\documentclass[traditabstract]{aa}

\usepackage{graphicx}
\usepackage{natbib}
\usepackage{longtable,lscape, booktabs}
\usepackage{xtab}
\usepackage{rotating}
\usepackage{fmtcount}
\usepackage{amsmath}
\usepackage{adjustbox}
\usepackage{txfonts}
\usepackage{hyperref}
\usepackage{threeparttable}
\usepackage{subfig}
\usepackage{multirow}
\usepackage{float}
\usepackage{ulem}
\usepackage{color}

\newcommand{\Msun}{\mbox{$\rm M_\odot\,$}}
\newcommand{\Lsun}{\mbox{$\rm L_\odot\,$}}

\newcommand{\mum}{\mbox{$\rm \mu m\,$}}

\def\arcmin{\hbox{$^\prime$}}
\def\arcsec{\hbox{$^{\prime\prime}$}}

\begin{document}

        \title{Photometric variability of massive young stellar objects. I.\thanks{Tables \ref{tab:sources_and_lcs},  \ref{tab:all_targets_sed_input_summary_first_filters}, and \ref{tab:all_targets_sed_fit_summary} are only available in electronic form at the CDS via anonymous ftp to cdsarc.u-strasbg.fr (130.79.125.5) or via http://cdsweb.u-strasbg.fr/cgi-bin/qcat?J/A+A/.}}
        \subtitle{}
        \author{G. D. C. Teixeira \inst{1,2}, M.S.N. Kumar\inst{1,3}, L. Smith\inst{4,3},  P. W. Lucas\inst{3}, C. Morris\inst{3},
                 J. Borissova\inst{5,6}, M. J. P. F. G. Monteiro \inst{1,2}, A. Caratti o Garatti\inst{7}, C. Contreras Pe\~{n}a\inst{8},  D. Froebrich \inst{9}, \and J.F. Gameiro\inst{1,2}
                }       
        \offprints{G. D. C. Teixeira; email:gteixeira@astro.up.pt}
        \institute{ Instituto de Astrof\'isica e Ci\^encias do Espa\c{c}o, Universidade do Porto, CAUP, Rua das Estrelas, 4150-762 Porto, Portugal
                \and Departamento de F\'isica e Astronomia, Faculdade de Ci\^encias, Universidade do Porto, Rua do Campo Alegre 687, PT4169-007 Porto, Portugal
                \and Centre for Astrophysics Research, University of Hertfordshire, Hatfield, AL10 AB, UK
                \and Institute of Astronomy, University of Cambridge, Madingley Road, Cambridge, CB3 0HA, UK    
                \and Instituto de Fisica y Astronom\'{i}a, Universidad de Valpara\'{i}so, Gran Breta\~{n}a 1111, Playa Ancha, Valpara\'{i}so, Chile
                \and Millennium Institute of Astrophysics, Vicu\~na Mackenna 4860, 7820436 Macul, Santiago, Chile
                \and Dublin Institute for Advanced Studies, Astronomy \& Astrophysics Section, 31 Fitzwilliam Place, Dublin 2, Ireland
                \and Department of Physics and Astronomy, University of Exeter, Stocker Road, Exeter, Devon EX4 4SB, UK
                \and Centre for Astrophysics and Planetary Science, School of Physical Sciences, University of Kent, Canterbury, CT2 7NH, UK
        }

        \authorrunning{Teixeira et al.}
        \titlerunning{MYSOs variability in VVV}
        \date{\today}

        \abstract{The Vista Variables in the Via Lactea (VVV) survey has allowed for an unprecedented number of multi-epoch observations of the southern Galactic plane. In a recent paper, 13 massive young stellar objects (MYSOs) have already been identified within the  highly variable ($\Delta K_s >1$ mag) YSO sample of another published work. This study aims to understand the general nature of variability in MYSOs. Here we present the first systematic study of variability in a large sample of candidate MYSOs. We examined the data for variability of the putative driving sources of all known Spitzer extended green objects (EGOs) (270) and bright 24 \mum sources coinciding with the peak of 870  \mum detected ATLASGAL clumps (448), a total of 718 targets. Of these, 190 point sources (139 EGOs and 51 non-EGOs) displayed variability ($IQR>0.05$, $\Delta K_s > 0.15$ mag). 111 and 79 light-curves were classified as periodic and aperiodic respectively. Light-curves have been sub-classified into eruptive, dipper, fader, short-term-variable and long-period-variable-YSO categories. Lomb-Scargle periodogram analysis of periodic light-curves was carried out. 1 - 870 \mum spectral energy distributions of all the variable sources were fitted with YSO models to obtain the representative properties of the variable sources. 41\% of the variable sources are represented by $>4 \ \Msun$ objects, and only 6\% were modelled as $>8 \ \Msun$ objects. The highest-mass objects are mostly non-EGOs, and deeply embedded, as indicated by nearly twice the extinction when compared with EGO sources. By placing them on the HR diagram we show that most of the lower mass, EGO type objects are concentrated on the putative birth-line position, while the luminous non-EGO type objects group around the zero-age-main-sequence track. Some of the most luminous far infrared (FIR) sources in the massive clumps and infrared quiet driving sources of EGOs have been missed out by this study owing to an uniform sample selection method. A high rate of detectable variability in EGO targets (139 out of 153 searched) implies that near-infrared variability in MYSOs is closely linked to the accretion phenomenon and outflow activity.}
        
        
        \maketitle
        %
        
\section{Introduction} \label{sec:intro}

The paradigm of accretion in young stellar objects (YSO) has shifted from a
model of constant mean accretion rate to that favouring short events of intense
accretion \citep{vorbas06,vorbas15,zhu09}.
 This shift is largely to address the issue of the `protostellar luminosity problem' \citep{keny90, ken95, dun14}. A variety of models including turbulent or competitive accretion, accretion regulated by core, disk, and feedback, are invoked to understand the deviation from the idealized case of isothermal sphere (
        \citet{keny90}, \citet{mcke10}, \citet{myer10}, \citet{vorbas08}, \citet{dun12}, \citet{dun14} and
        references therein). However, most of these models share the variable accretion component,  albeit differing at various mass regimes. The accumulated observational evidence appears to favour variable
        accretion instead of constant mean scenarios \citep{dun14}.
Photometric variability of YSOs can be related to their natal environment,
accretion physics or a combination of both \
(\citet{contreras17}, \citet{kesseli16}, \citet{meyerMNRAS2017} and references therein). Some of
the variability can be caused by cold and hot spots formed on the surface of
the YSO by infalling material from the disc. Dust clumps in the stellar medium
surrounding the YSO can cause variable extinction of star-light as it passes
along the observers line of sight (e.g. \citet{herbst99}, \citet{eiro02} among others).

The FUors (FU Orionis) and EXors (EX Lupi) examples of high amplitude
photometric variability result from variable accretion. Respectively, they last from a
few years to a few months. These objects are known to be
low-mass YSOs, although similar counterparts in the higher mass range have
been found \citep{kumar2016,garat17}.
\citet{kumar2016} uses highly variable light curves (LCs) of massive young
stellar objects (MYSOs) candidates from the Vista Variables in the Via Lactea
(VVV) survey \citep{vvv2010}, arguing that they were signposts of ongoing episodic
accretion. Photometric and spectroscopic variability in a
20 \Msun MYSO was used by \citet{garat17} to conclude that disk-mediated accretion bursts are a
common mechanism across stellar masses. ALMA observations were used by
\citet{hunter17} as evidence that sudden accretion is responsible for the
growth of a massive protostar. These findings suggest that episodic accretion
maybe a common mechanism in star formation, independent of mass. Computational
models predict luminous flares in MYSOs, which are morphologically similar to FUors and
EXors \citep{meyerMNRAS2017}.

The findings in \citet{kumar2016} raise the question of the overall nature of
variability in massive YSOs. In this paper, we attempt to examine the
variability phenomena in known extended green objects (EGOs) \citep{cyg08} and
IR sources, deeply embedded in clumps identified by the APEX Telescope Large
Area Survey of the Galaxy (ATLASGAL) \citep{schull09}.
They represent unbiased large
samples of point-like massive young stellar candidates, therefore, allowing
us to use the point source photometry to examine variability. We surmise that
the RMS and UCHII regions represent an important MYSO sample, however, it
requires larger aperture photometry of extended objects to examine
variability, which we postpone to a different study.

Employing point source photometry requires that the targets are point like in
MIPS, have associated high mass star forming
signposts, and finally they are also point-like in the $K_s$ band. The
selection of such targets is described in Section \ref{sec:surveys}.
In Sect. \ref{sec:results} we describe the results obtained and discuss their implications
in Sect. \ref{sec:discussion}.

\section{Target sample, data, and methods}\label{sec:surveys}

Identification of point-like MYSO targets is based on the Spitzer
GLIMPSE and MIPSGAL surveys \citep{car09}, the ATLASGAL survey \citep{schull09}, and
the VVV survey \citep{vvv2010}. These different surveys are highly complementary, covering
much of the same area but at different wavelengths (from $\sim 1.2 - 870$ \mum). We searched for: a) driving
sources of EGOs \citep{cyg08, chen2013, chen2013pt2}
and b) luminous MIPS $24 \ \mum$ point sources embedded in ATLASGAL
clumps.
The two samples are expected to roughly represent two early evolutionary phases of massive stars; the EGOs, with an active phase of mass ejection, and non-EGOs which are likely yet to begin outflow activity.

\subsection{MYSO sample}\label{sec:iding}

\subsubsection{EGO sample}
 EGOs are objects with extended emission in the Spitzer $4.5 \mum$ band
(IRAC 2). This band is of particular significance since it contains both
$H_2$, and $CO$ lines, which can be excited by shocks when outflows and jets
interact with the interstellar medium (ISM). This is particularly the case
when the extended emission in the $4.5 \ \mum$ band is in excess with respect to
emission in the other IRAC bands. 

They were first catalogued by \citet{cyg08},
and later the catalogue was extended by \citet{chen2013,chen2013pt2}.
EGOs are thought to represent the $H_{2}$ flows driven by MYSOs \citep{cyg08} 
or MYSO outflow cavities \citep{taka12}.
A total of 270 unique EGO targets have been catalogued so far. By original classification \citep{cyg08} these targets have a MIPS 24  \mum detection, usually representing the driving source of the outflow.
In order to find the near-infrared counterpart of these driving sources we
searched for 2 $\mu$m sources in the VVV catalogue with a search radius of 1.0\arcsec and 0.5\arcsec
from the known EGO positions. We find 187 and 153 driving sources. We allowed for sources classified as both point-like and extended to be selected, even though, 80\% of the detected sources were point-like. Young stellar objects with disk and outflow activity are often surrounded by circumstellar nebulae in the near-infrared, leading to a classification as extended. These objects were kept in the sample list. Additionally,
three colour composites, shown in Fig. \ref{fig:erup1}, were used to
visually examine whether the identified point sources are good representations
of an outflow driving source. This examination led us to retain the 153 sources
which clearly represent $2 \ \mum$ counterparts of the $24 \ \mum$ source, hence the 
putative driving source of the EGO target.

\subsubsection{Non-EGO sample} 

\citet{kumar2016} identified a sample of highly variable VVV objects and
found MYSO counterparts in ATLASGAL clumps \citep{contreras13}. Here an inverse
approach is used.
Using ATLASGAL, \citet{contreras13} and \citet{urquhartcsc14} built the Compact
Source Catalogue (CSC) which identified $\sim 10000$ dense clumps.
The mass, density, and distance to these clumps are provided by \citet{urqu17}
and they are believed to represent active sites of high-mass star formation.
Assuming that ATLASGAL clumps host MYSOs, we searched
for MIPSGAL point-like
sources that matched with ATLASGAL CSC sources within a radius of 5\arcsec.
This ensured that we matched red point-like sources in $24 \ \mu m$ band with
the peak emission in the 870 $\mu m$ observations of ATLASGAL. 873 point sources
were found with this search.
When there were multiple matches we chose the object with the closest centroid distance.
The MIPS FWHM is equal to 6\arcsec, therefore, a further search of the 873 targets with a
matching radius of 5\arcsec was performed with the VVV catalogue, allowing us to
find 574 $K_s$-band targets. These 574 targets display more than one Ks band source within the 5\arcsec radius.
In the next step, the point source closest to the MIPS peak was searched for, by constraining the search radius to
1.0\arcsec, only for the 574 targets. This retrieved a list of 2171 sources from the VVV source
catalogue. Out of these, any source which had less than ten non-saturated epochs (over the full five-year period)
was removed. This led us to find 367 single
detections and 147 multiple detections in the centroid search with $r \leq 1 \arcsec$.
The multiple detection targets were examined visually considering the source magnitude, colour and centroid distance, based on which 66 of the 147 sources were rejected, retaining 81 sources.
These 448 (367 + 81) sources are, therefore, the
$K_s$-band point sources representing the MYSO candidate at the peak of an
ATLASGAL clump with a spectral energy distribution (SED) that can be
assembled from at least $2 \ \mum$ unto $870 \ \mum$.

The final MYSO sample we produced to study the variability is, therefore, composed of
153 EGO and 448 non-EGO sources, resulting in 601 targets. We note that
66 of the 153 EGO targets also lie within the ATLASGAL clumps,  the non-EGO
sources being exclusively those that coincide with the peak of ATLASGAL clumps.

\subsection{VVV survey data}

\begin{figure*}
	
	\includegraphics[width=\textwidth]{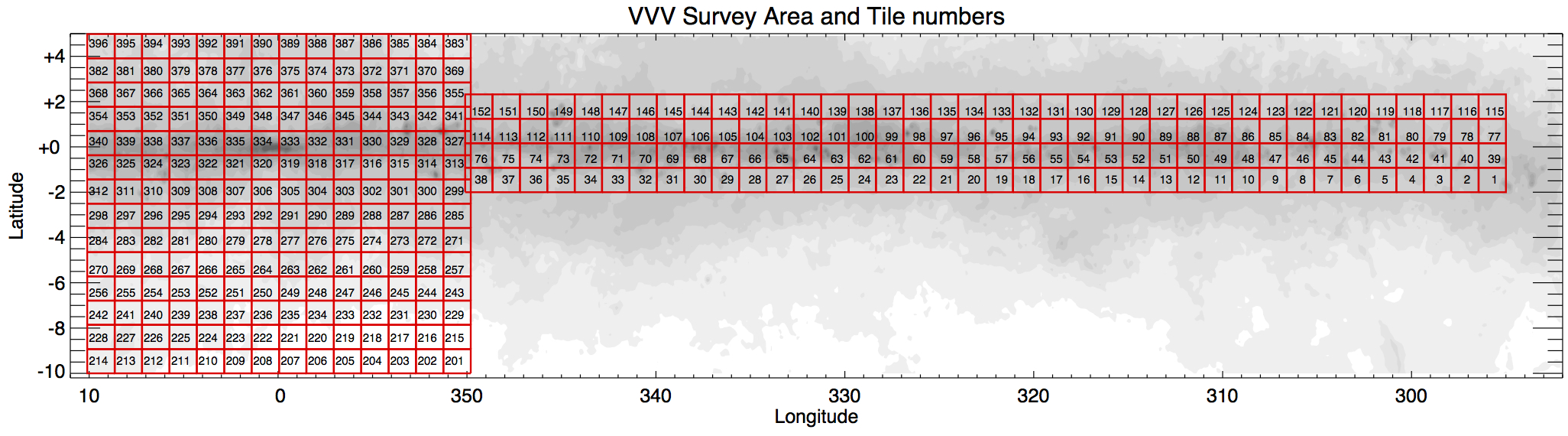}
	\caption{VVV survey area.}
	\label{fig:galaxy}
\end{figure*}

The VISTA Variables in the Via Lactea (VVV) survey has obtained
photometric observations in the near-infrared (NIR) passbands (0.9-2.5 \mum),
covering multiple epochs, spread over five years (from 2010 to 2014) and covering a 520 $deg^2$ area
of the inner Galactic plane (see Fig. \ref{fig:galaxy}) \citep{vvv2010}.        
The survey data is made publicly available through the Cambridge Survey
Astronomical Unit (CASU), which is the photometry obtained on the final
combined `tile' images. A tile image incorporates multiple `pawprint' that
are single exposures on sky. The pawprint data (available to the VVV team and
also made public at the VISTA Science Archive in Edinburgh),
is the basic product of the observations which often holds better photometric
and seeing information, as they are better calibrated and tend to have sharper
image profiles than the tile data. In this work we have exploited this
full potential by using the pawprint photometry.

\subsection{Processing of the pawprint photometry}

The pawprint photometry and photometric classification used are standard
pipeline products from the Cambridge Astronomical Survey Unit (CASU), as
detailed in \citet{lewis2010}. Matching and combining detections between multiple
pawprints were made following the approach detailed by \citet{smith2017}.
Sources are classified according to their morphology and flagged as 
1, 0, -1, -2, -3, -7, -9, respectively as; a galaxy, noise, stellar,
probably stellar, probable galaxy, bad pixel within 2\arcsec aperture, and
saturated.

The pawprint observing pattern of dithers and
overlaps, implies that each source might have between two and six image frames for
the same observing epoch, they can also be detected only once along certain edges of the survey. 
Since these observations are close in
time, we chose to bin them together and compute the median magnitude of
all observations in intervals of half a day. This binning 
prevents the detection of variations with timescales smaller than 
half a day, but reduces the
level of scatter in short-periods. The gain in photometric sensitivity will
thus be a factor of the number of observations binned, and scales according
to $\sqrt{n}$ where $n$ is the number of binned observations. The
typical error in the photometry will be $K\_err \leq 0.05$ mag which allows
for the detection of low-level variability. Following the reasoning explained
in \citet{smith2017} we employ the aperMag2 ($r \sim 0.71\arcsec$) as the $K_s$
magnitude for all analysis.

For each source we assembled a database that contains: a unique identification, median co-ordinates in the
ICRS, median magnitude (over half a day) in the K-band, the median absolute
deviation (MAD), the standard deviation, the inter-quartile range (IQR),
the number of pawprints in which the source was observed, the number of total
observed epochs, the modal class, the number of epochs classified with each
flag, the K-band magnitude, the quality classifier of the photometry, and the
modified julian date (MJD) of the observation.

The median of the co-ordinates and magnitude, and the modal class were computed for all
pawprint observations. The MAD and IQR were computed for each source
as they are robust statistical indicators to
measure the amplitude and dispersion of the variability
\citep{hamp74, upton96,soko17}. These parameters are less sensitive to
outliers than the standard deviation. A high value of the MAD or IQR
can be a good indicator of the inherent variability of the source.
The IQR measures the amplitude of the difference between the third  and first quartiles (Q3 and Q1) of a distribution, in this case the distribution of
magnitudes
\begin{equation}
IQR = Q3-Q1.
\end{equation}
The MAD is the
median of the absolute differences between each data point and the median, as
shown by the following equation:
\begin{equation}
MAD = median(\left | K_i-median(K) \right |)
\end{equation}
in which, $K_i$ is an observation and $K$ represents all the
observations.

\subsection{Light curves and their reliability}\label{subsec:lcs}

The light-curves (LCs) of 448 non-EGO and 153 EGO targets were produced in the following
way by querying the database assembled above.
First, we queried a set of co-ordinates and search radius on the database. Secondly,
	we built a list with all observations that matched the query. Thirdly, we 
	excluded all saturated observations (modal class = -9). Fourthly, we produced 
	a LC for each target using the difference from median
	($K_{median}-Kmag_i$).

\begin{figure}
 
        \includegraphics[width=\columnwidth]{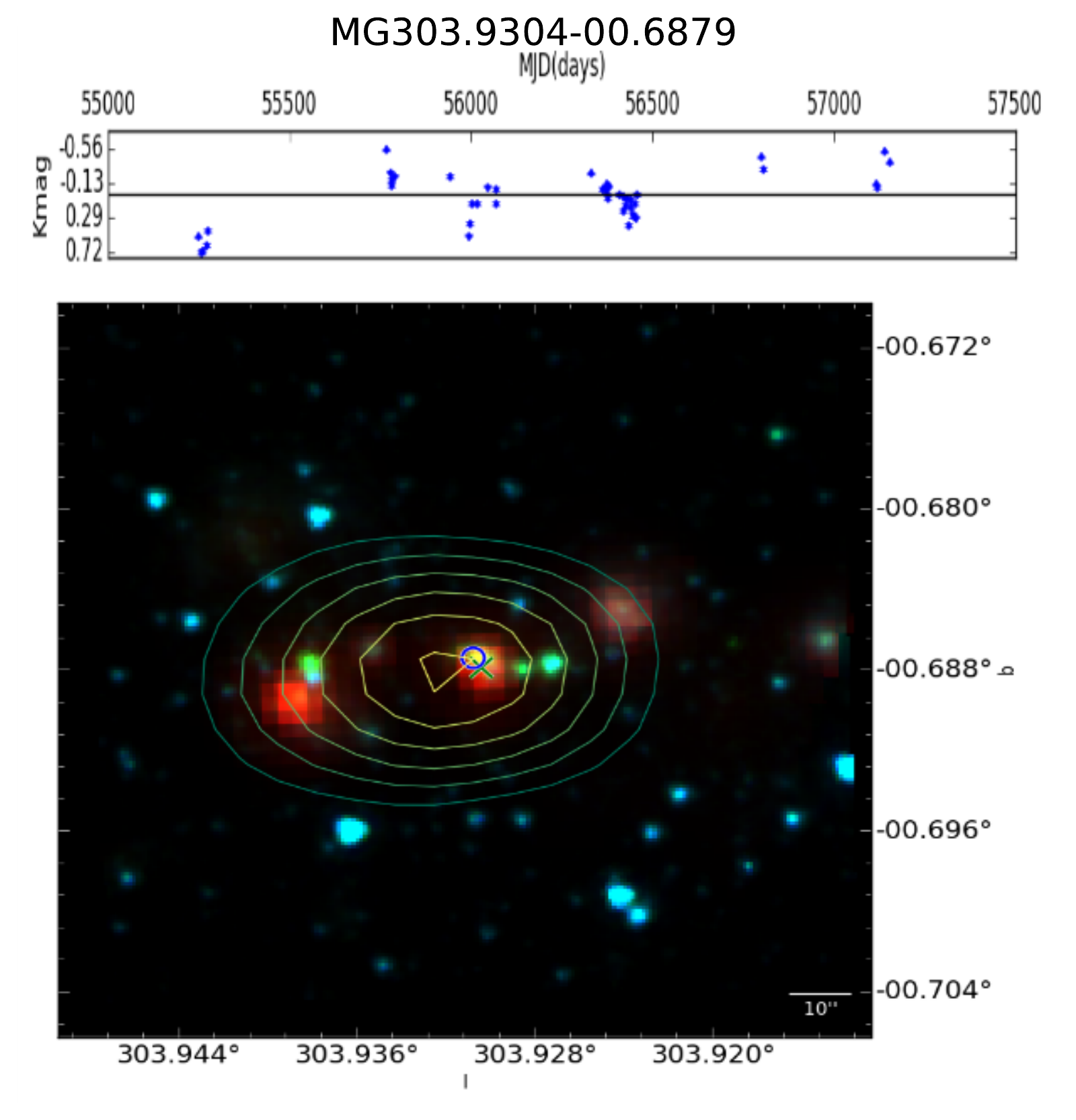}
        \caption{LC of an eruptive event. Top panel shows the LC of the source, the
                error bars represent MAD($\Delta S_{i_{mjd}}$), the bottom plot shows the
                RGB image of the source using the Spitzer IRAC 3.6 \mum, IRAC 4.0 \mum, and the 24\mum MIPS band
                as blue, green and red, respectively. The VVV source is indicated by the blue
                circle and the green cross represents the MIPS co-ordinates. The contours of
                the RGB are in the interval of [Peak-$5\sigma$, Peak] from the ATLASGAL
                observation at $850$ \mum.}
        \label{fig:erup1}
\end{figure}

In the top panel of Fig. \ref{fig:erup1} we show an example of a LC. To ascertain that the
photometry of the target at a given time epoch is not affected by poor
observing conditions, flat field errors, improper photometry, or poor seeing,
we performed a few tests.

\subsubsection{Identifying the variable source}

        Stellar sources within two annuli defined by r=1\arcsec  and r=60\arcsec from
        the target were selected. Typically 100-200 sources were found by this
        selection. For each such source $S_i$, the magnitude deviation
        ($\Delta S_{i_{mjd}}$) from its median value ($\widetilde{S_{i_{mjd}}}$) over all epochs
        was computed. The median value $\widetilde{\Delta S_{i_{mjd}}}$ for all sources 
        in the annulus, is a representation of the photometric deviation (if any) of the
        the individual epoch over the time-line. 
        for each source, at each epoch, the offset $\widetilde{\Delta S_{i_{mjd}}}$ was added to 
        $S_{i_{mjd}}$ to produce the corrected light curve. 
        The MAD of the deviations for all selected sources, MAD($\Delta S_{i_{mjd}}$), is used 
        as an approximation of the $1\sigma$ photometric error of a 1\arcmin \ field around the target
        for a given epoch. And is shown as the error bars in Fig. \ref{fig:erup1}.

        \begin{figure}
                \includegraphics[width=\columnwidth]{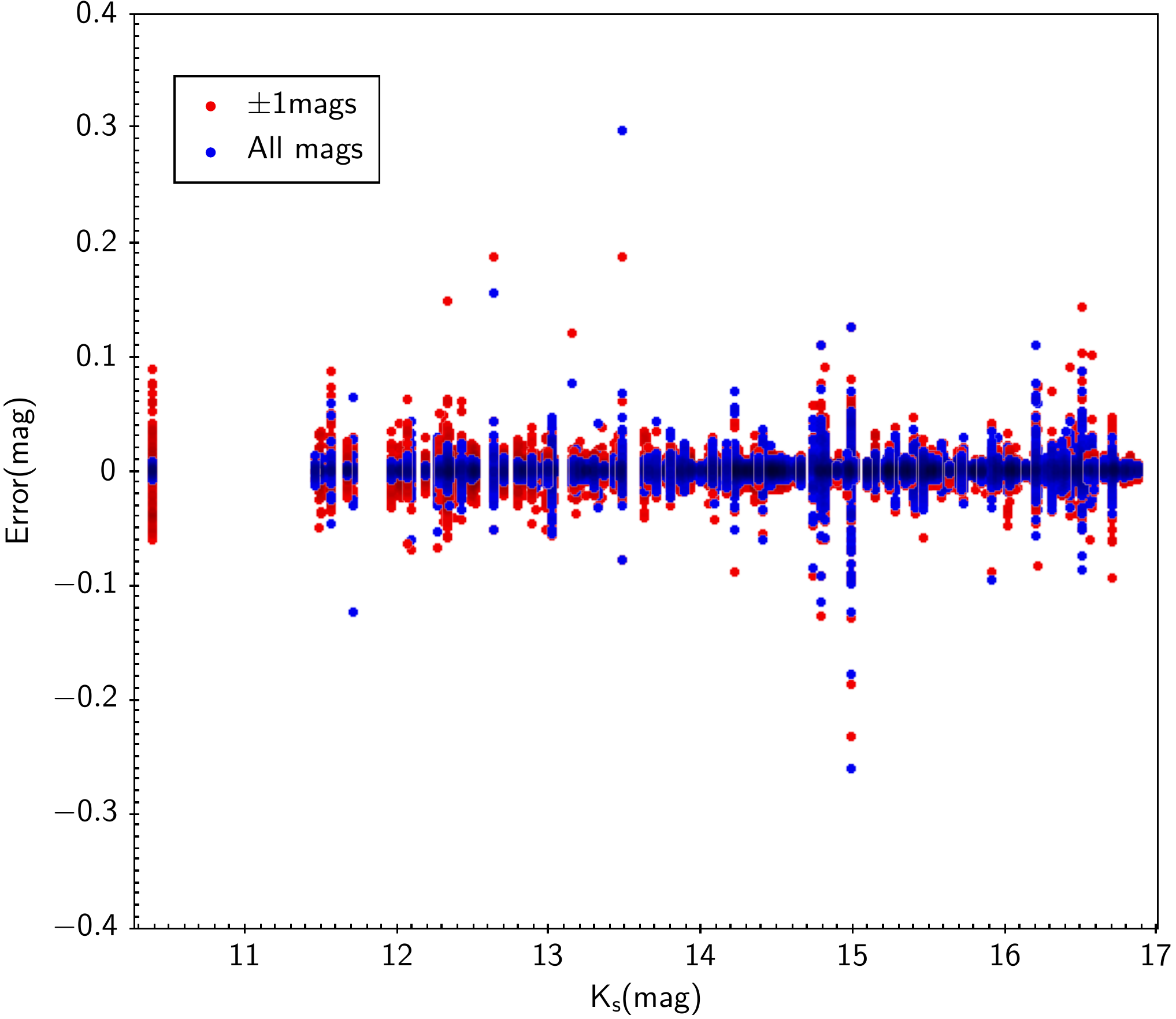}
                \caption{Systematic errors as a function of the magnitude of each target.
                        The red points represents the case where we only consider the stellar sources with $\pm 1$ mag around our targets.
                        The blue points represent the case in which we consider all stellar sources in the vicinity of our targets.}
                \label{fig:2d_mags}
        \end{figure}
        
\subsubsection{Influence of magnitude on variability}
        Next we assessed the influence of using any and all, or, only sources
        with magnitudes comparable to that of the target inside the annuli for
        computing the $1\sigma$ error. For this purpose we filtered the sources
        within a magnitude range $\pm 1$ mag to that of the target, which
        decreased the number of sources by a factor of approximately ten. Figure
        \ref{fig:2d_mags} illustrates the results of this test. It can be seen
        that the difference in $1\sigma$ error by using the two comparison samples
        is $K_s \sim 0.0018 -0.0031$ mag, which is 1-2 orders of magnitude below
        the typical $1\sigma$ errors in the target fields.

\subsubsection{Control field test}      
        The targets of study are found in the midst of star forming regions, often deeply
        embedded in dark clouds, leading to reduced and non-uniform source distribution.
        Also YSOs (in general) are known to be variable objects, so, many sources in a 
        given field may be variable. To address the influence of these effects, we used a
        control field region randomly selected to be 5\arcmin away from the target field.
        For each control field the steps explained in the two subsections above were
        executed. We find that the control field variability was very similar to the 
        MAD($\Delta S_{i_{mjd}}$) computed above.

\subsection{Periodograms, false alarm probability, and their aliases}

Once the LCs are assembled, we computed the Lomb Scargle periodogram, identify
the max power frequency component, and use it to produce a phase-folded LC.
\citet{scargle82} defined the false alarm probability (FAP) as a measurement
of the probability of a signal without any periodic component to have a peak
amplitude. The predictive power of the FAP decreases in the presence of
correlated noise, non-gaussian errors, and, highly non-sinusoidal variability.
The $90\%$, $95\%$, and $99\%$ FAP levels have been computed for periodograms
of each target.

A given periodicity can, by a compound effect of binning, observational
window, and noise, produce harmonics of itself, which appear in the
periodograms as additional peaks, or aliases \citep{vanderplas17}. In an
effort to verify if the peaks determined were in fact real signals or their
aliases, an additional verification step was added. The highest peak of the
periodograms and the following 10 highest peaks were identified.
Aliases were searched by examining: a) multiples in the frequency range; b) multiples
in the period range; and c) solving the following equation:
\begin{equation}
f_i = f_t + n * f_w
\end{equation}
where $f_i$ is the frequency of the alias, $f_t$ is the true frequency, $n$ is
an integer, and $f_w$ is a frequency window, using the windows of 1 year
($0.0027 \ day^{-1}$), one day ($1 \ day^{-1}$), and a sidereal day
($1.0027 \ day^{-1}$), as these are the most common aliases for Earth-based
telescopes \citep{vanderplas17}.

\subsection{SED analysis}\label{sec:sedanalysis}

The target samples are generally considered to represent MYSOs based on signposts
of high mass star formation and survey shallowness. To better understand the 
nature of the sources studied here in detail for variability, we have analysed their
$1.2 \ \mum$ - $870 \ \mum$ spectral energy distributions (SEDs).

\begin{table}
        \centering
        \caption{Filters and apertures used for building the SEDs.}
        \label{tab:sed_apertures}
        \begin{tabular}{lcc}
                \hline
                \hline
                Filter & Wavelength & Aperture \\
                & ($\mum$) & (\arcsec)\\
                \hline
                J        & 1.235 & 3   \\
                H        & 1.662 & 3   \\
                $K_{s}$  & 2.159 & 3   \\
                IRAC1    & 3.6   & 4   \\
                IRAC2    & 4.5   & 4   \\
                IRAC3    & 5.8   & 4   \\
                IRAC4    & 8.0   & 4   \\
                MIPS24   & 24    & 6   \\
                PACS70   & 70    & 5.6 \\
                PACS160  & 160   & 10.7\\
                SPIRE250 & 250   & 17  \\
                SPIRE350 & 350   & 24  \\
                SPIRE500 & 500   & 35  \\
                AGAL870  & 870   & 19.2\\
                \hline
                
        \end{tabular}
\end{table}

The Python version of the SED fitting tool \citep{rob10} was used to 
model SEDs of the target sources. The photometric bands, filter, and apertures
used to construct the SEDs can be found on Table \ref{tab:sed_apertures}. 
The photometric data used to construct the SEDs was obtained from querying the public
online archives of 2MASS, SPITZER, ATLASGAL, and Herschel \citep{masssurv,car09,schull09,herschel10}.
Our SED fitting follows the method detailed in \citet{grave09}. 
An uniform photometric error
of $10\%$  was assumed.
Longer wavelength data were usually set as upper limits, because their large beams include emission from multiple sources, sometimes small clusters, even those that are well resolved at $24 \ \mum$ and below. Data at wavelengths 
shorter than $24 \ \mum$ are set as data points. 
However, for the EGO sample, the $4.5 \ \mum$
IRAC band data was set as upper limit by default, as their main characteristic is 
to have excess emission in that band.

We used a range of extinction of Av = 0-50 mag for all targets. 
Distances are available \citep{urqu17} for 105 targets, non-EGO and EGOs, 
they were used with an uncertainty of $\pm 1$ kpc while fitting. 
For the remaining 102 targets we allowed a full plausible range of d=1-13 kpc.

For each target all the models which have a $\chi^2 - \chi_{best}^2 <3$ were
used, and the parameters of the source were computed by performing a weighted
mean, weighted by the inverse $\chi^2$ as described in \citet{grave09}.
The observational data used to construct the SEDs is listed in the Table 
\ref{tab:all_targets_sed_input_summary_first_filters} .

\section{Results}\label{sec:results}

The LCs of 601 (448 non-EGO + 153 EGO) were visually examined and compared with the 
source IQR, while considering the deliberation made in Sect. \ref{subsec:lcs}. 
We consistently find that an $IQR>0.05$ is associated with visually $>$20\% 
of the data-points in the light-curve that are above the $1\sigma$ error of the 
field, as shown by the error bars for each source.

\begin{figure*}
        
        \includegraphics[width=\textwidth]{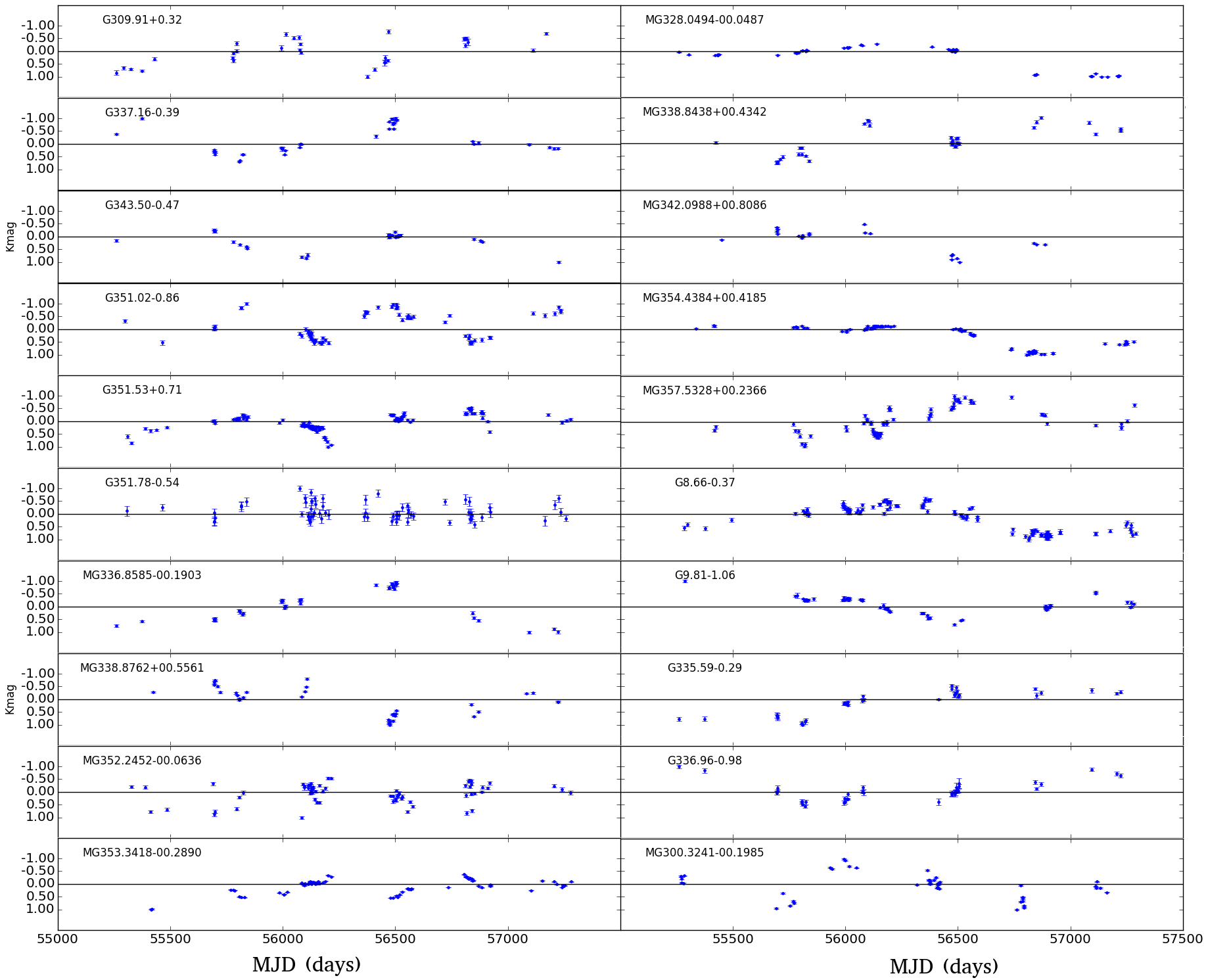}
        \caption{Some of the clearer LCs, periodic (left column) and aperiodic (right column). Each figure shows the LC of the source, error bars represent
                MAD($\Delta S_{i_{mjd}}$). The vertical axis represents the variability from the median normalized by $max(\left | K_i-median(K) \right | )$.}
        \label{fig:fullpageoflcs}
\end{figure*}

\begin{table*}
        \centering
        \caption{Source co-ordinates, photometric data and variability}
        \label{tab:sources_and_lcs}
        \begin{tabular}{lcccccccc}
                \hline
                \hline
                Source & RA & DEC & $K_s$ & MAD & IQR & $\Delta K_s$ & Class & Period\\
                & (deg) & (deg) & (mag) & (mag) & (mag) & (mag) &     & (day)\\
                
                \hline
                MG303.9304-00.6879 & 195.10156 & -63.54177 & 15.21 & 0.15 & 0.33 & 1.28 & Erup     & NA     \\
                
                MG328.0494-00.0487 & 238.7064 & -53.7280 & 12.28 & 0.149 & 0.278 & 1.83 & Fad      & NA     \\
                
                MG352.2452-00.0636 & 261.5178 & -35.5005 & 15.95 & 0.079 & 0.166 & 0.53 & STV      & 29.4   \\
                
                MG354.4384+00.4185 & 262.5086 & -33.4088 & 14.66 & 0.091 & 0.523 & 0.89 & Dip      & NA     \\
                
                G309.91+0.32       & 207.7246 & -61.7394 & 13.65 & 0.204 & 0.383 & 0.81 & LPV-yso  & 545.9  \\
                
                G335.59-0.29       & 247.7437 & -48.7308 & 13.16 & 0.097 & 0.348 & 0.61 & low-Erup & NA     \\
                
                G351.78-0.54       & 261.6775 & -36.1536 & 14.46 & 0.06 & 0.12 & 0.38 & STV      & 18.3   \\
                
                G343.50-0.47       & 255.3267 & -42.8267 & 15.38 & 0.10 & 0.18 & 0.86 & LPV-yso  & 1156.3  \\
                
                \hline
                
        \end{tabular}
        \begin{tablenotes}
                \item For full table check the online data.
        \end{tablenotes}
\end{table*}

This selection  criteria resulted in 51 (of the 448) non-EGO and 139 (of the 153)
EGO targets to be classified as variable sources. They are listed in 
Table \ref{tab:sources_and_lcs}, along with
the LC classification. In Fig. \ref{fig:fullpageoflcs} we display some of the clear LCs of both periodic and aperiodic nature. For each source (see Fig. \ref{fig:lpv}), the LC, periodogram, phase-folded LC, and a three colour composite image
of the target is made available.

\subsection{Light curve classification}\label{subs:classification}

Light curves can be classified based on their behaviour and, often, such classification
represents a close connection with certain physical processes. A classification scheme similar to the one used in
\citet{contreras17} was followed here, and LCs were divided into : a) long period
variables (LPV-yso); b) short timescale variables (STV); c) dippers and faders;
d) eruptive. In
defining periodic variables, \citet{contreras17} only included periods of
the highest power, while we include all significant periods.
Four LCs for which we have only a short time coverage were considered to be unclassified.

\begin{figure*}
        \begin{tabular}{cc}
                
                \includegraphics[width=0.5\textwidth]{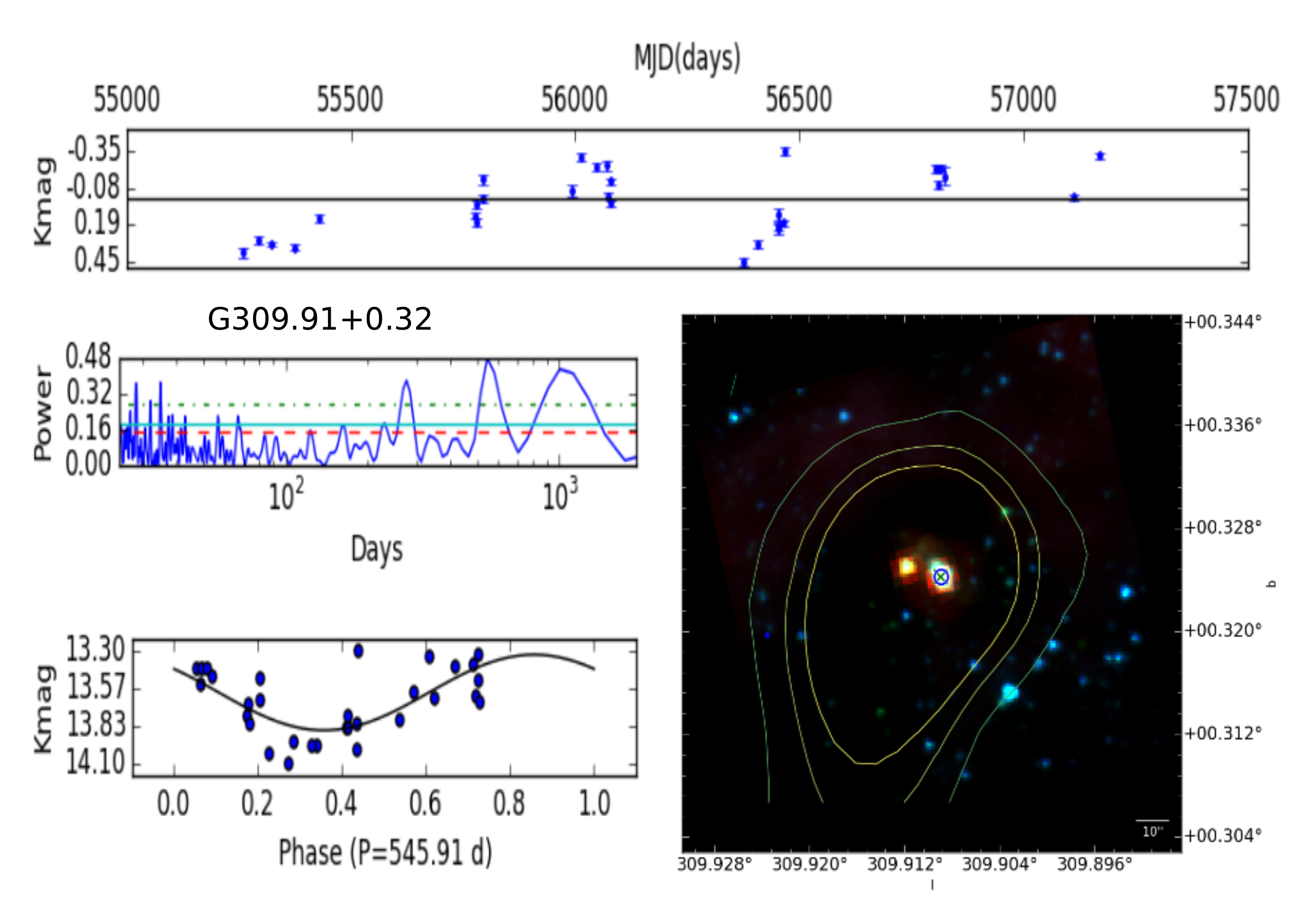} &
                \includegraphics[width=0.5\textwidth]{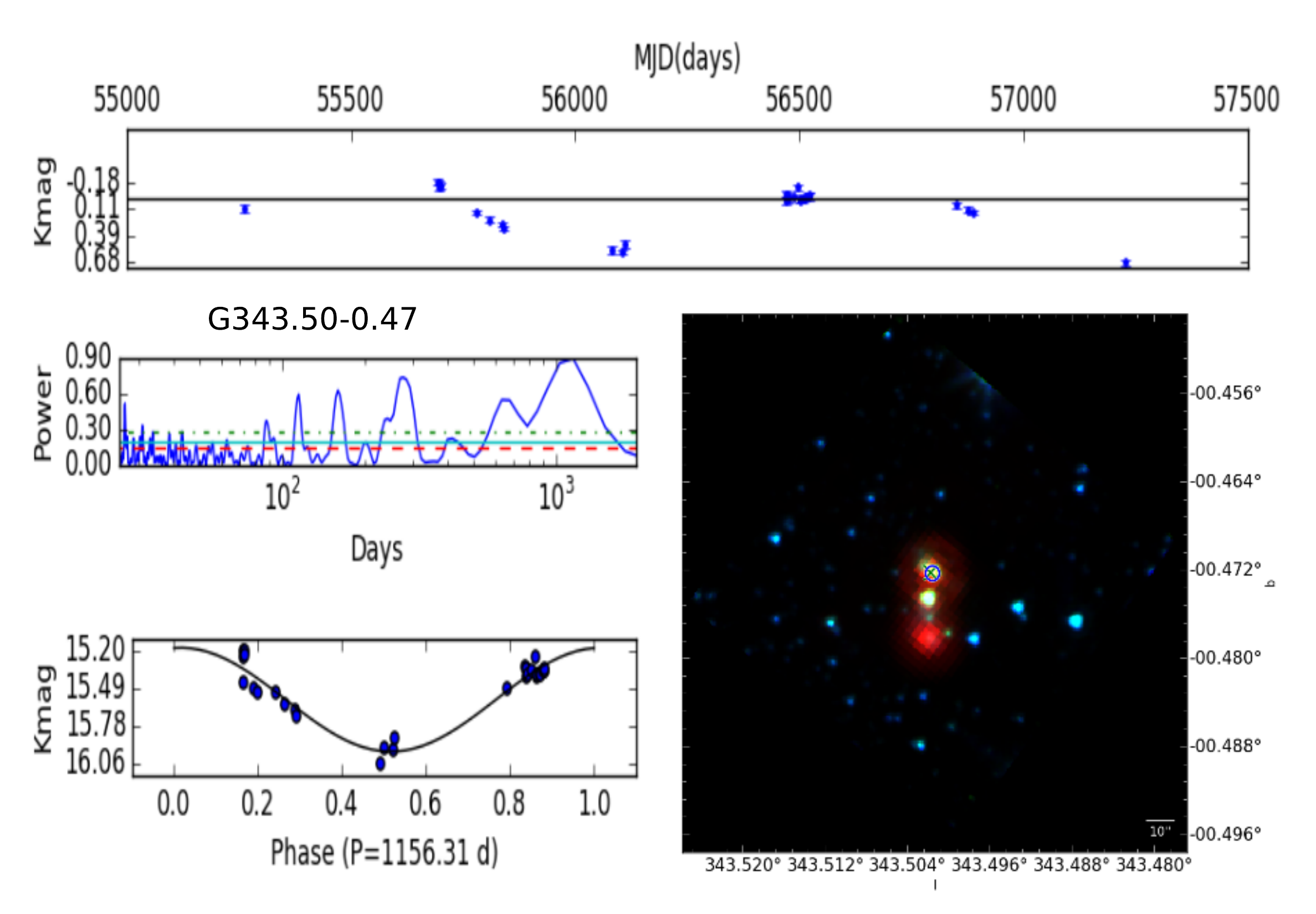} \\

        \end{tabular}
        \caption{Prototypical LPV-yso sources: Top panel for each figure shows the LC of the source, error bars represent
                MAD($\Delta S_{i_{mjd}}$), the left middle panel shows the
                corresponding periodogram in logarithmic scale (also plotted are the $99\%$, $95\%$, and $90\%$ false probability levels,
                respectively: the green dot-dashed line, the cyan full line, and the red dashed line), the bottom left panel shows the phase-folded
                light curve of the source using the best period fitted (also shows the corresponding value in days), the bottom right
                plot shows the 
                RGB image of the source using the Spitzer IRAC 3.6 \mum, IRAC 4.0 \mum, and the 24\mum MIPS band
                as blue, green and red, respectively. The VVV source is indicated by the blue circle and the green cross represents the MIPS co-ordinates.
                The contours of the RGB are in the interval of [Peak-$5\sigma$, Peak] from the ATLASGAL observation at $850$ \mum.}
        \label{fig:lpv}
\end{figure*}

Long period variables (LPV-yso) are defined in \citet{contreras17} as sources
with periodic photometric variability and periods larger than $P>100$ days.
LPV-ysos have periods larger than the stellar rotation or
inner disc orbits of young stellar objects, which are typically $P<15$ 
days. Figure \ref{fig:lpv} shows the example of two LPV-ysos,
source G309.91+0.32 and G343.50-0.47, which have periods of $\sim 545 $ days, and
$\sim 1156$ days.
The RGB image of source G309.91+0.32 reveals distinct extended green
emission, a signpost of the presence of an outflow, its periodogram shows a
prominent signal well above the $99\%$ FAP level. The source
has a median brightness of Ks = 13.65 mag in the VVV and the amplitude
between the brightest and dimmest point of its LC is $\sim 0.81$ mag.
The other prototypical LPV-yso selected, G343.50-0.47, and is part of a complex of three
MIPS bright sources. It is a source with Ks=15.38 mag, the amplitude of its variability is
close to $\sim 0.86$ mag and, the periodogram of the source shows a distinct
peak well above the $99\%$ FAP level. There are no aliases in the periodograms of either of these 
sources.

\begin{figure*}
        \begin{tabular}{cc}
                \includegraphics[width=0.5\textwidth]{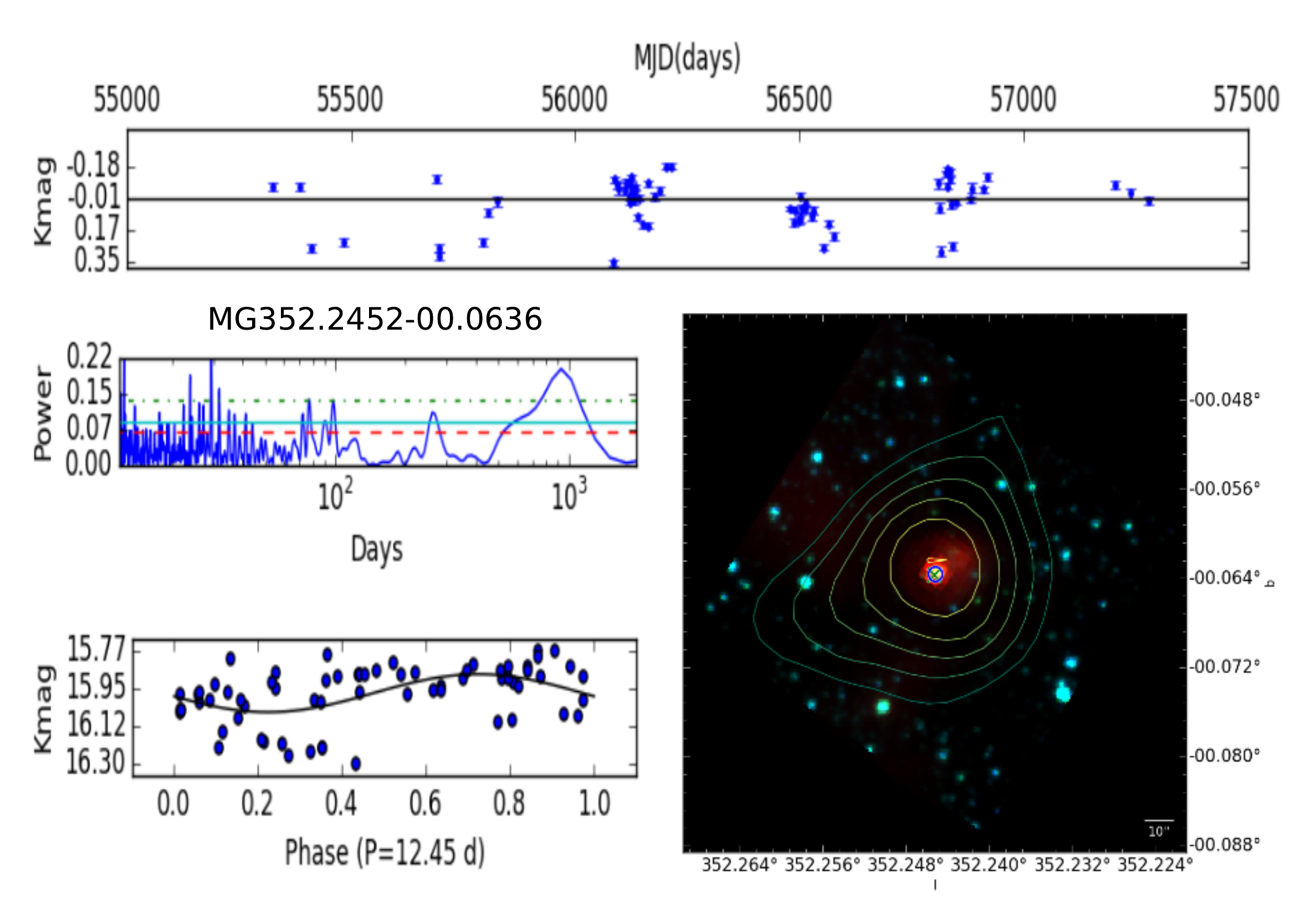} &
                \includegraphics[width=0.5\textwidth]{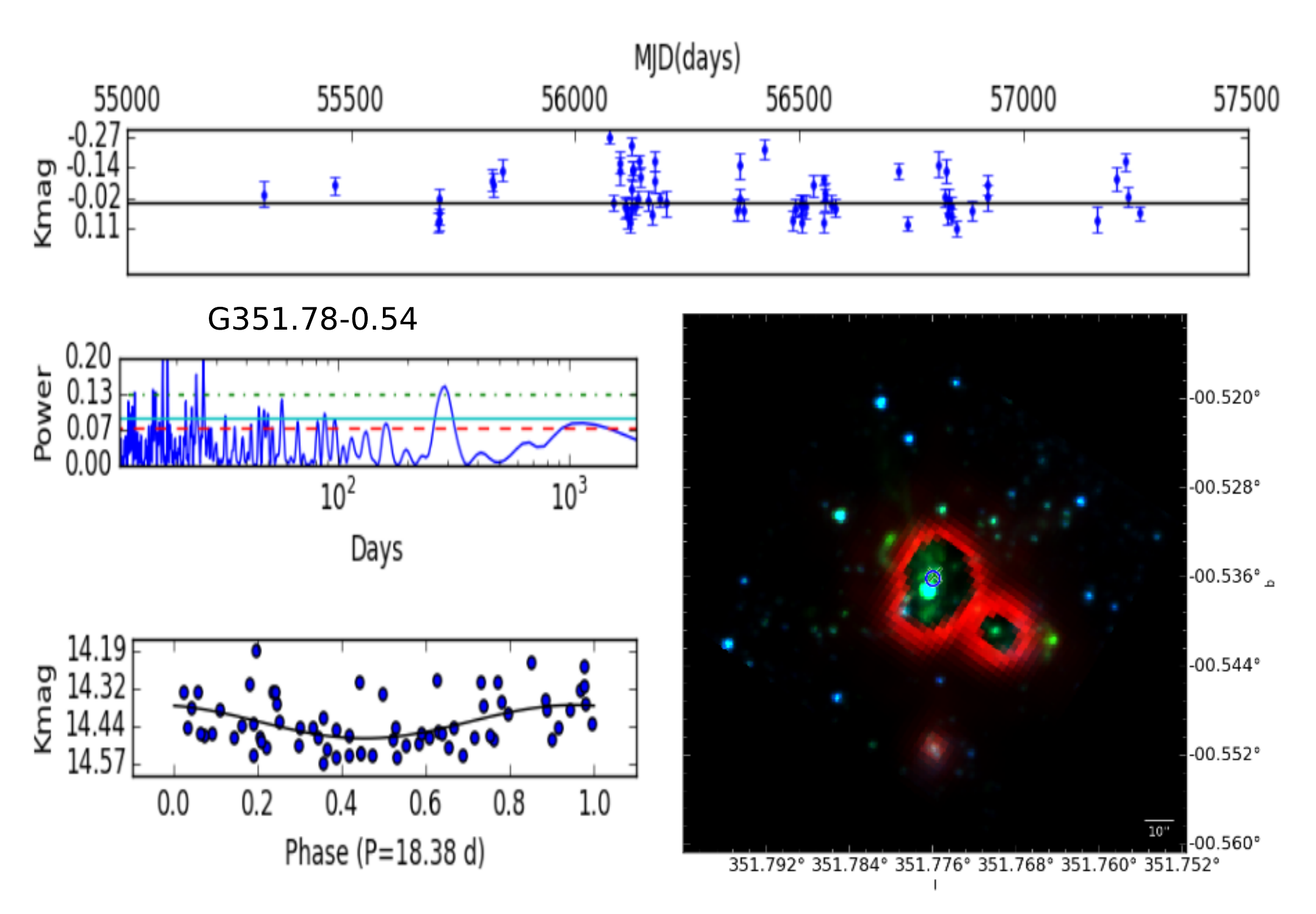}\\
                
        \end{tabular}
        \caption{Prototypical STV sources: Top panel for each figure shows the LC of the source, error bars represent MAD($\Delta S_{i_{mjd}}$), the left middle panel shows the
                corresponding periodogram in logarithmic scale (also plotted are the $99\%$, $95\%$, and $90\%$ false probability levels,
                respectively: the green dot-dashed line, the cyan full line, and the red dashed line), the bottom left panel shows the phase-folded
                light curve of the source using the best period fitted (also shows the corresponding value in days), the bottom right
                plot shows the RGB image of the source using the Spitzer IRAC 3.6 \mum, IRAC 4.0 \mum, and the 24\mum MIPS band
                as blue, green and red, respectively. The VVV source is indicated by the blue circle and the green cross represents the MIPS co-ordinates.
                The contours of the RGB are in the interval of [Peak-$5*\sigma$, Peak] from the ATLASGAL observation at $850\mum$.}
        \label{fig:stv}
\end{figure*}

Short timescale variables are objects with short timescales of periodic variability 
($P<100$ days), or without an
apparent periodicity. Periods larger than the stellar rotation or inner disc
orbits, $15<P<100$ days can be explained by phenomena such as obscuration from
circumbinary disc or by variable accretion \citep{contreras17, bouv03}.
Sources MG352.2452-00.0636 and G351.78-0.54 are typical examples of STVs, shown
in Fig. \ref{fig:stv}, with typical periods of approximately 29 and 18 days, respectively. Both sources match well ($r<2\arcsec$) with the
$870 \ \mum$ emission peak, additionally MG352.2452-00.0636 coincides with an 
IRDC filament, and G351.78-0.54 is close ($r<2\arcsec$) to the VLA1a source studied by \citet{zapa2008} as part of a compact cluster of MYSOs. G351.78-0.54 also coincides with an highly variable maser
studied by \citet{goed2014}.

\begin{figure}
        \includegraphics[width=\columnwidth]{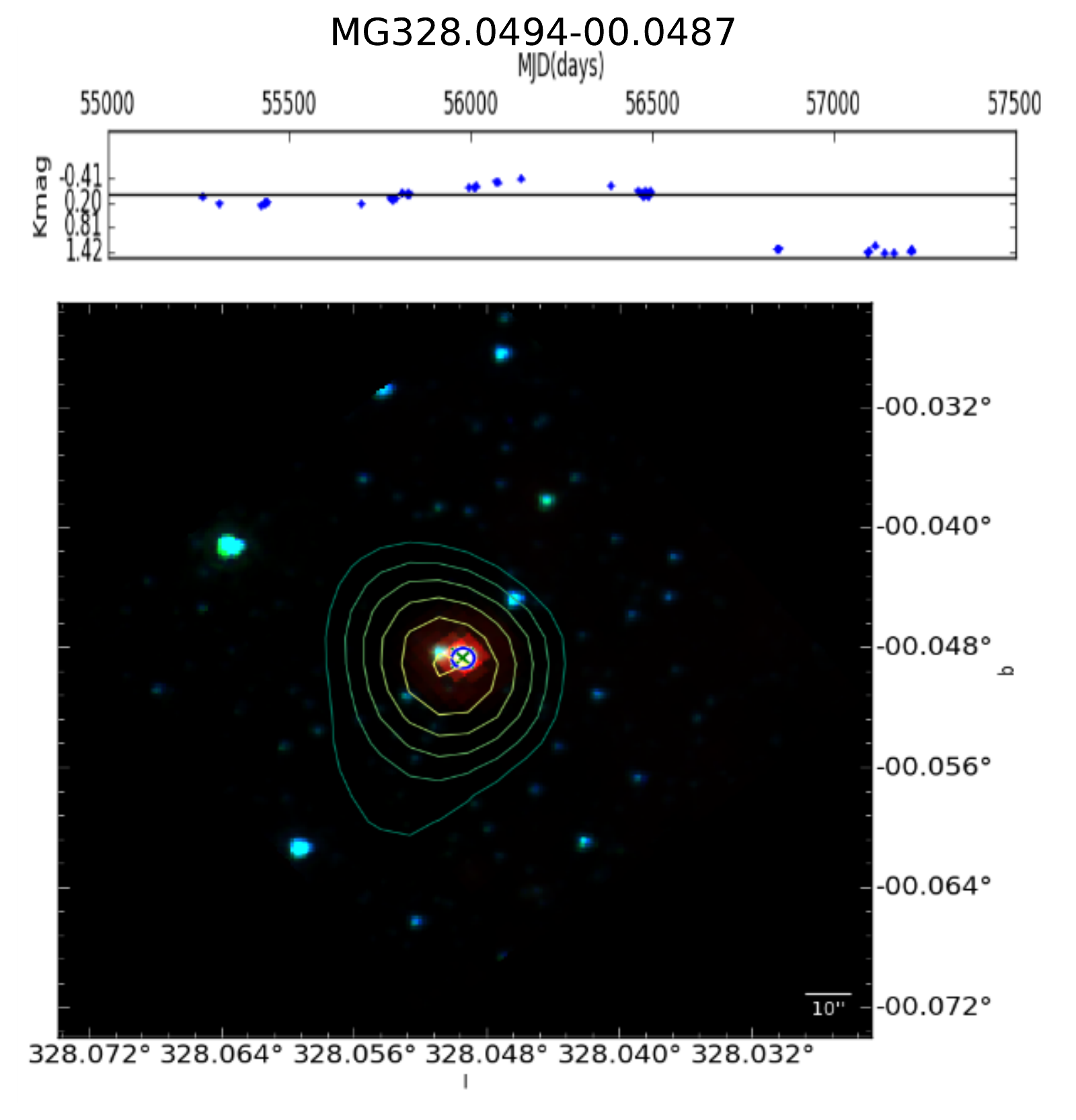}
        \caption{Typical Fader event.Colours and symbols are the same as in Fig. \ref{fig:lpv}.}
        \label{fig:fader}
\end{figure}

\begin{figure}
        \includegraphics[width=\columnwidth]{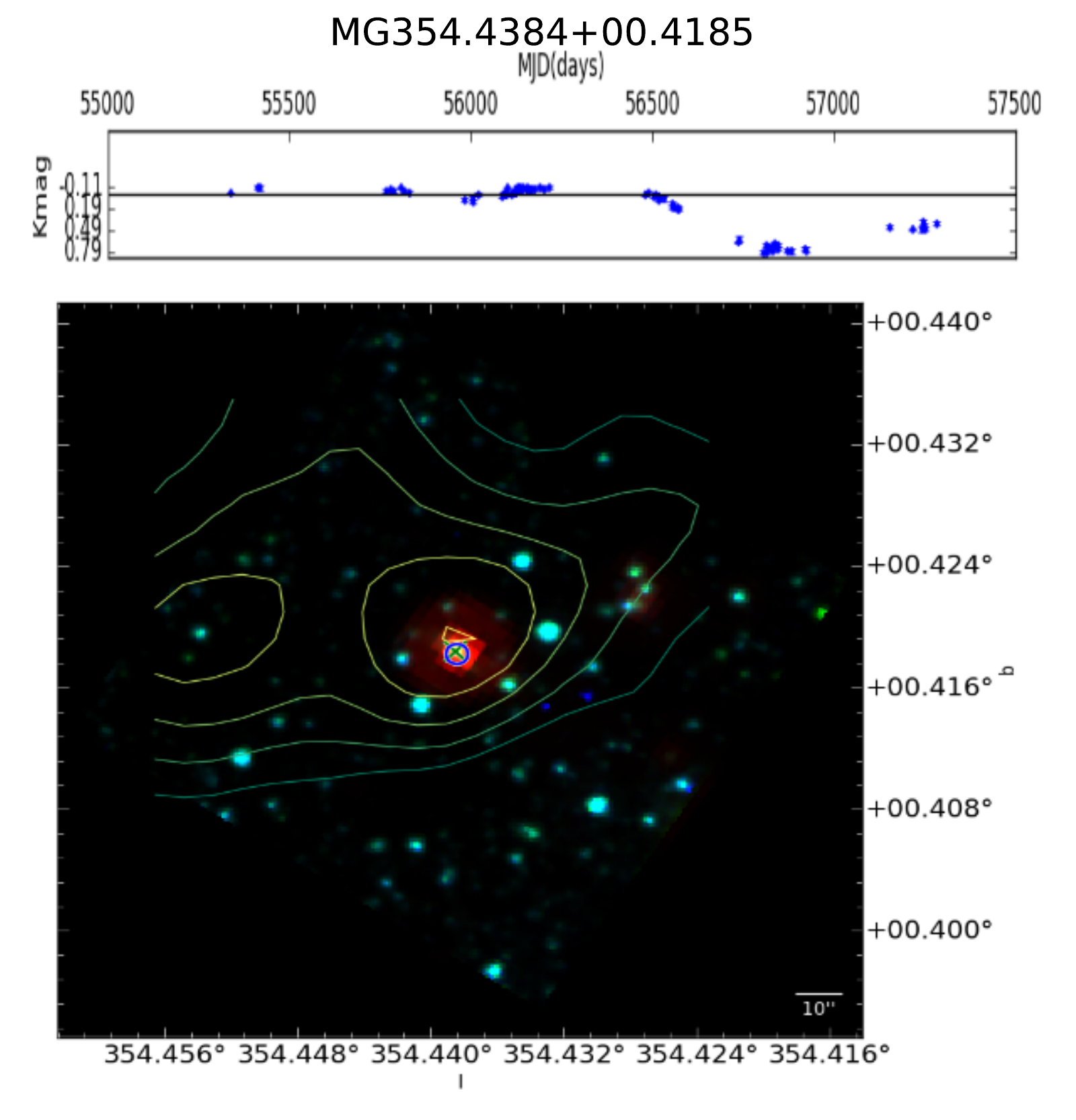}
        \caption{LC of a dipper event. Colours and symbols are the same as in Fig. \ref{fig:lpv}.}
        \label{fig:dipper}
\end{figure}

Faders and dippers are two classes of LCs with aperiodic
photometric variability. Dippers are characterized by long-lasting (lasting months to years) dimming events
followed by a return to normal brightness, while the same terminology is found in works of the YSOVAR team \citep{moralesysovar} they use it to classify phenomena on shorter timescales (hours to days). Faders show light-curves
that slowly decline over time, or a
period of continuous brightness followed by a sudden decrease sustained 
over a year. Dippers are often associated with increased extinction from surrounding
material. Faders can be caused by either a return to a quiescent accreting
phase or a long lasting increase in extinction. It should be noted that both
dippers and faders share common LC morphologies and can be easily
mistaken for one another. A snapshot of a dipper event not returning
to normal brightness can likely be mistaken as a fader.
Source MG328.0494-00.0487 (Fig. \ref{fig:fader}) is the prototypical example
of a fader event. It matches the peak emission of the ATLASGAL observations, and is
an extended object in the $8 \ \mum$ \textit{Spitzer} band, it has a close-by
companion. There is no clear peak in its periodogram and the LC shows some
periodic variability until around MJD 56500, at which point there is a drop in
brightness of close to $\Delta K \sim 1.4$ mag.
The morphology of Dipper events is typified by source MG354.4384+00.4185 which
is plotted in Fig. \ref{fig:dipper}. There is no clear peak in its periodogram, and its
light curve could be considered as non-variable, except for an abrupt drop in
brightness of $\Delta K \sim 0.8$ until the target recovers more than half its
brightness about 750 days later.

Eruptive LCs are also aperiodic, but they are characterized by outbursts and
increases in brightness, typically over periods of months or years but, in
some cases, lasting a few weeks. Objects with increases in their luminosity,
likely from ongoing accretion events, will produce such LCs. 
FUors or EXors are classic examples showing
eruptive morphologies.
Similarly to \citet{med2018} we employ a subdivision of the
eruptive class, `low amplitude eruptives' for sources with
$\Delta K<1.0 $ mag. This distinction is made to emphasize that such
variations are much less extreme than those in FUors and EXors in the optical
wavelengths, and more similar to common short term variability in their
amplitude. Nevertheless, for certain disk geometries and high extinction it is
possible for a FUor or EXor-like eruption to appear as low-amplitude
variability in the NIR. A low-amplitude eruptive LC can either correspond to a
low-amplitude variable source or to a high-amplitude source with a
geometry + extinction combination such that it appears as low-amplitude in the
$K_s$ band. Overall, there are 26 low-amplitude eruptives and 15 normal eruptives
in our samples. While this identification is presented in Table \ref{tab:sources_and_lcs}, for the analysis they were not considered as separate classes.

\begin{figure}
        
        \includegraphics[width=\columnwidth]{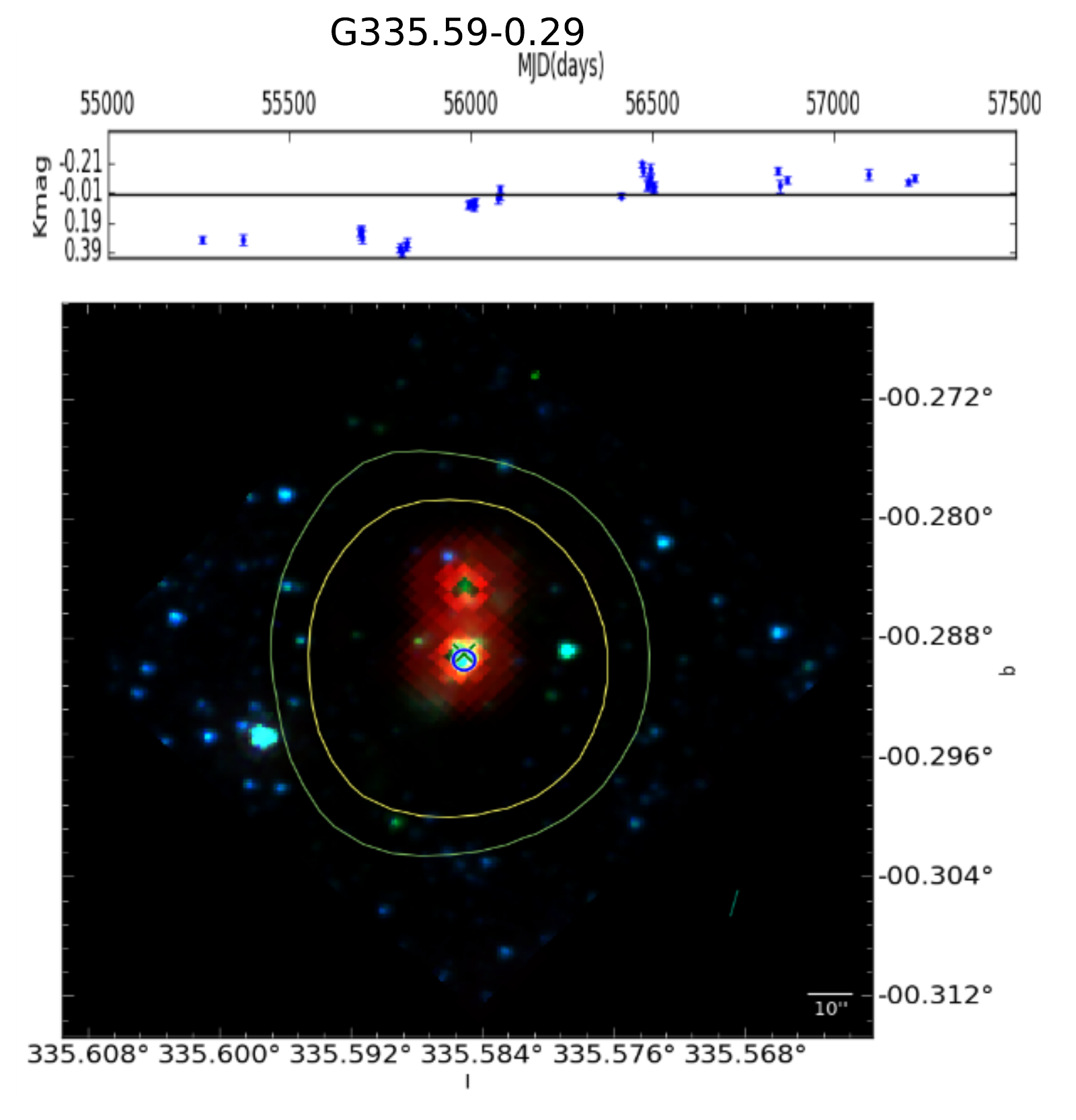}
        \caption{LC of the example of a low-amplitude eruptive event. Colours and symbols are the same as in Fig. \ref{fig:lpv}.}
        \label{fig:erup2}
\end{figure}

One source that features an ongoing eruptive event is MG303.9304-00.6879, plotted
in Fig. \ref{fig:erup1}, showing multiple stages of increased brightness
over the entire time-line, with two large amplitude brightness changes over years. 
Figure \ref{fig:erup2} shows an example of low-amplitude eruptive LC. 
This source, G335.59-0.29, displays C-shaped green emission,
characteristic of jet emission. The main feature of this
LC is its sustained increase in brightness over time, with a total amplitude
of $\Delta K \sim 0.68$ mag.
Overall, the variable sources can be split into
the periodic category, composed of LPV-yso and STVs, or the aperiodic category, 
including faders, dippers, and eruptive sources.
Their detailed distributions represented by the different MYSO samples can be found in
Table \ref{tab:egosandagalsummary_lc}.
Analysis of periodogram aliases (see Sect. \ref{subsec:lcs}) indicates that 1 and 15 members of the 
non-EGO + EGO samples, respectively, could be classified differently (see also Sect. \ref{sec:discussion}).

\begin{table}[]
        \centering
        \caption{Observed parameters of LC classes, for both EGO and non-EGO samples.}
        \label{tab:egosandagalsummary_lc}
        \begin{tabular}{lccc}
                \hline 
                \hline
                LC classification & EGO   & non-EGO & Total \\
                \hline 
                Periodic          & 90 ($\sim65\%$) & 21 ($\sim41\%$)   & 111\\ 
                Aperiodic         & 49 ($\sim35\%$) & 30 ($\sim59\%$)   & 79\\
                LPV-yso           & 53 ($\sim38\%$) & 9  ($\sim18\%$)   & 62\\
                STV               & 37 ($\sim27\%$) & 12 ($\sim23\%$)   & 49\\
                Dipper            & 15 ($\sim11\%$) & 5  ($\sim10\%$)   & 20\\
                Fader             & 13 ($\sim9\%$)  & 5  ($\sim10\%$)   & 18\\
                Eruptive          & 21 ($\sim15\%$) & 20 ($\sim39\%$)   & 41\\
                \hline
        \end{tabular}
\end{table}

The sources MG300.3241-00.1985, MG322.4833+00.6447, MG342.3189+00.5876 have also been
studied as highly variable objects \citep{contreras17,kumar2016}. 
Of these, MG300.3241-00.1985 was studied spectroscopically by \citet{contrerasspec}
and classified as an eruptive MNor, an object with a mixture of characteristics 
from FUors and EXors.
We note that other $\Delta K >1$ mag sources listed here were not found in 
\citet{contreras17} because they were not highly variable in the 2010-2012 period.

\subsection{Variable source SEDs}       

\begin{figure*}
        \begin{tabular}{ccc}
                \subfloat{\includegraphics[width = 0.3\textwidth]{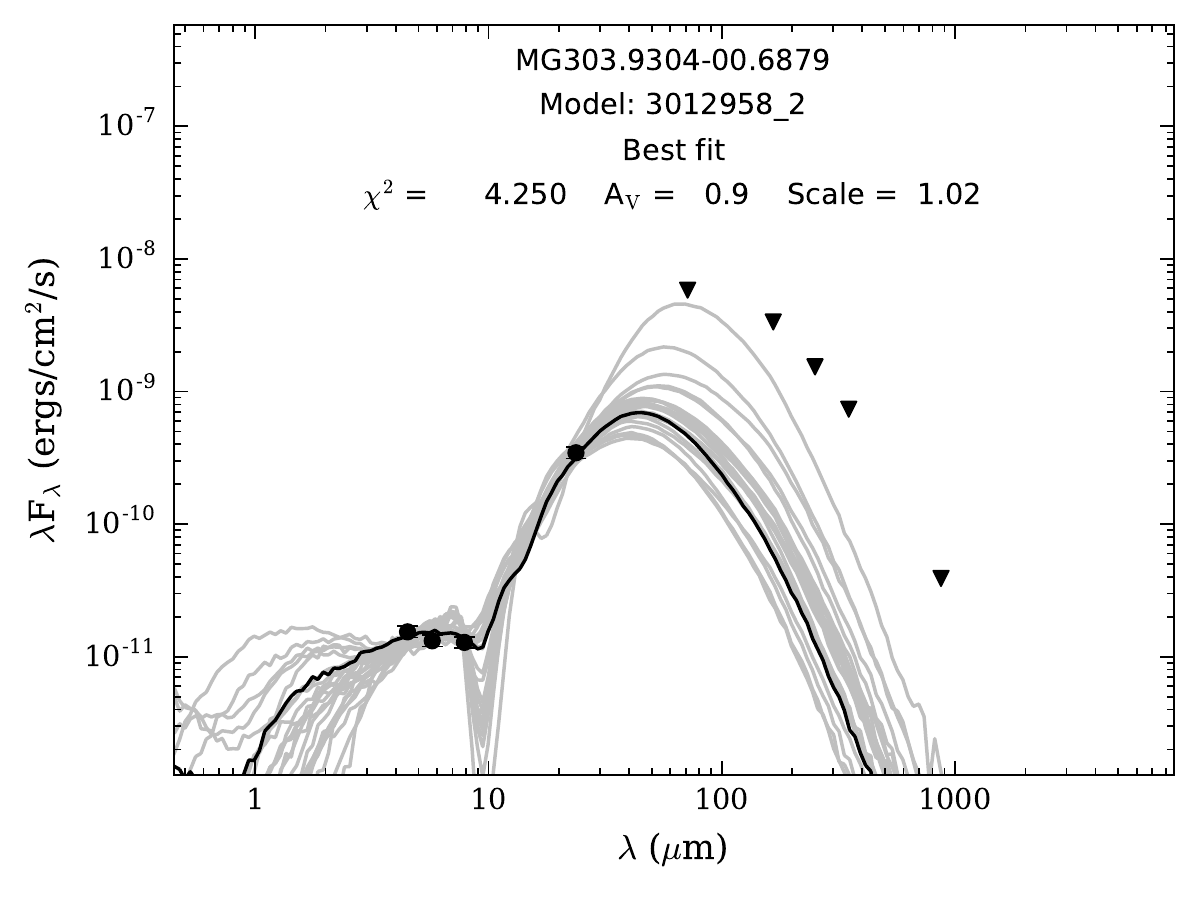}} &
                \subfloat{\includegraphics[width = 0.3\textwidth]{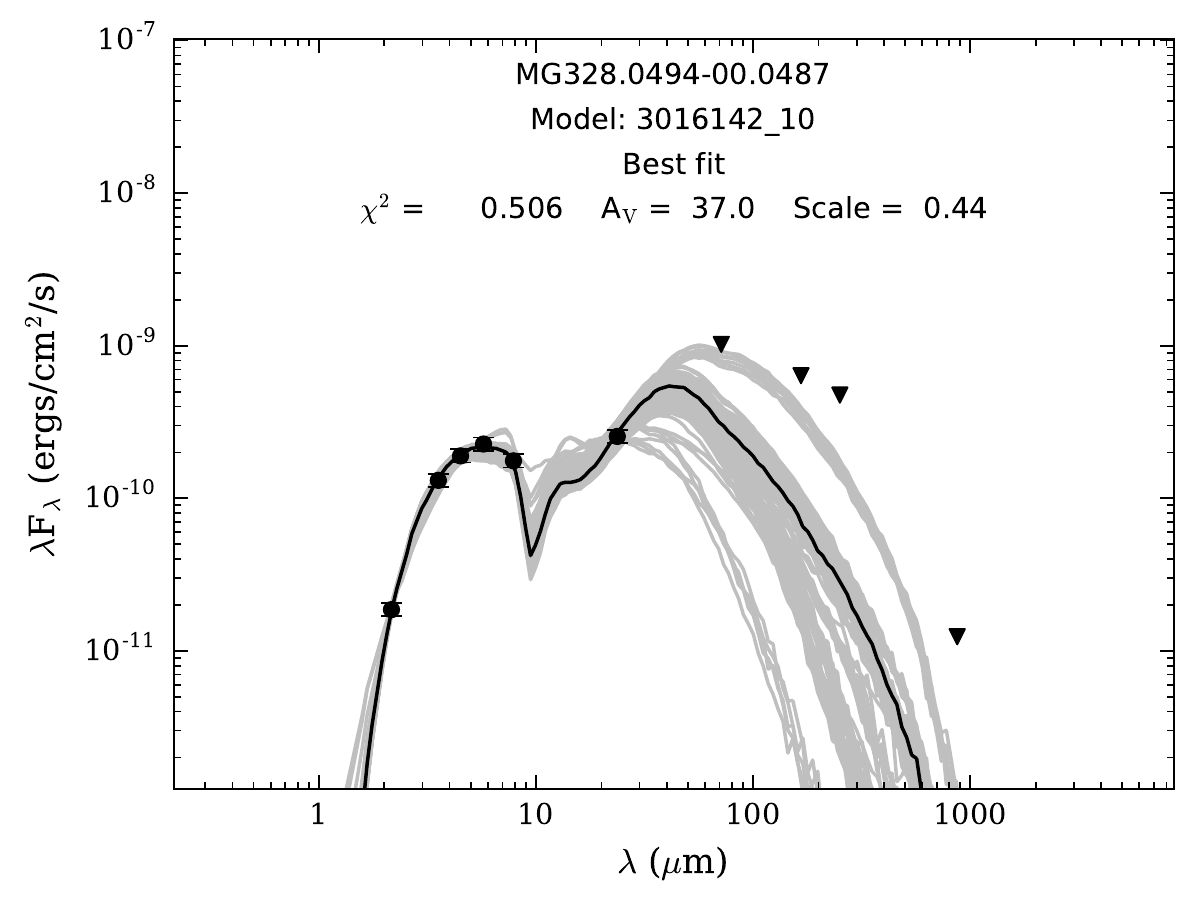}} &
                \subfloat{\includegraphics[width = 0.3\textwidth]{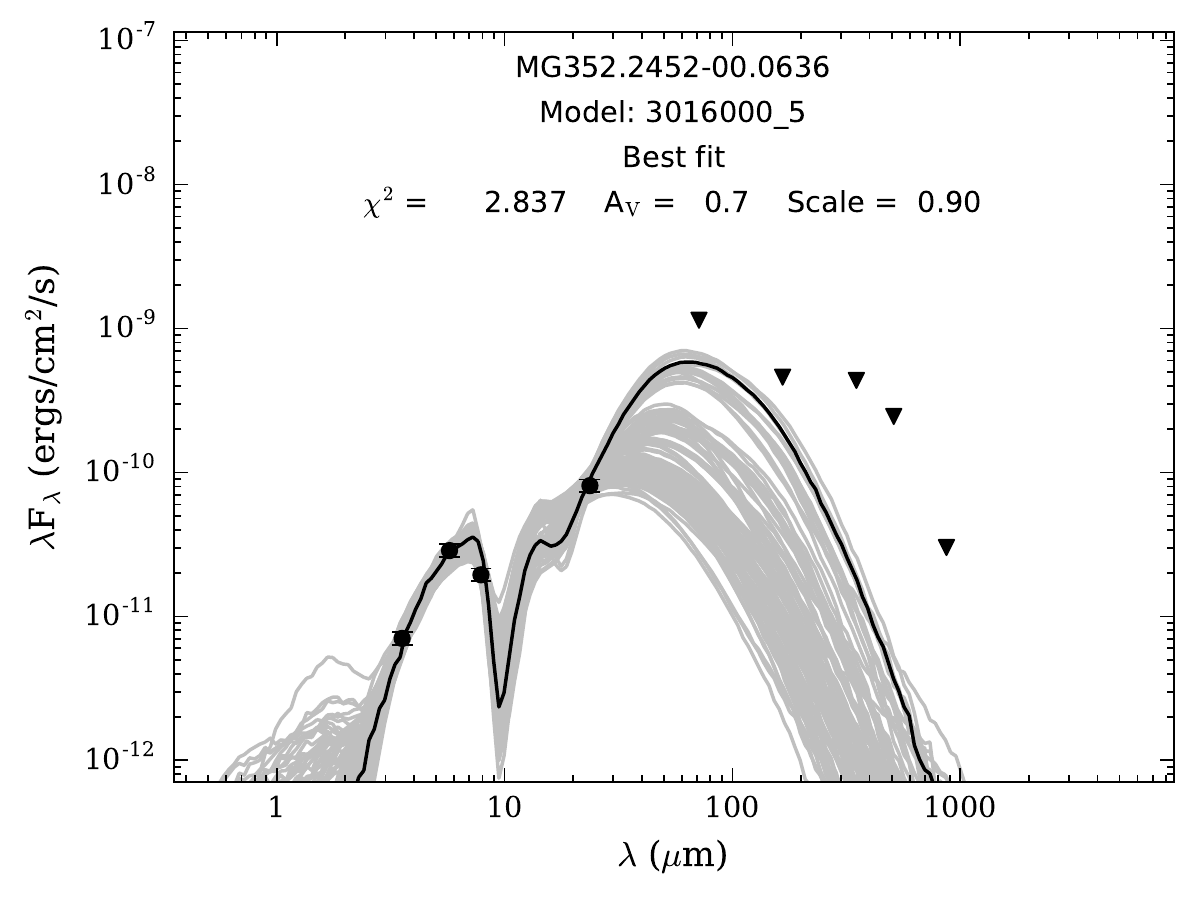}} \\
                \subfloat{\includegraphics[width = 0.3\textwidth]{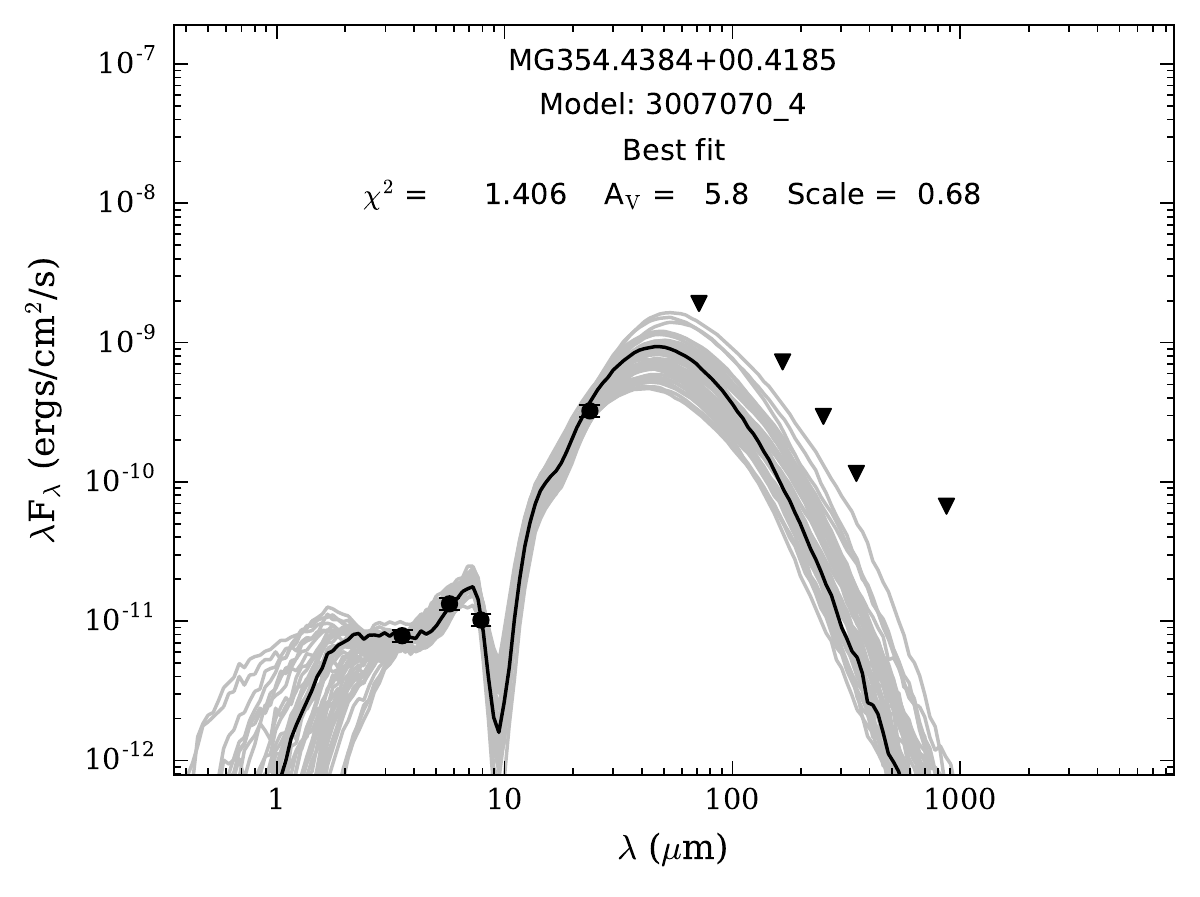}} &
                \subfloat{\includegraphics[width = 0.3\textwidth]{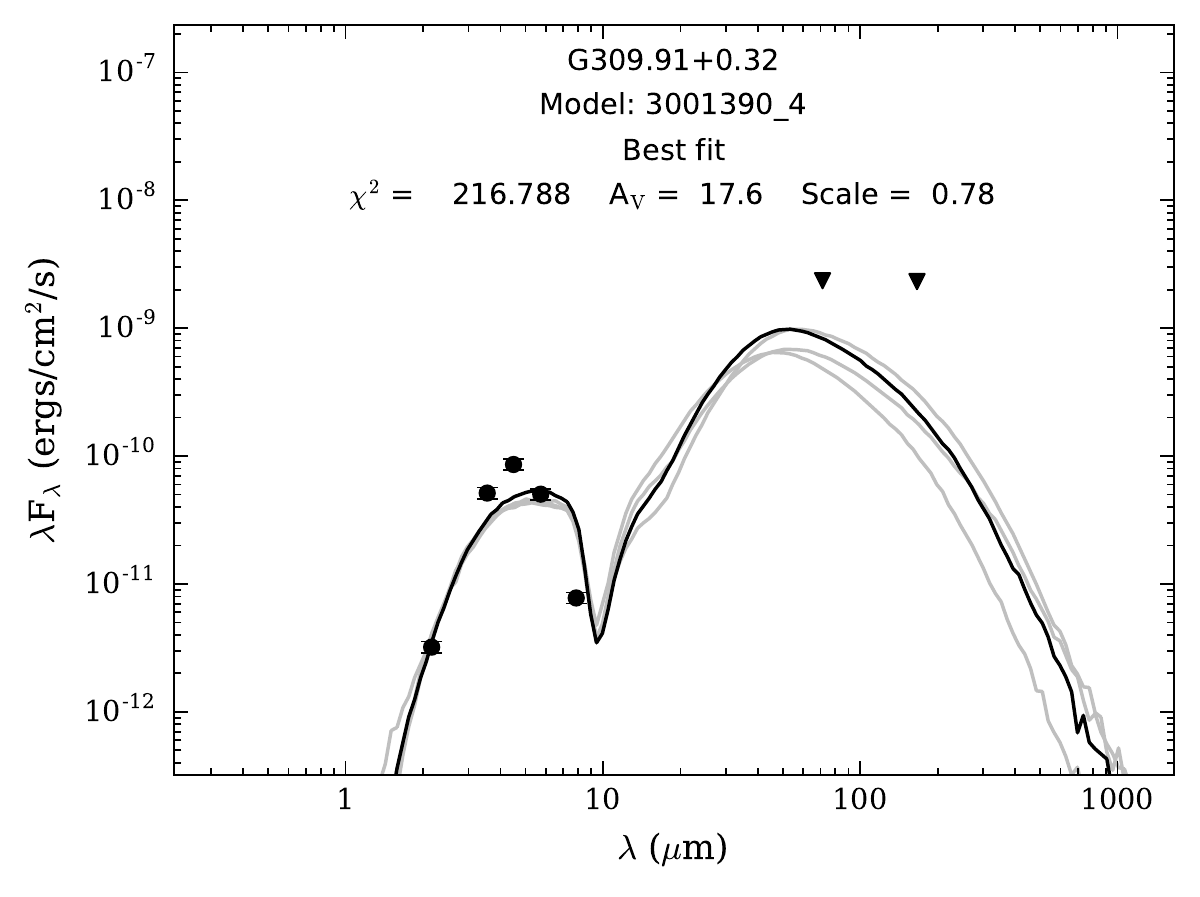}} &
                \subfloat{\includegraphics[width = 0.3\textwidth]{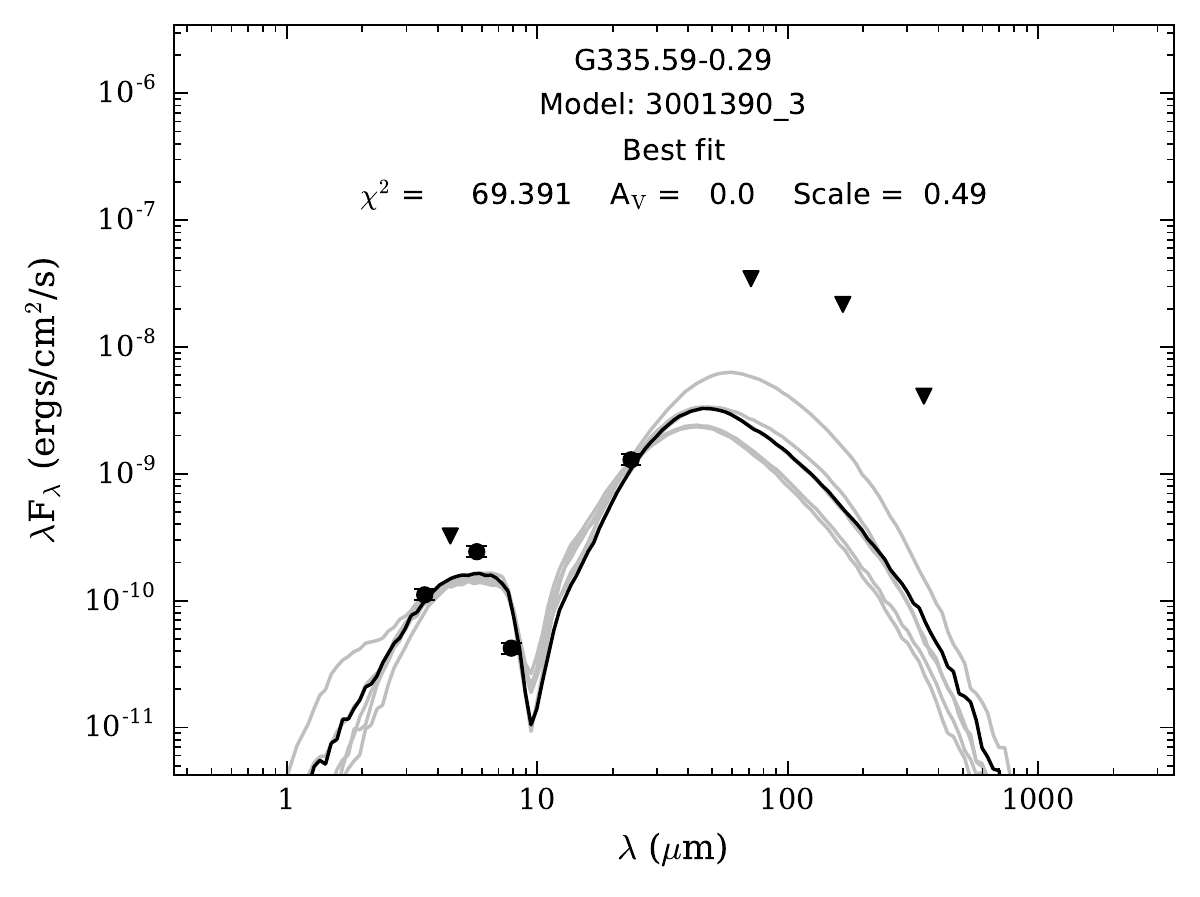}} \\
                \subfloat{\includegraphics[width = 0.3\textwidth]{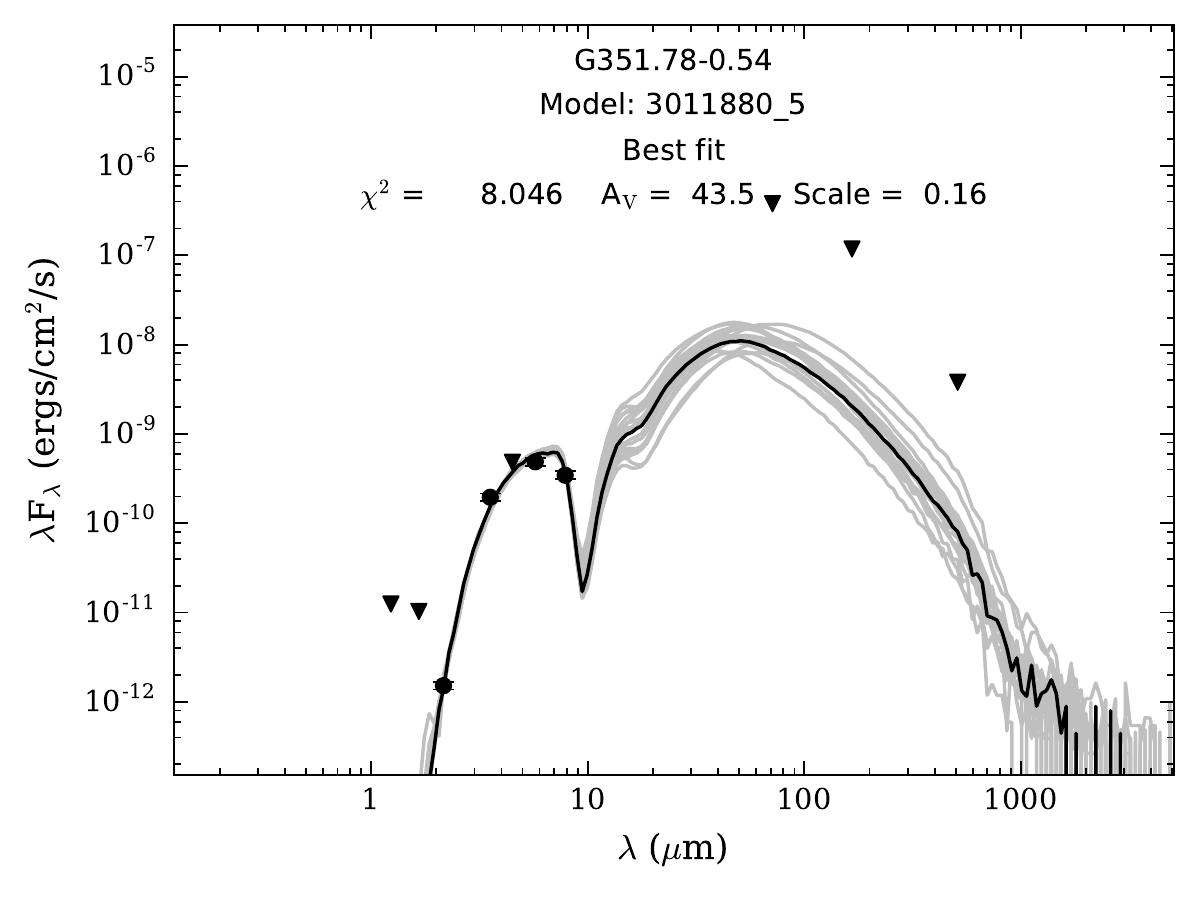}} &
                \subfloat{\includegraphics[width = 0.3\textwidth]{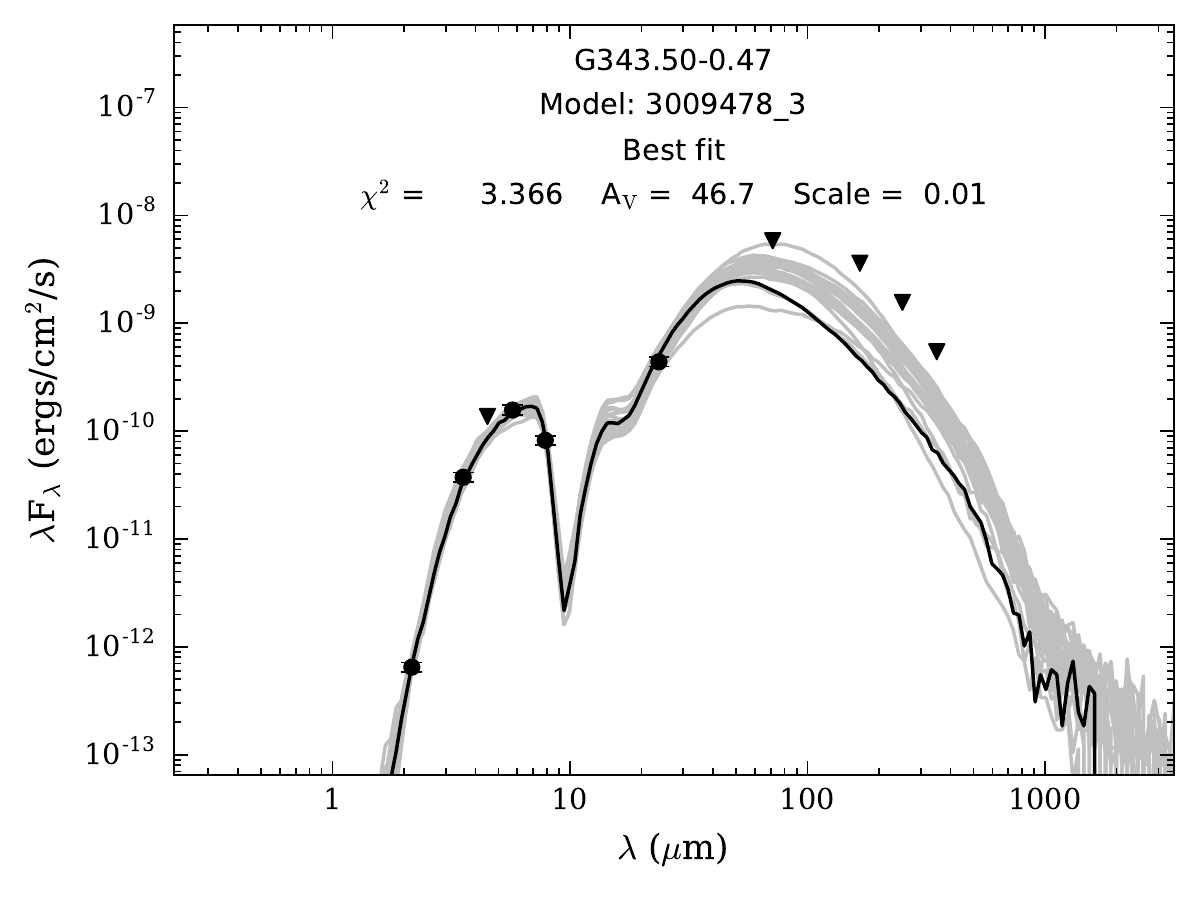}} &
        \end{tabular}
        \caption{Grid of SEDs for our prototypical sources. The dark line corresponds to the best fit model. The grey lines correspond to other $\chi^2 - \chi_{best}^2 <3$ models.}
        \label{fig:best_seds}
\end{figure*}

The goal of SED fitting was to test if the variable targets indeed represent MYSOs.  
The SEDs of the variable sources were fitted by YSO models \citep{robmodels} (see Sect. \ref{sec:sedanalysis}) 
allowing us to constrain the 
properties of these objects. The results of this fitting procedure can be
found in table \ref{tab:all_targets_sed_fit_summary}.
This table contains the full sample of variables with 190 entries. However,
as mentioned in Sect. \ref{sec:sedanalysis} only 105 targets have known distances, 
where the SEDs can be reasonably constrained. We note that in 
those cases where distances are not available, fitting with the full range of 
1-13 kpc has resulted in some model fits that outputs 
sub-stellar masses. This result is likely to be a consequence of
unknown distance rather than the true nature of the source because 
the indicators of 
high-mass star formation used in the
original selection are more reliable. 
In Fig.  \ref{fig:best_seds}, the data and model fits can be visualized
for the example targets with different LC classes mentioned in the previous
section. The masses of these example targets range from $1.84$ to
$10.30$ \Msun, with luminosities between $57$ and $6918 \Lsun$, representing evolutionary ages
between $10^4$ to $10^6$ yrs.
Table \ref{tab:sed_mass_bins} summarizes the SED results by listing various 
properties of the sources grouped in mass ranges roughly separating the 
low, intermediate, and high-mass sources. It can be seen that about $\sim 35\%$
of the targets are modelled in the 4-8 \Msun range and only $6\%$ representing
$\geq 8 \ \Msun$ objects. A large fraction ($\sim 60\%$) are fitted with YSO models representing sources
with $M<4\Msun$.

\begin{table*}
        \centering
        \caption{SED results by mass bin.}
        \label{tab:sed_mass_bins}
        \begin{tabular}{lccccccccc}
                \hline
                \hline
                M       & Sources & L       & L       & $\dot{M}_{env}$           & $\dot{M}_{env}$           &  $\dot{M}_{disk}$    & $\dot{M}_{disk}$    & $A_{{\rm V}_{circum}}$ & $A_{{\rm V}_{circum}}$ \\
                (\Msun) & ($\%$)  & (\Lsun) & (\Lsun) & ($\Msun \ yr^{-1}$) & ($\Msun \ yr^{-1}$) &  ($\Msun \ yr^{-1}$) & ($\Msun \ yr^{-1}$) &            &  \\
                Range   &  Ratio  & Range   & Median  & Range               & Median              &  Range               & Median              & Range      & Median \\
                \hline
                $M<4$     & $\sim 59$ & [4.0E-1,9.0E2] & 5.0E1 & [0,4E-4]    & 1.3E-5 & [-8E-3,4E-5] & 2E-7  & [6E-1,6E5] & 74  \\
                $4\leq M < 6$  & $\sim 21$ & [8.8E1,1.2E3]  & 2.9E2 & [0,4E-4]    & 7.8E-5 & [-4E-2,9E-6] & 6E-7  & [2E0,2E4]  & 56  \\
                $6\leq M < 8$  & $\sim 14$ & [2.9E2,5.1E3]  & 9.3E2 & [0,6E-4]    & 2.0E-4 & [-2E-1,3E-5] & 2E-6  & [5E0,1E5]  & 66  \\
                $8\leq M$ & $\sim 6$  & [1.3E3,3.7E4]  & 3.0E3 & [1E-4,4E-3] & 2.8E-4 & [-1E0,4E-6]  & -2E-3 & [4E1,4E5]  & 228 \\
                \hline

        \end{tabular}
        
\end{table*}

The 4-8 \Msun sources display $\dot{M}_{env} \sim 10^{-4} \ \Msun yr^{-1}$, $\dot{M}_{Disk} \sim 10^{-6} \ \Msun yr^{-1}$ and a few hundred solar luminosities. The number of EGO and non-EGO sources fitted as
low, intermediate, and high-mass stars are 87, 45, 10 and 25, 21, 1 respectively. 
It is worth noting that all but one of the sources fitted by models $\geq 8 \Msun$ are EGO objects.
These SEDs are well-fitted by MYSO models similar to those represented in \citet{grave09}.
Four of the 11  objects ($\geq 8 \ \Msun$ ) are included in the 6.7 GHz class II methanol maser surveys
and they show  emission. These four also show class I methanol maser emission.

\section{Discussion}\label{sec:discussion}

\begin{figure}
        \begin{tabular}{cc}
                \includegraphics[width=0.45\columnwidth]{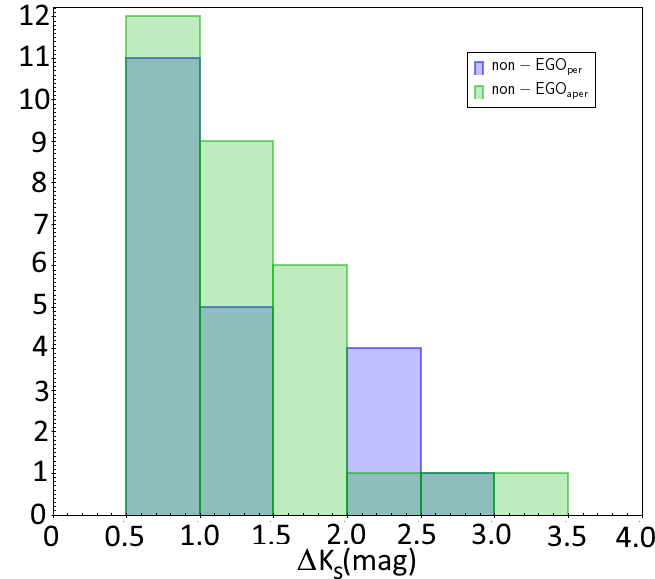} &

                \includegraphics[width=0.45\columnwidth]{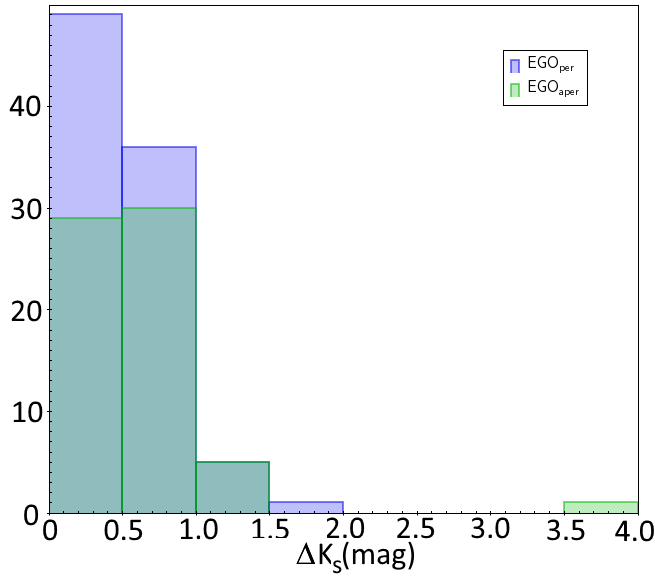} \\
                
        \end{tabular}
        \caption{Histogram of $\Delta K$ divided by sample and periodicity. 
                EGO and non-EGO sources are shown, respectively, on the right and left plots.}
        
        \label{hist:deltaK}
\end{figure}

The results show that 139 of 156 (91\%) EGO sample presents variability in contrast to 51 of the 433  (12\%) non-EGO targets, implying that variability is strongly correlated with the outflow activity in MYSOs. Table 3 summarizes the variability statistics. More than half (64\%) of the variable EGOs are classified as periodic contrasting more than half (59\%) of the non-EGO sample that are classified as aperiodic. Table \ref{tab:egosandagalsummary_sed} allows us to discern the differences between EGO and non-EGO samples, and sources classified as periodic or otherwise. It can be seen from Fig. \ref{hist:deltaK} and Table \ref{tab:egosandagalsummary_sed}, that the amplitude range of variation in non-EGOs is roughly twice as much as that of EGOs. Of the modelled parameters, the circumstellar extinction ($A_{{\rm V}_{circum}}$) for non-EGO targets clearly stand out as twice the median value for EGOs. Also, it appears that non-EGO variable sources may simply be more luminous objects located in slightly farther away targets. Together, the $\Delta K_s$, Av and L comparison indicate that the non-EGO variable sources are relatively more embedded objects when compared to EGOs.

The results of the search for aliases among the ten frequencies with     greater
power, found aliases for the highest peak of the periodogram in five non-EGO targets
($\sim 9\%$), and 22 ($\sim 15\%$) EGO targets. Of these, only 1 ($\sim 2\%$) of the non-EGO
targets would change their classification from LPV-yso to STV, while, for the
EGO sample 15 ($\sim 10\%$) of the targets could change from LPV-YSO to STV or
vice-versa. Therefore, these aliases would not change any periodic to aperiodic
source, as
period length is not the only condition defining a periodic source (LC
morphology is also one of the main factors).

\begin{table}
        \centering 
        \caption{Summary of the median fit parameters, for both EGO and non-EGO samples divided by periodicity.}
        \label{tab:egosandagalsummary_sed}              
        \begin{tabular}{lcccc}
                \hline
                \hline
                Parameter                        & EGO & non-EGO & Periodic & Aperiodic \\
                \hline 
                $\Delta K_s$ (mag)               & 0.52 & 1.02 & 0.58 & 0.69 \\
                Period (days)                    & 312  & 416  & 126  & - \\
                M (\Msun)                        & 3.2  & 3.8  & 3.2  & 3.6 \\ 
                $\dot{M}$ ($\Msun yr^-1$)        & 4E-5 & 6E-6 & 4E-5 & 2E-5\\
                $\dot{M}_{disk}$ ($\Msun yr^-1$) & 3E-7 & 7E-7 & 3E-7 & 6E-7 \\
                L (\Lsun)                        & 125  & 212  & 125  & 190 \\
                Age (Myr)                        & 5.0  & 5.6  & 5.0  & 5.0 \\
                T (K)                            & 4841 & 7795 & 4857 & 5990 \\
                $A_{{\rm V}_{circum}}$           & 61   & 125  & 71   & 54 \\
                \hline 
        \end{tabular}
\end{table}

\begin{figure}
        
        \includegraphics[width=\columnwidth]{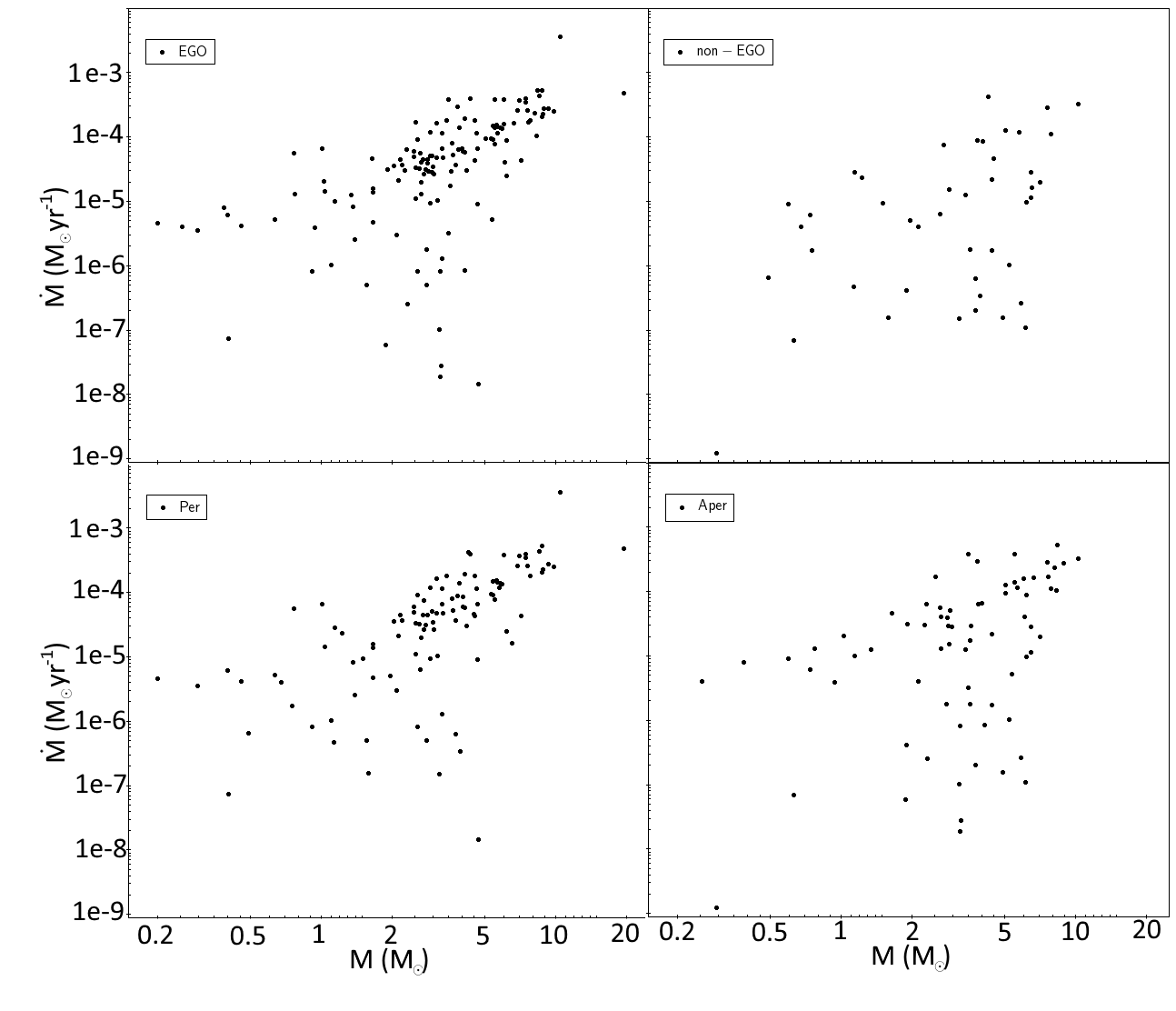}
        \caption{Mass versus envelope-accretion rate for the fitted SEDs of EGO and non-EGO sources, in logarithmic scale. EGO, non-EGO, periodic, and aperiodic, are plotted at the top left, top right, bottom left, and 
                bottom right, respectively.}
        \label{fig:MvsMdot}
\end{figure}

It can be seen from Fig. \ref{fig:MvsMdot} that the envelope accretion rate (see also Table \ref{tab:egosandagalsummary_sed}) for non-EGO sources is an order of magnitude smaller than that for the EGO sources. The same effect can be noticed for aperiodic sources (bottom panel). We note that the non-EGOs are dominated by aperiodic sources, that should be indicative of the differences observed in the top and bottom panels of Fig. \ref{fig:MvsMdot}. Aperiodic LCs, represented by eruptive, dippers and fader classes, are thought to trace objects with low level of quiescent accretion, that will undergo short periods of intense accretion. The lower level of accretion found in these objects can therefore be explained by this behaviour.

\begin{figure}
        
        \includegraphics[width=\columnwidth]{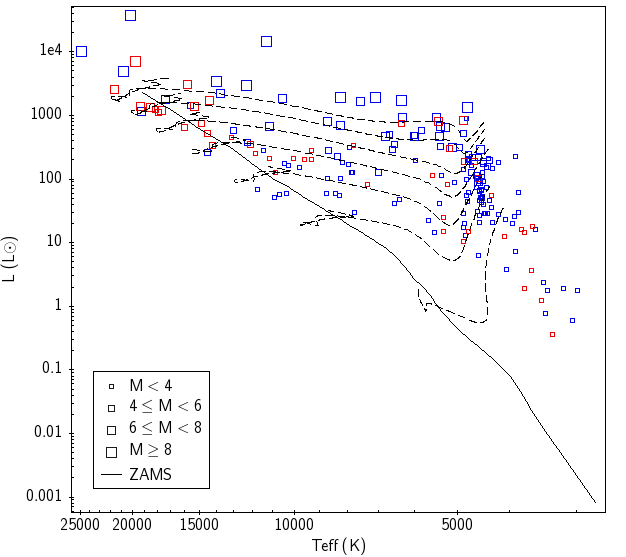}
        \caption{HR diagram for our sources. Symbol size corresponds to $M<4$, $4\leq M <6$, $6\leq M <8$, $M \geq 8 $ \Msun, from smaller to larger, respectively. The dashed lines(from bottom to top) are the PMS tracks for 1, 2 , 3, 4, 5, 6, and 7 \Msun, the filled line is the ZAMS. Blue and red symbols are, respectively, EGOs and non-EGOs. }
        \label{fig:hrdiagram}
\end{figure}

We compared the SED fitted model properties with the amplitude of variation and did not find any correlations to understand the variability as a function of mass, accretion rate, luminosity or temperature.
Figure \ref{fig:hrdiagram} shows an HR diagram, by plotting the luminosity versus temperature of all variable sources derived from the SED fitting. The zero-age-main-sequence (ZAMS) curve \citep{siess00} is shown by a solid curve. The seven dashed curves display the pre-main-sequence tracks (also from \citet{siess00}, for solar metallicity) for objects of 1-7 \Msun in steps of 1 \Msun.
EGOs are concentrated closer to the putative birth-line position of the massive stars and are also largely lower mass objects ($<4$ \Msun). The precursor to a high mass star is considered to be a lower mass object which continues to accrete material for more than half of its life until it contracts on the main sequence. In view of that conjecture, it is not surprising that a majority of the EGO driving sources are modelled by young low to intermediate mass stars.
        Furthermore, the HR diagram, with the associated PMS tracks, validates the fitted masses.

Most parameters resulting from SED fitting are model
        dependent, with known correlations within the model grid between the age, mass and accretion rates \citep{robmodels}. However, the observed data is scaled to match the luminosity and
        temperature of the selected models from the grid, therefore, these are relatively
        more reliable fitted parameters. Unlike the lower mass stars, the luminosity and temperature differences prominently distinguish the sparsely populated massive stellar models in the grid. These two parameters are used for comparison in this analysis, to ensure that the inferences made are reasonably free of biases in the grid models.

There is an apparent concentration of non-EGO objects, probably with slightly more higher mass objects on the ZAMS. It was previously noted that the non-EGO targets are significantly more embedded objects displaying larger $\Delta K_s$ compared to EGOs. The objects located closer to the ZAMS may therefore be candidate sources to test the hypothesis of bloated and pulsating young massive stars. \citet{hosok10} argue that high-mass stars are bloated objects. Such objects are also thought to be pulsationally unstable, or at least go through a period of significant pulsations as they settle down on the ZAMS \citep{ina13}.
\citet{contreras17} indicates that eruptive variable behaviour 
        is more common or recurs more frequently at earlier stages of stellar PMS. 
        The analysis in this work show that $\sim 70\%$ eruptive variables are concentrated on the birthline and the ZAMS in nearly half proportion. Protostellar envelopes are a prominent feature of objects located on the birthline, therefore suggesting that most eruptive MYSOs are indeed the result of envelope accretion. High-mass protostellar objects ingesting a burst of accreting matter enter a `bloated phase' before re-adjustment and contraction \citep{hosok10}. This could be the case for those eruptive sources located on the ZAMS.

In Table \ref{tab:masers}, all the variable sources (32 targets) with known 6.7 GHz class II methanol maser detection are listed, along with, the simultaneous detection of class I methanol maser. Those sources with only class I methanol maser detections are not listed. The detection of class II methanol maser is considered as a strong sign-post of high-mass star formation, especially massive outflow activity \citep{devill15}. Of the 32 sources, only two are non-EGOs, therefore, reinstating the association of class II methanol masers with MYSO outflow activity. \citet{goed2014} have studied variability of methanol masers, and two of the infrared variable sources presented here, G351.78-0.54 and G298.26+0.74, were analysed in that study and G351.78-0.54
is considered as an highly variable maser, while G298.26+0.74 does not present maser variability above instrumental noise.

Our selection criteria for non-EGO sources, 24 \mum MIPS sources matching ATLASGAL CSC objects ($r<5$\arcsec) might lead us to miss some of the most important sources in the clumps. 
Since the most luminous source inside each clump can be offset by more than $r<5$\arcsec we  have missed many of the MYSOs in these regions. The criteria used ensures that the targets are good MYSO candidates but, the most luminous
FIR sources and their counterparts will be examined in a future work.

\section{Summary}

This study has investigated the nature of near-infrared variability in MYSOs, focusing on the driving
sources of EGOs and luminous 24 \mum point sources coinciding within 5\arcsec of the massive star forming clumps
mapped at 870 \mum by ATLASGAL. The search led us to examine the $K_s$-band light-curves of 718 point sources.

\begin{itemize}
        \item 190 sources (139 EGOs and 51 non-EGOs) were found to be variable with an IQR$>0.05$ and $\Delta K_s >0.15$.
        111 and 79 of these objects are classified as periodic + aperiodic, respectively.
        
        \item The 2\mum - 870\mum spectral energy distribution of the variable point sources were assembled and fitted
        with YSO models. 47 and 6 sources were modelled as $\geq4$ \Msun and $\geq8$ \Msun, respectively.
        
        \item On an HR diagram, most lower mass EGO sources concentrate along a putative birth-line.
        
        \item A high rate of detectable variability in EGO targets (139 out of 153 searched) implies that 
        near-infrared variability in MYSOs is closely linked to the accretion phenomenon and outflow activity.
        
\end{itemize} 

        Further to the discovery of a dozen high-amplitude variable MYSOs \citep{kumar2016}, this is the first large scale systematic study of near-infrared variability in MYSOs. The variable sources identified in this work are excellent targets with which to undertake follow-up studies to understand the circumstellar environment of MYSOs in detail.

\begin{table*}
        \centering
        \caption{EGO and non-EGO MYSO candidates with nearby methanol masers.}
        \label{tab:masers}
                \begin{threeparttable}
                        \begin{tabular}{lccccccc}
                                \hline
                                \hline
                                
                                Source             &  $\widetilde{K\_mag}$ & IQR    & Distance & Class & ClassII  & ClassI \\
                                &  (mag)                & (mag)  & (kpc)      &       & Maser    & Maser  \\
                                \hline
                                MG003.5016-00.2020 &    16.07   &       0.23    & 5.0   & Erup          & Y &      \\
                                MG006.9222-00.2512 &    14.38   &       0.26    & 3.0   & Erup          & Y &     Y\\
                                MG332.3652+00.6046 &    14.17   &       0.09    & 2.7   & Fad           & Y &     Y\\
                                MG333.0294-00.0149 &    15.24   &       0.18    & 4.0   & Dip           & Y &     N\\
                                MG339.2939+00.1387 &    15.63   &       0.41    & 4.8   & STV           & Y &      \\
                                MG339.5843-00.1282 &    13.16   &       0.16    & 2.6   & Dip           & Y &     Y\\
                                MG345.5764-00.2252 &    15.33   &       0.3         & 7.9   & Erup        & Y &      \\
                                MG352.6040-00.2253 &    15.38   &       0.22    & 7.6   & Erup          & Y &      \\
                                MG358.4604-00.3929 &    16.03   &       0.16    & 5.0   & LPV-yso     & Y &       Y\\
                                G9.62+0.20             &        14.38   &       0.11    & 5.2   & STV         & Y &       Y\\
                                G6.19-0.36             &        14.52   &       0.09    & 5.1   & STV         & Y &       Y\\
                                G5.62-0.08             &        15.43   &       0.07    & 5.1   & LPV-yso     & Y &       Y\\
                                G359.44-0.10       &    14.99   &       0.13    &       & LPV-yso     & Y &       Y\\
                                G358.84-0.74       &    13.82   &       0.12    & 6.8   & LPV-yso     & Y &       Y\\
                                G358.46-0.39(b)    &    15.45   &       0.16    & 2.9   & STV         & Y &       Y\\
                                G358.39-0.48       &    13.93   &       0.19    & 2.4   & Erup        & Y &       Y\\
                                G358.26-2.06       &    12.26   &       0.08    & 3.0   & Fad         & Y &        \\
                                G355.54-0.10       &    14.08   &       0.15    & 3.0   & LPV-yso     & Y &       Y\\
                                G355.18-0.42       &    14.98   &       0.08    & 1.2   & Erup        & Y &       Y\\
                                G353.46+0.56       &    13.18   &       0.1         & 11.2  & LPV-yso     & Y &  Y\\
                                G352.63-1.07       &    14.56   &       0.14    & 0.9   & STV         & Y &       Y\\
                                G352.58-0.18       &    15.62   &       0.09    & 5.1   & LPV-yso     & Y &        \\
                                G352.13-0.94       &    12.79   &       0.1         & 2.3   & LPV-yso     & Y &  Y\\
                                G351.78-0.54       &    14.46   &       0.12    & 0.7   & STV         & Y &       Y\\
                                G351.69+0.17       &    14.91   &       0.05    & 12.1  & STV         & Y &        \\
                                G351.38-0.18       &    15.8    &       0.07    & 5.6     & STV         & Y &     N\\
                                G351.16+0.69       &    10.4    &       0.15    & 1.8     & STV         & Y &     Y\\
                                G350.52-0.35       &    15.02   &       0.17    & 3.1     & Erup        & Y &     N\\
                                G350.36-0.07       &    14.31   &       0.09    & 11.2    & Fad         & Y &      \\
                                G2.54+0.20             &        12.71   &       0.09    & 4.0     & LPV-yso         & Y & N\\
                                G2.14+0.01             &        13.03   &       0.03    & 11.2    & Non-var     & Y &      \\
                                G0.09-0.66             &        13.87   &       0.08    & 8.2     & STV         & Y &     Y\\
                                \hline
                        \end{tabular}                        
                \end{threeparttable}                
\end{table*}

\begin{acknowledgements}
        G.D.C.T. is supported by an FCT/Portugal PhD grant PD/BD/113478/2015.
        MSNK acknowledges the support from Funda\c{c}\~ao para a Ci\^encia e Tecnologia (FCT)
        through Investigador FCT contracts IF/00956/2015/CP1273/CT0002, and the H2020 Marie-Curie
        Intra-European Fellowship project GESTATE (661249).
        Support for JB is provided by the Ministry of Economy, Development, and 
        Tourism's Millennium Science Initiative through grant IC120009,  awarded to 
        The Millennium Institute of Astrophysics, MAS.
        A.C.G. has received funding
        from the European Research Council (ERC) under the European Union’s Horizon 2020 research and innovation programme (grant agreement No. 743029).
        PWL acknowledges the support of consolidated grants (ST/R000905/1
        and ST/M001008/1) funded by the UK Science and Technology Facilities Research Council.
        CCP acknowledges support from the Leverhulme Trust.
        JFG is supported by Funda\c{c}\~ao para a Ci\^encia e a Tecnologia (FCT) through national funds (UID/FIS/04434/2013)
        and by FEDER through COMPETE2020 (POCI-01-0145-
        FEDER-007672).
\end{acknowledgements}

\onecolumn
\begin{landscape}

\begin{longtable}{lcccccccccccccccc}
                \hline
                \hline
                SOURCE & $\chi^2$  & Av   & M    & $\dot{M}$       & $M_{disk}$ & $\dot{M}_{disk}$        &  L   & Age     \\
                &           &  mag  & \Msun & $\log(\Msun yr^-1)$ & $\log(\Msun)$  & $\log(\Msun yr^-1)$ & $\log(\Lsun)$ & $\log(yr)$ \\
                \hline
                MG303.9304-00.6879 & $11.69  \pm 3.10$ & $2.7 \pm 4.6$ & $10.3 \pm 2.1$ & $-3.5 \pm -3.35$ & $-0.86 \pm -0.75$ & $-5.33 \pm -4.89$ & $3.84 \pm 3.78$ & $5.0 \pm 0.4$ \\
                MG328.0494-00.0487 & $9.53  \pm 5.06$ & $38.23 \pm 5.49$  & $5.58 \pm 0.94$ & $-5.89 \pm -5.28$ & $-1.61 \pm -1.40$ & $-6.25 \pm -5.46$ & $2.91 \pm 2.73$ & $5.94 \pm 0.42$ \\
                MG352.2452-00.0636 & $7.97  \pm 2.81$ & $26.17 \pm 16.92$ & $1.84 \pm 1.32$ & $-4.53 \pm -4.37$ & $-1.60 \pm -1.40$ & $-5.76 \pm -5.36$ & $1.76 \pm 2.10$ & $4.21 \pm 0.71$ \\
                MG354.4384+00.4185 & $6.33  \pm 3.51$ & $14.70 \pm 14.42$ & $3.65 \pm 1.03$ & $-4.15 \pm -4.33$ & $-1.40 \pm -1.23$ & $-6.13 \pm -5.77$ & $2.06 \pm 1.70$ & $4.63 \pm 0.38$ \\
                
                G309.91+0.32 & $38.03 \pm 2.33$ & $0.00 \pm 0.02$ & $5.39 \pm 0.78$ & $-3.84 \pm -3.57$ & $-1.55 \pm -1.29$ & $-6.60 \pm -6.26$ & $2.36 \pm 2.00$ & $4.78 \pm 0.35$\\
                G335.59-0.29 & $33.02 \pm 2.66$ & $0.07 \pm 0.16$ & $8.53 \pm 0.66$ & $-3.34 \pm -3.07$ & $-0.88 \pm -0.66$ & $-5.66 \pm -5.61$ & $3.15 \pm 2.76$ & $4.53 \pm 0.33$\\
\end{longtable}

\end{landscape}
\twocolumn

\appendix

\section{Tables}

\onecolumn
\begin{landscape}
        \begin{longtable}{lcccccccccccccc}
                
                \caption{Input data to build the SEDs.}
                \label{tab:all_targets_sed_input_summary_first_filters} \\

                \hline
                \hline
                
                Source   &   J   & H     & $K_s$   &  [3.6]  & [4.5]  & [5.8]  & [8.0]  &[24]  & 70  & 160  & 250  & 350 & 500  & 870    \\
                & (mJy) & (mJy) & (mJy)   &  (mJy)  & (mJy)  & (mJy)  & (mJy)  &(mJy) &(mJy)   &(mJy)    & (mJy)    &(mJy)    &(mJy)     &(mJy)      \\
                \hline

                \endfirsthead
                \multicolumn{15}{c}{\tablename~\thetable~ (continued)} \\
                \addlinespace

                \hline
                \hline
                
                Source   &   J   & H     & $K_s$   &  [3.6]  & [4.5]  & [5.8]  & [8.0]  &[24]  & 70  & 160  & 250  & 350 & 500  & 870    \\
                & (mJy) & (mJy) & (mJy)   &  (mJy)  & (mJy)  & (mJy)  & (mJy)  &(mJy) &(mJy)   &(mJy)    & (mJy)    &(mJy)    &(mJy)     &(mJy)      \\
                \hline

                \endhead
                \midrule
                \addlinespace[-2pt]
                \multicolumn{15}{r@{}}{\footnotesize (To be continued)}
                \endfoot
                \bottomrule
                \endlastfoot
                MG002.5577-00.7510 &  &  &  & 3.73 & 7.91 & 10.15 & 13.15 & 88.48 & 931.42 & 4905.53 &  &  &  & 1220\\
                MG003.5016-00.2020 &  &  &  &  &  &  &  & 53.51 &  &  &  &  &  & 2380\\
                MG006.2185-00.5837 &  &  & 7.18 & 121.61 & 224.36 & 367.37 & 551.90 & 3801.94 & 29697.80 & 28396.90 & 26346 &  &  & 4330\\
                MG006.9222-00.2512 &  &  &  & 20.80 & 90.98 & 232.43 & 329.81 & 882.17 & 12654.10 & 38301.50 & 53000 & 45976 &  & 8180\\
                MG300.3241-00.1985 &  & 0.86 & 7.18 & 53.97 & 123.87 & 189.28 & 268.57 & 712.49 & 4438.84 & 11381.30 & 22095 & 11587 &  & 2280\\
                MG303.9304-00.6879 &  &  &  &  & 23.19 & 25.32 & 33.84 & 2739.57 & 138174 & 185408 & 128313 & 85812 &  & 11300\\
                MG305.5101+00.3661 &  &  & 1.82 & 9.52 & 22.56 & 38.40 & 42.64 & 162.77 & 2620.13 & 7523.15 &  & 53309 &  & 6680\\
                MG306.1352+00.1330 &  &  &  &  & 3.05 &  &  & 172.43 & 27512.50 & 60414.60 &  &  & 23380 & 5290\\
\hline
                
        \end{longtable}
\end{landscape}

\begin{landscape}

\small
        \begin{longtable}{lcccccccccccccccc}
                
                \caption{Results of the weighted averaging of the fit SEDs.}
                \label{tab:all_targets_sed_fit_summary} \\

                \hline
                \hline
                
                SOURCE & $\chi^2$ & $\sigma_{\chi^2}$ & Av  & $\sigma_{Av }$ & M    & $\sigma_{M}$ & $\dot{M}$            & $\sigma_{\dot{M}}$        & $M_{disk}$ & $\sigma_{M_{disk}}$ & $\dot{M}_{disk}$        & $\sigma_{\dot{M}_{disk}}$  & L    & $\sigma_{L}$ & Age     & $\sigma_{Age}$ \\
                &        &        &  mag        &             & \Msun &   & $\log(\Msun yr^-1)$ &  & $\log(\Msun)$  &       & $\log(\Msun yr^-1)$ &  & $\log(\Lsun)$ &   & $\log(yr)$ &  \\
                
                \hline
                \endfirsthead
                \multicolumn{17}{c}{\tablename~\thetable~ (continued)} \\
                \addlinespace
                \hline
                \hline

                SOURCE & $\chi^2$ & $\sigma_{\chi^2}$ & Av  & $\sigma_{Av }$ & M    & $\sigma_{M}$ & $\dot{M}$            & $\sigma_{\dot{M}}$        & $M_{disk}$ & $\sigma_{M_{disk}}$ & $\dot{M}_{disk}$        & $\sigma_{\dot{M}_{disk}}$  & L    & $\sigma_{L}$ & Age     & $\sigma_{Age}$ \\
                &        &        &  mag        &             & \Msun &   & $\log(\Msun yr^-1)$ &  & $\log(\Msun)$  &       & $\log(\Msun yr^-1)$ &  & $\log(\Lsun)$ &   & $\log(yr)$ &  \\
                
                \hline
                \endhead
                \midrule
                \addlinespace[-2pt]
                \multicolumn{17}{r@{}}{\footnotesize (To be continued)}
                \endfoot
                \bottomrule
                \endlastfoot

                MG358.0590-00.4698 & 6.98 & 3.21 & 12.18 & 16.83 & 2.14 & 1.83 & -5.40 & -5.67 & -1.58 & -1.58 & -5.63 & -5.35 & 1.91 & 2.13 & 4.62 & 0.95\\
                
                MG356.5212+00.2141 & 10.40 & 2.95 & 42.94 & 11.85 & 3.20 & 1.69 & -6.83 & -6.41 & -3.54 & -2.97 & -9.32 & -8.75 & 2.31 & 2.35 & 5.98 & 0.27\\
                
                MG354.4384+00.4185 & 7.45 & 3.72 & 8.38 & 9.13 & 7.87 & 0.91 & -3.96 & -4.24 & -1.00 & -0.74 & -5.30 & -4.86 & 2.93 & 2.58 & 4.31 & 0.40\\
                
                MG352.6040-00.2253 & 8.75 & 4.29 & 39.18 & 8.79 & 7.60 & 0.75 & -3.55 & -3.70 & -1.02 & -0.76 & -5.57 & -5.26 & 2.91 & 2.43 & 4.53 & 0.45\\
                
                MG352.2452-00.0636 & 10.11 & 2.99 & 19.40 & 18.02 & 5.74 & 1.85 & -3.93 & -3.77 & -1.08 & -0.92 & - & - & 2.87 & 2.84 & 4.27 & 0.78\\
                
                MG351.8098+00.6433 & 10.70 & 3.01 & 47.02 & 6.49 & 1.89 & 1.11 & -7.23 & -6.61 & -1.76 & -1.87 & -6.88 & -6.73 & 1.61 & 1.84 & 6.45 & 0.43\\
                
                MG347.7561+00.2323 & 8.34 & 3.14 & 44.67 & 8.74 & 6.53 & 0.76 & -4.78 & -3.77 & -1.48 & -1.46 & -6.15 & -5.79 & 3.15 & 2.76 & 6.31 & 0.38\\
                
                MG345.5764-00.2252 & 12.24 & 2.98 & 38.91 & 7.03 & 4.40 & 0.53 & -5.76 & -4.63 & -1.40 & -1.42 & -6.24 & -5.62 & 2.53 & 2.06 & 6.42 & 0.36\\
\hline
                
        \end{longtable}
\end{landscape}

\twocolumn
\section{Figures}
\onecolumn

\begin{figure}
        \begin{tabular}{cc}
                \includegraphics[width=0.48\textwidth]{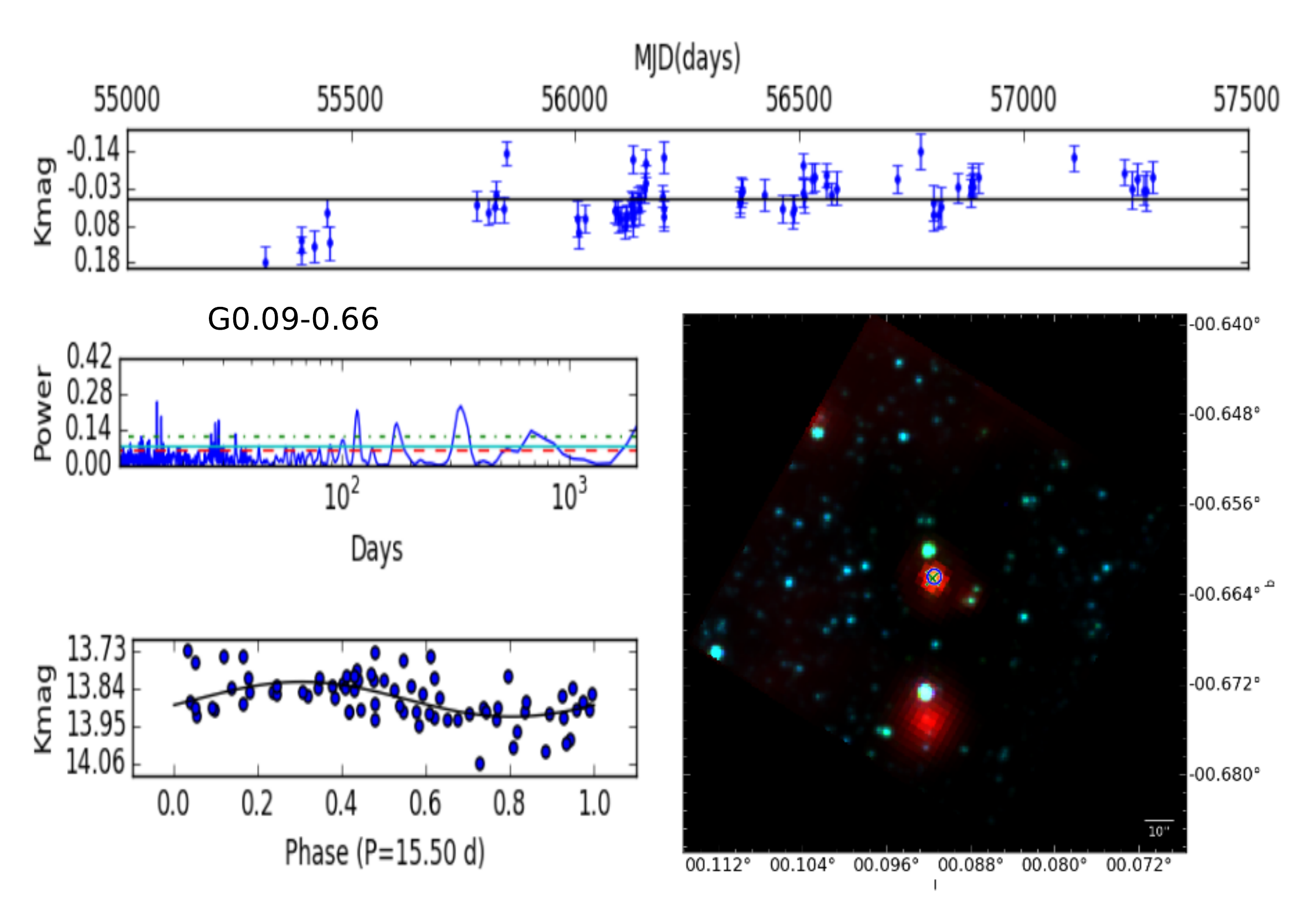}&
                \includegraphics[width=0.48\textwidth]{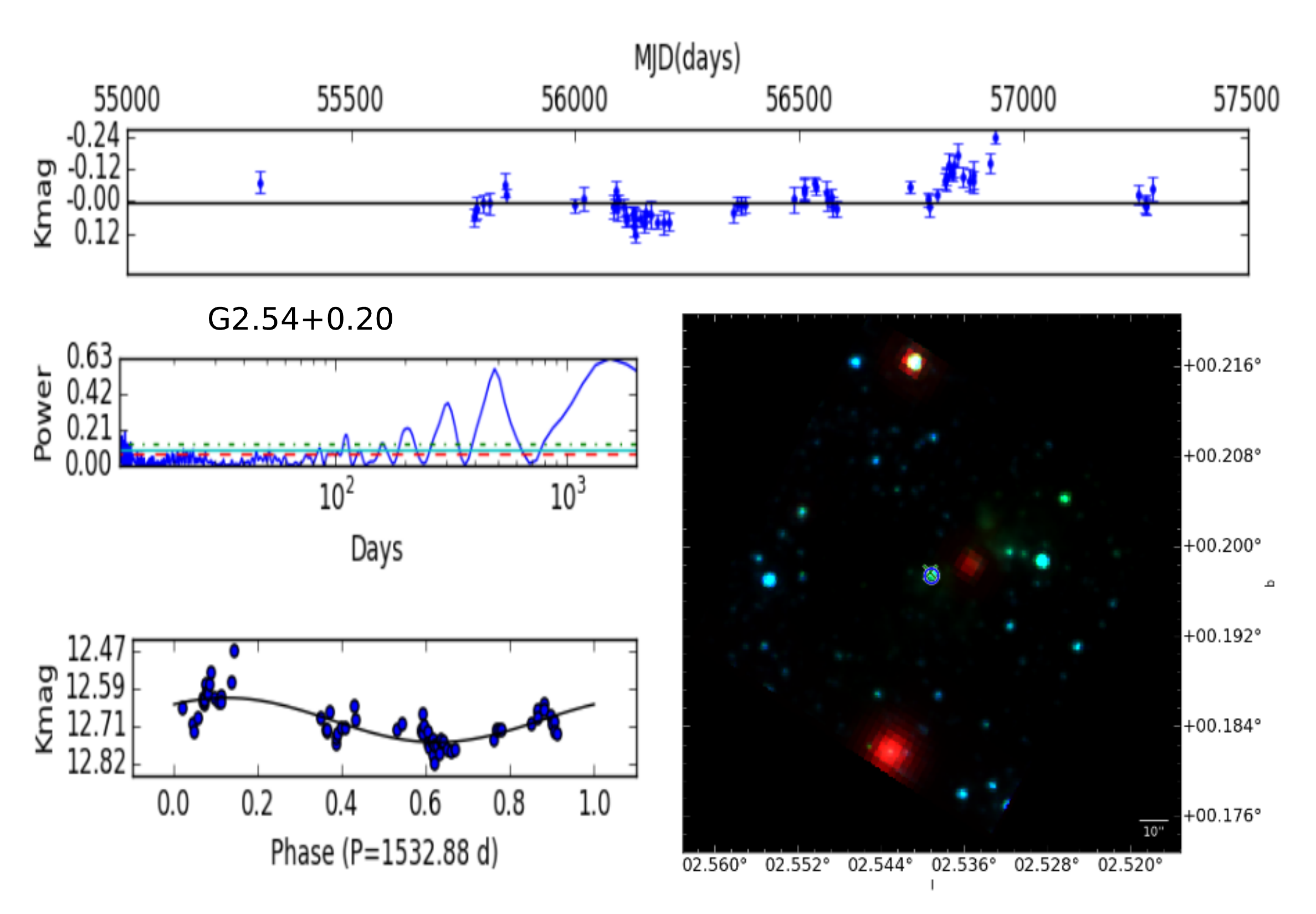}\\
                \includegraphics[width=0.48\textwidth]{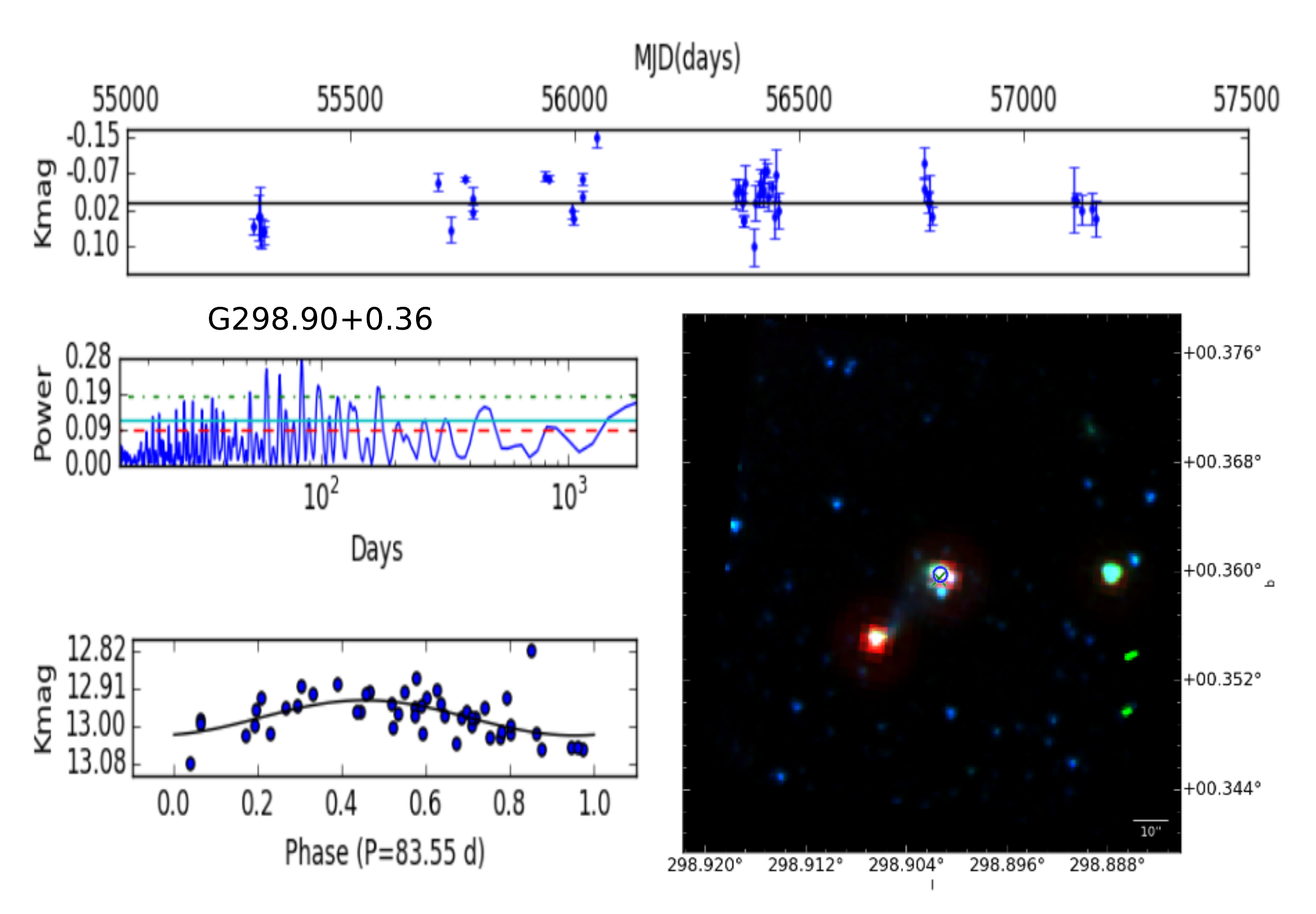}&
                \includegraphics[width=0.48\textwidth]{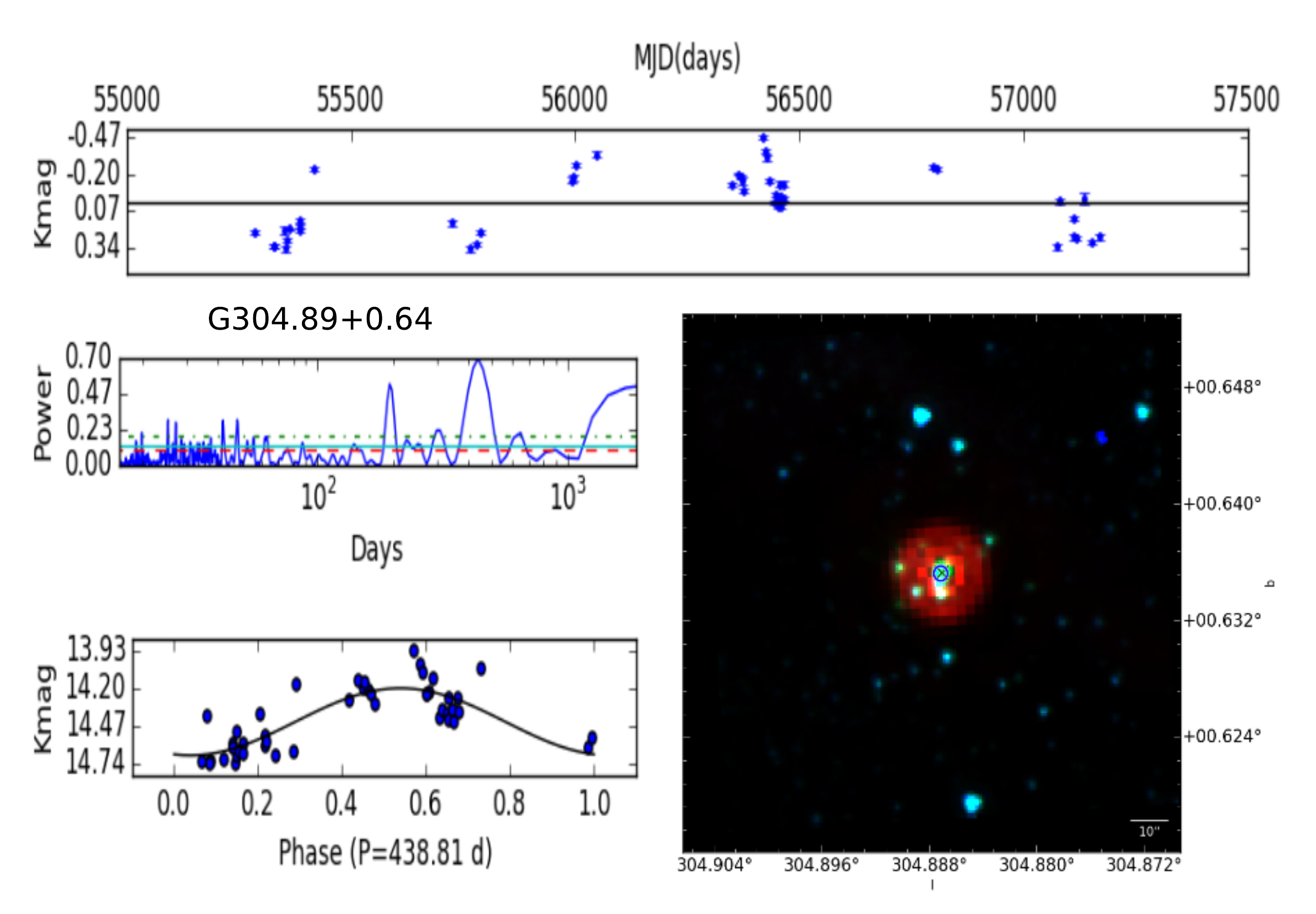}\\
                \includegraphics[width=0.48\textwidth]{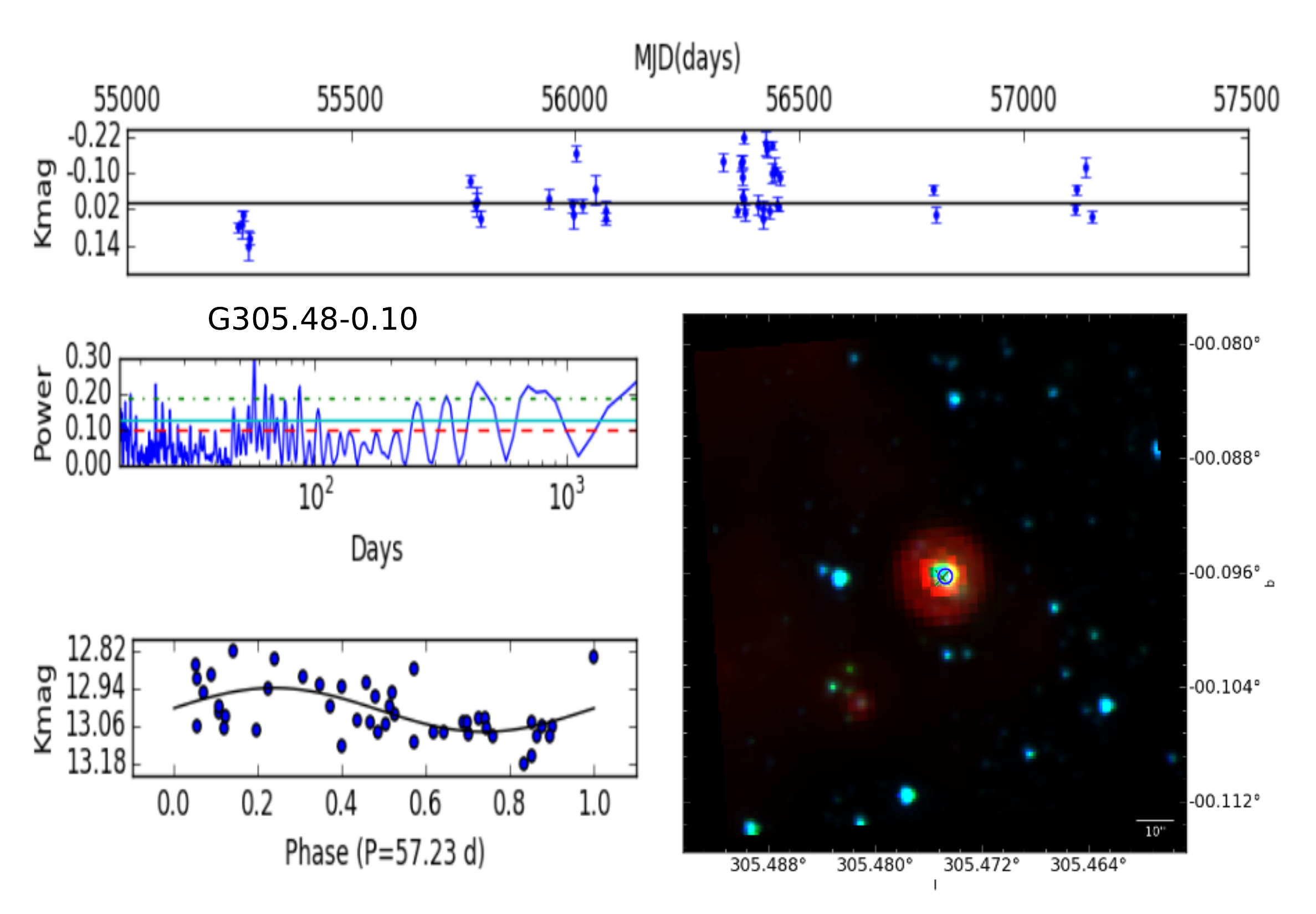}&
                \includegraphics[width=0.48\textwidth]{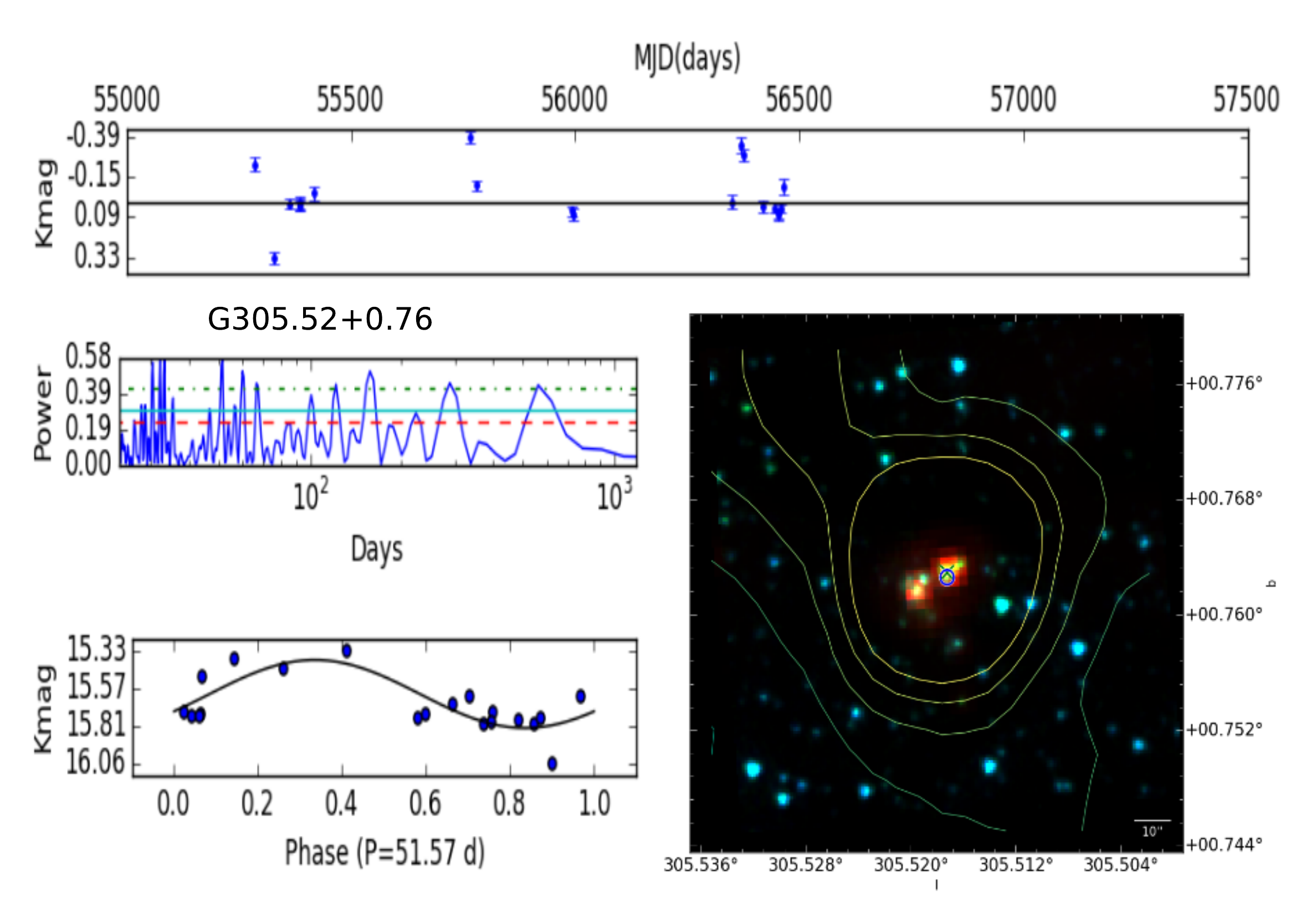}\\
                \includegraphics[width=0.48\textwidth]{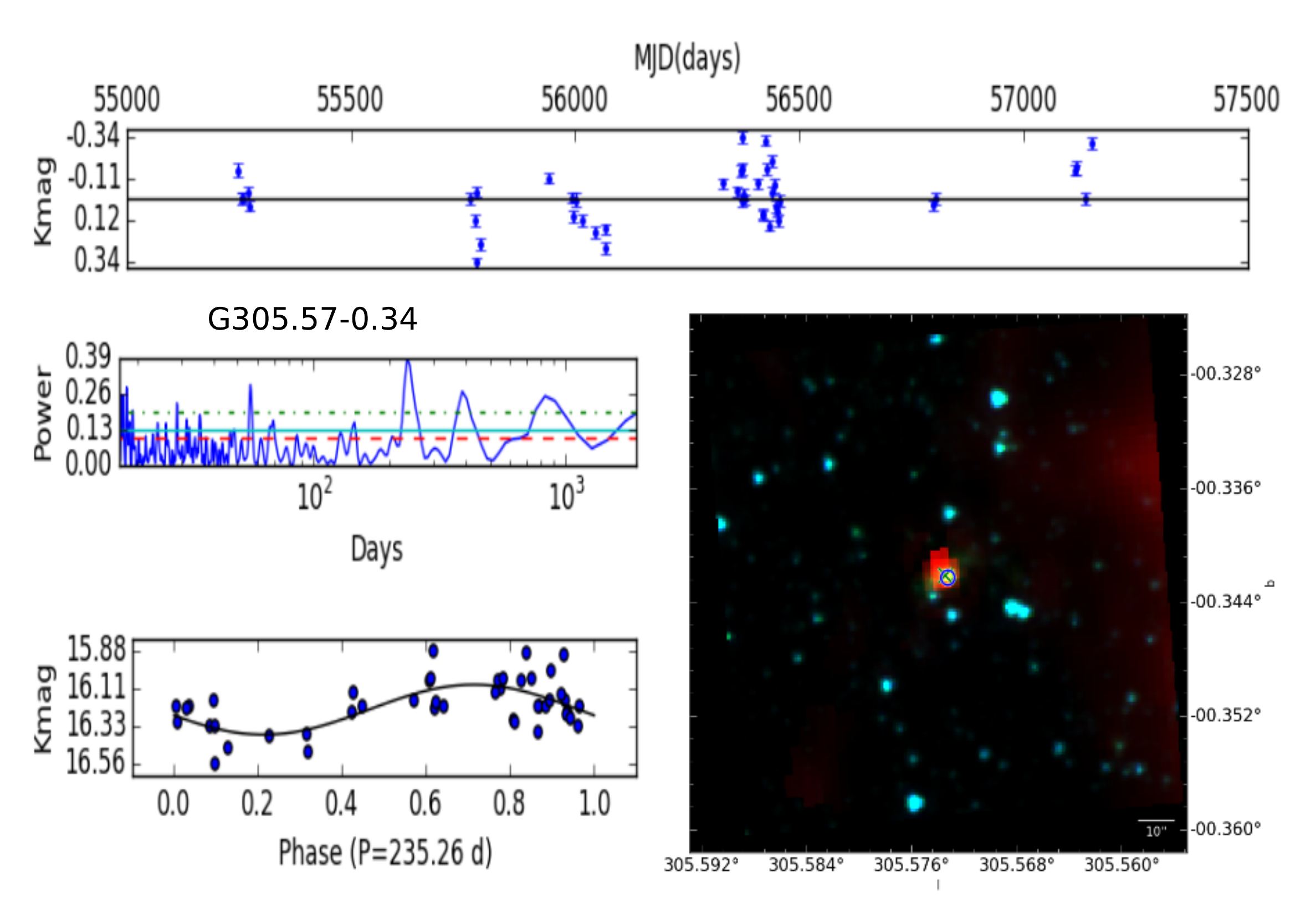}&
                \includegraphics[width=0.48\textwidth]{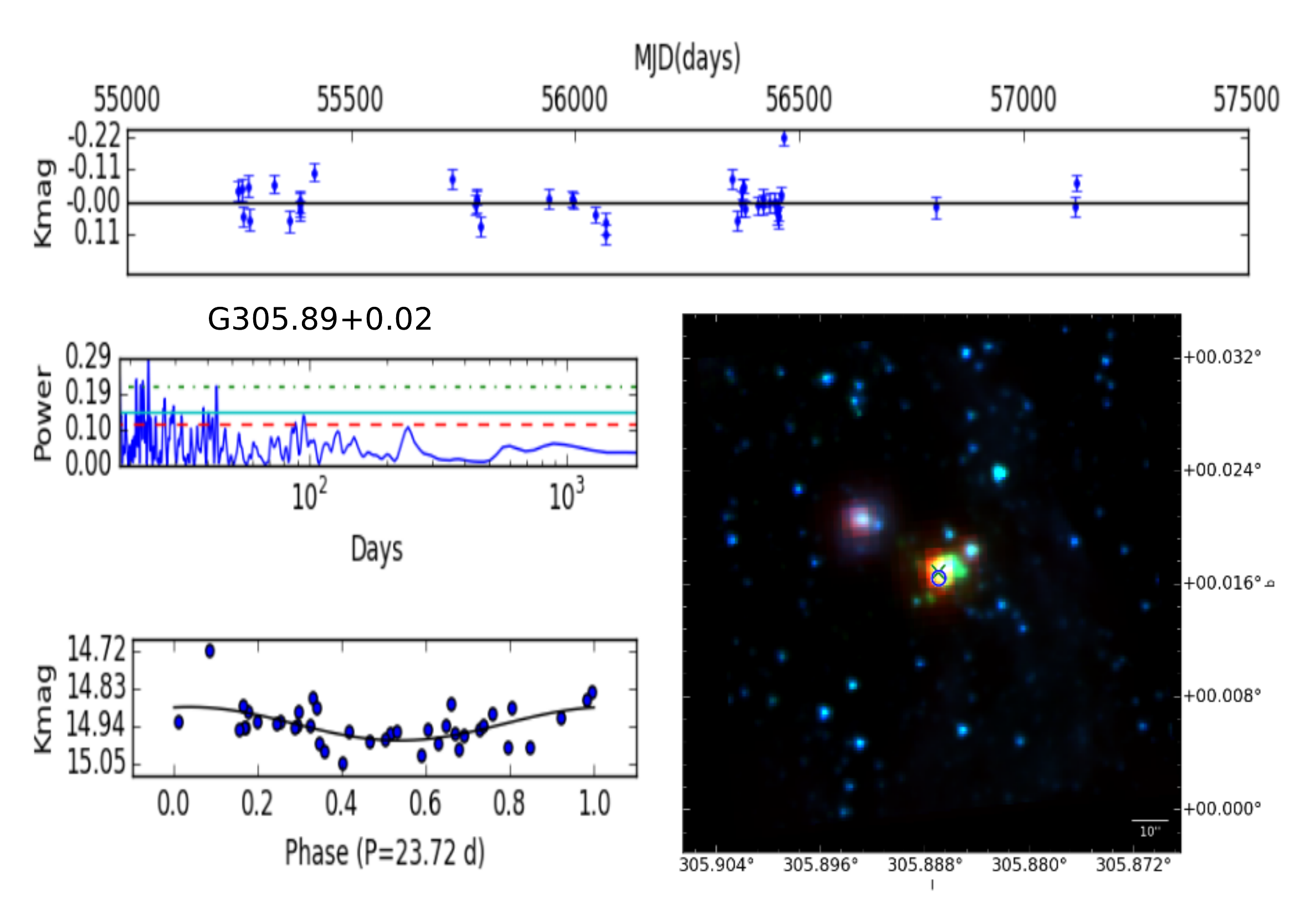}\\

        \end{tabular}
        \caption{LC of the source, with error bars representing
                MAD($\Delta S_{i_{mjd}}$), periodograms (also plotted are the $99\%$, $95\%$, and $90\%$ false probability levels,
                respectively: the green dot-dashed line, the cyan full line, and the red dashed line), the phase-folded
                LC using the best period fitted, the 
                RGB image of the source using the Spitzer IRAC 3.6 \mum, IRAC 4.0 \mum, and the 24\mum MIPS band
                as blue, green and red, respectively. The VVV source is marked by the blue circle and the green cross represents the MIPS co-ordinates. The contours of the RGB are in the interval of [Peak-$5\sigma$, Peak] from the ATLASGAL observation at $850$ \mum.}
        \label{fig:testgrid}
\end{figure}

\begin{figure}
\begin{tabular}{cc}
                \includegraphics[width=0.48\textwidth]{{./G309.91+0.32_combined}.pdf}&
                \includegraphics[width=0.48\textwidth]{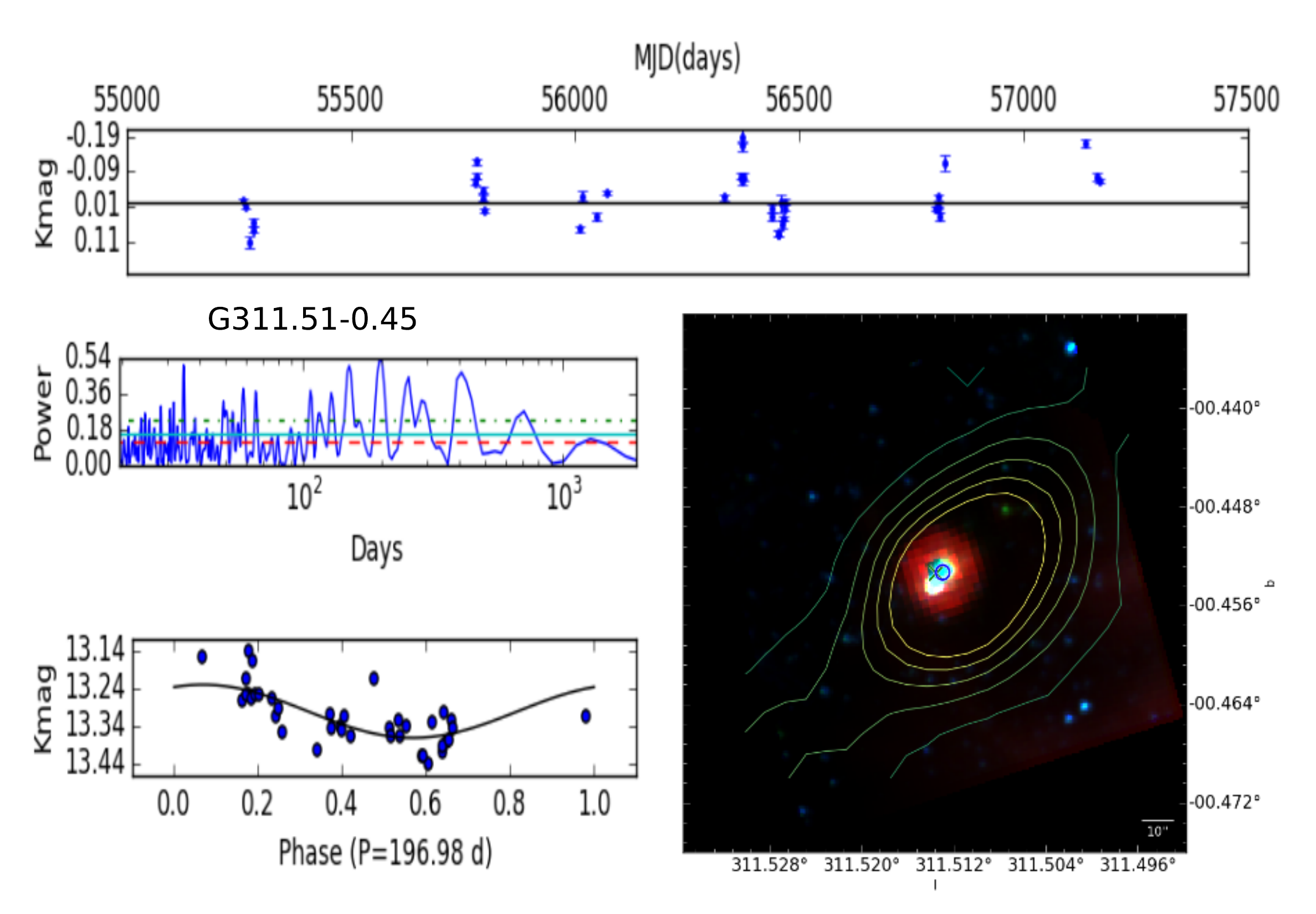}\\
                \includegraphics[width=0.48\textwidth]{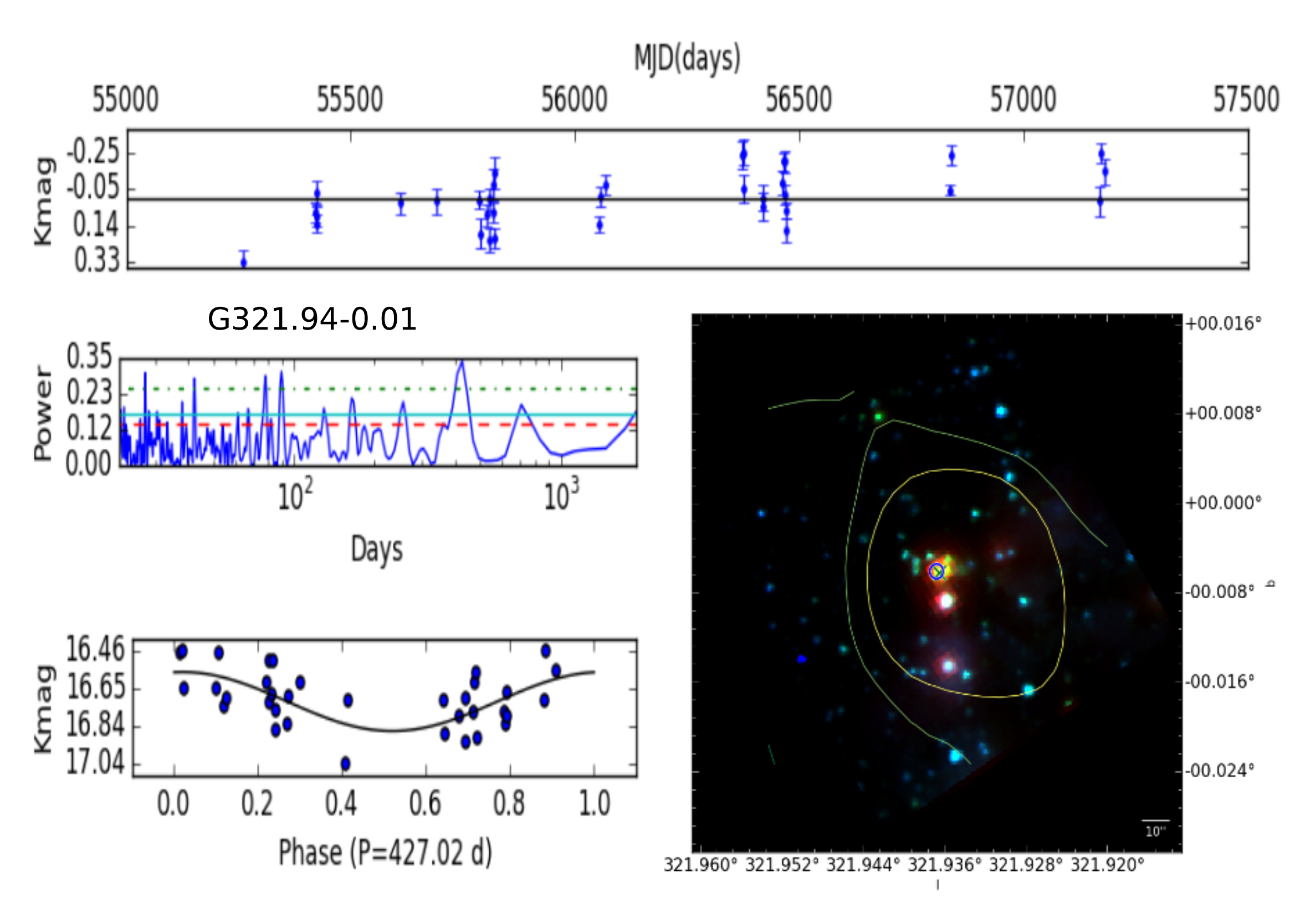}&
                \includegraphics[width=0.48\textwidth]{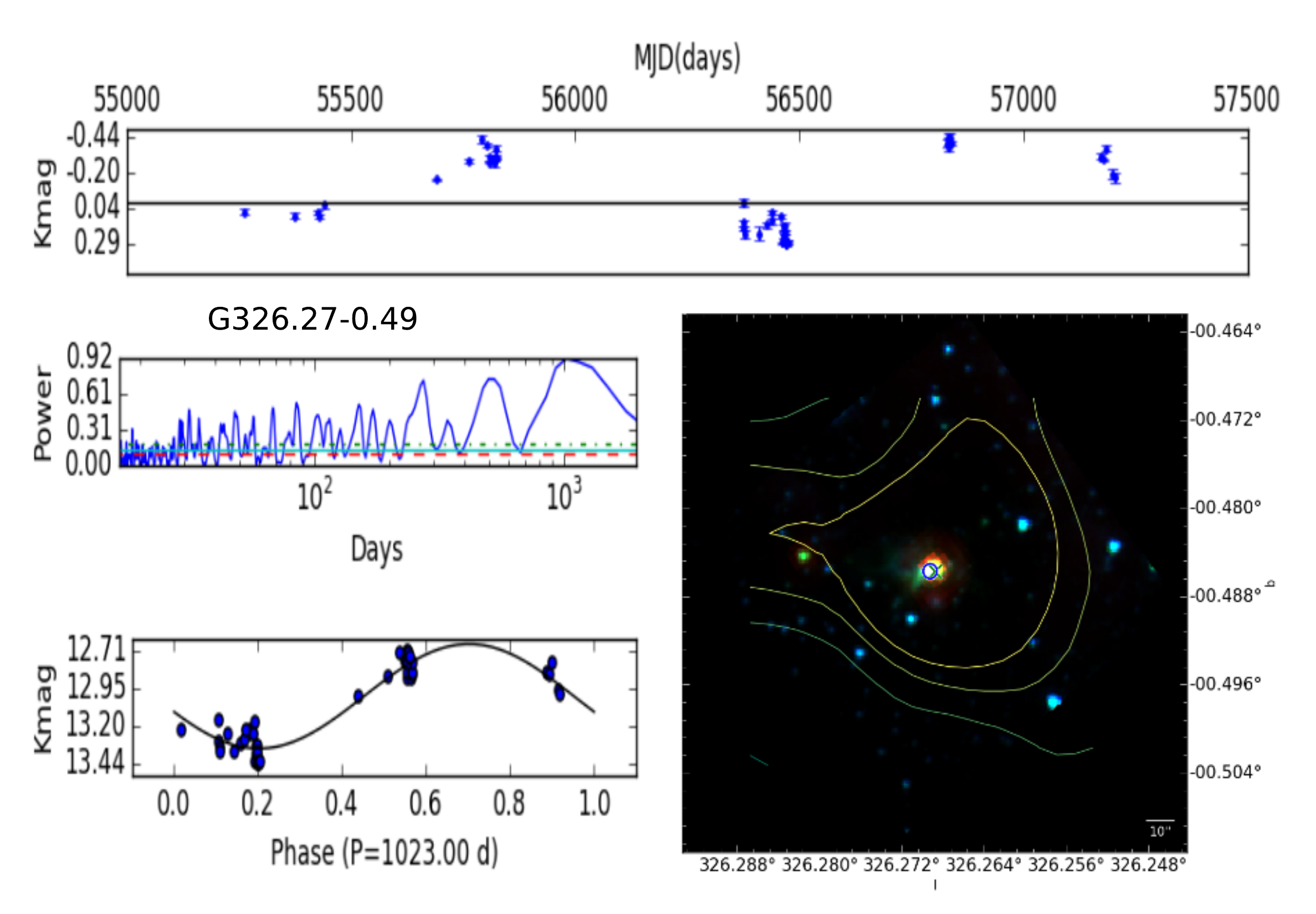}\\
                \includegraphics[width=0.48\textwidth]{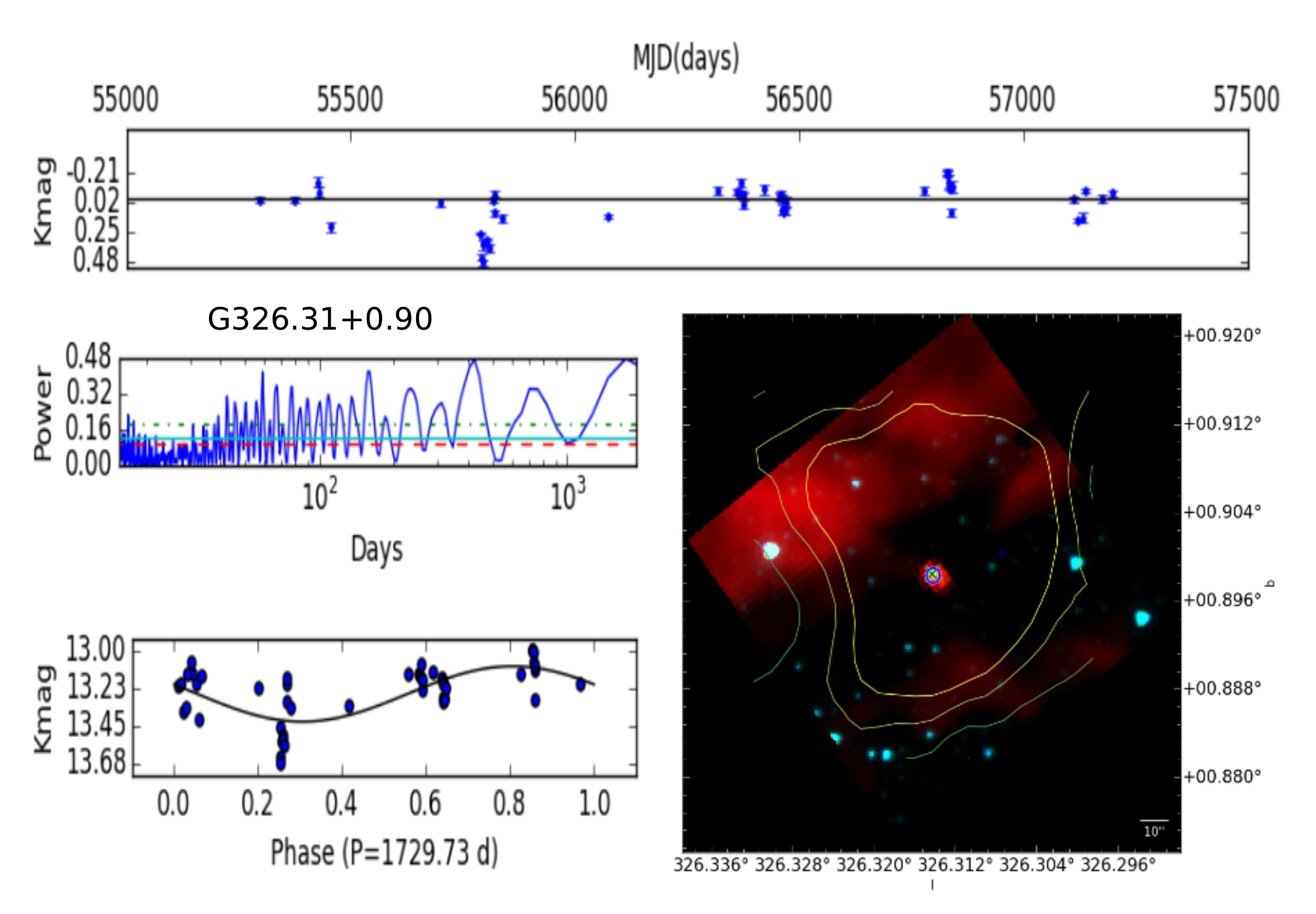}&
                \includegraphics[width=0.48\textwidth]{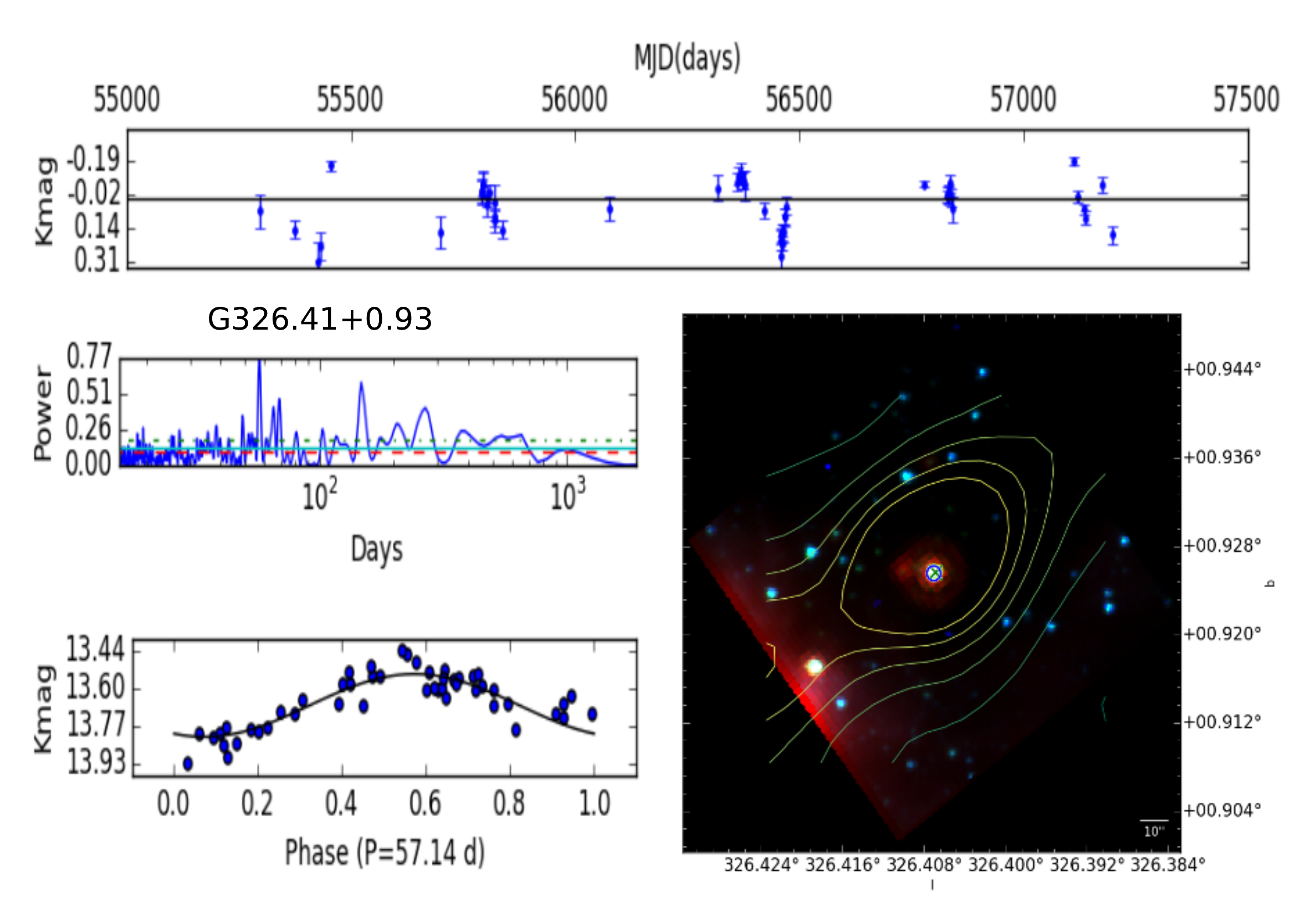}\\
                \includegraphics[width=0.48\textwidth]{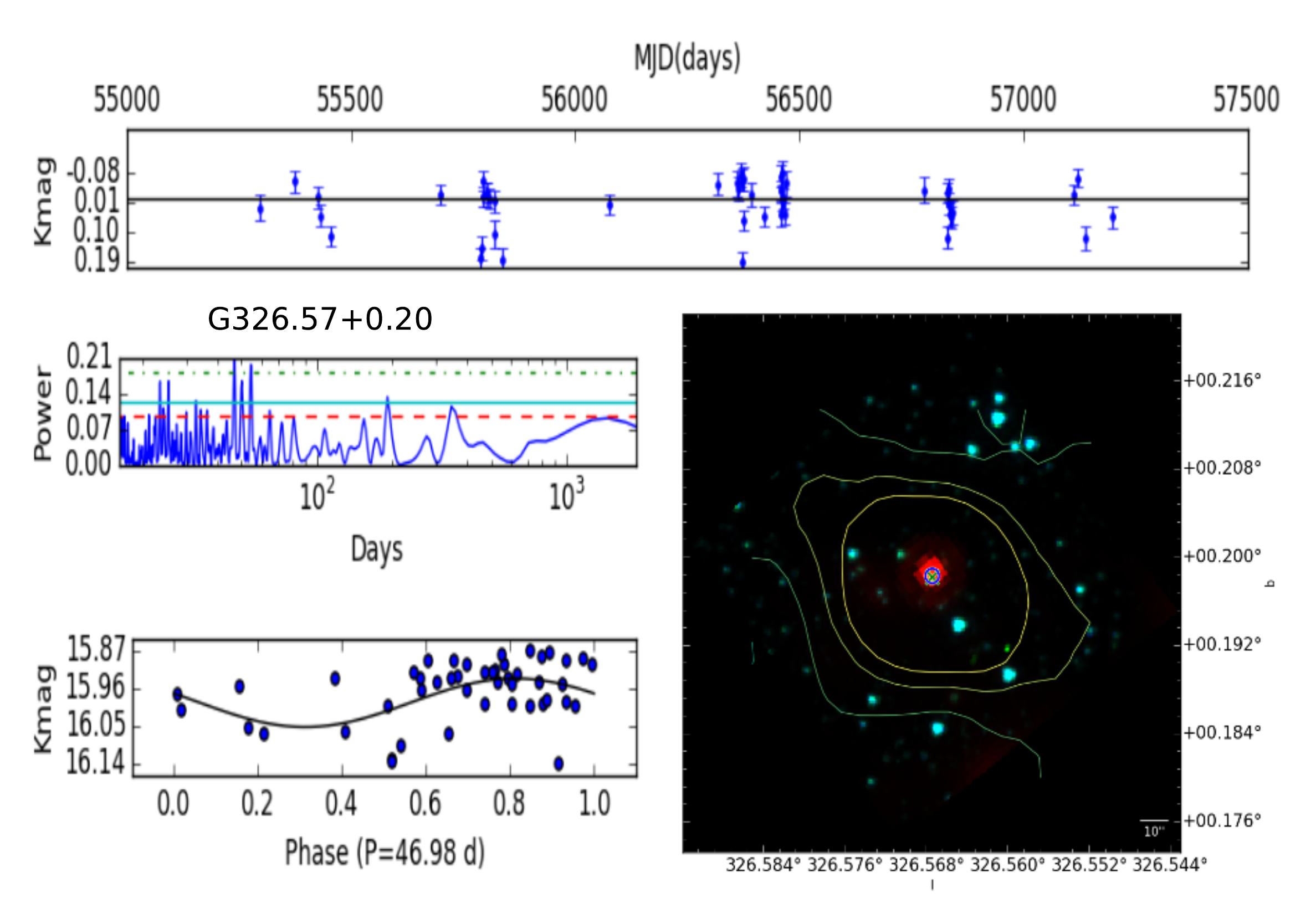}&
                \includegraphics[width=0.48\textwidth]{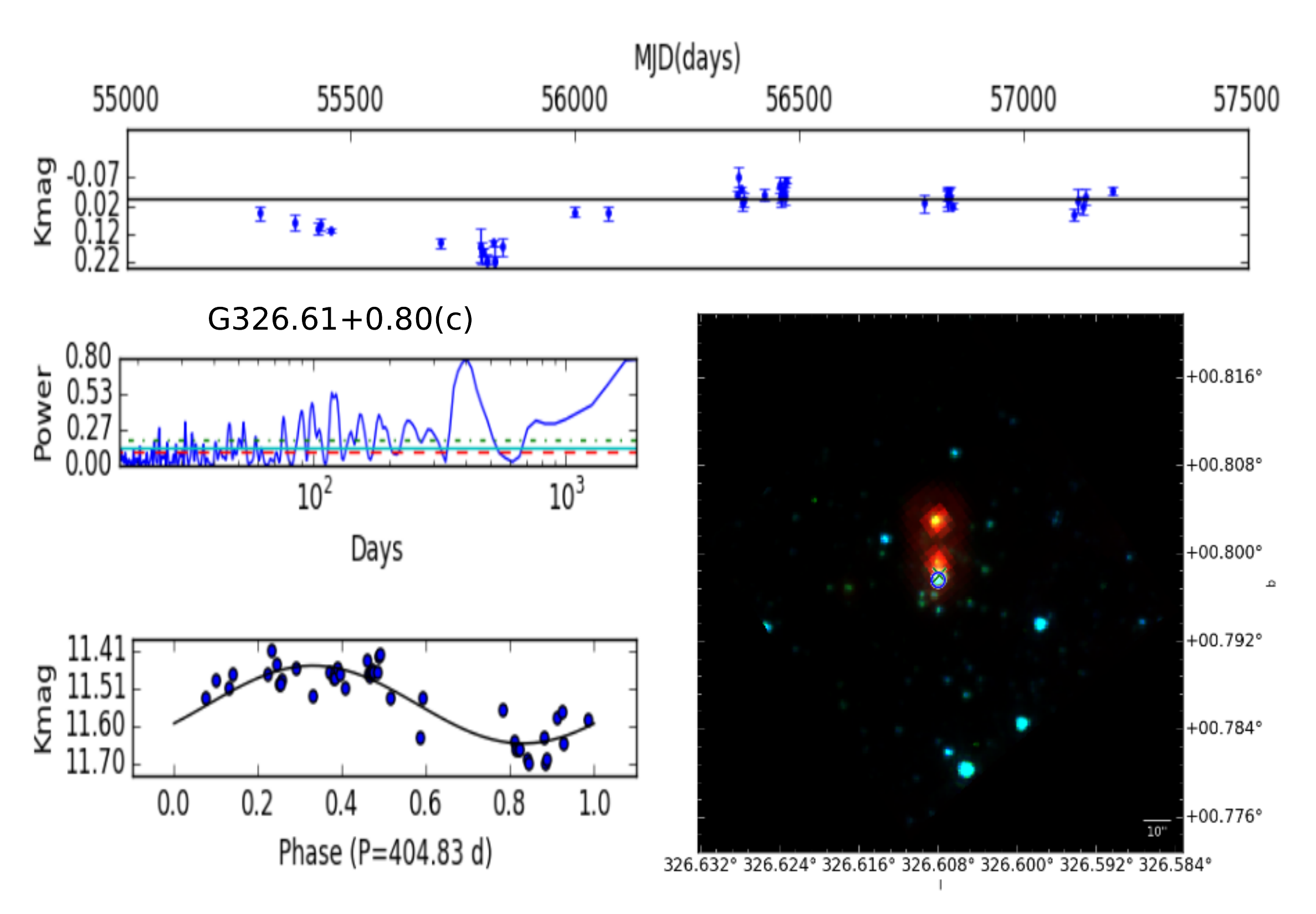}\\
        \end{tabular}
        \caption{Continuation of Fig. \ref{fig:testgrid}.}
\end{figure}

\begin{figure}
        \begin{tabular}{cc}
                \includegraphics[width=0.48\textwidth]{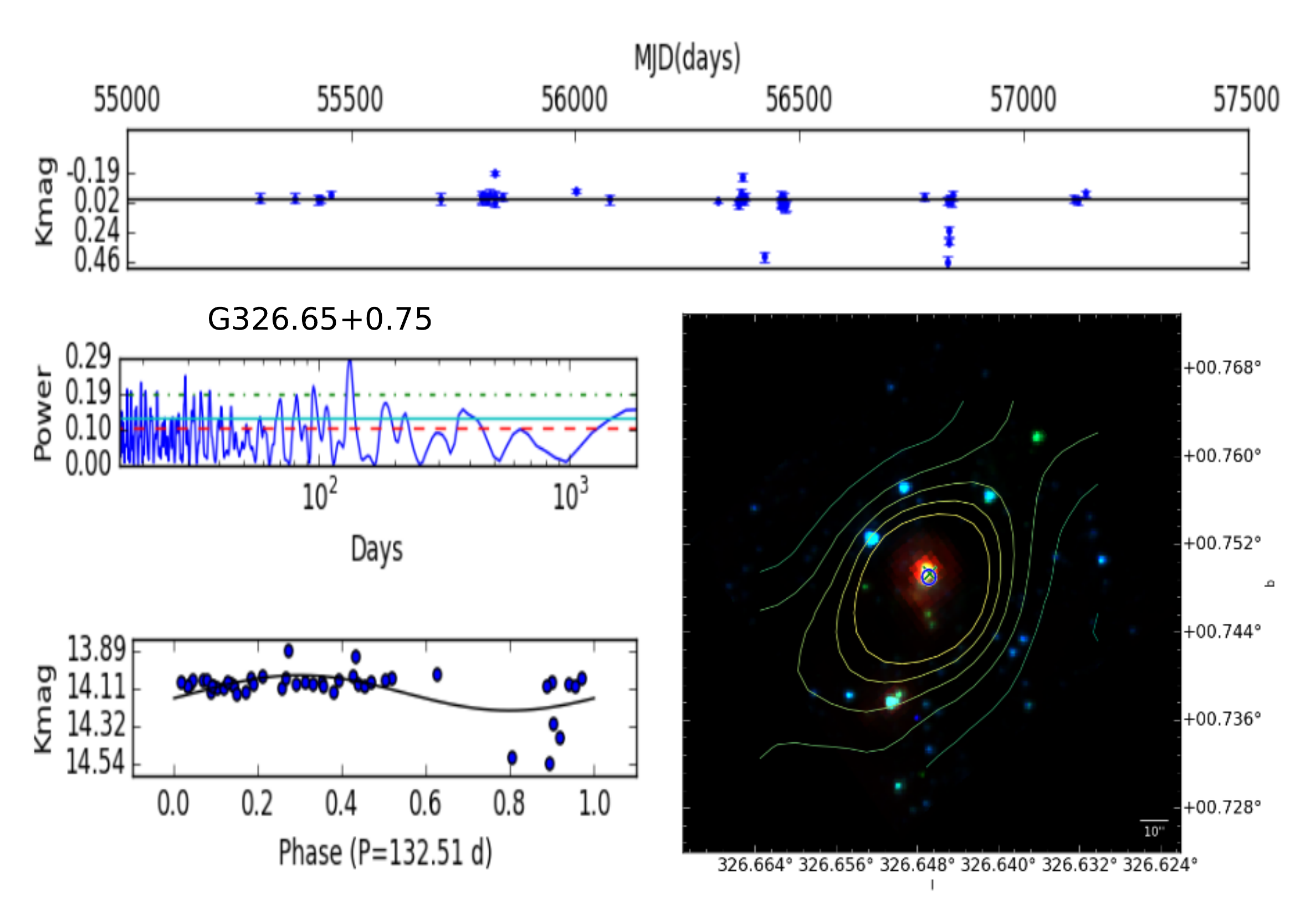}&
                \includegraphics[width=0.48\textwidth]{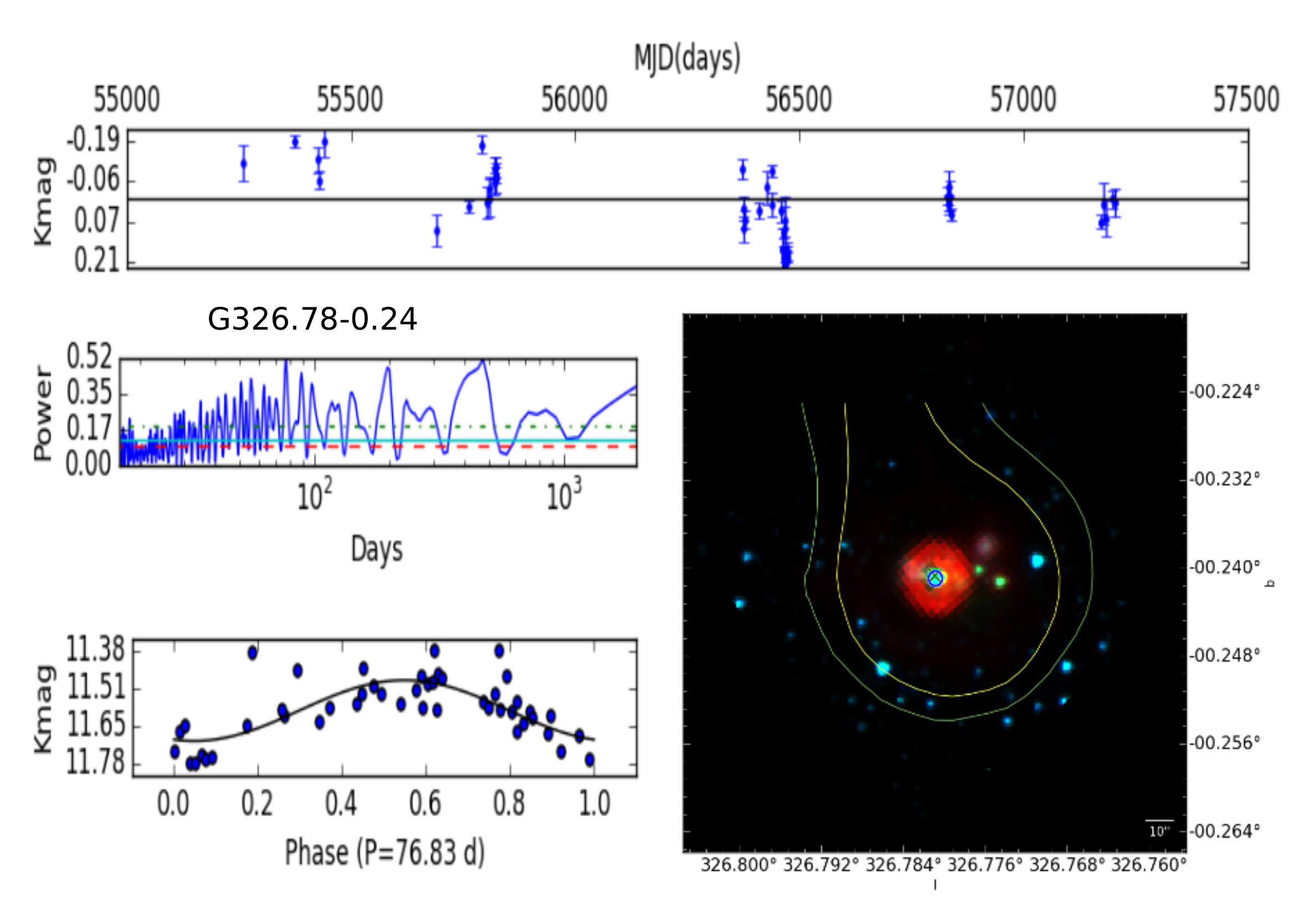}\\
                \includegraphics[width=0.48\textwidth]{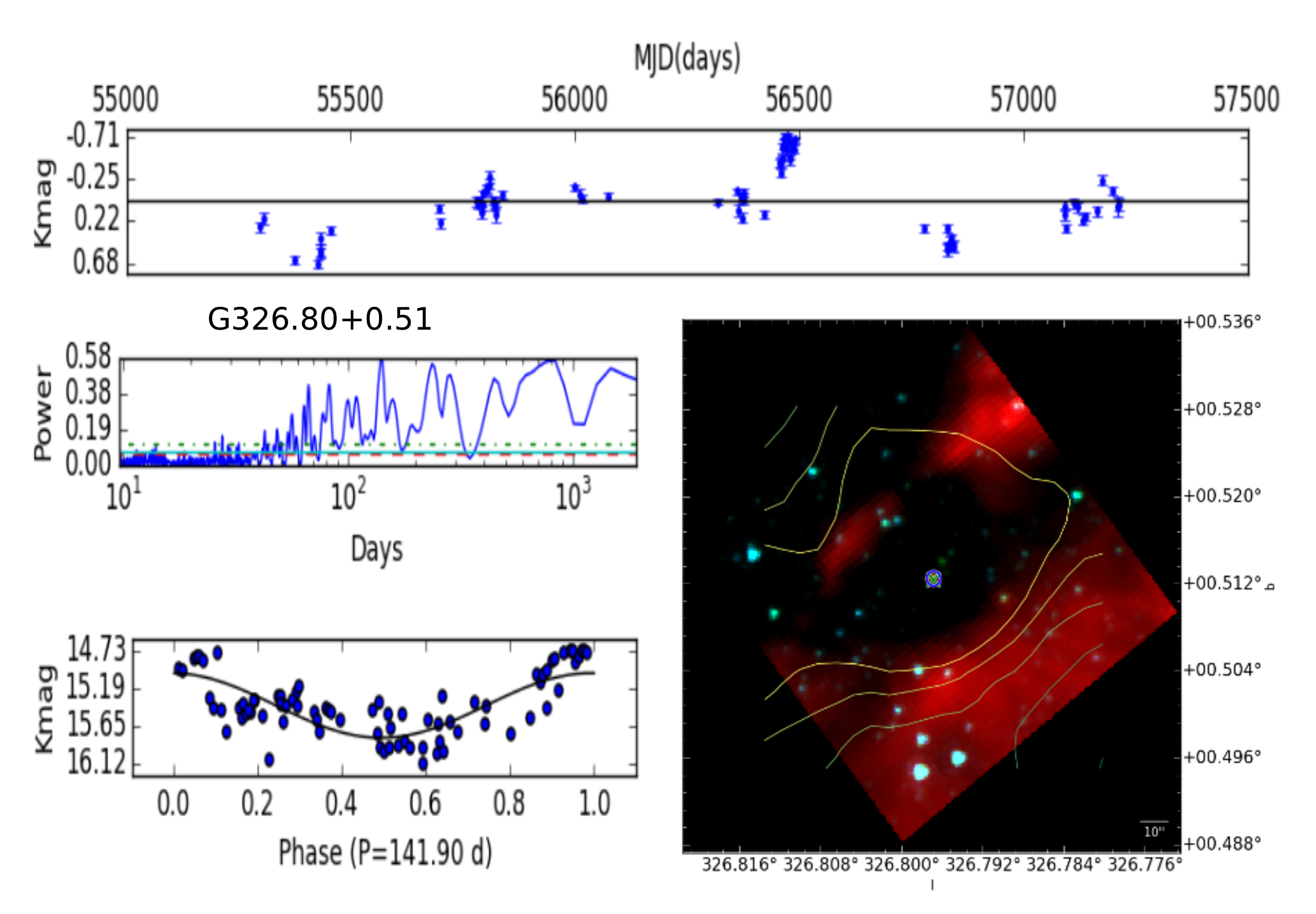}&
                \includegraphics[width=0.48\textwidth]{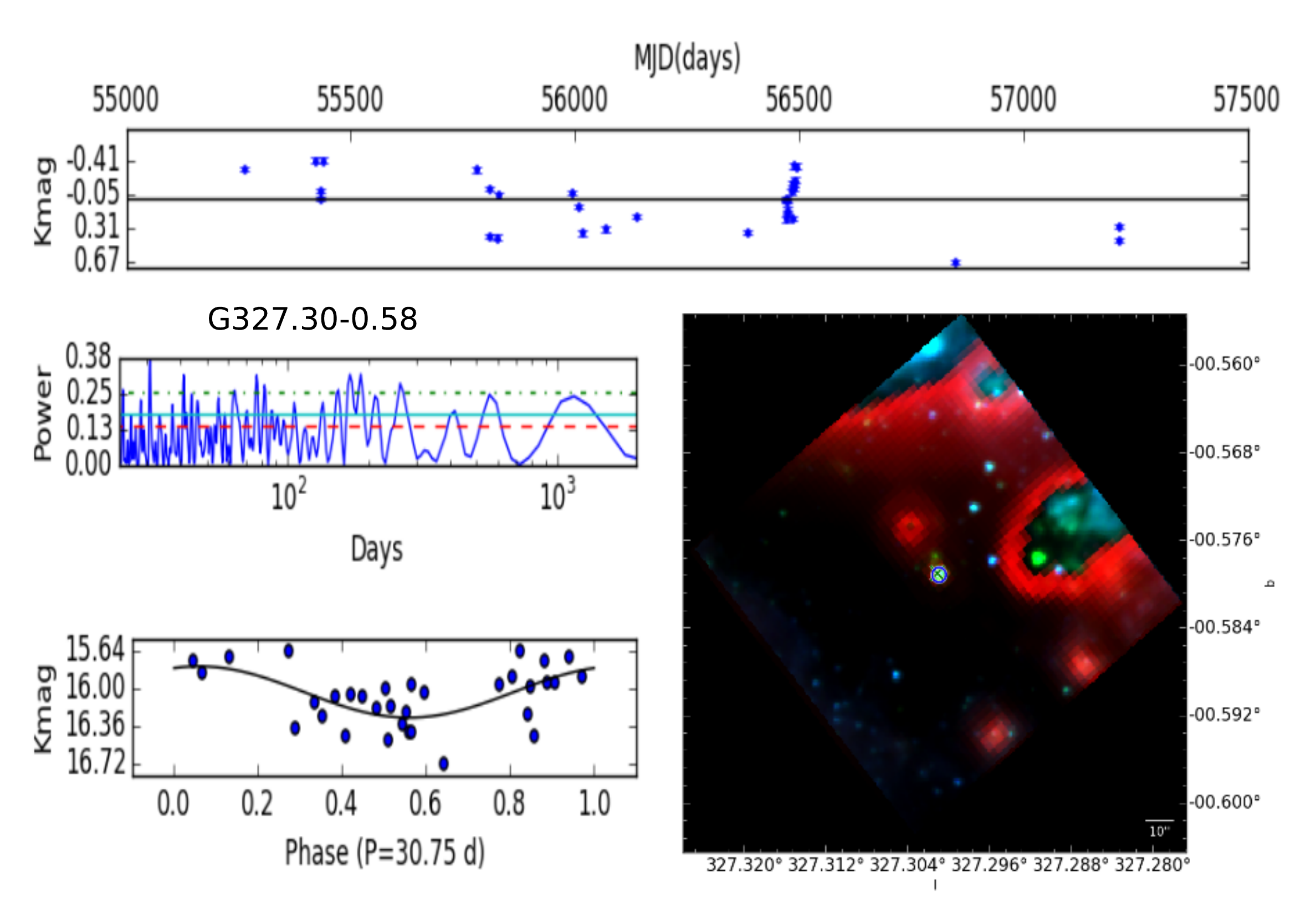}\\
                \includegraphics[width=0.48\textwidth]{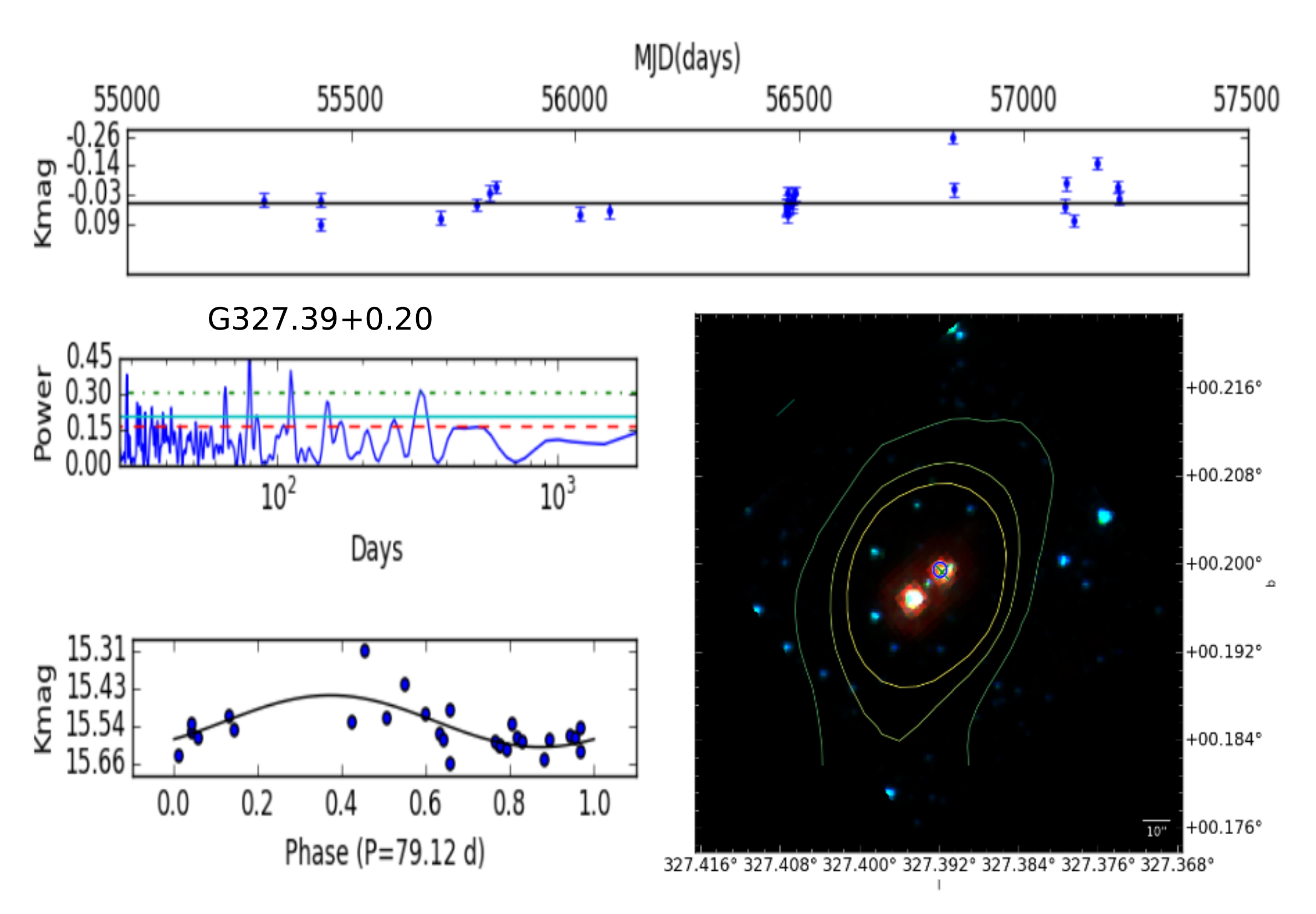}&
                \includegraphics[width=0.48\textwidth]{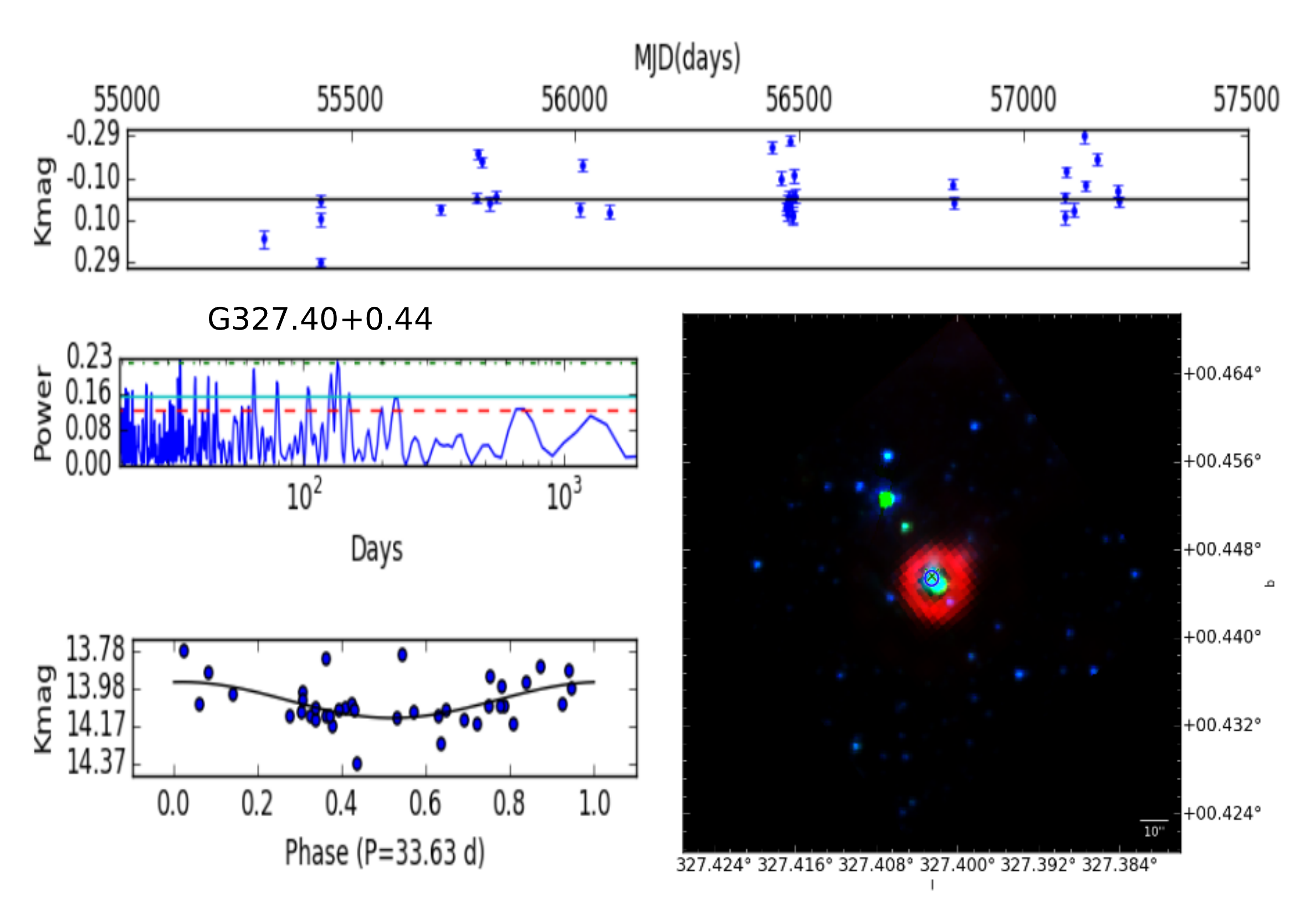}\\
                \includegraphics[width=0.48\textwidth]{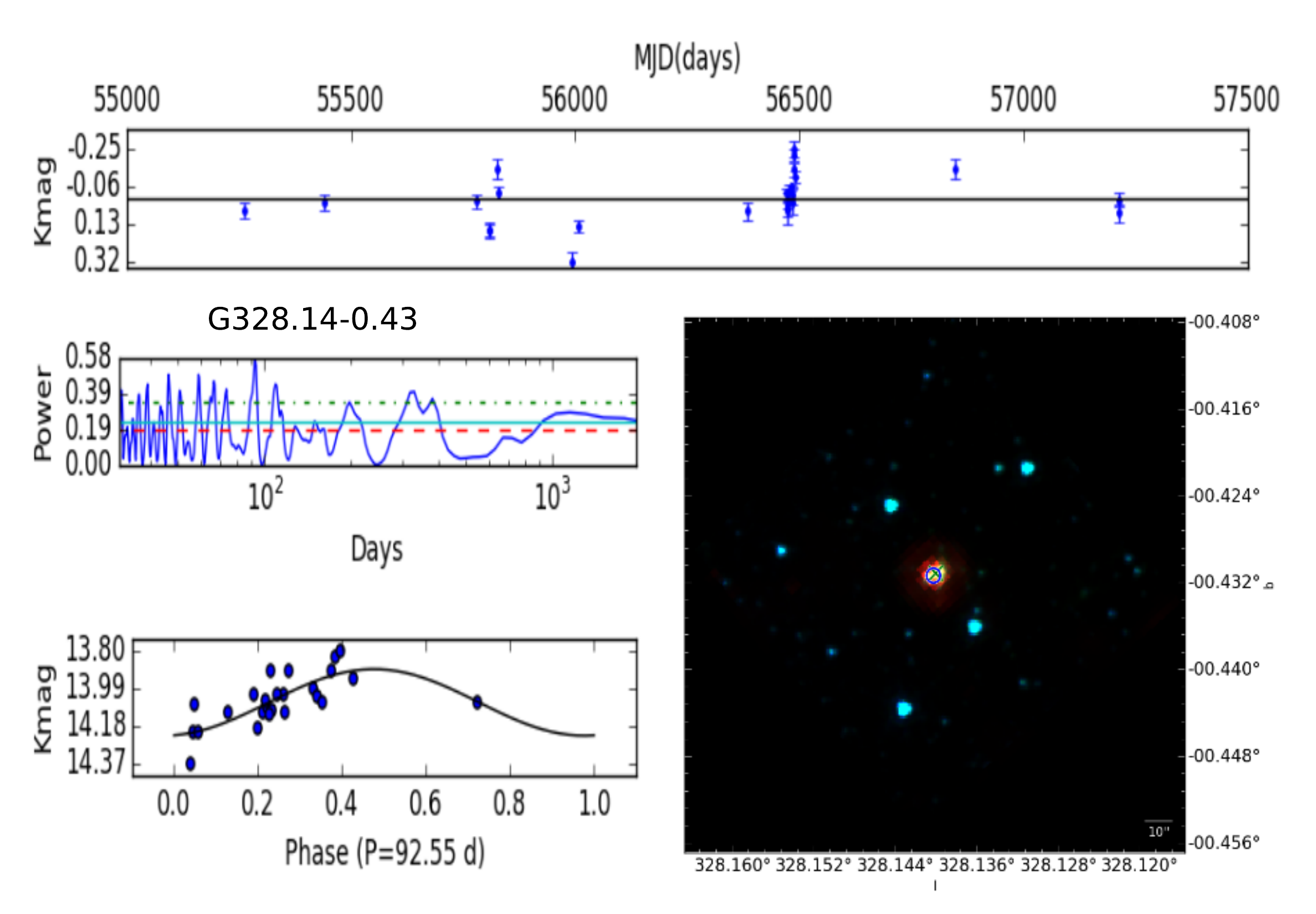}&
                \includegraphics[width=0.48\textwidth]{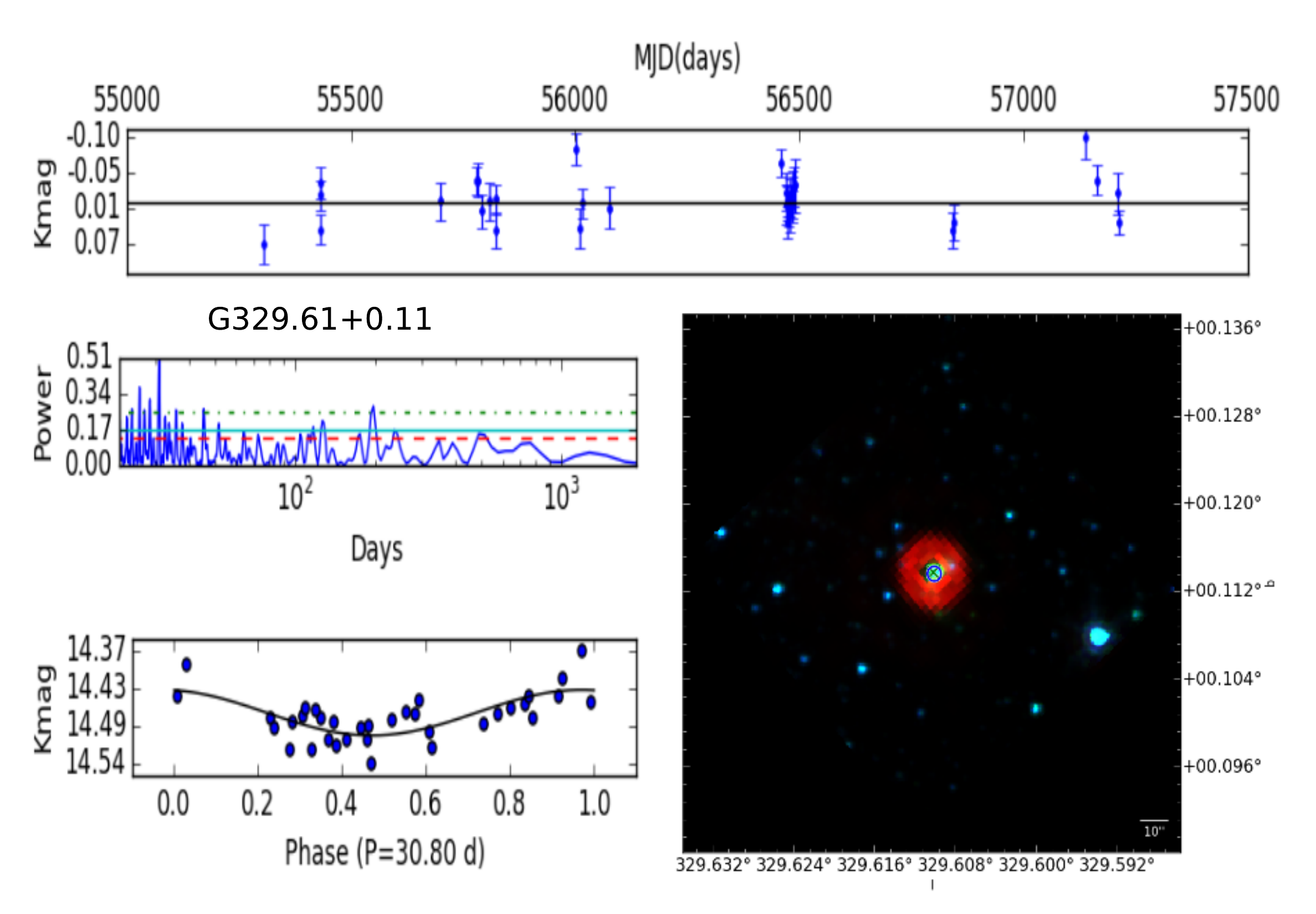}\\

        \end{tabular}
        \caption{Continuation of Fig. \ref{fig:testgrid}.}
\end{figure}

\begin{figure}
        \begin{tabular}{cc}
                \includegraphics[width=0.48\textwidth]{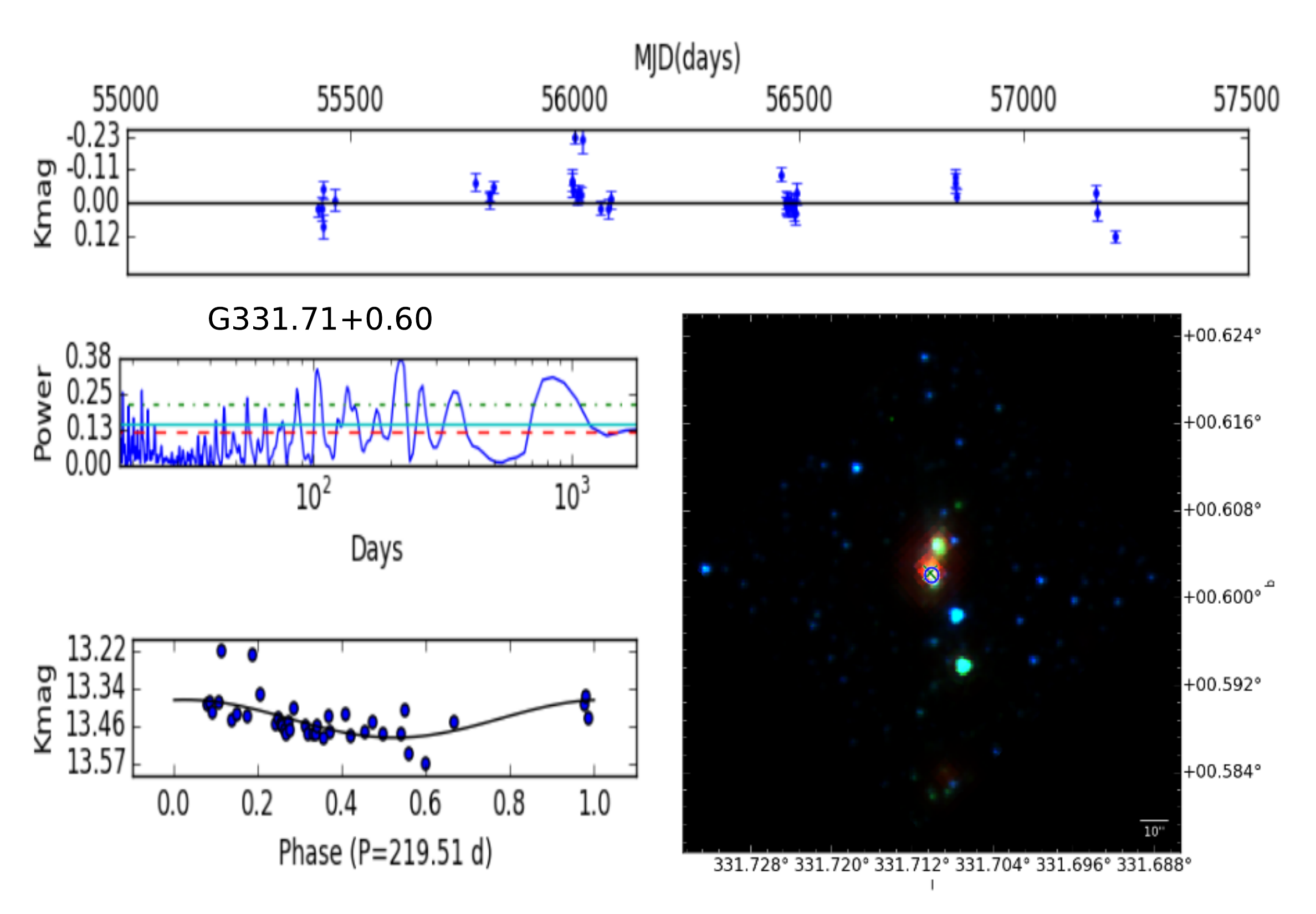}&
                \includegraphics[width=0.48\textwidth]{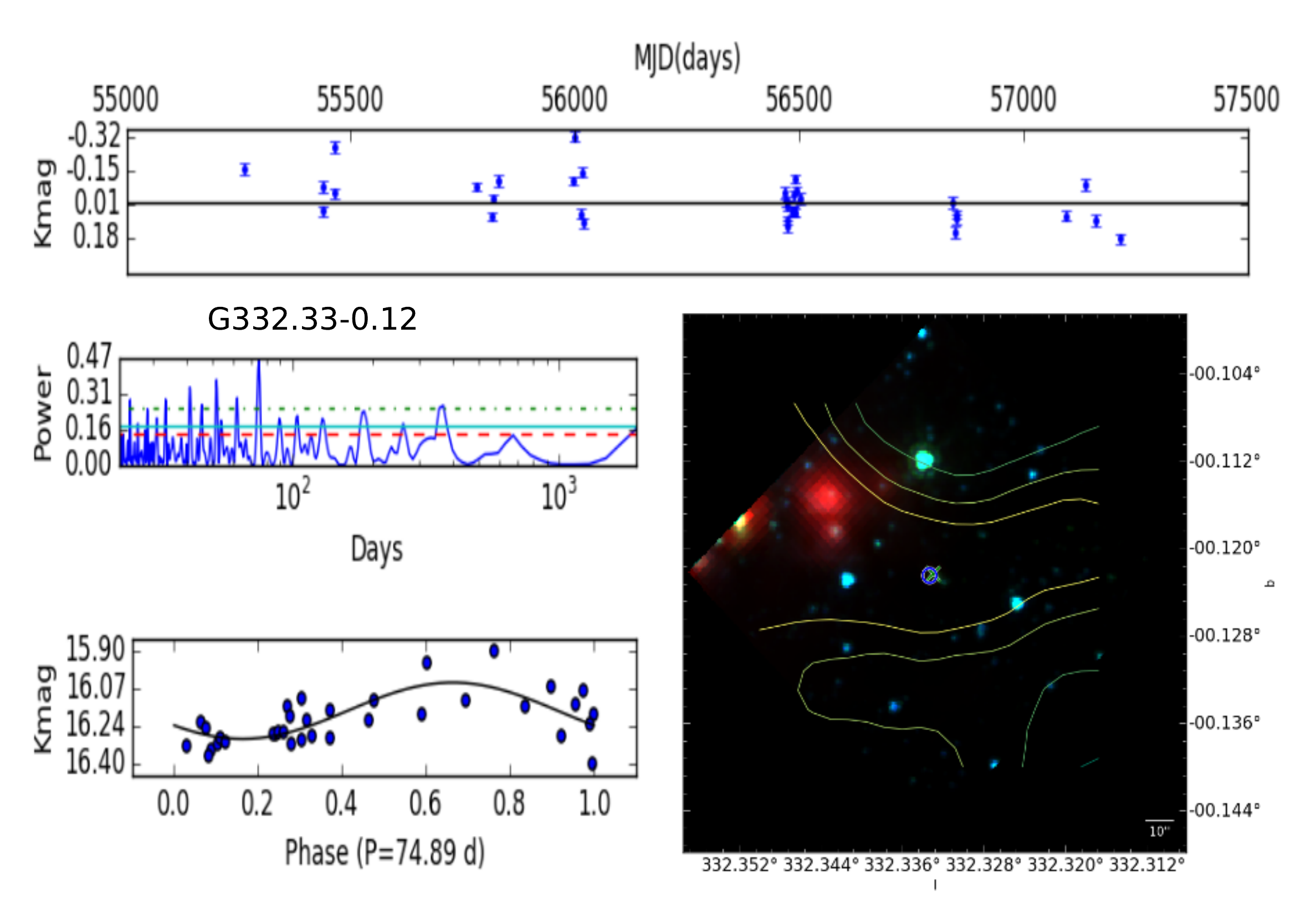}\\
                \includegraphics[width=0.48\textwidth]{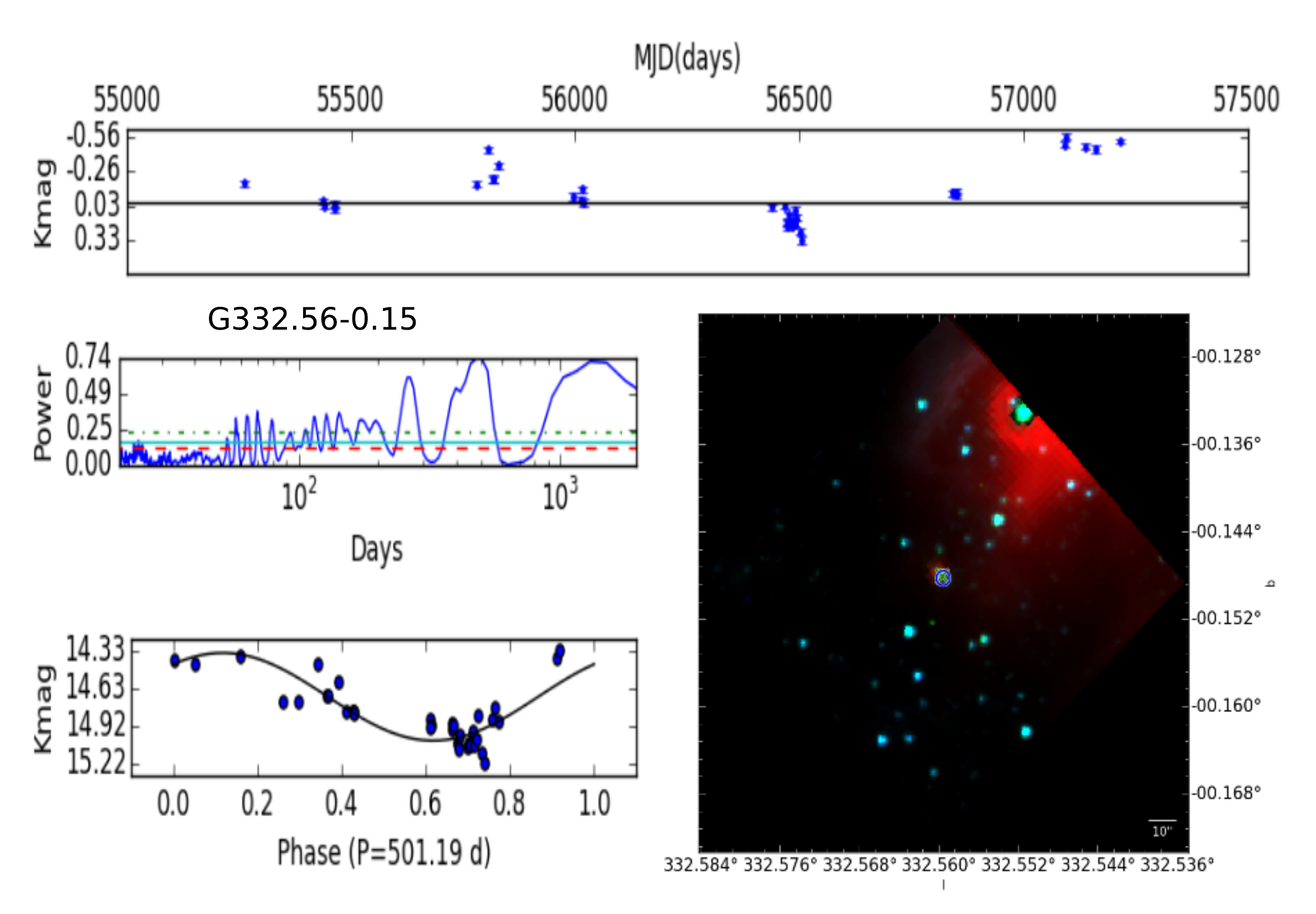}&
                \includegraphics[width=0.48\textwidth]{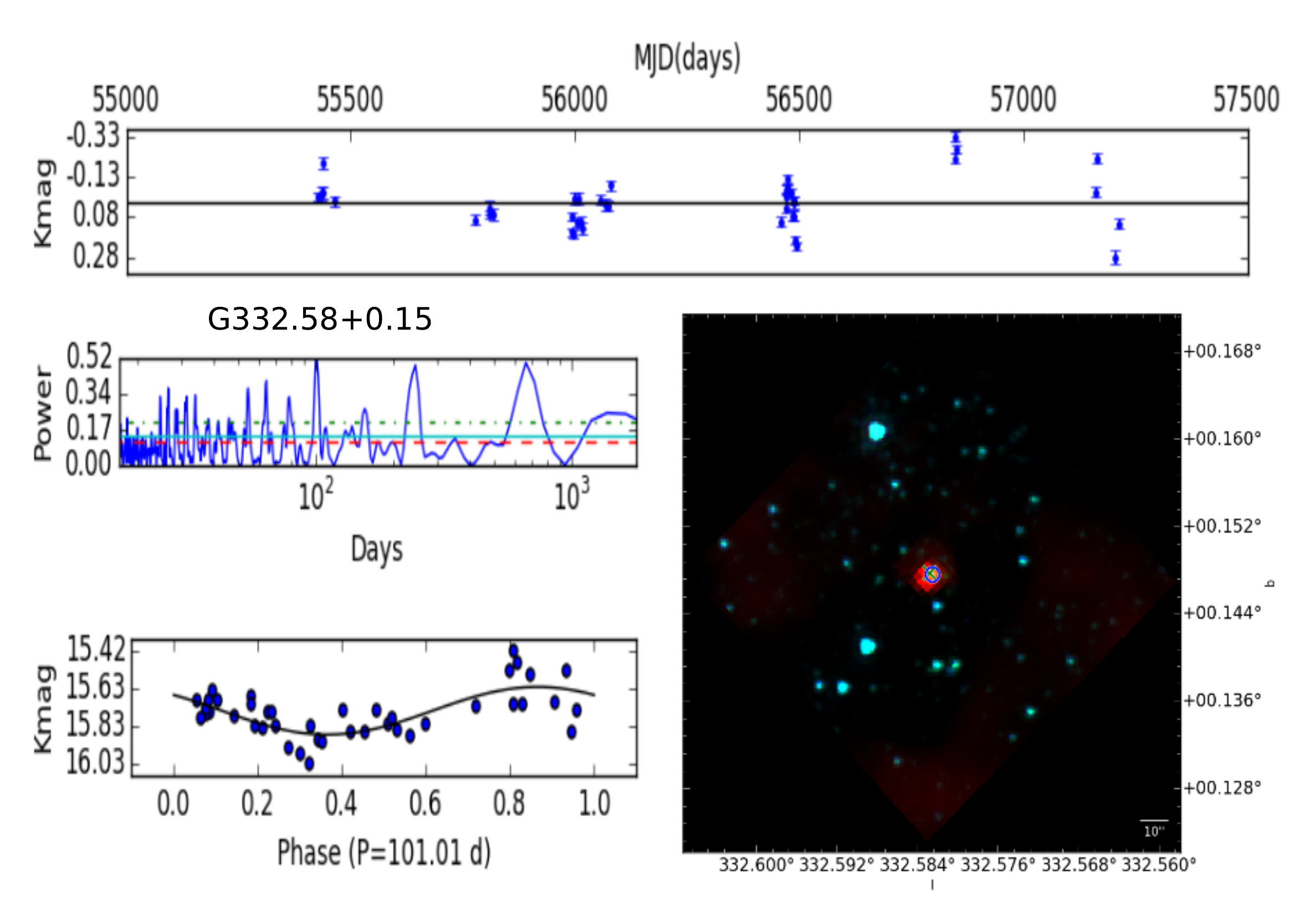}\\
                \includegraphics[width=0.48\textwidth]{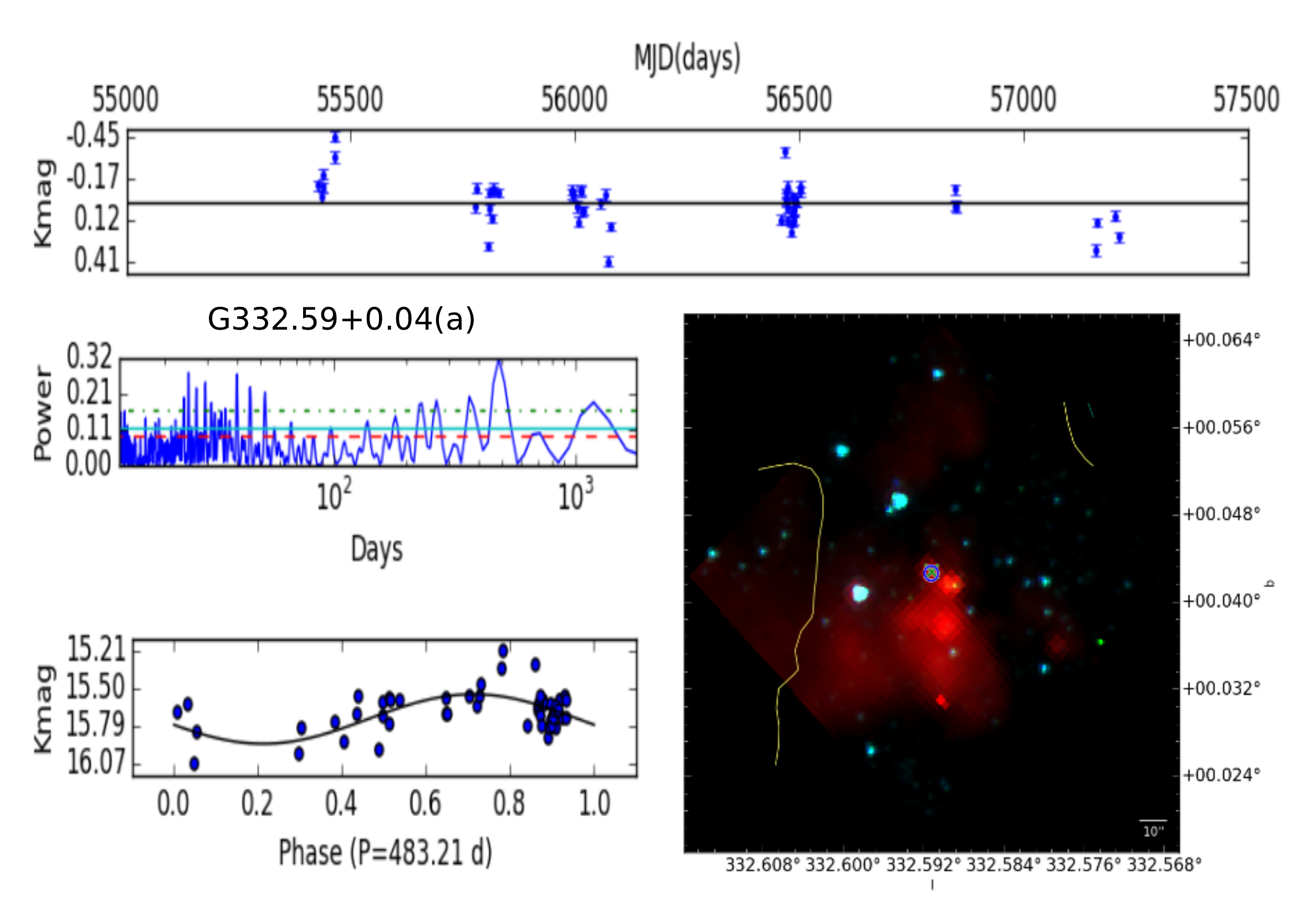}&
                \includegraphics[width=0.48\textwidth]{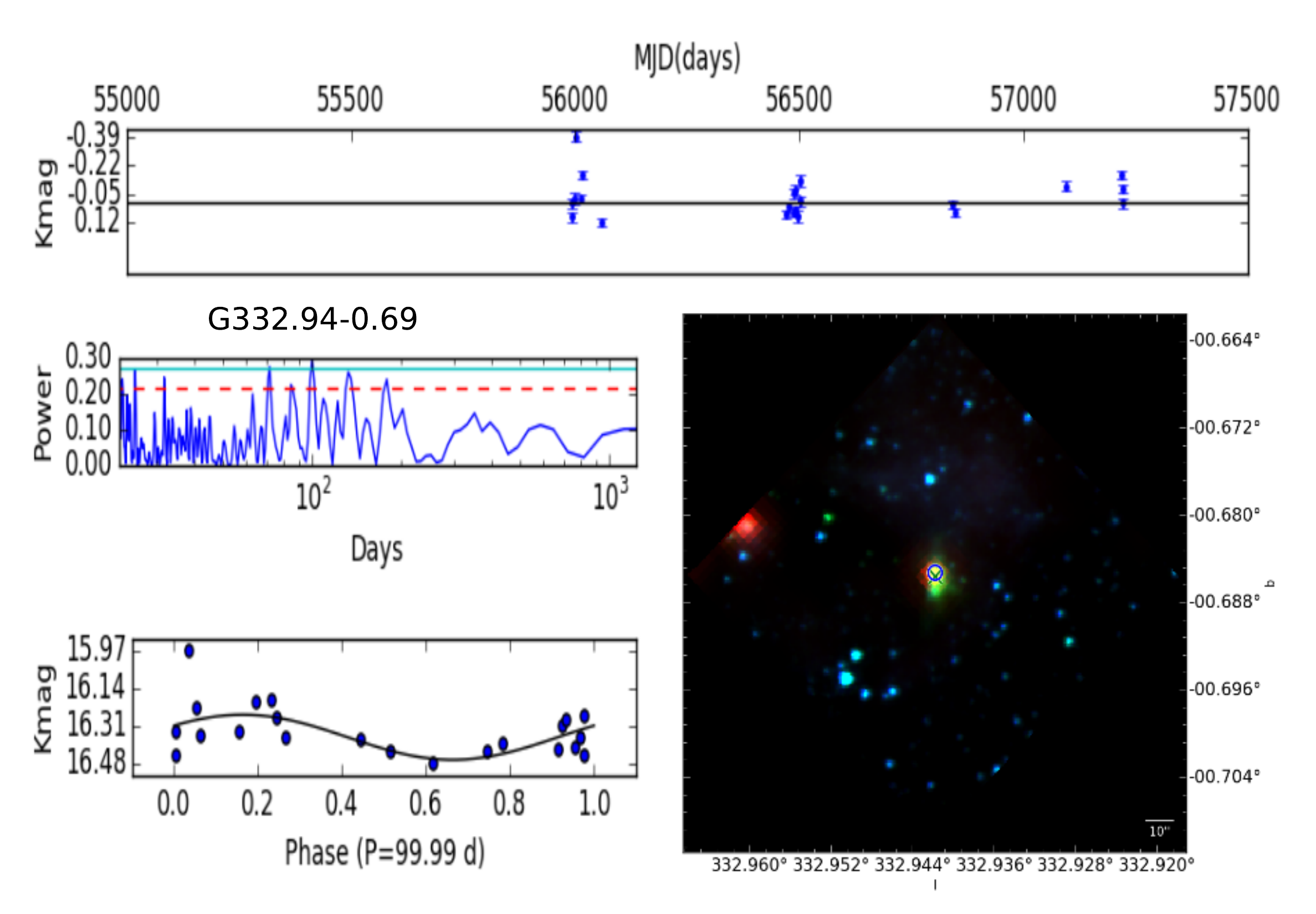}\\    
                \includegraphics[width=0.48\textwidth]{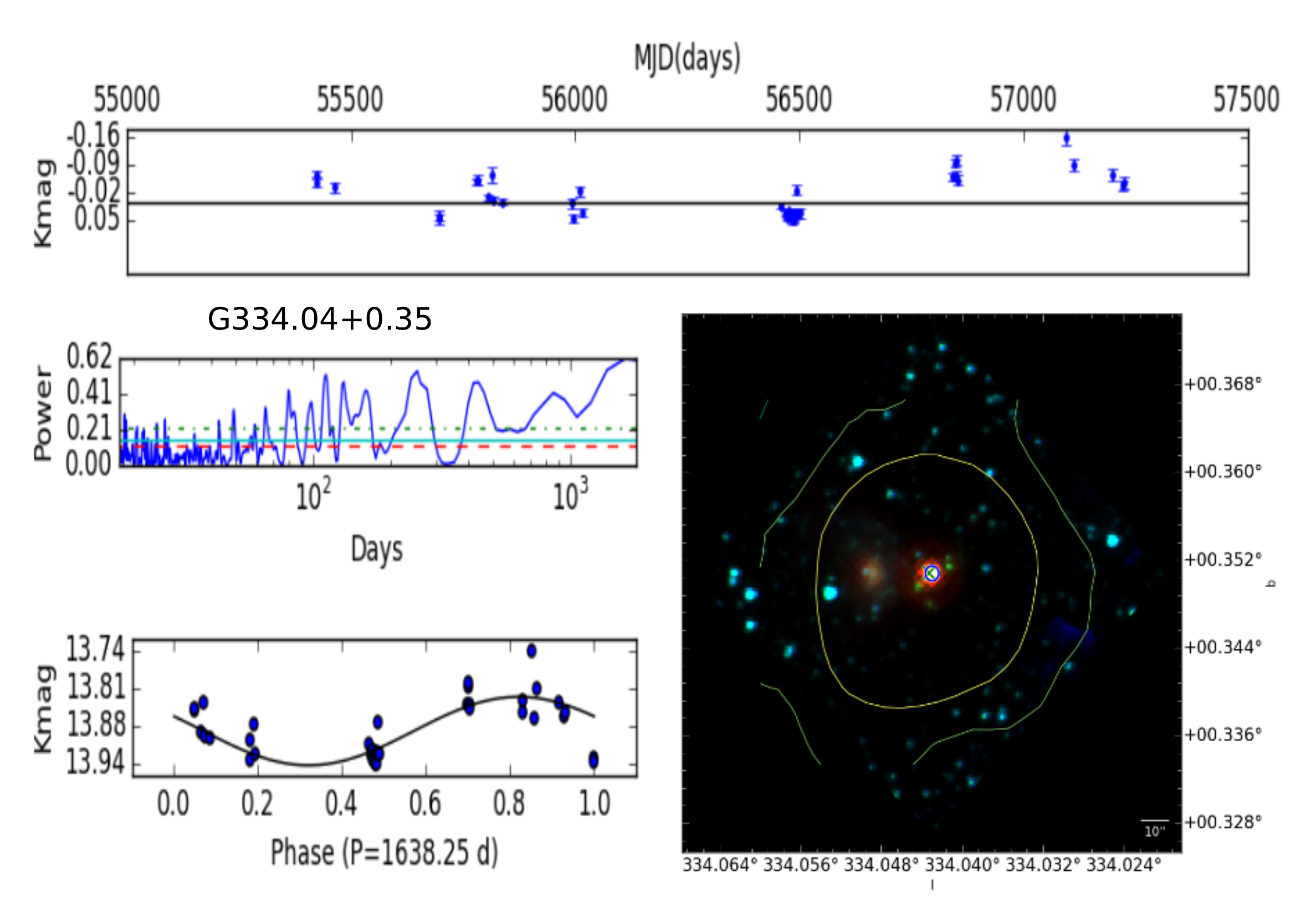}&
                \includegraphics[width=0.48\textwidth]{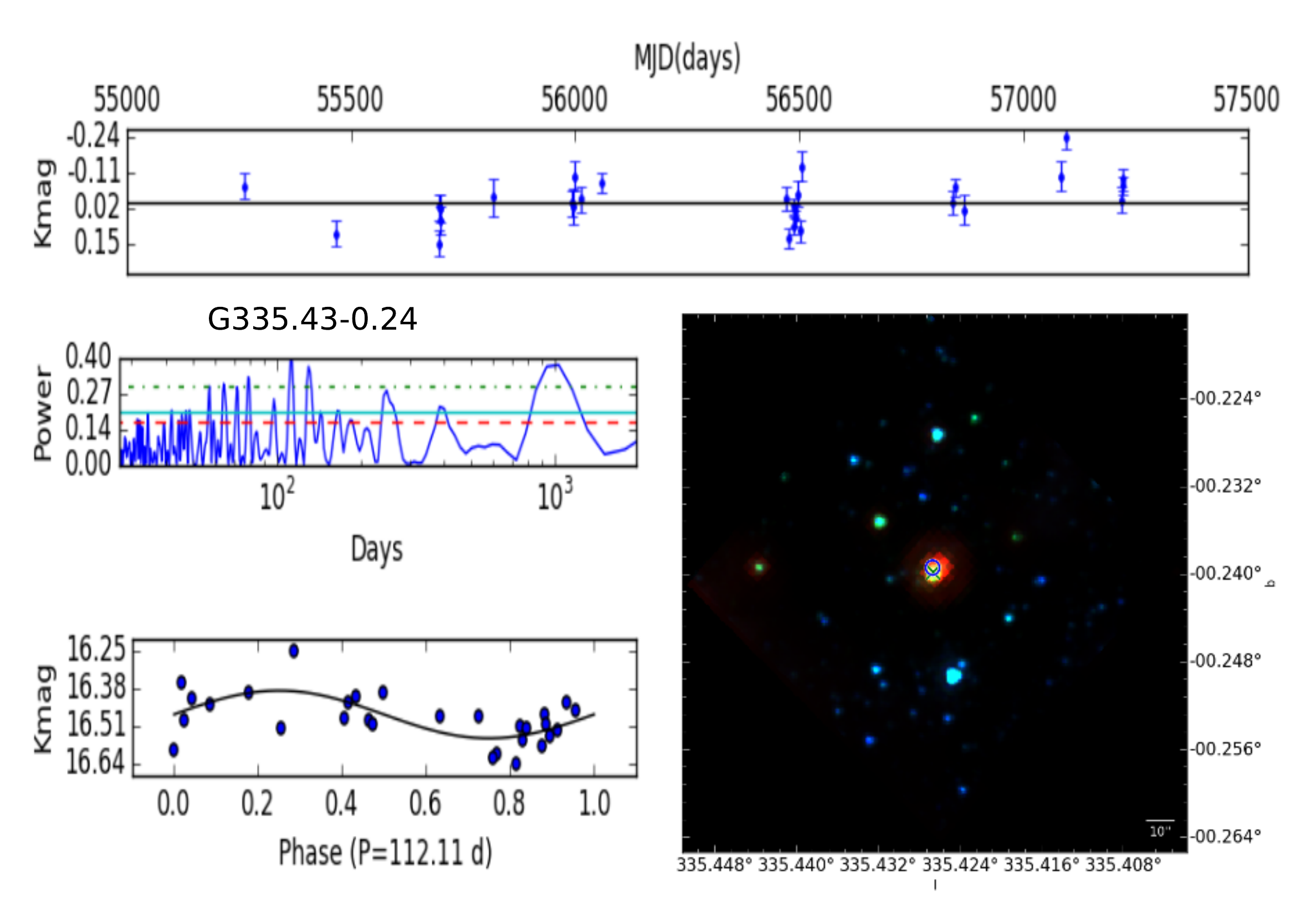}\\
                
        \end{tabular}
        \caption{Continuation of Fig. \ref{fig:testgrid}.}
\end{figure}

\begin{figure}
        \begin{tabular}{cc}

                \includegraphics[width=0.48\textwidth]{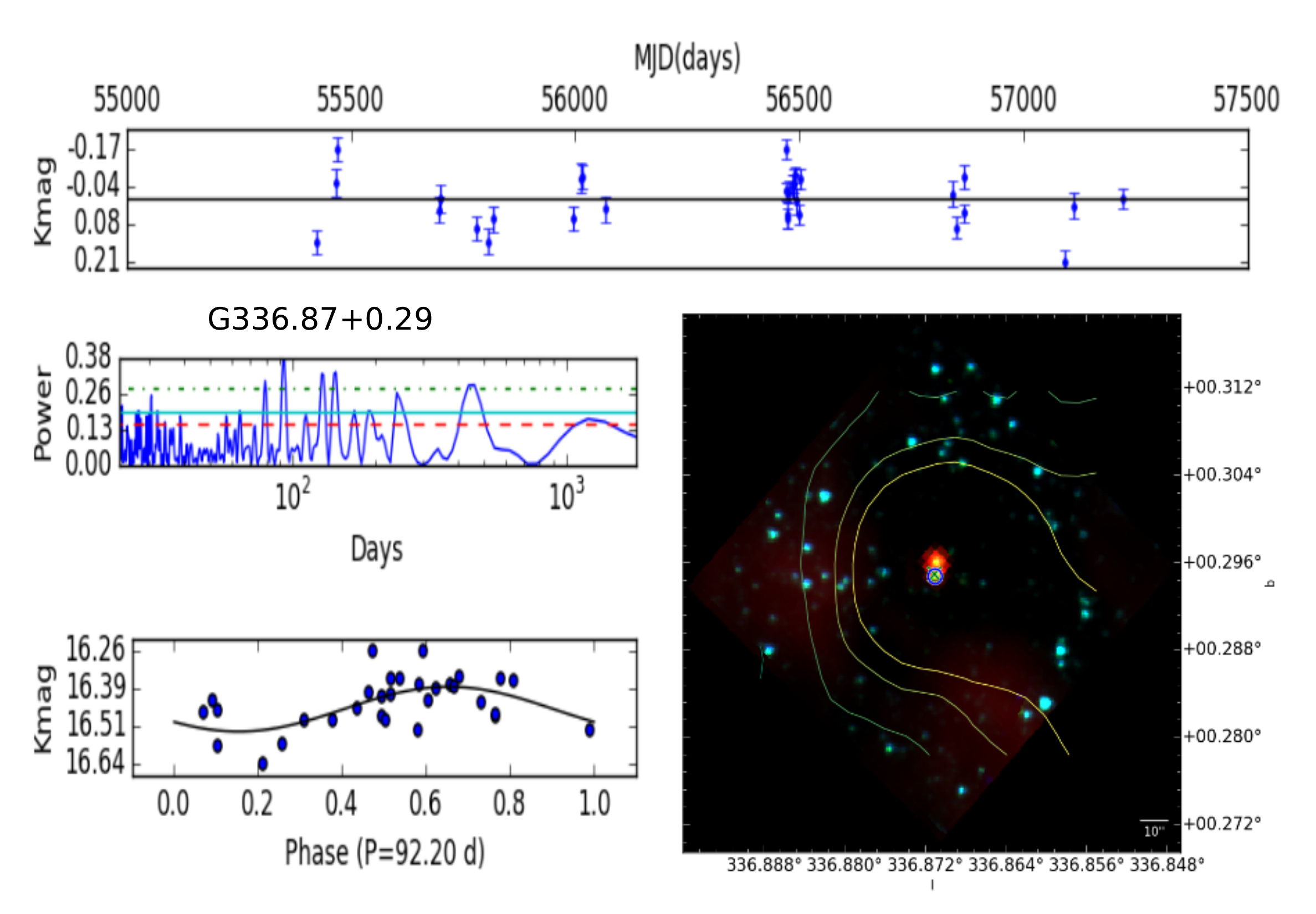}&
                \includegraphics[width=0.48\textwidth]{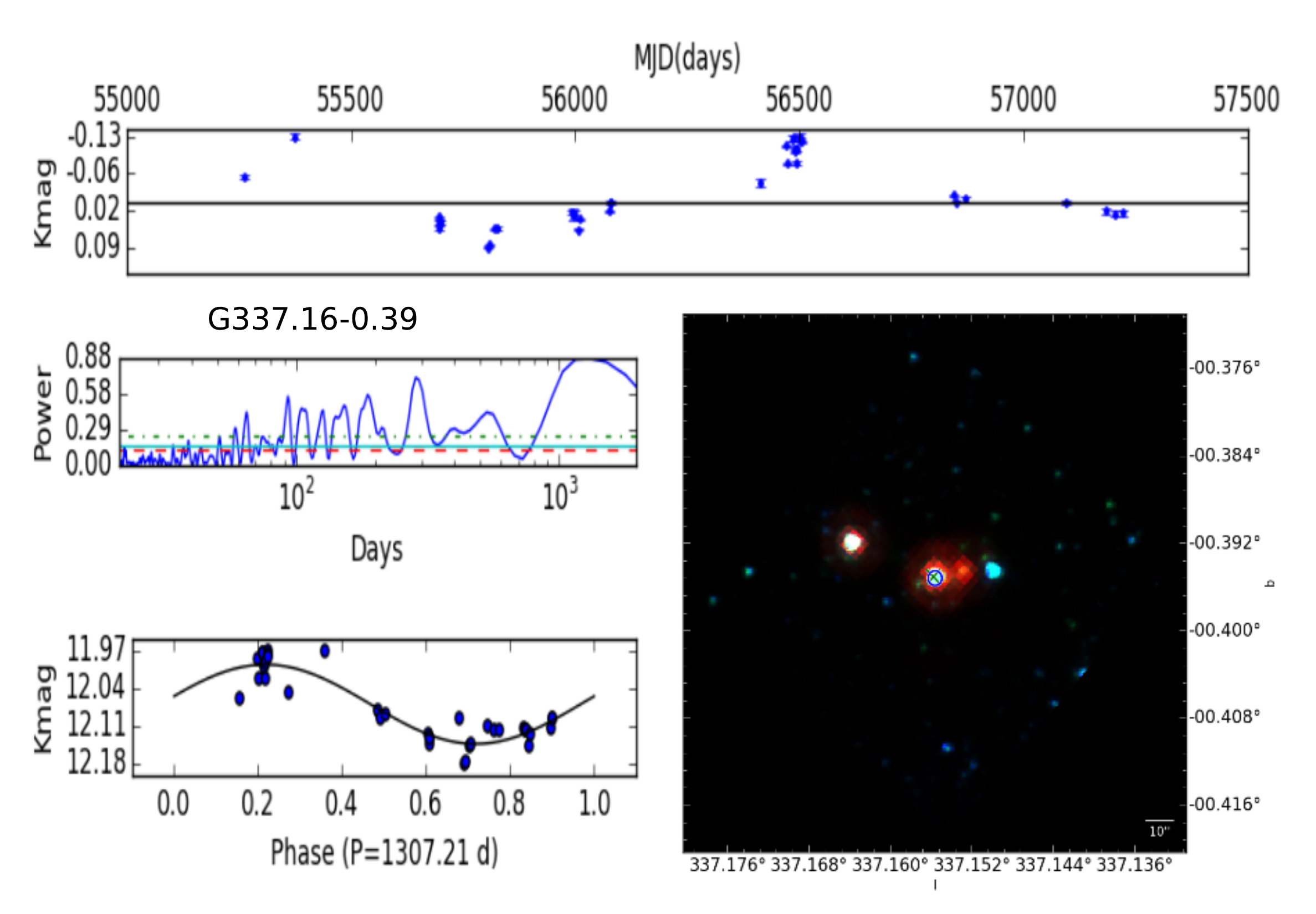}\\
                \includegraphics[width=0.48\textwidth]{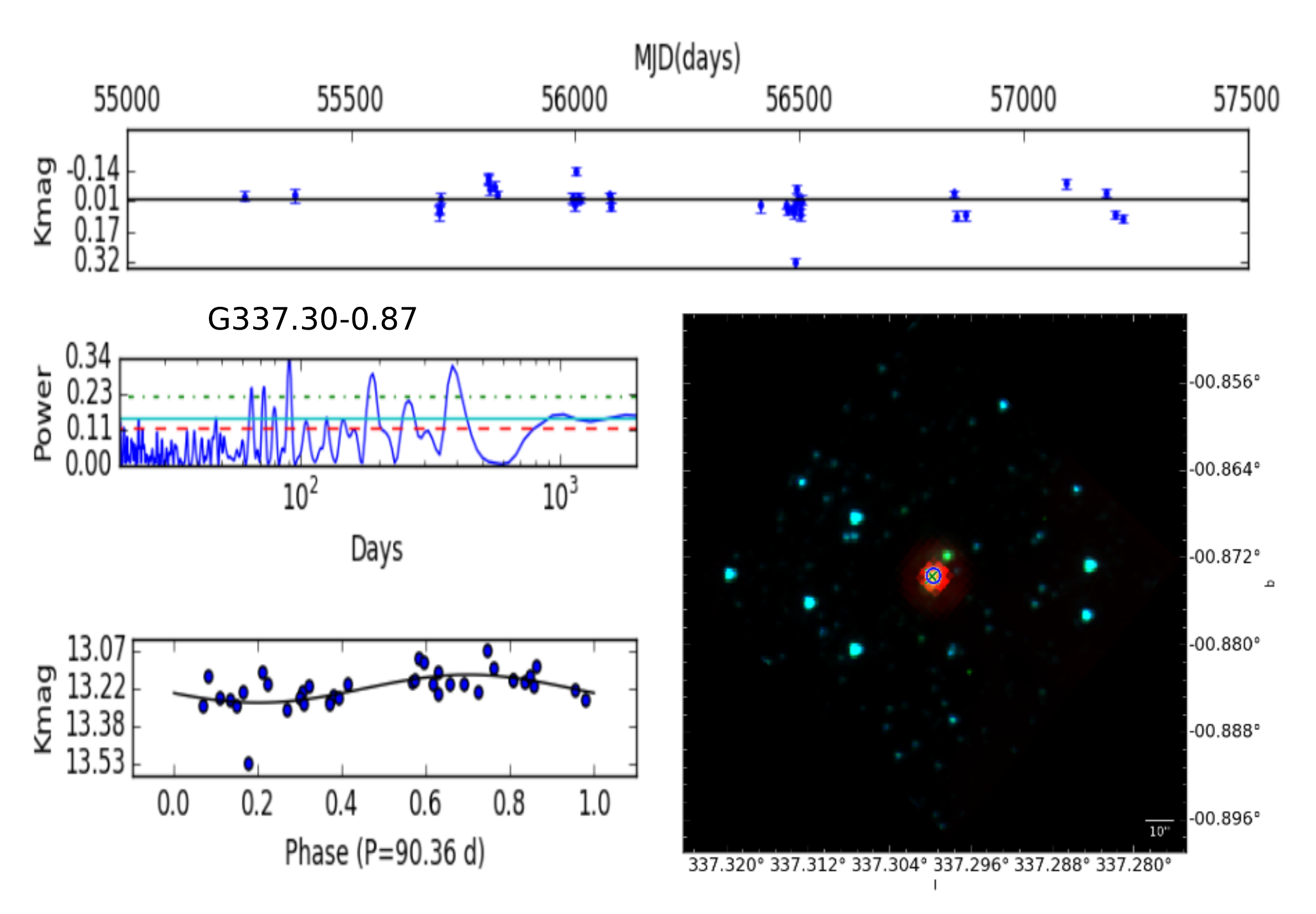}&
                \includegraphics[width=0.48\textwidth]{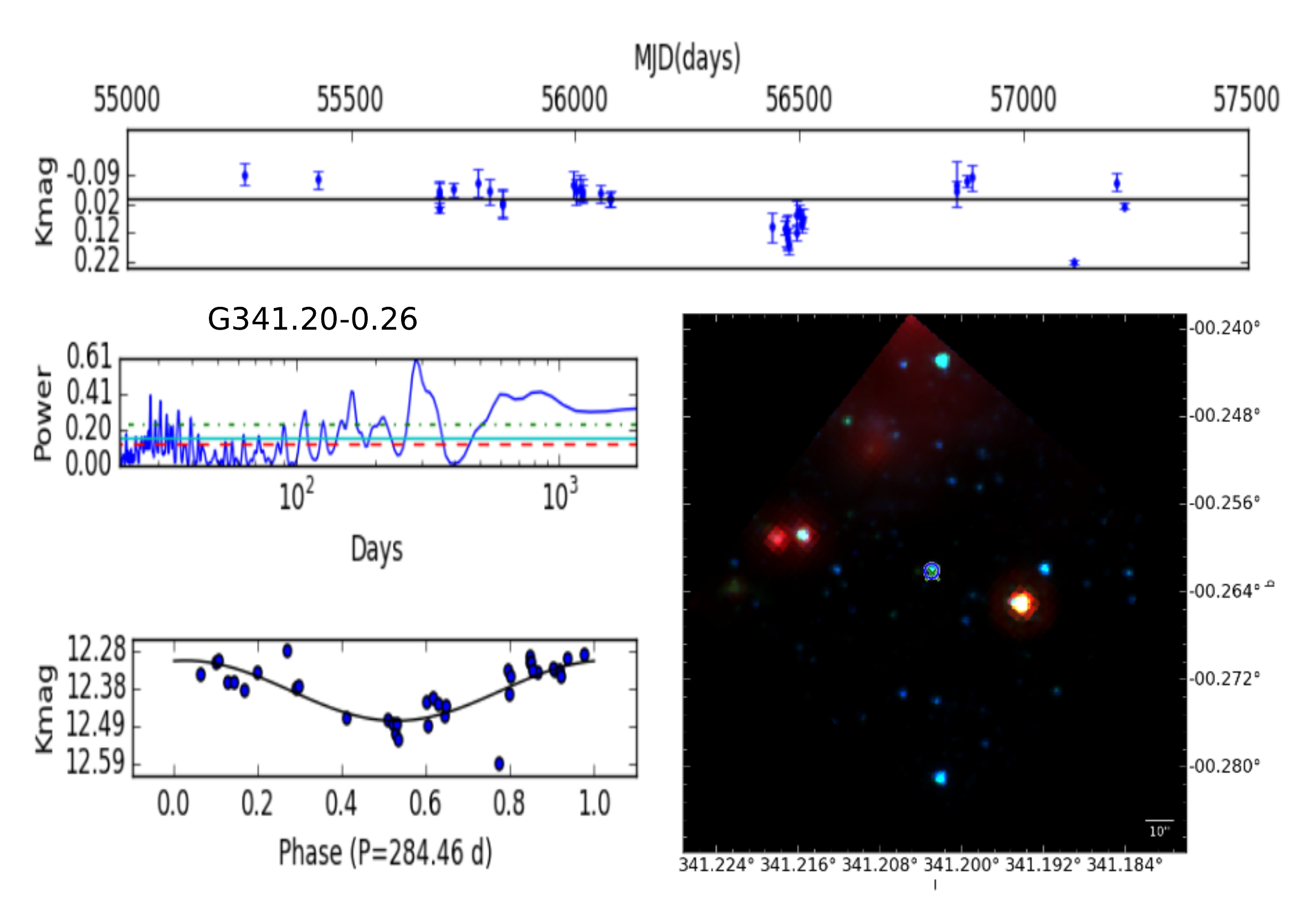}\\
                \includegraphics[width=0.48\textwidth]{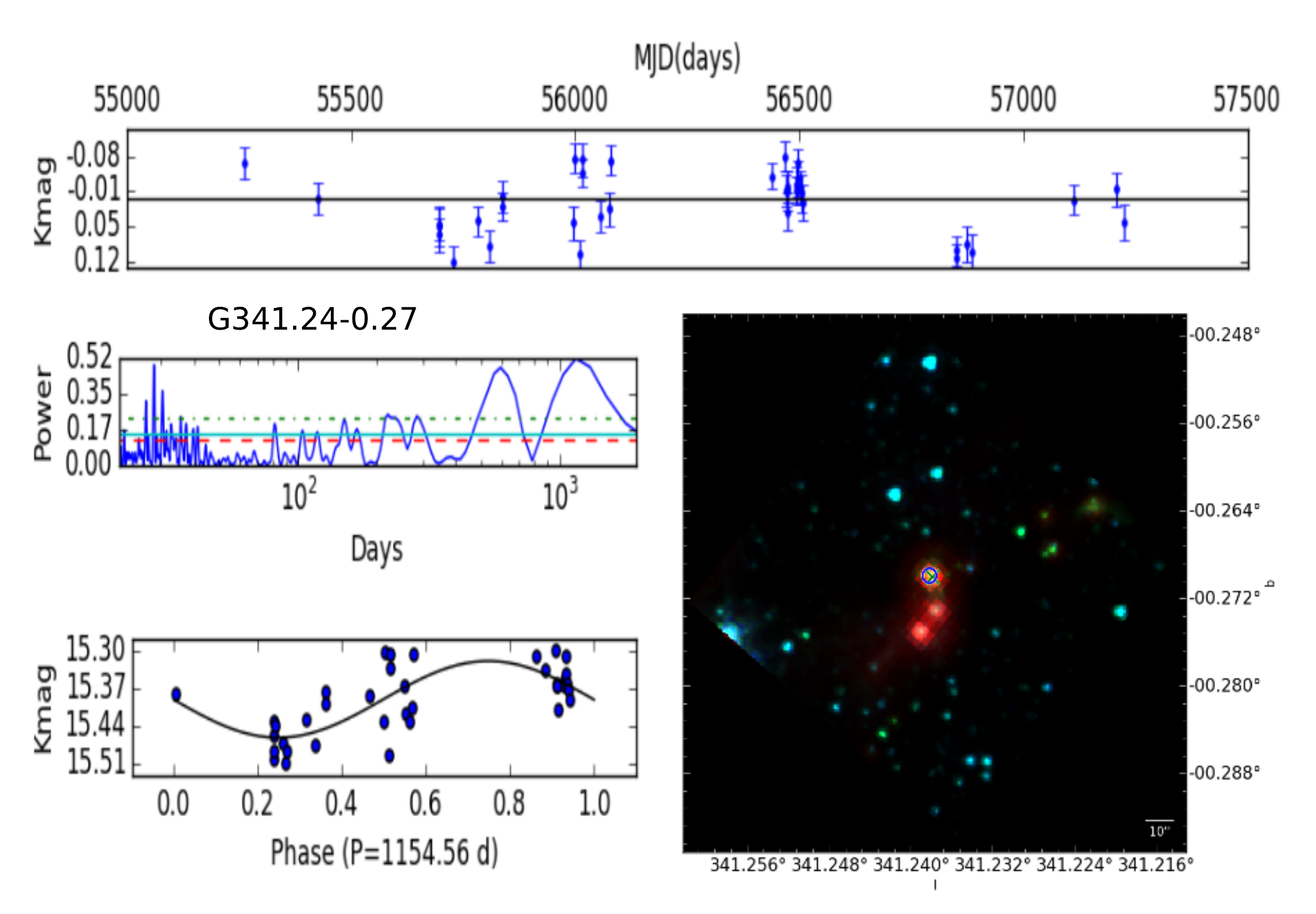}&
                \includegraphics[width=0.48\textwidth]{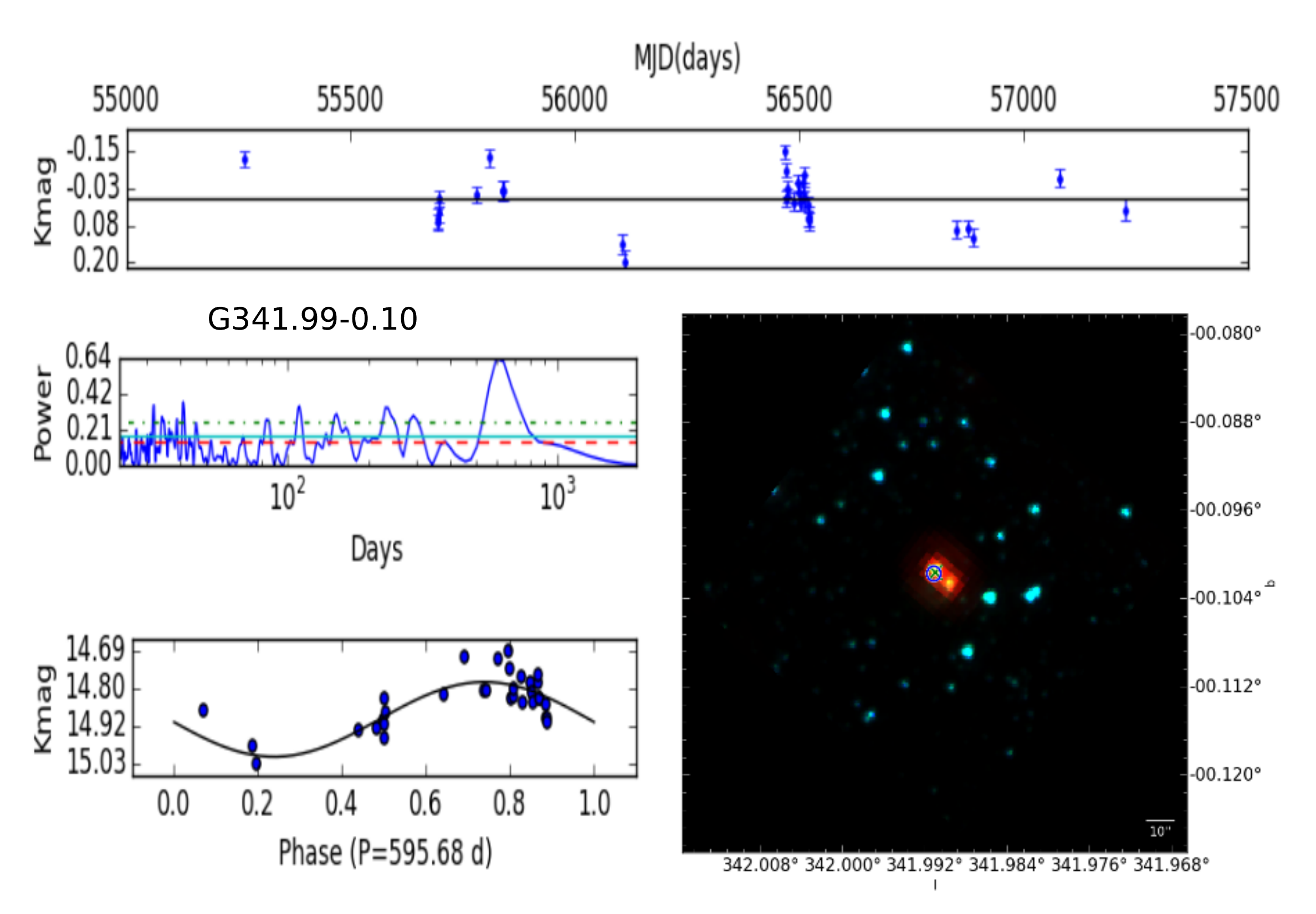}\\
                \includegraphics[width=0.48\textwidth]{{./G343.50-0.47_combined}.pdf}&
                \includegraphics[width=0.48\textwidth]{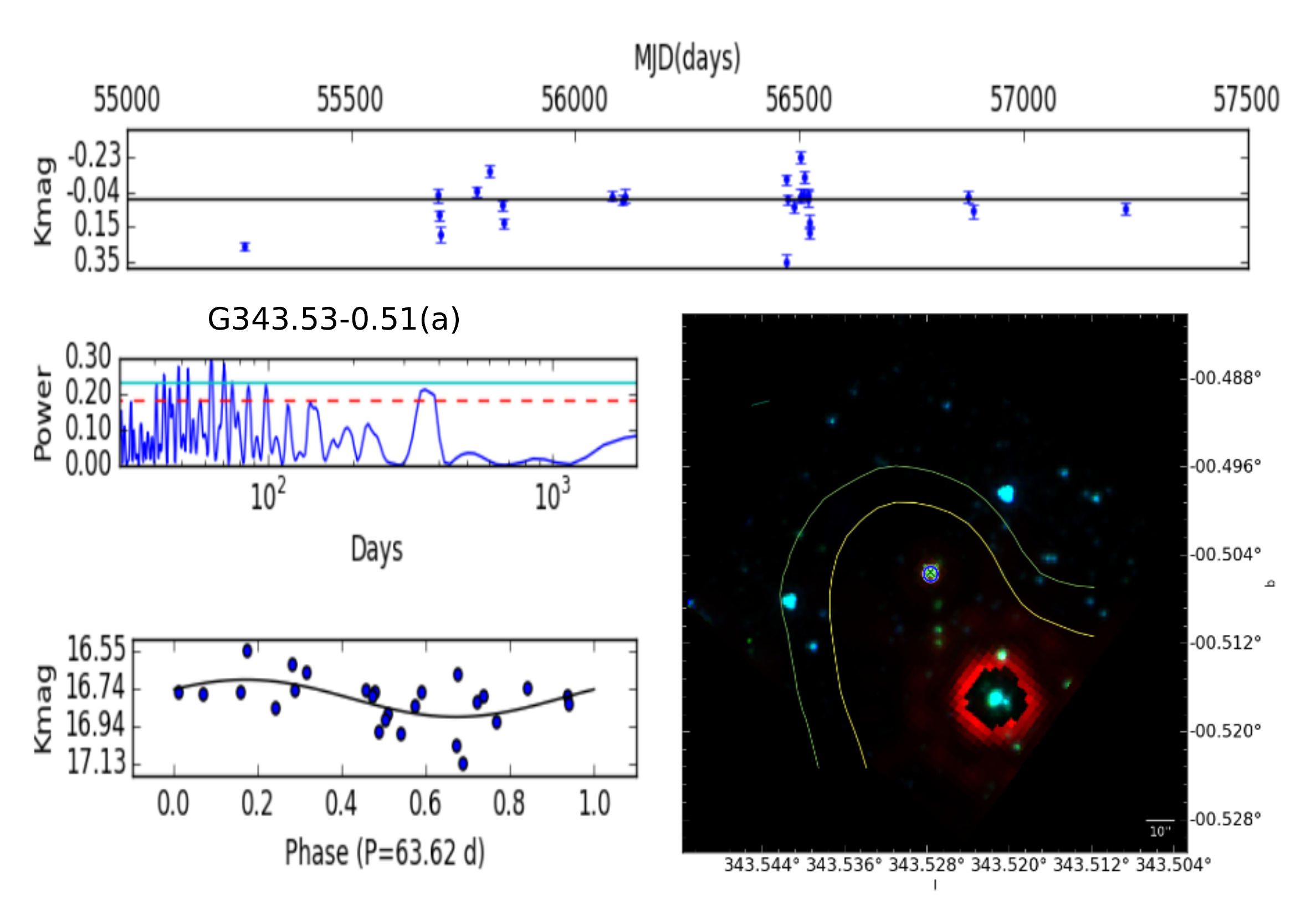}\\
                
        \end{tabular}
        \caption{Continuation of Fig. \ref{fig:testgrid}.}
\end{figure}

\begin{figure}
        \begin{tabular}{cc}
                
                \includegraphics[width=0.48\textwidth]{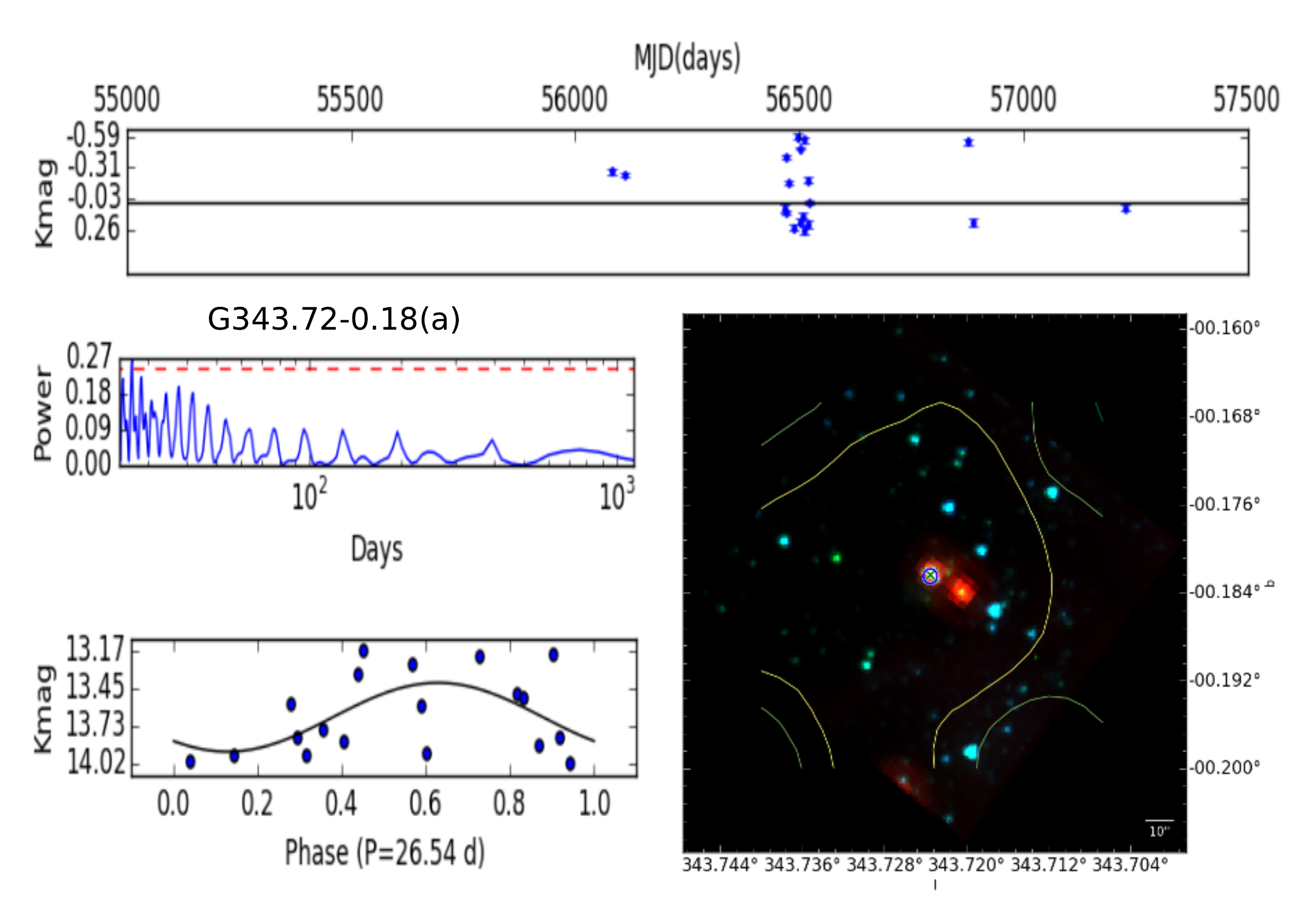}&
                \includegraphics[width=0.48\textwidth]{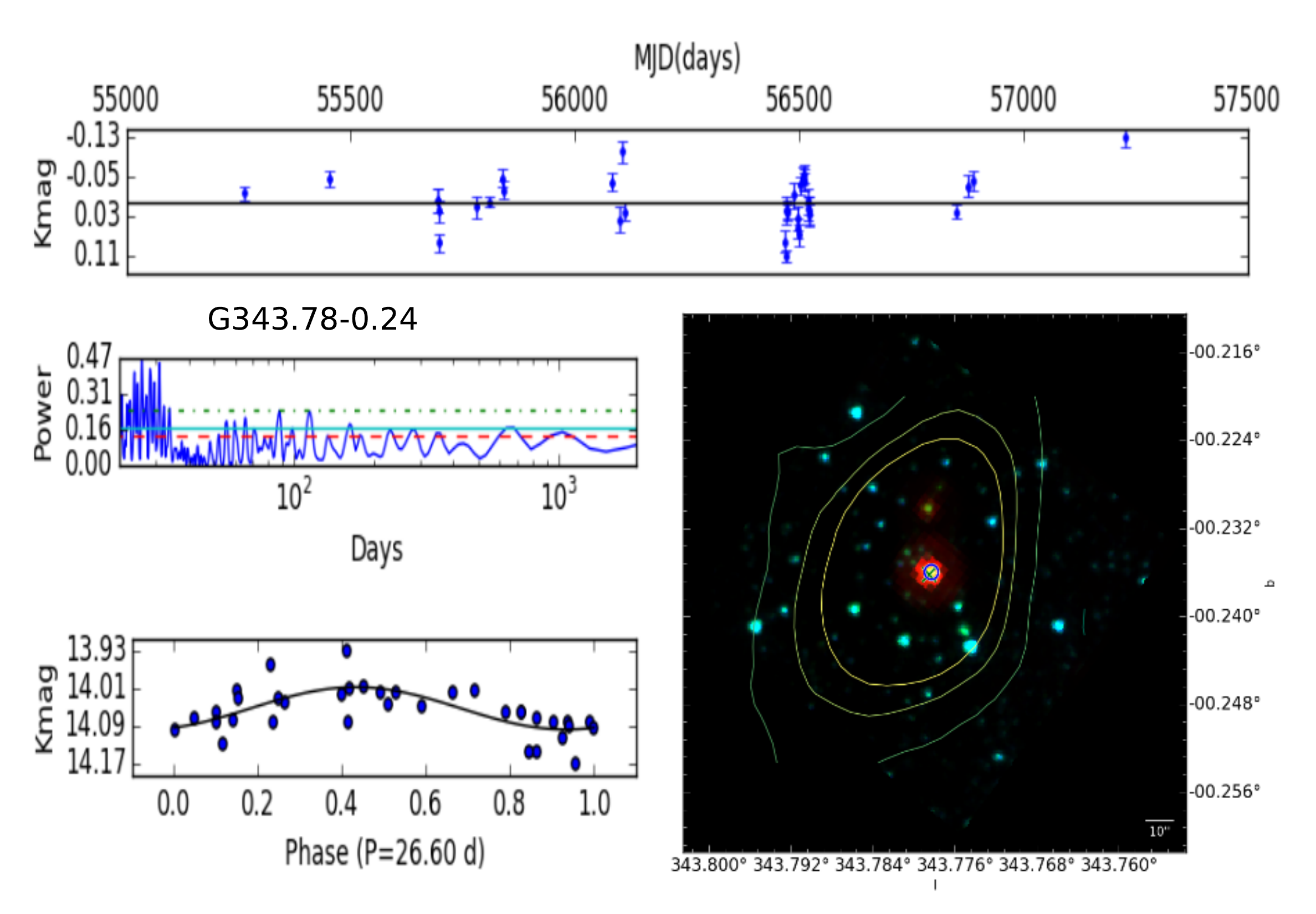}\\
                \includegraphics[width=0.48\textwidth]{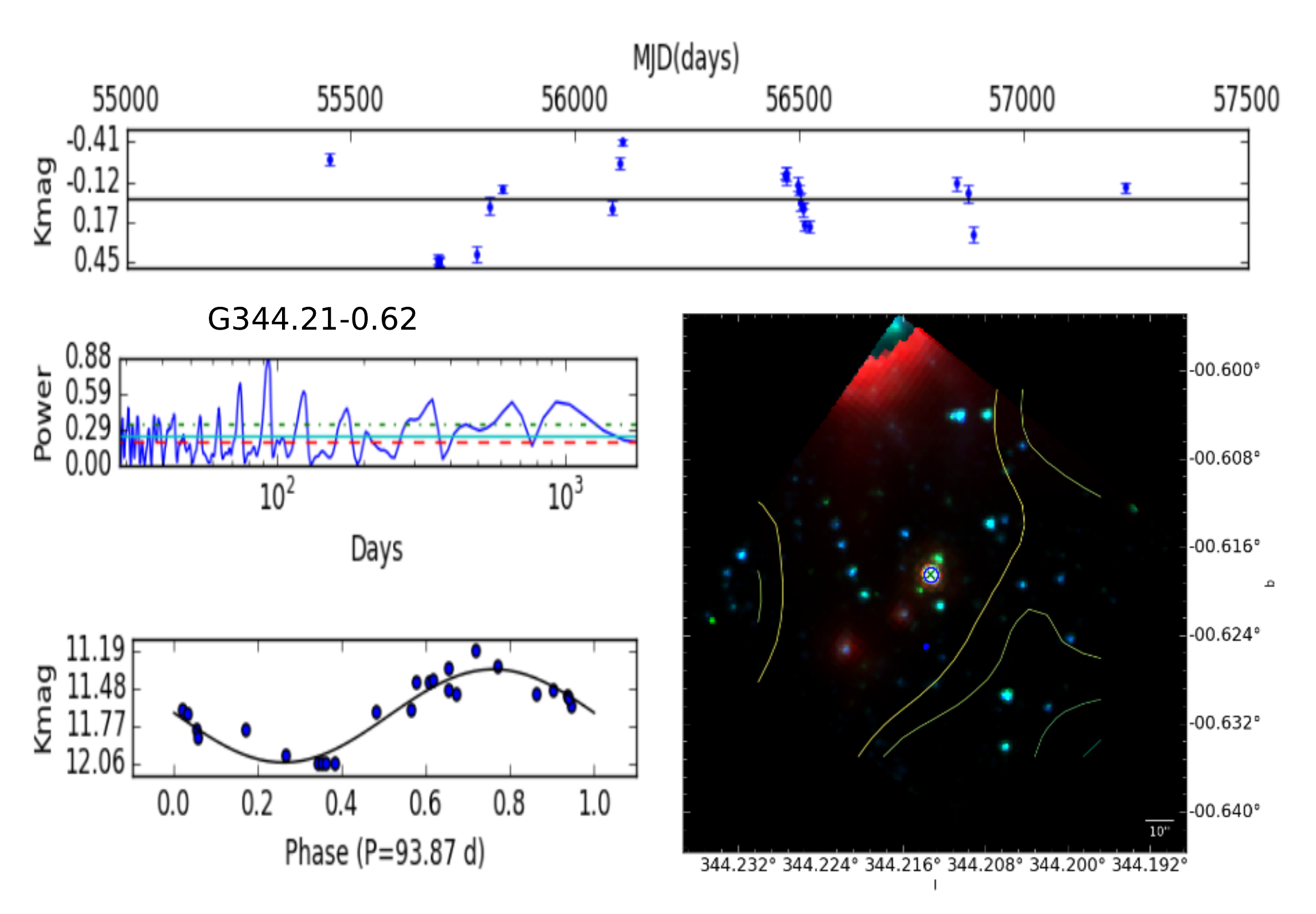}&
                \includegraphics[width=0.48\textwidth]{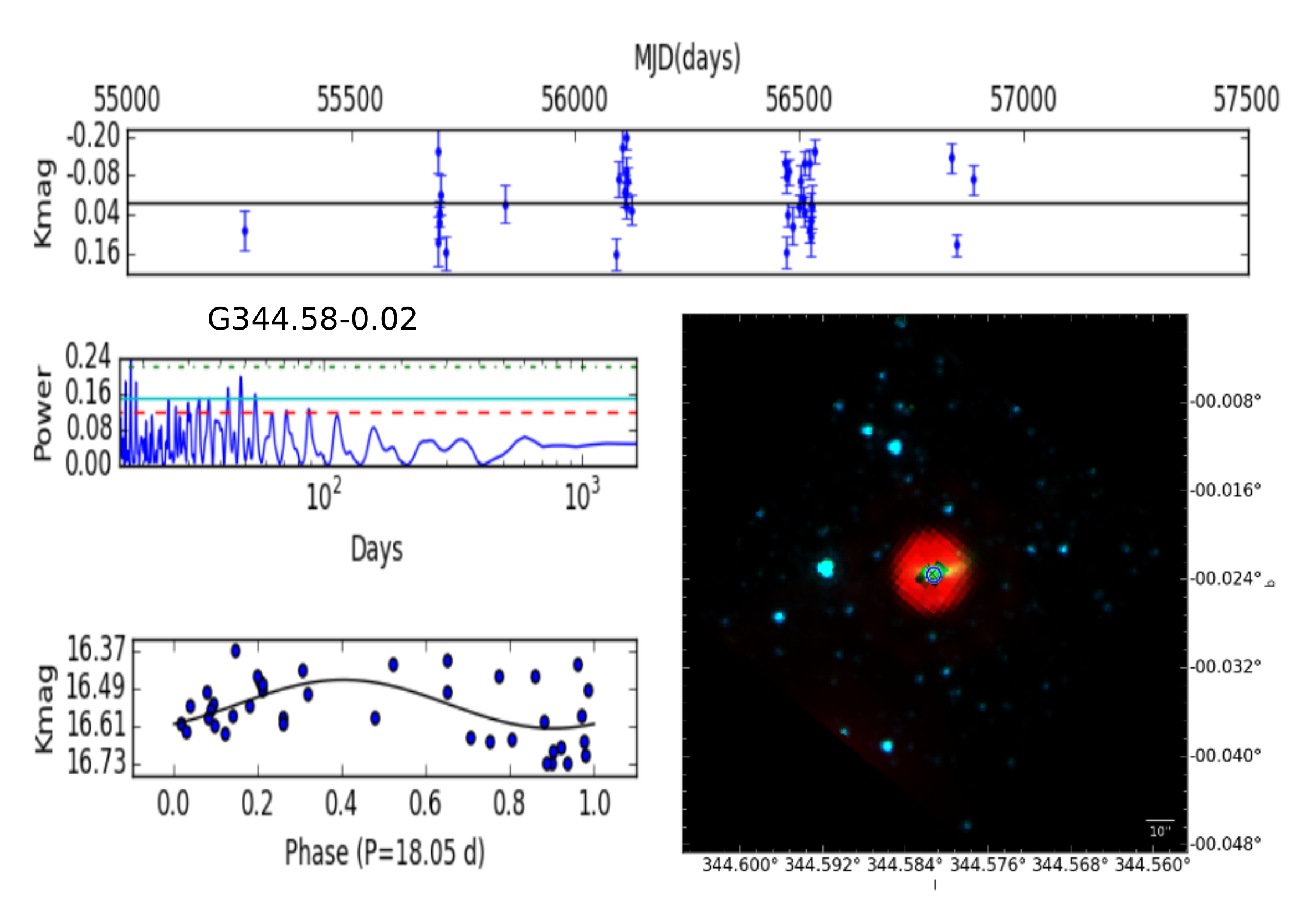}\\
                \includegraphics[width=0.48\textwidth]{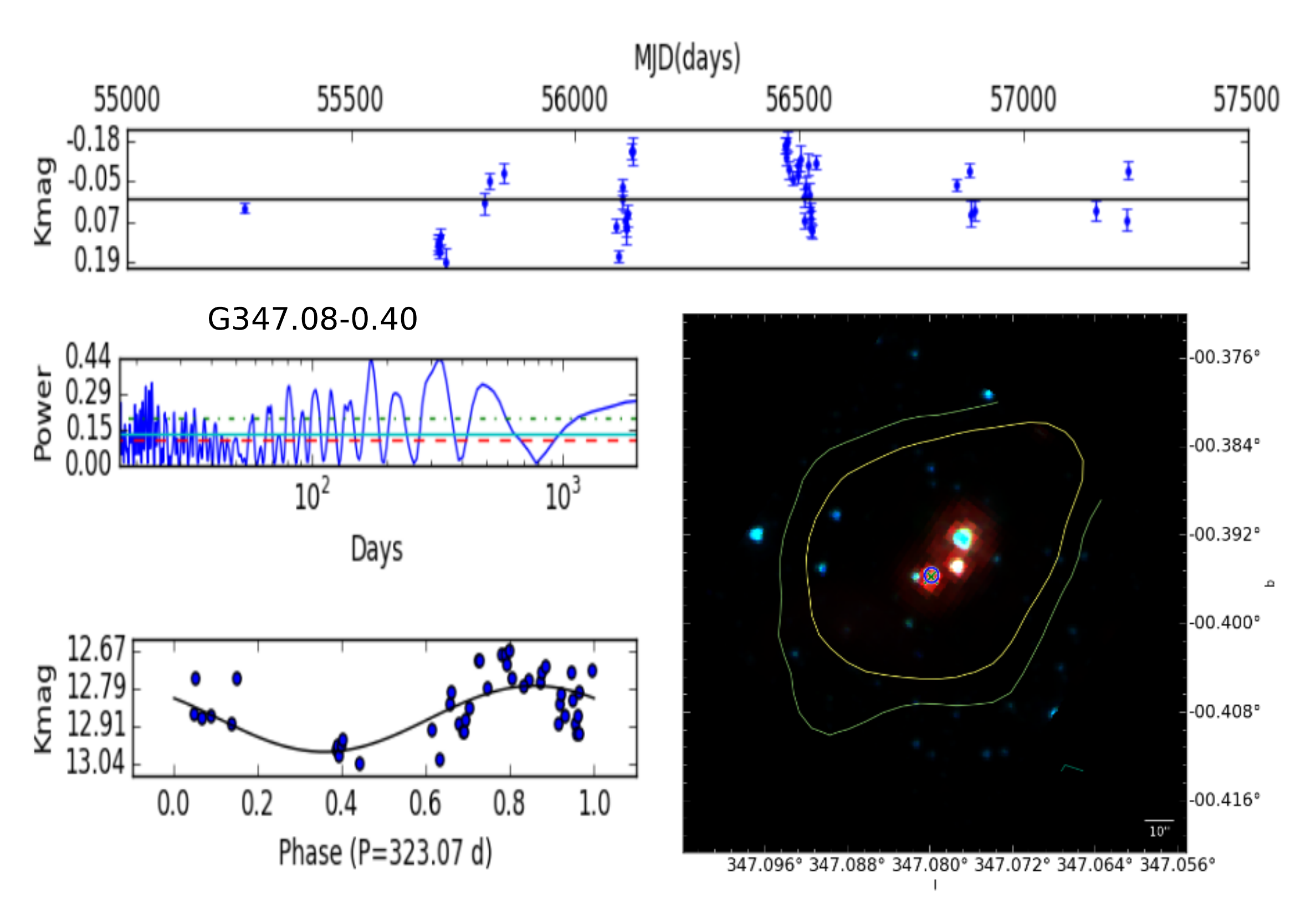}&
                \includegraphics[width=0.48\textwidth]{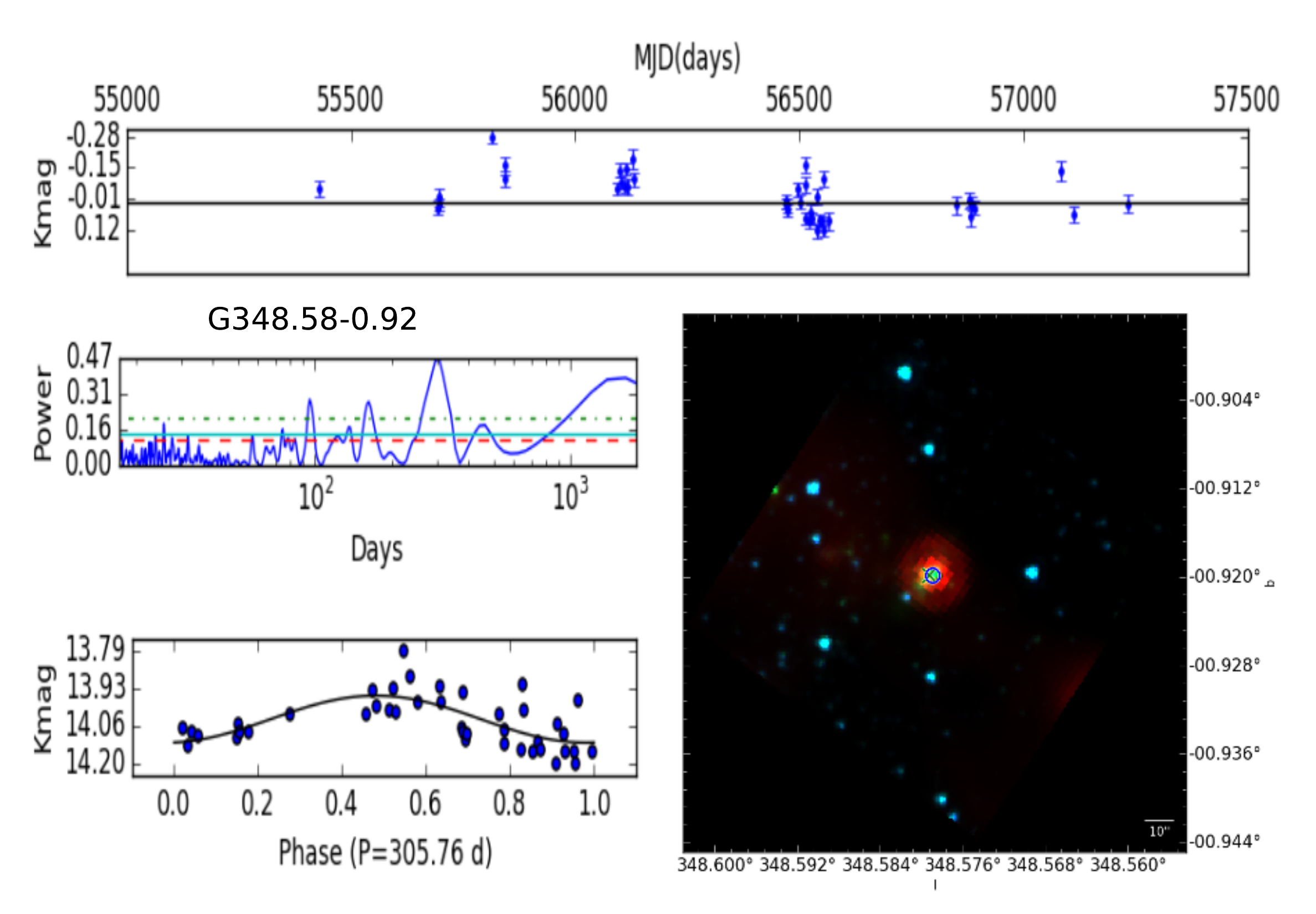}\\
                \includegraphics[width=0.48\textwidth]{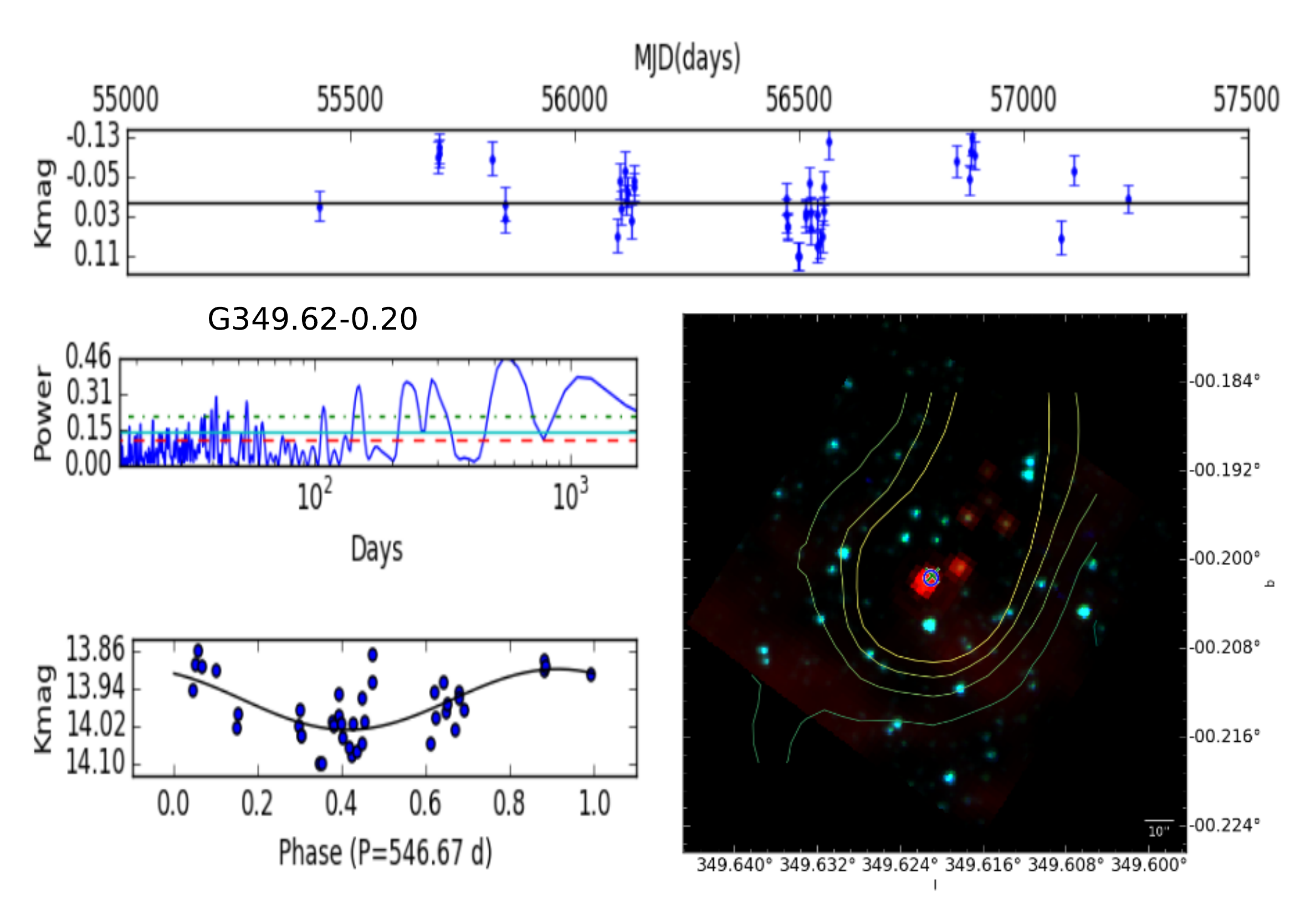}&
                \includegraphics[width=0.48\textwidth]{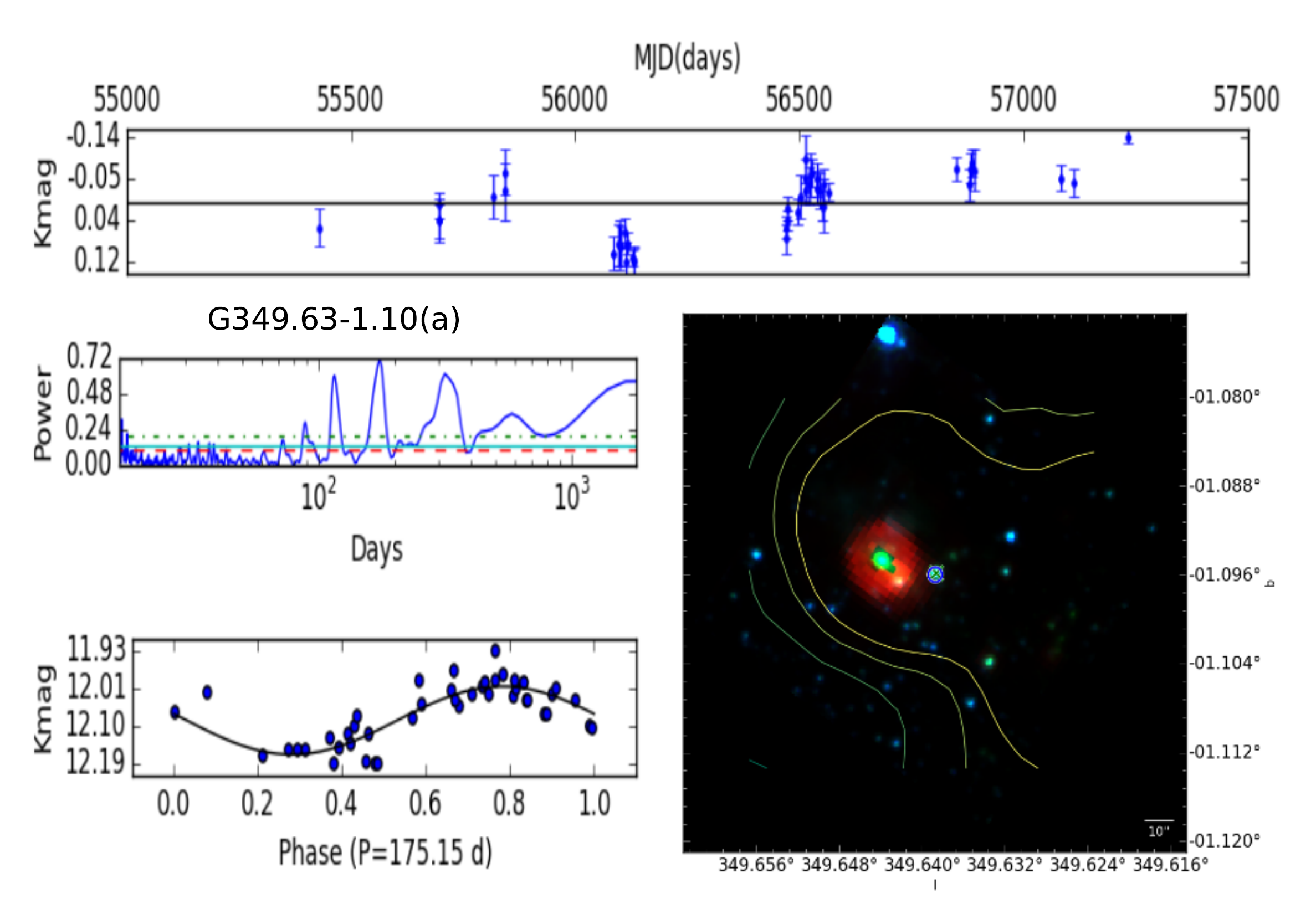}\\
                
        \end{tabular}
        \caption{Continuation of Fig. \ref{fig:testgrid}.}
\end{figure}

\begin{figure}
        \begin{tabular}{cc}

                \includegraphics[width=0.48\textwidth]{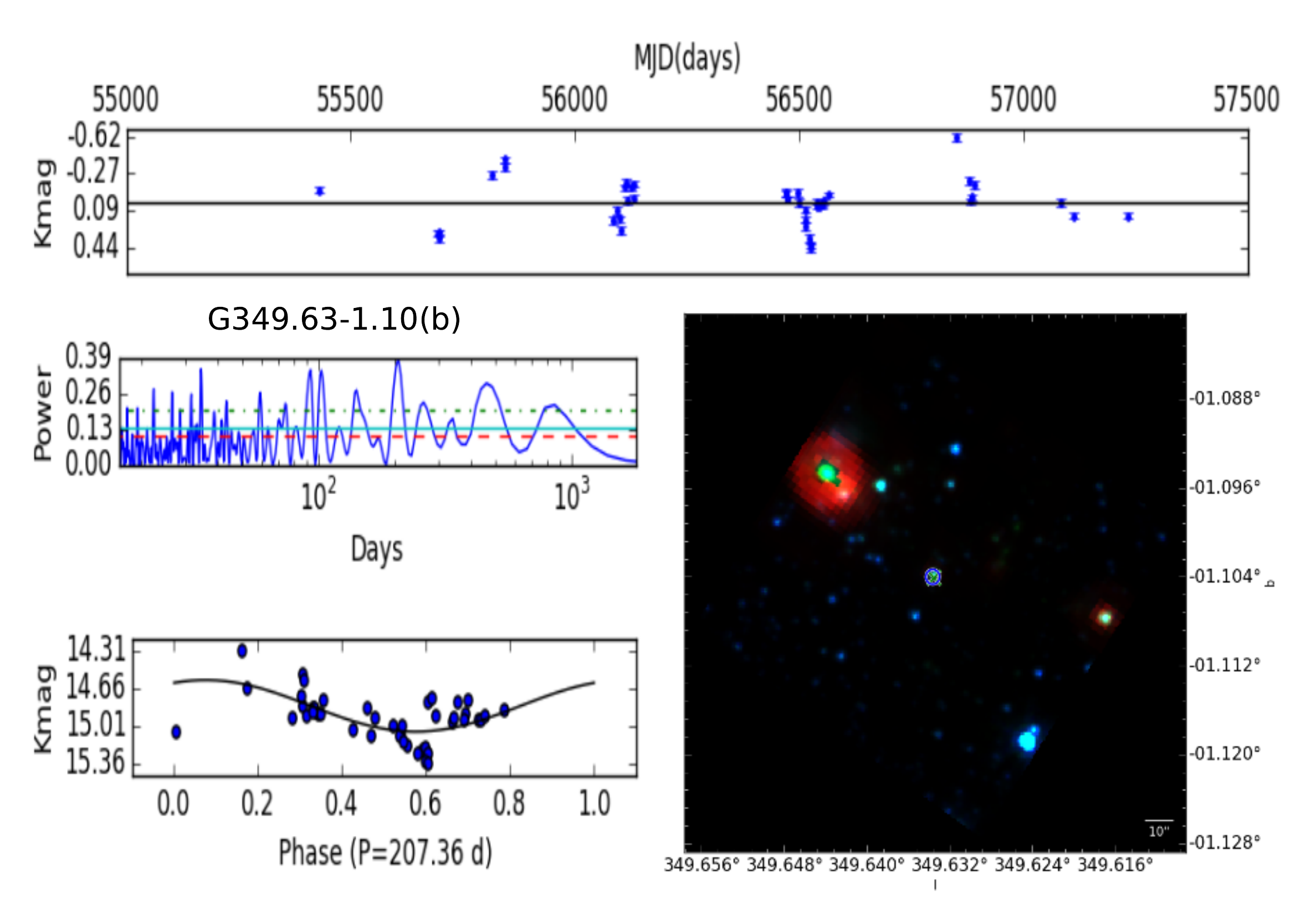}&
                \includegraphics[width=0.48\textwidth]{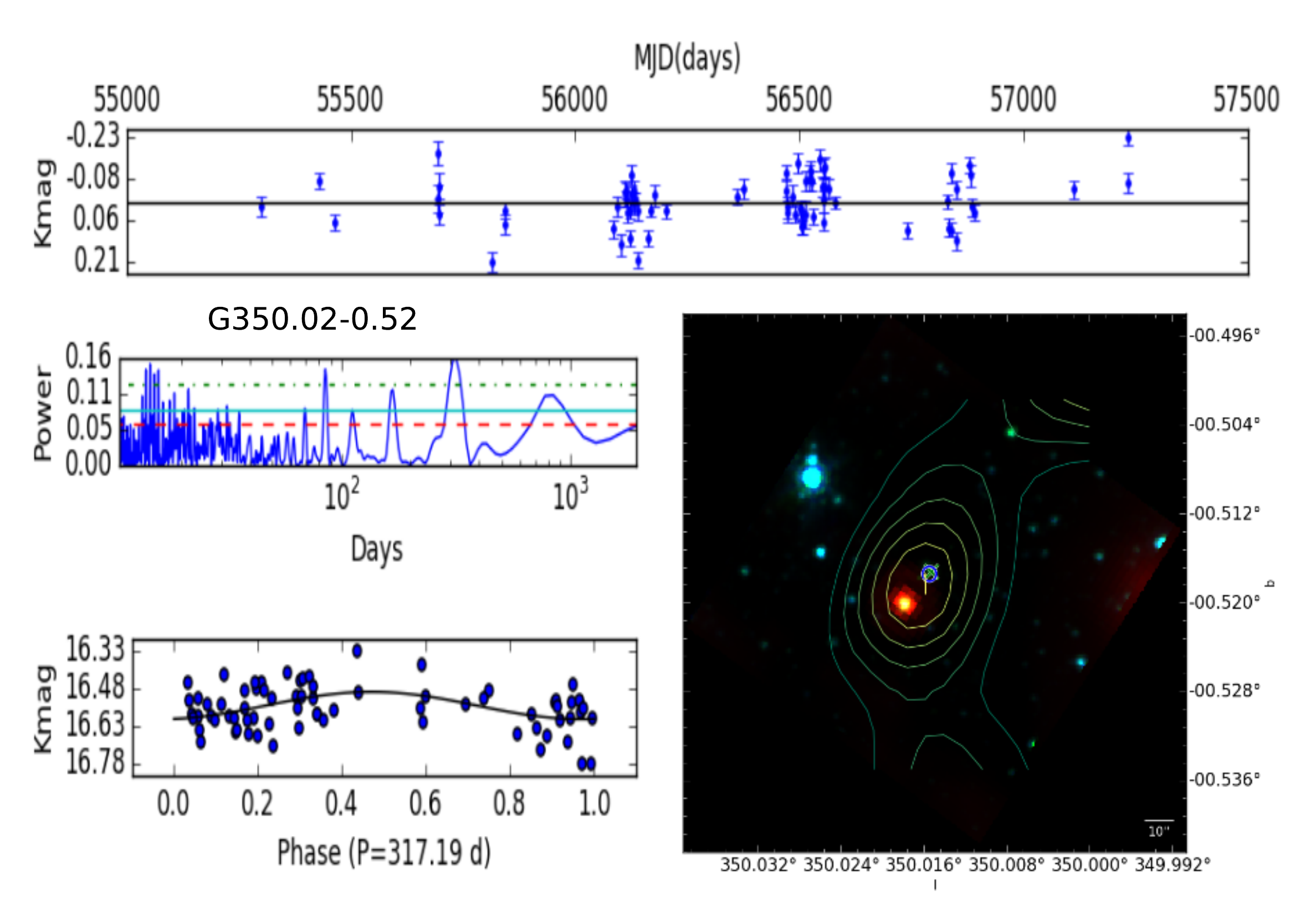}\\
                \includegraphics[width=0.48\textwidth]{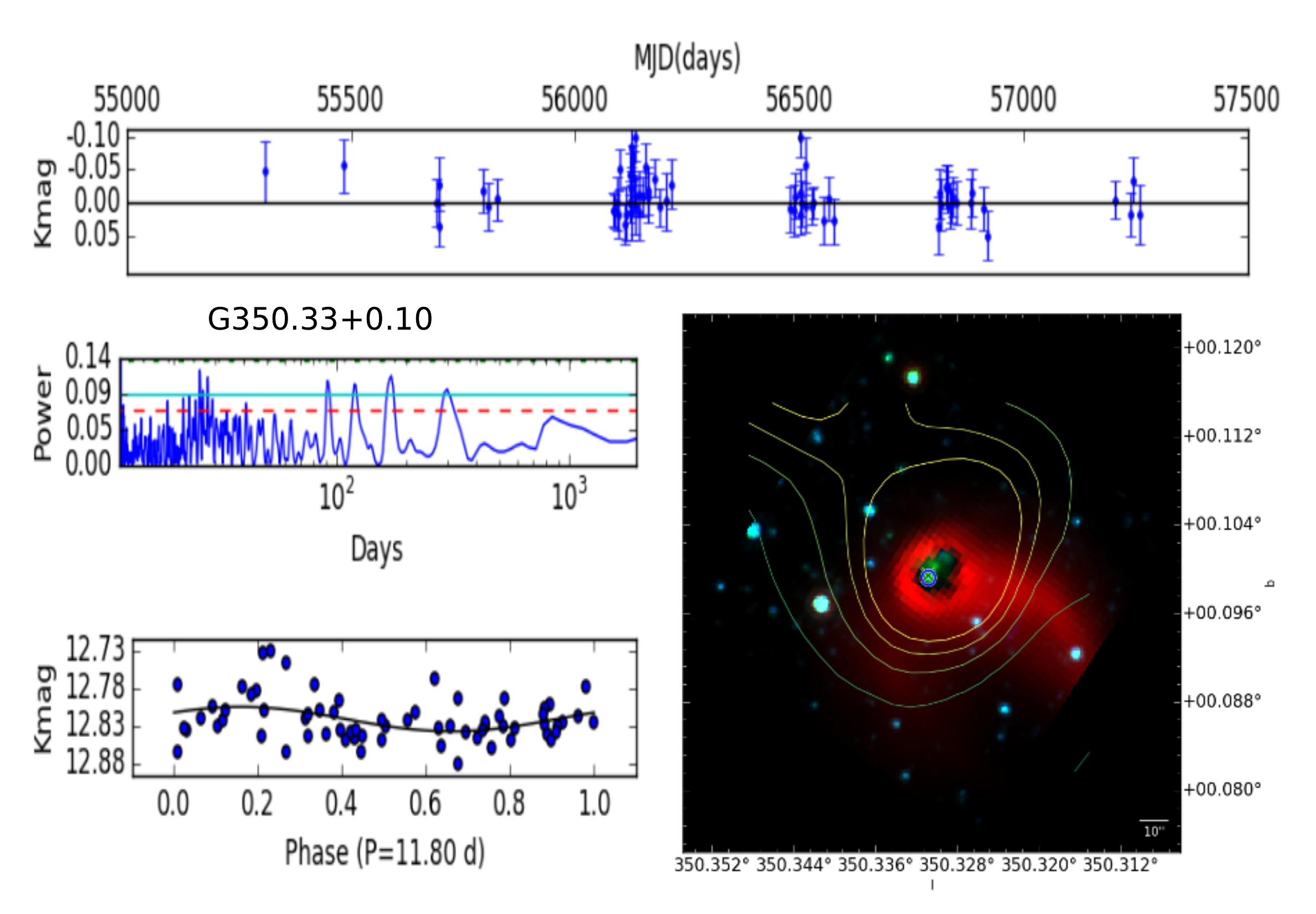}&
                \includegraphics[width=0.48\textwidth]{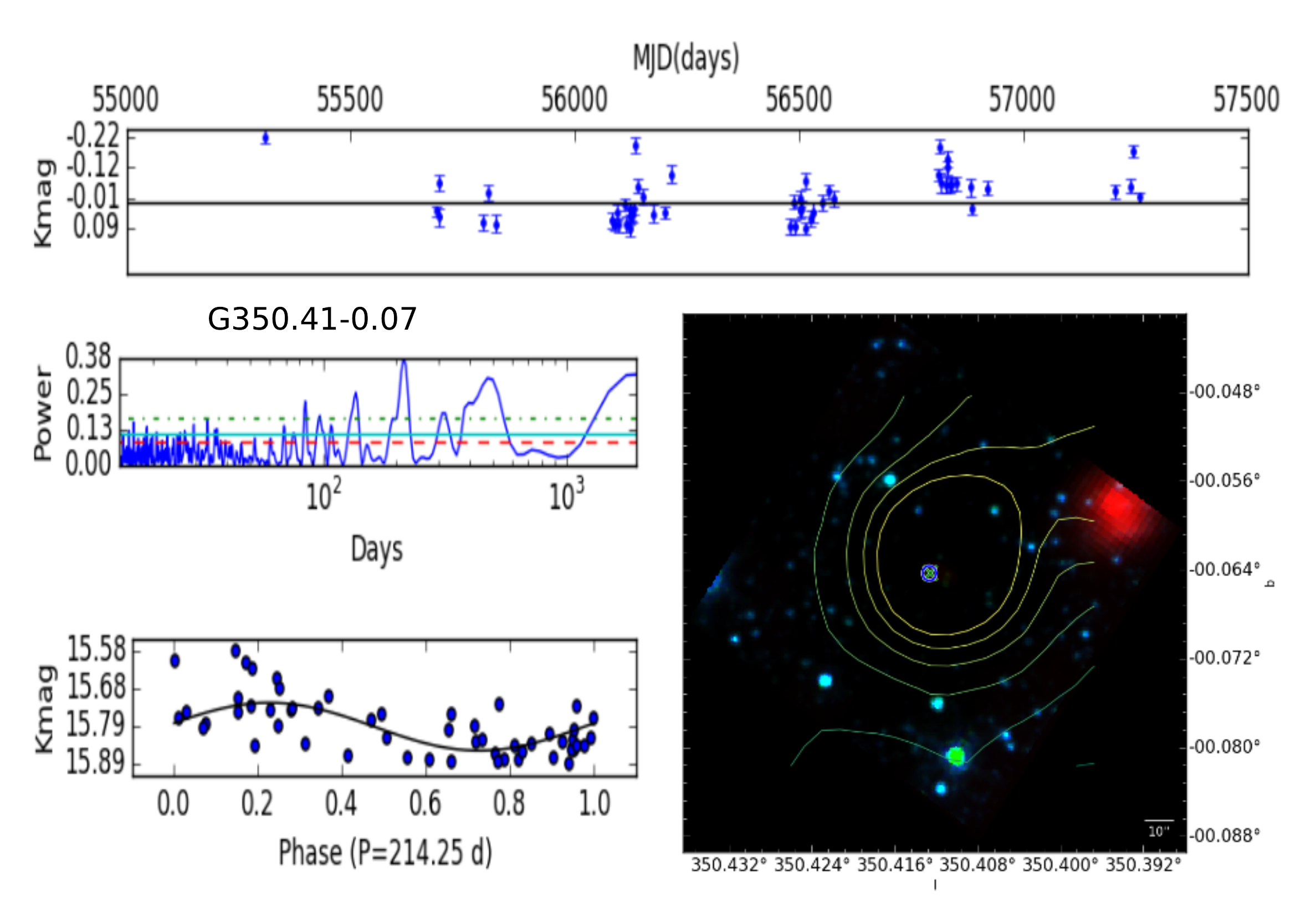}\\
                \includegraphics[width=0.48\textwidth]{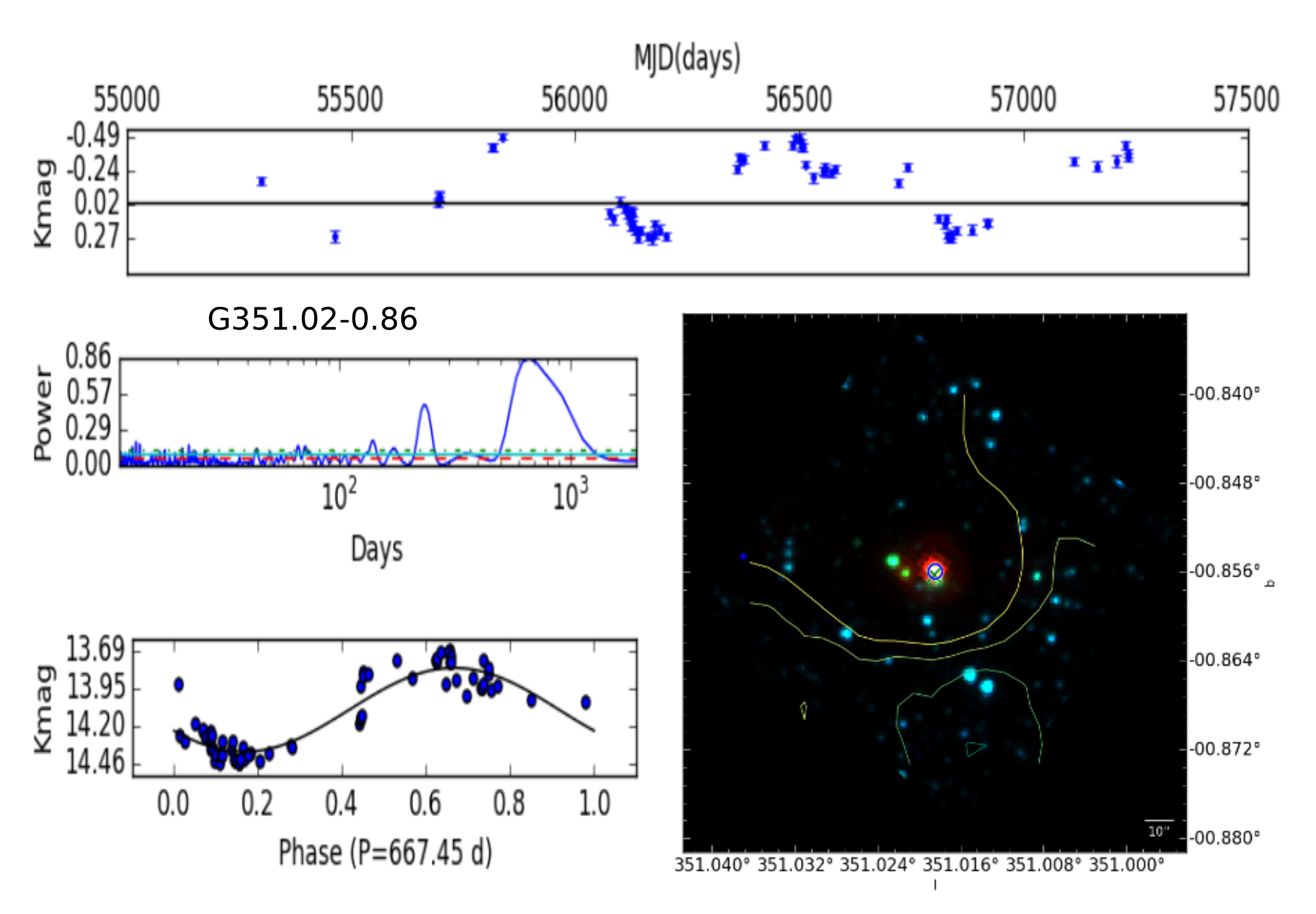}&
                \includegraphics[width=0.48\textwidth]{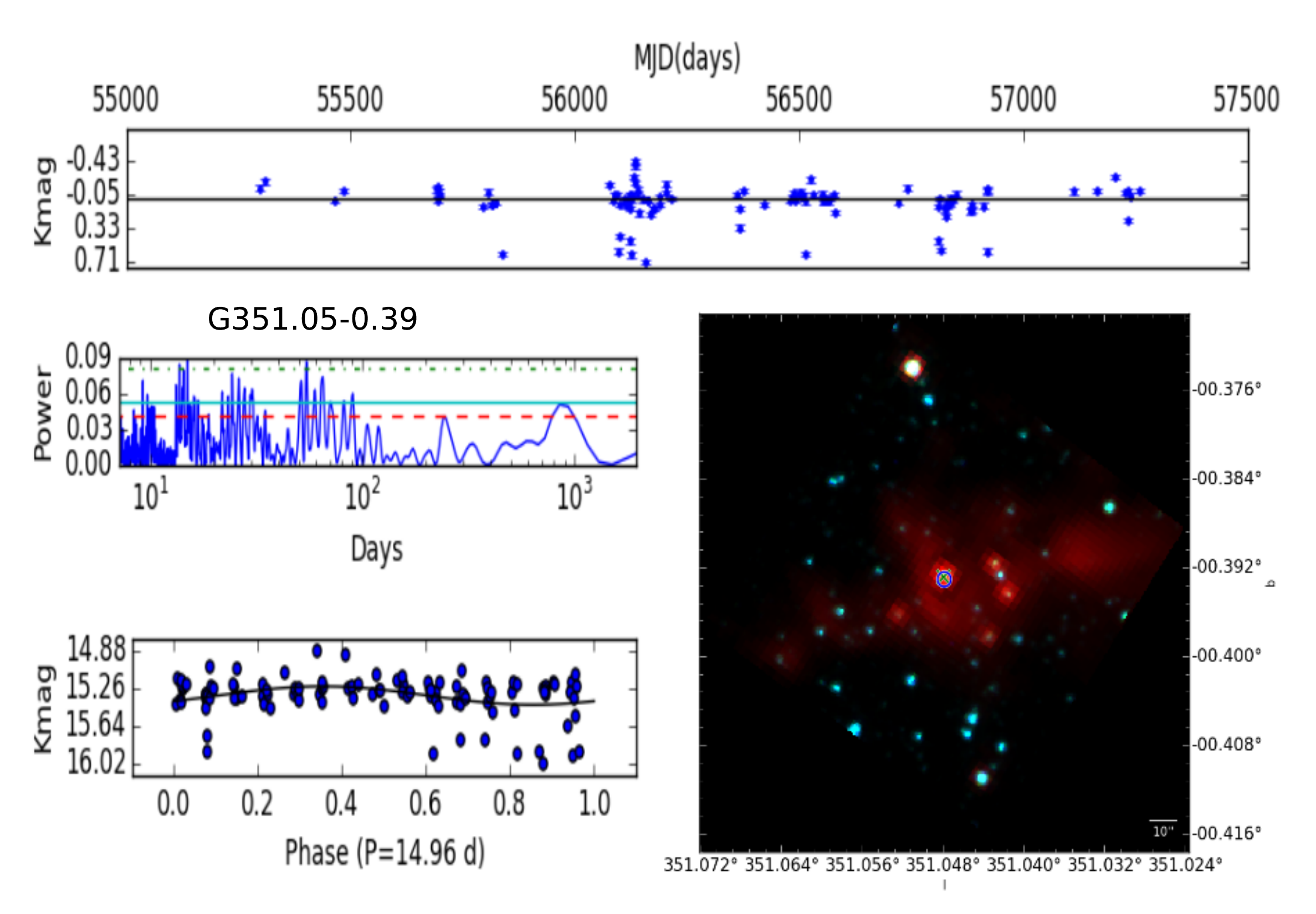}\\
                \includegraphics[width=0.48\textwidth]{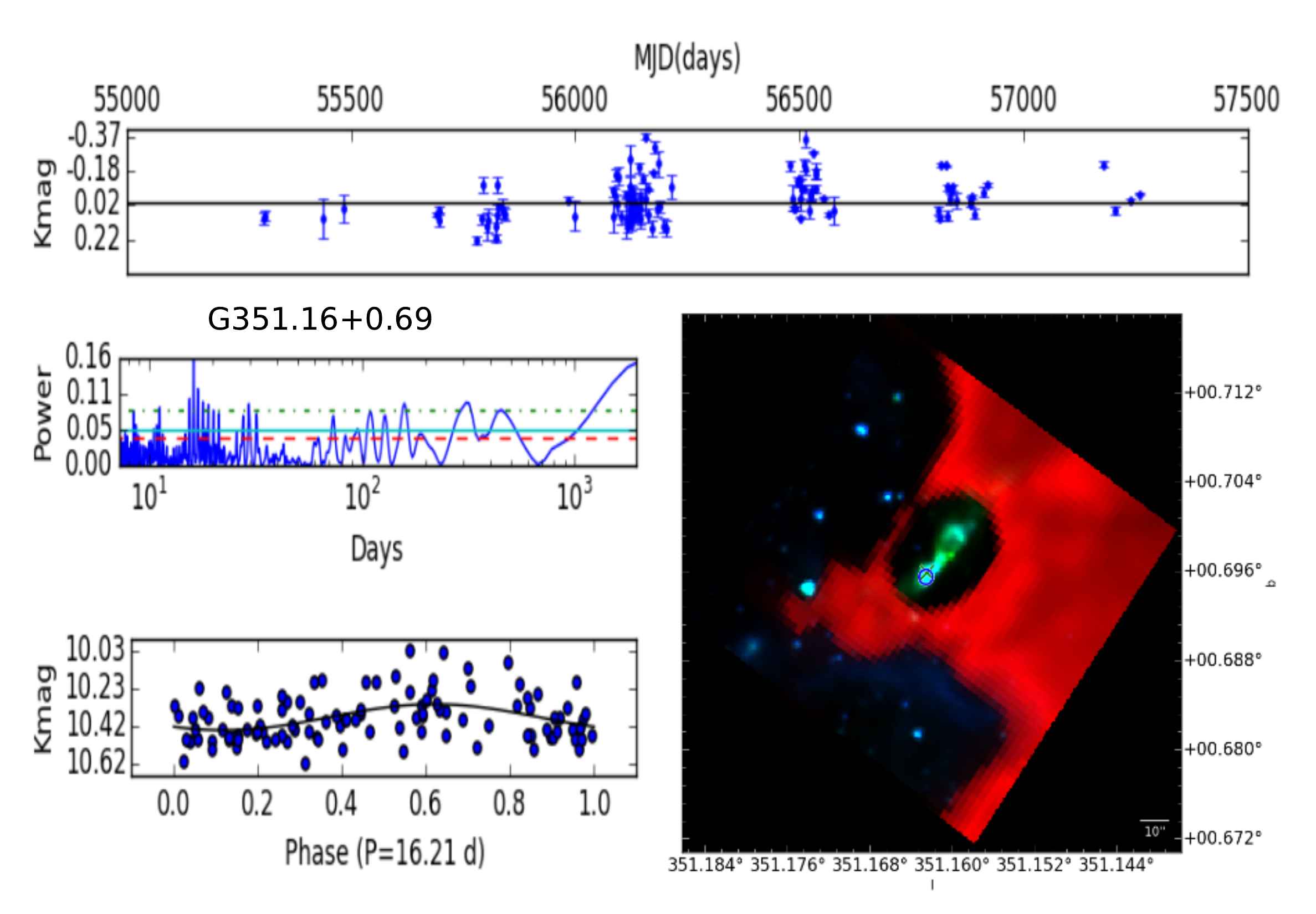}&
                \includegraphics[width=0.48\textwidth]{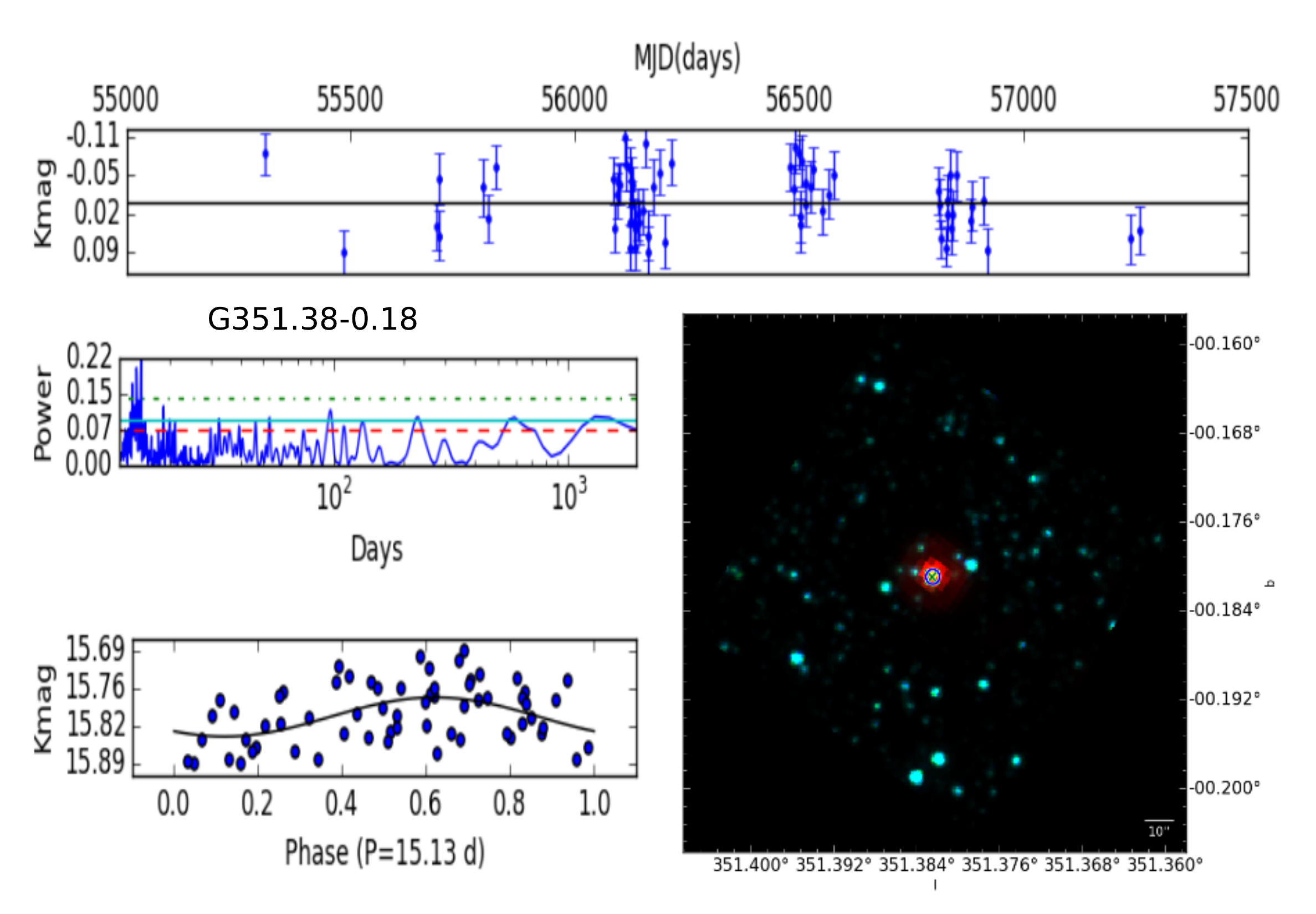}\\

        \end{tabular}
        \caption{Continuation of Fig. \ref{fig:testgrid}.}
\end{figure}

\begin{figure}
        \begin{tabular}{cc}
                \includegraphics[width=0.48\textwidth]{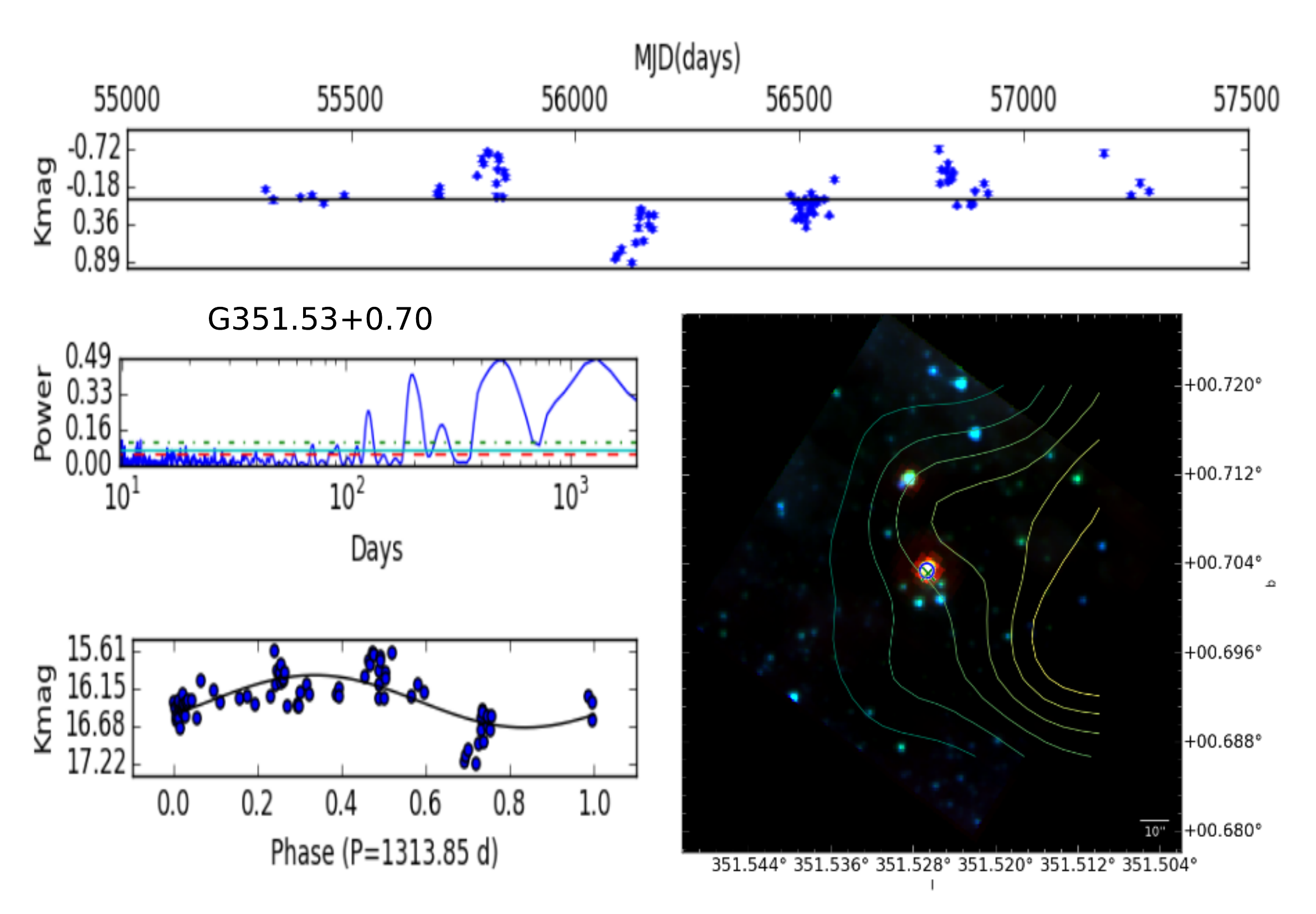}&
                \includegraphics[width=0.48\textwidth]{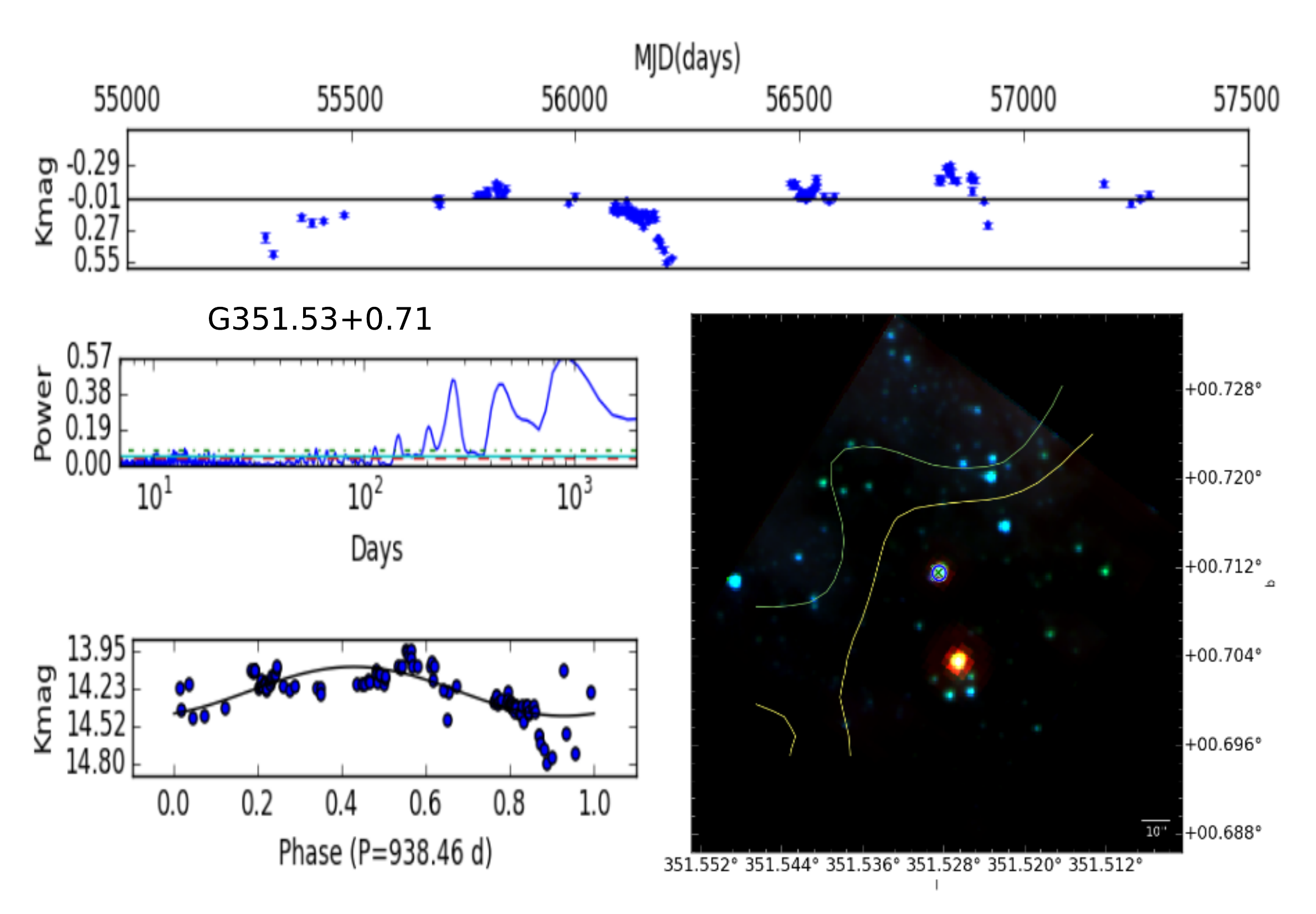}\\
                \includegraphics[width=0.48\textwidth]{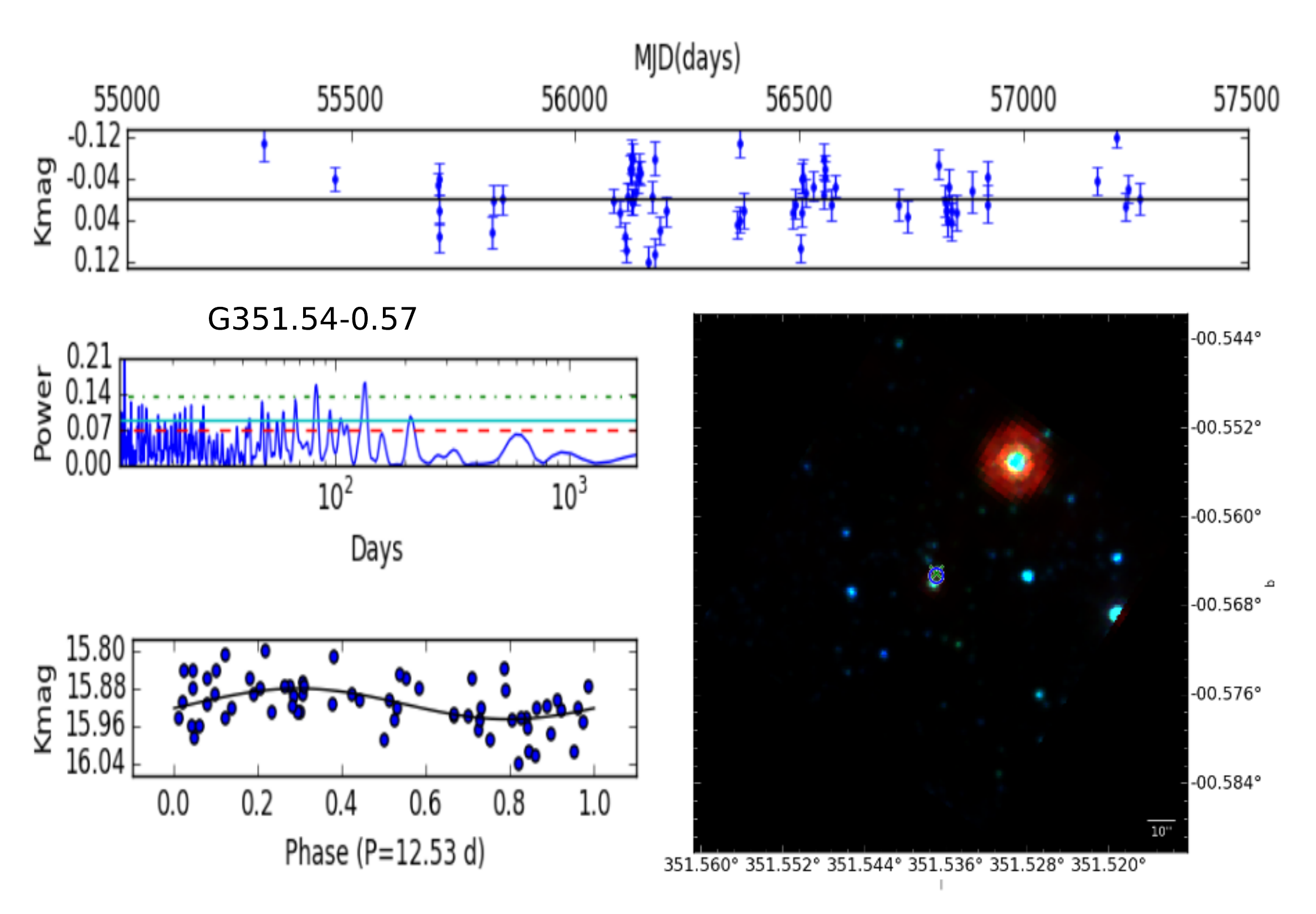}&
                \includegraphics[width=0.48\textwidth]{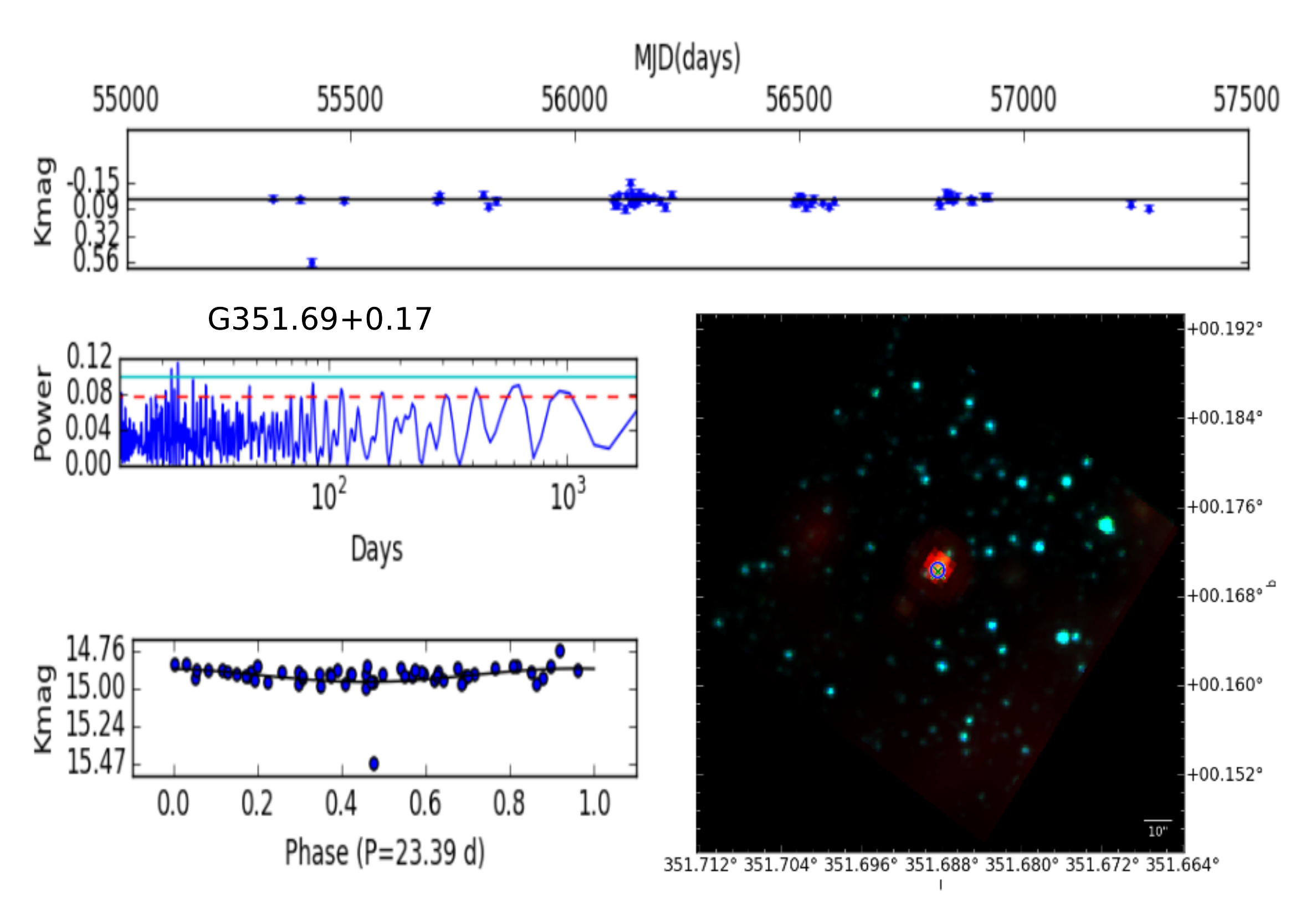}\\
                \includegraphics[width=0.48\textwidth]{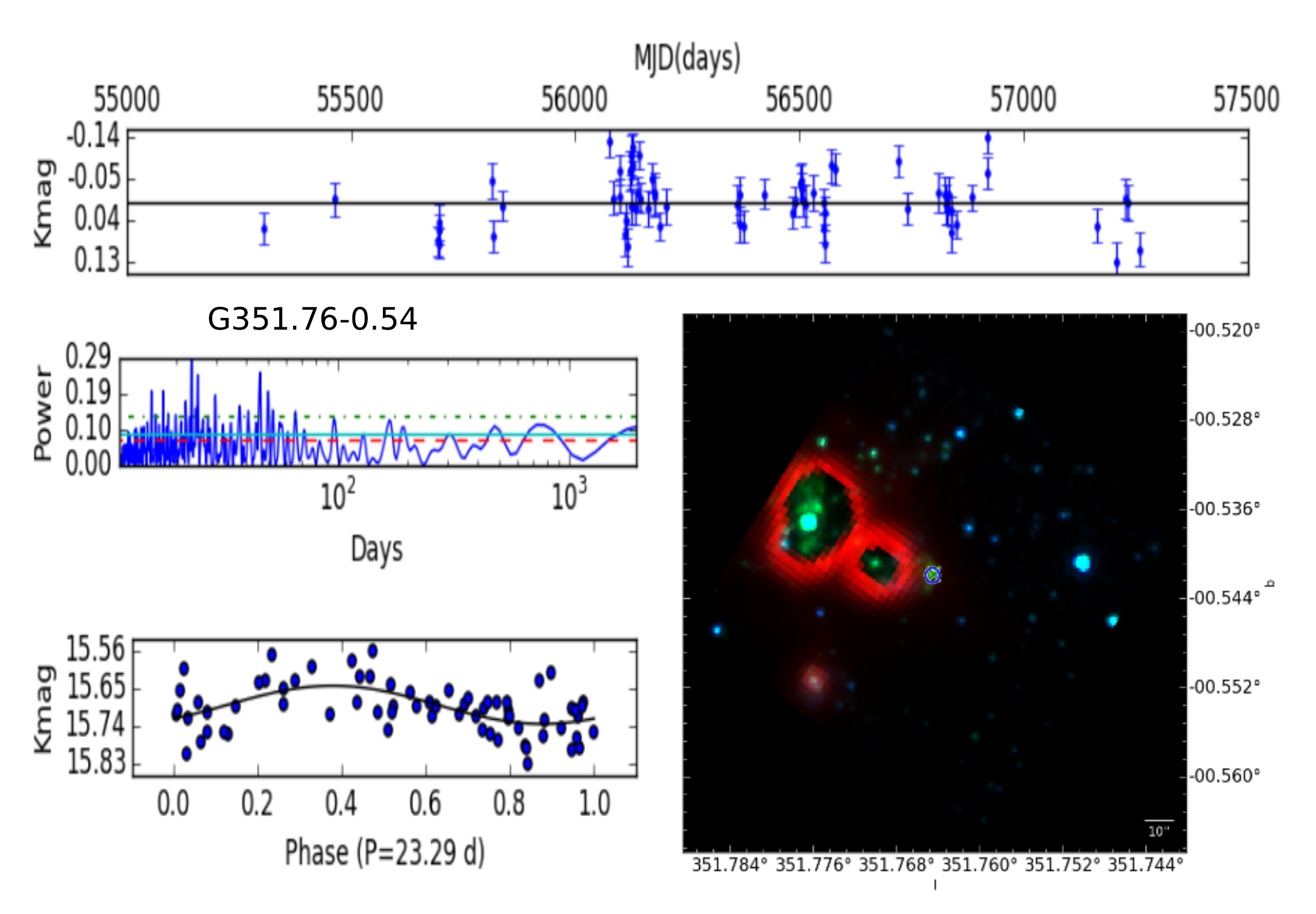}&
                \includegraphics[width=0.48\textwidth]{{./G351.78-0.54_combined}.pdf}\\
                \includegraphics[width=0.48\textwidth]{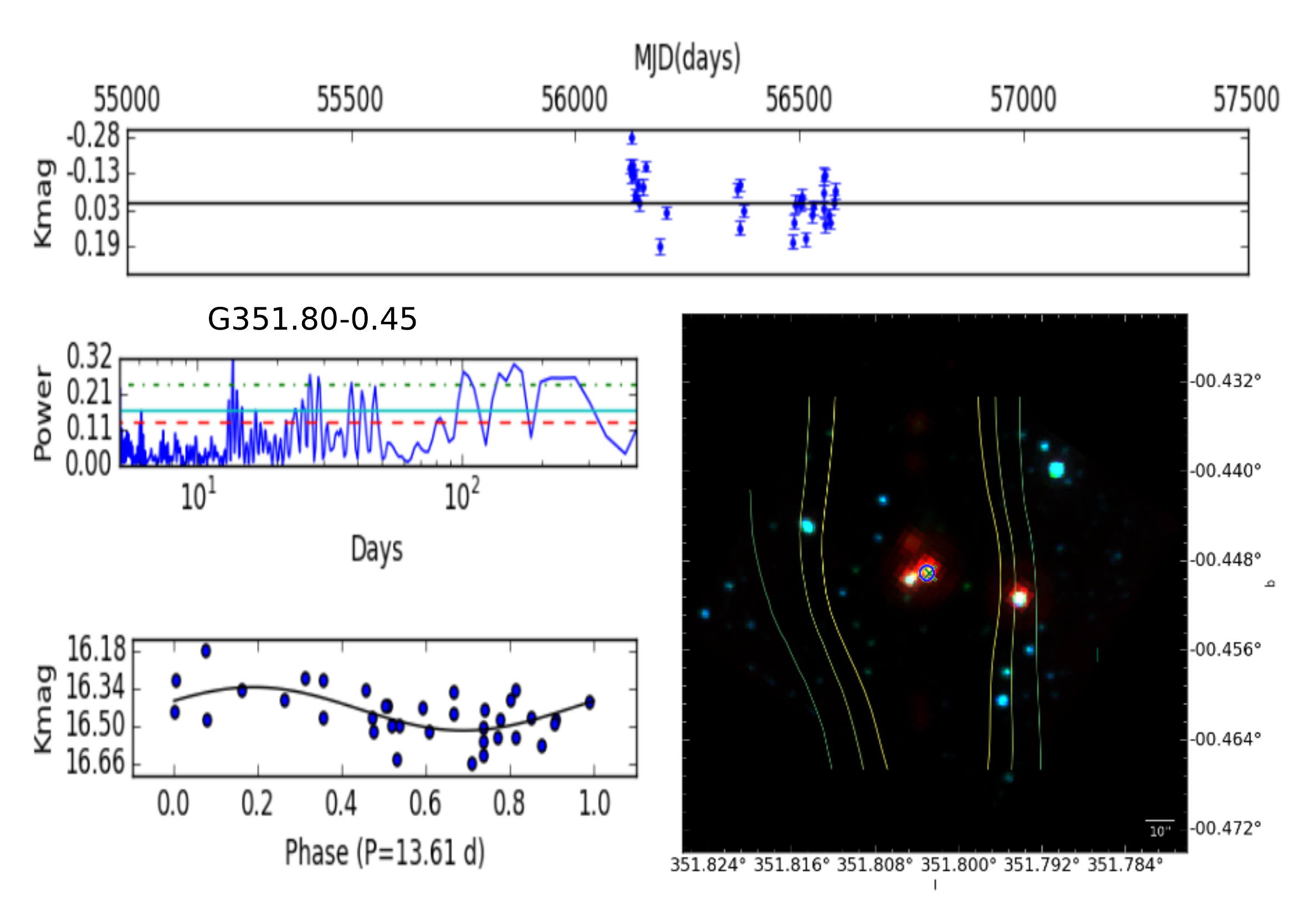}&
                \includegraphics[width=0.48\textwidth]{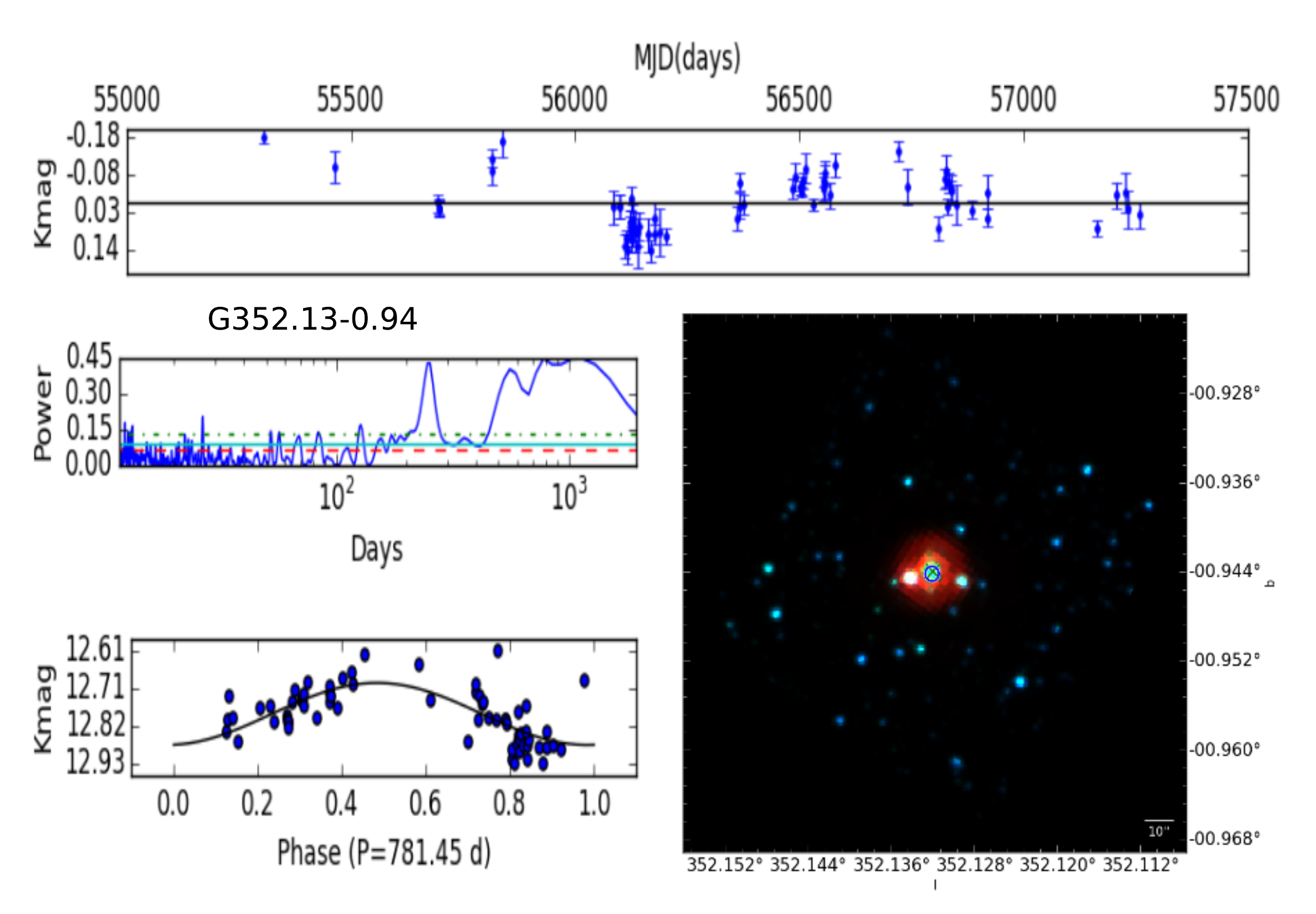}\\
        \end{tabular}
        \caption{Continuation of Fig. \ref{fig:testgrid}.}
\end{figure}

\begin{figure}
        \begin{tabular}{cc}

                \includegraphics[width=0.48\textwidth]{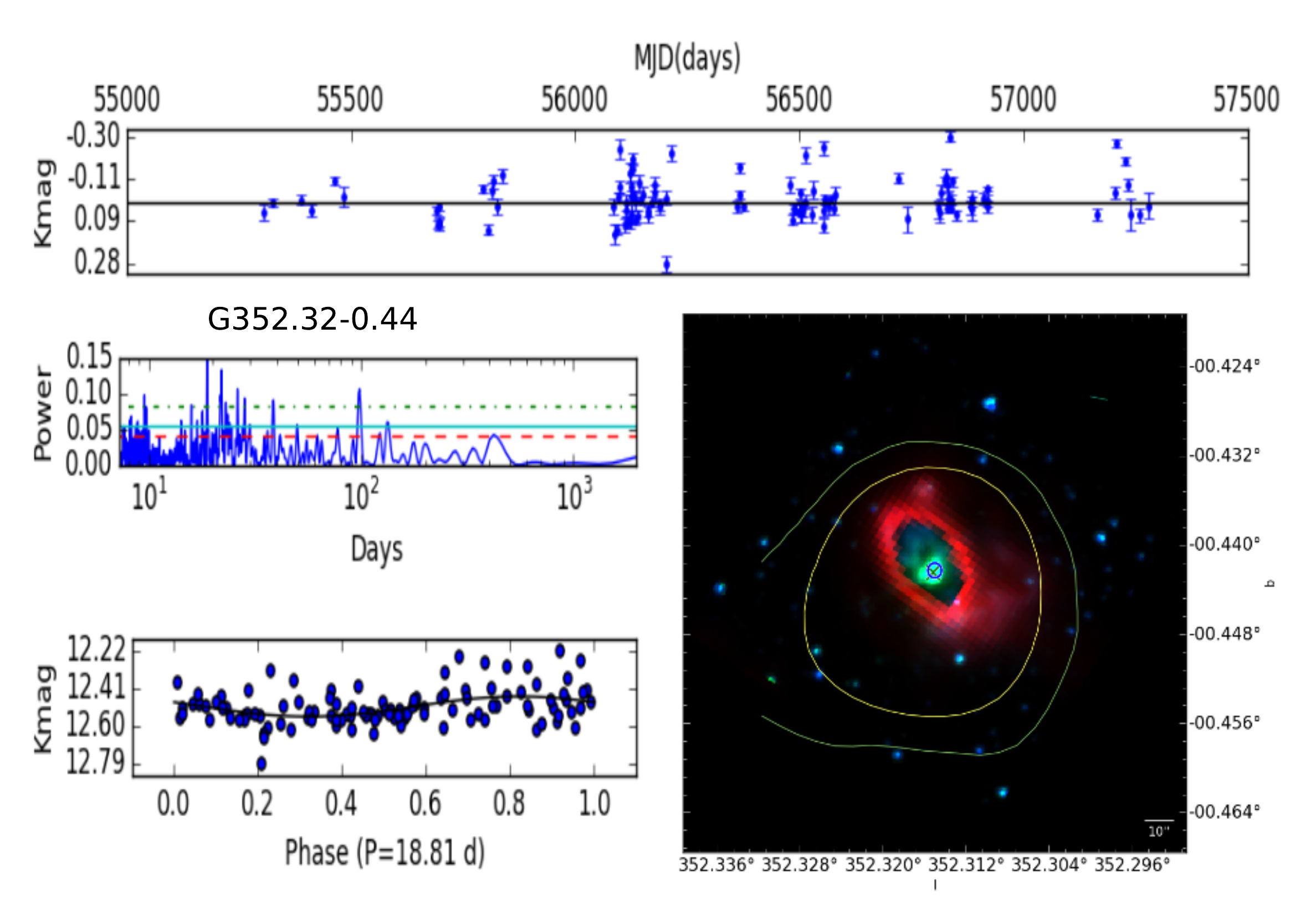}&
                \includegraphics[width=0.48\textwidth]{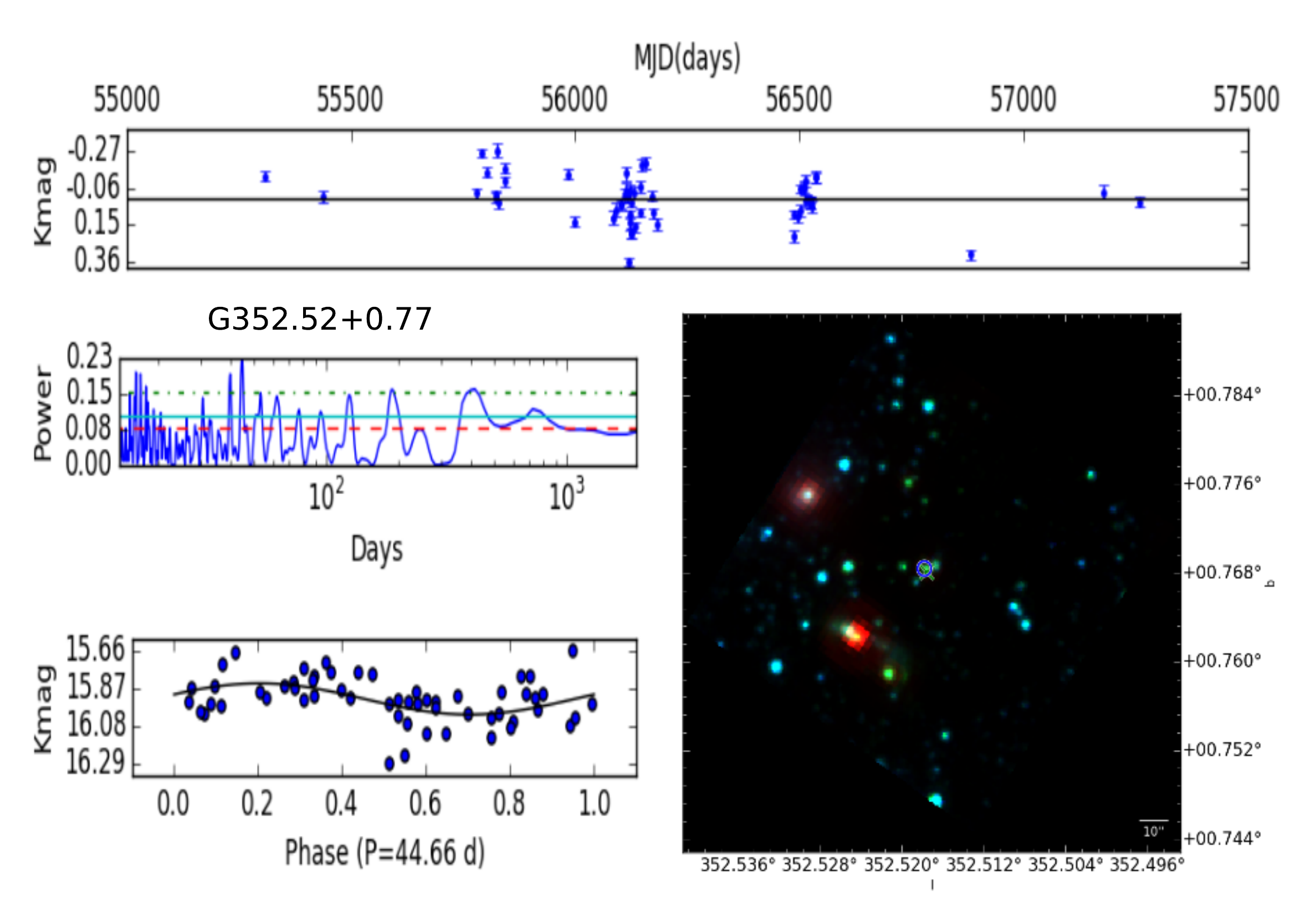}\\
                \includegraphics[width=0.48\textwidth]{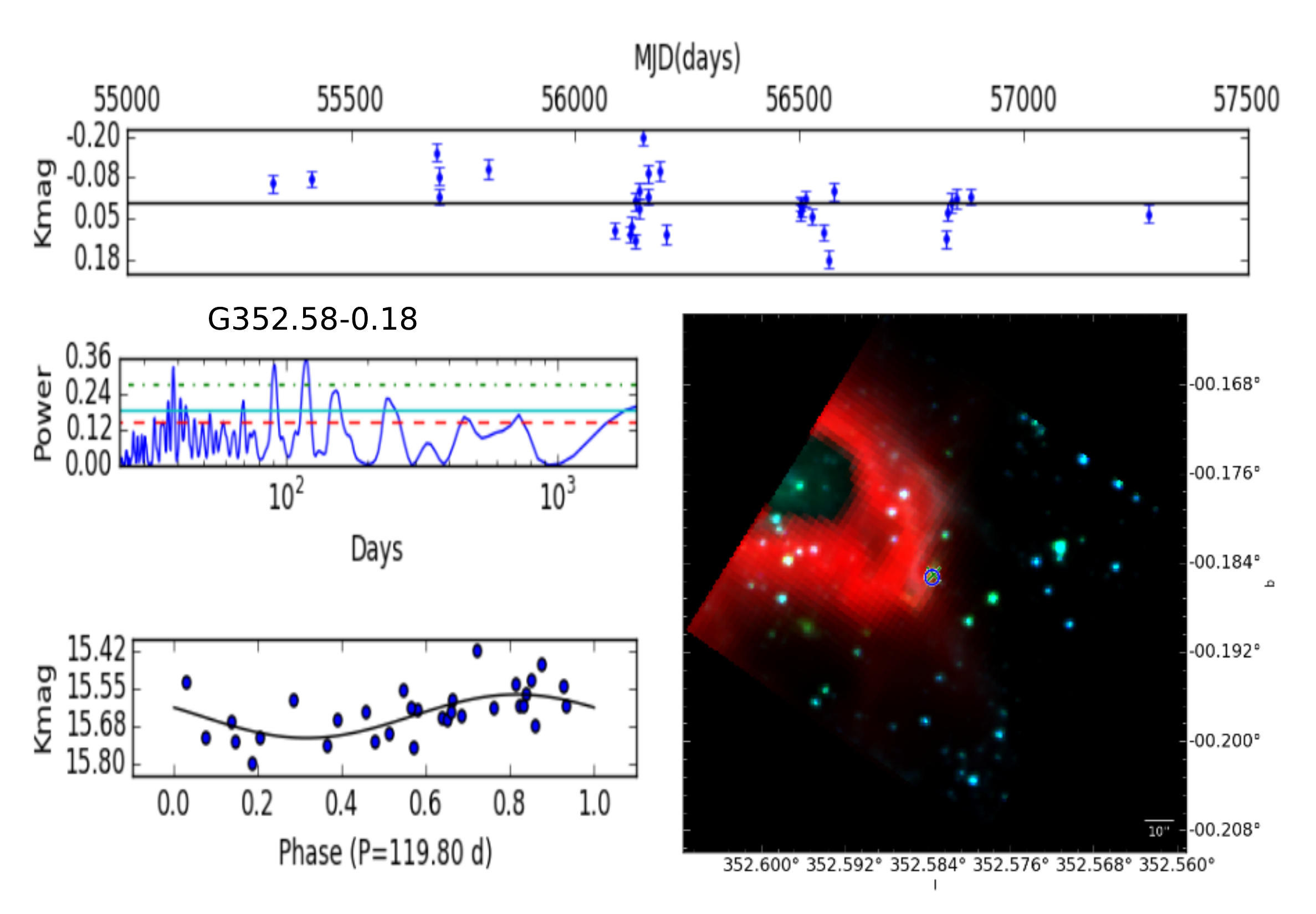}&
                \includegraphics[width=0.48\textwidth]{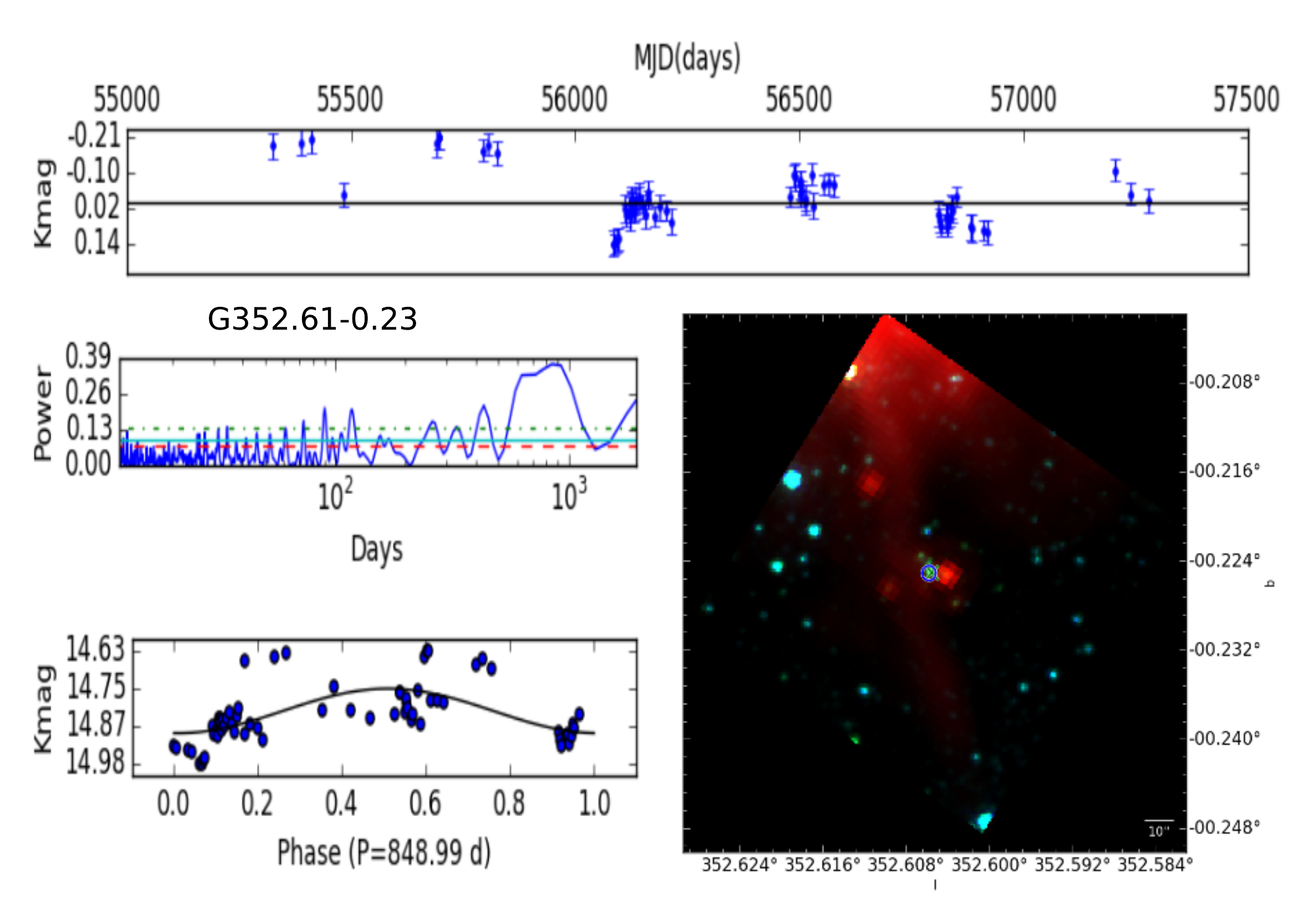}\\
                \includegraphics[width=0.48\textwidth]{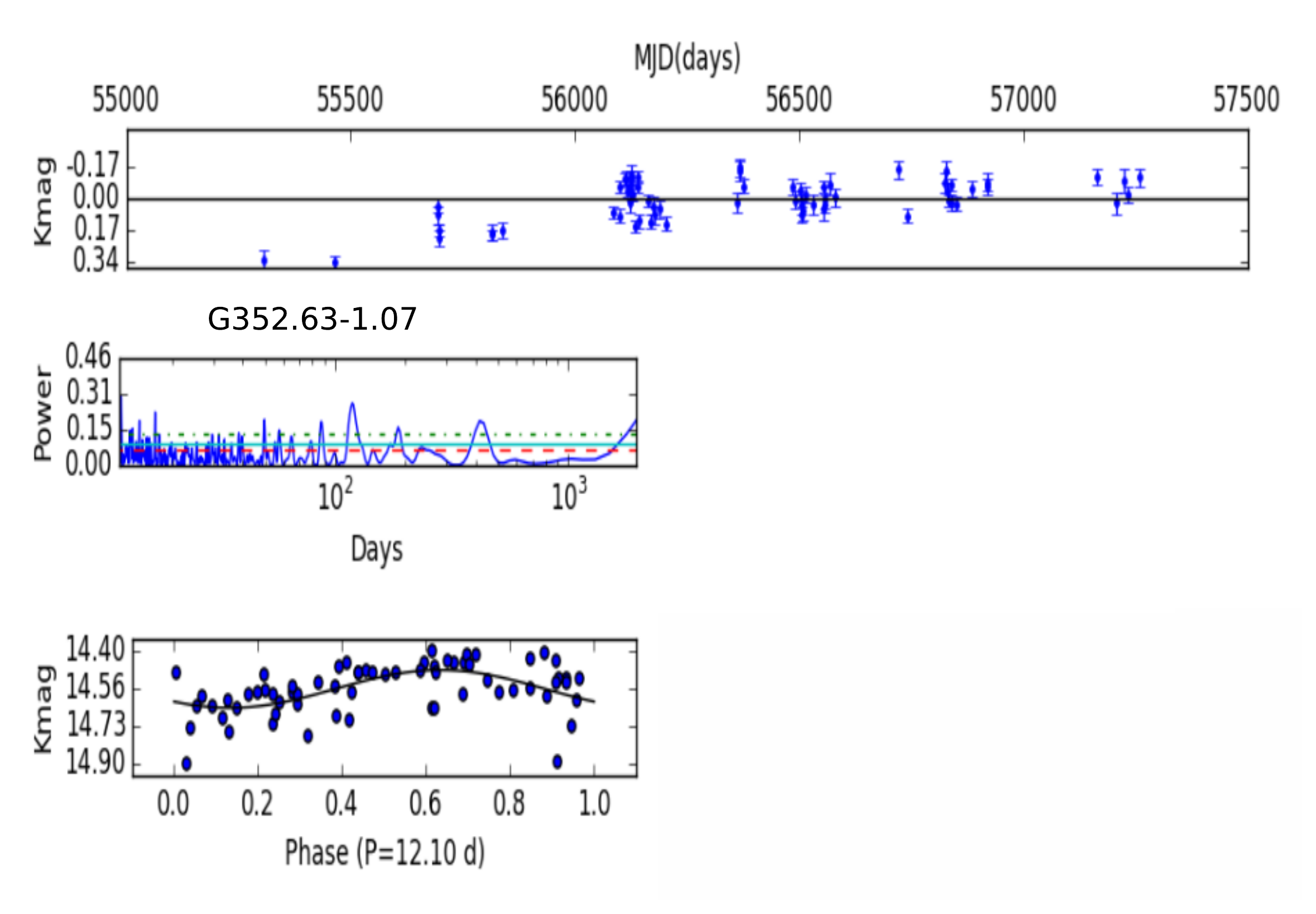}&
                \includegraphics[width=0.48\textwidth]{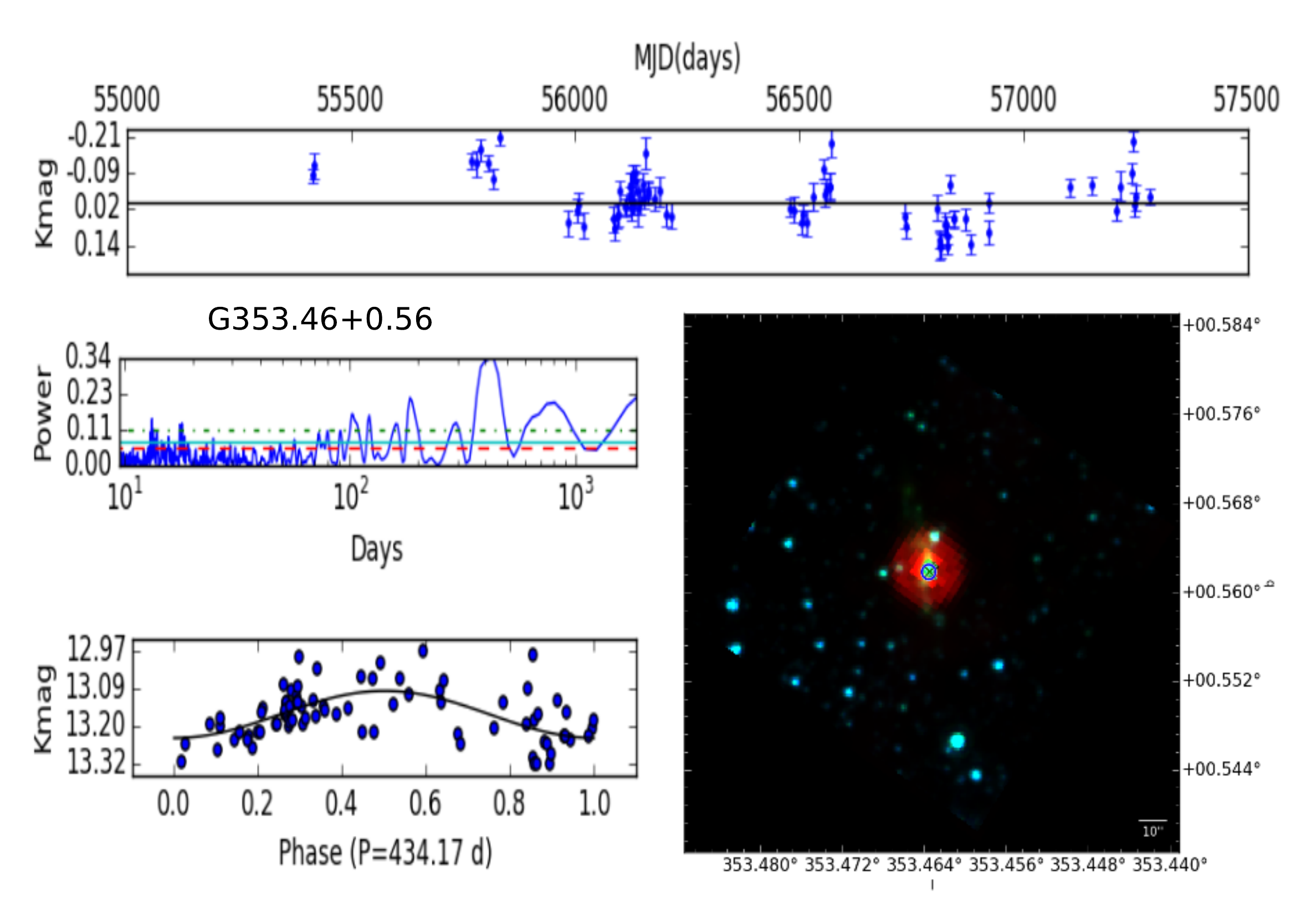}\\
                \includegraphics[width=0.48\textwidth]{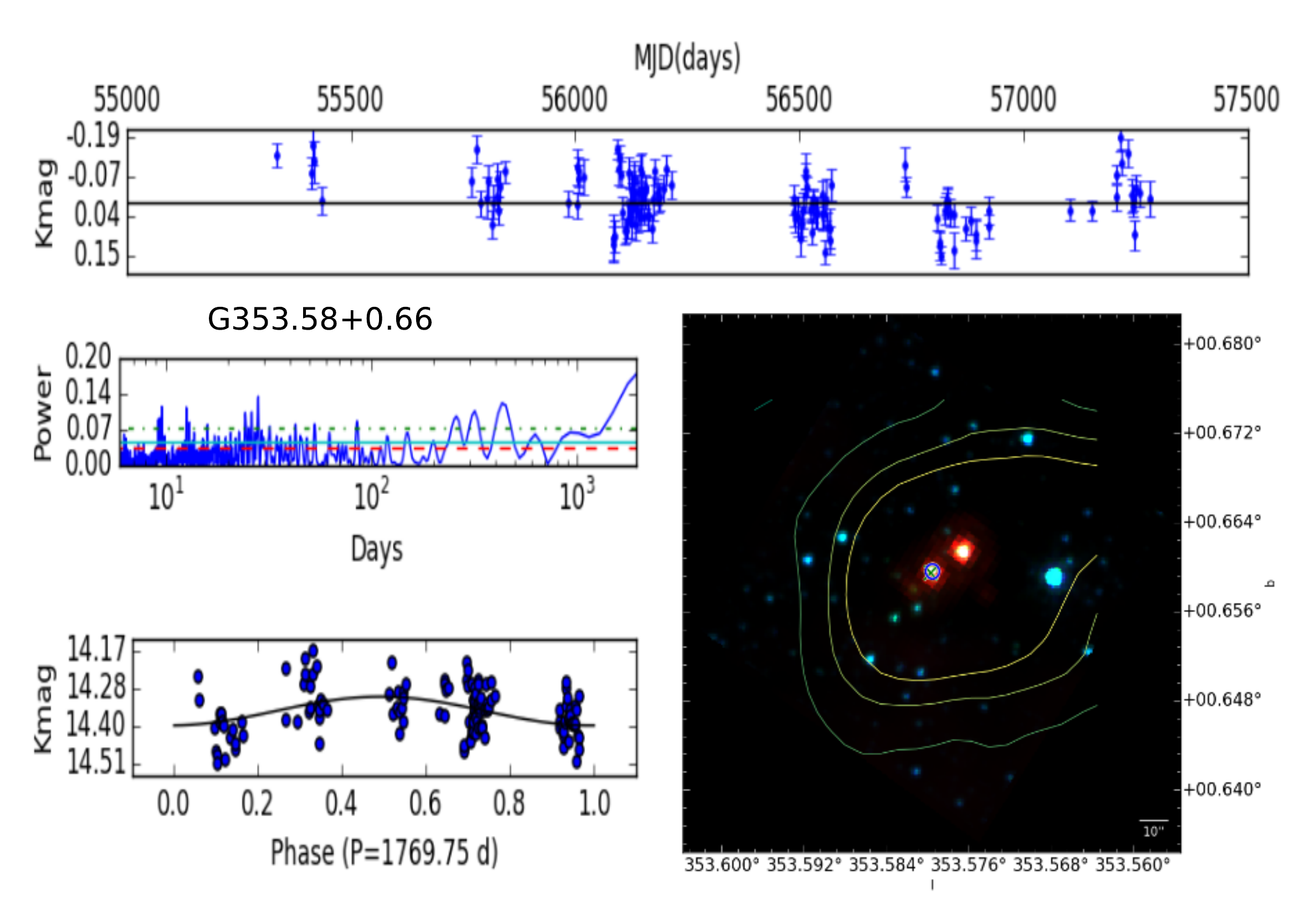}&            
                \includegraphics[width=0.48\textwidth]{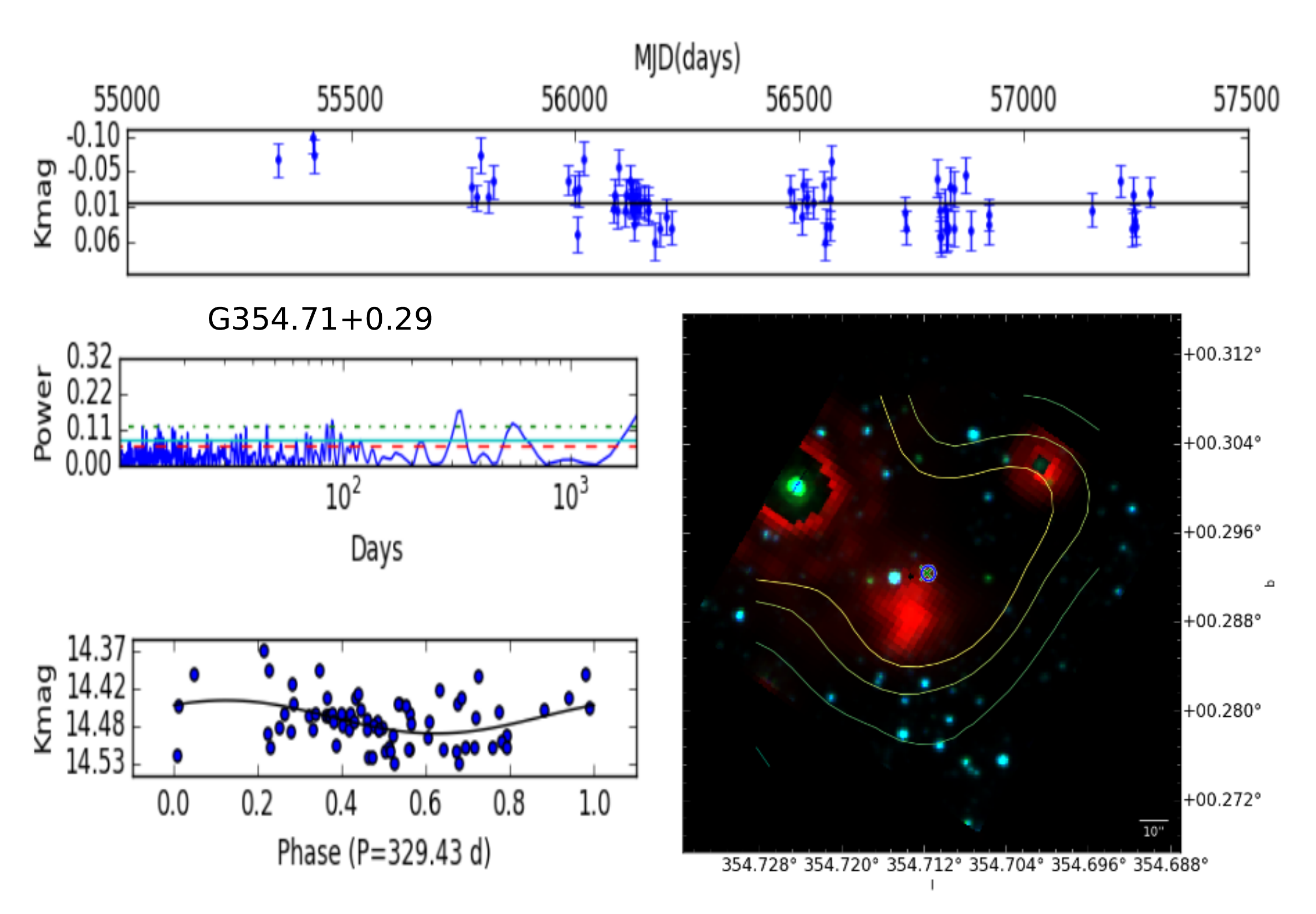}\\
                
        \end{tabular}
        \caption{Continuation of Fig. \ref{fig:testgrid}.}
\end{figure}

\begin{figure}
        \begin{tabular}{cc}
                \includegraphics[width=0.48\textwidth]{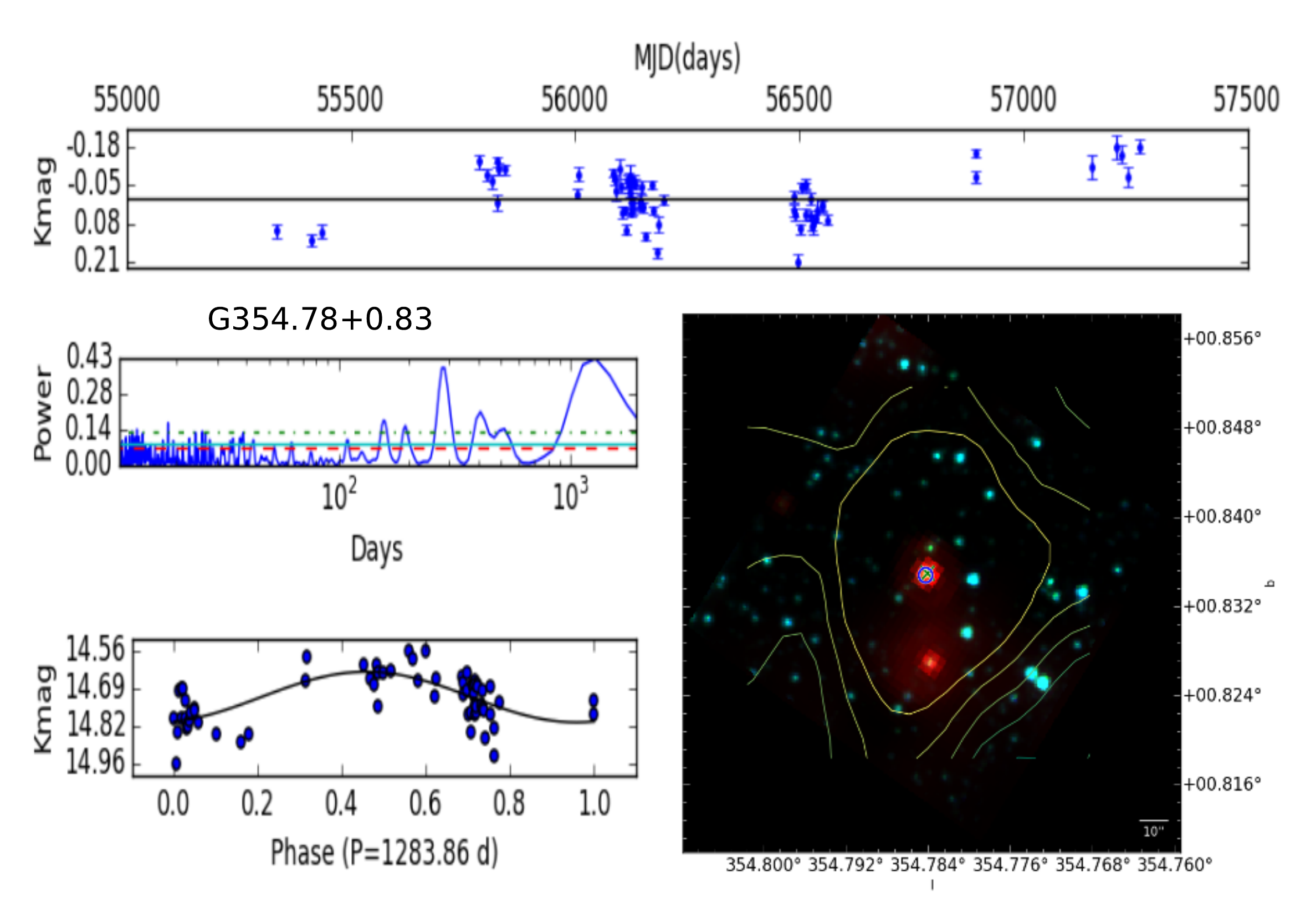}&
                \includegraphics[width=0.48\textwidth]{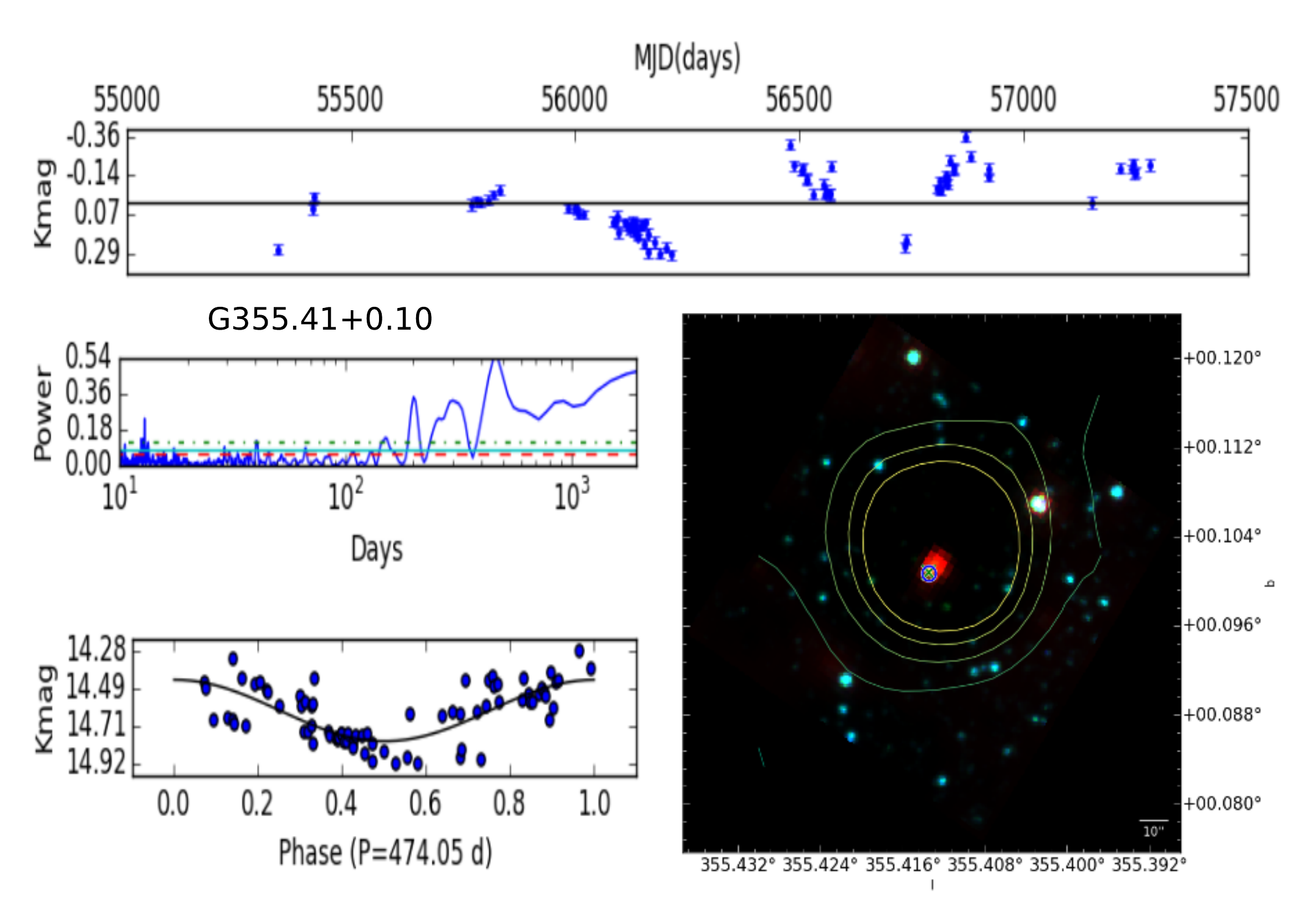}\\
                \includegraphics[width=0.48\textwidth]{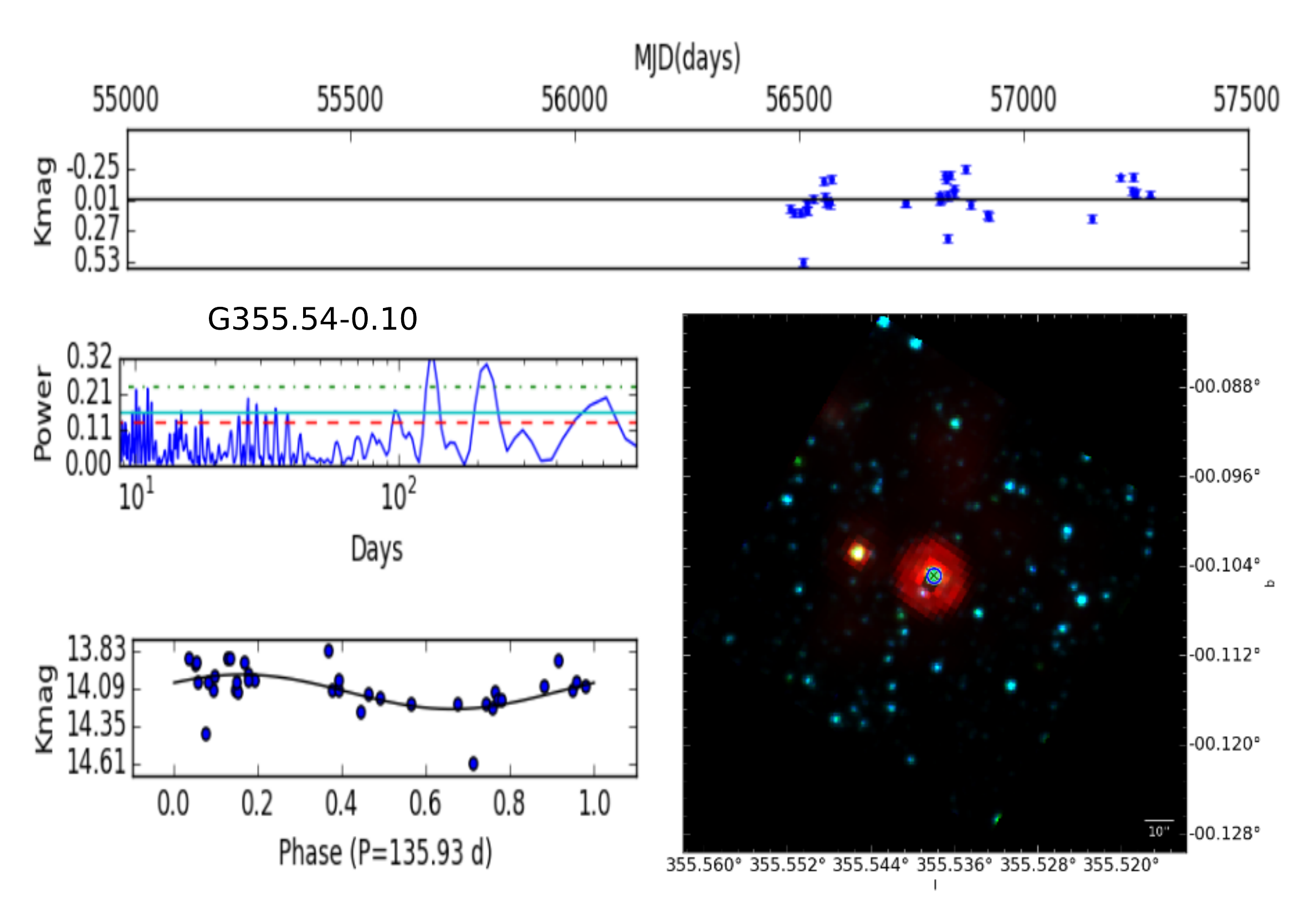}&
                \includegraphics[width=0.48\textwidth]{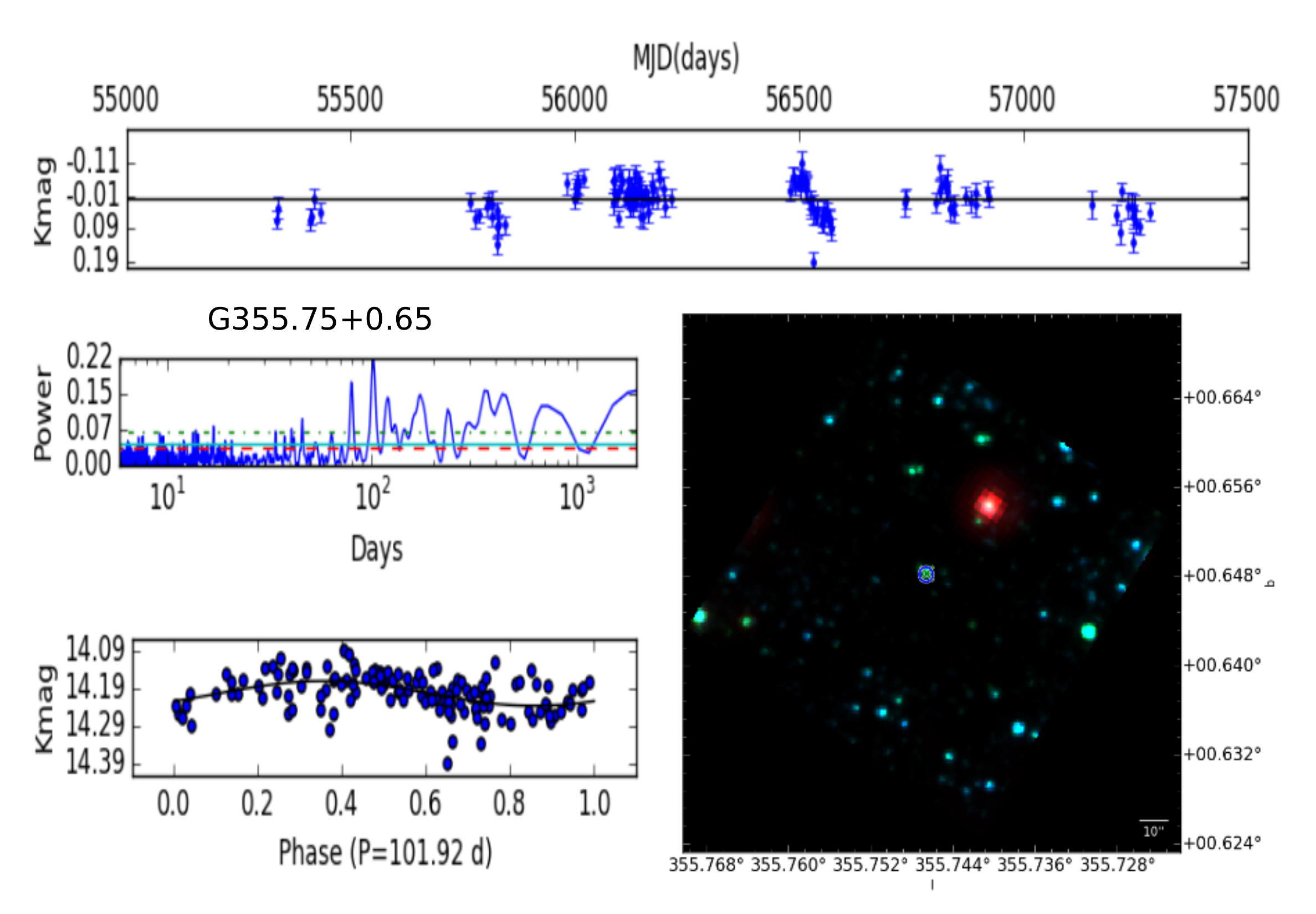}\\
                \includegraphics[width=0.48\textwidth]{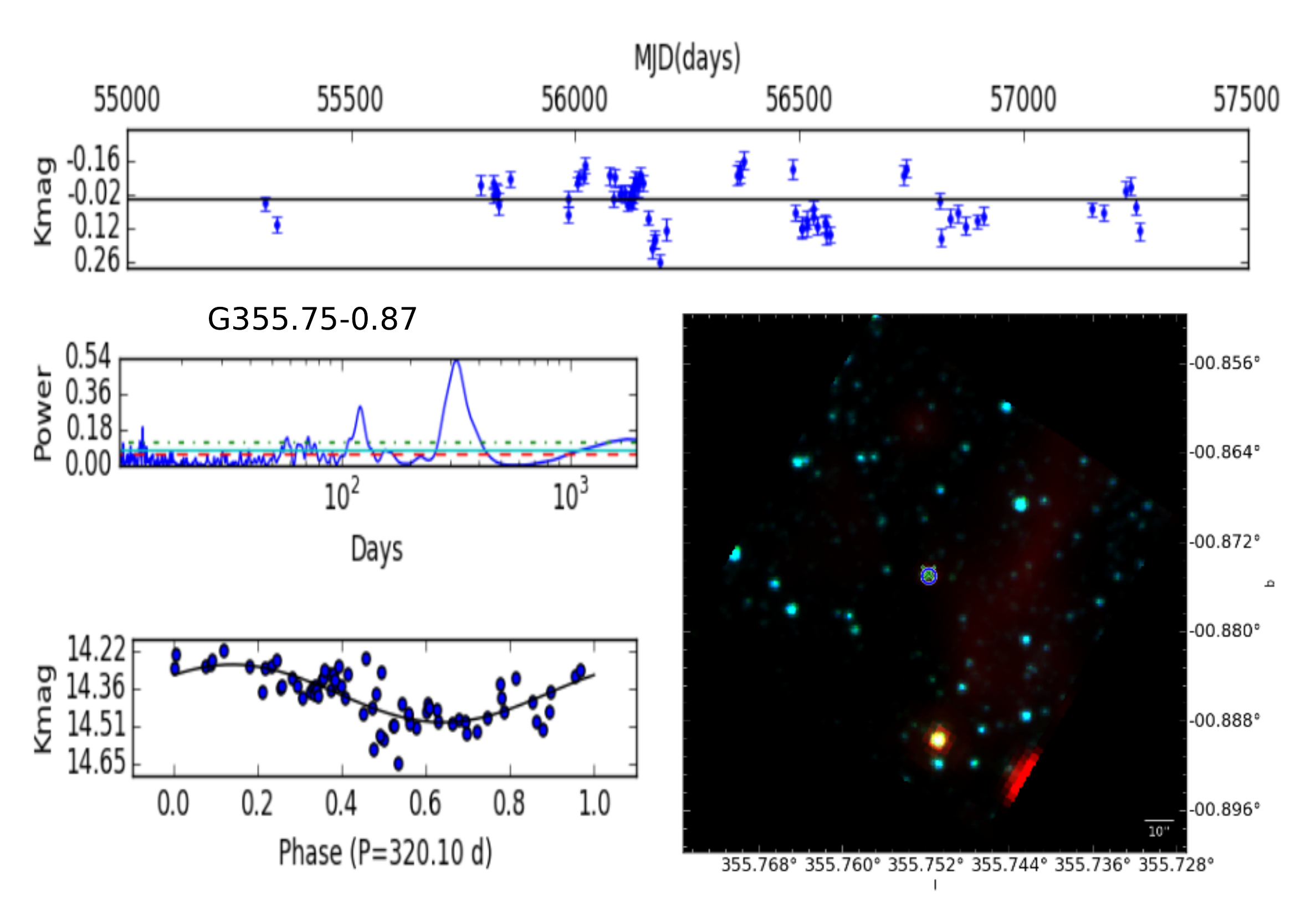}&
                \includegraphics[width=0.48\textwidth]{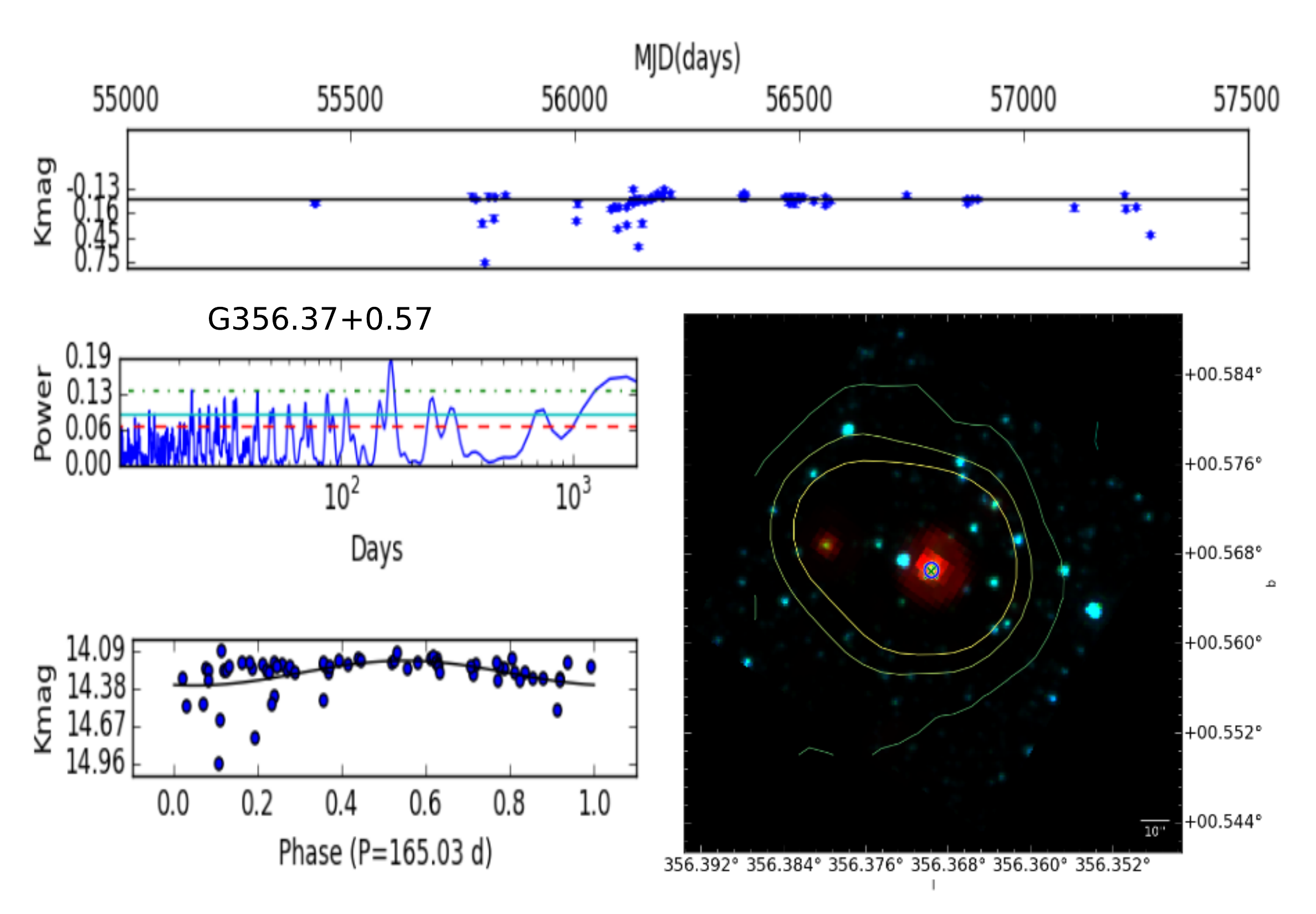}\\
                \includegraphics[width=0.48\textwidth]{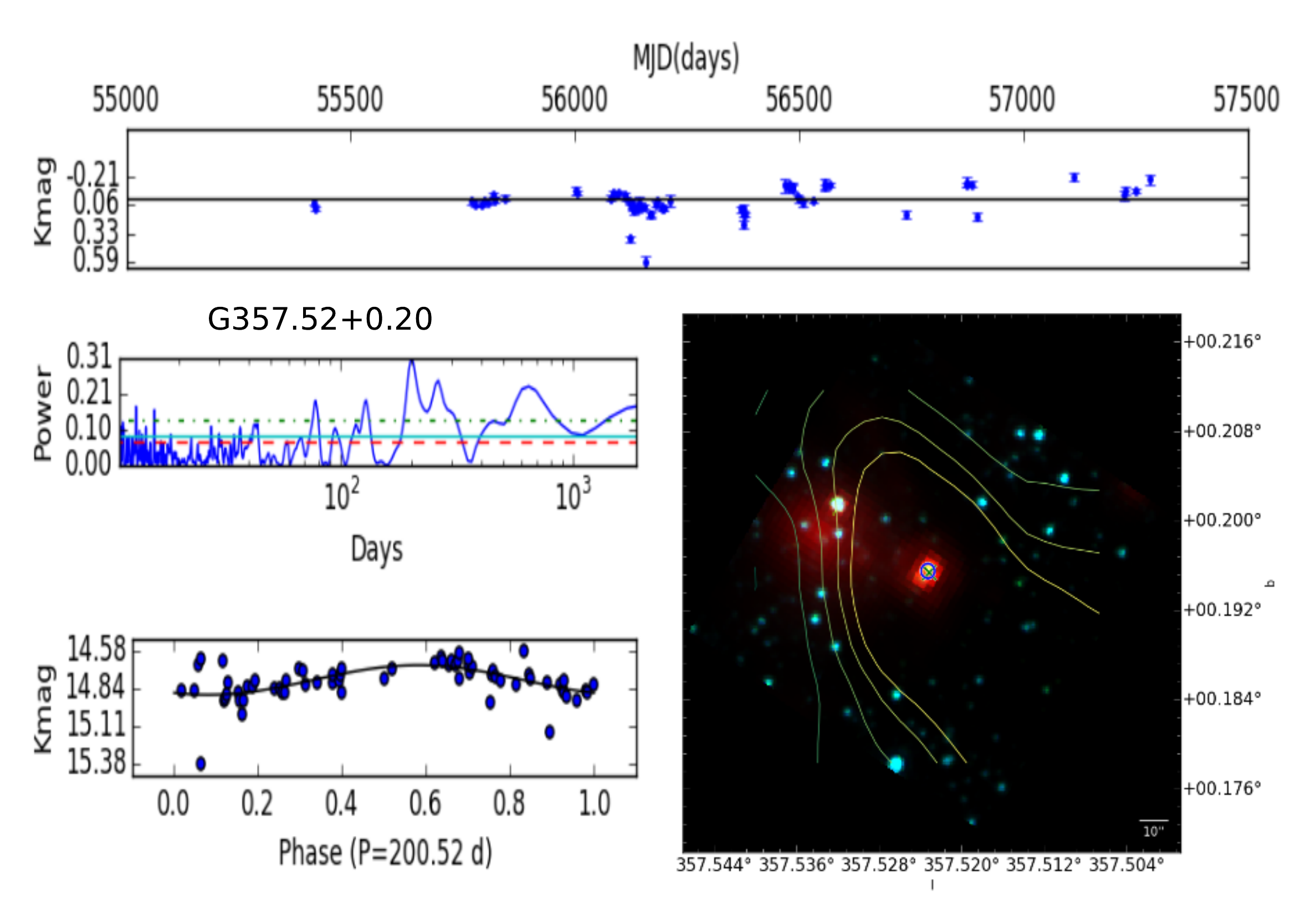}&
                \includegraphics[width=0.48\textwidth]{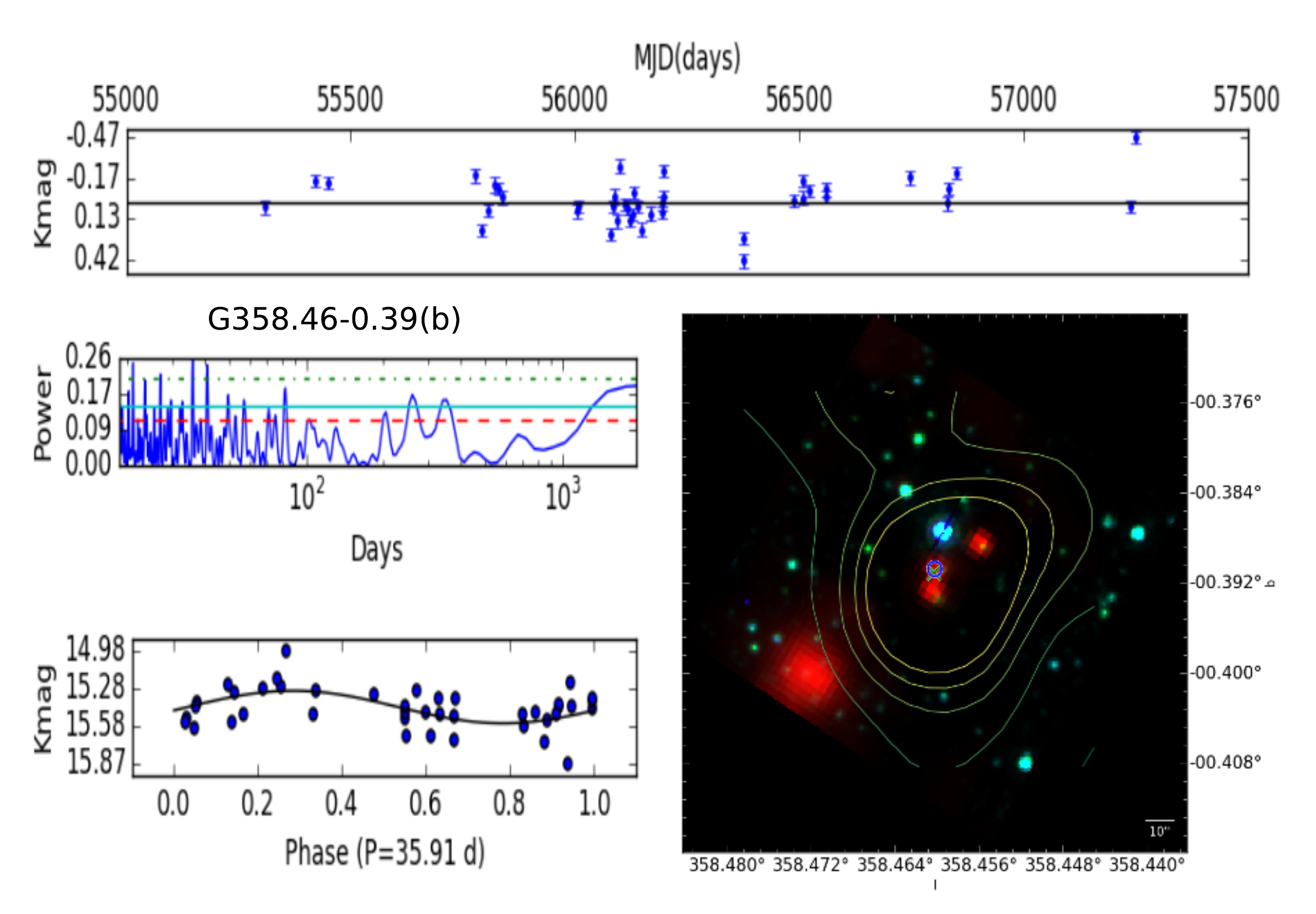}\\
        \end{tabular}
        \caption{Continuation of Fig. \ref{fig:testgrid}.}
\end{figure}    

\begin{figure}
        \begin{tabular}{cc}

                \includegraphics[width=0.48\textwidth]{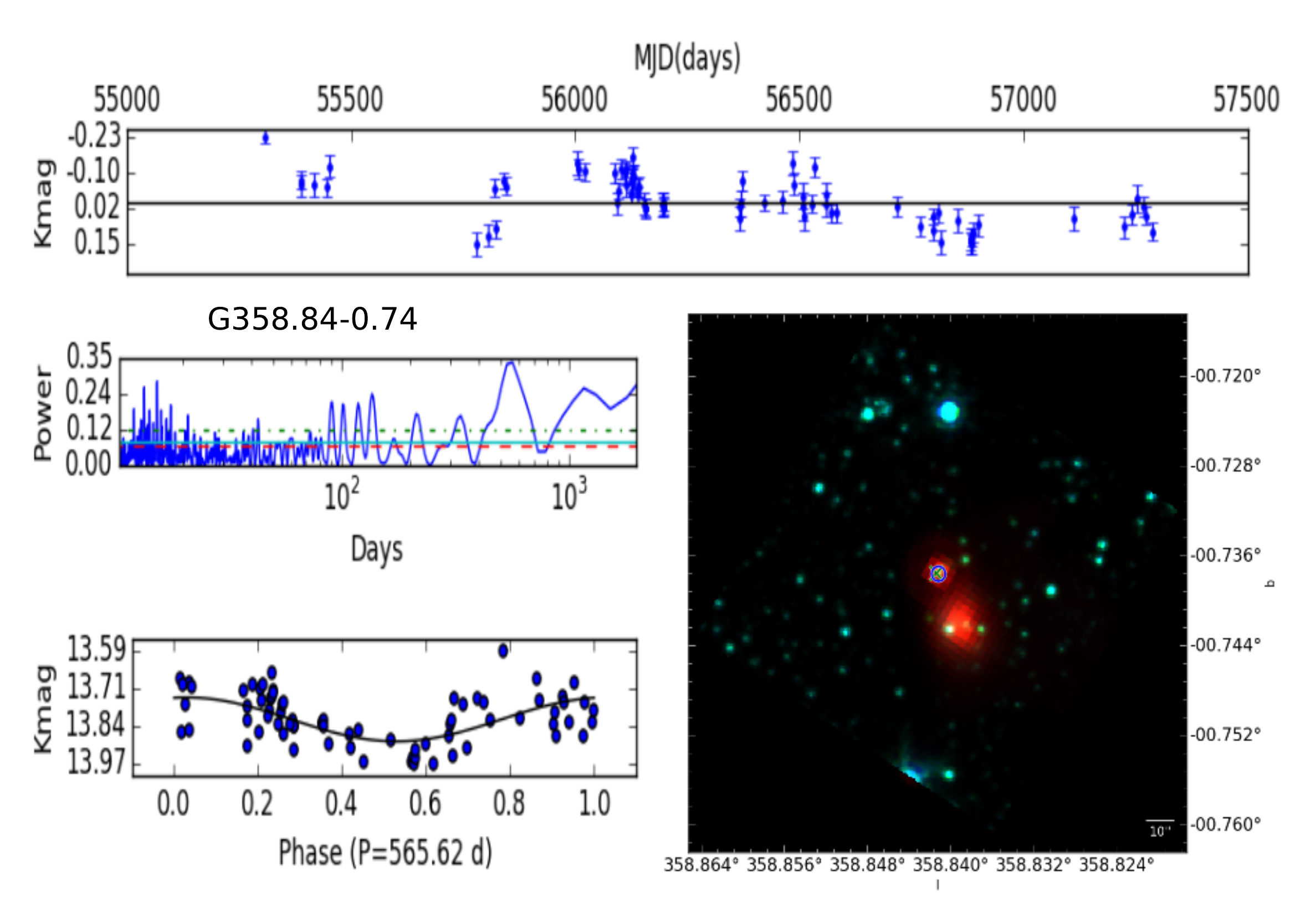}&
                \includegraphics[width=0.48\textwidth]{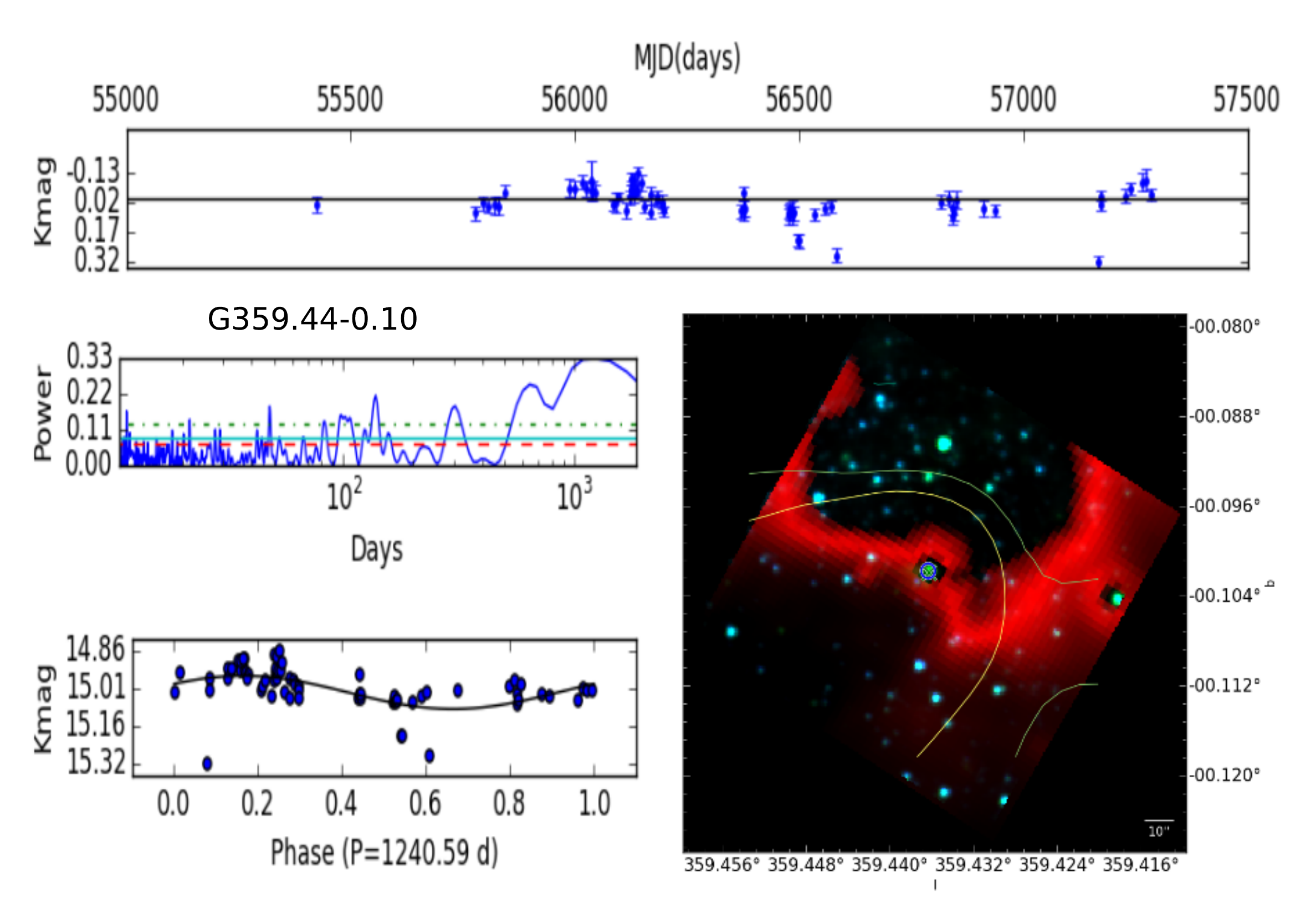}\\
                \includegraphics[width=0.48\textwidth]{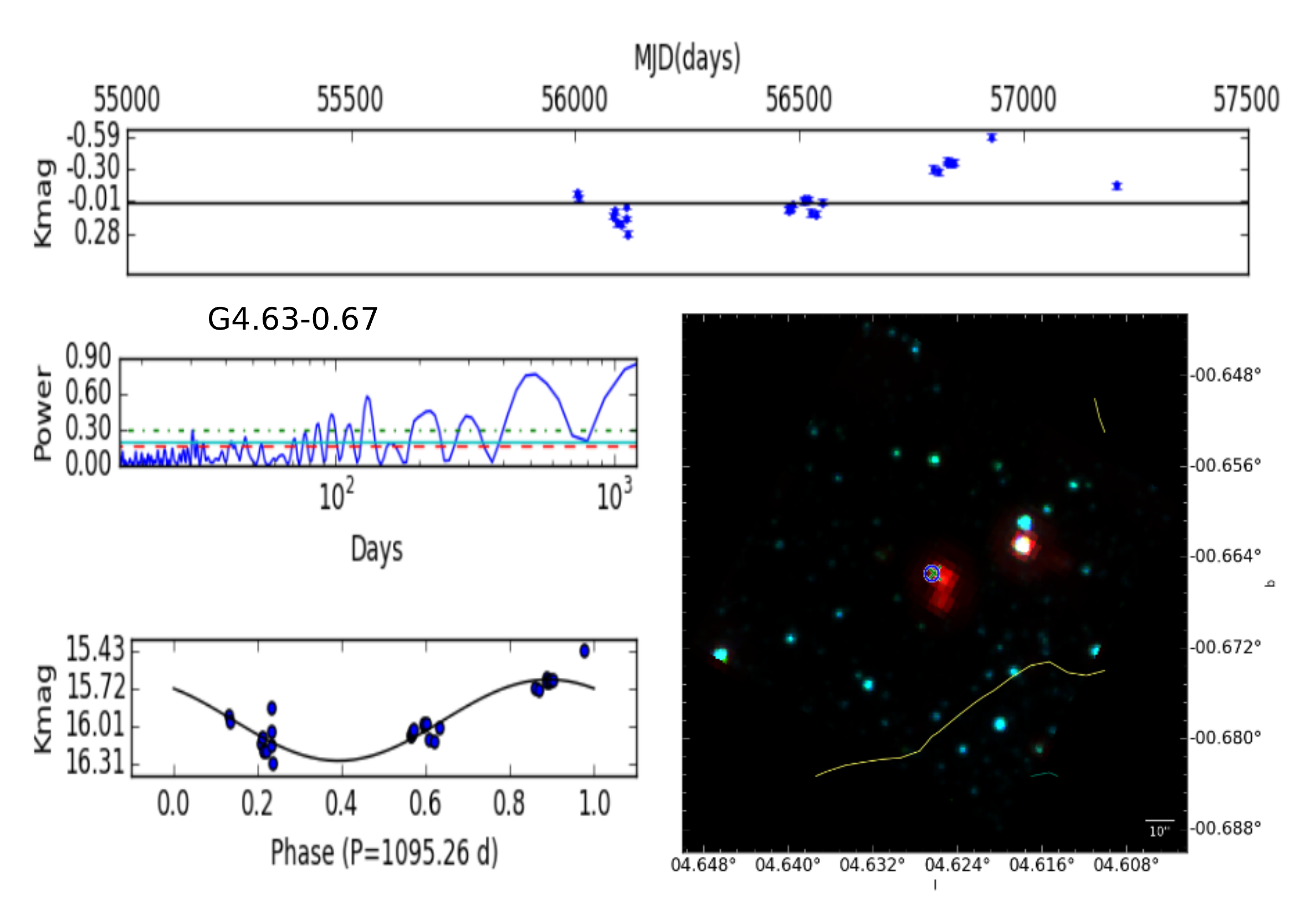}&
                \includegraphics[width=0.48\textwidth]{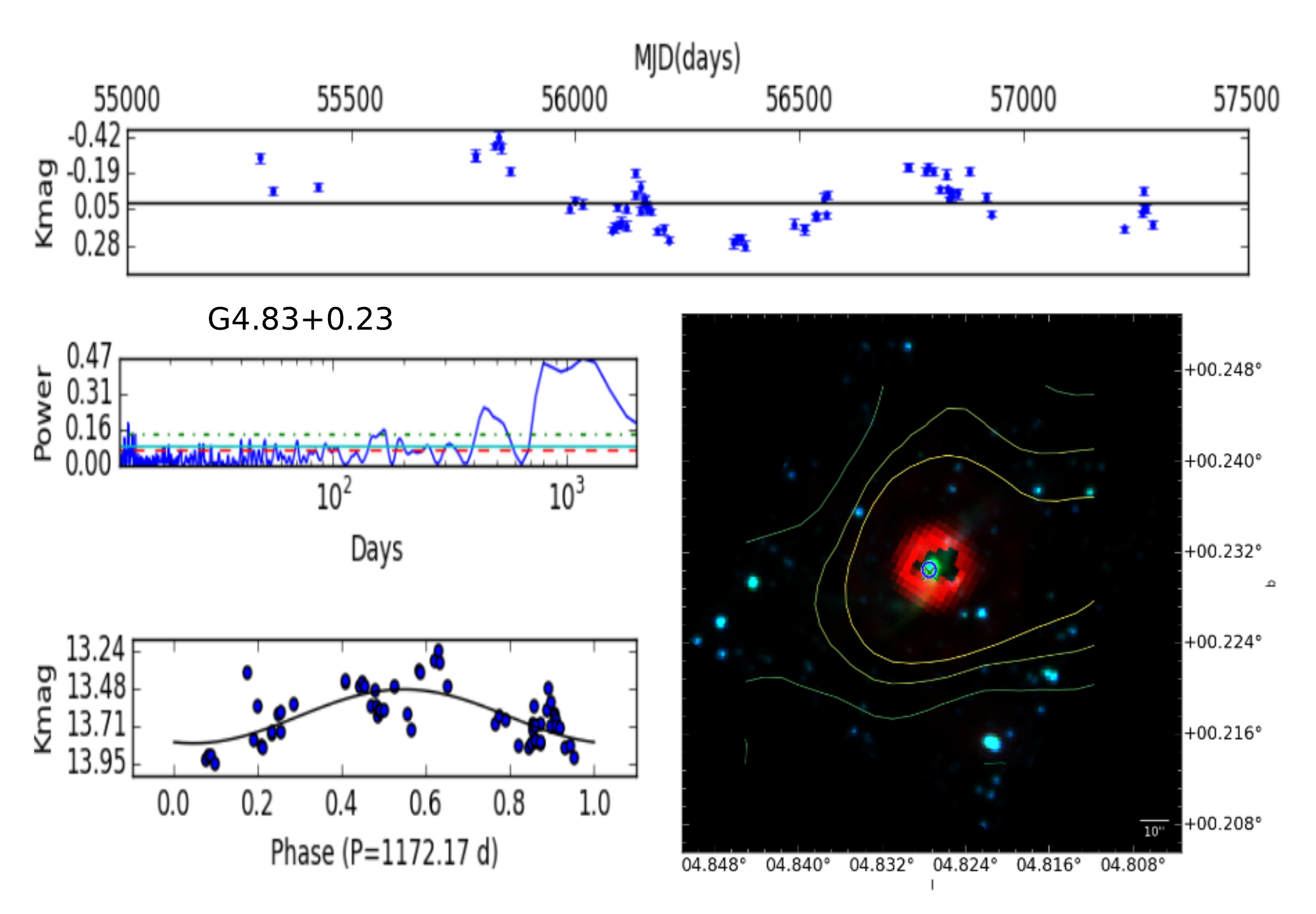}\\
                \includegraphics[width=0.48\textwidth]{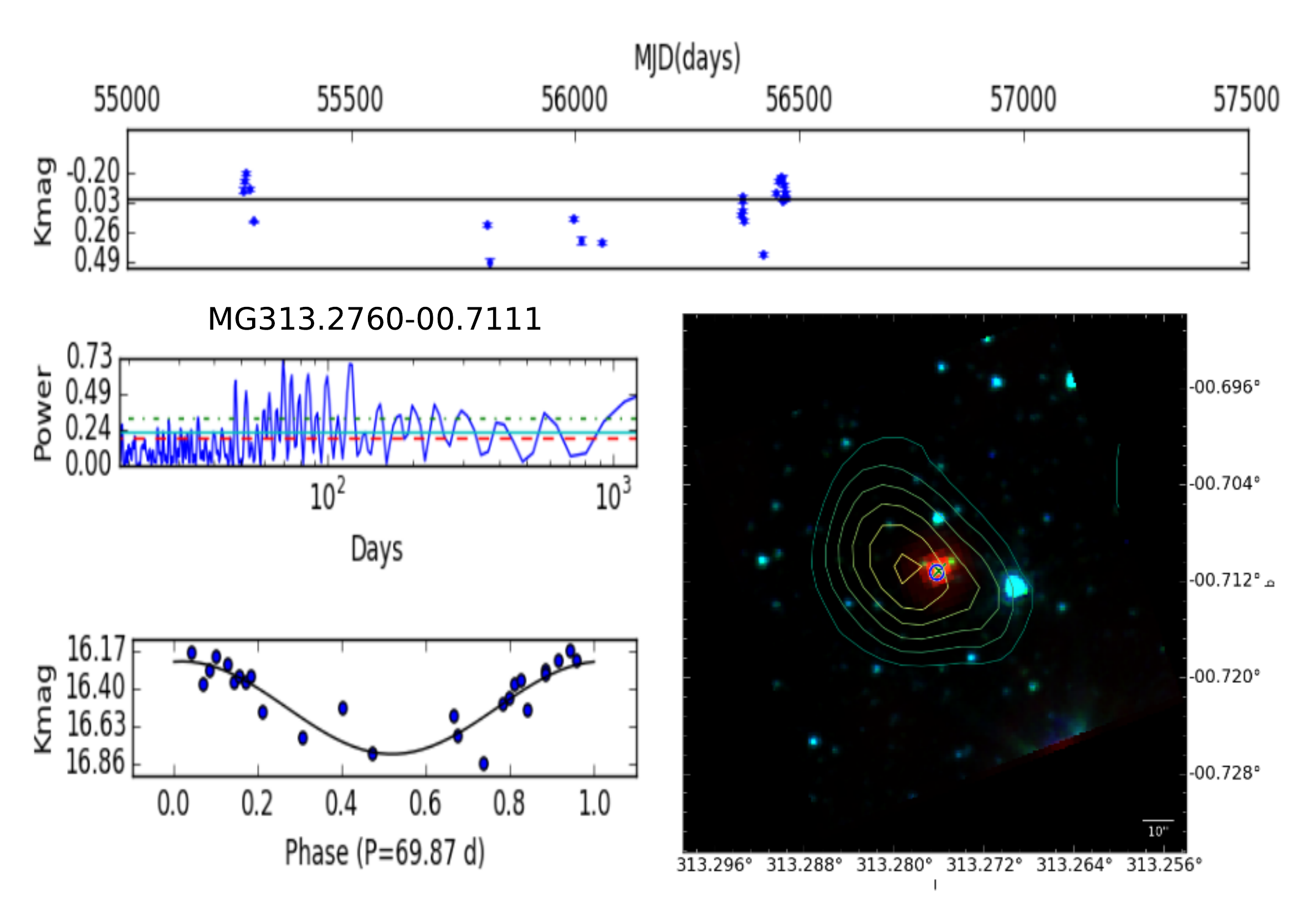}&
                \includegraphics[width=0.48\textwidth]{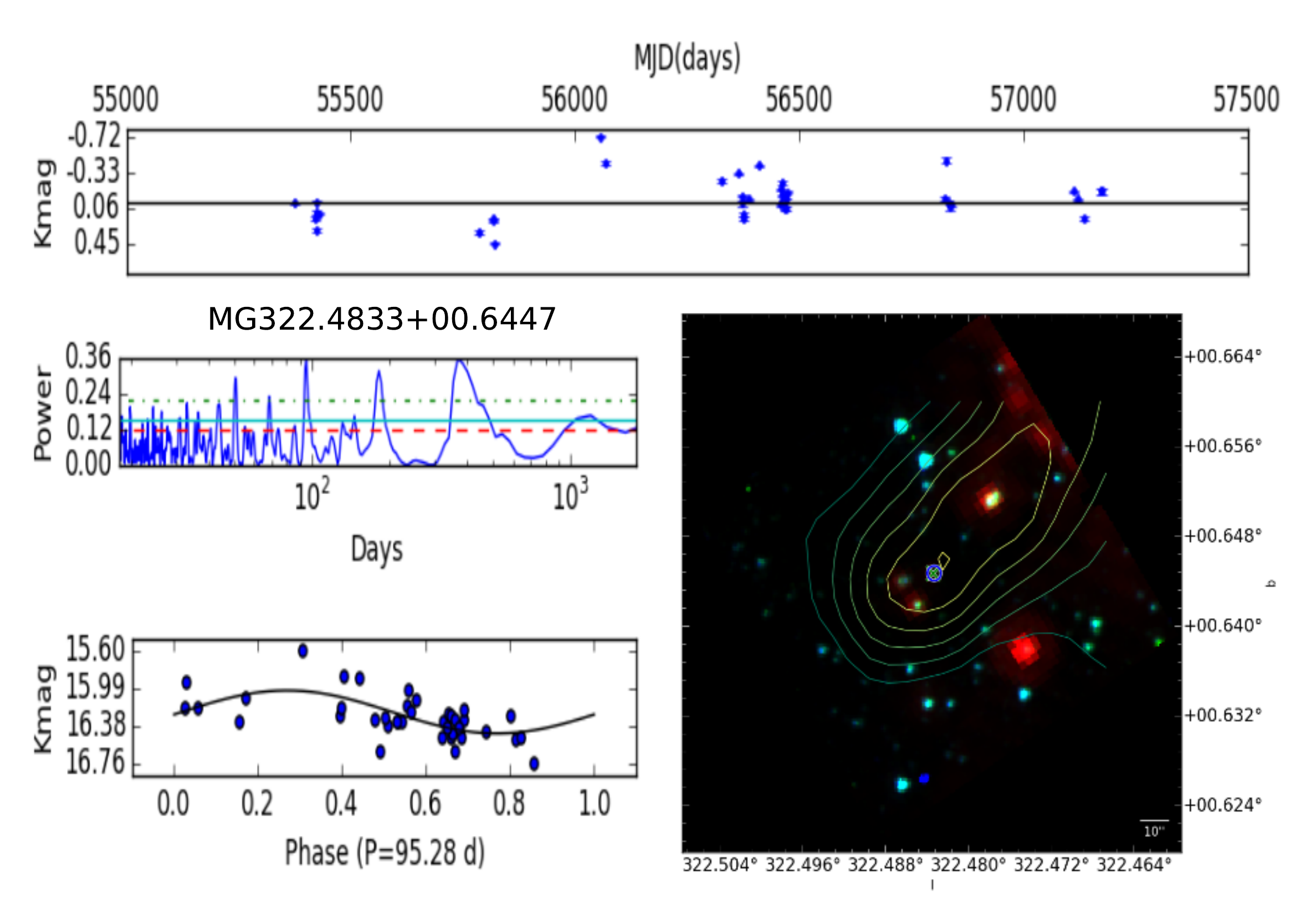}\\
                \includegraphics[width=0.48\textwidth]{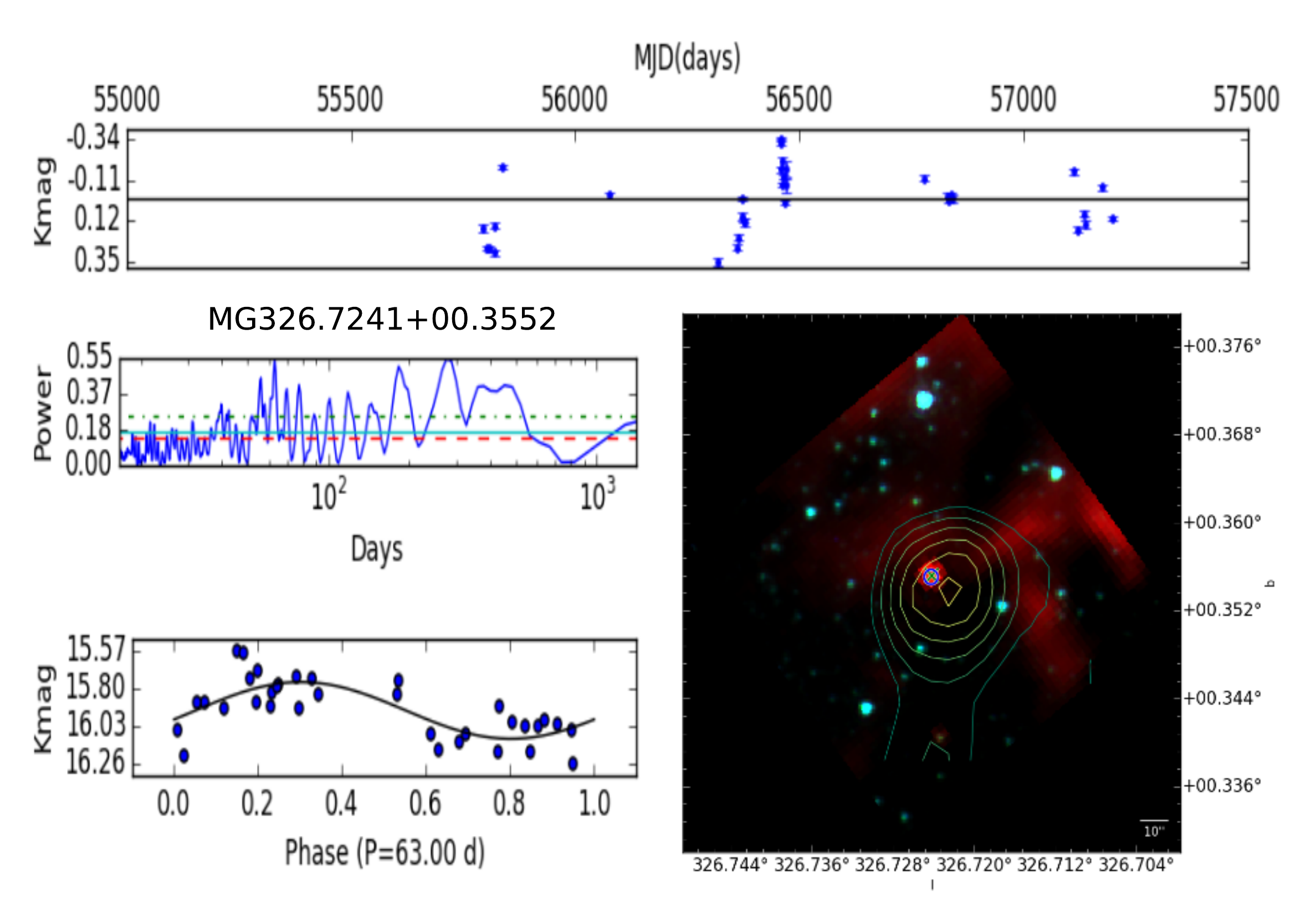}&
                \includegraphics[width=0.48\textwidth]{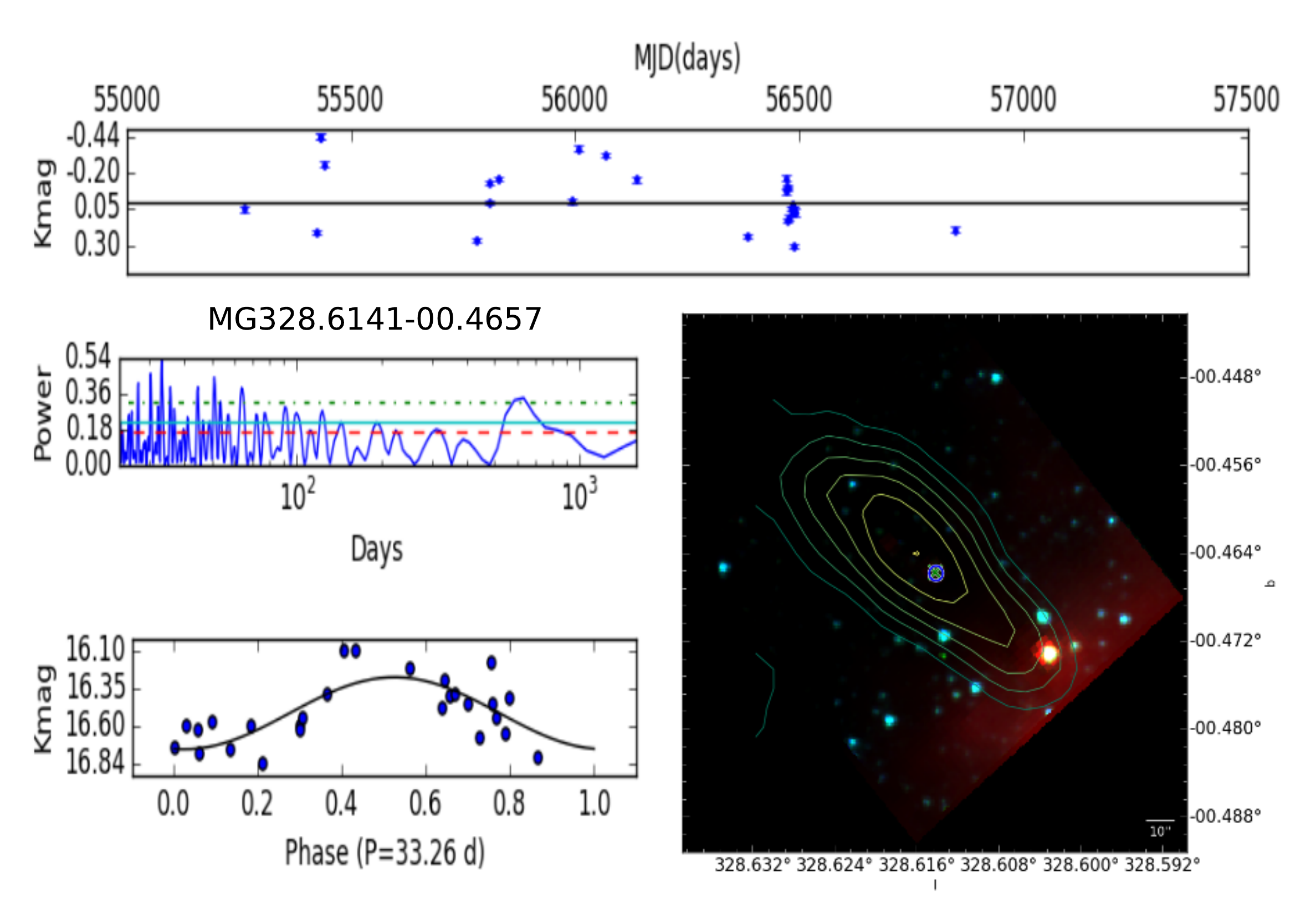}\\
                
        \end{tabular}
        \caption{Continuation of Fig. \ref{fig:testgrid}.}
\end{figure}

\begin{figure}
        \begin{tabular}{cc}
                \includegraphics[width=0.48\textwidth]{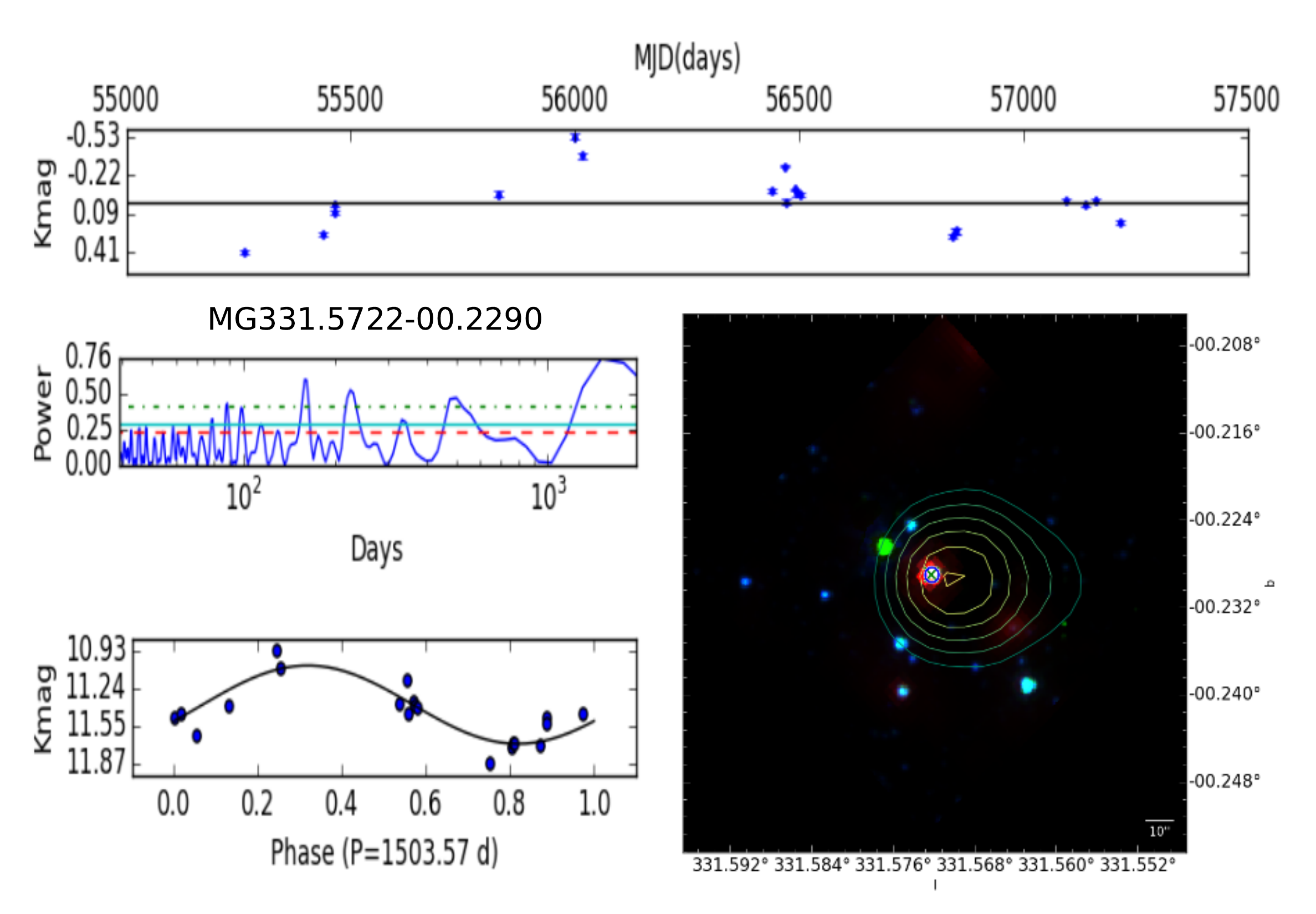}&
                \includegraphics[width=0.48\textwidth]{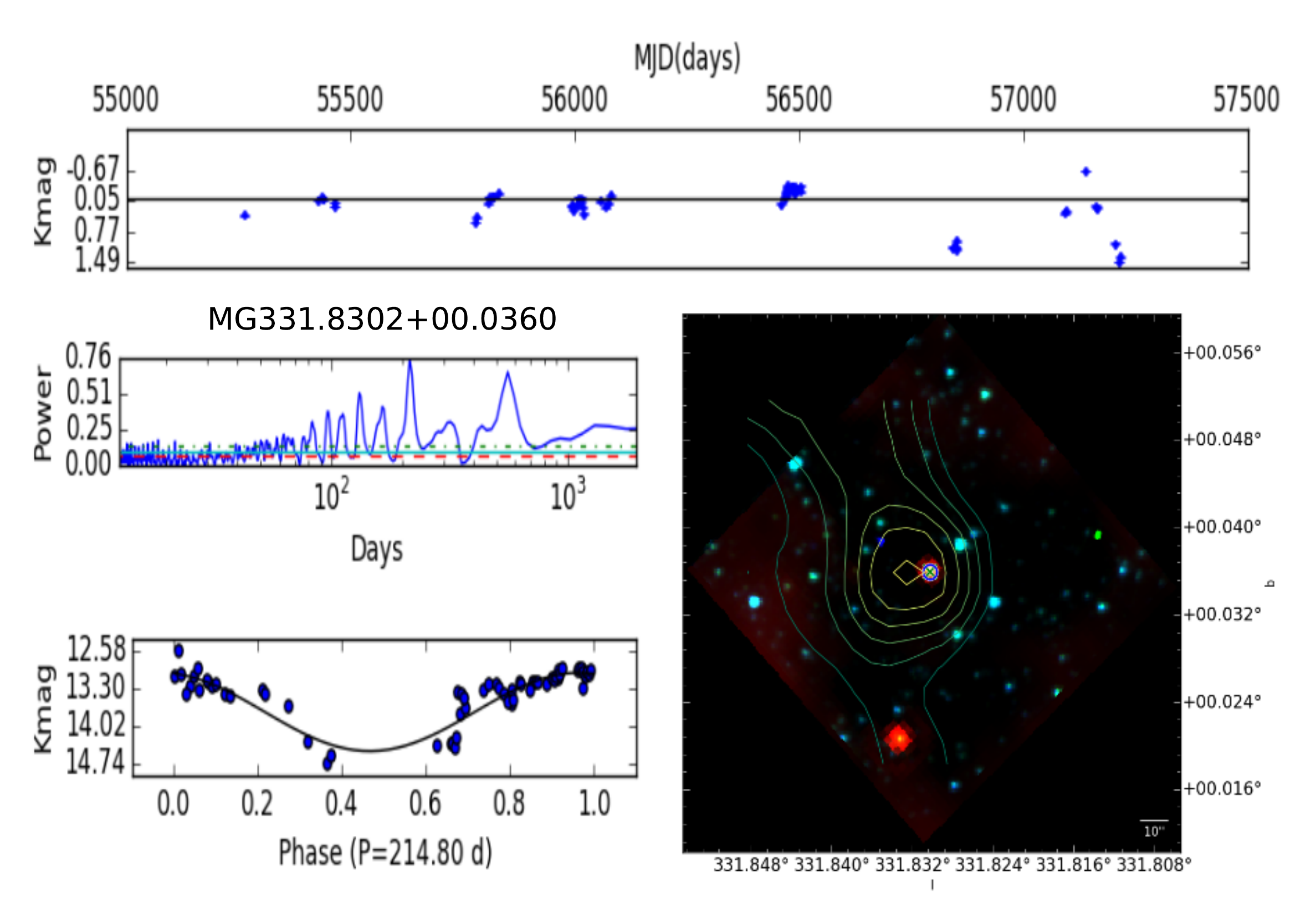}\\
                \includegraphics[width=0.48\textwidth]{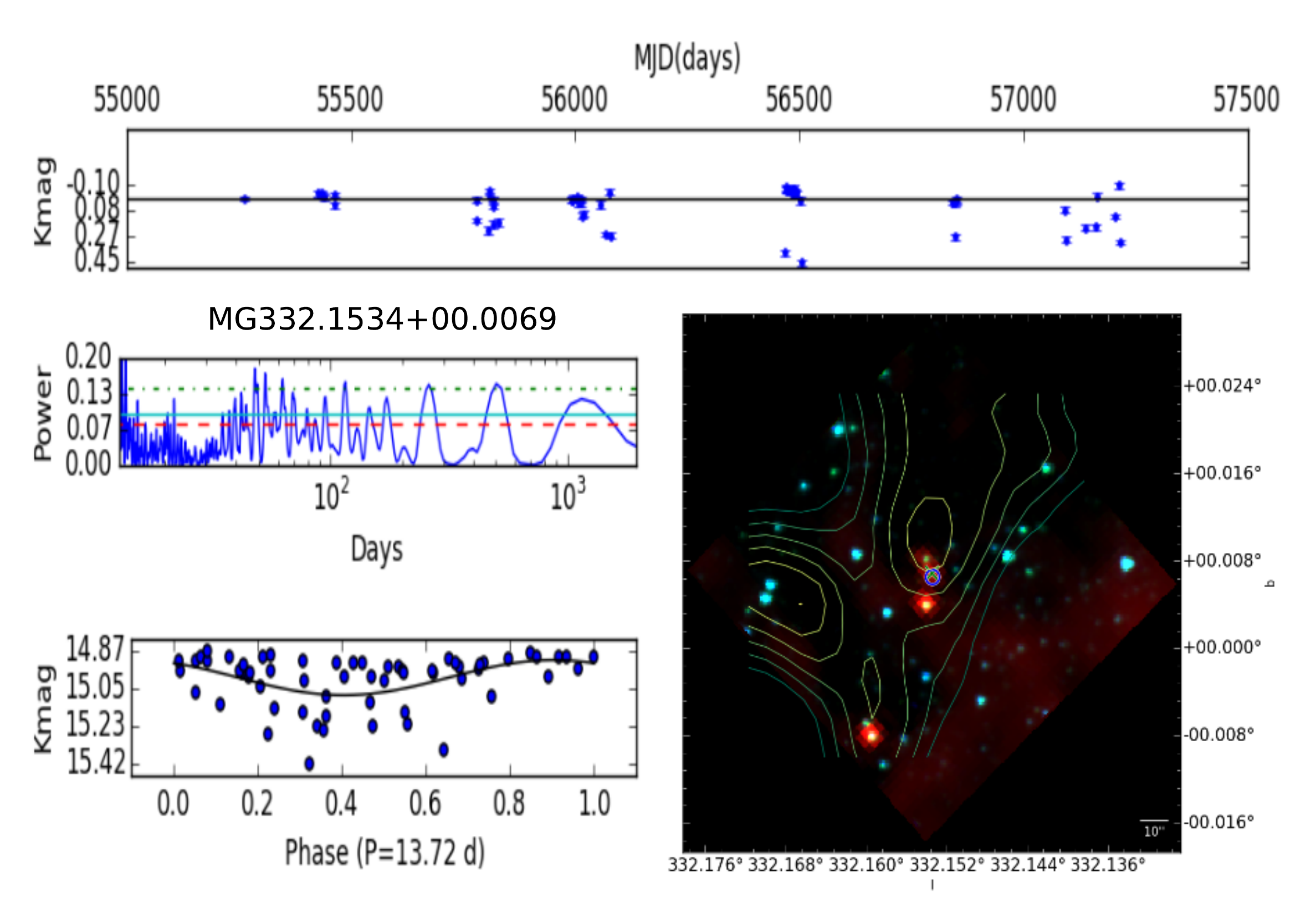}&
                \includegraphics[width=0.48\textwidth]{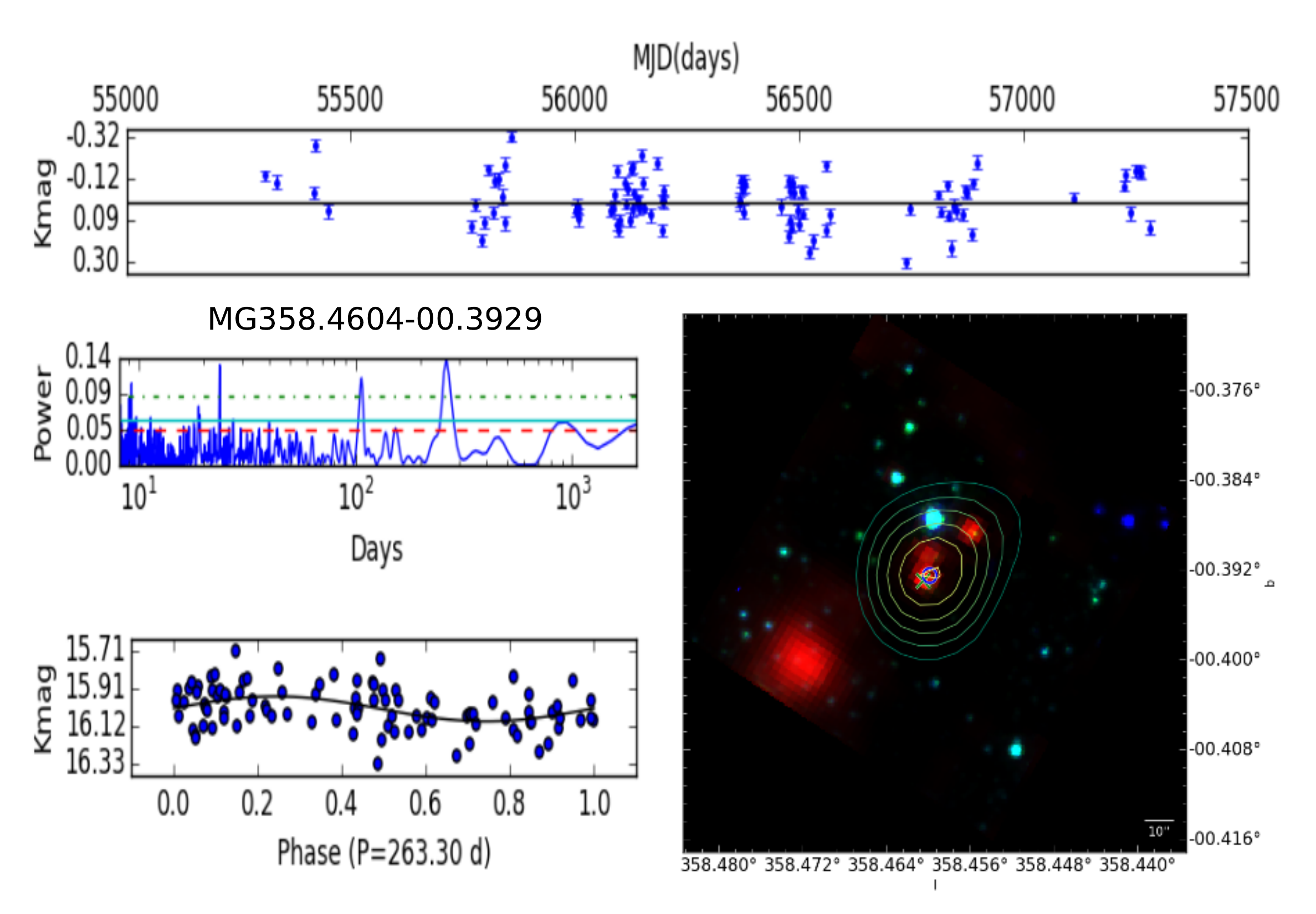}\\
                \includegraphics[width=0.48\textwidth]{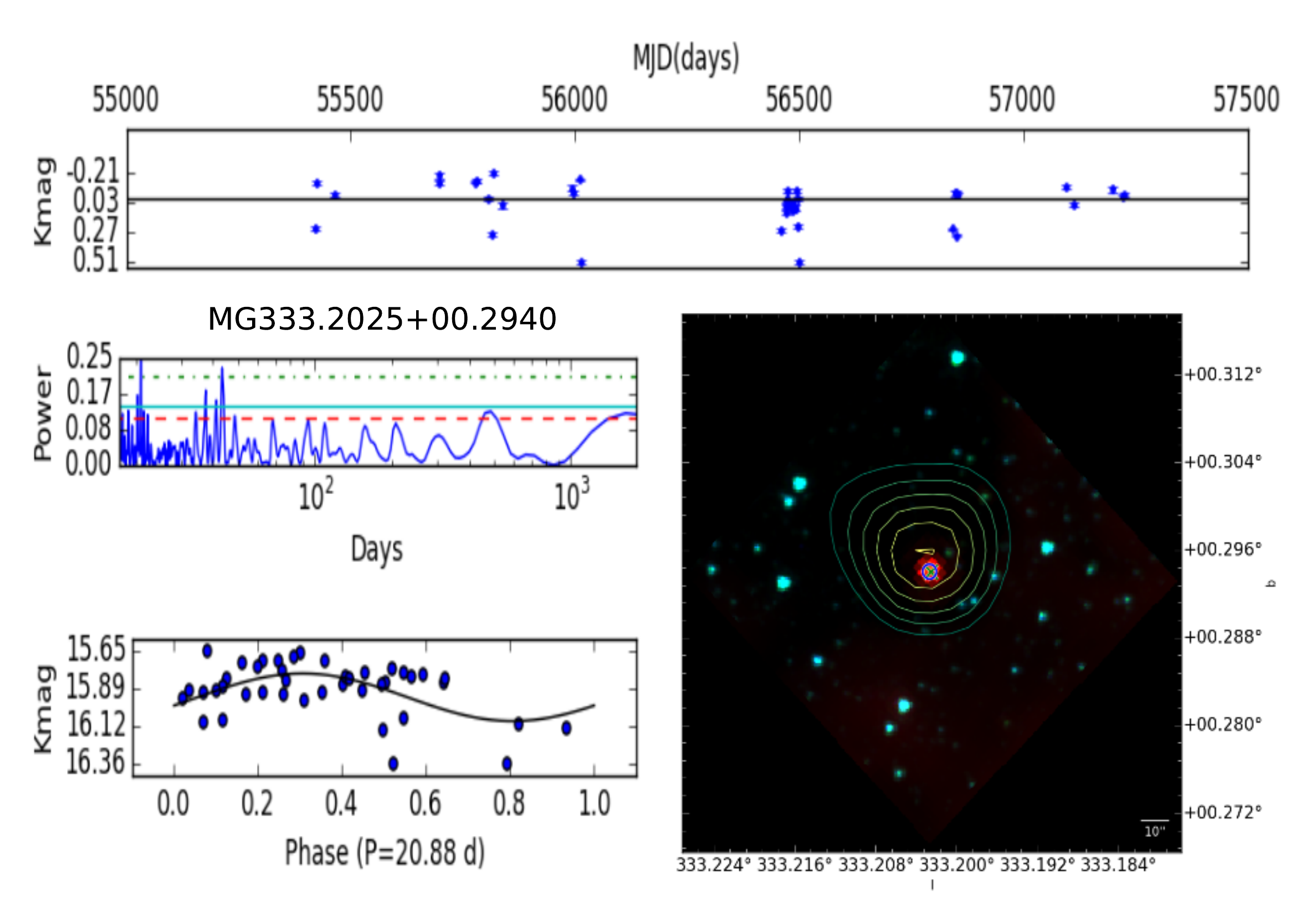}&
                \includegraphics[width=0.48\textwidth]{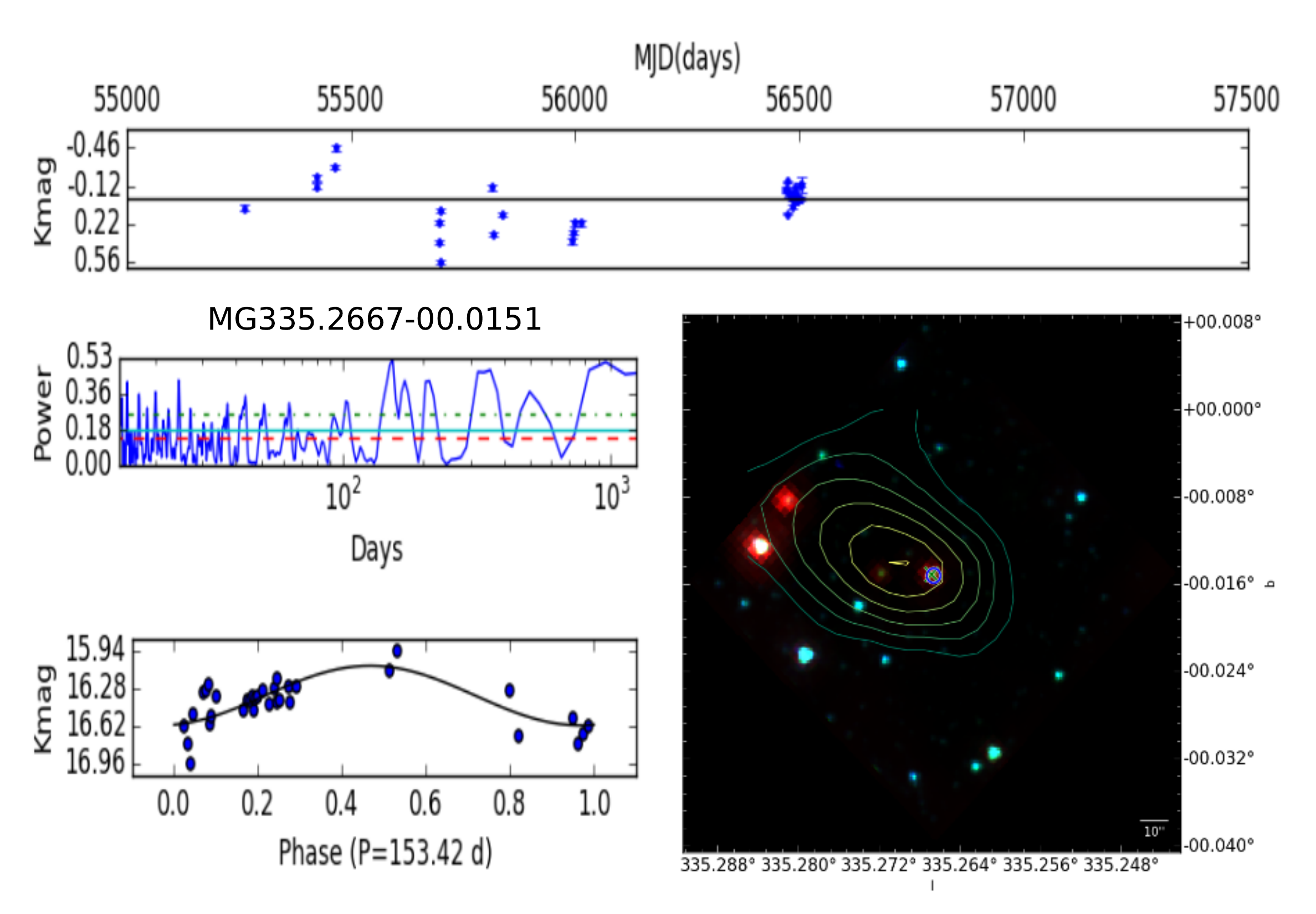}\\
                \includegraphics[width=0.48\textwidth]{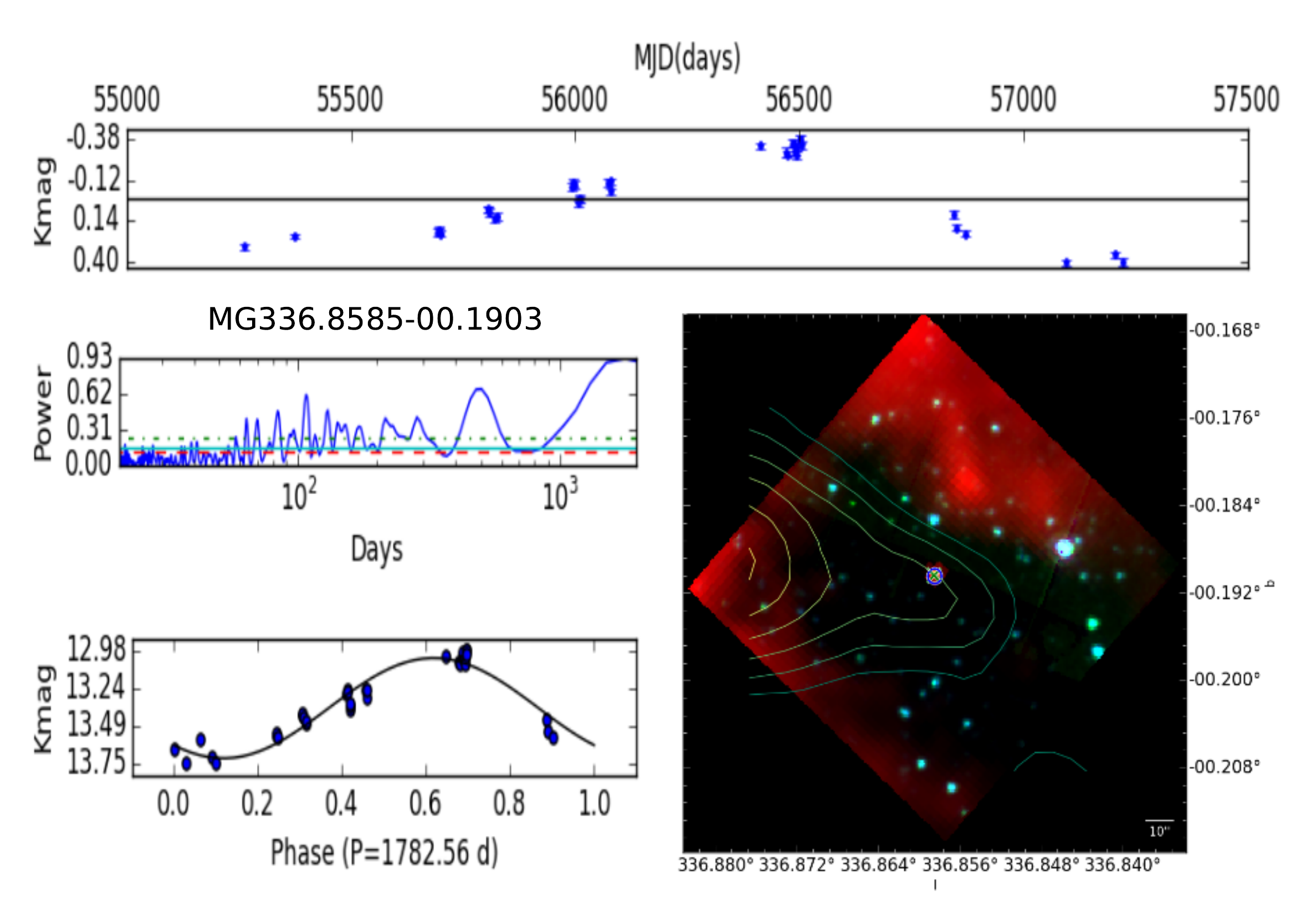}&
                \includegraphics[width=0.48\textwidth]{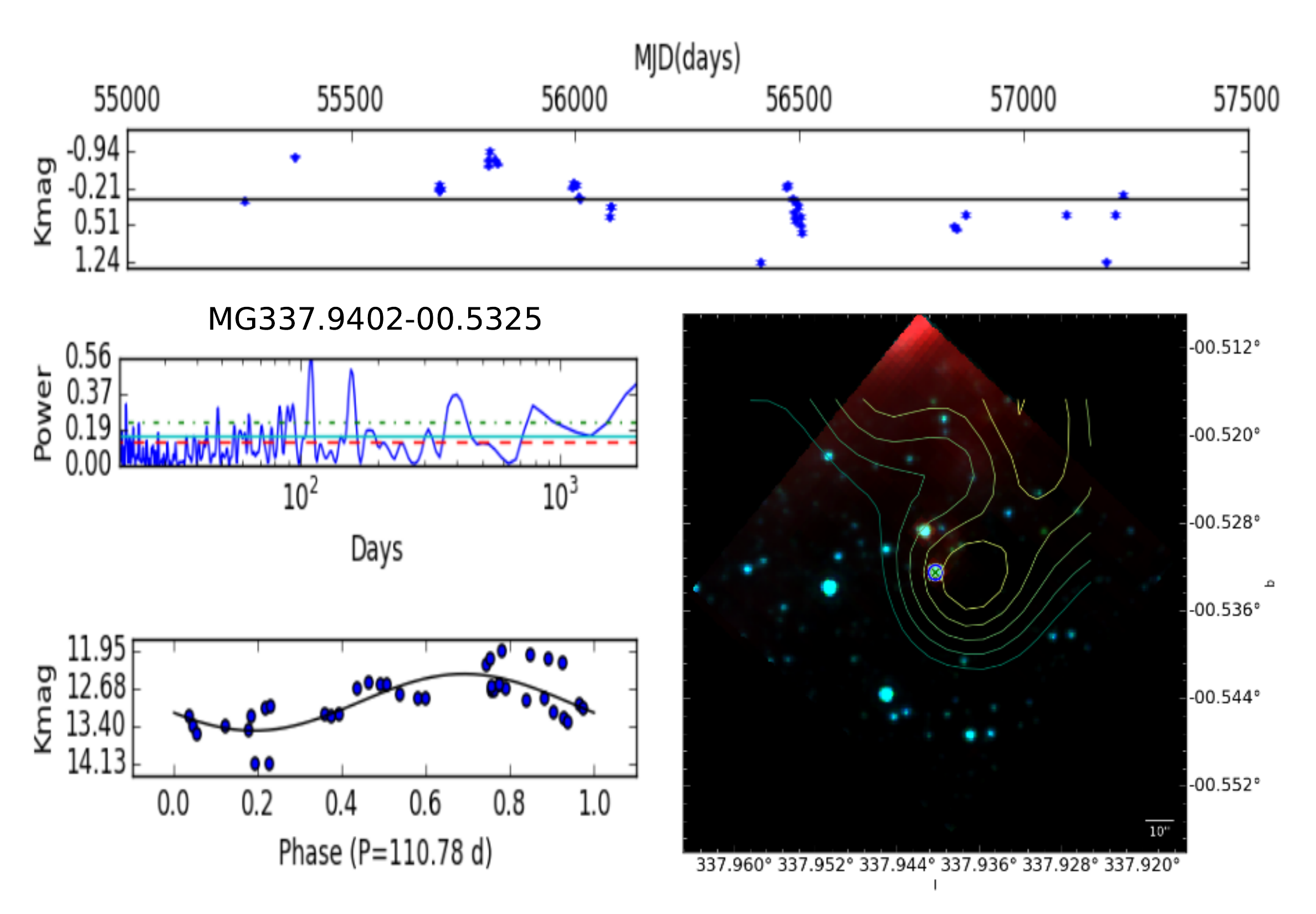}\\
        \end{tabular}
        \caption{Continuation of Fig. \ref{fig:testgrid}.}
\end{figure}

\begin{figure}
        \begin{tabular}{cc}

                \includegraphics[width=0.48\textwidth]{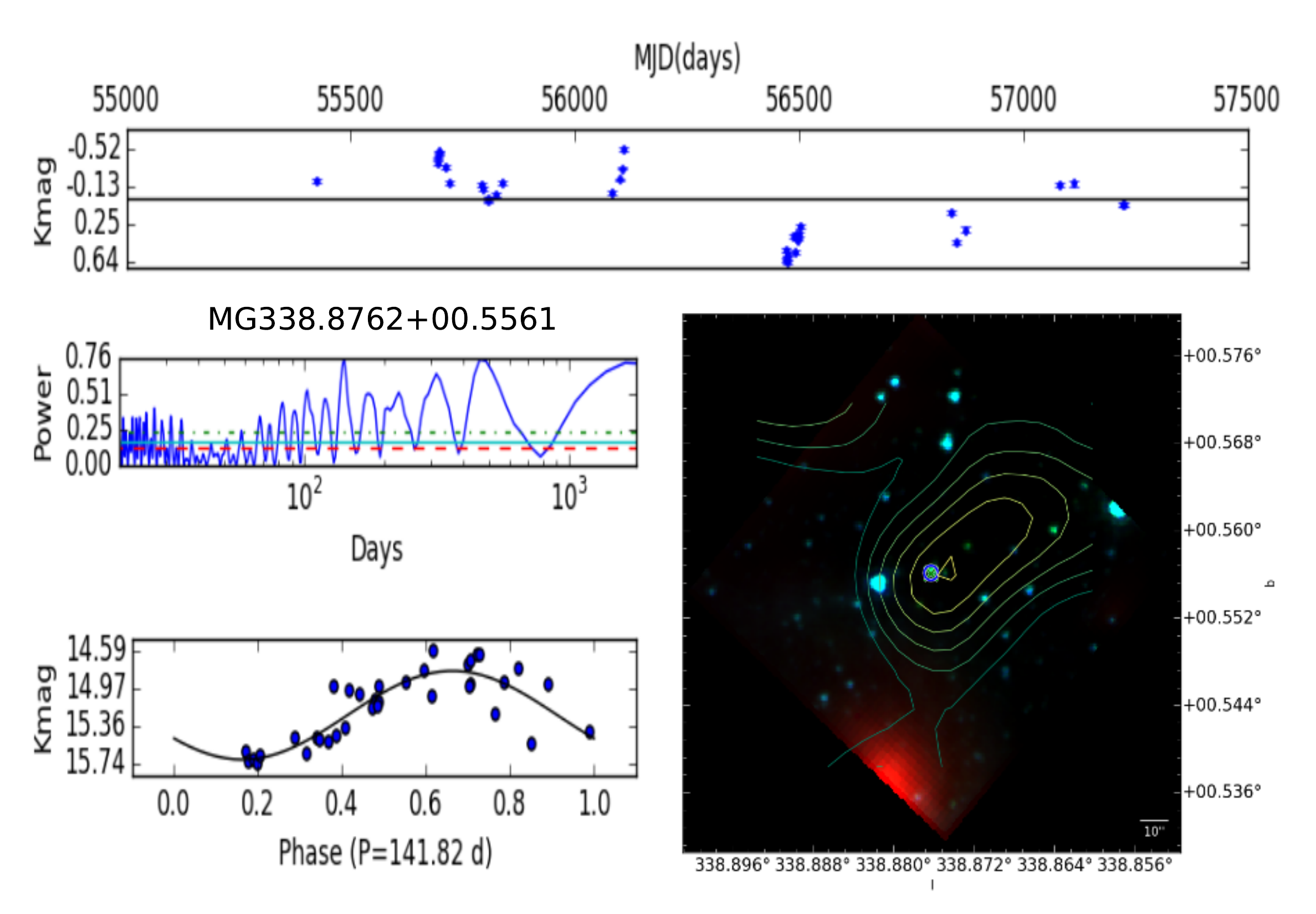}&
                \includegraphics[width=0.48\textwidth]{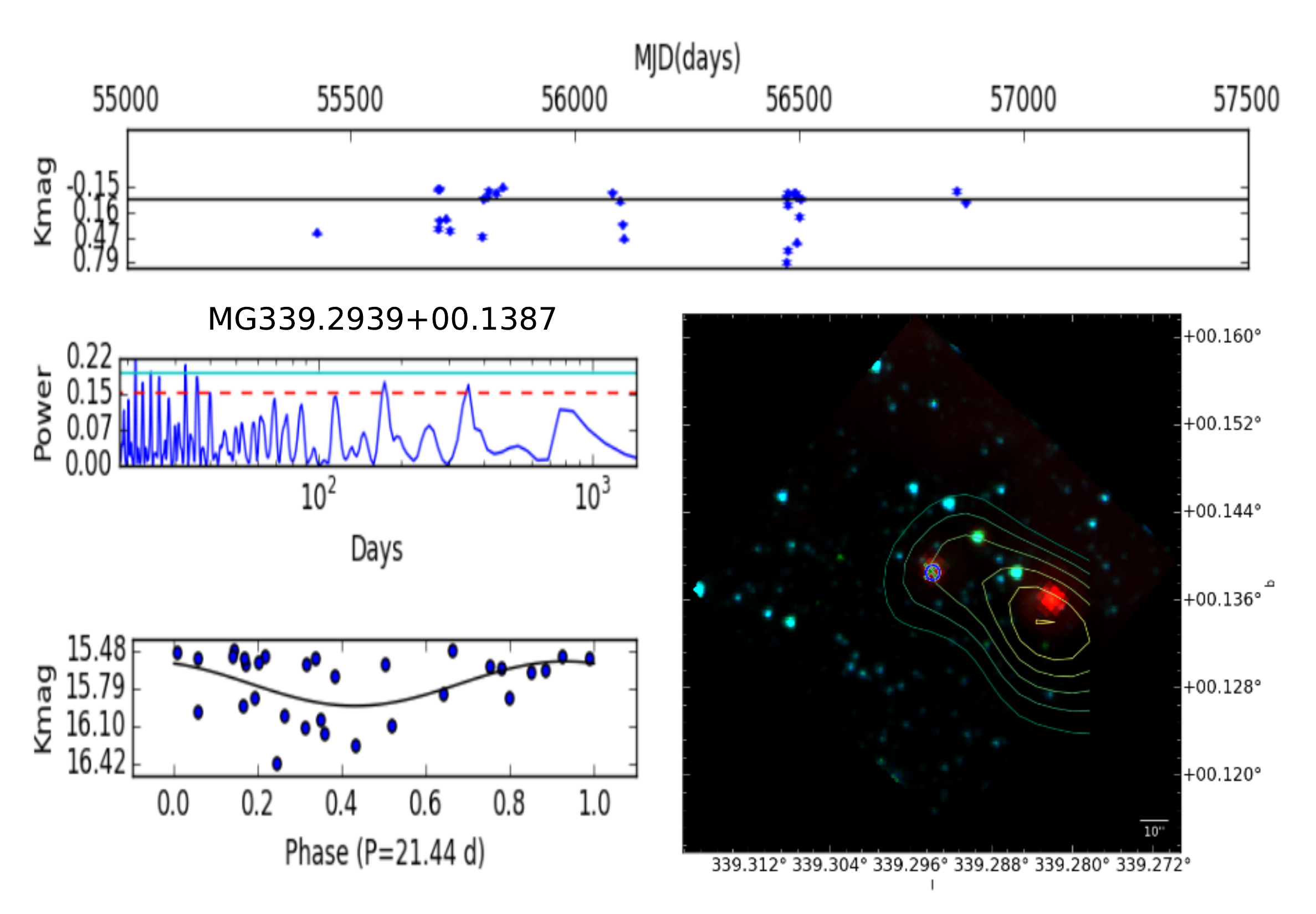}\\
                \includegraphics[width=0.48\textwidth]{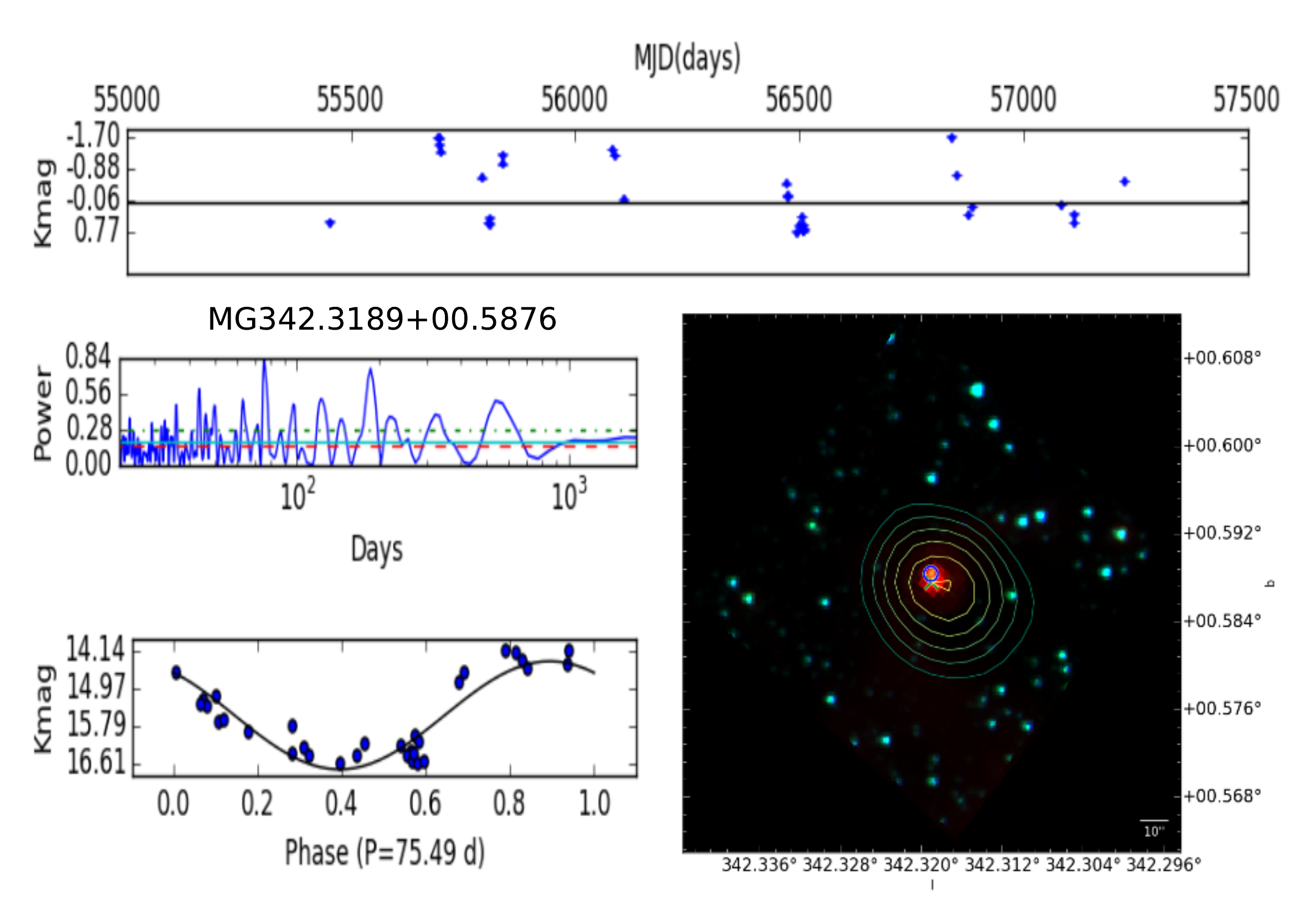}&
                \includegraphics[width=0.48\textwidth]{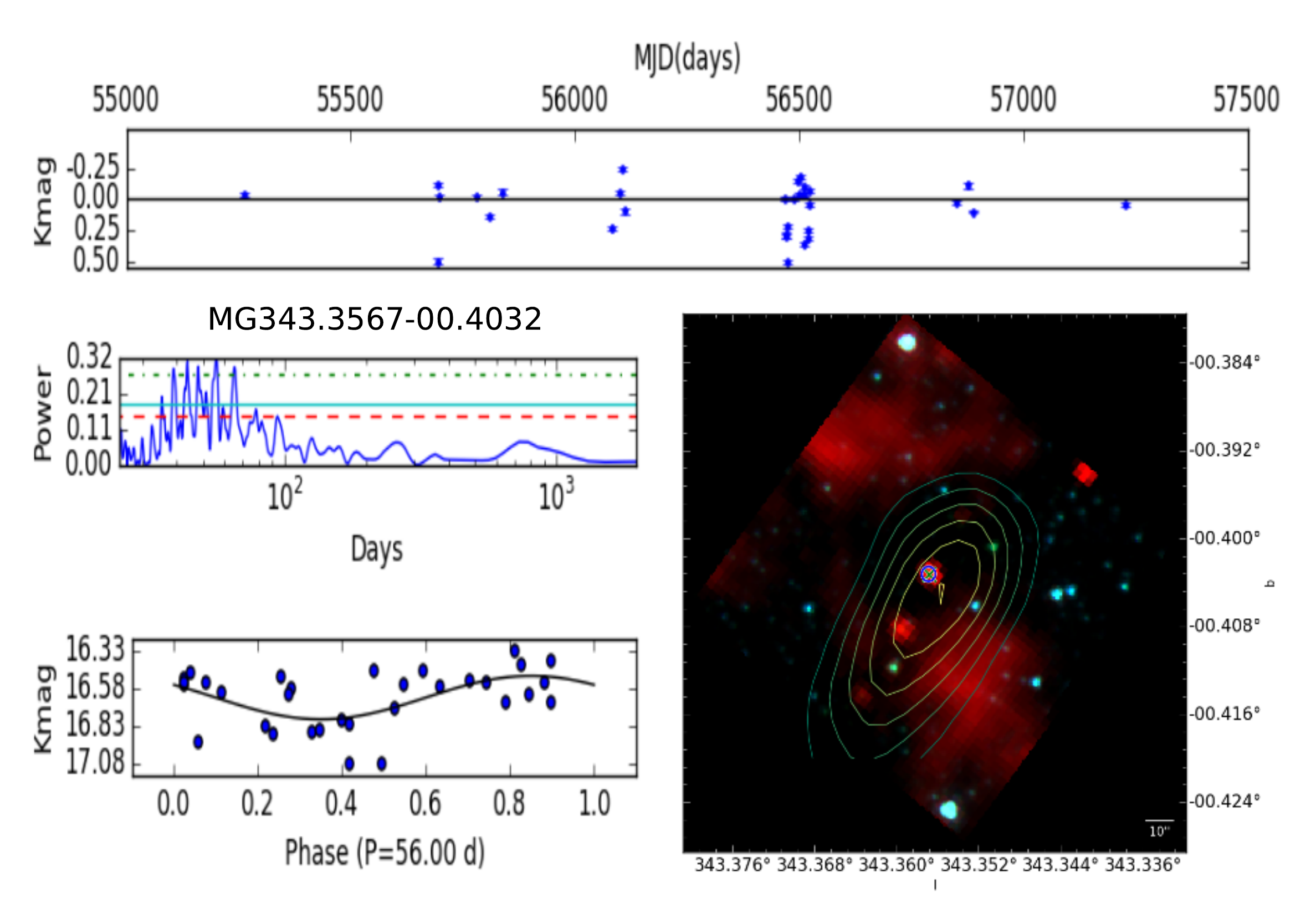}\\             
                \includegraphics[width=0.48\textwidth]{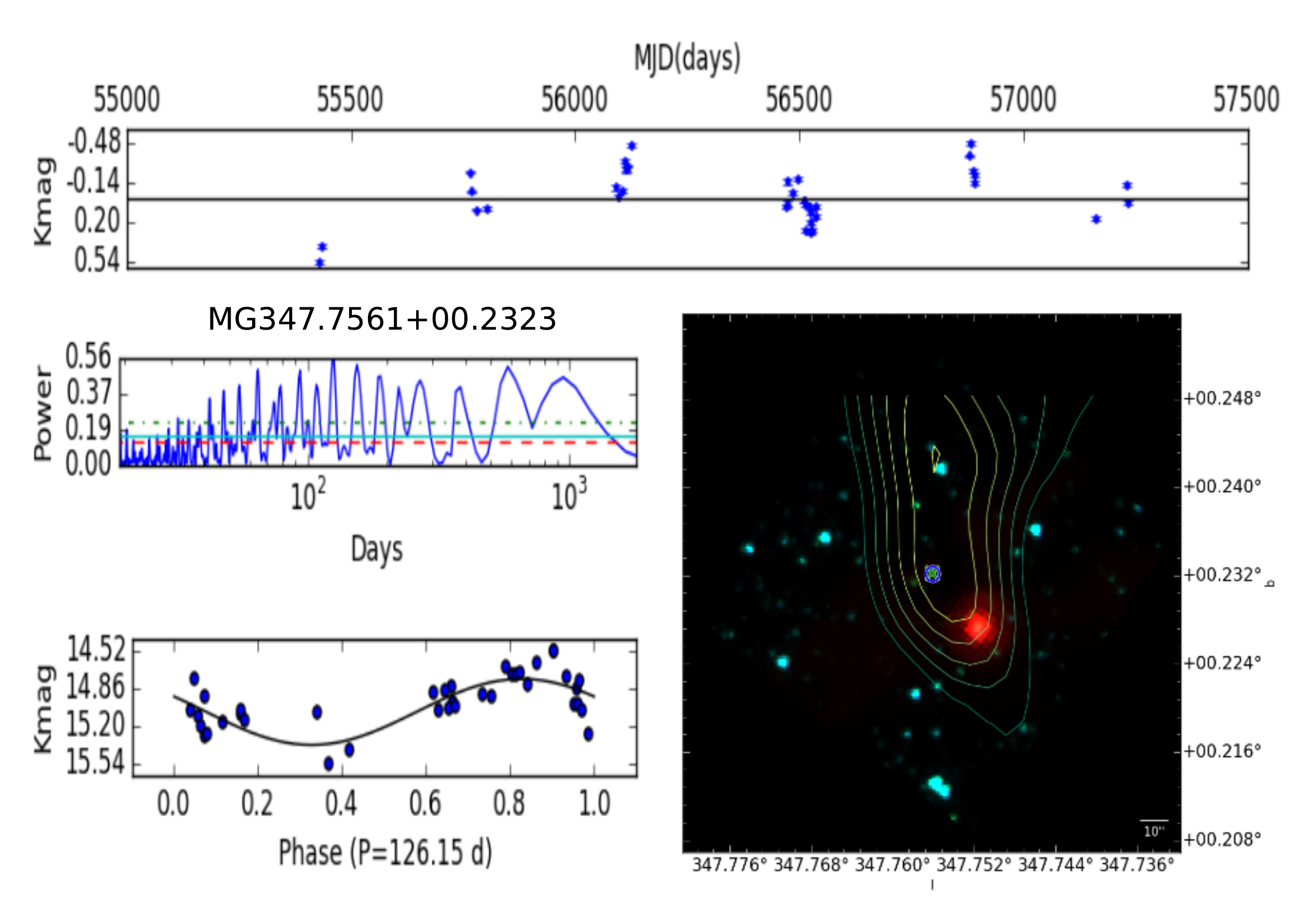}&
                \includegraphics[width=0.48\textwidth]{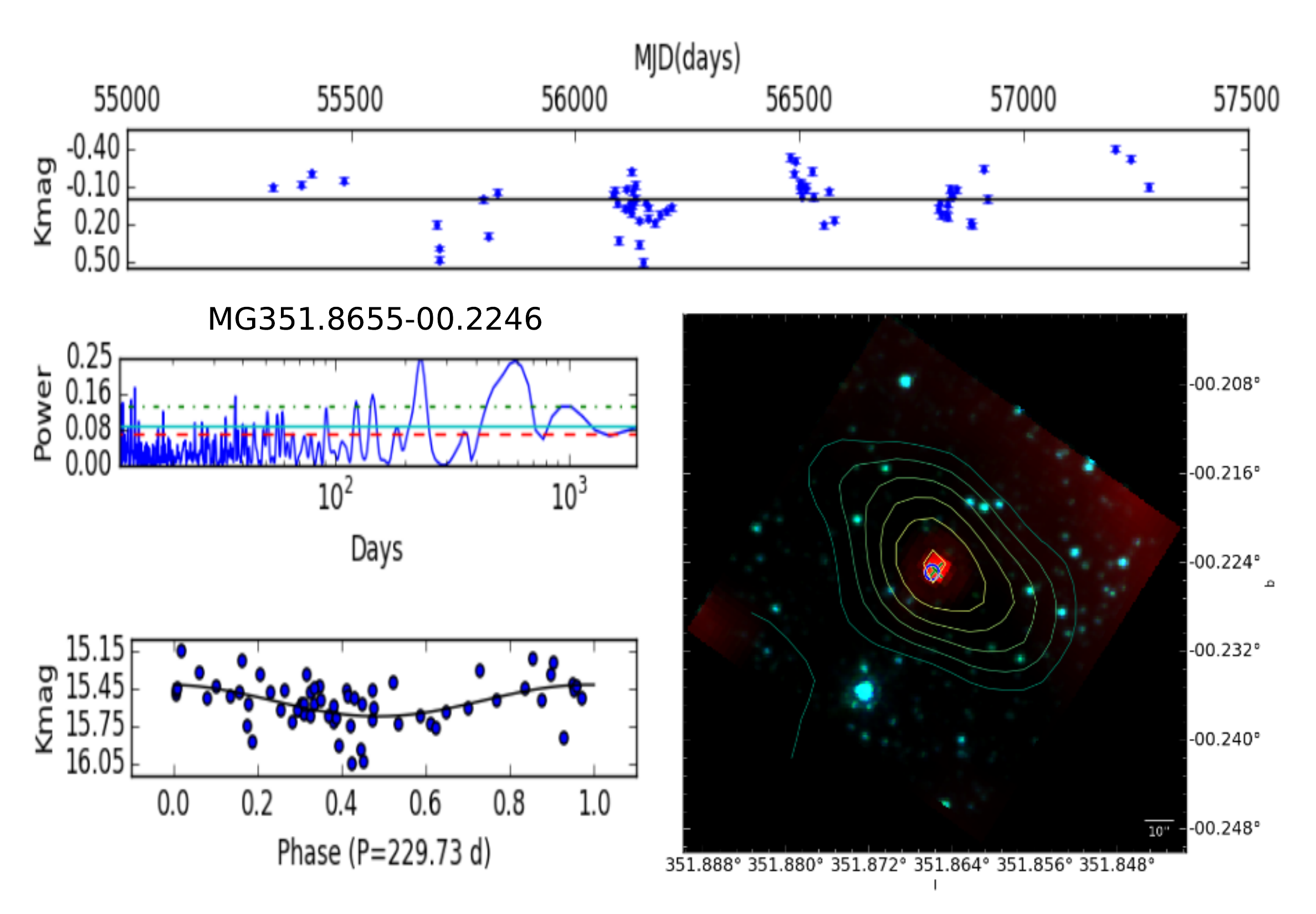}\\
                \includegraphics[width=0.48\textwidth]{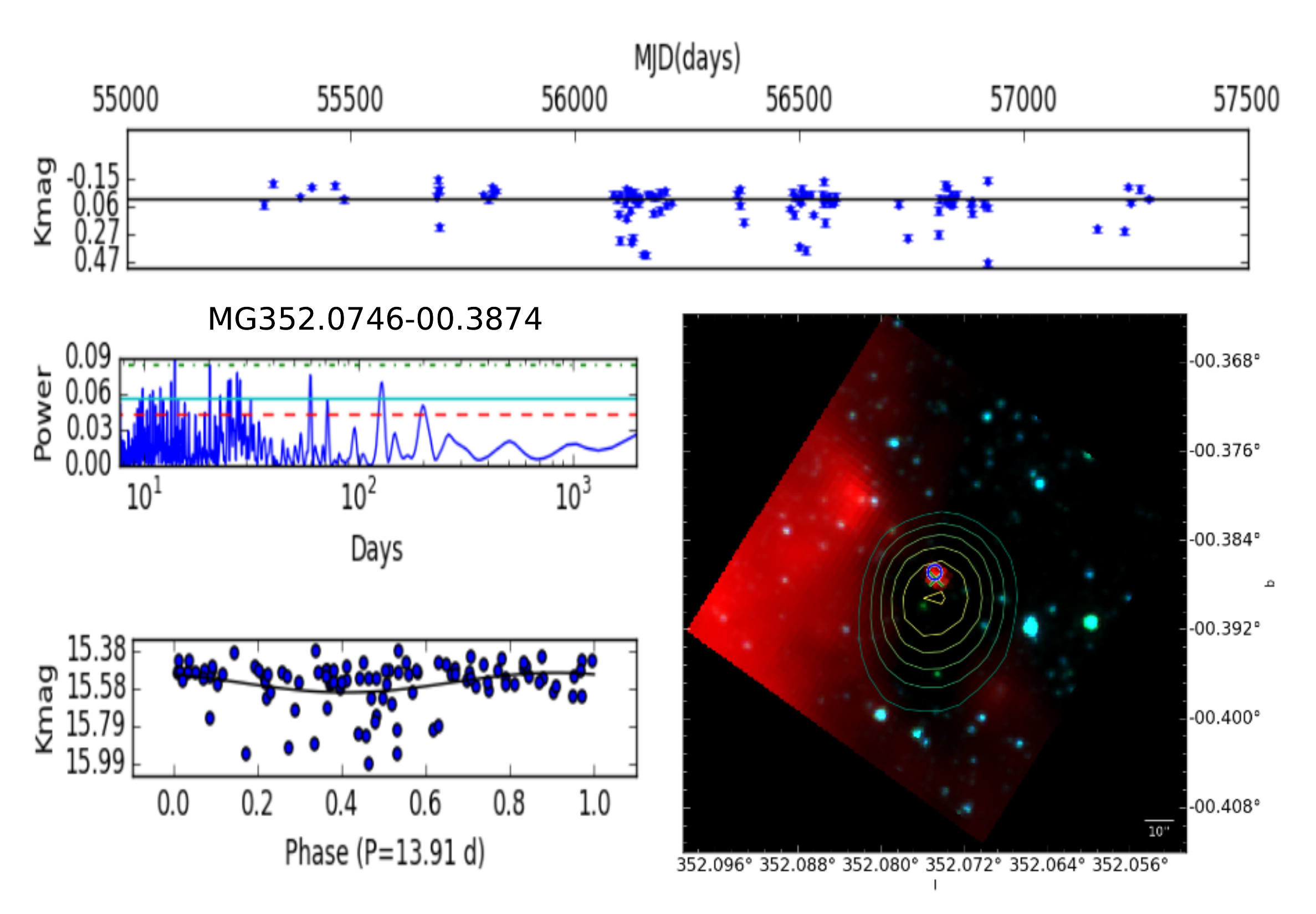}&
                \includegraphics[width=0.48\textwidth]{{./MG352.2452-00.0636_combined}.pdf}\\
                
        \end{tabular}
        \caption{Continuation of Fig. \ref{fig:testgrid}.}
\end{figure}

\begin{figure}
        \begin{tabular}{cc}
                
                \includegraphics[width=0.48\textwidth]{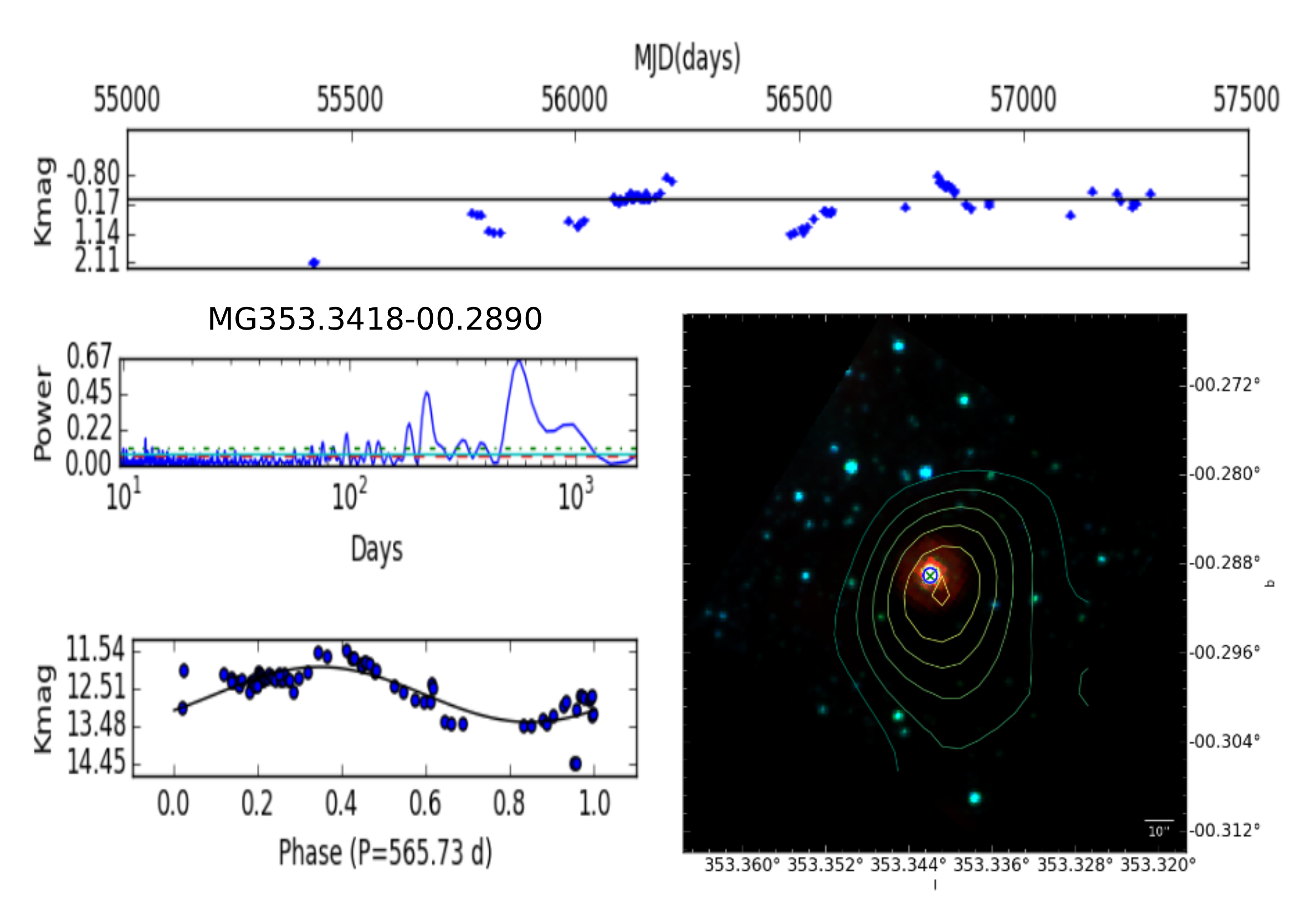}&
                \includegraphics[width=0.48\textwidth]{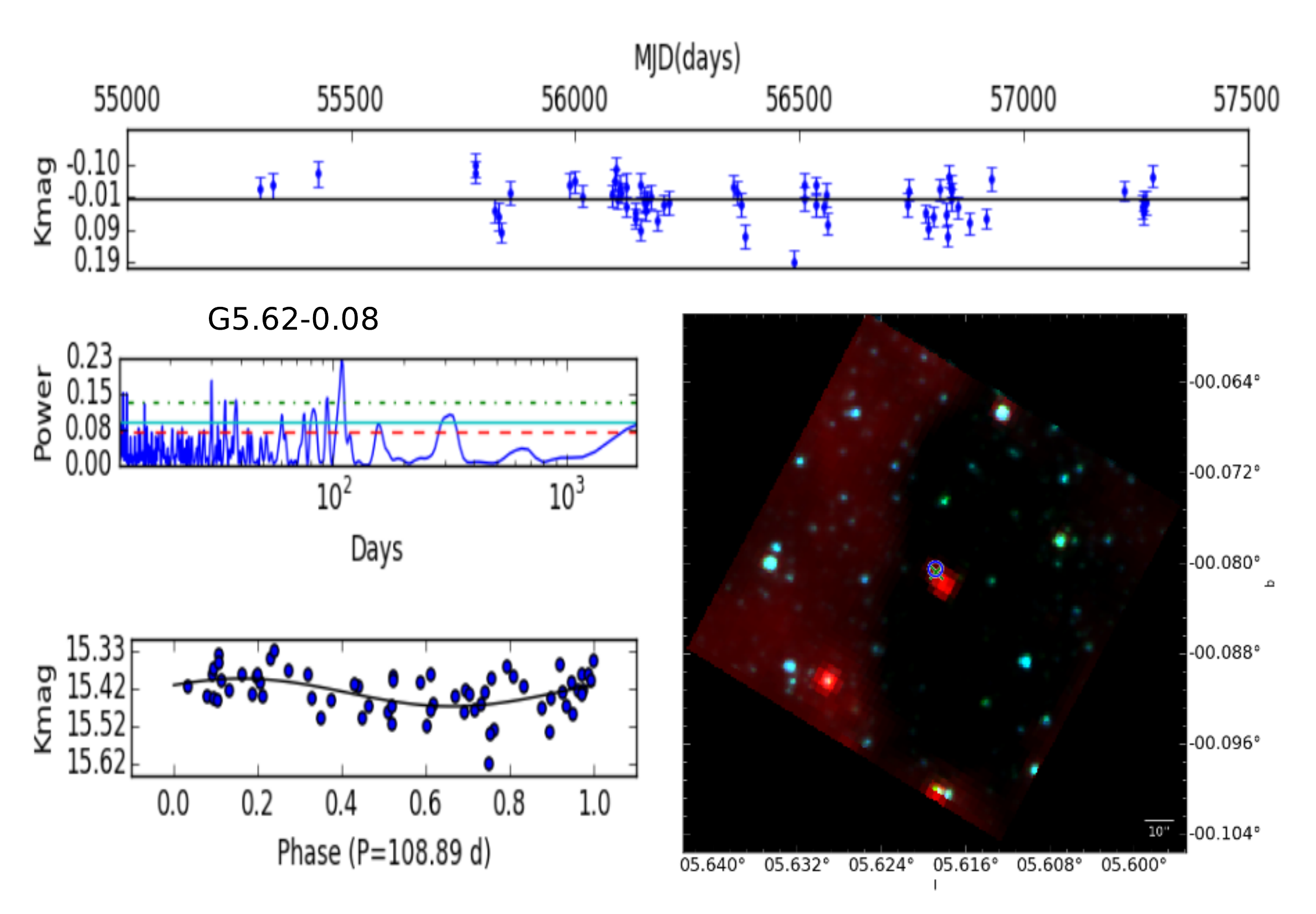}\\
                \includegraphics[width=0.48\textwidth]{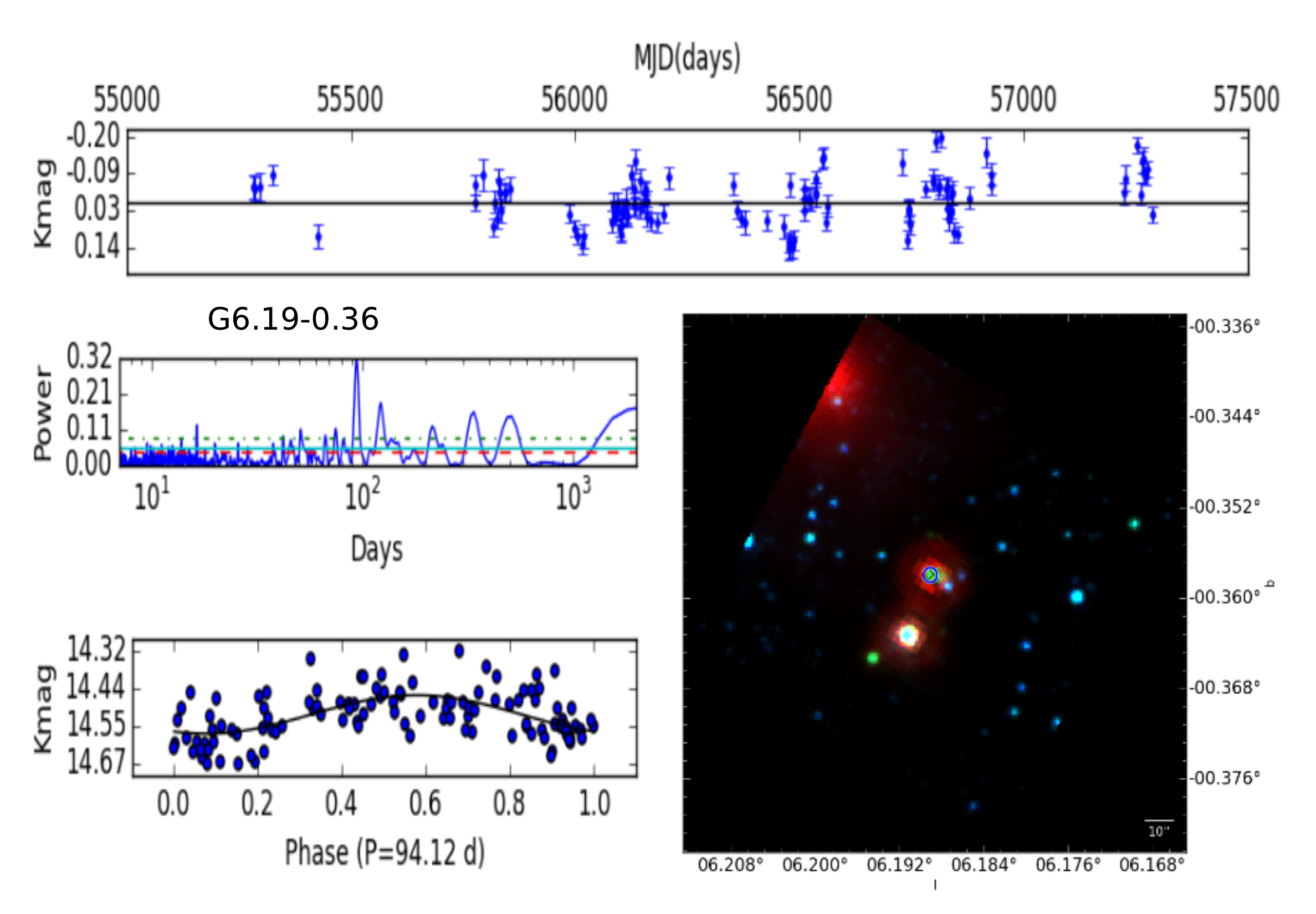}&
                \includegraphics[width=0.48\textwidth]{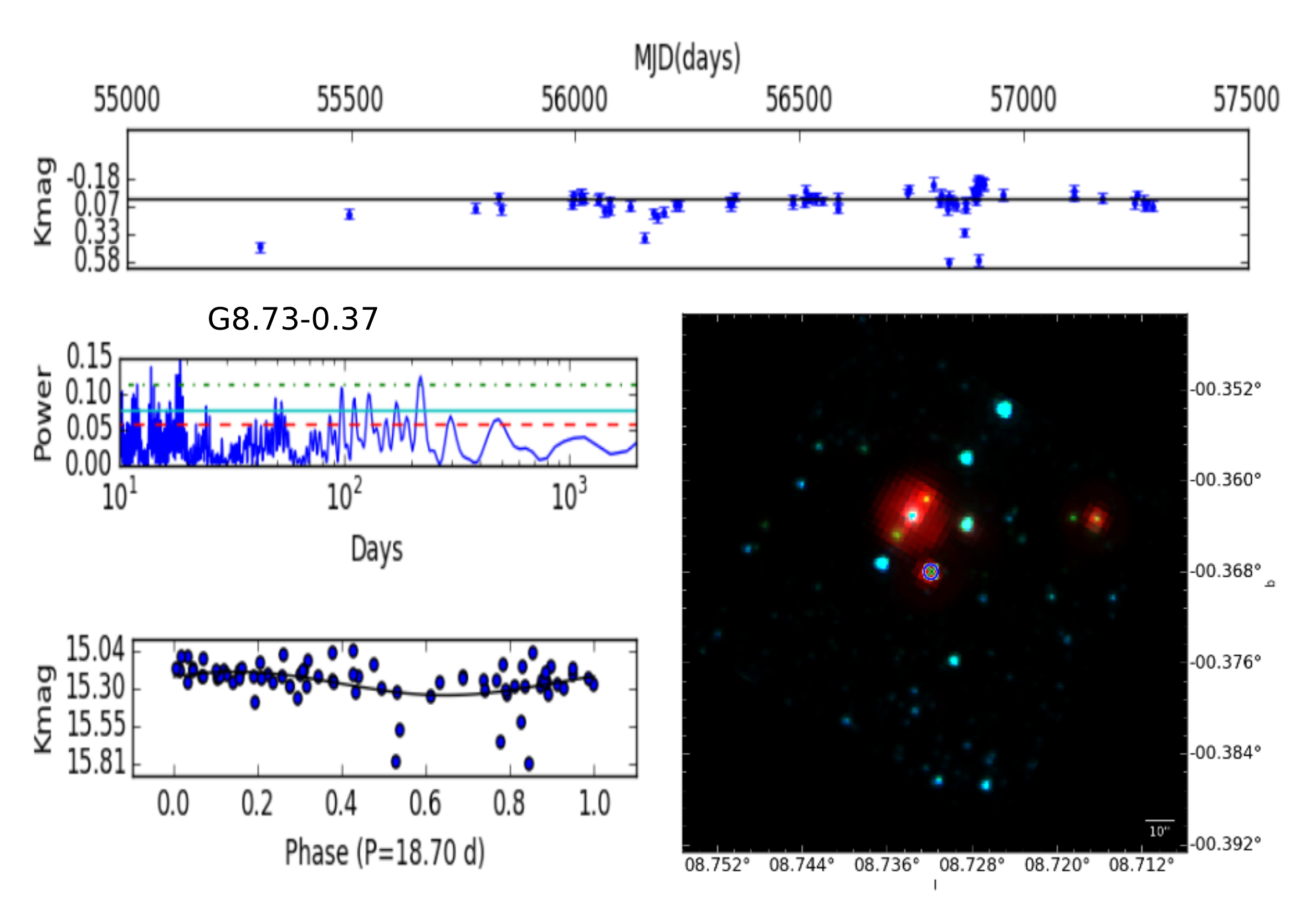}\\
                \includegraphics[width=0.48\textwidth]{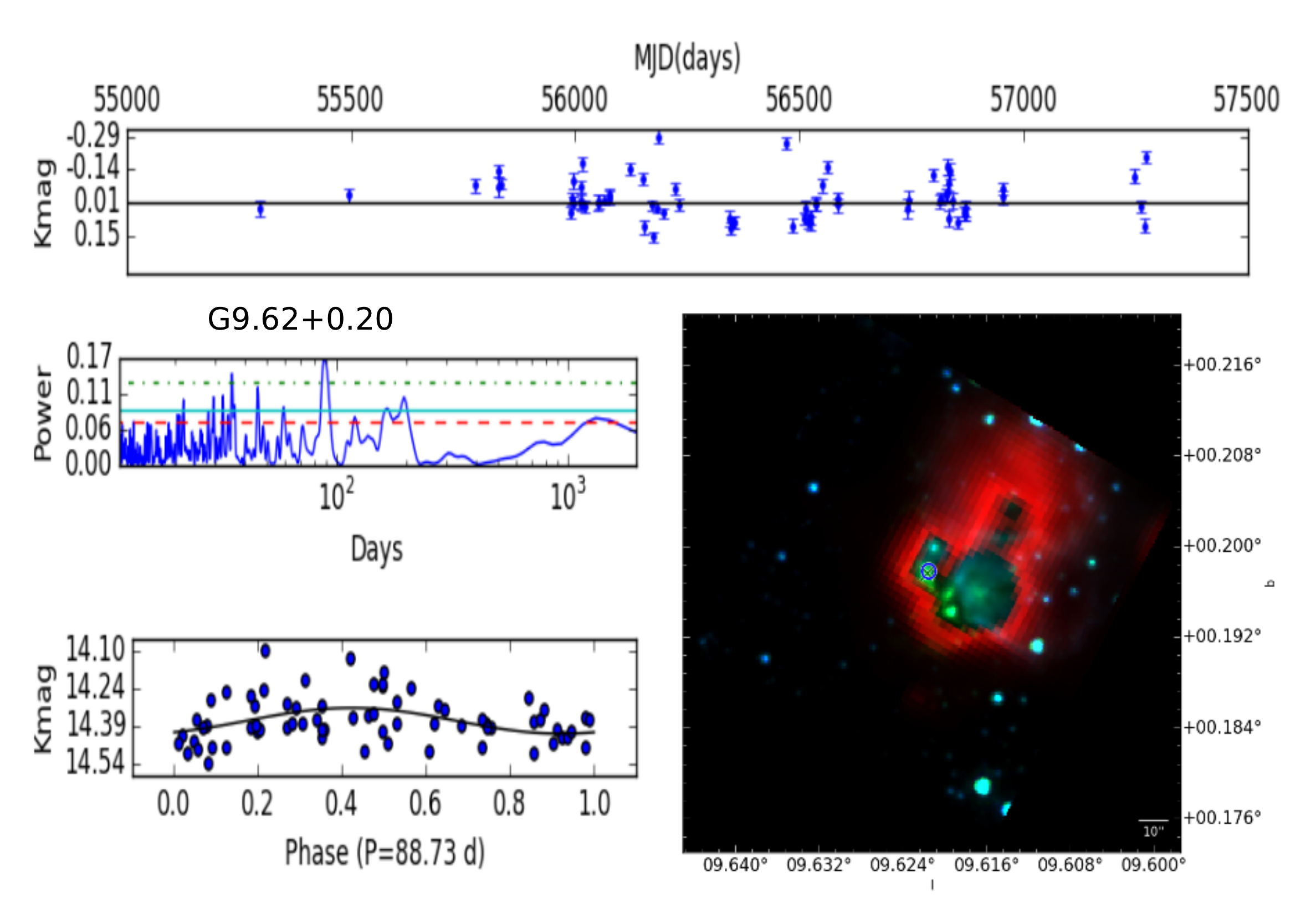}&
                \includegraphics[width=0.48\textwidth]{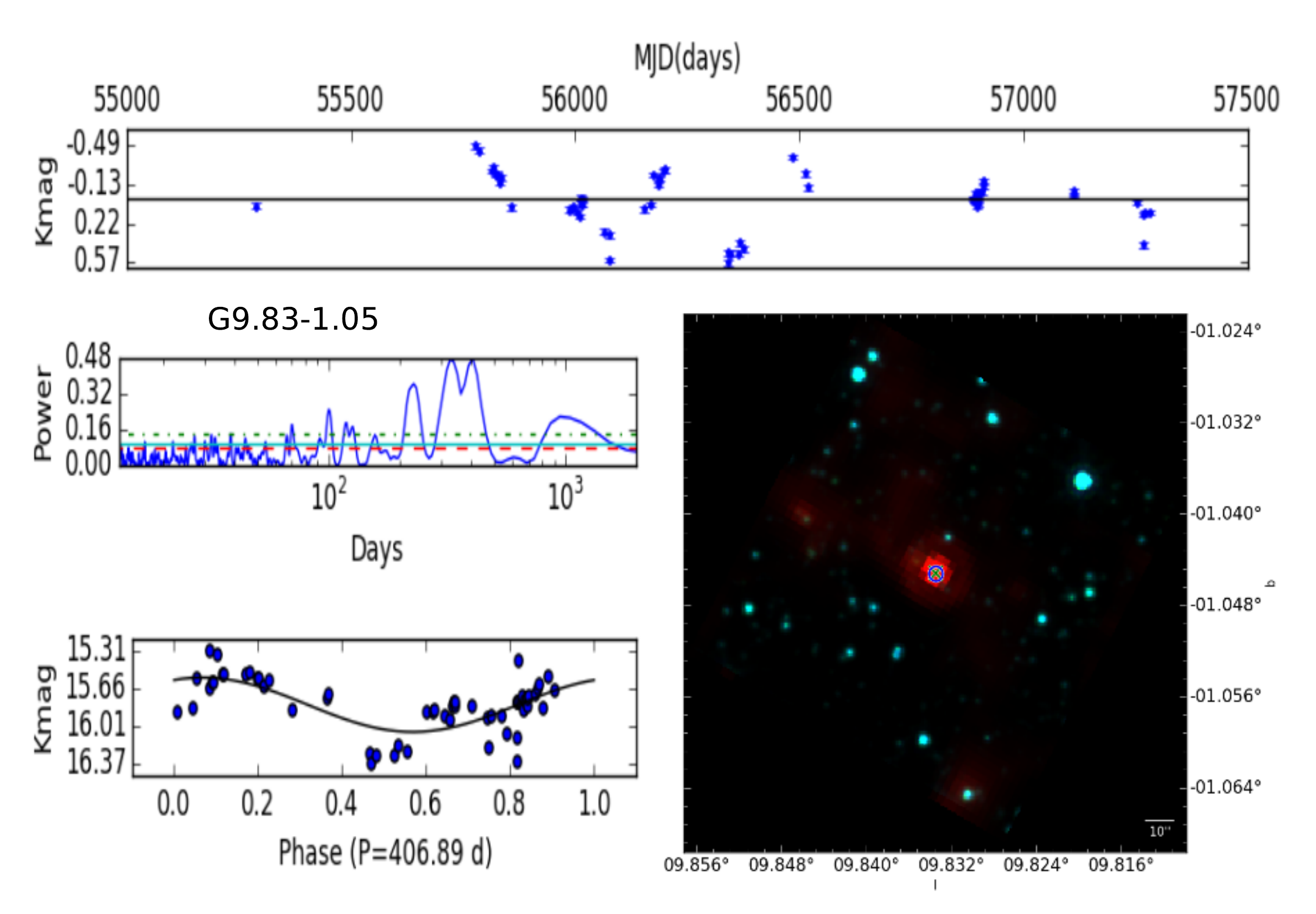}\\
                \includegraphics[width=0.48\textwidth]{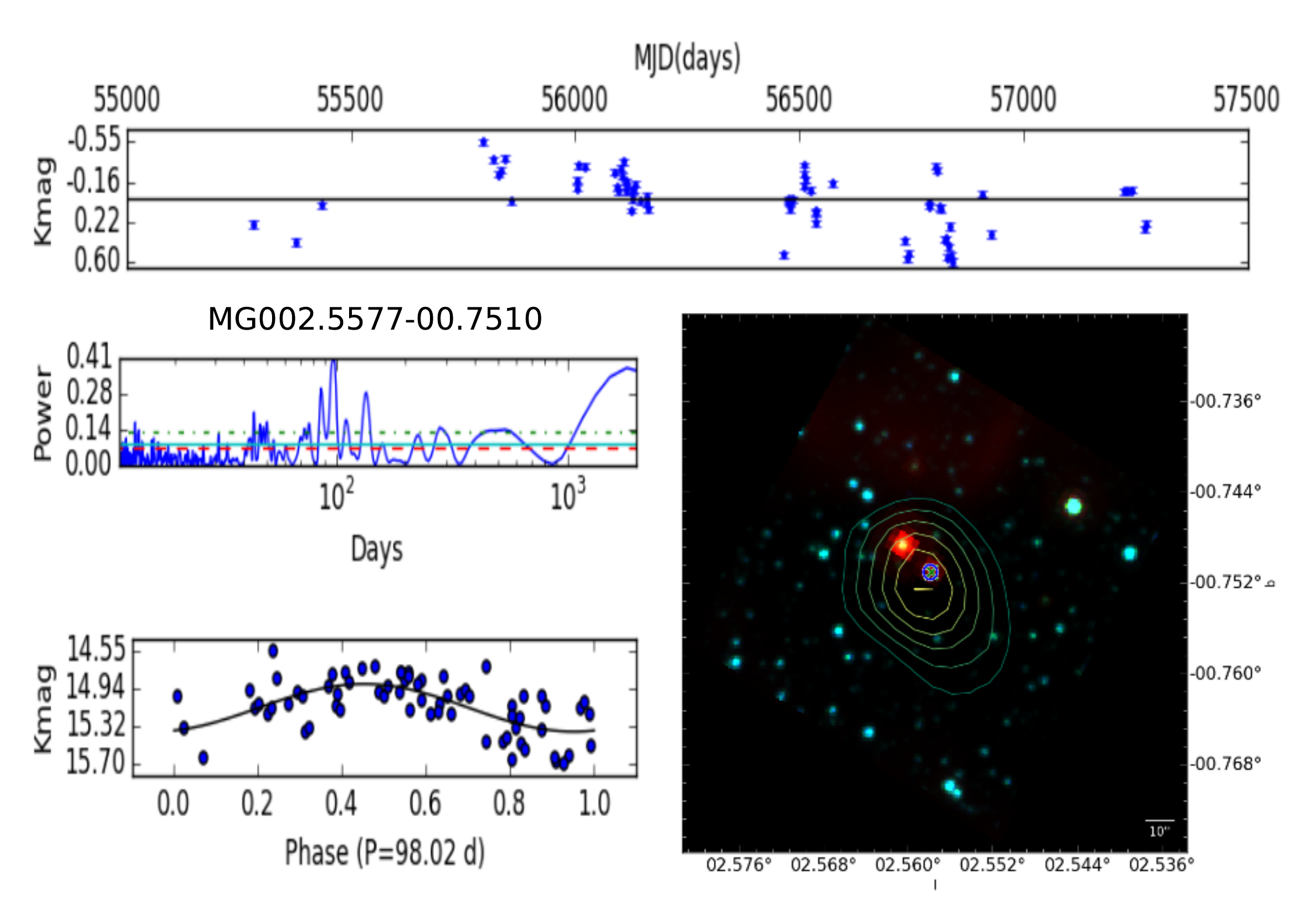}&
                \includegraphics[width=0.48\textwidth]{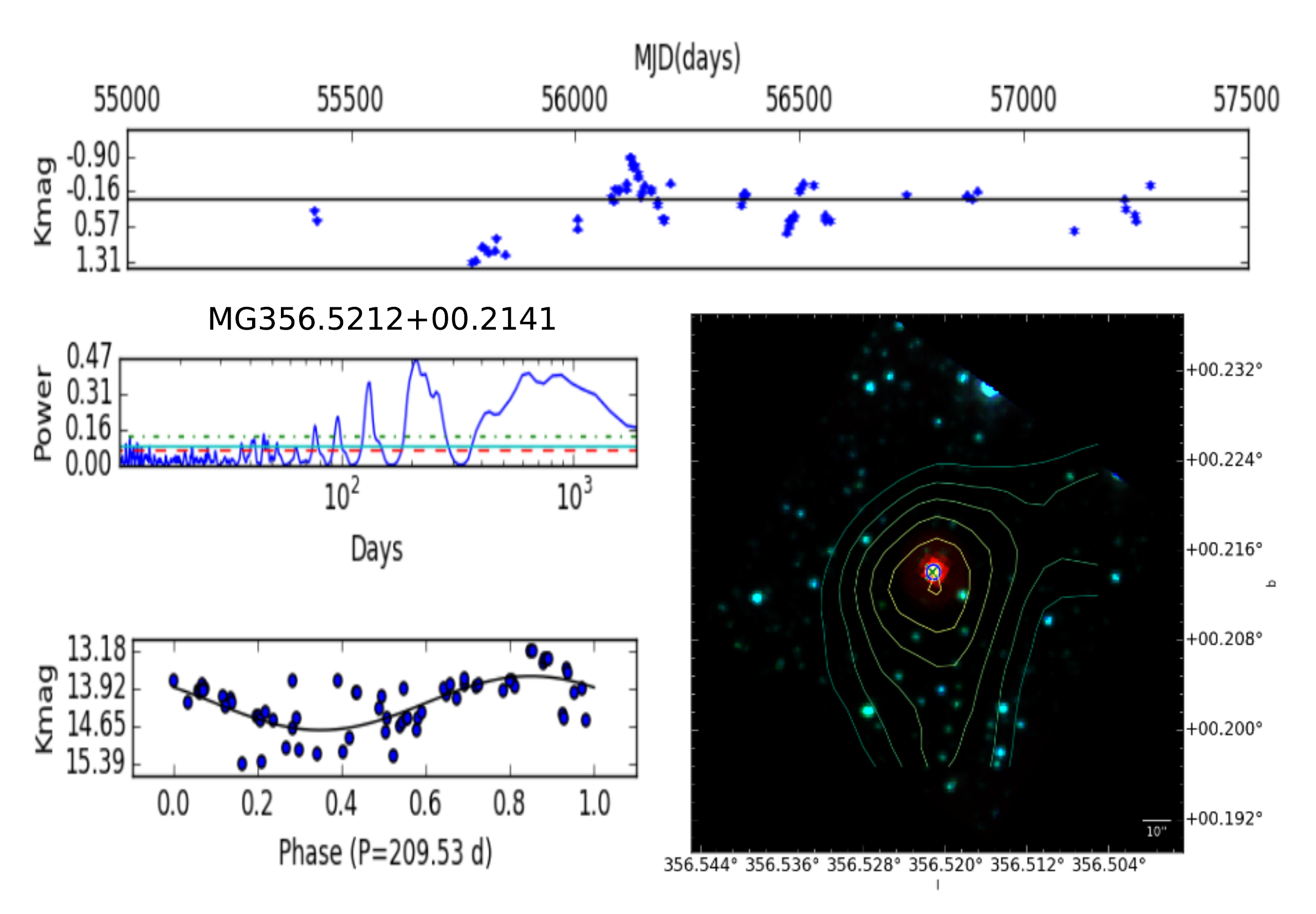}\\
                &
                \\
                
        \end{tabular}
        \caption{Continuation of Fig. \ref{fig:testgrid}.}
\end{figure}

\begin{figure}
        \begin{tabular}{ccc}

                \includegraphics[width=0.32\textwidth]{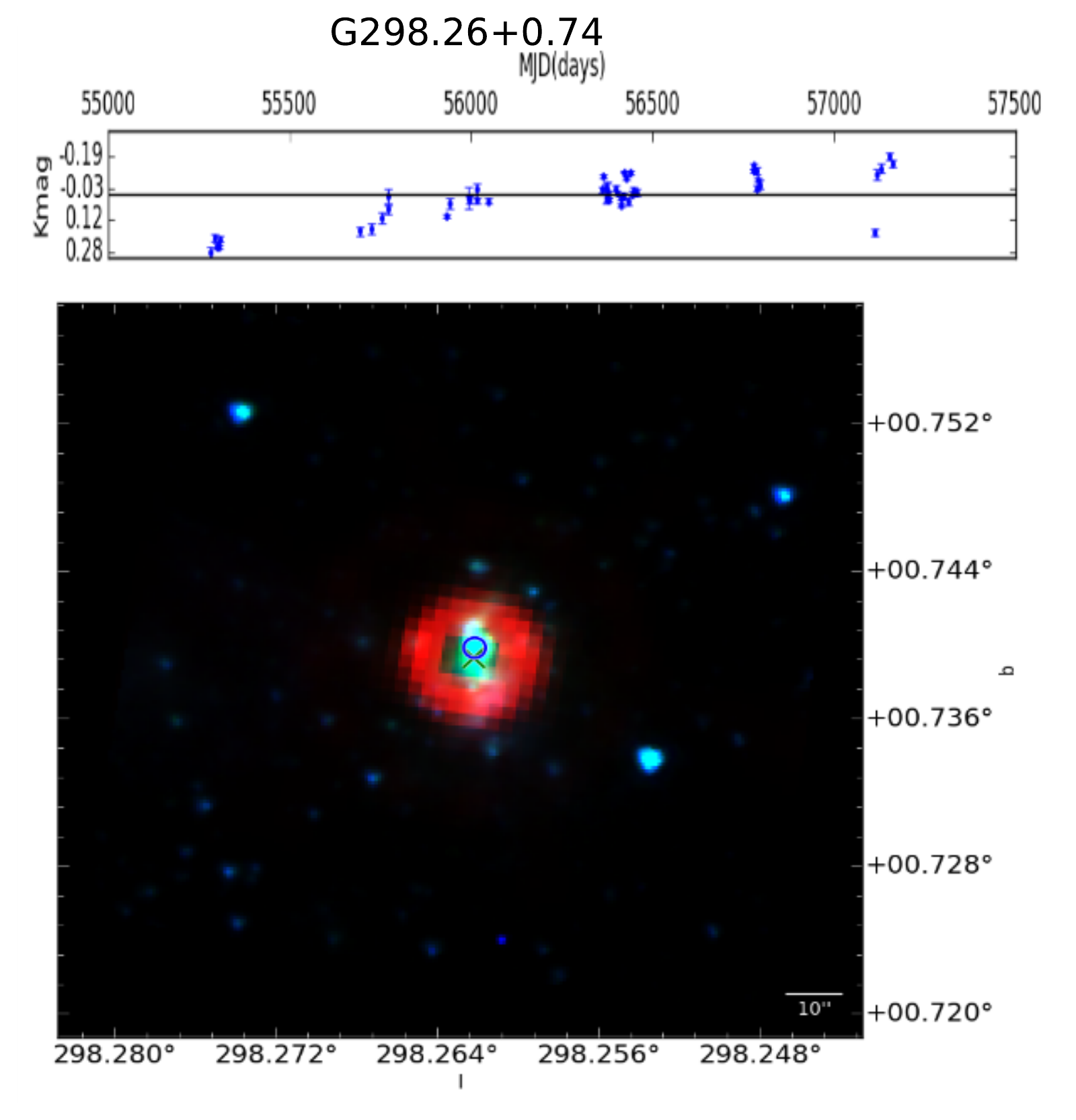}&
                \includegraphics[width=0.32\textwidth]{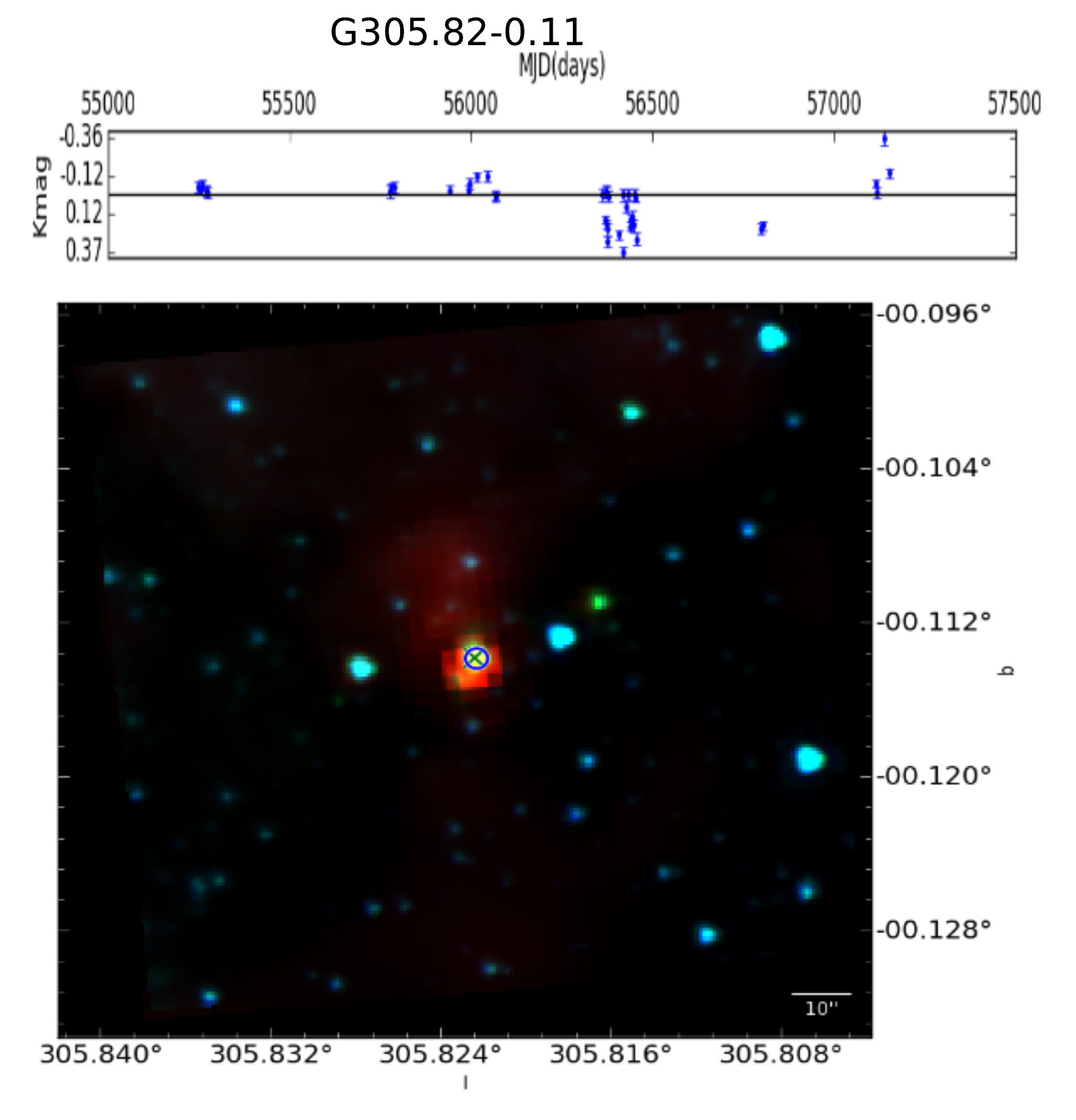}&
                \includegraphics[width=0.32\textwidth]{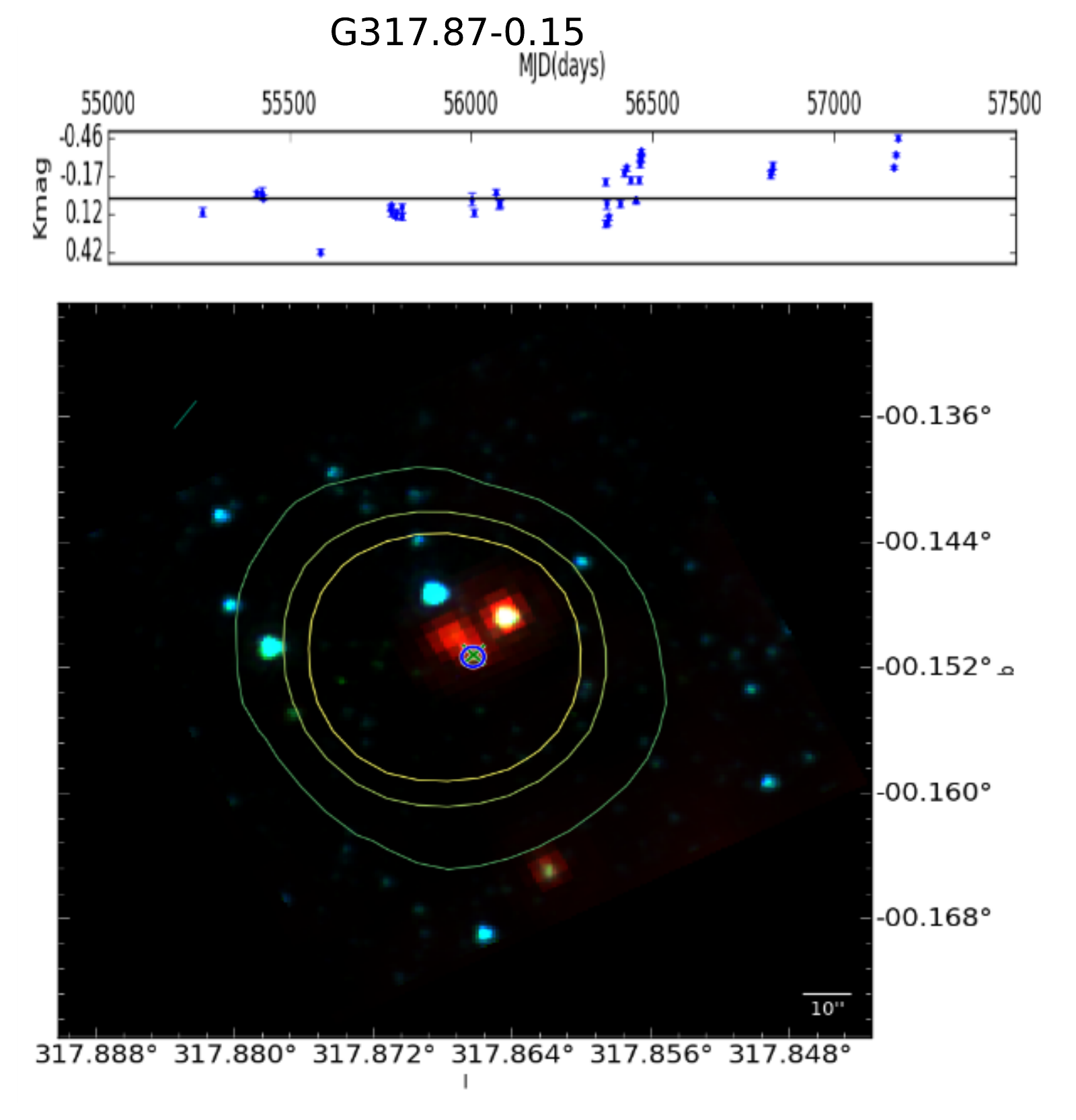}\\
                \includegraphics[width=0.32\textwidth]{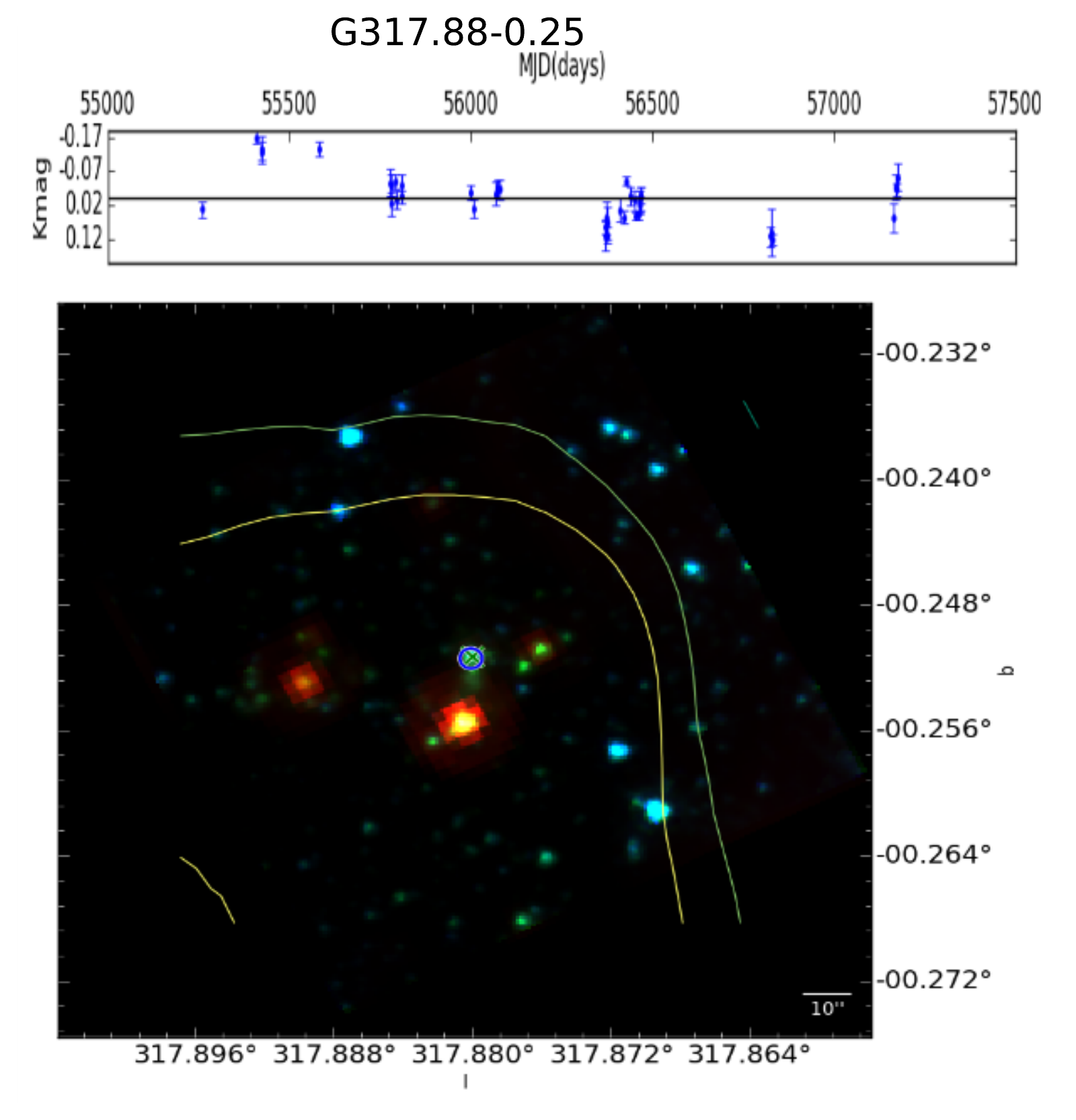}&
                \includegraphics[width=0.32\textwidth]{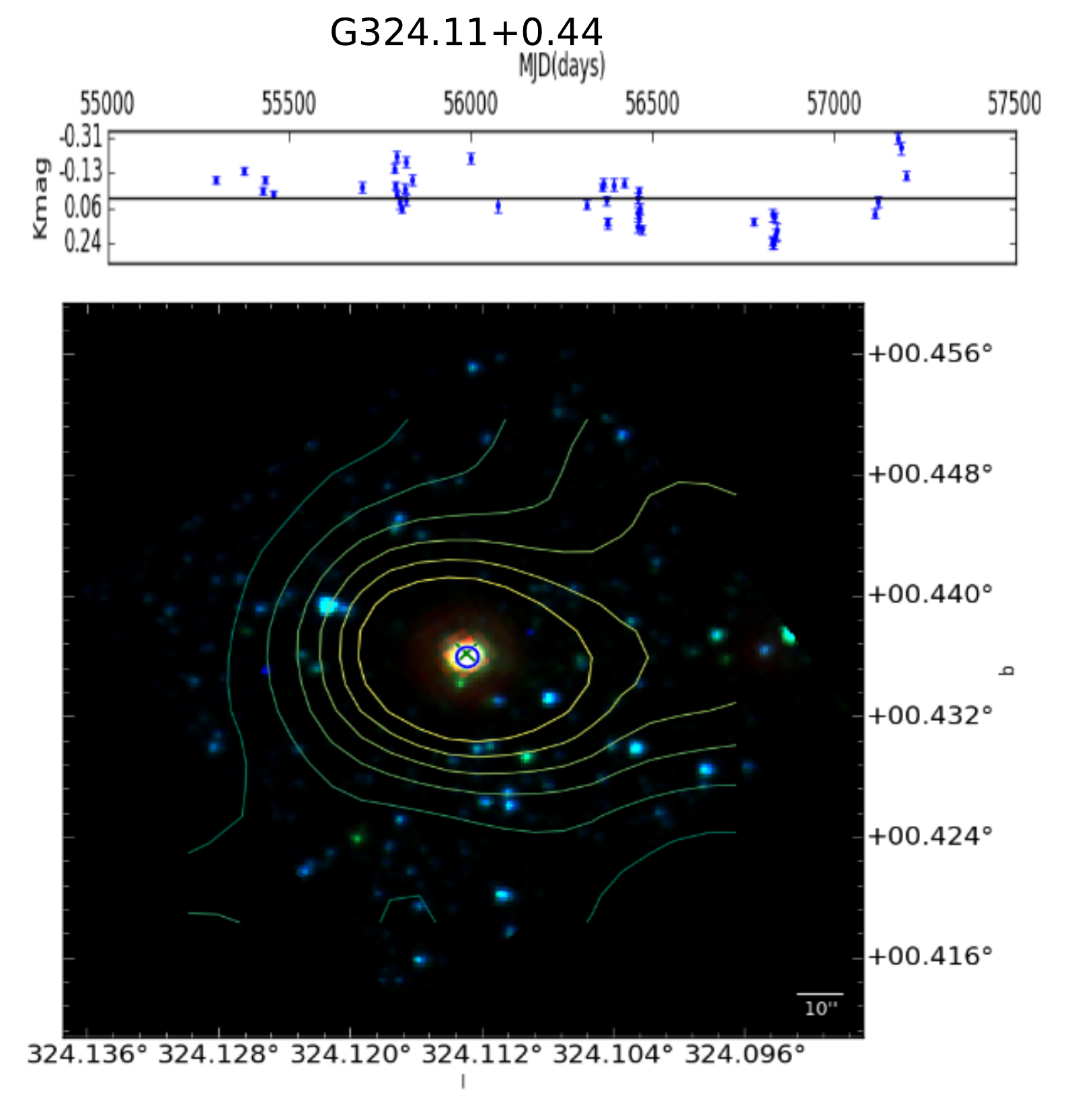}&
                \includegraphics[width=0.32\textwidth]{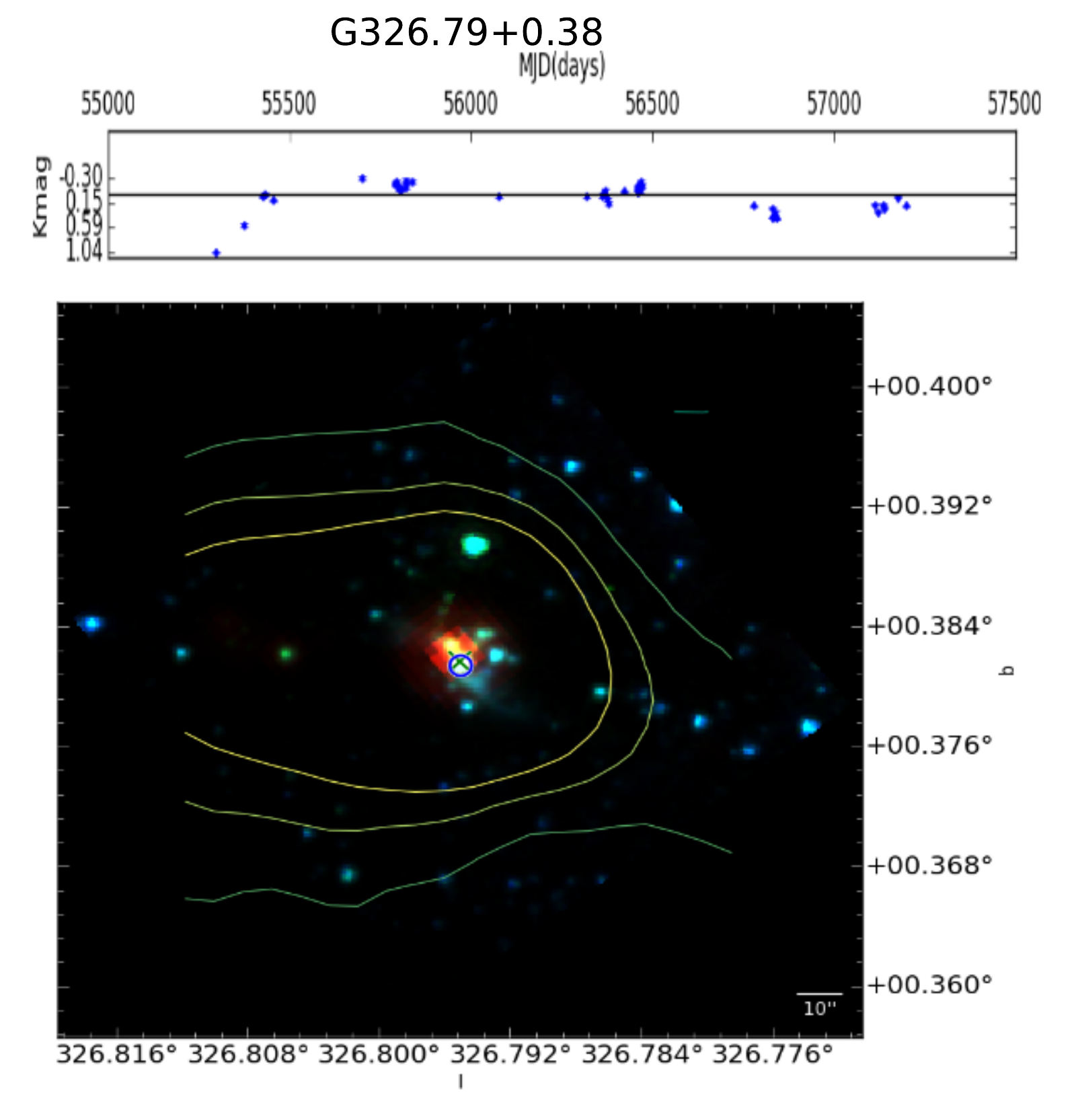}\\
                \includegraphics[width=0.32\textwidth]{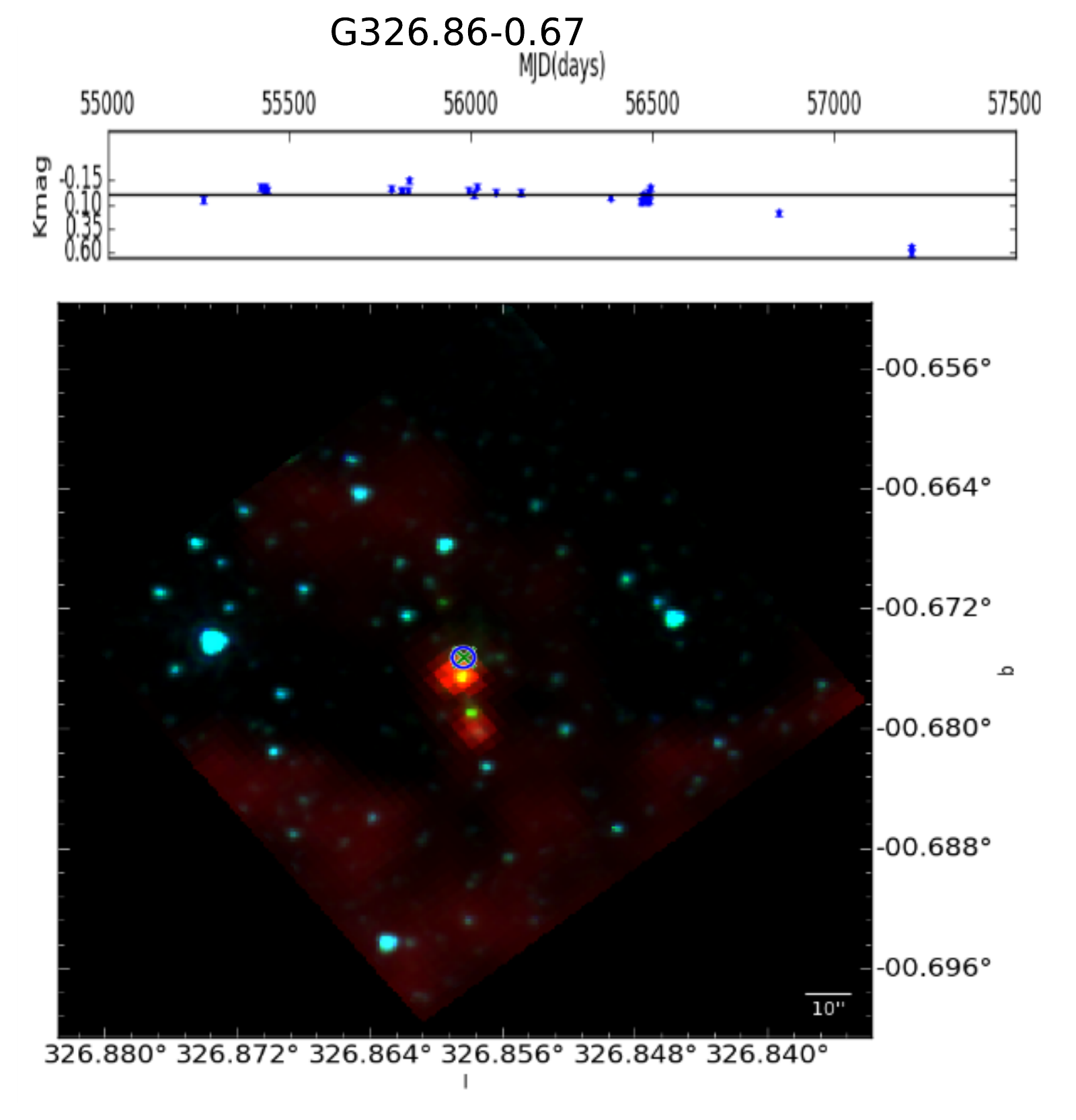}&
                \includegraphics[width=0.32\textwidth]{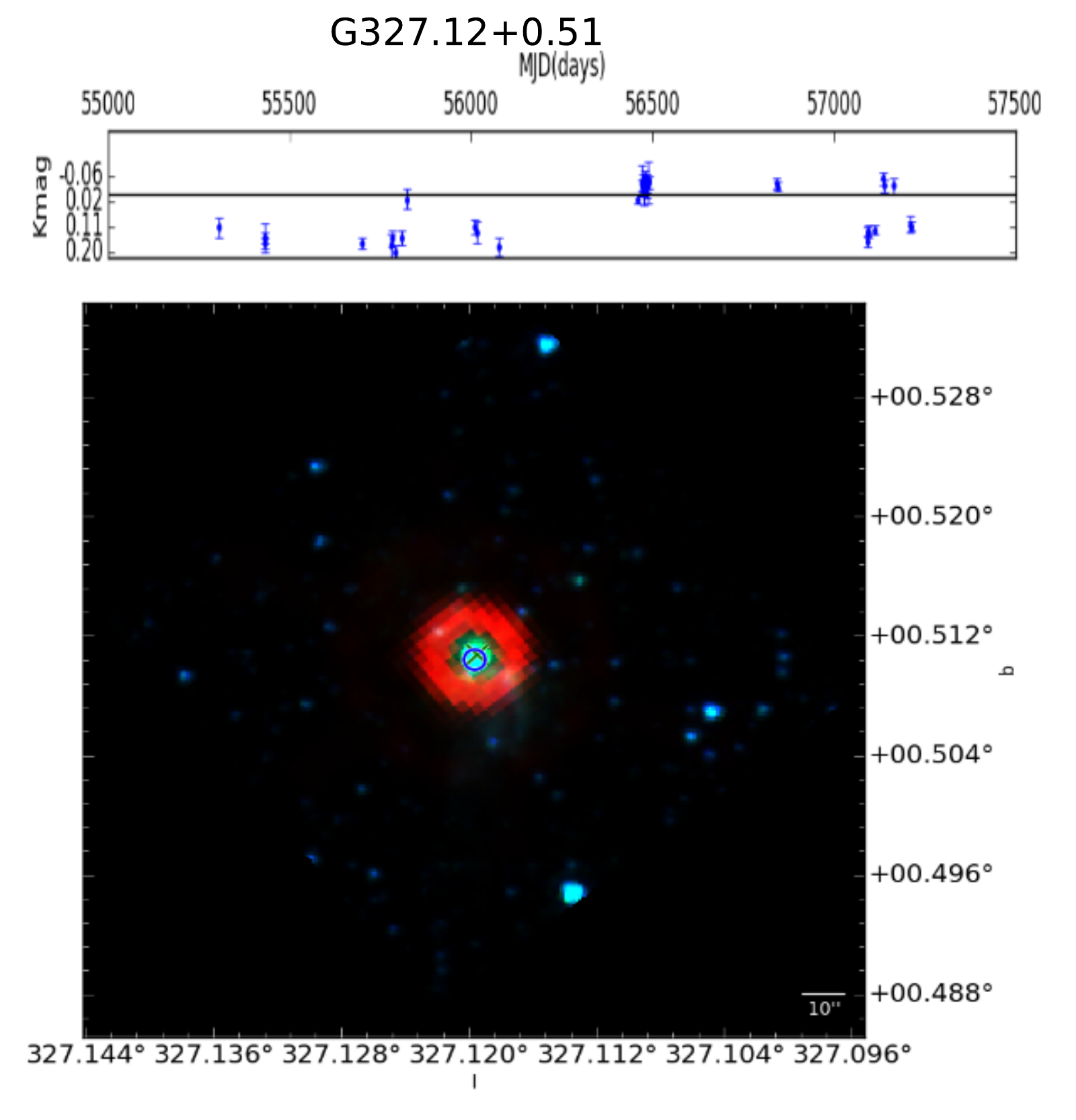}&
                \includegraphics[width=0.32\textwidth]{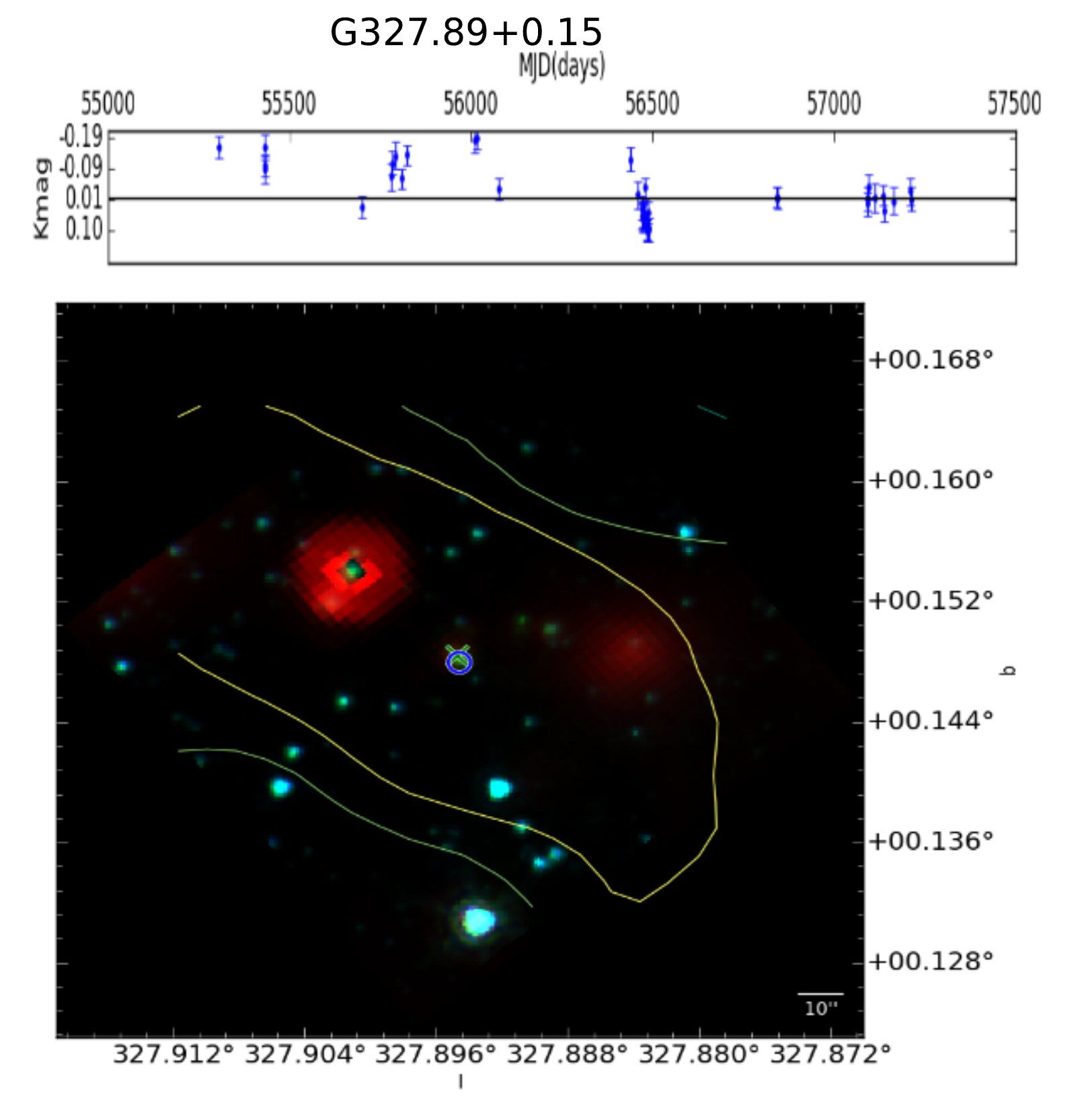}\\
                \includegraphics[width=0.32\textwidth]{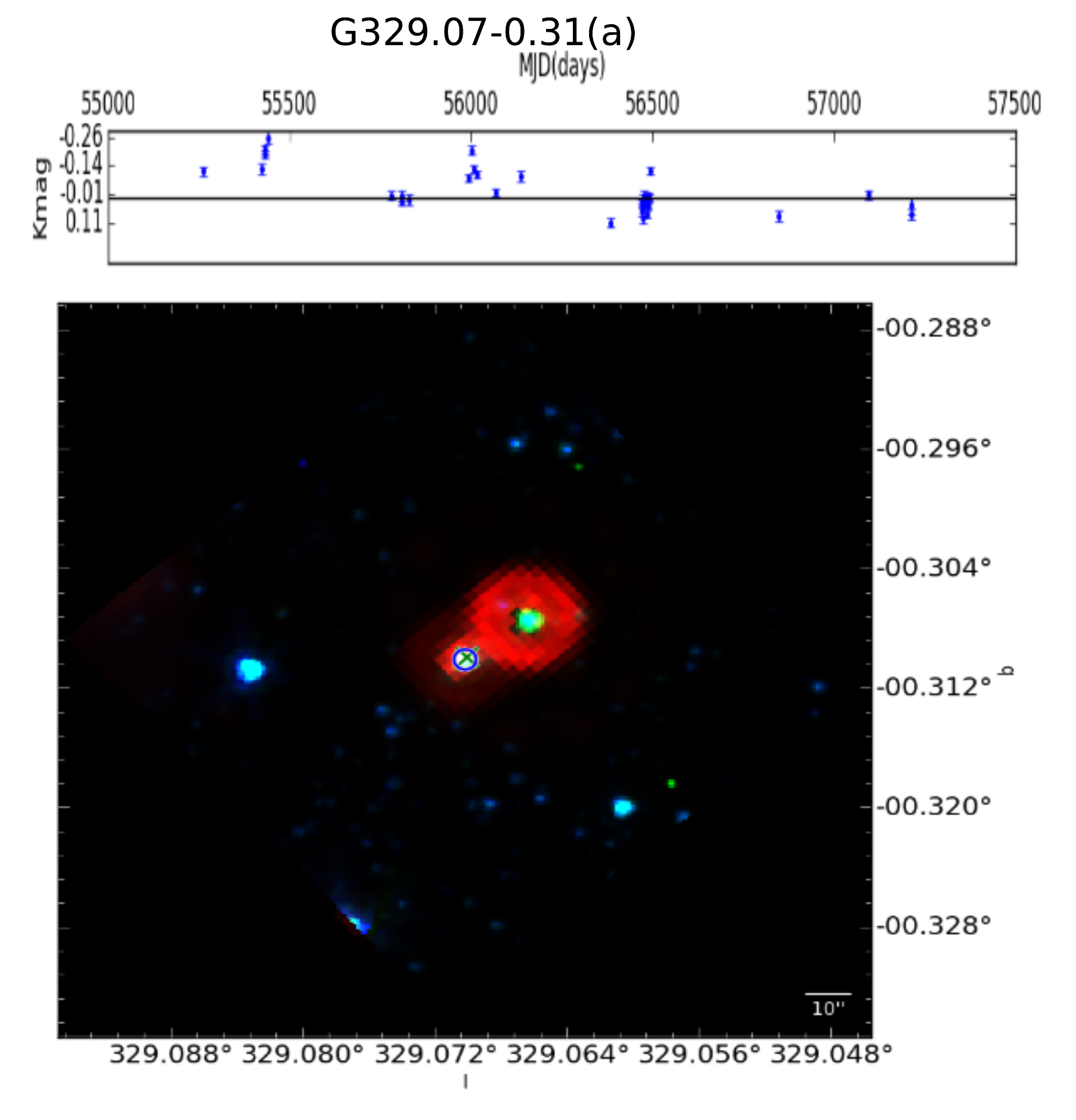}&
                \includegraphics[width=0.32\textwidth]{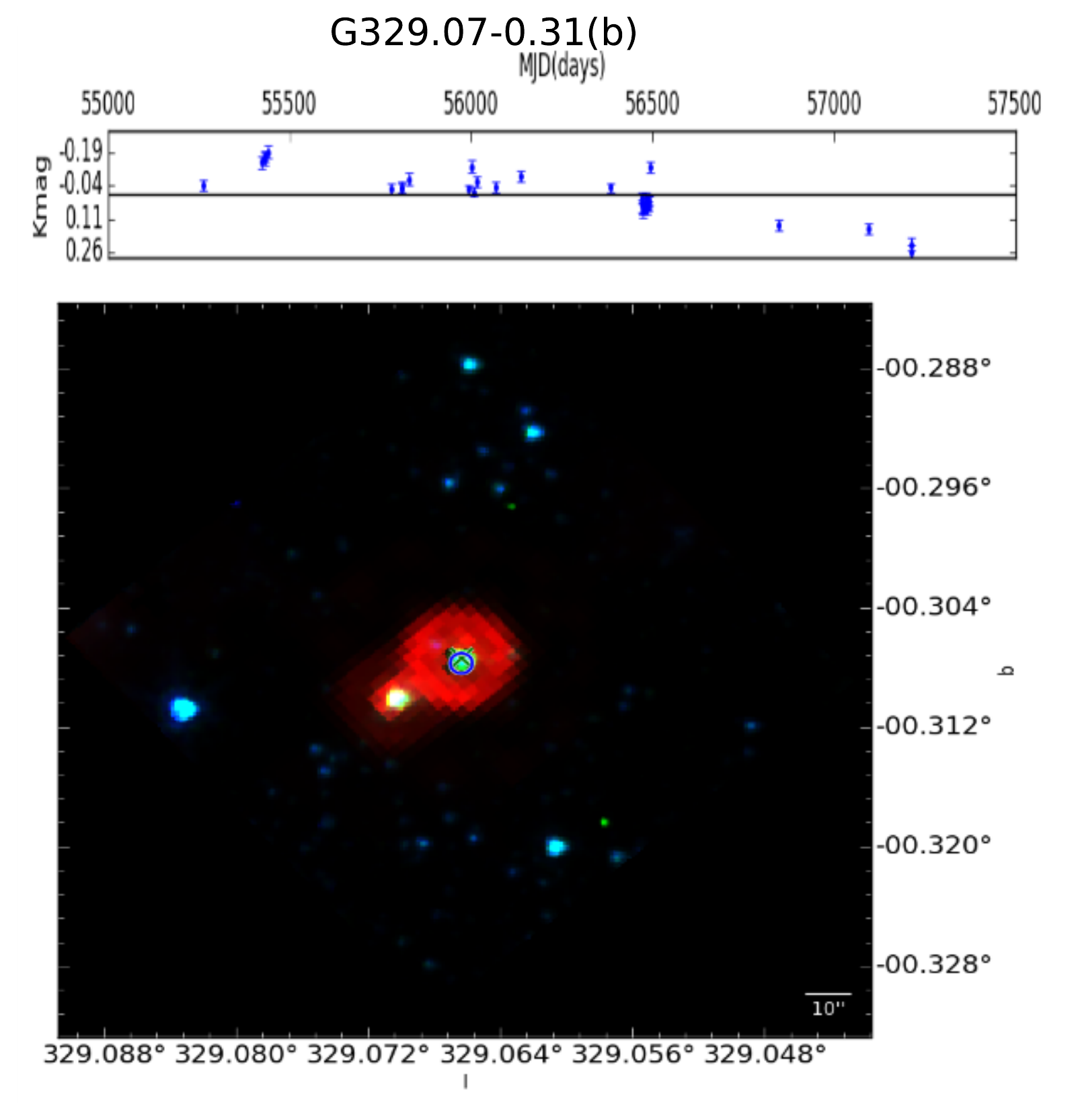}&
                \includegraphics[width=0.32\textwidth]{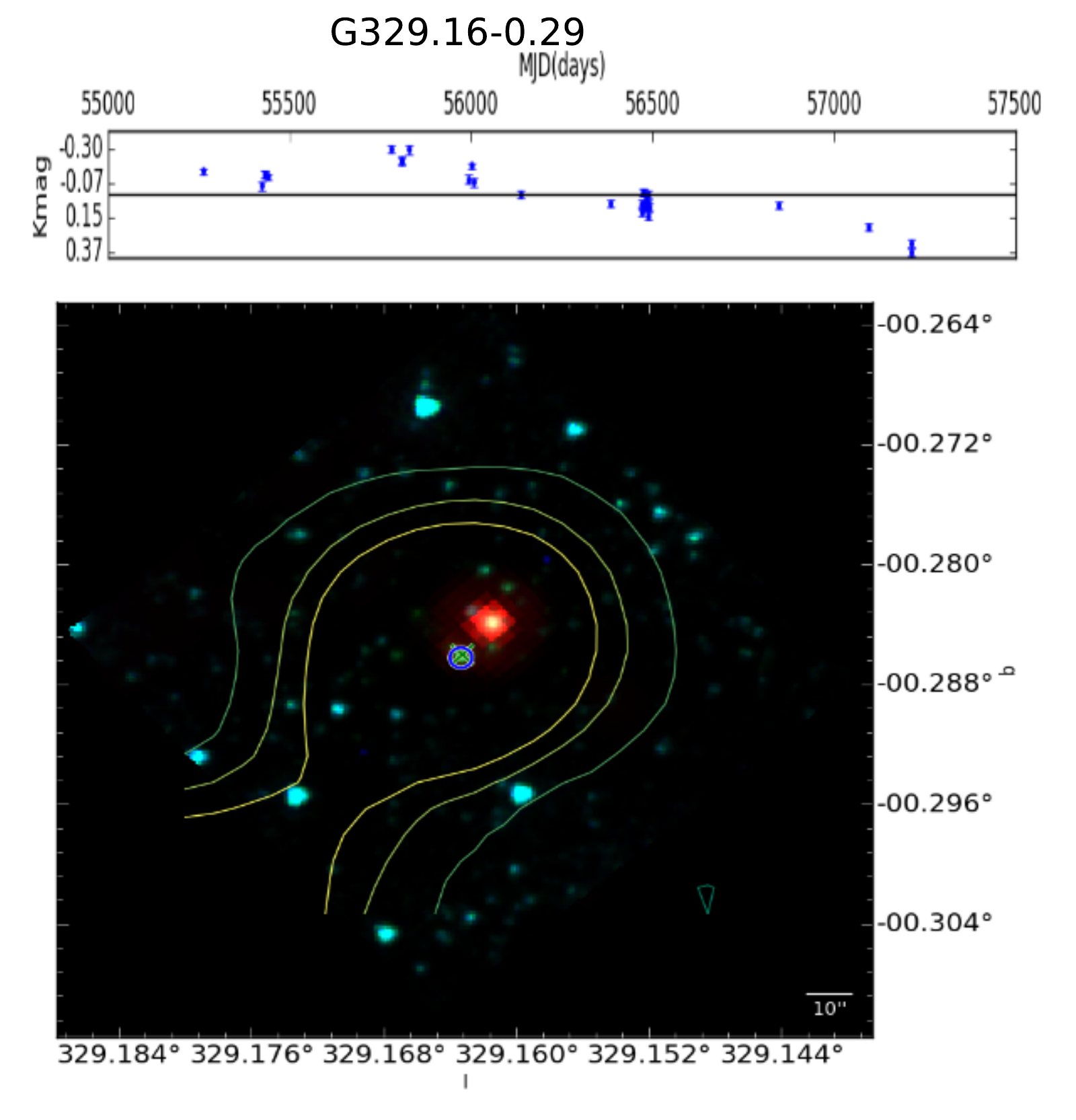}\\
                
        \end{tabular}
        \caption{Continuation of Fig. \ref{fig:testgrid}.}
\end{figure}

\begin{figure}
        \begin{tabular}{ccc}

                \includegraphics[width=0.32\textwidth]{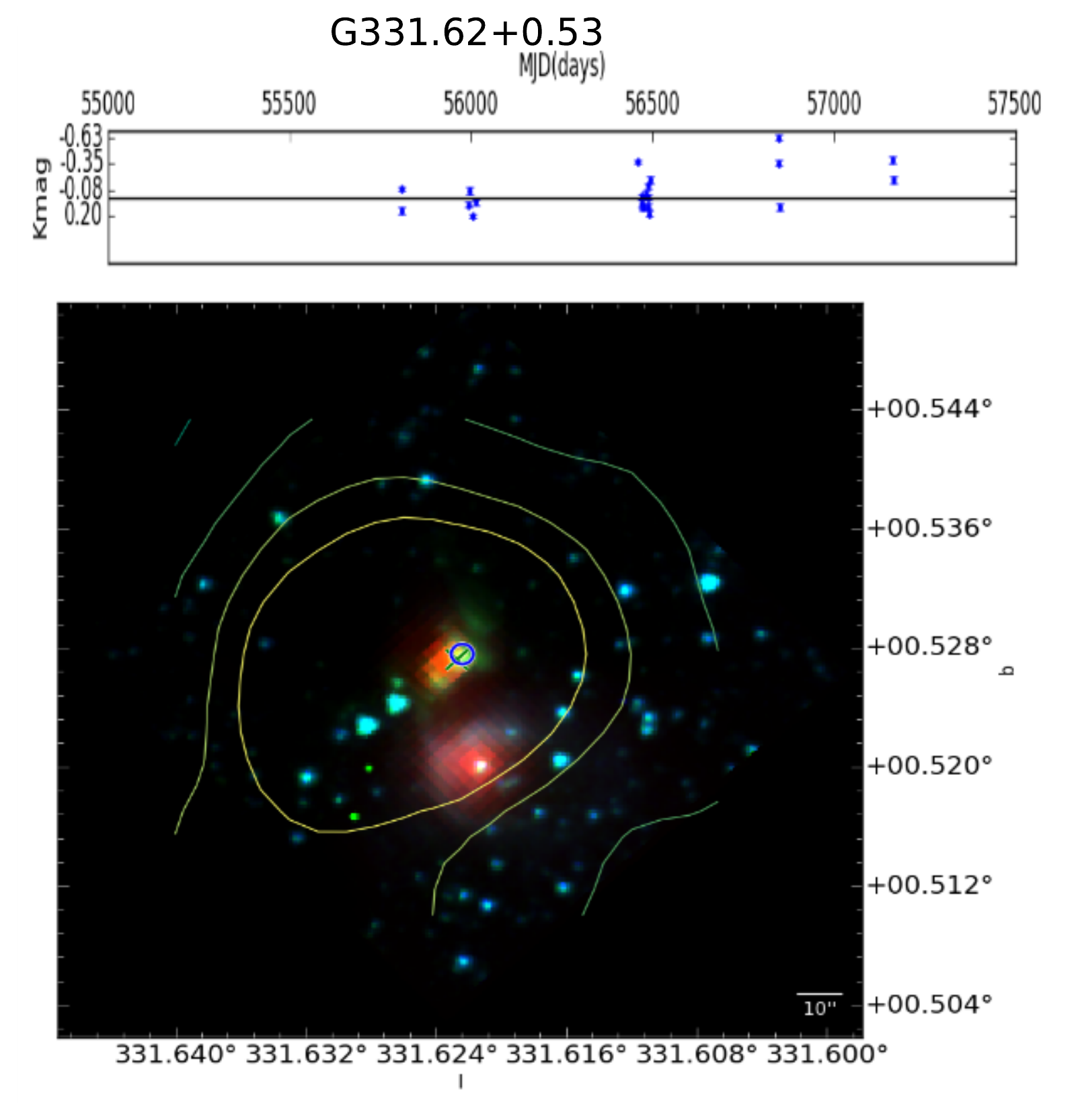}&
                \includegraphics[width=0.32\textwidth]{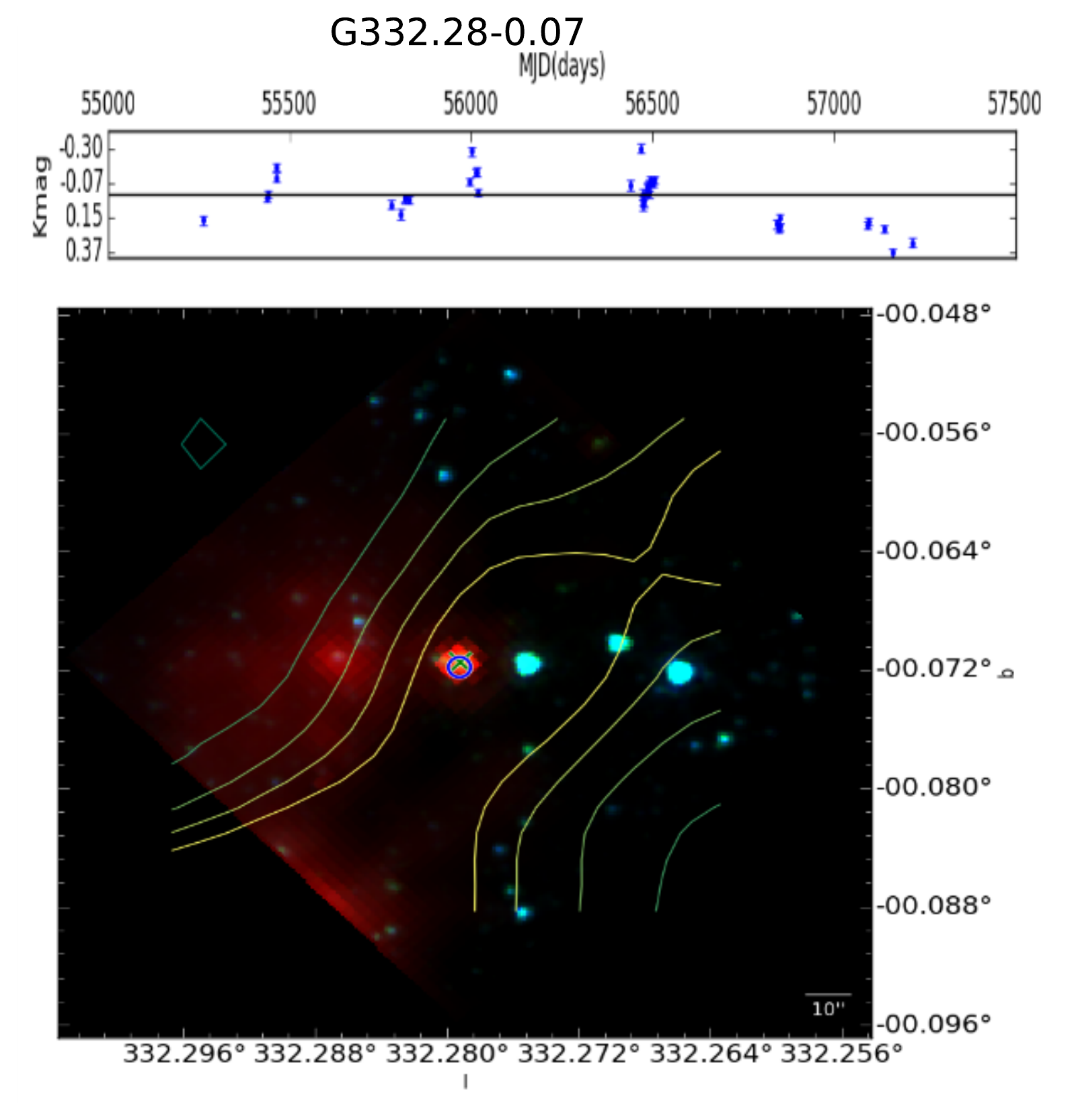}&
                \includegraphics[width=0.32\textwidth]{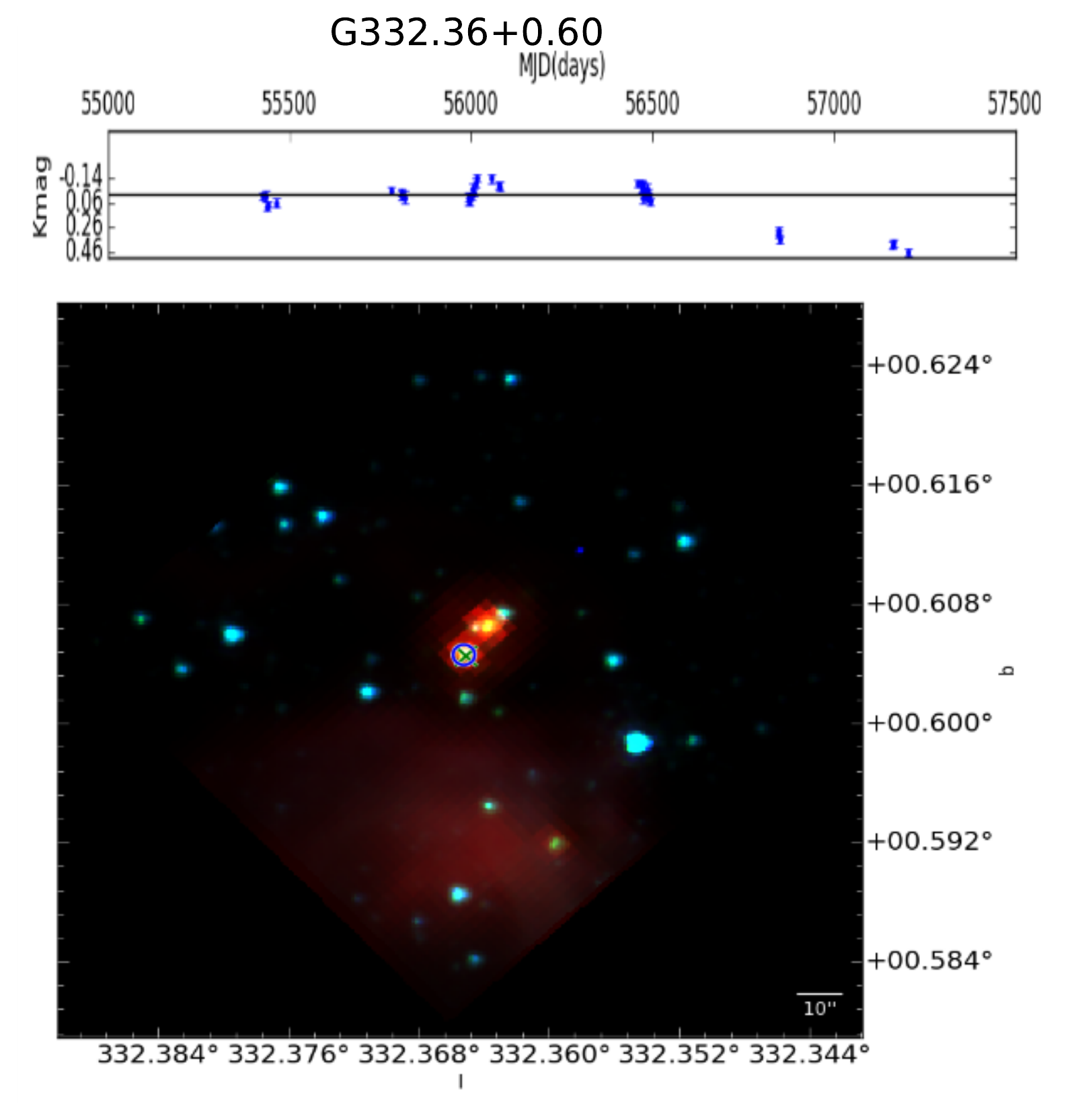}\\
                \includegraphics[width=0.32\textwidth]{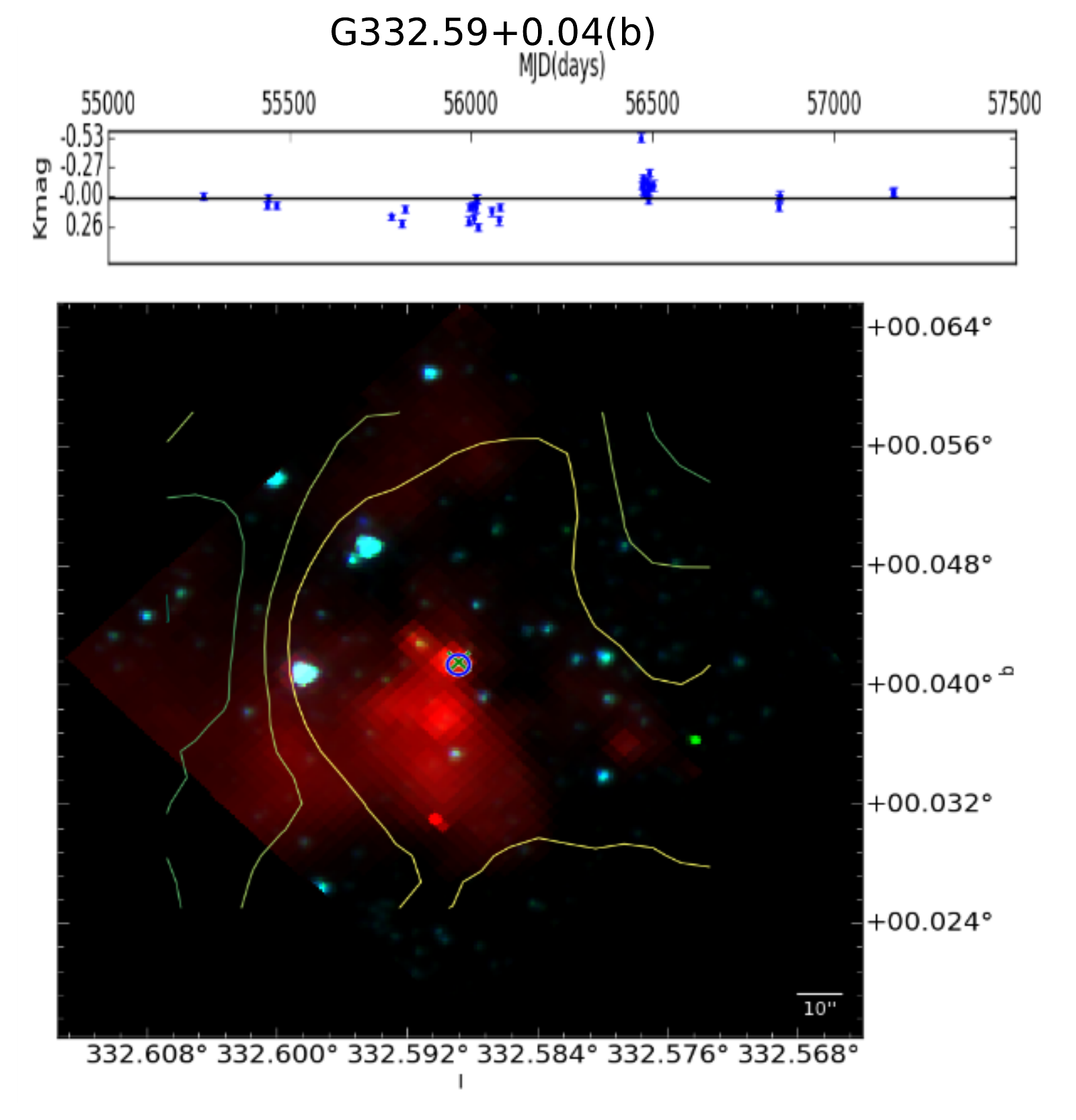}&
                \includegraphics[width=0.32\textwidth]{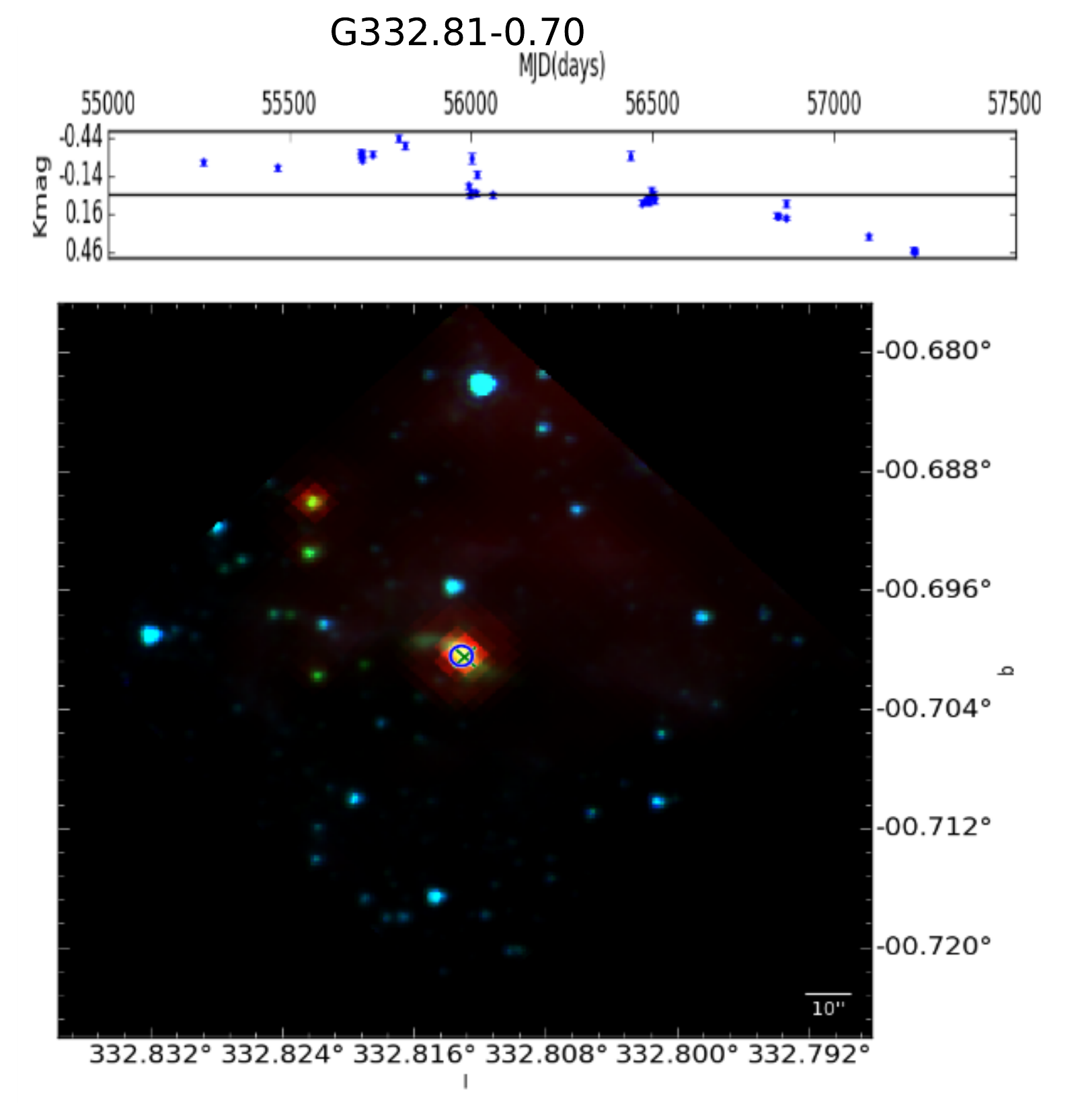}&
                \includegraphics[width=0.32\textwidth]{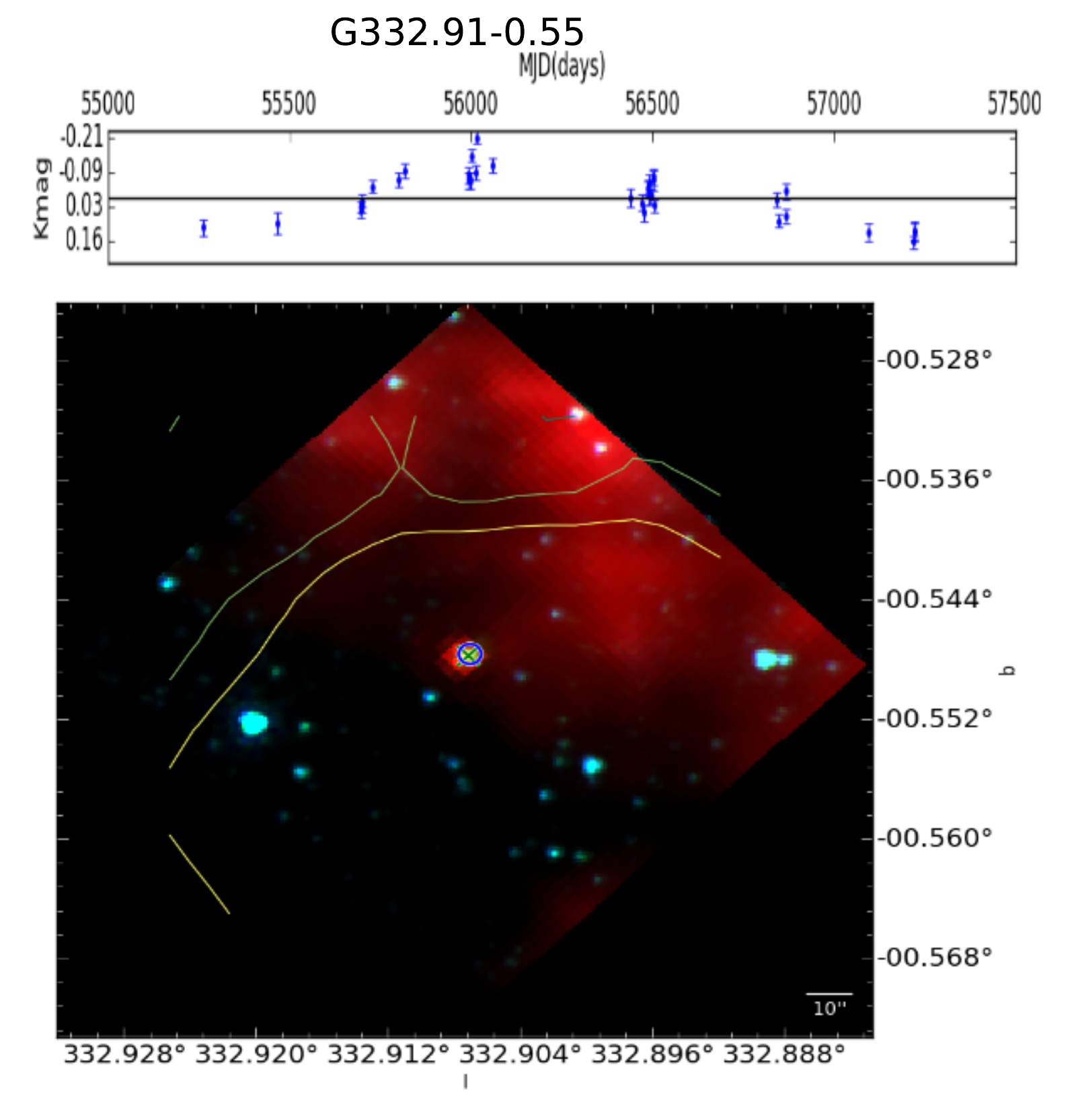}\\
                \includegraphics[width=0.32\textwidth]{{./G335.59-0.29_combined}.pdf}&
                \includegraphics[width=0.32\textwidth]{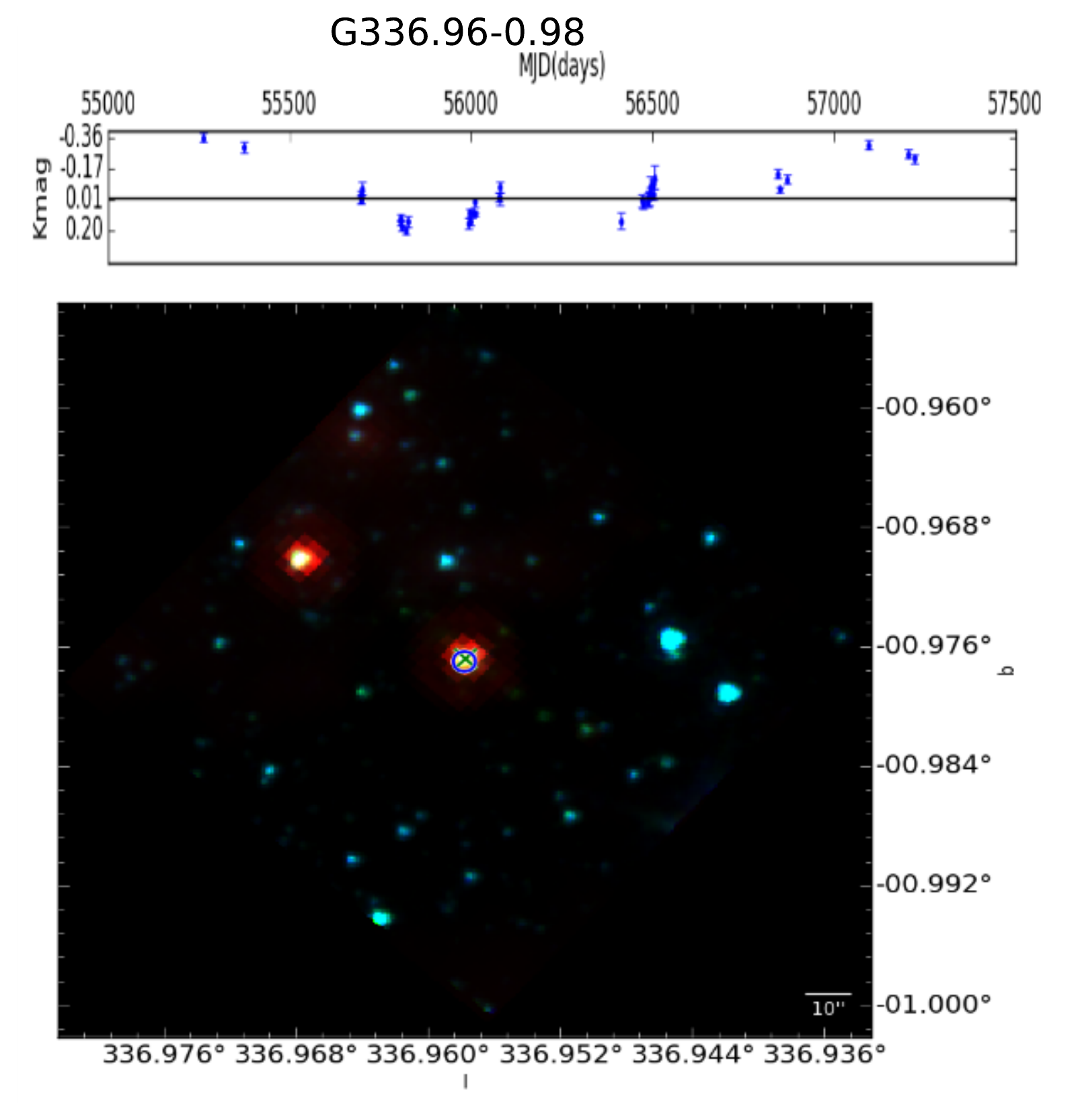}&
                \includegraphics[width=0.32\textwidth]{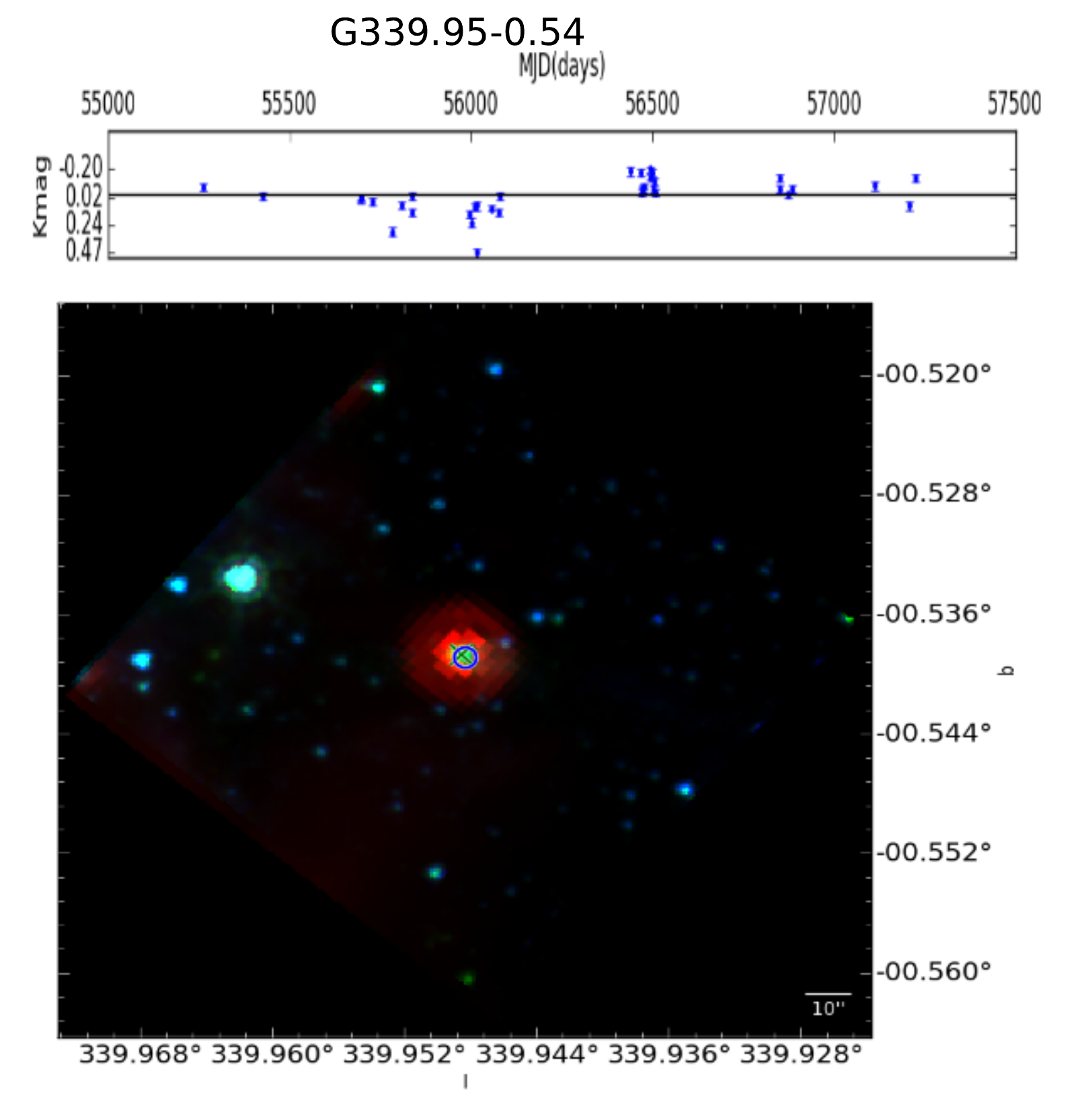}\\
                \includegraphics[width=0.32\textwidth]{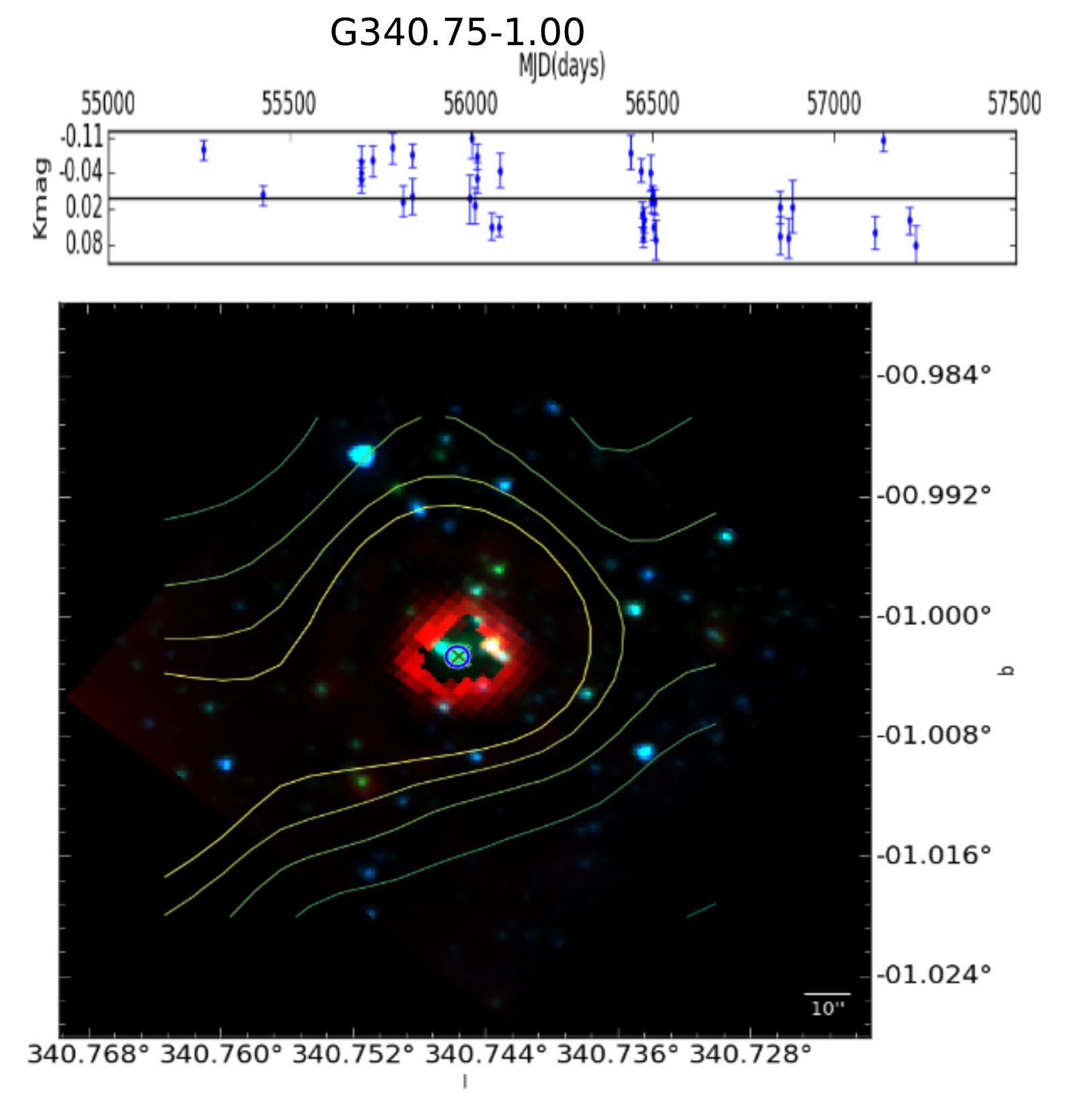}&
                \includegraphics[width=0.32\textwidth]{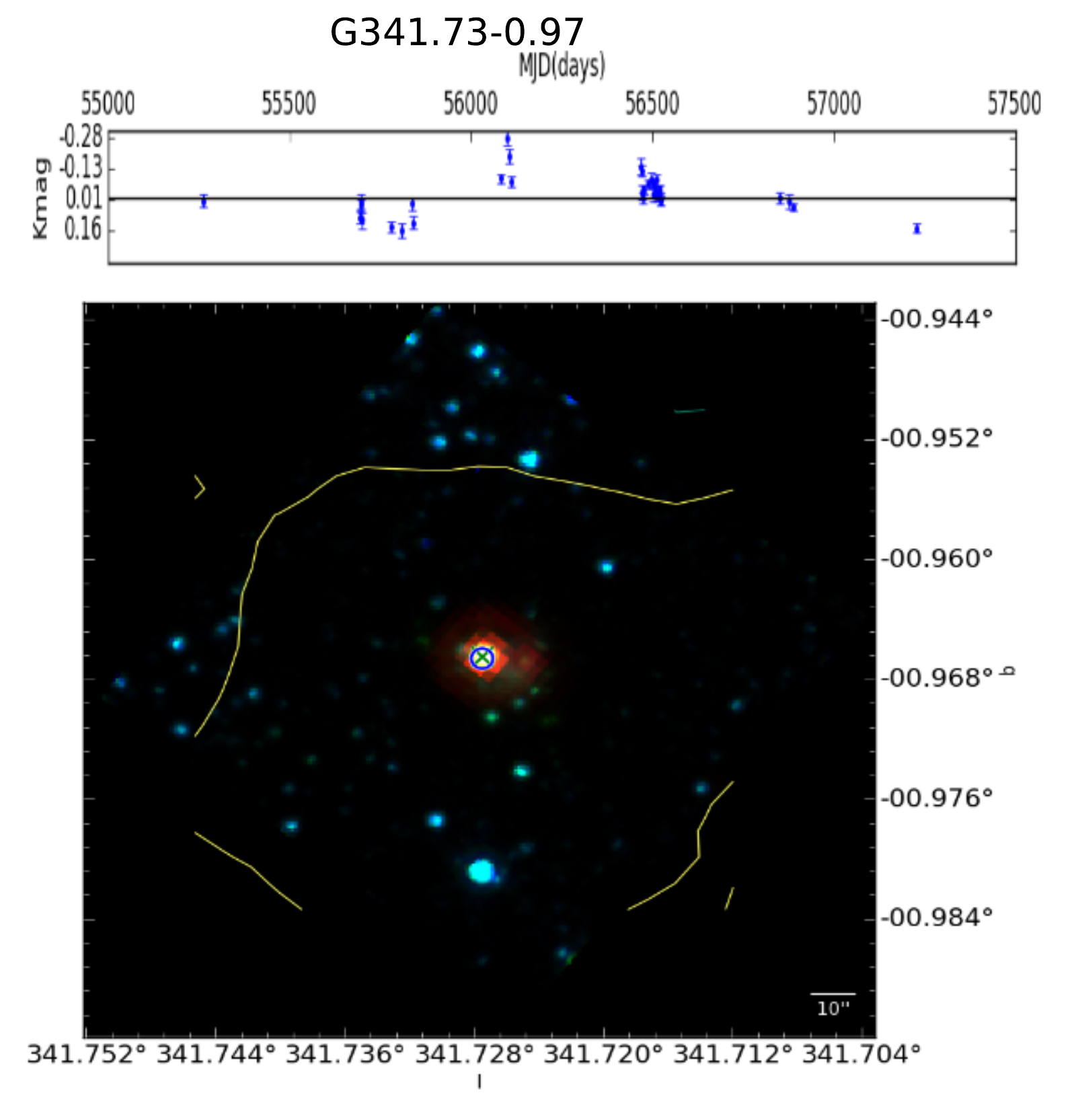}&
                \includegraphics[width=0.32\textwidth]{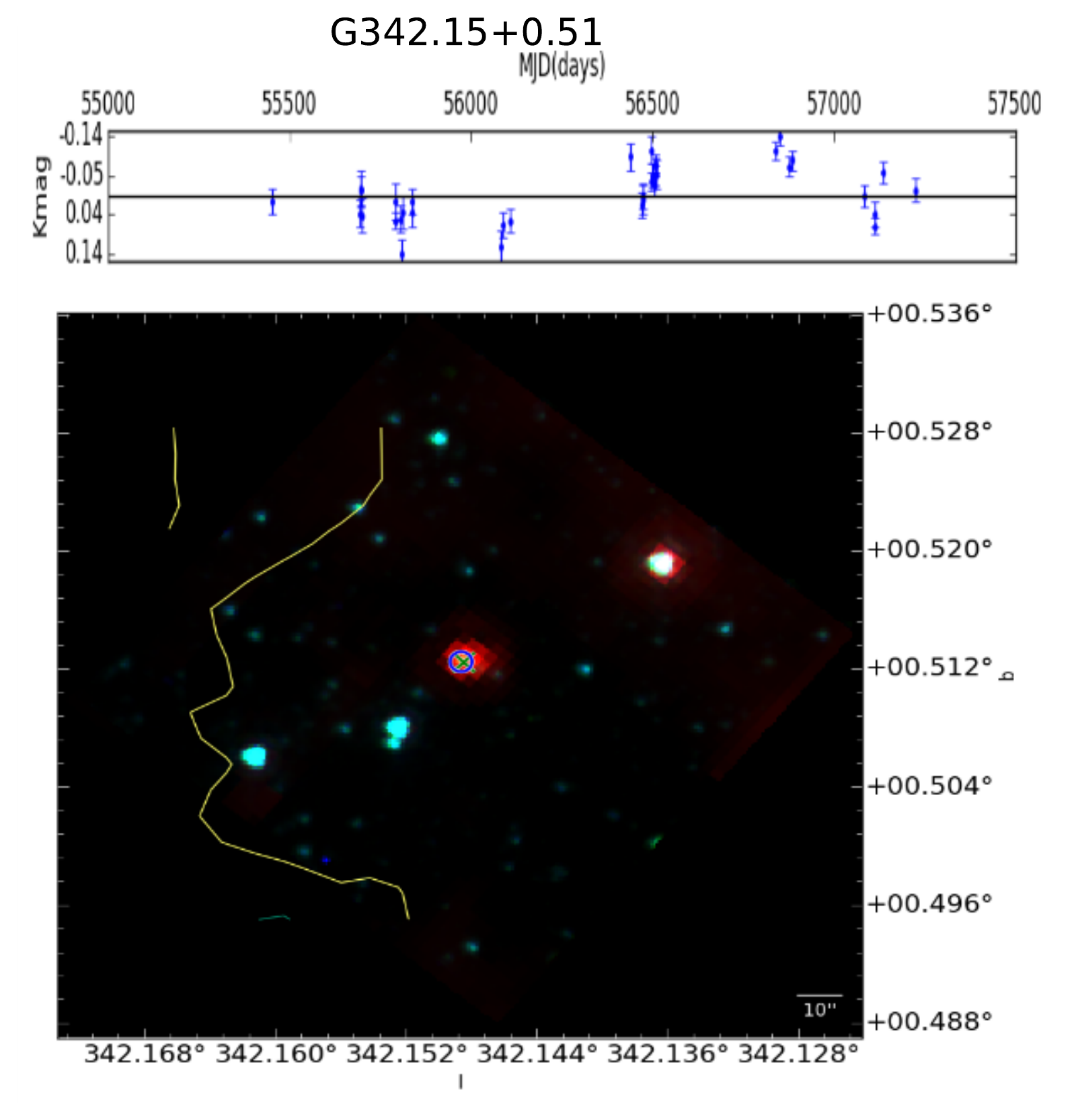}\\
        \end{tabular}
        \caption{Continuation of Fig. \ref{fig:testgrid}.}
\end{figure}

\begin{figure}
        \begin{tabular}{ccc}

                \includegraphics[width=0.32\textwidth]{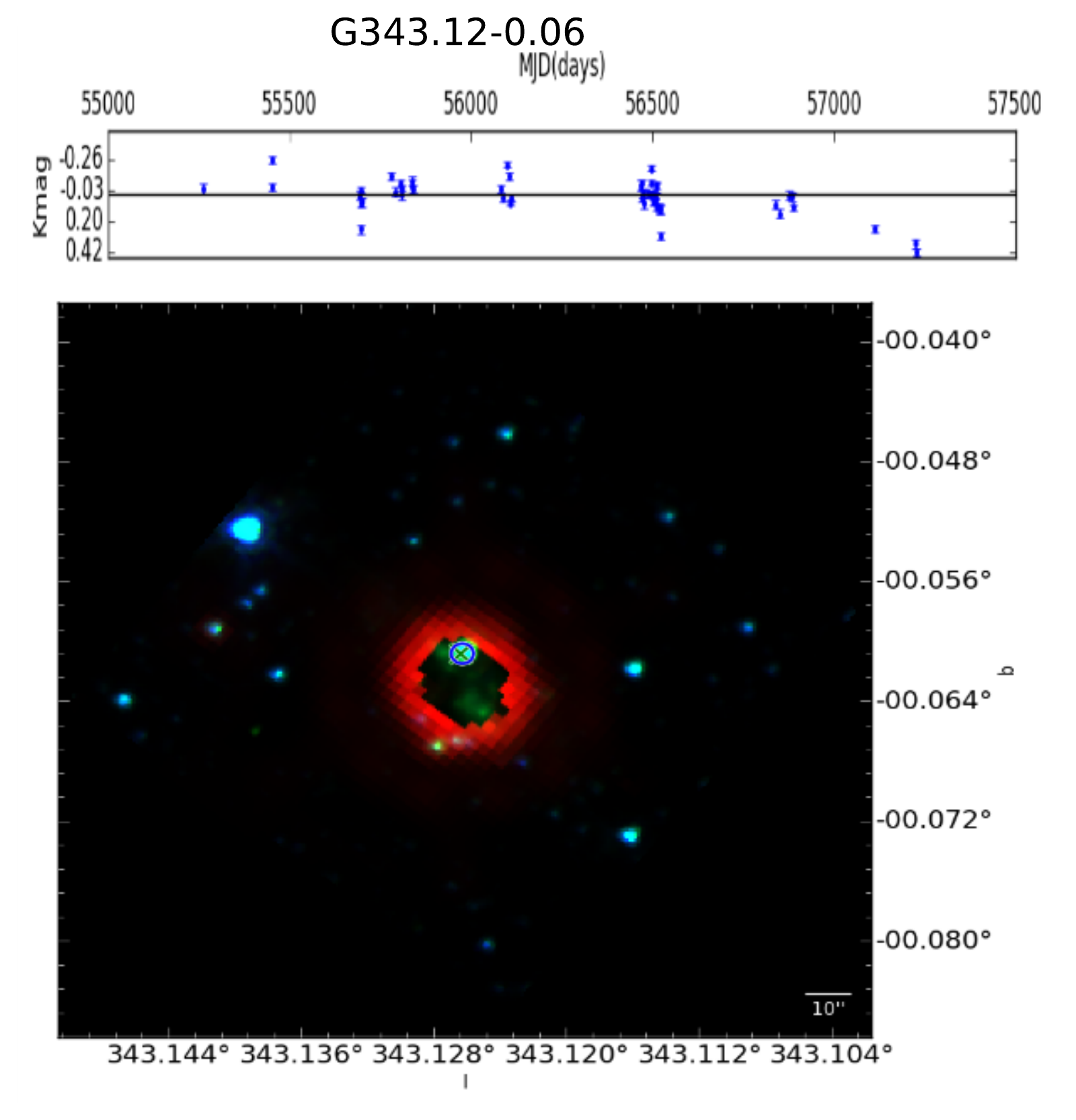}&
                \includegraphics[width=0.32\textwidth]{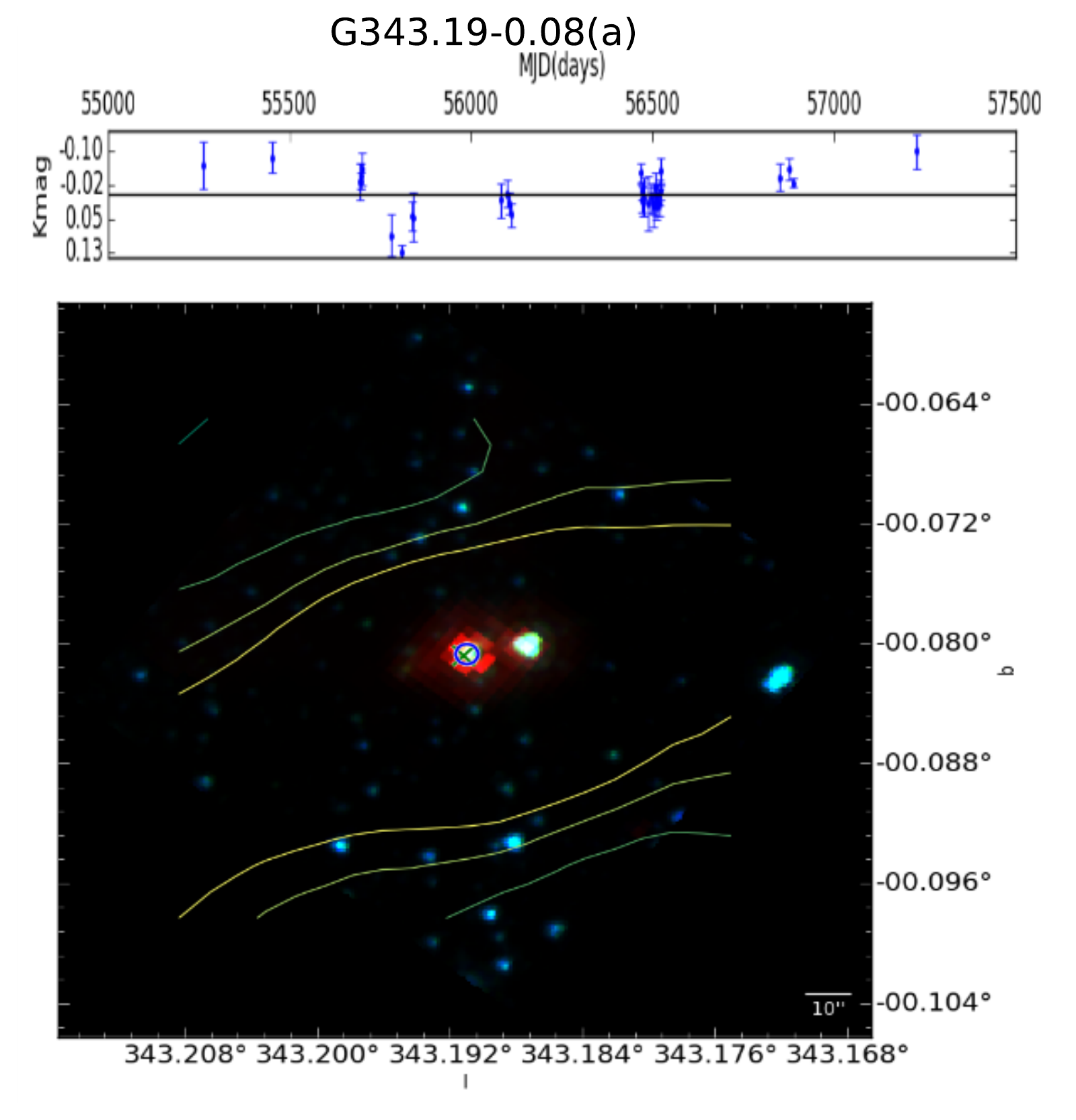}&
                \includegraphics[width=0.32\textwidth]{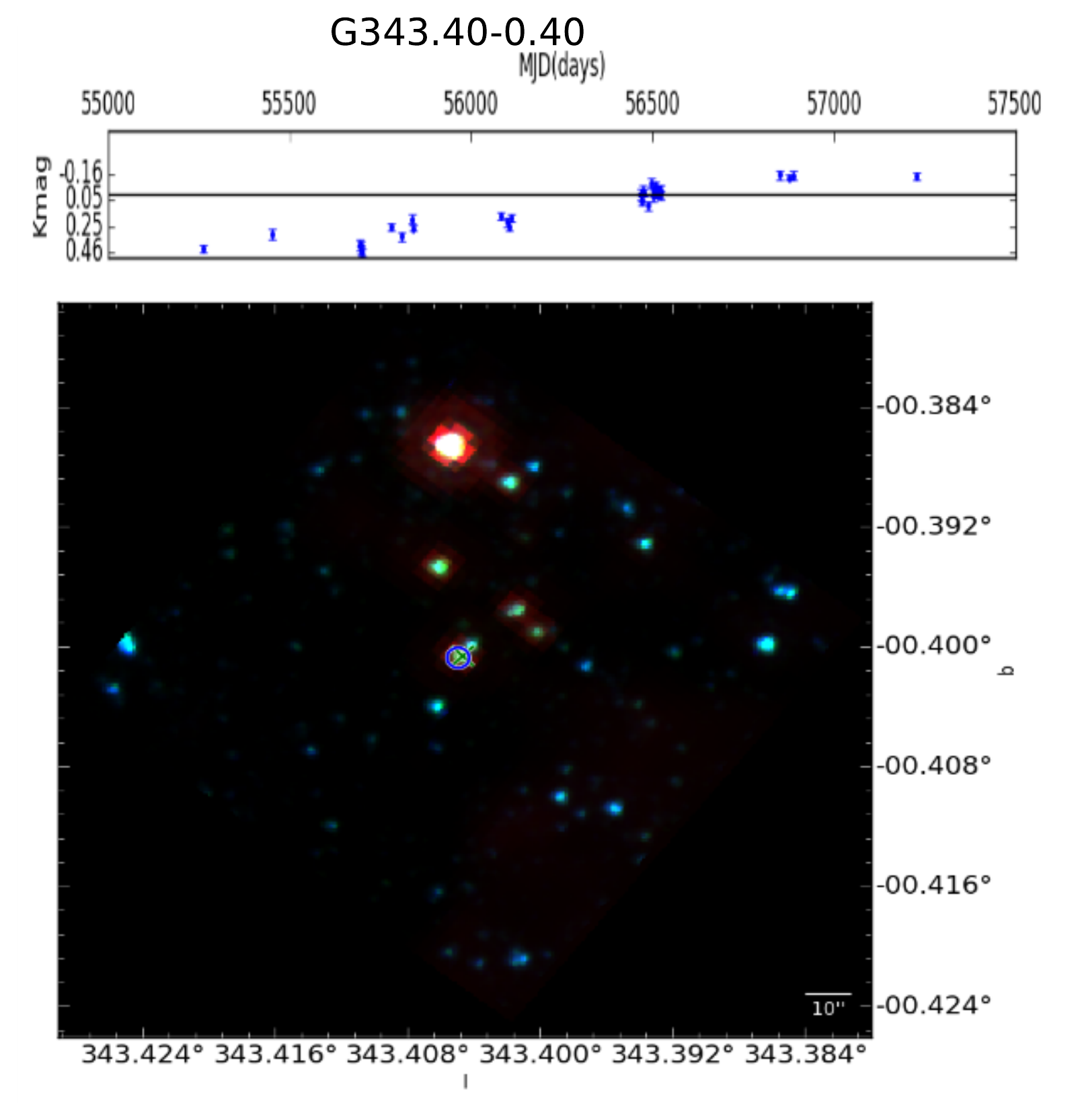}\\
                \includegraphics[width=0.32\textwidth]{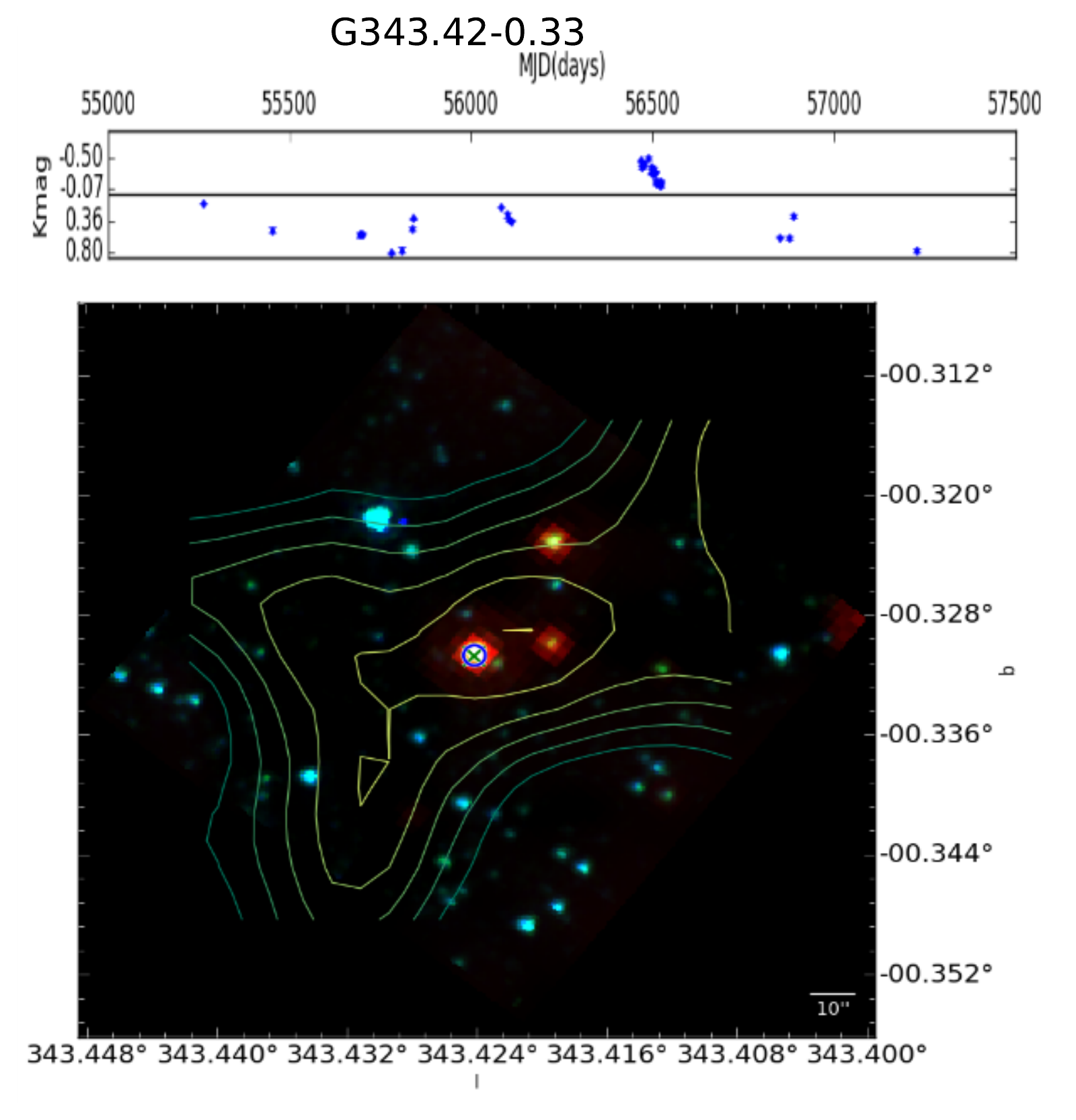}&
                \includegraphics[width=0.32\textwidth]{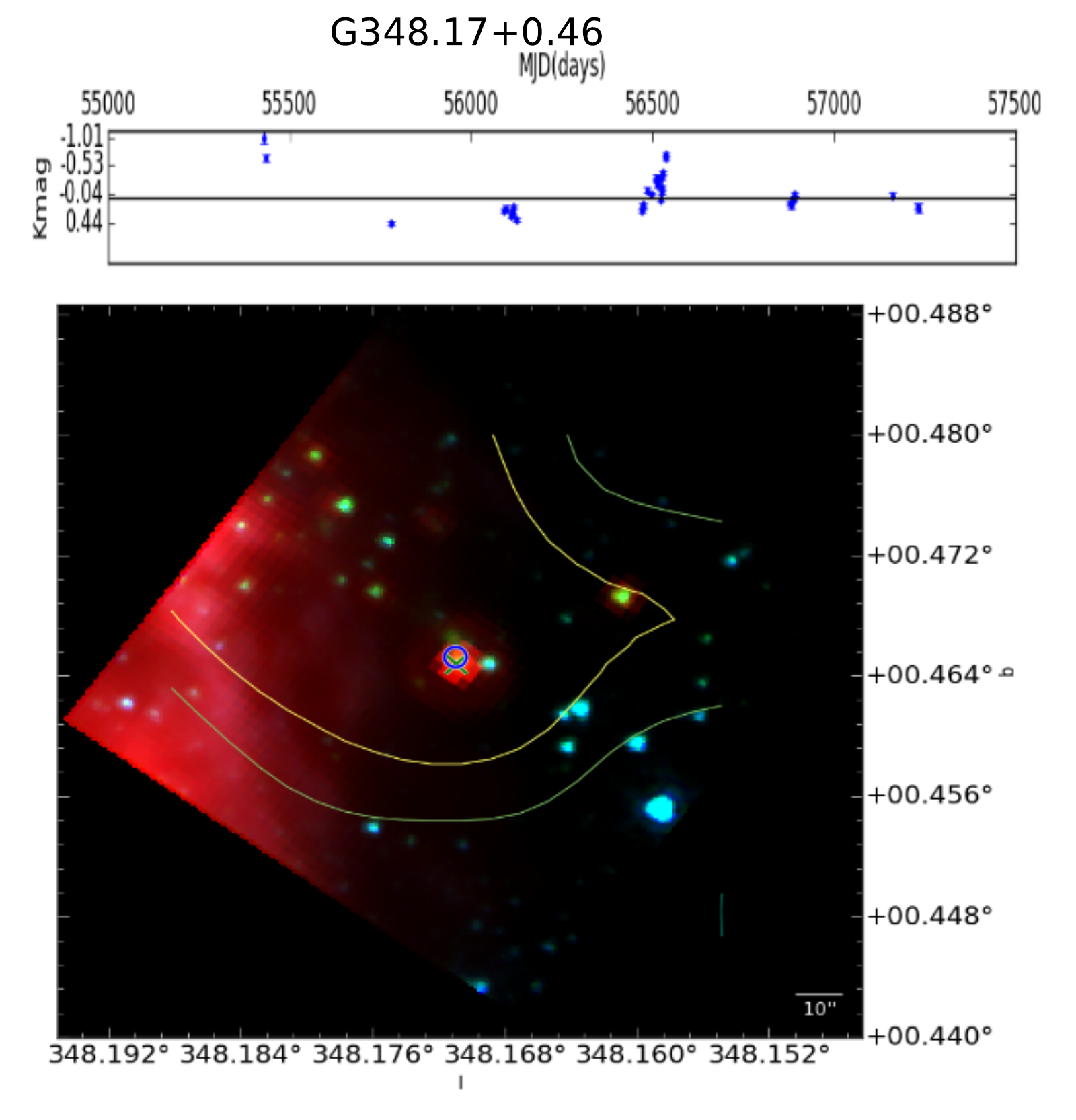}&
                \includegraphics[width=0.32\textwidth]{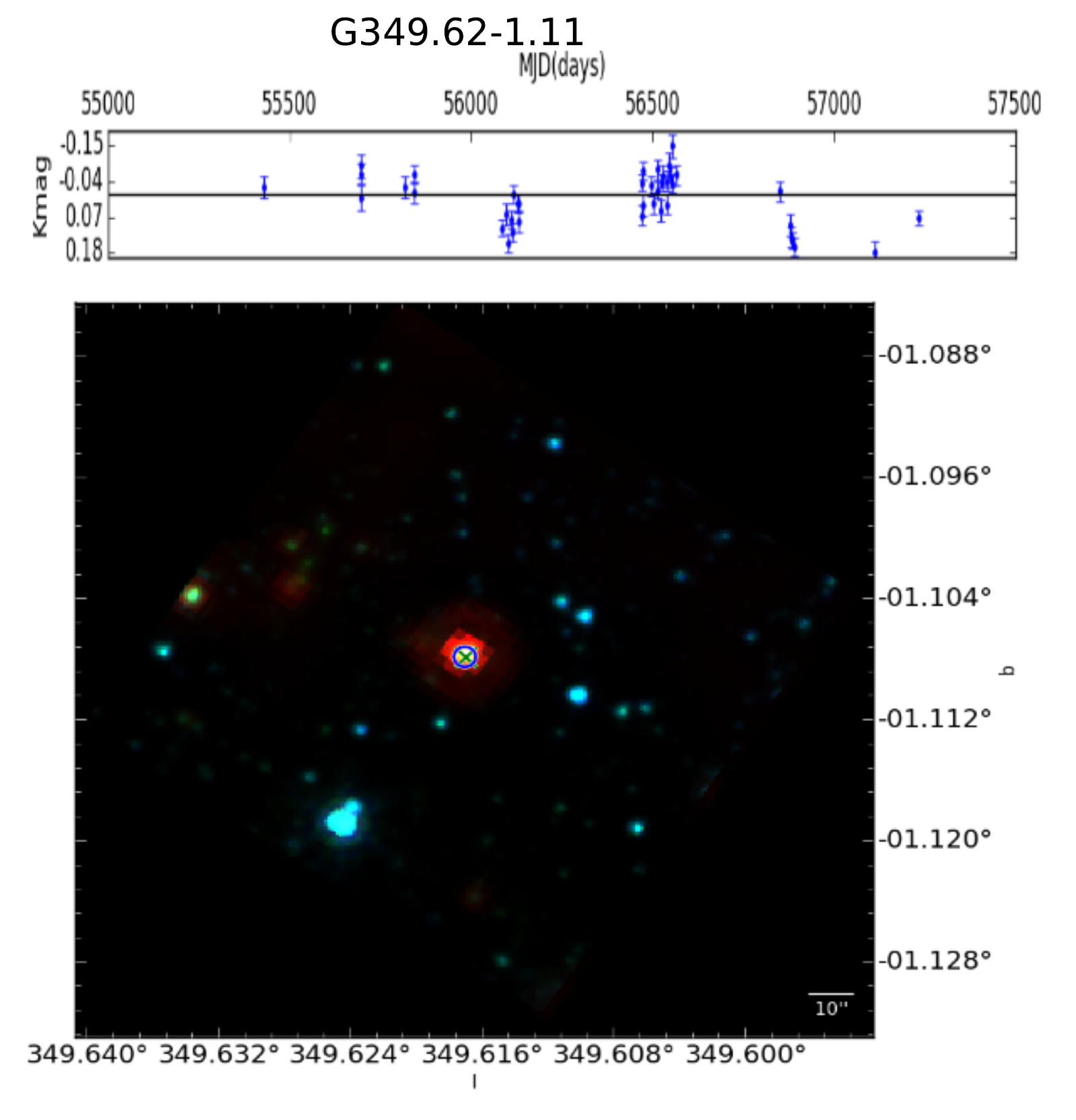}\\
                \includegraphics[width=0.32\textwidth]{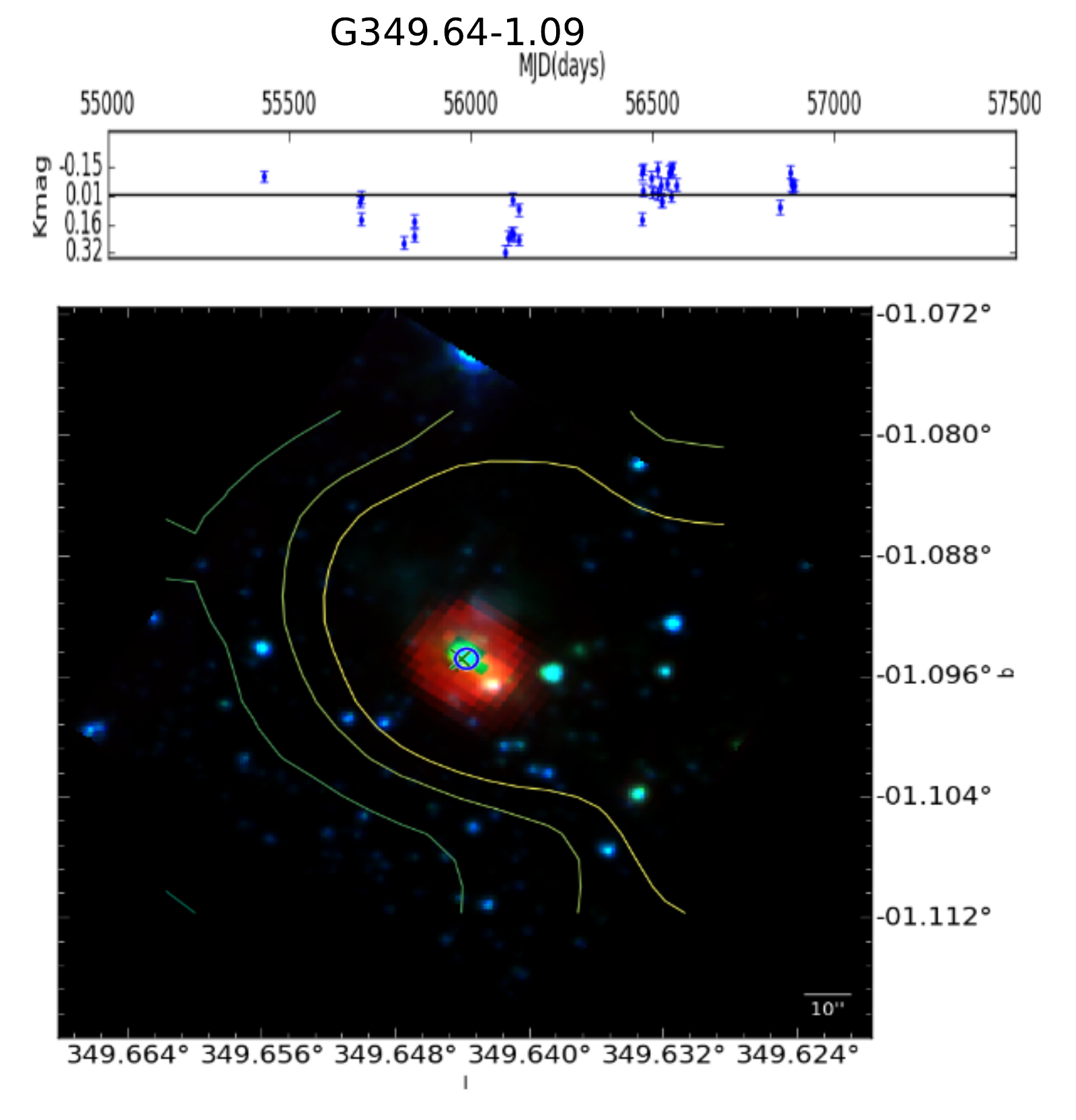}&
                \includegraphics[width=0.32\textwidth]{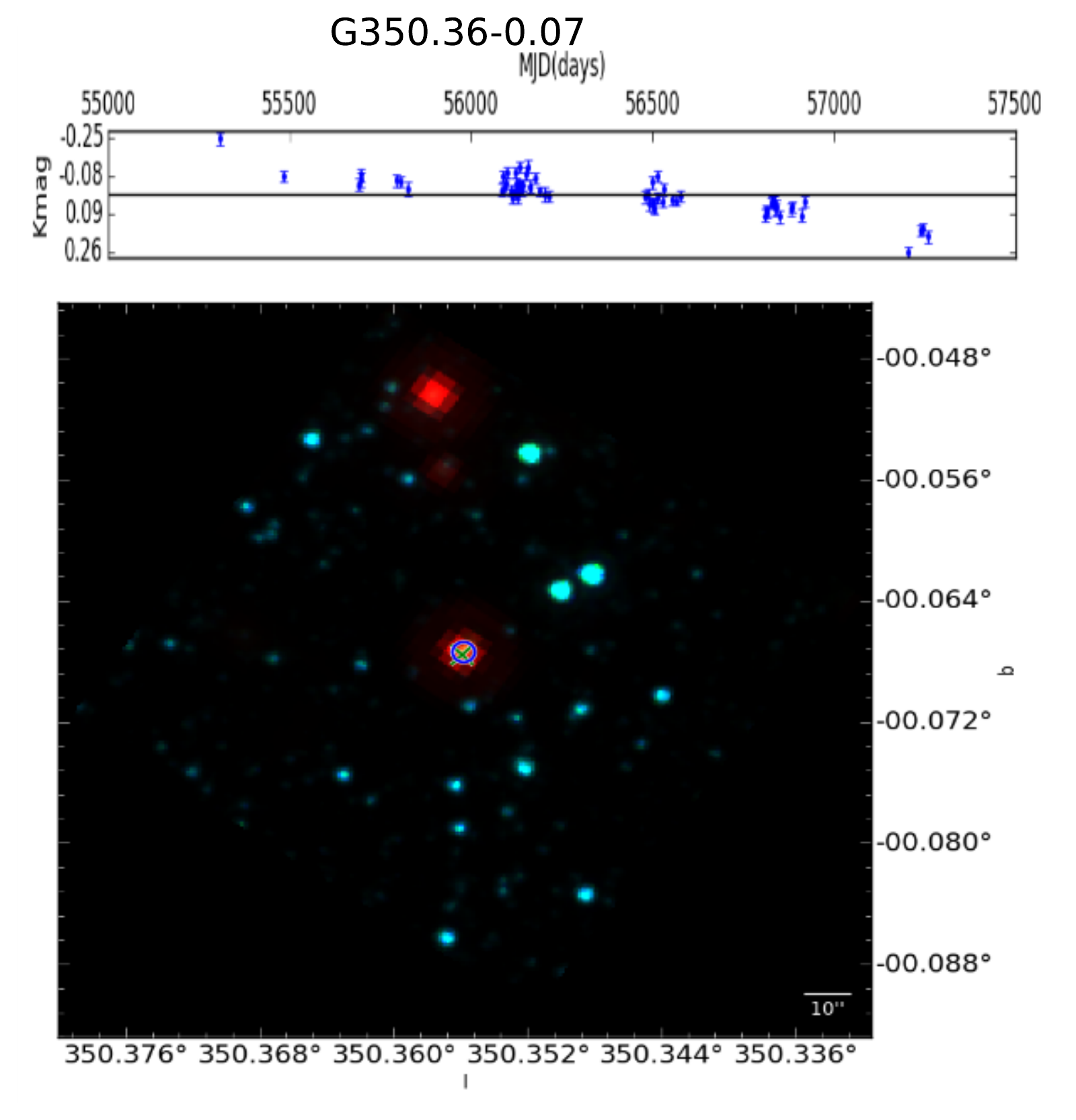}&
                \includegraphics[width=0.32\textwidth]{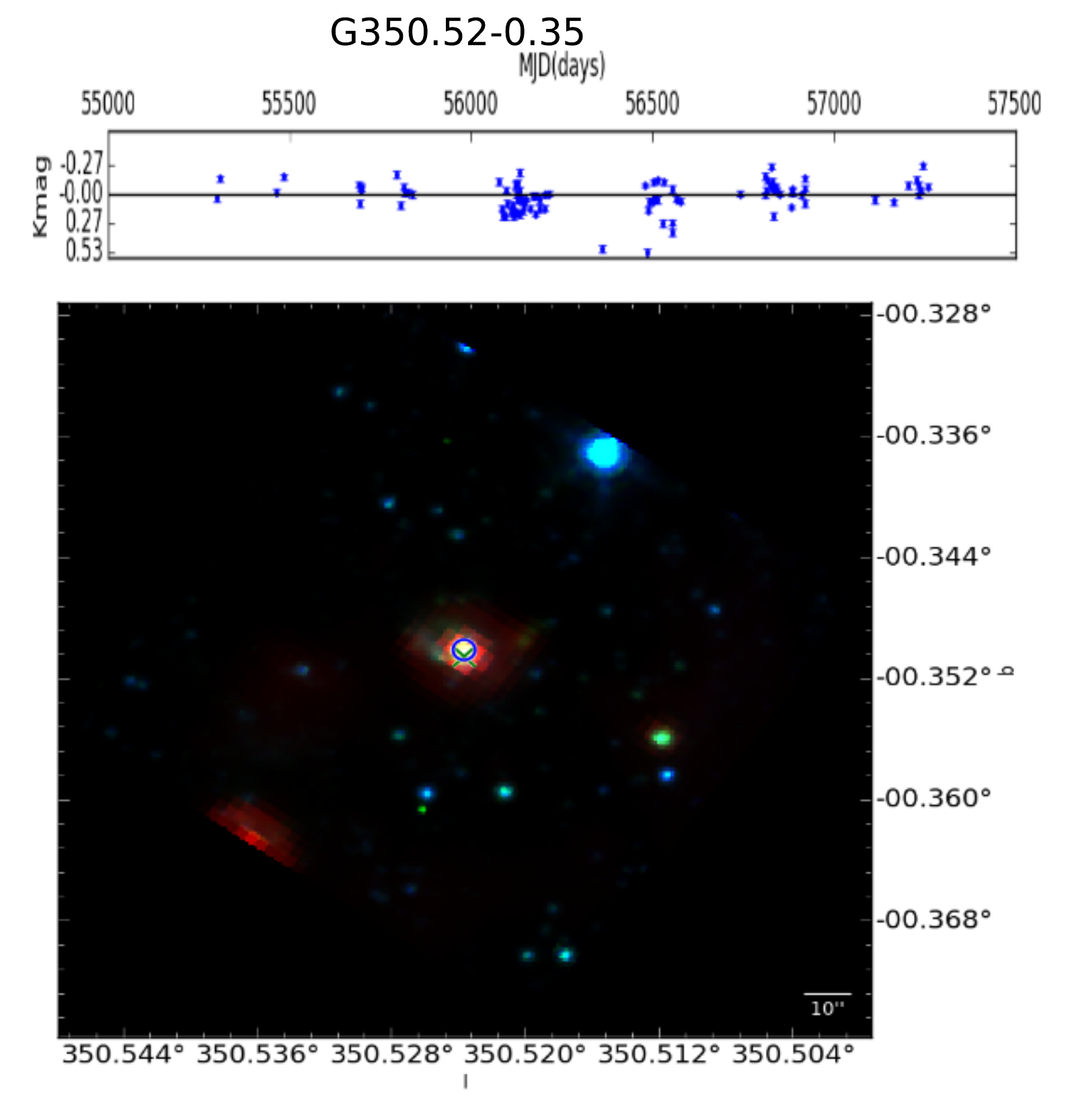}\\
                \includegraphics[width=0.32\textwidth]{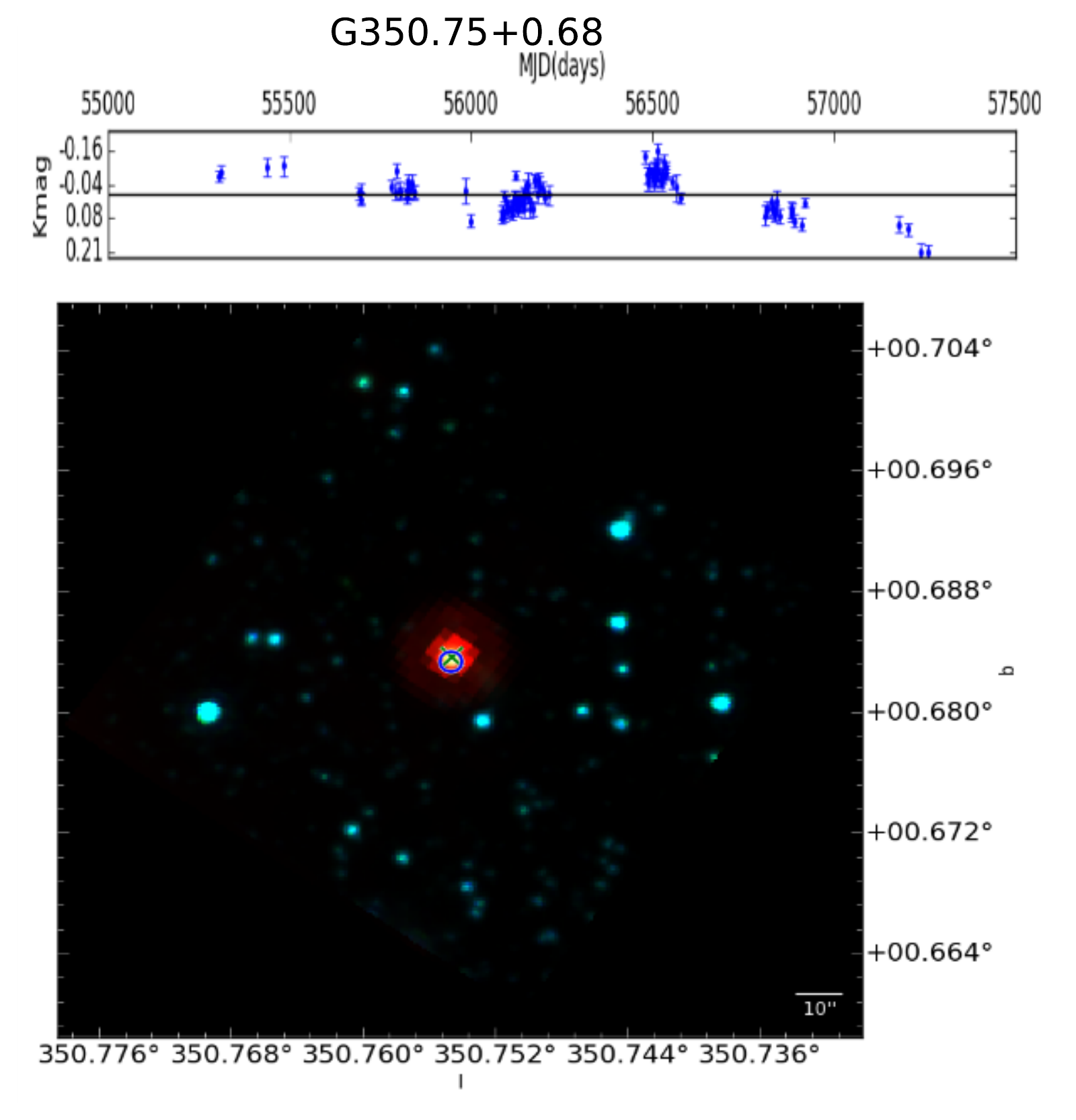}&
                \includegraphics[width=0.32\textwidth]{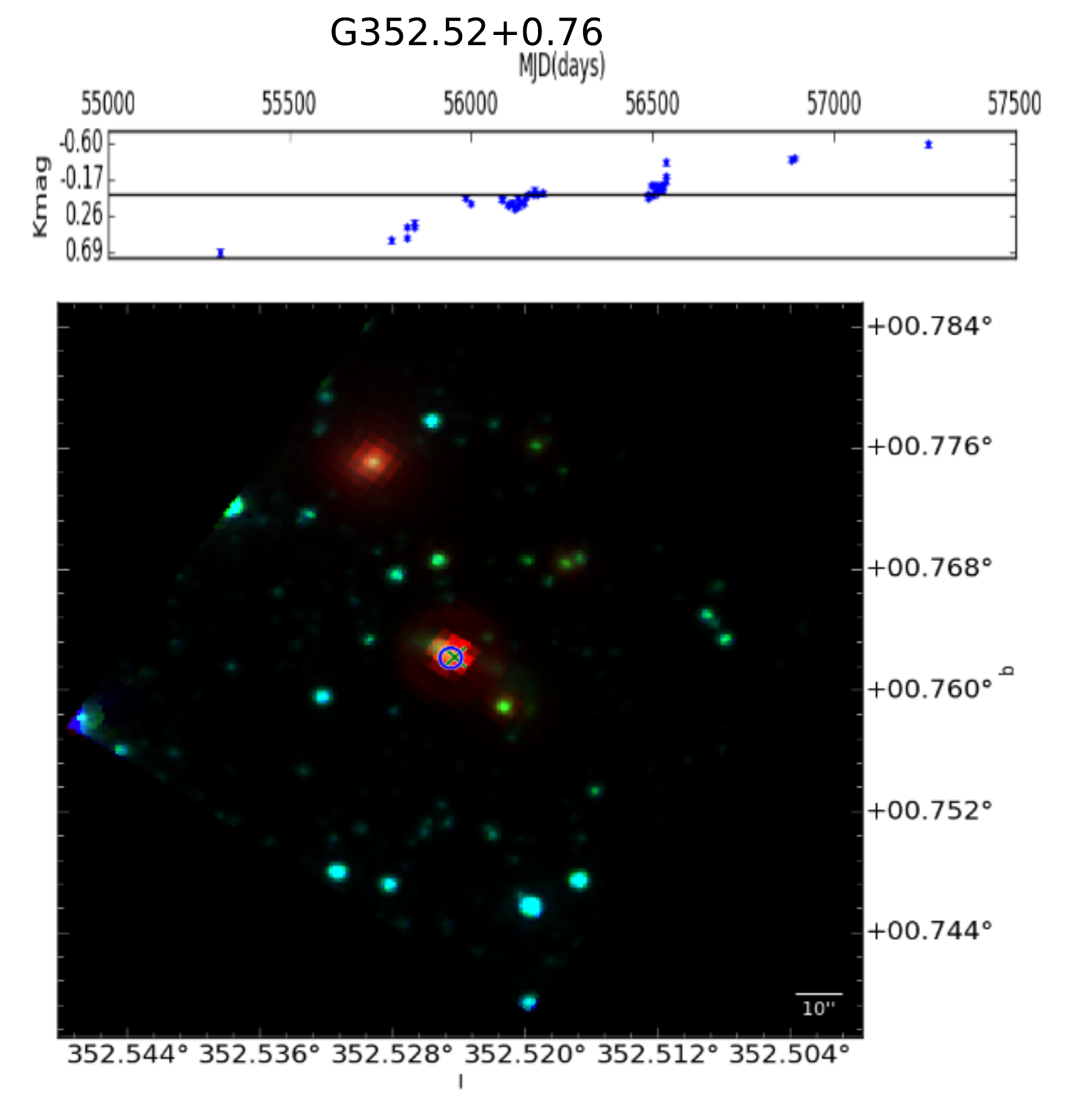}&
                \includegraphics[width=0.32\textwidth]{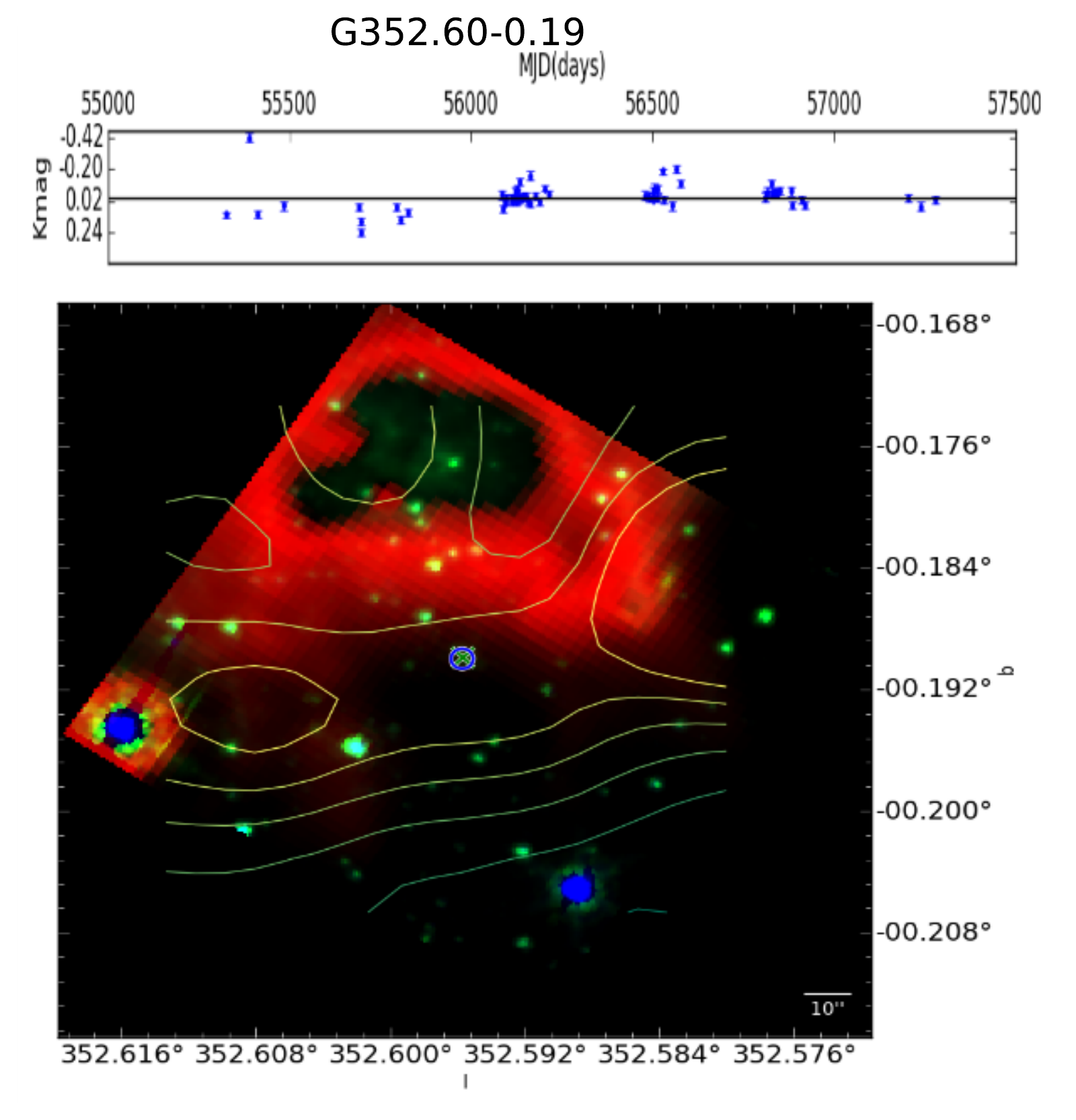}\\
        \end{tabular}
        \caption{Continuation of Fig. \ref{fig:testgrid}.}
\end{figure}

\begin{figure}
        \begin{tabular}{ccc}

                \includegraphics[width=0.32\textwidth]{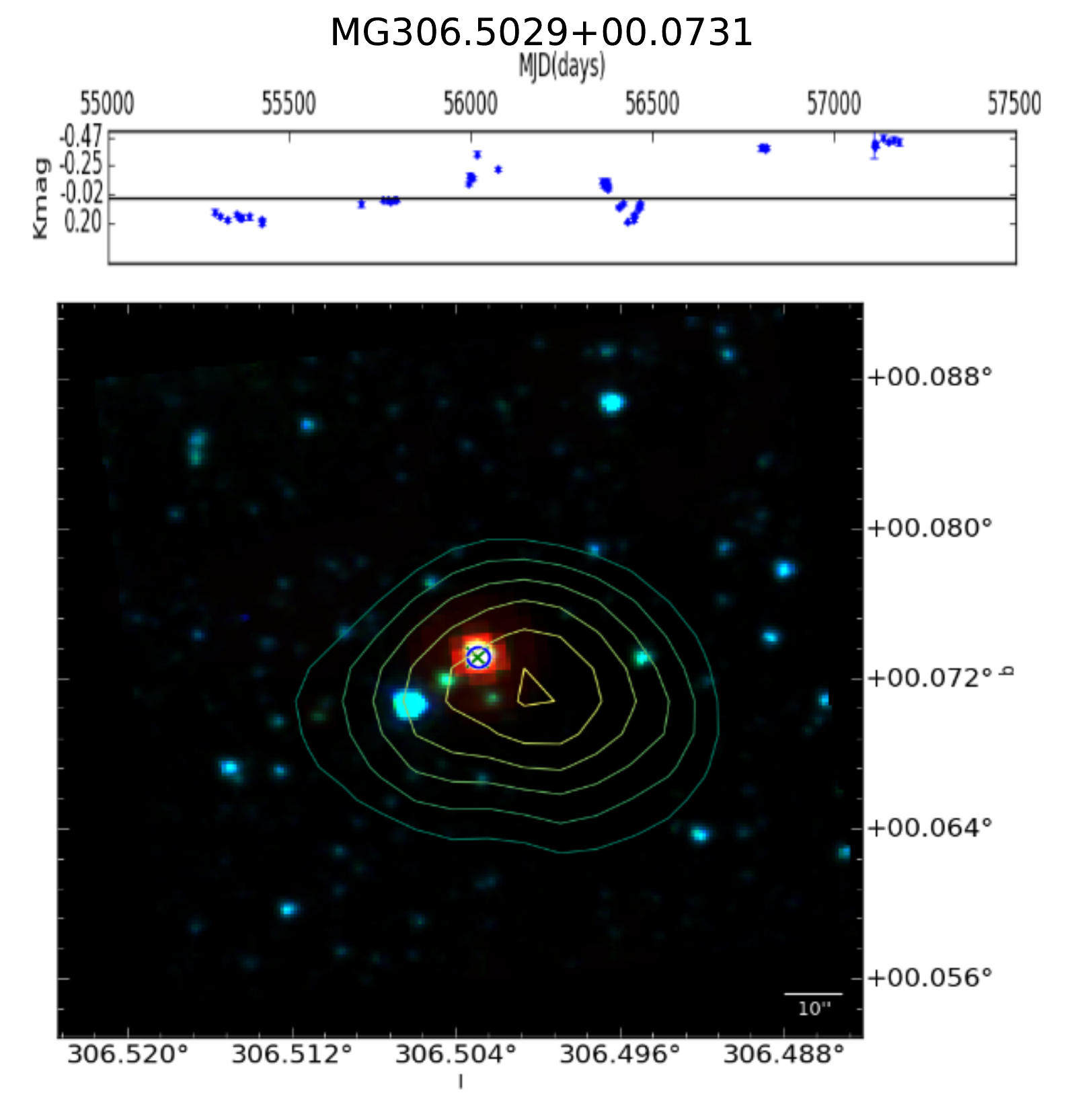}&
                \includegraphics[width=0.32\textwidth]{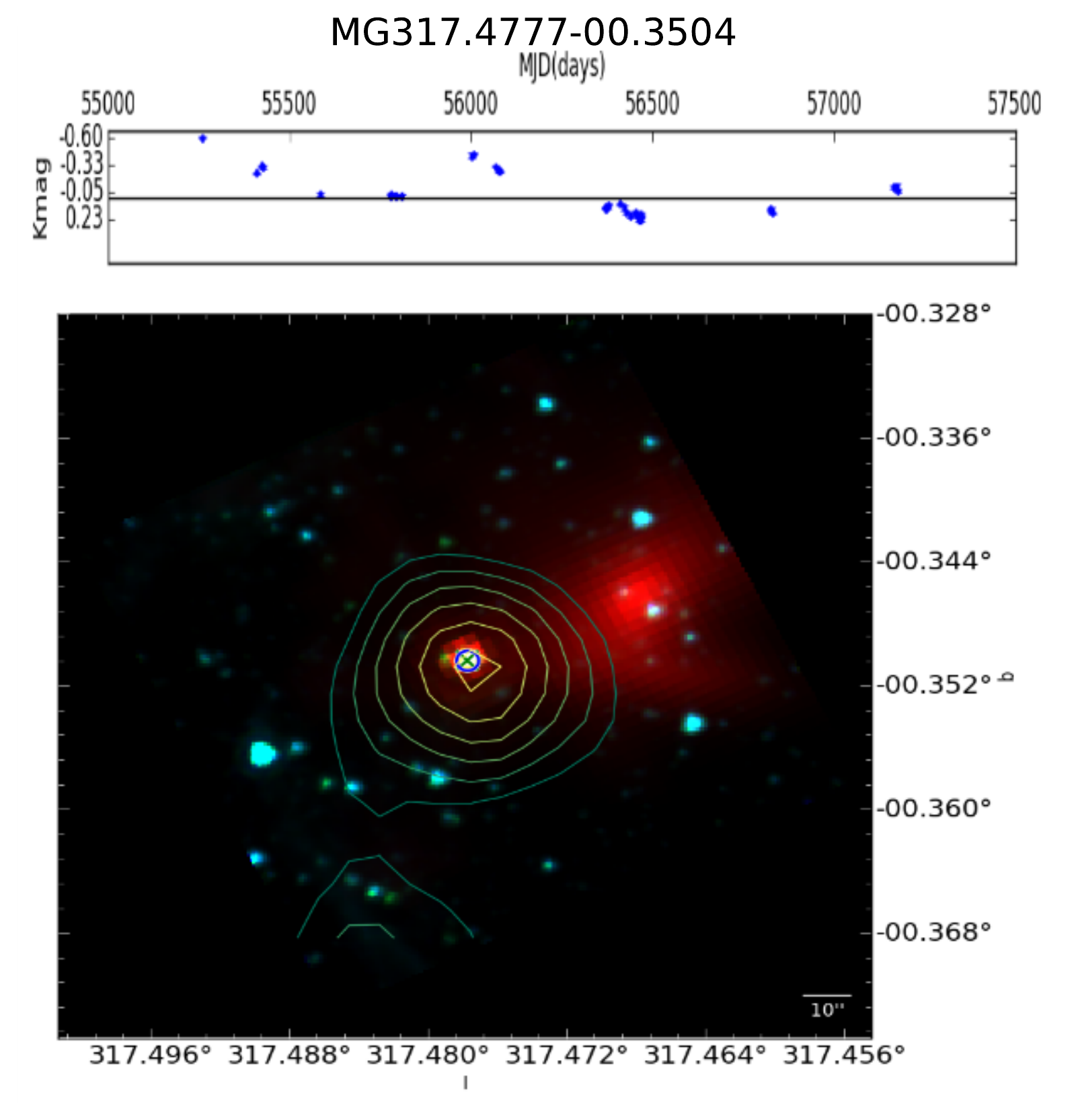}&
                \includegraphics[width=0.32\textwidth]{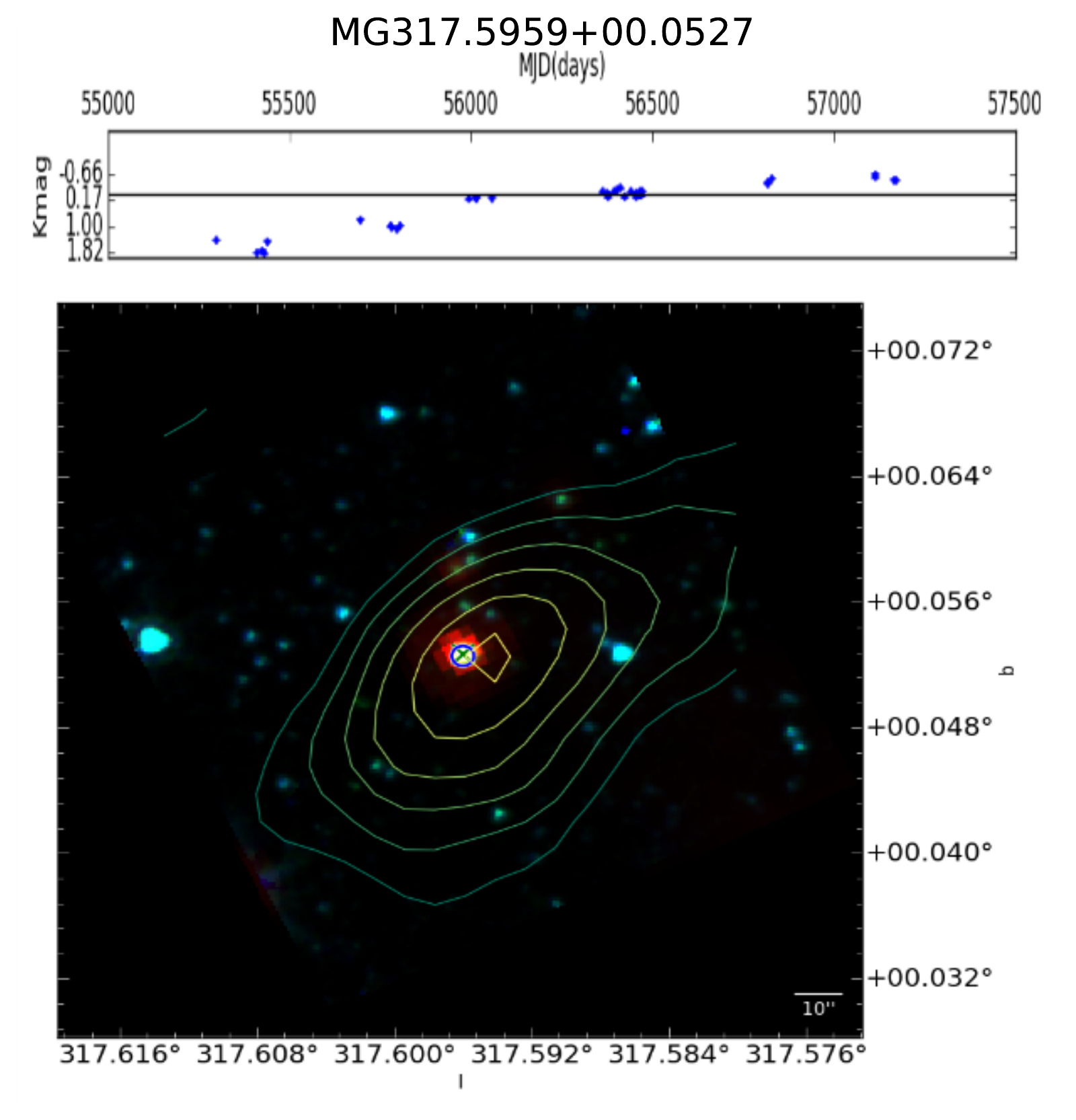}\\
                \includegraphics[width=0.32\textwidth]{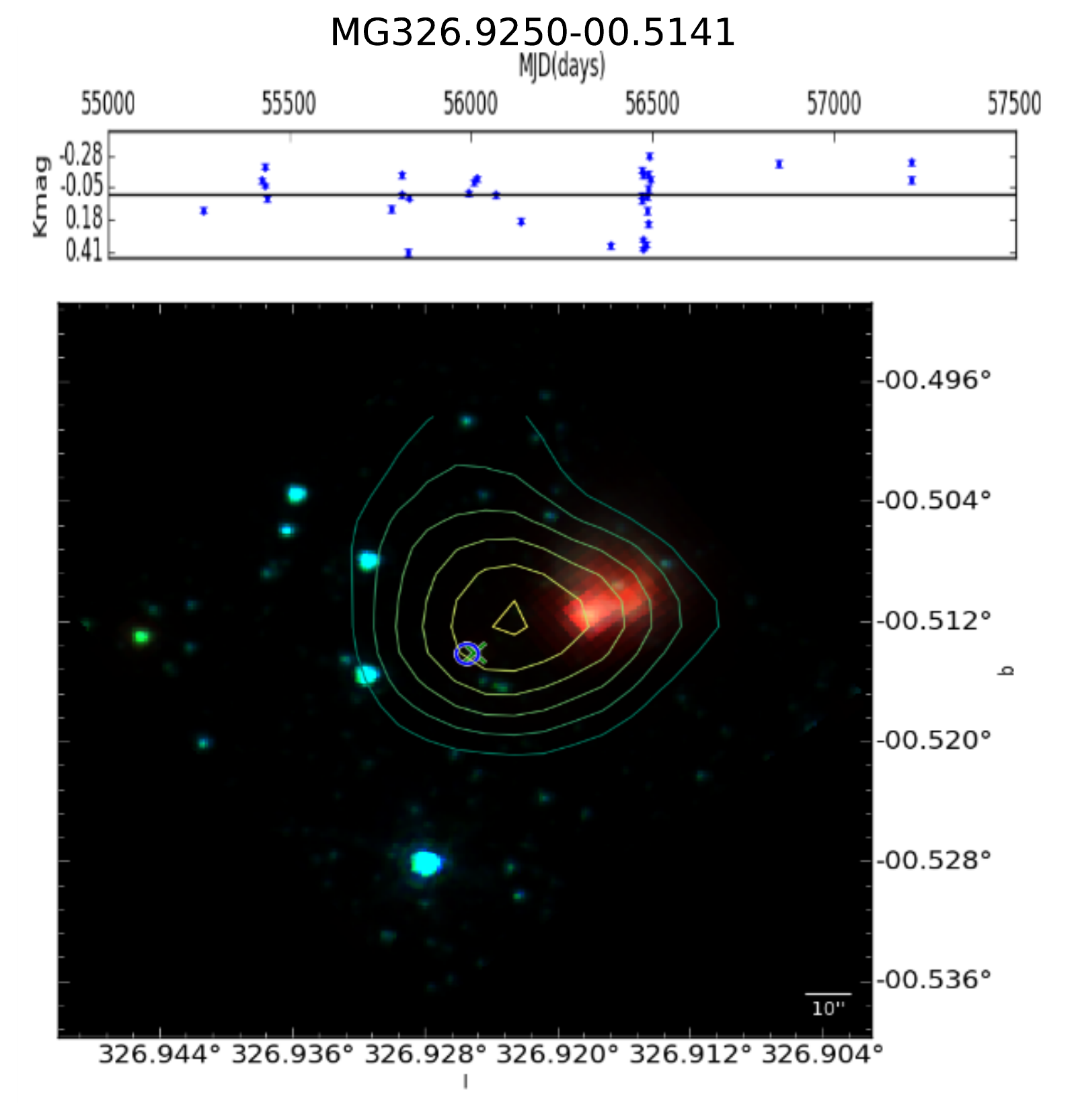}&
                \includegraphics[width=0.32\textwidth]{{./MG328.0494-00.0487_combined}.pdf}&
                \includegraphics[width=0.32\textwidth]{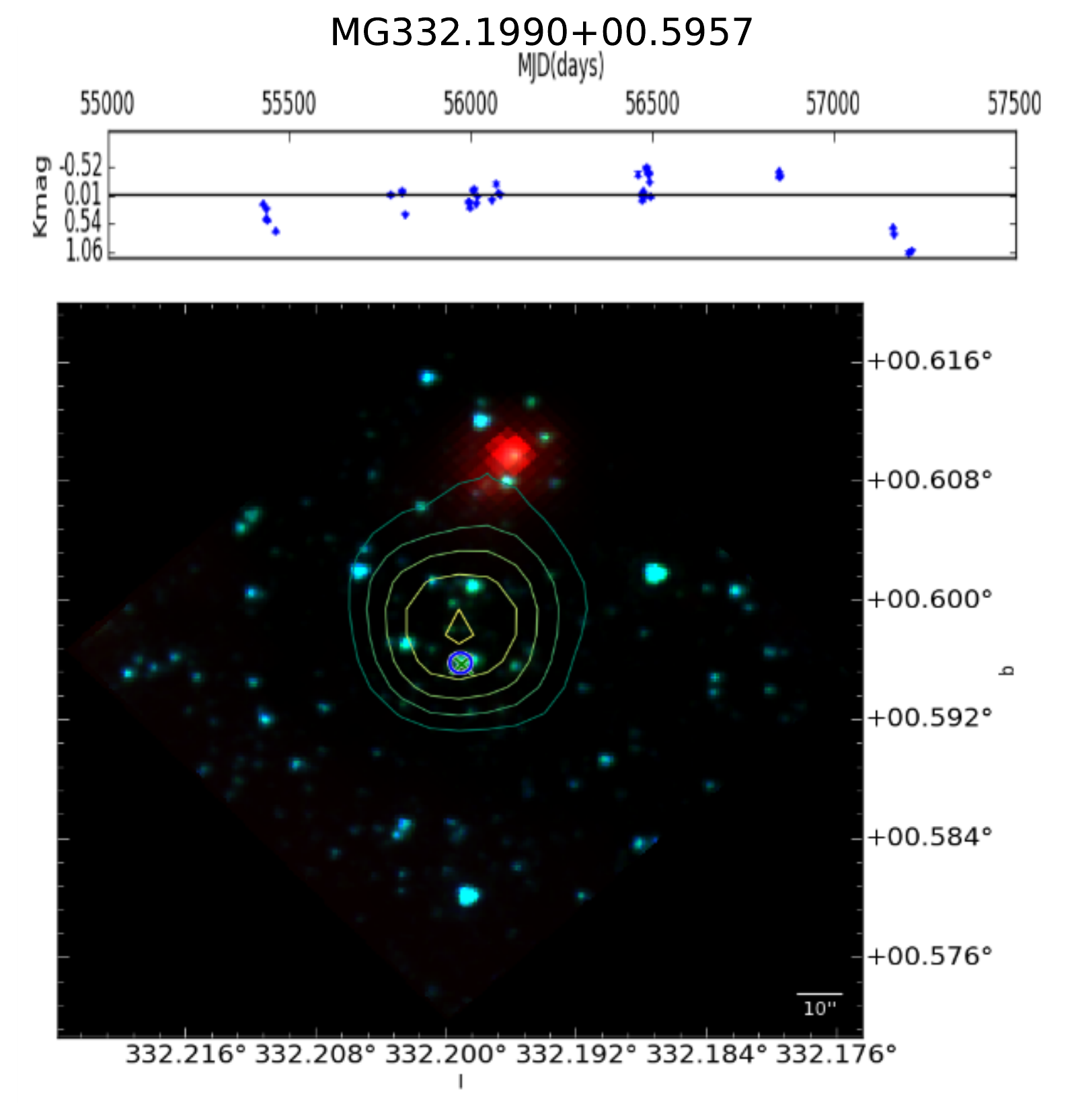}\\
                \includegraphics[width=0.32\textwidth]{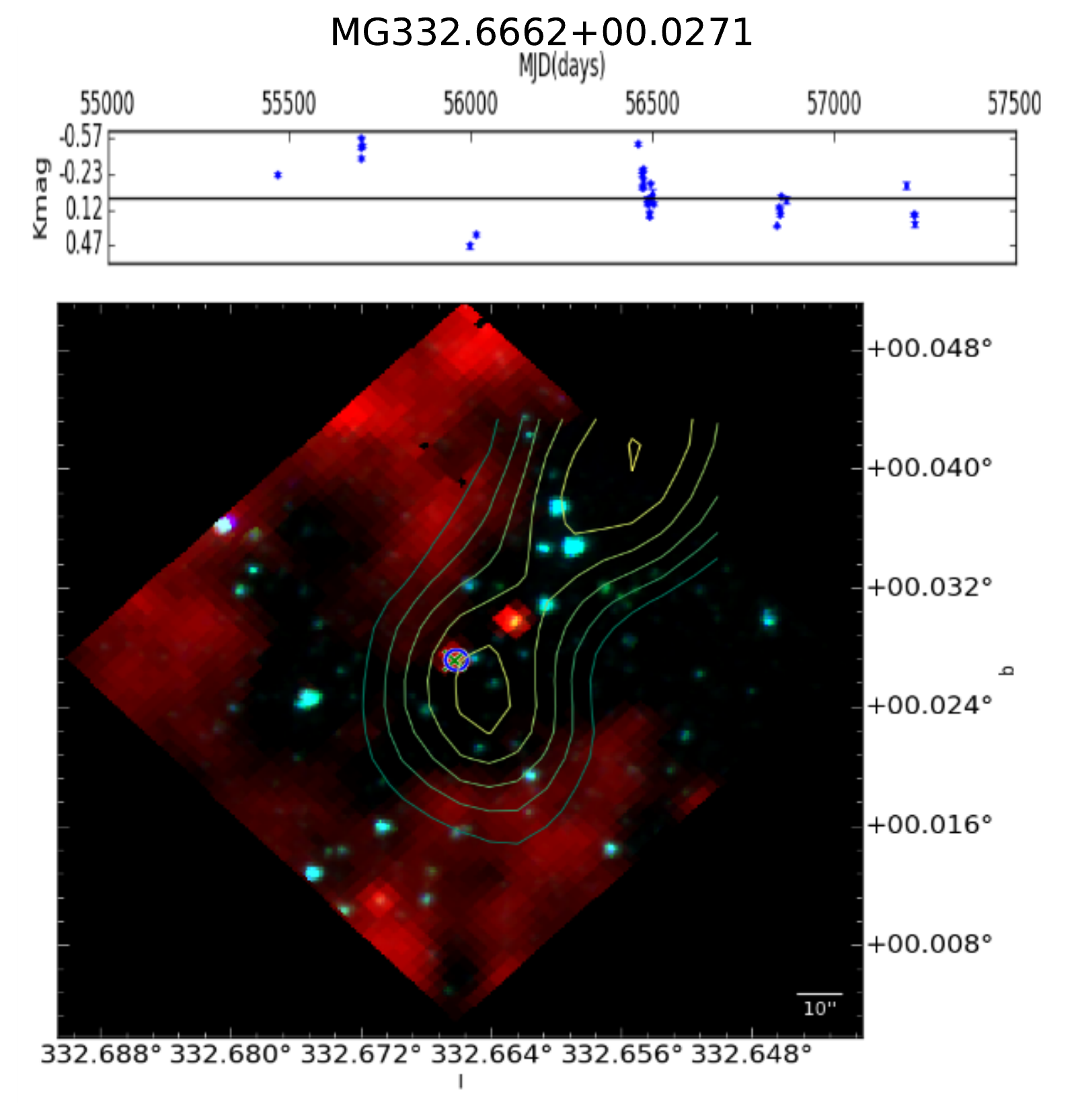}&
                \includegraphics[width=0.32\textwidth]{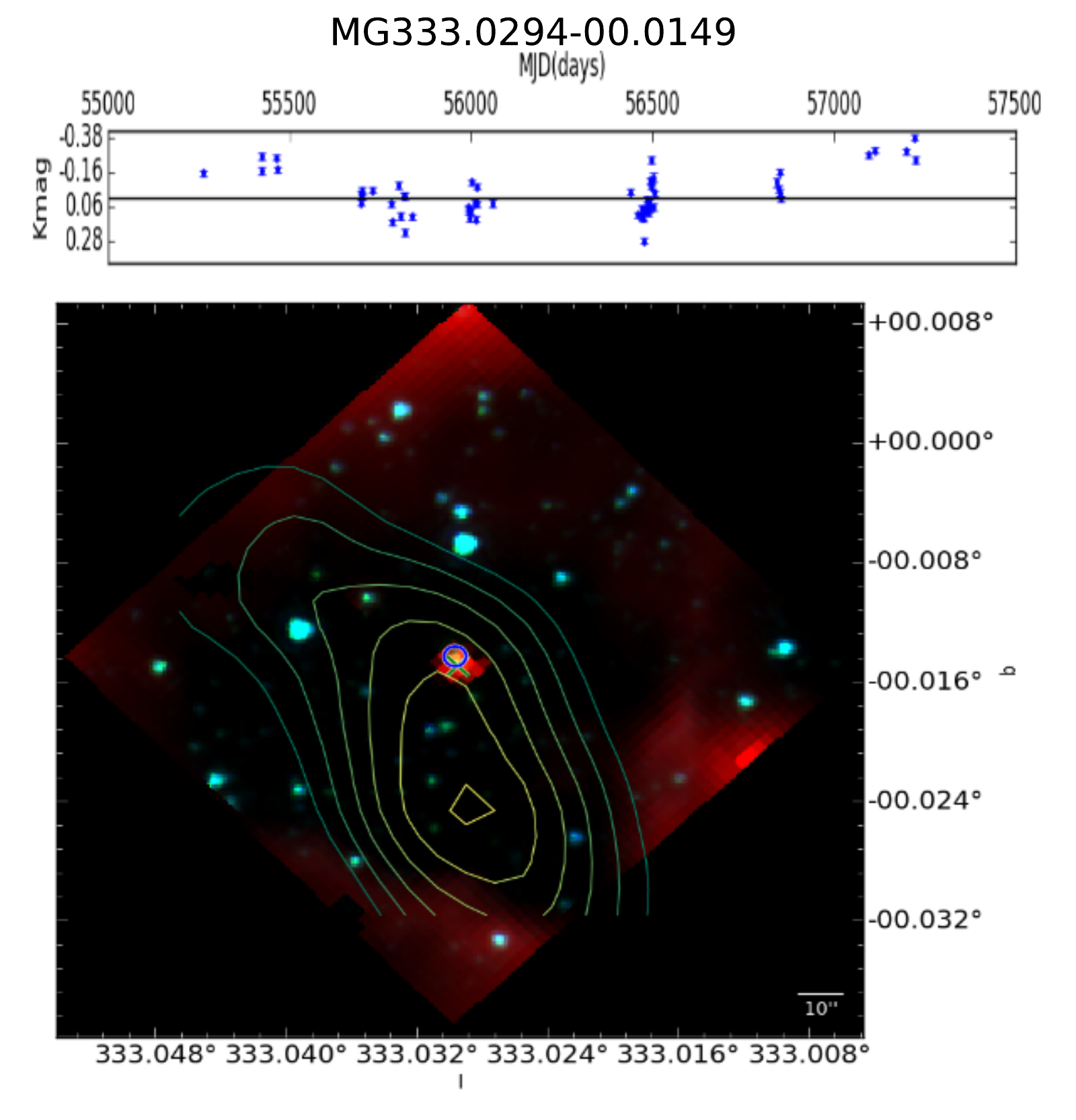}&
                \includegraphics[width=0.32\textwidth]{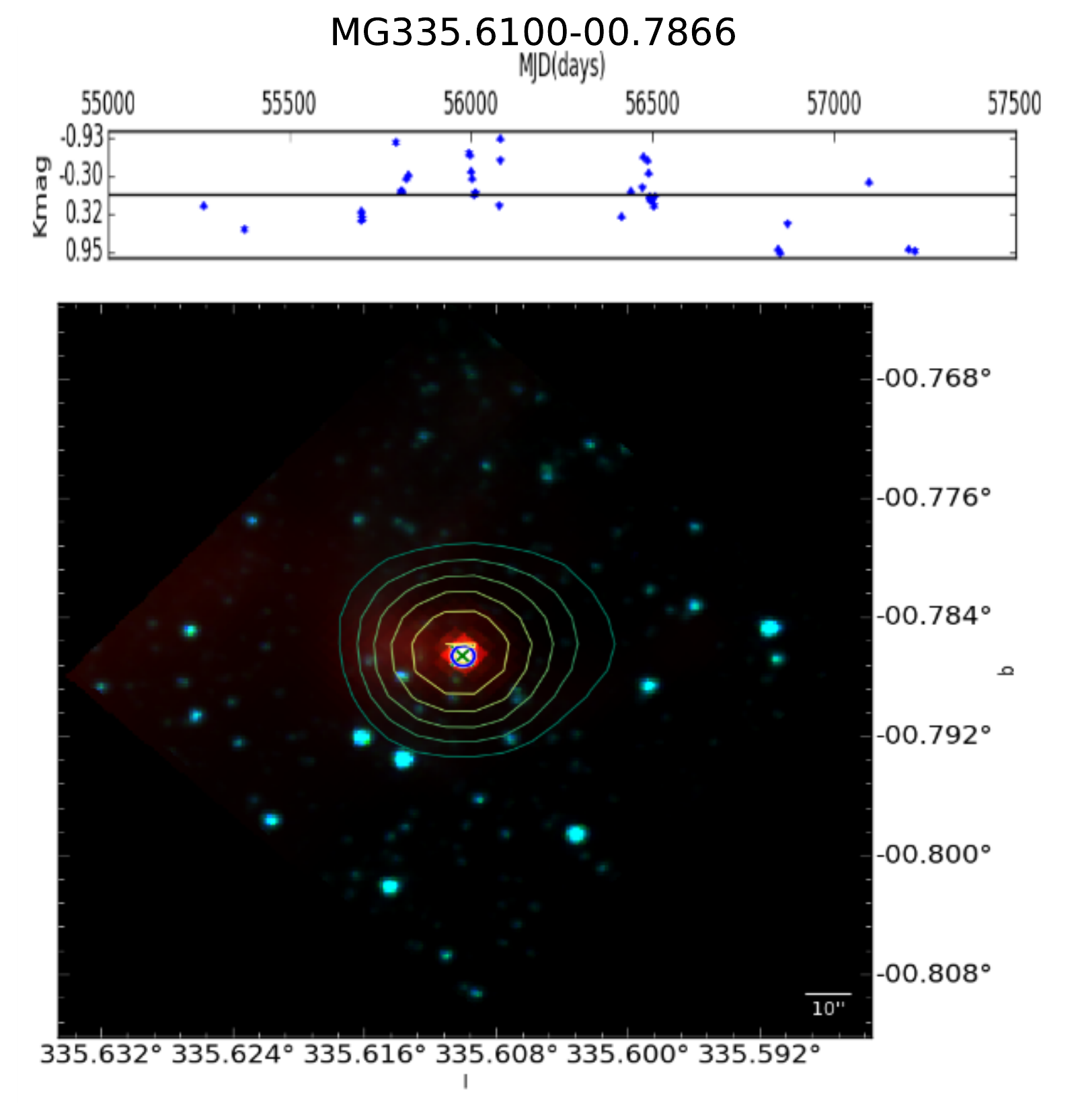}\\
                \includegraphics[width=0.32\textwidth]{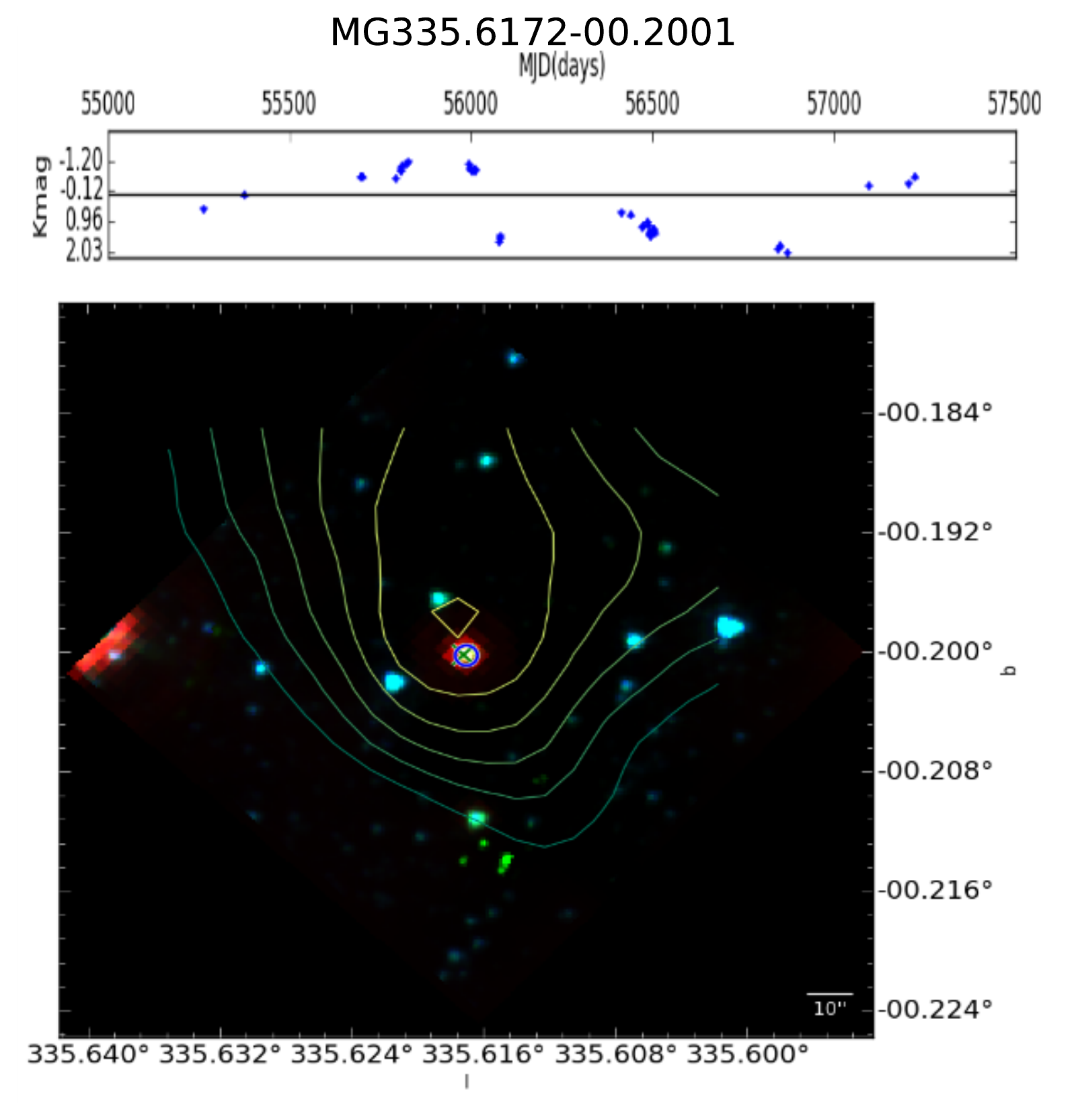}&
                \includegraphics[width=0.32\textwidth]{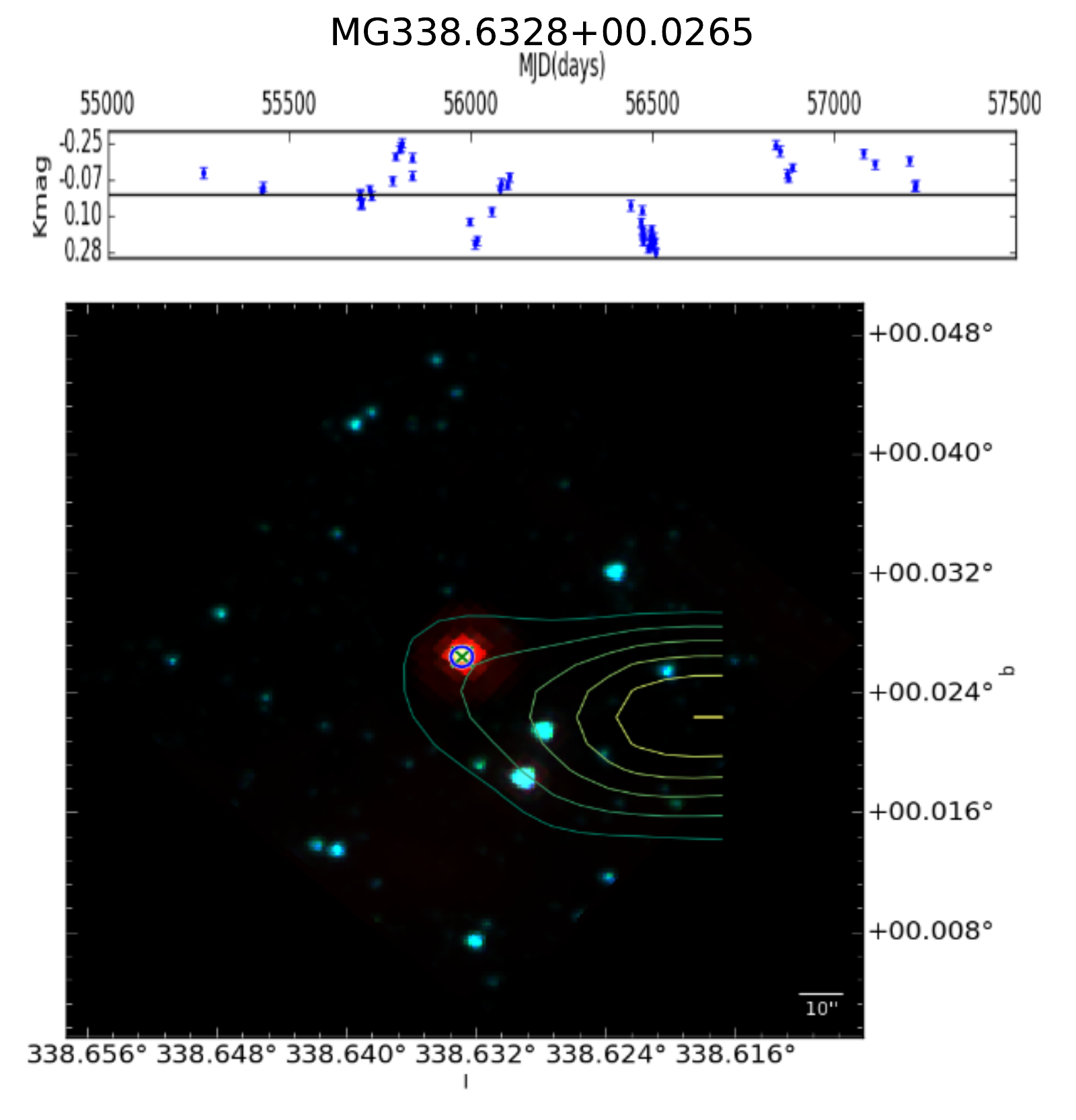}&
                \includegraphics[width=0.32\textwidth]{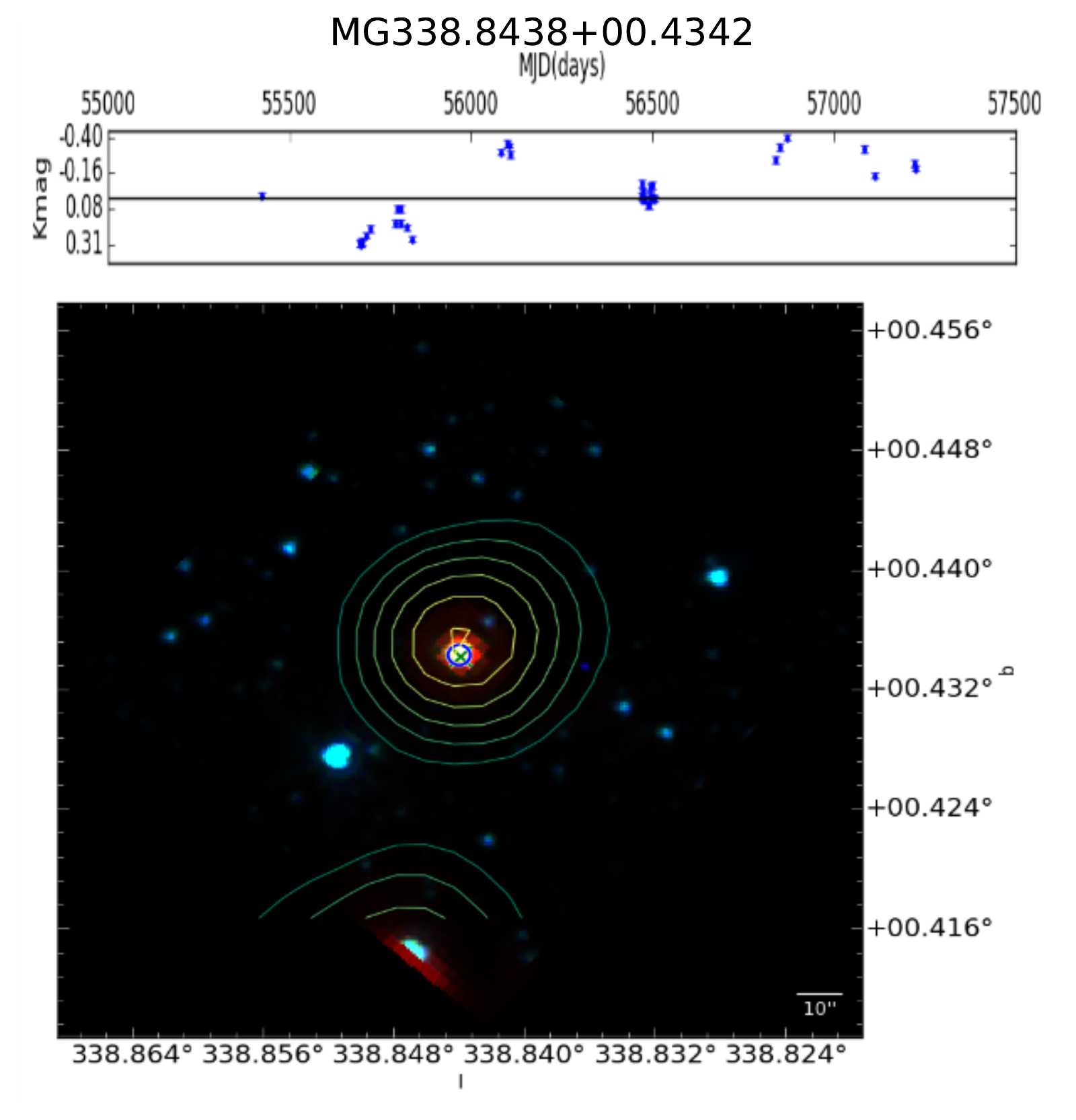}\\
        \end{tabular}
        \caption{Continuation of Fig. \ref{fig:testgrid}.}
\end{figure}

\begin{figure}
        \begin{tabular}{ccc}

                \includegraphics[width=0.32\textwidth]{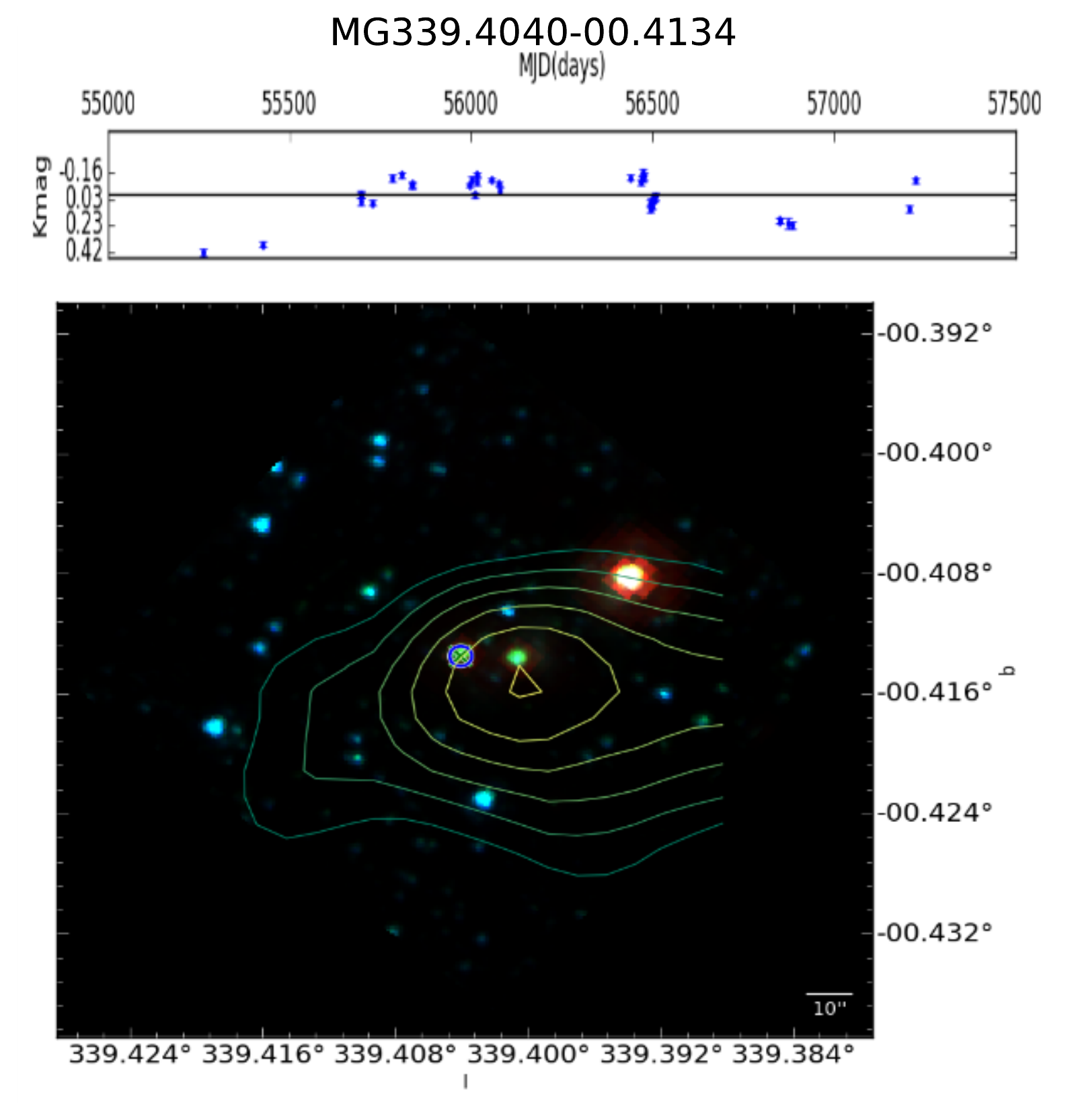}&
                \includegraphics[width=0.32\textwidth]{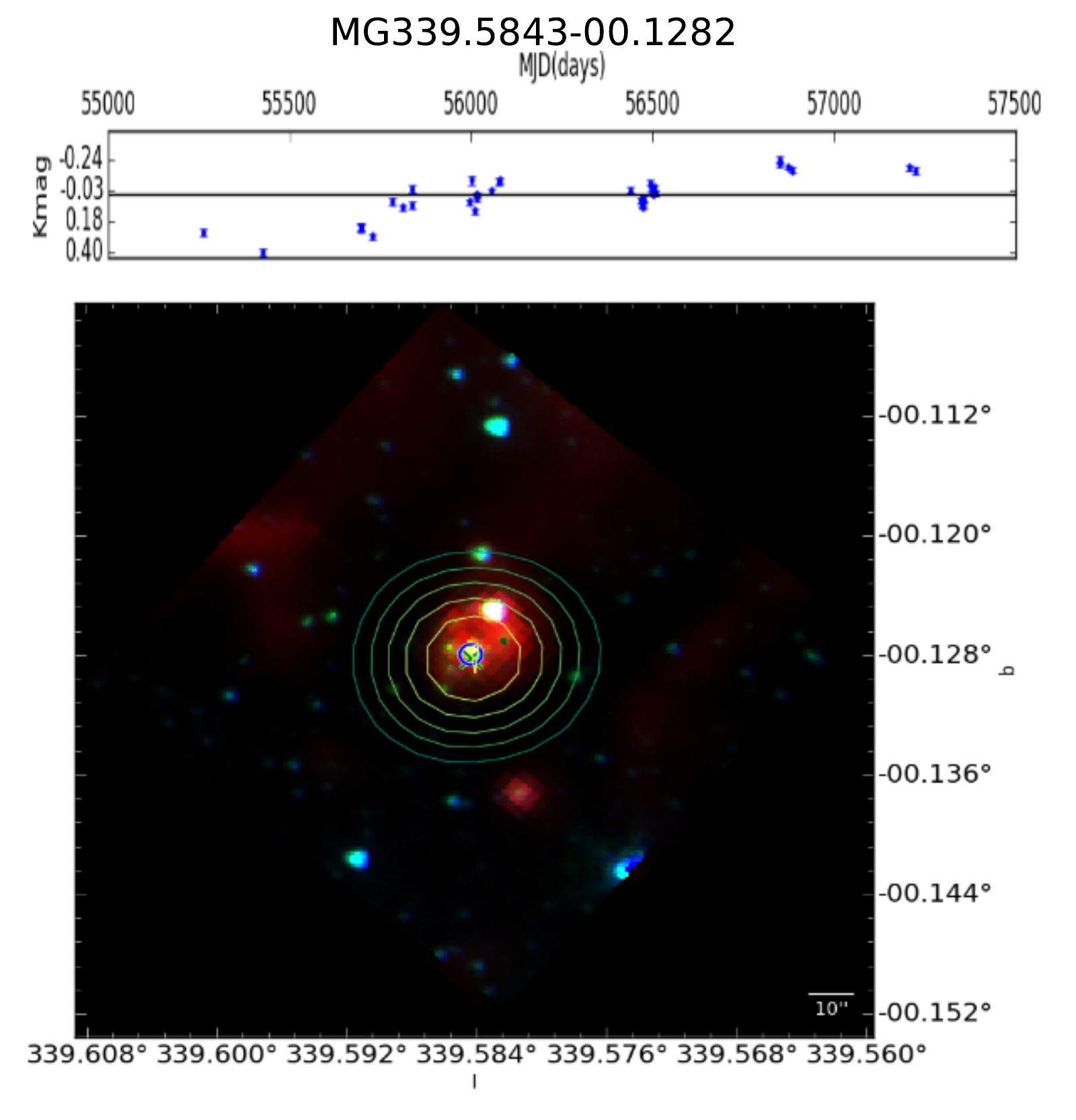}&
                \includegraphics[width=0.32\textwidth]{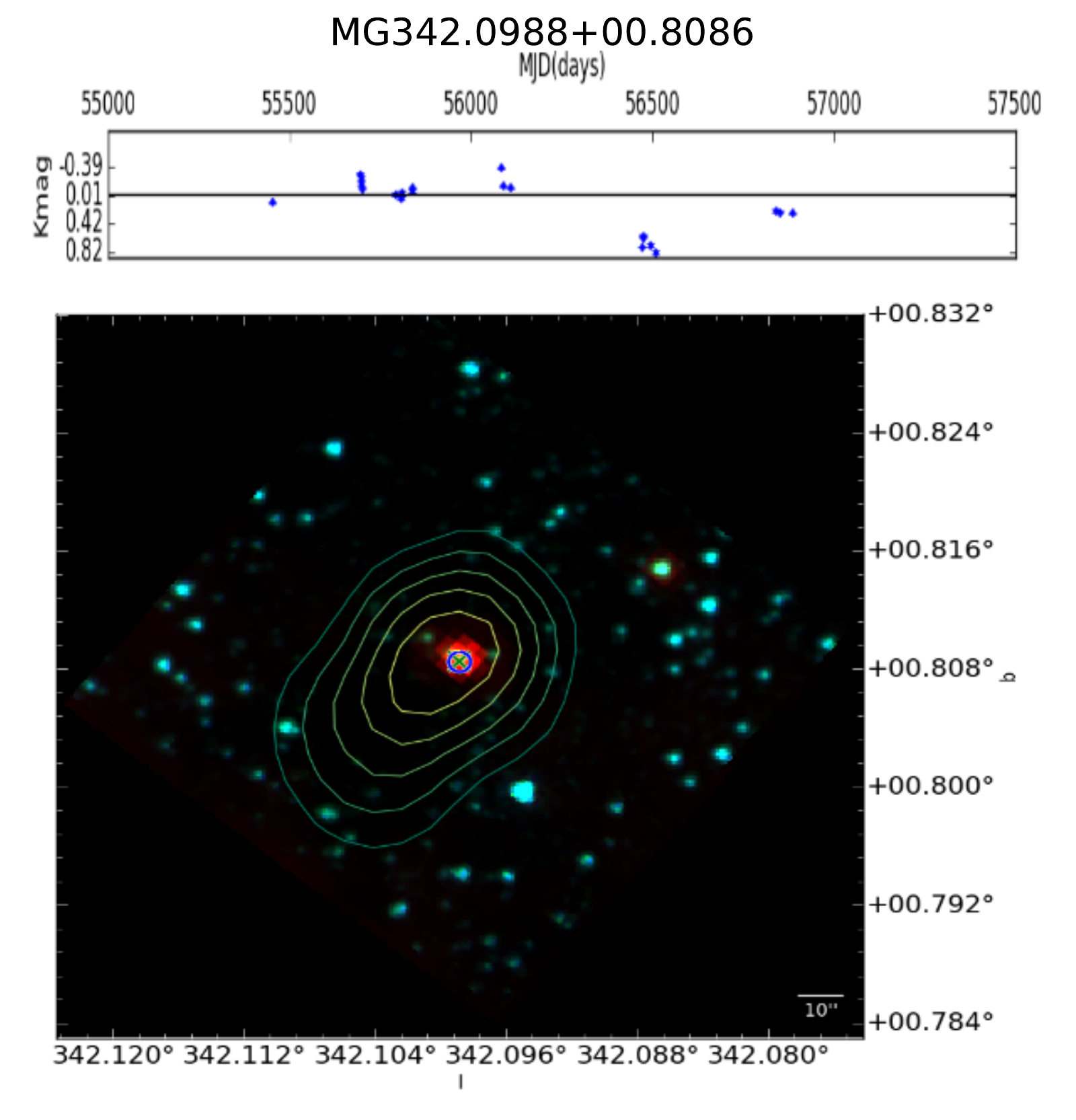}\\
                \includegraphics[width=0.32\textwidth]{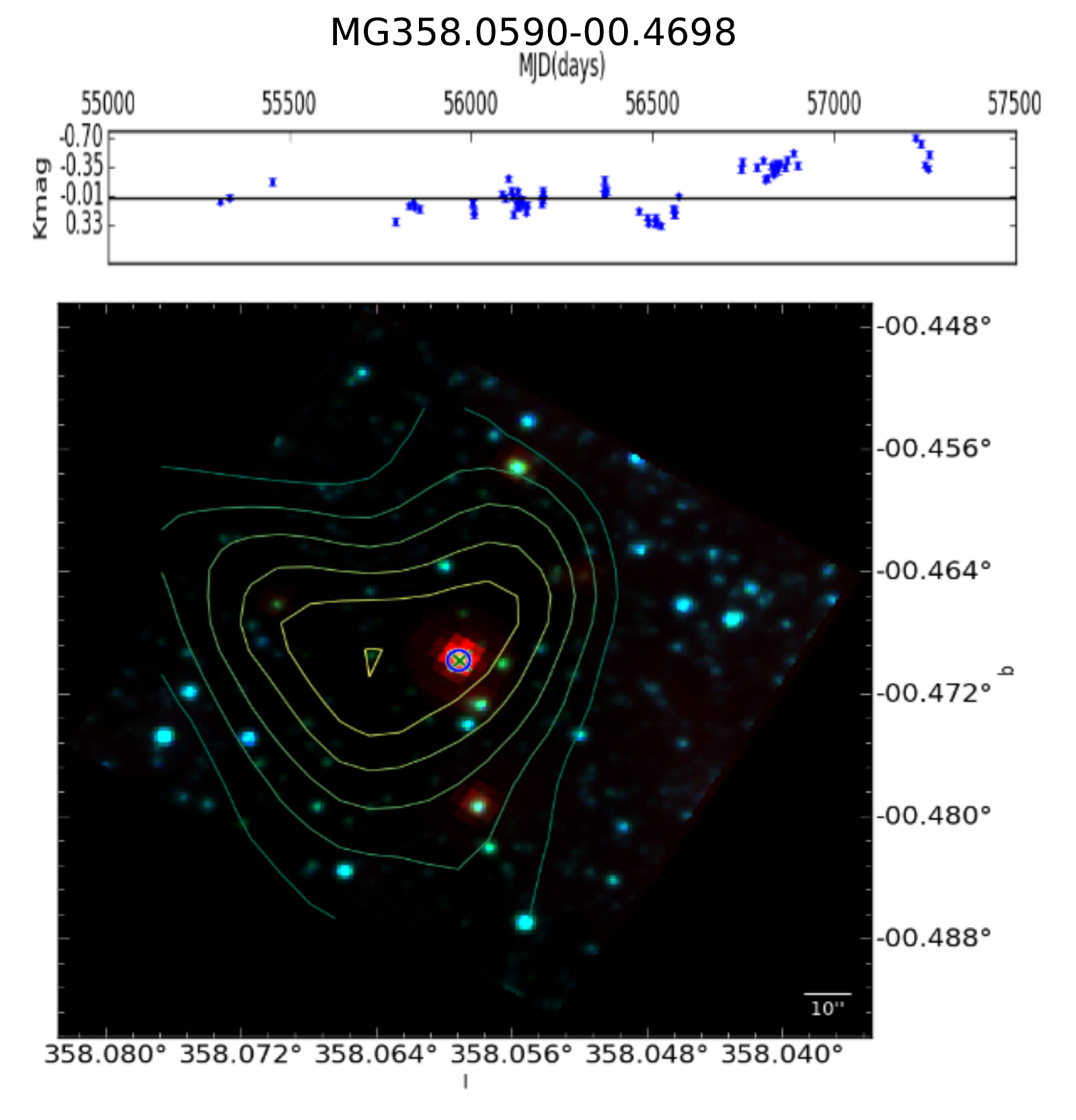}&
                \includegraphics[width=0.32\textwidth]{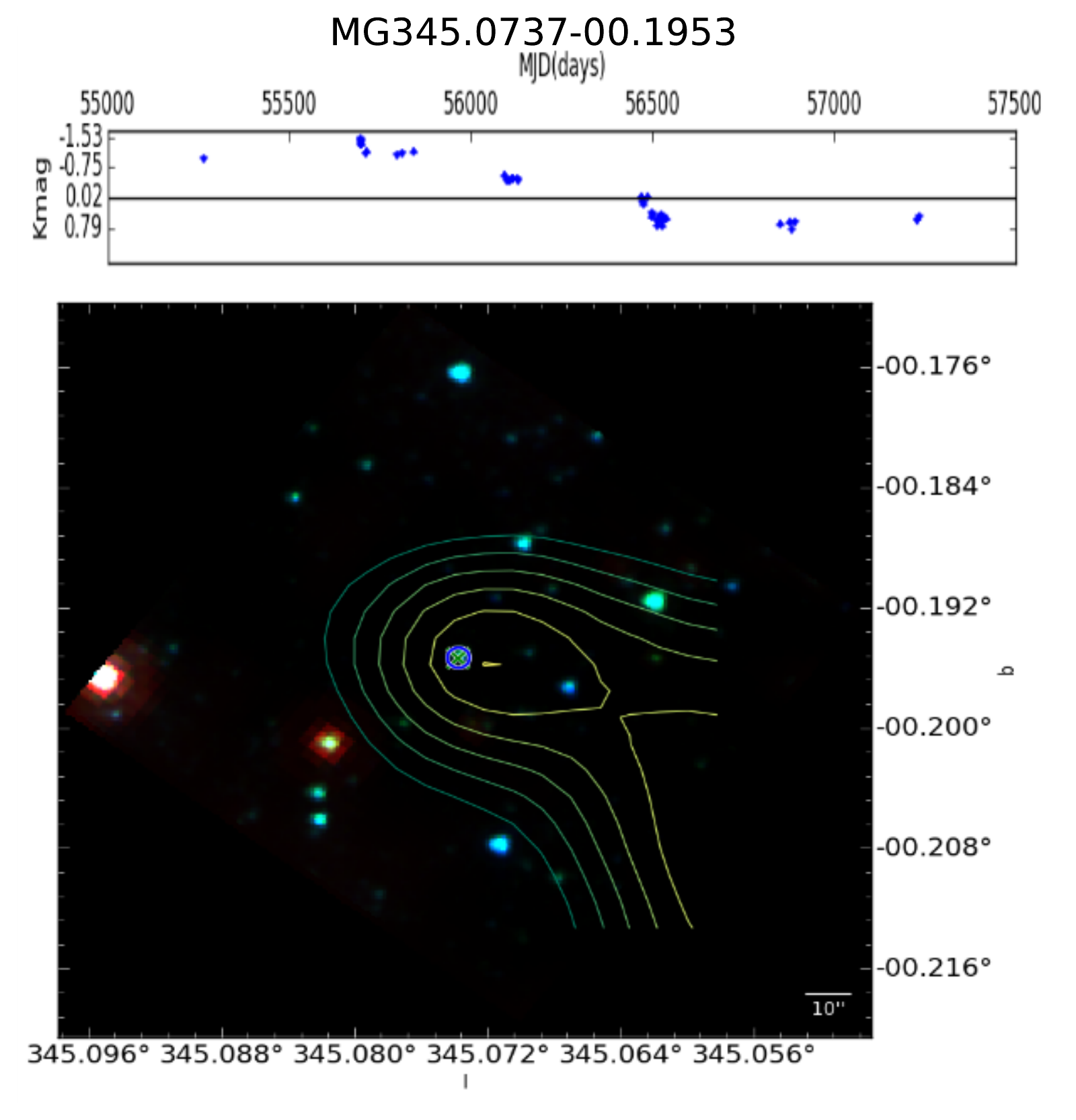}&
                \includegraphics[width=0.32\textwidth]{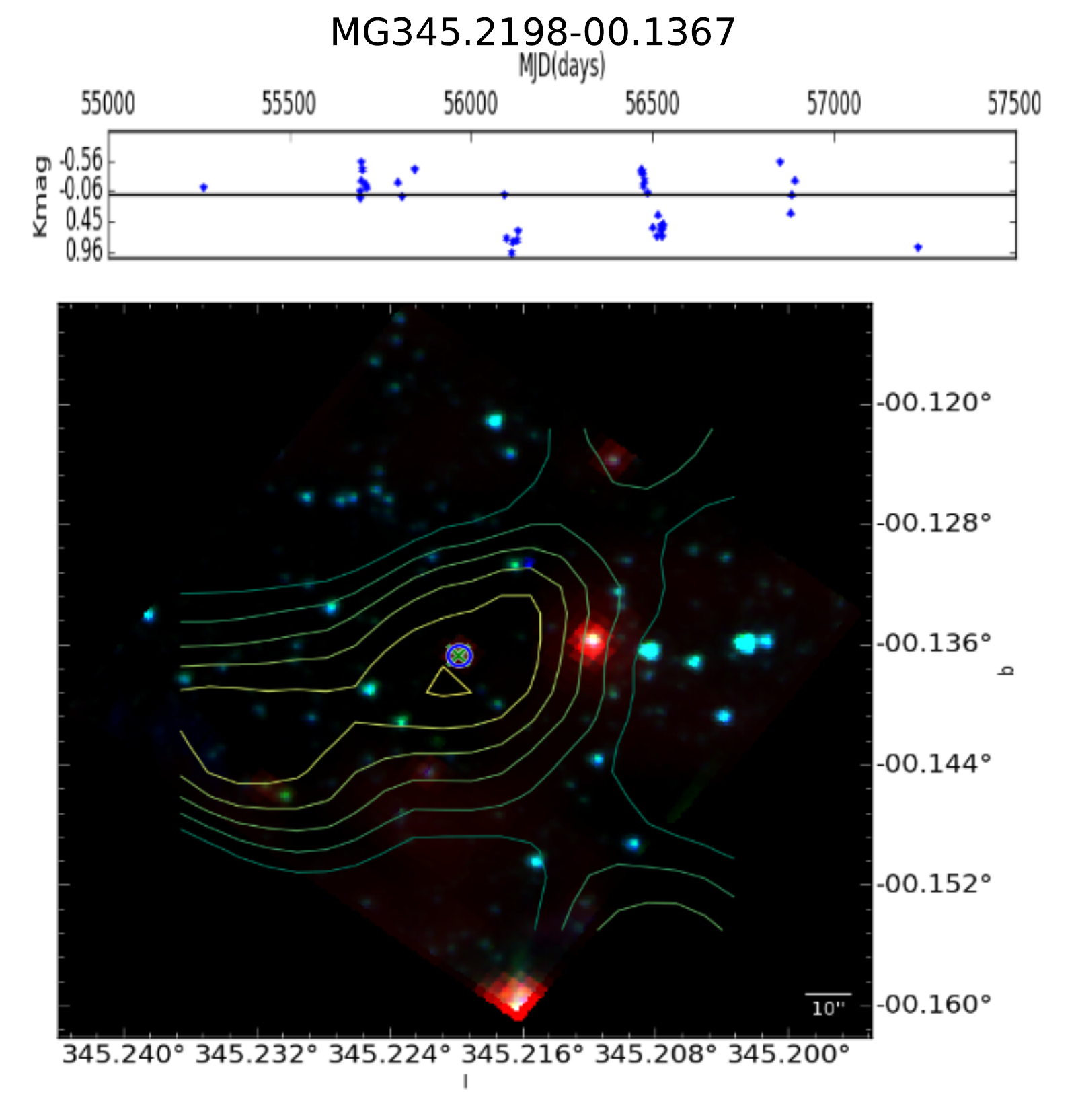}\\
                \includegraphics[width=0.32\textwidth]{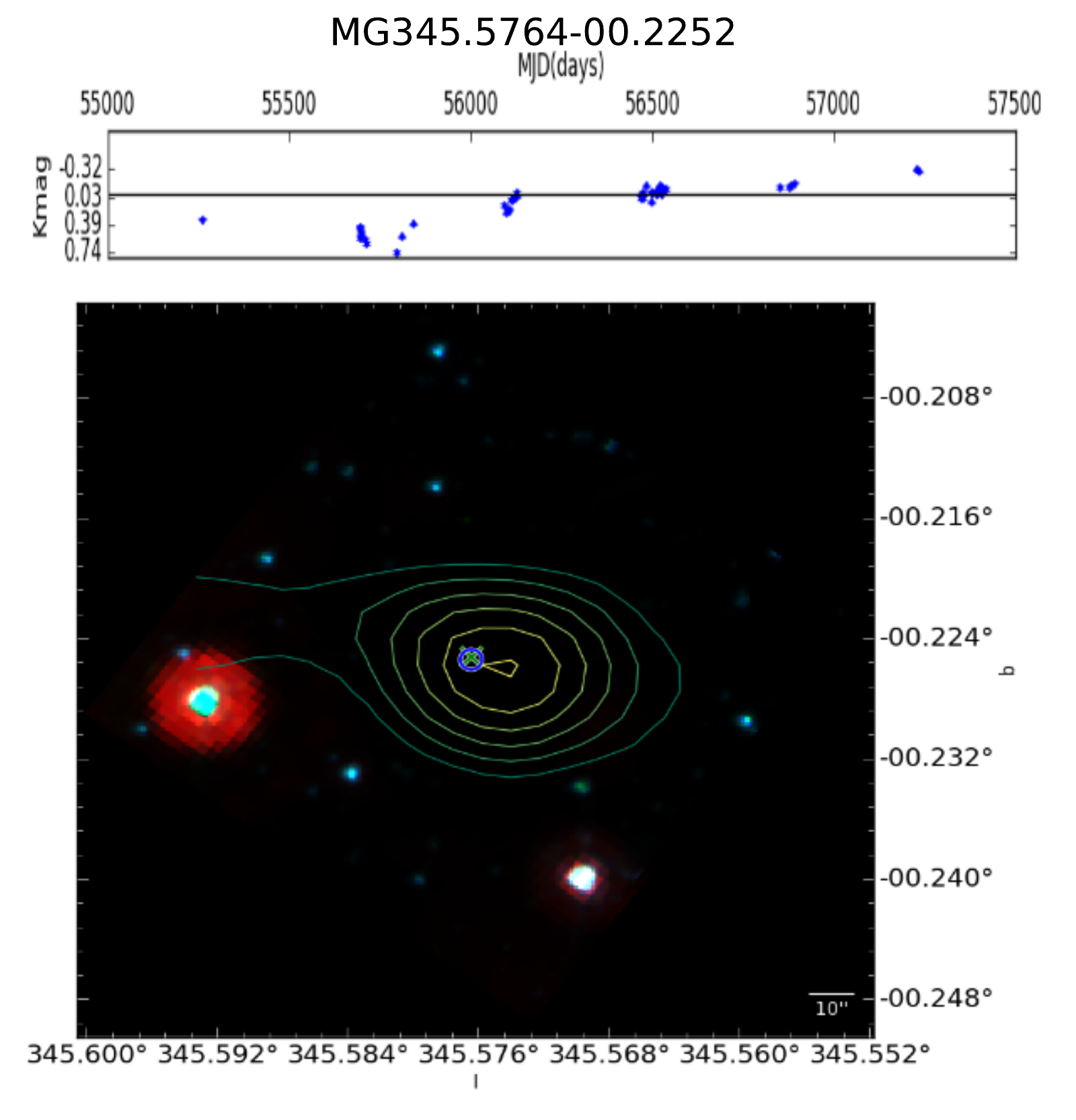}&
                \includegraphics[width=0.32\textwidth]{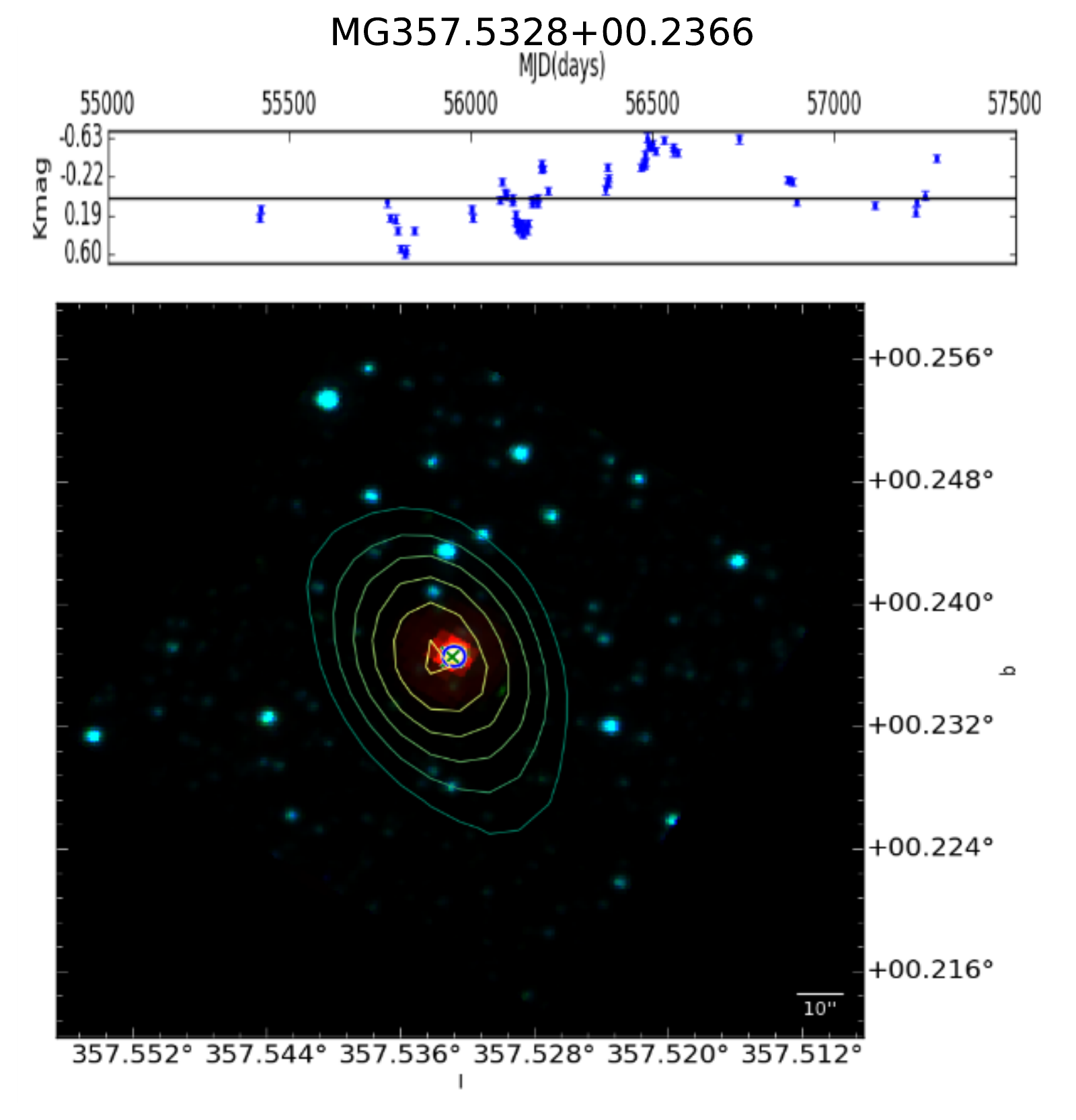}&
                \includegraphics[width=0.32\textwidth]{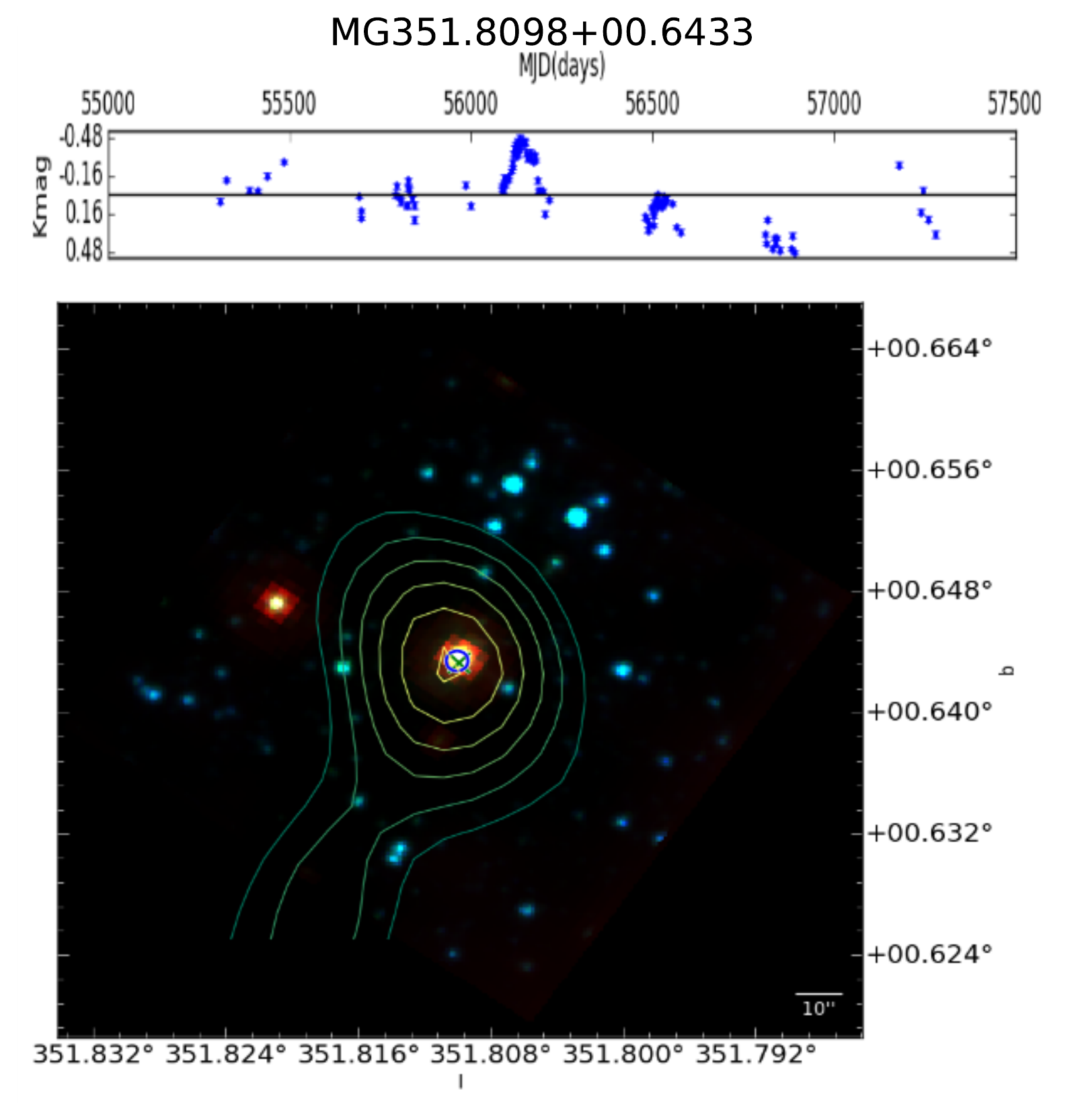}\\
                \includegraphics[width=0.32\textwidth]{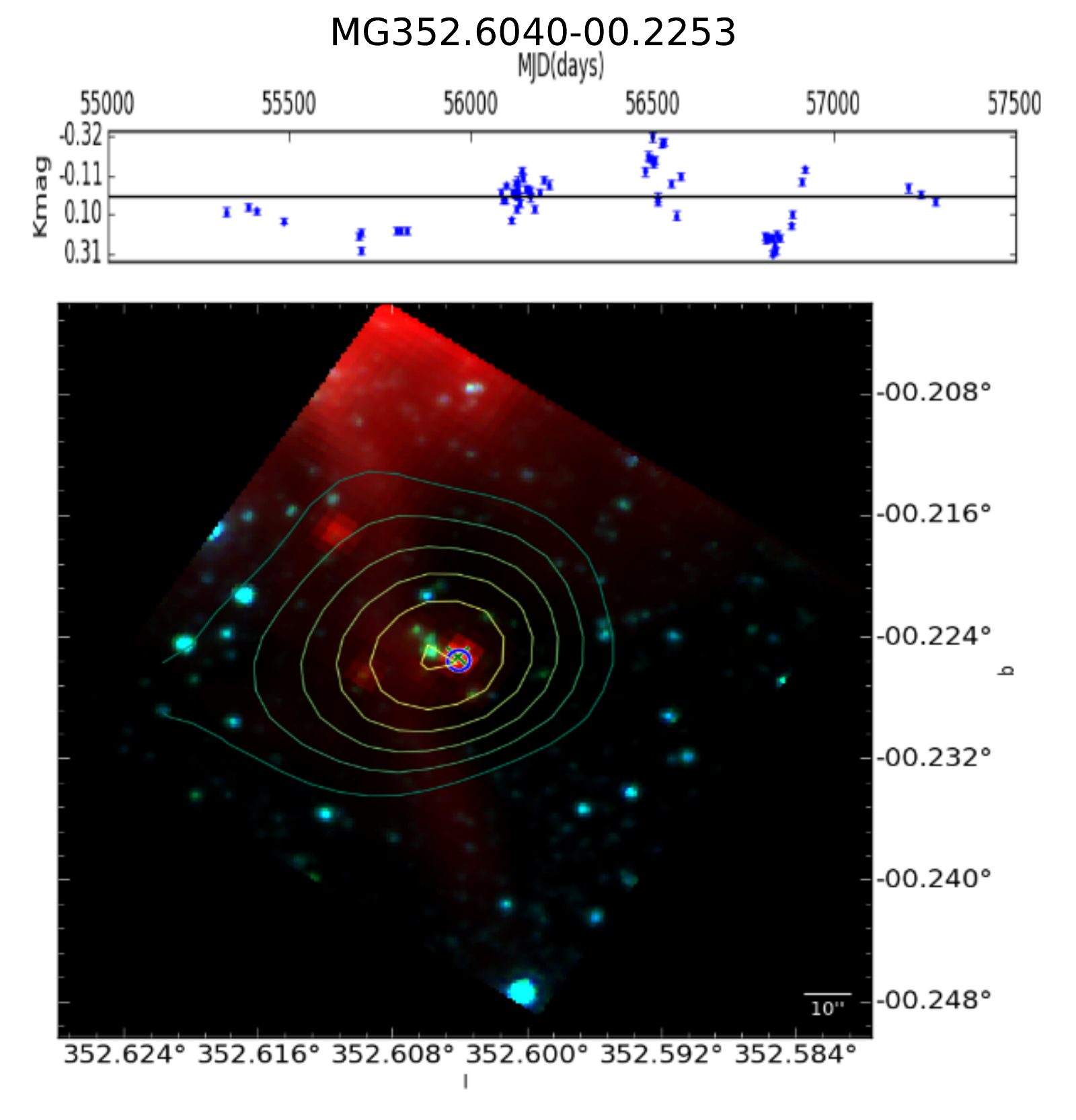}&
                \includegraphics[width=0.32\textwidth]{{./MG354.4384+00.4185_combined}.pdf}&
                \includegraphics[width=0.32\textwidth]{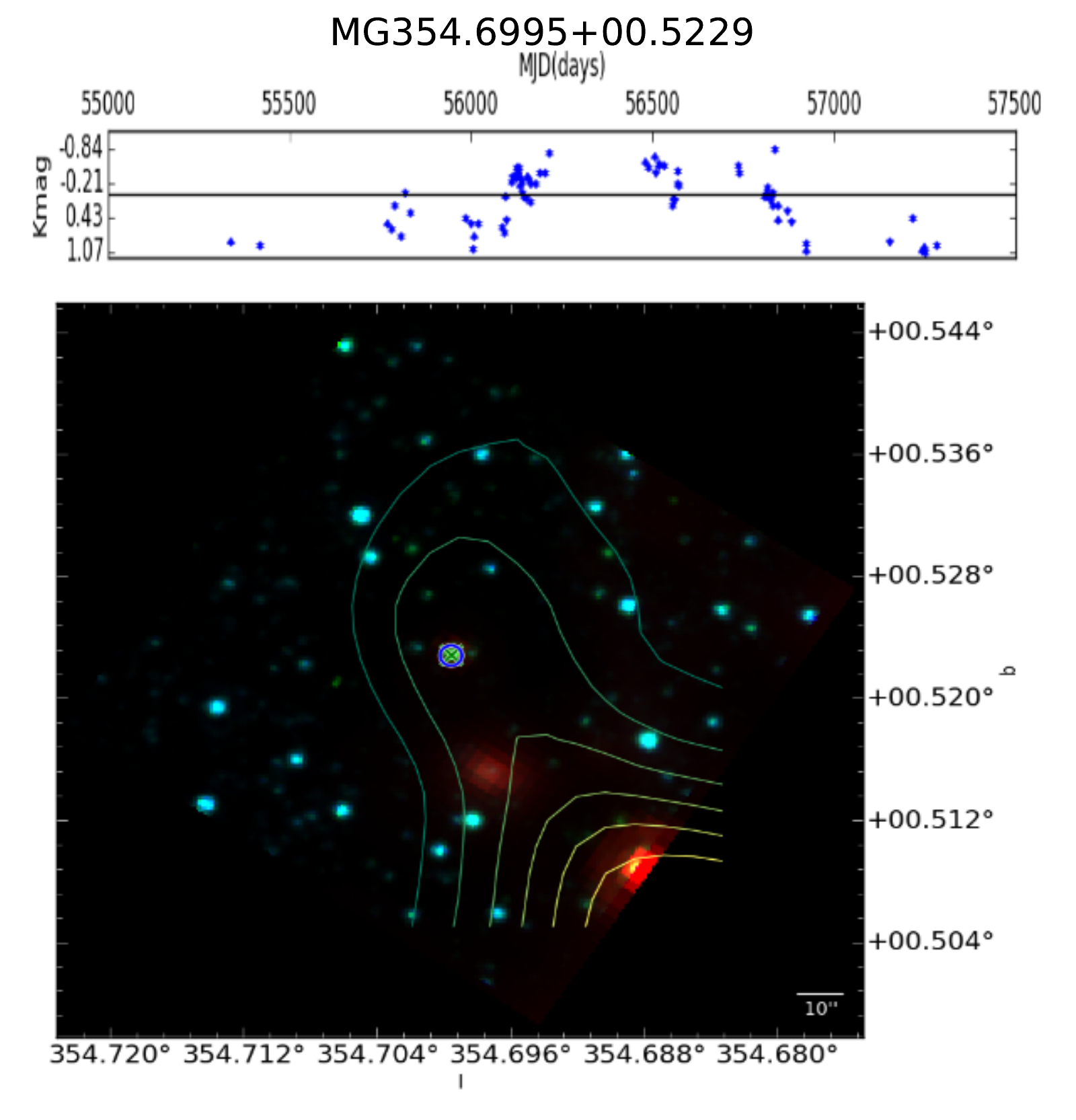}\\
        \end{tabular}
        \caption{Continuation of Fig. \ref{fig:testgrid}.}
\end{figure}

\begin{figure}
        \begin{tabular}{ccc}

                \includegraphics[width=0.32\textwidth]{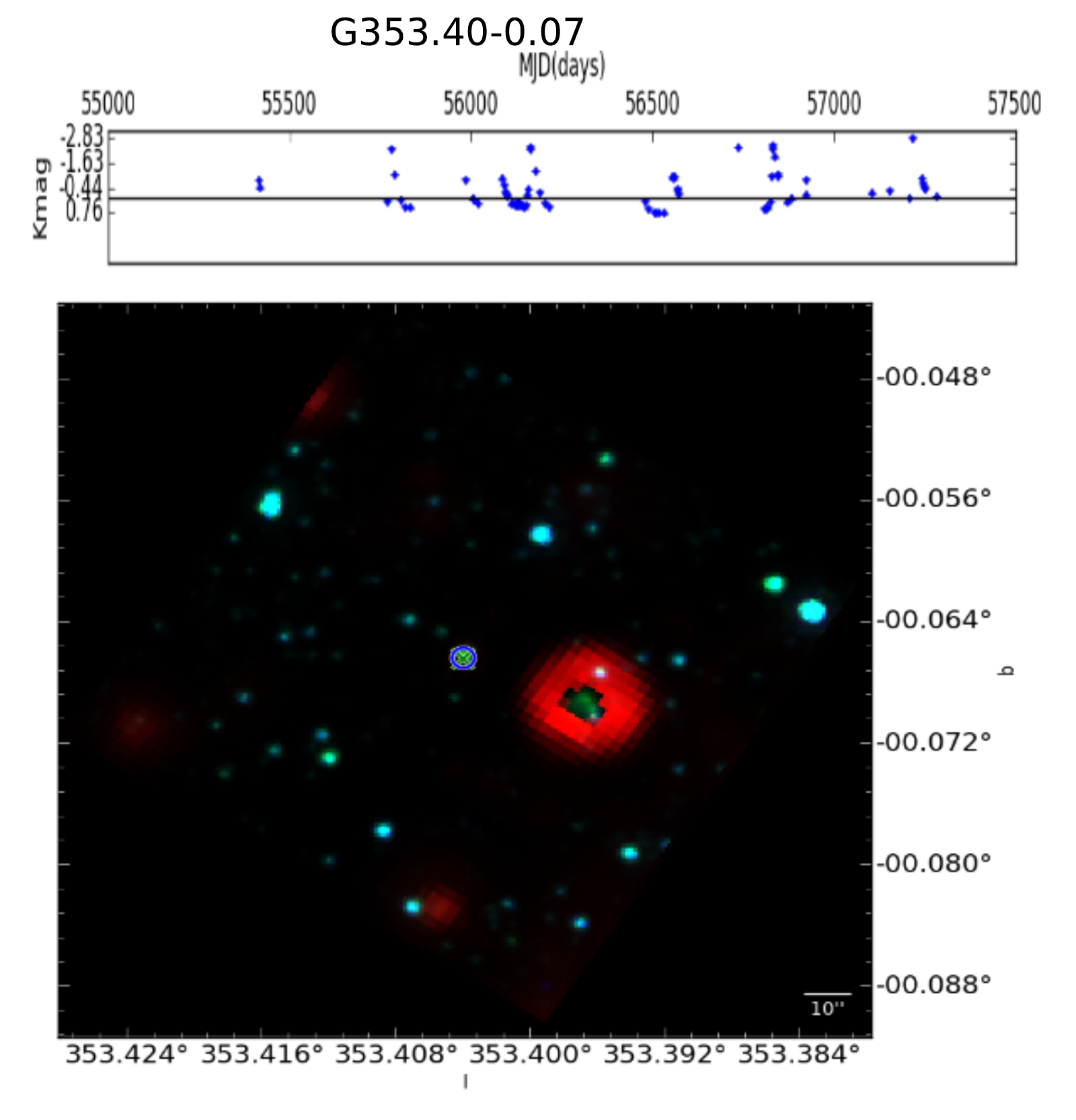}&
                \includegraphics[width=0.32\textwidth]{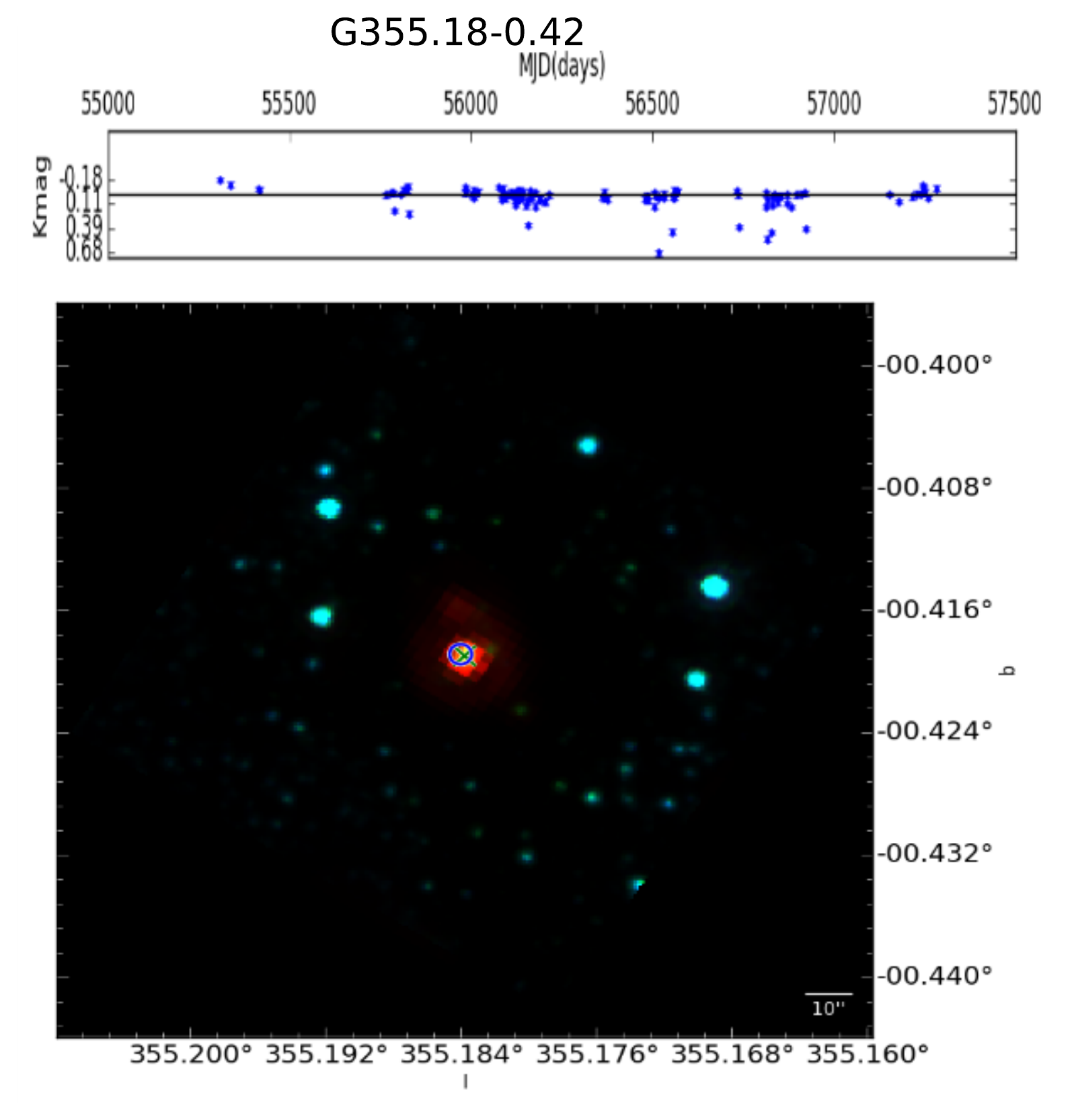}&
                \includegraphics[width=0.32\textwidth]{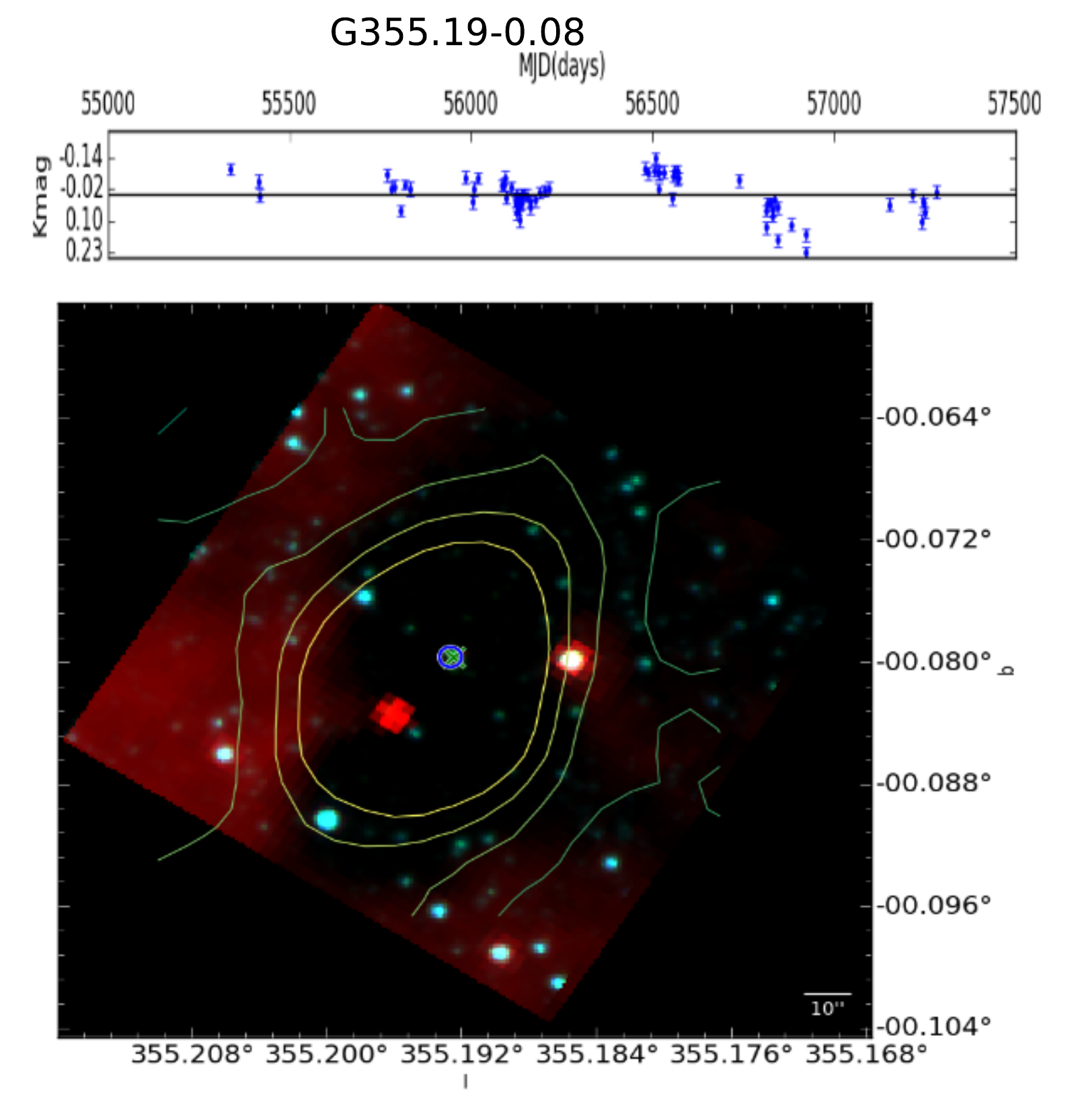}\\
                \includegraphics[width=0.32\textwidth]{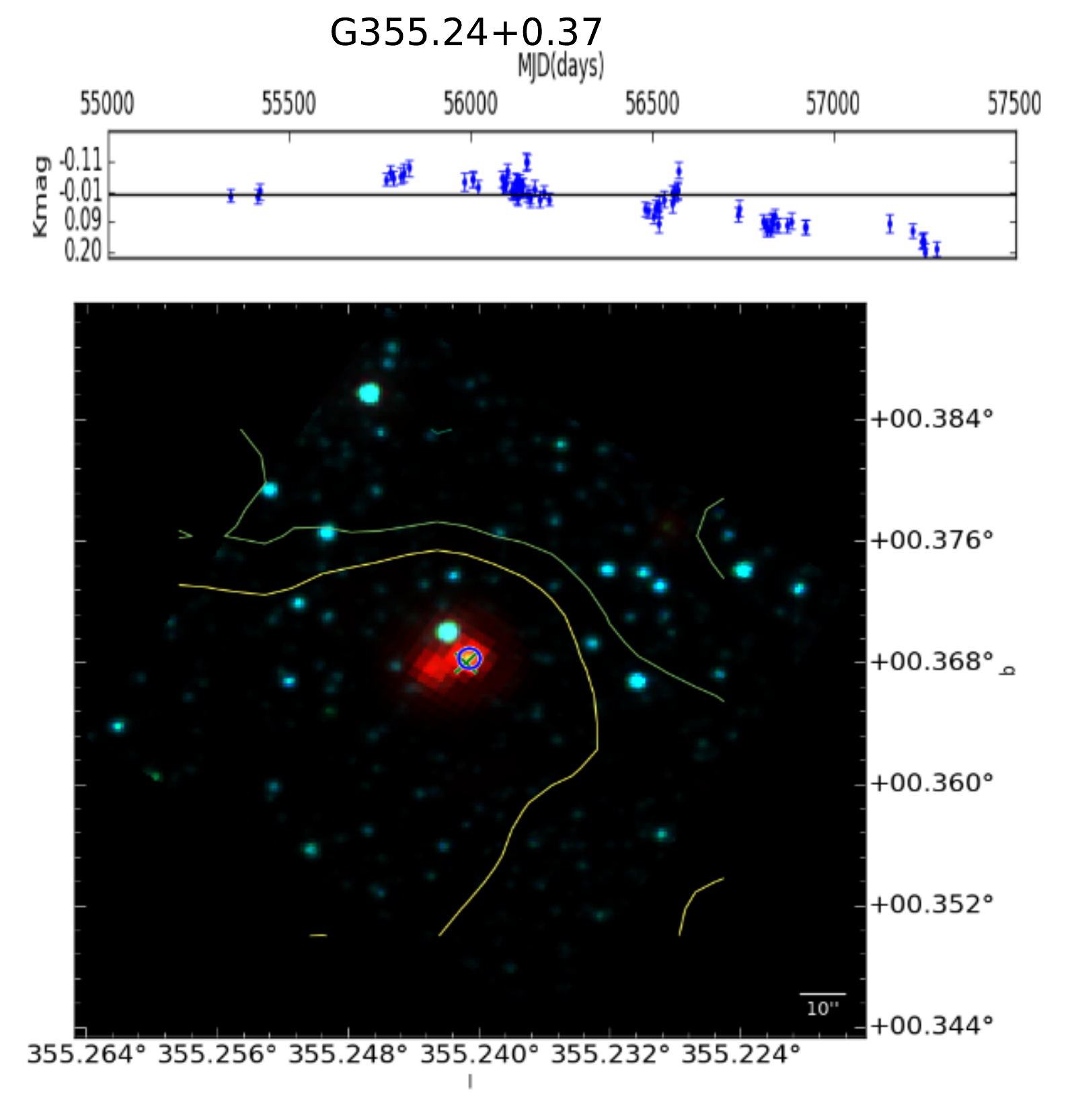}&
                \includegraphics[width=0.32\textwidth]{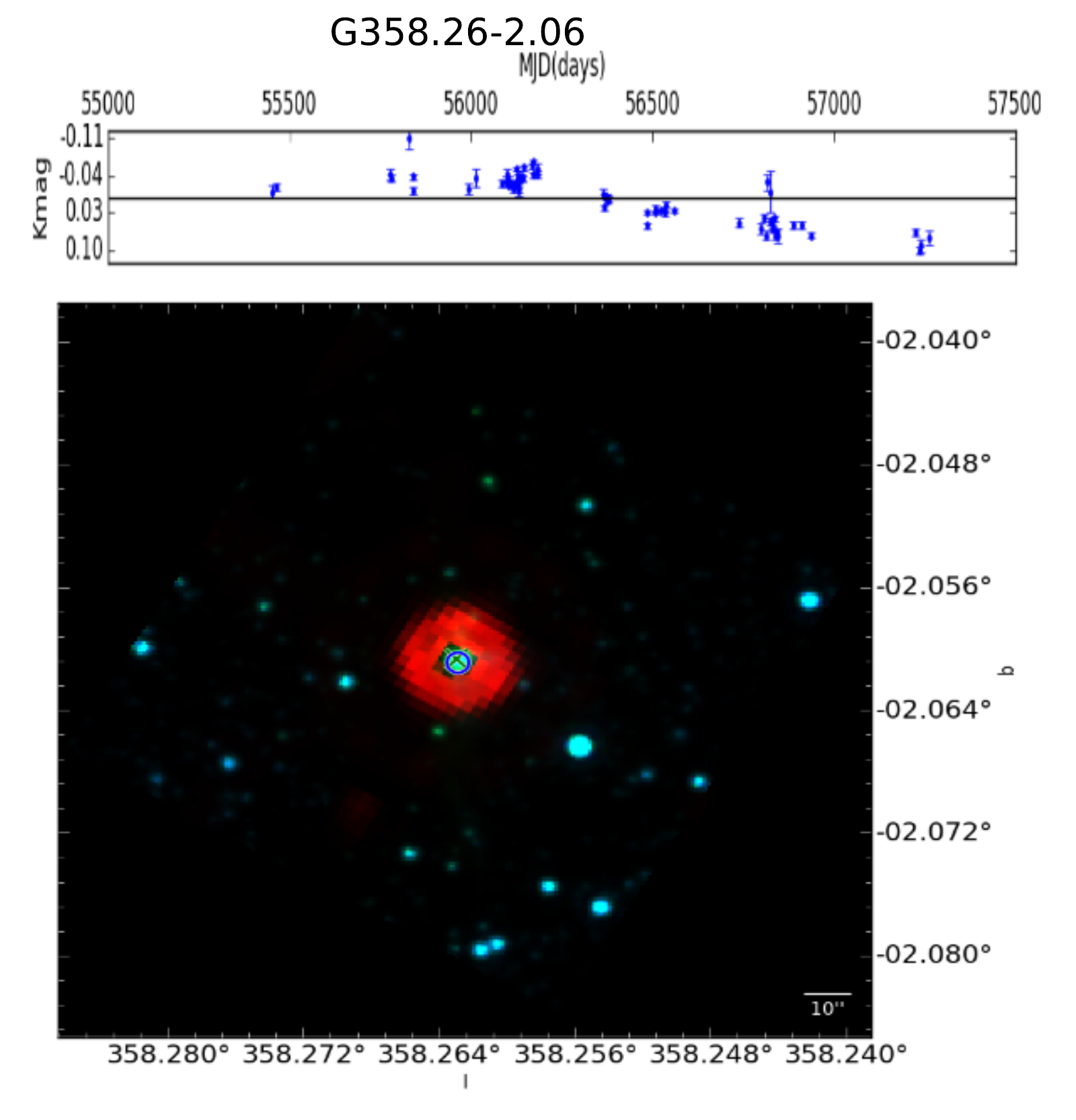}&
                \includegraphics[width=0.32\textwidth]{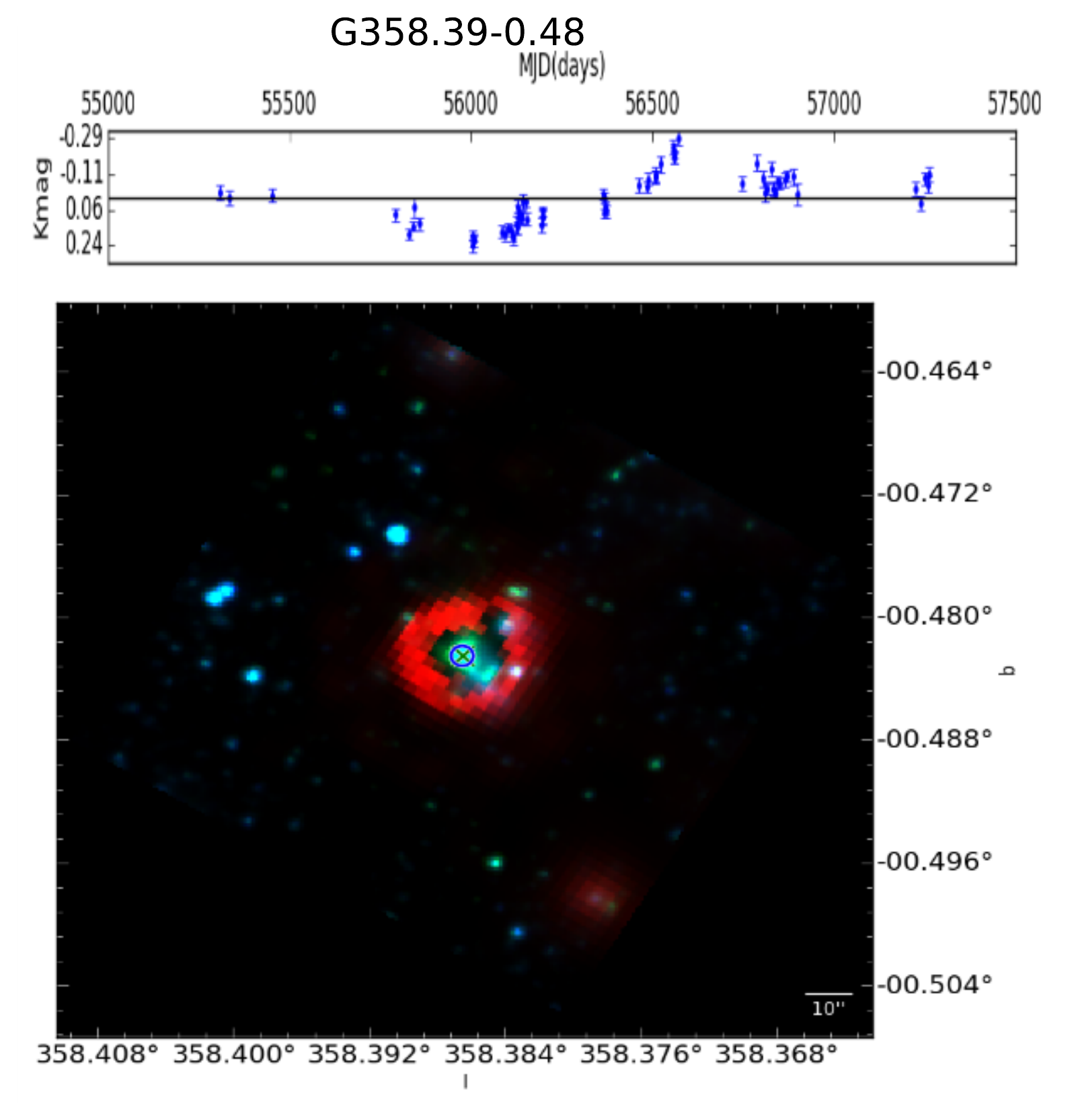}\\
                \includegraphics[width=0.32\textwidth]{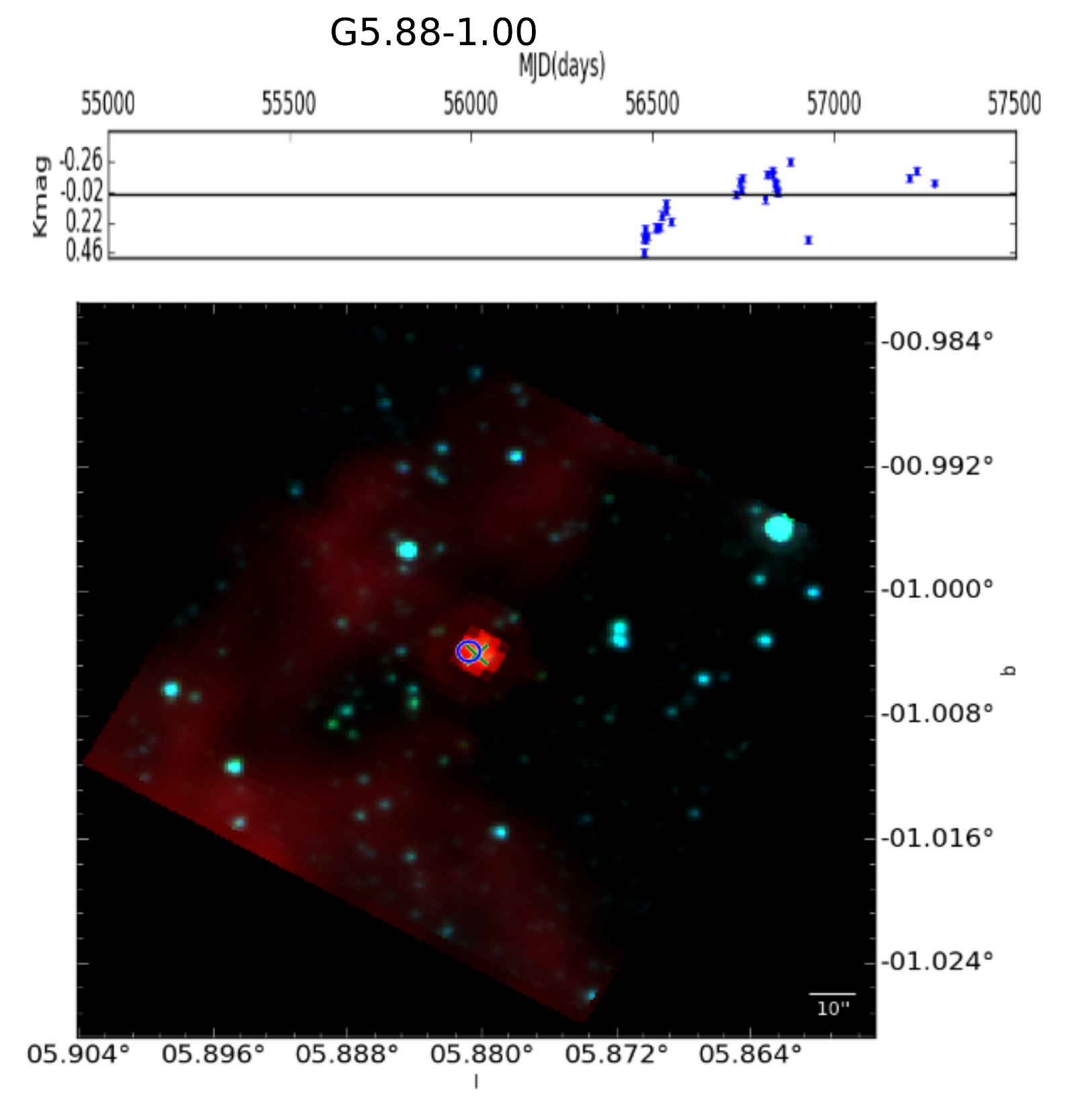}&
                \includegraphics[width=0.32\textwidth]{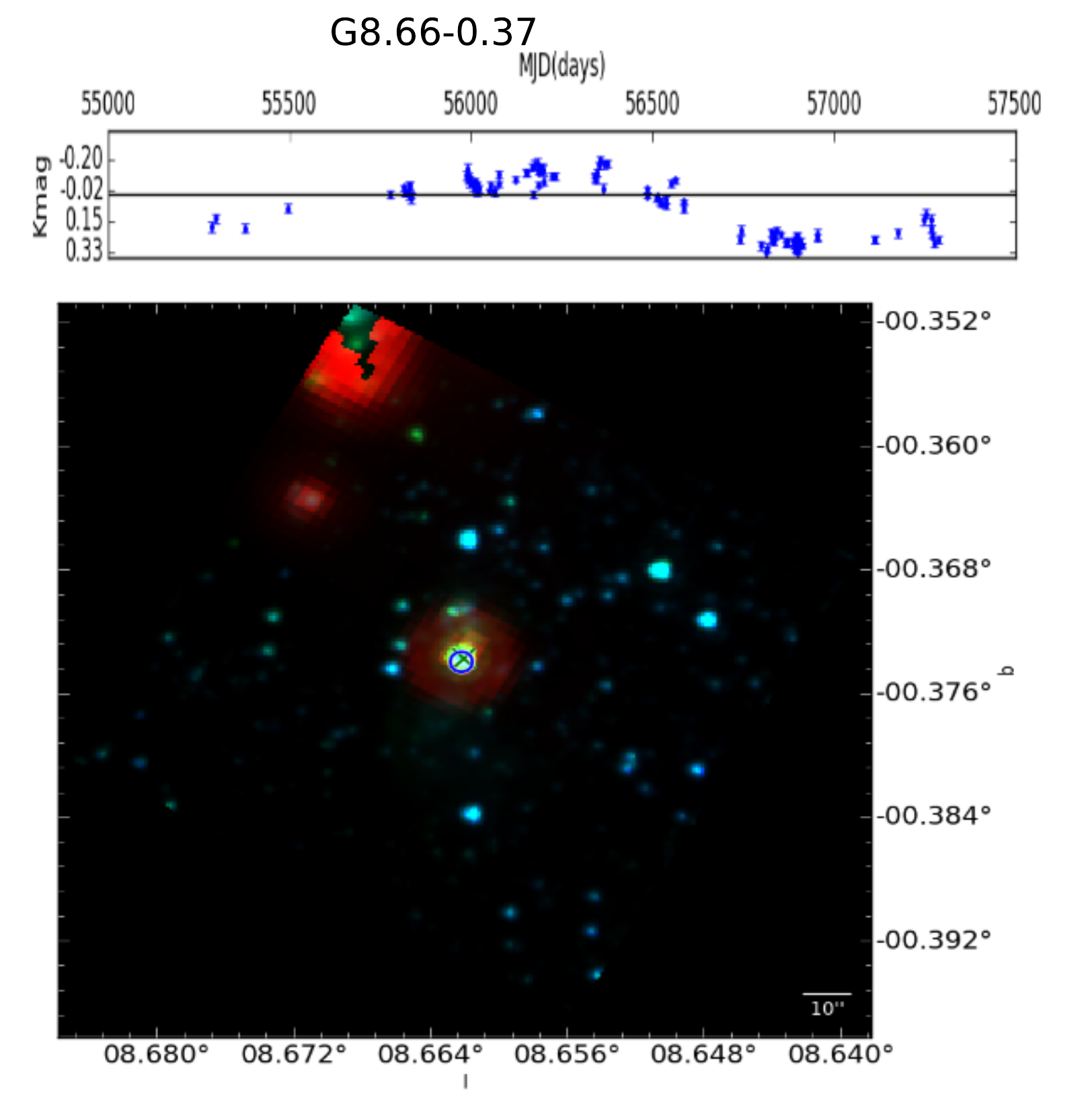}&
                \includegraphics[width=0.32\textwidth]{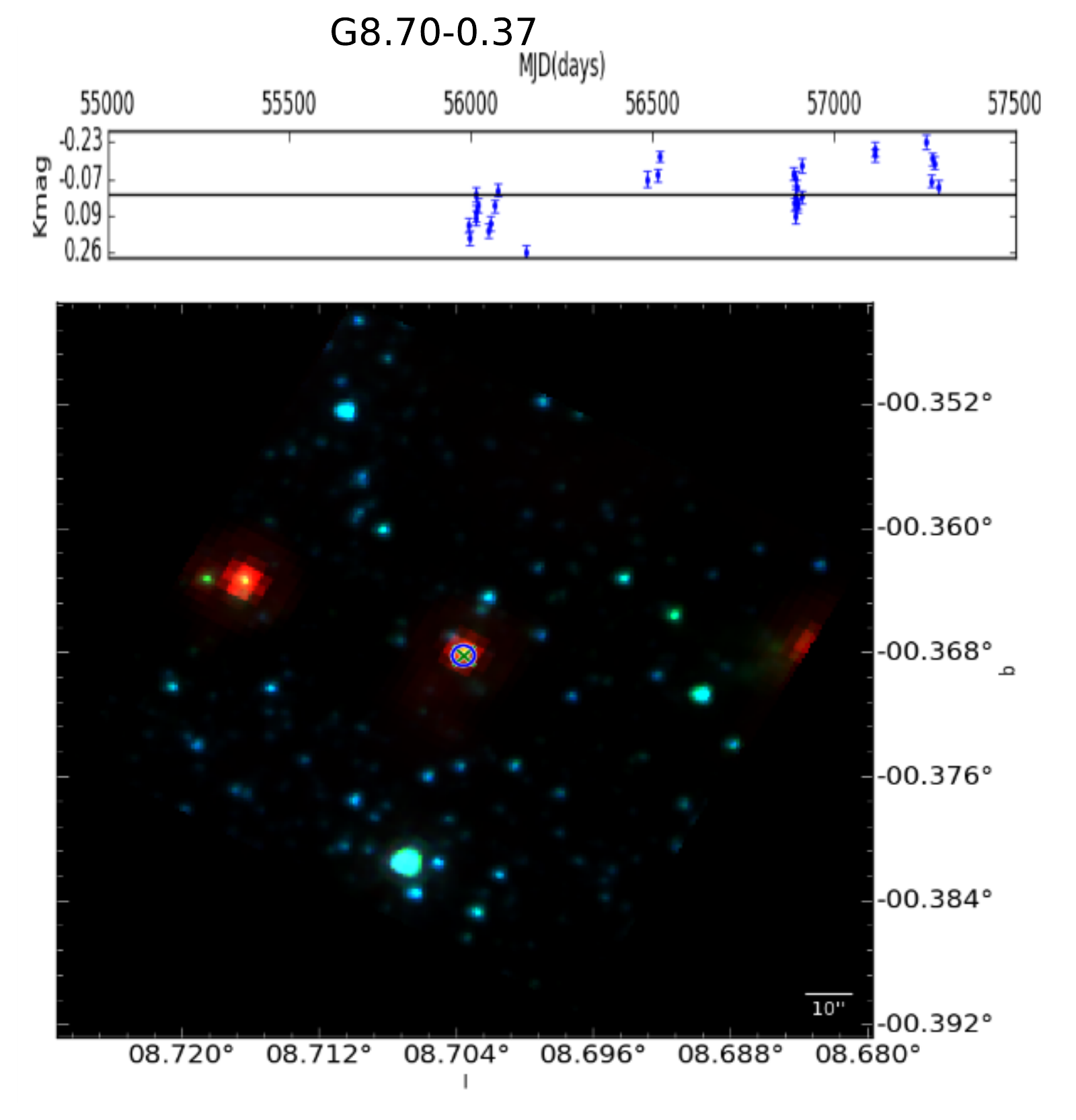}\\
                \includegraphics[width=0.32\textwidth]{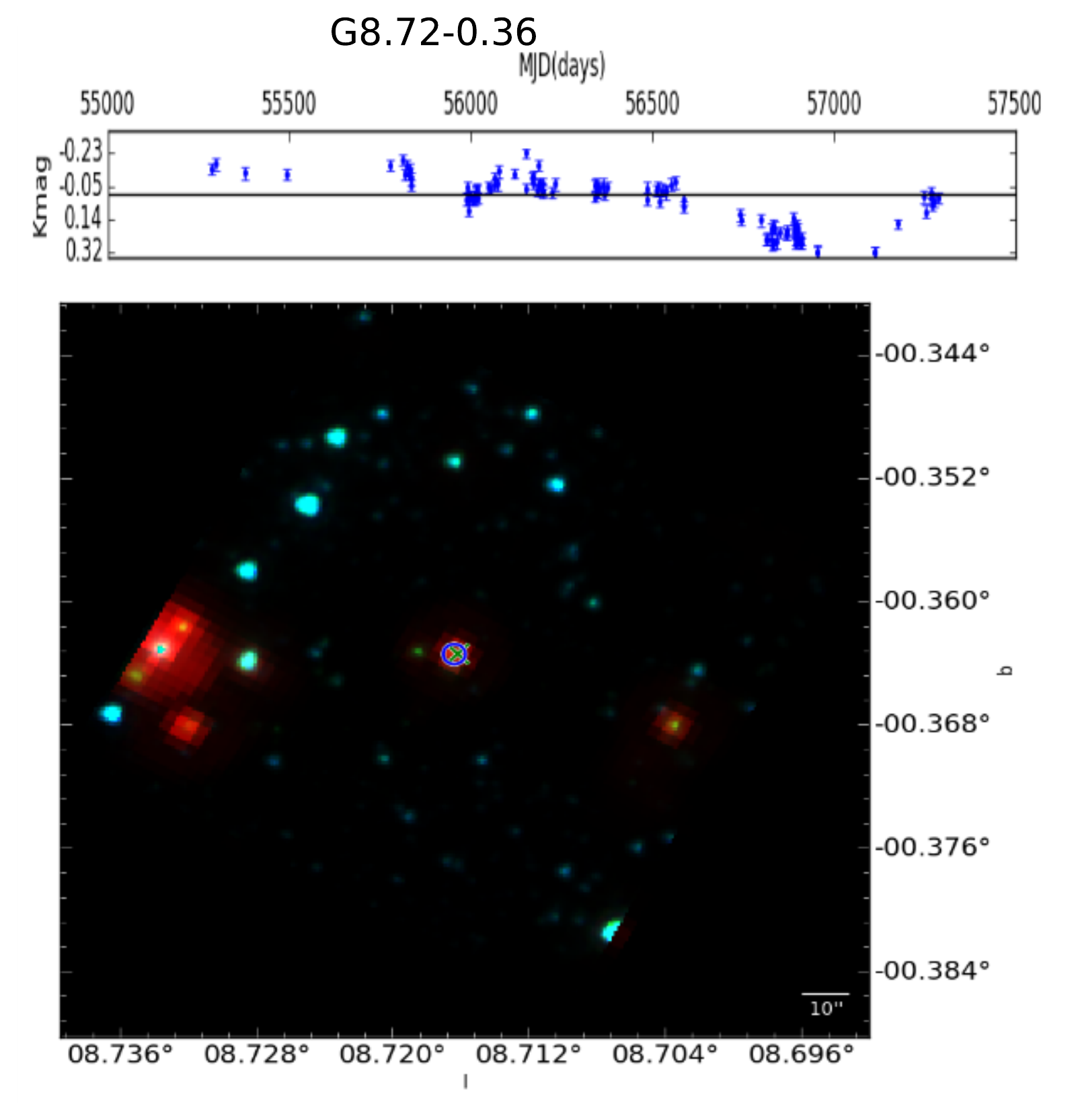}&
                \includegraphics[width=0.32\textwidth]{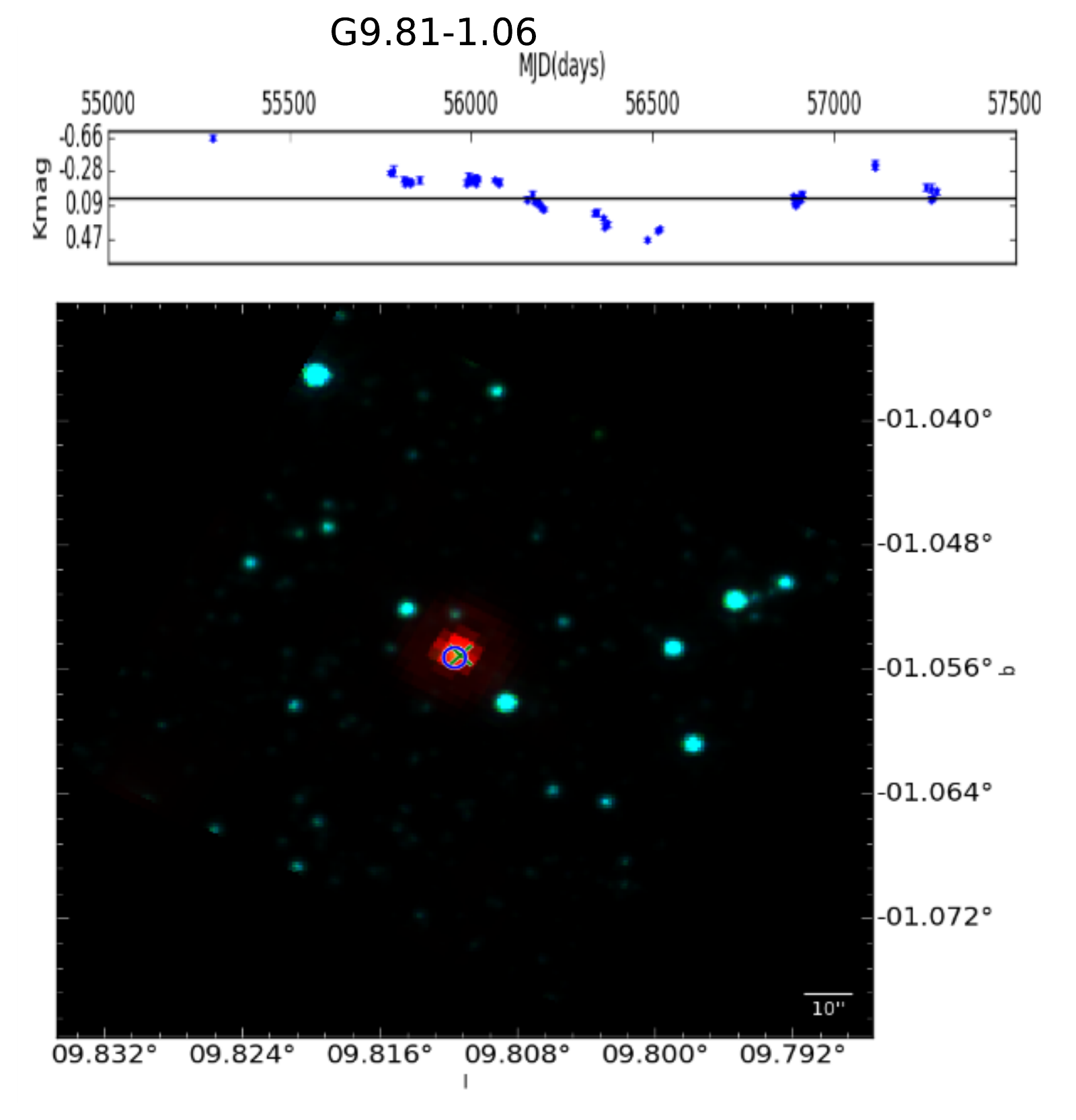}&
                \includegraphics[width=0.32\textwidth]{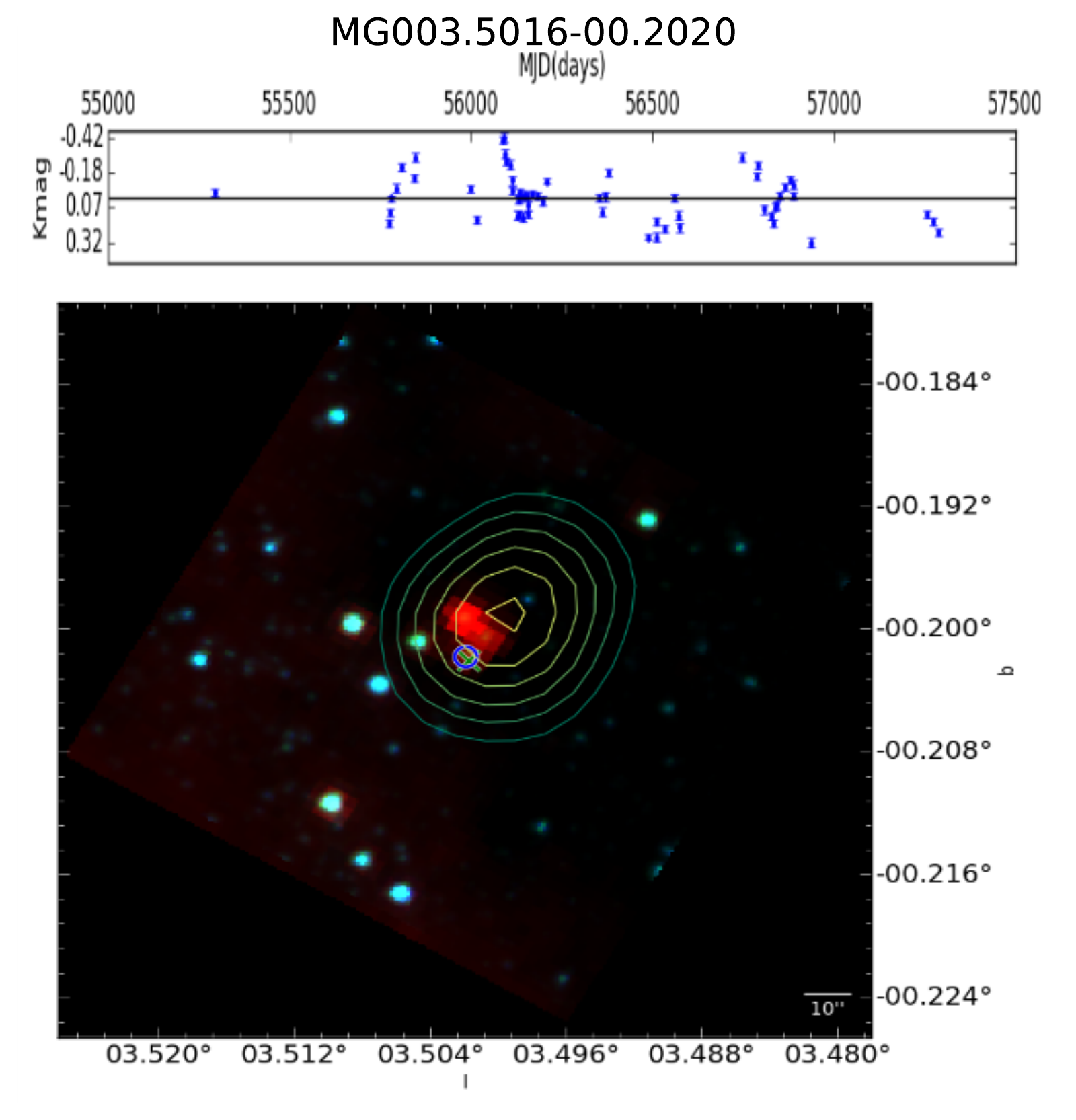}\\
        \end{tabular}
        \caption{Continuation of Fig. \ref{fig:testgrid}.}
\end{figure}

\begin{figure}
        \begin{tabular}{ccc}

                \includegraphics[width=0.32\textwidth]{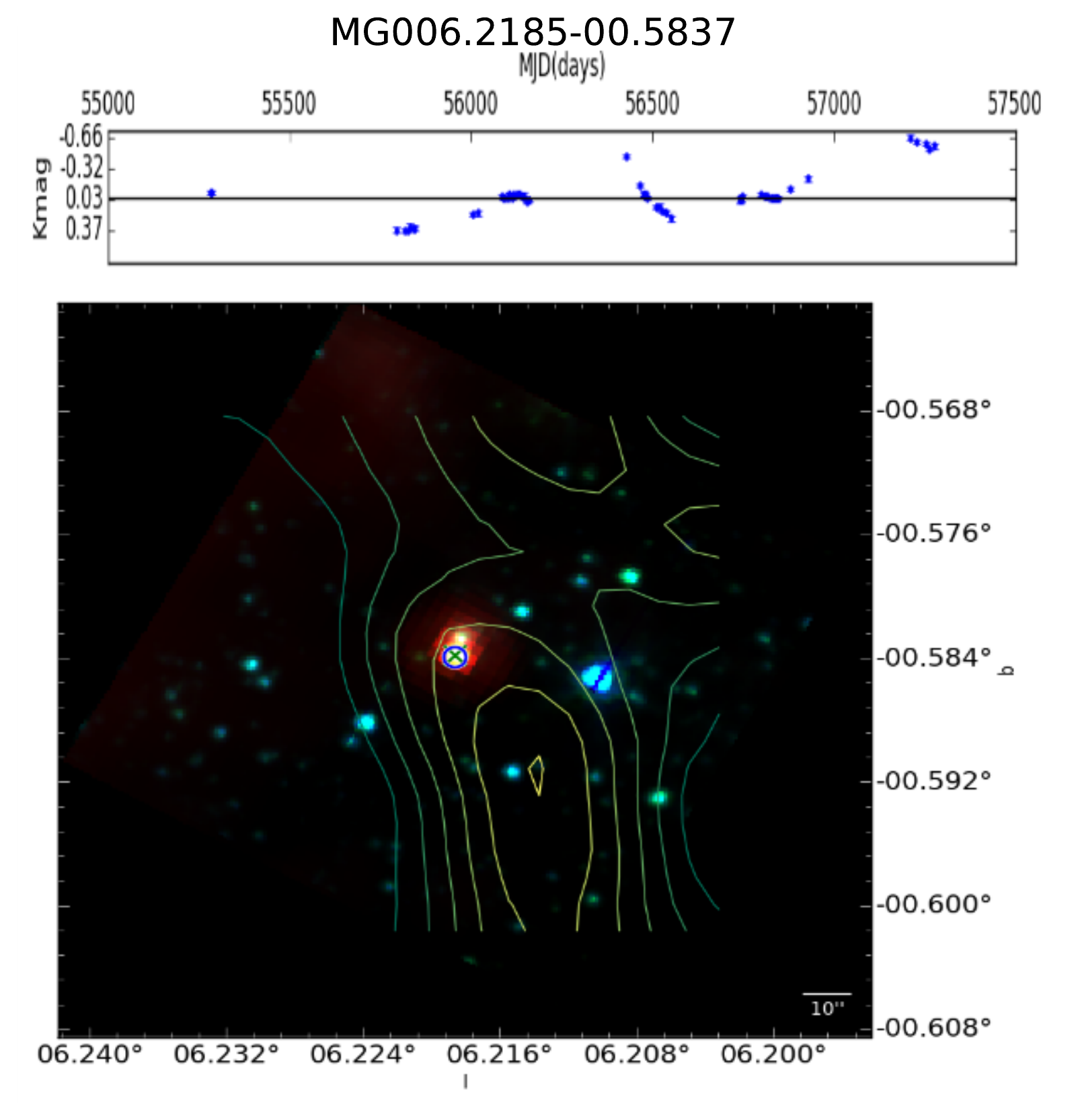}&
                \includegraphics[width=0.32\textwidth]{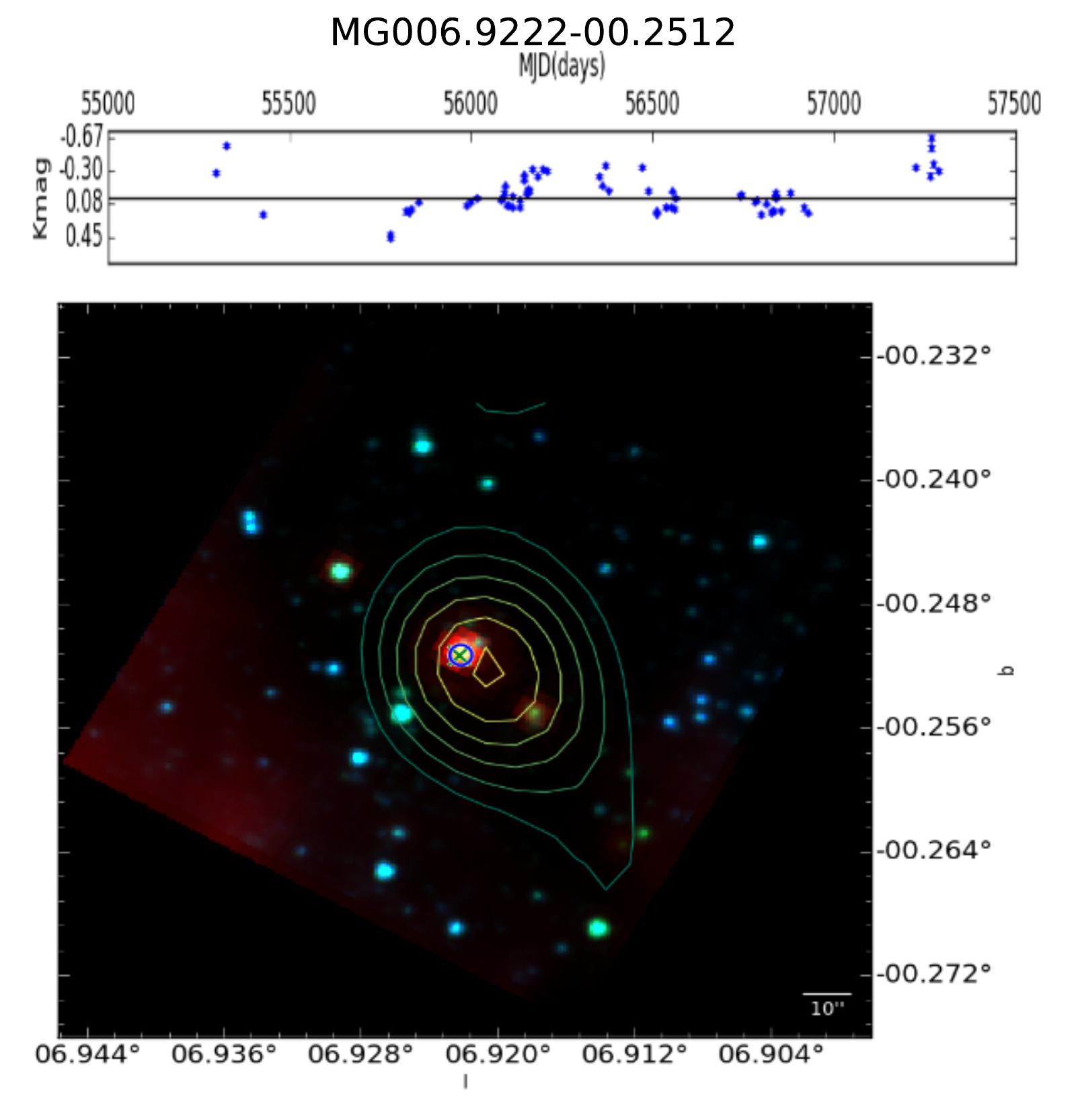}&
                \includegraphics[width=0.32\textwidth]{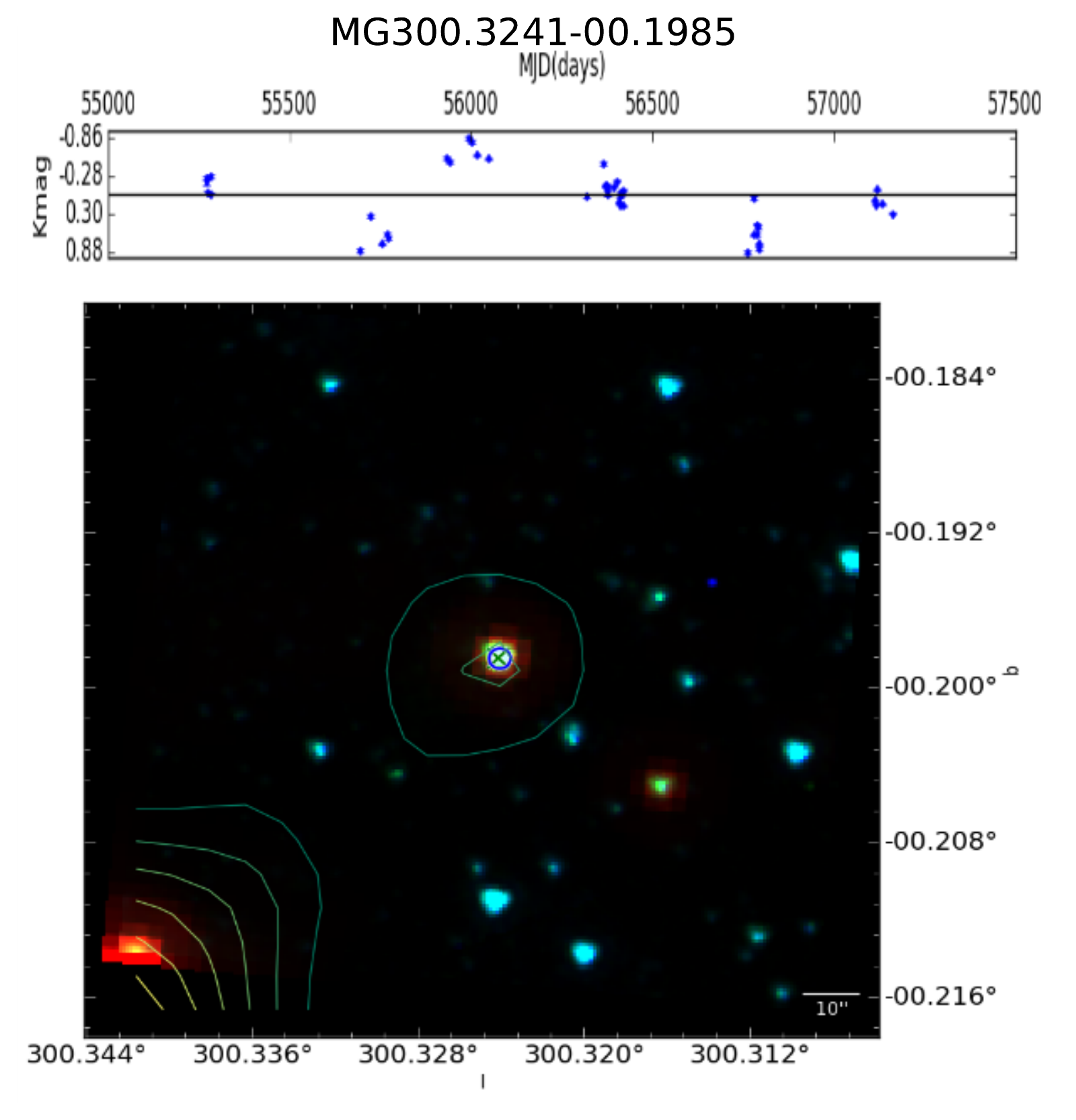}\\
                \includegraphics[width=0.32\textwidth]{{./MG303.9304-00.6879_combined}.pdf}&
                \includegraphics[width=0.32\textwidth]{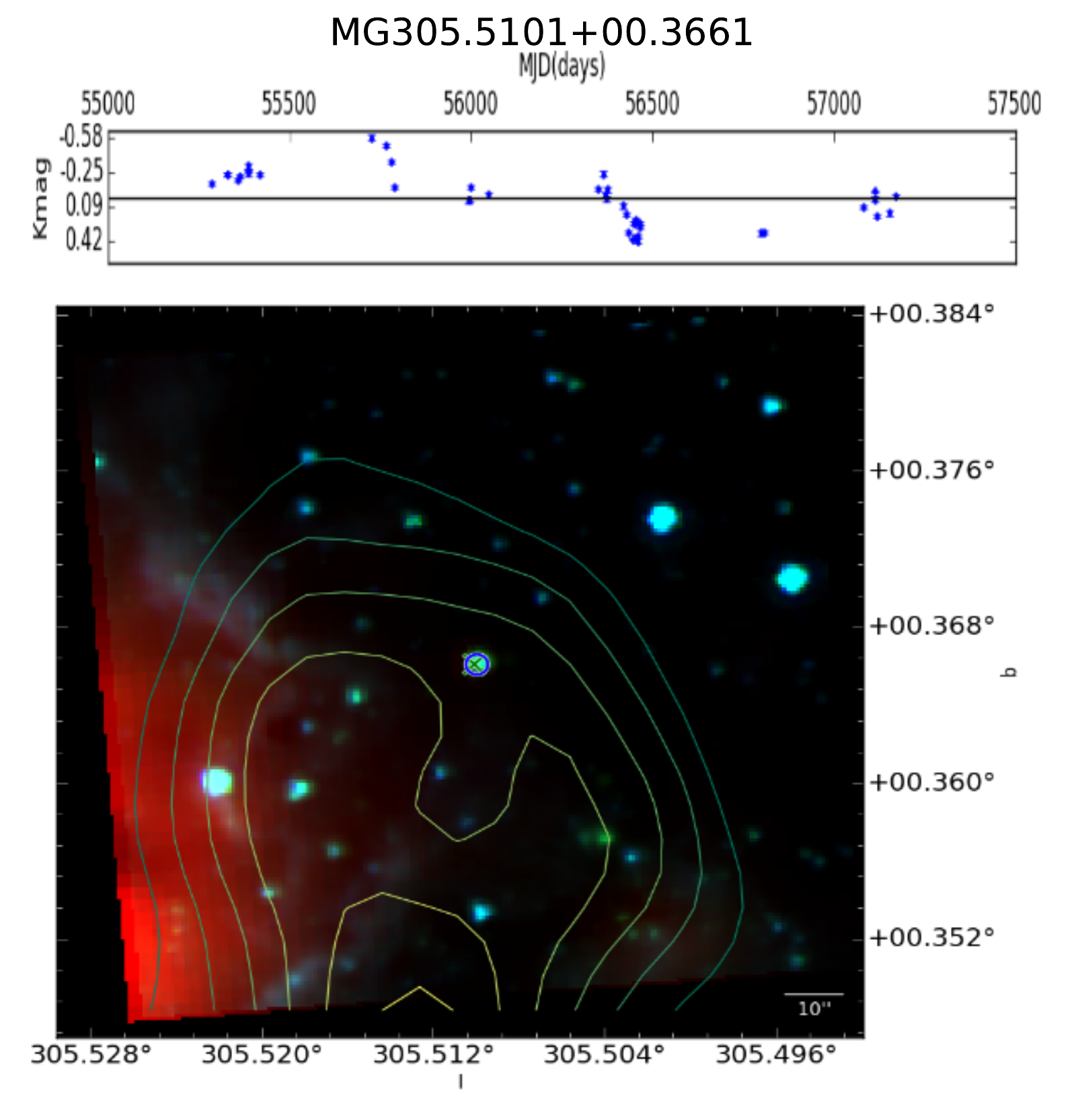}&
                \includegraphics[width=0.32\textwidth]{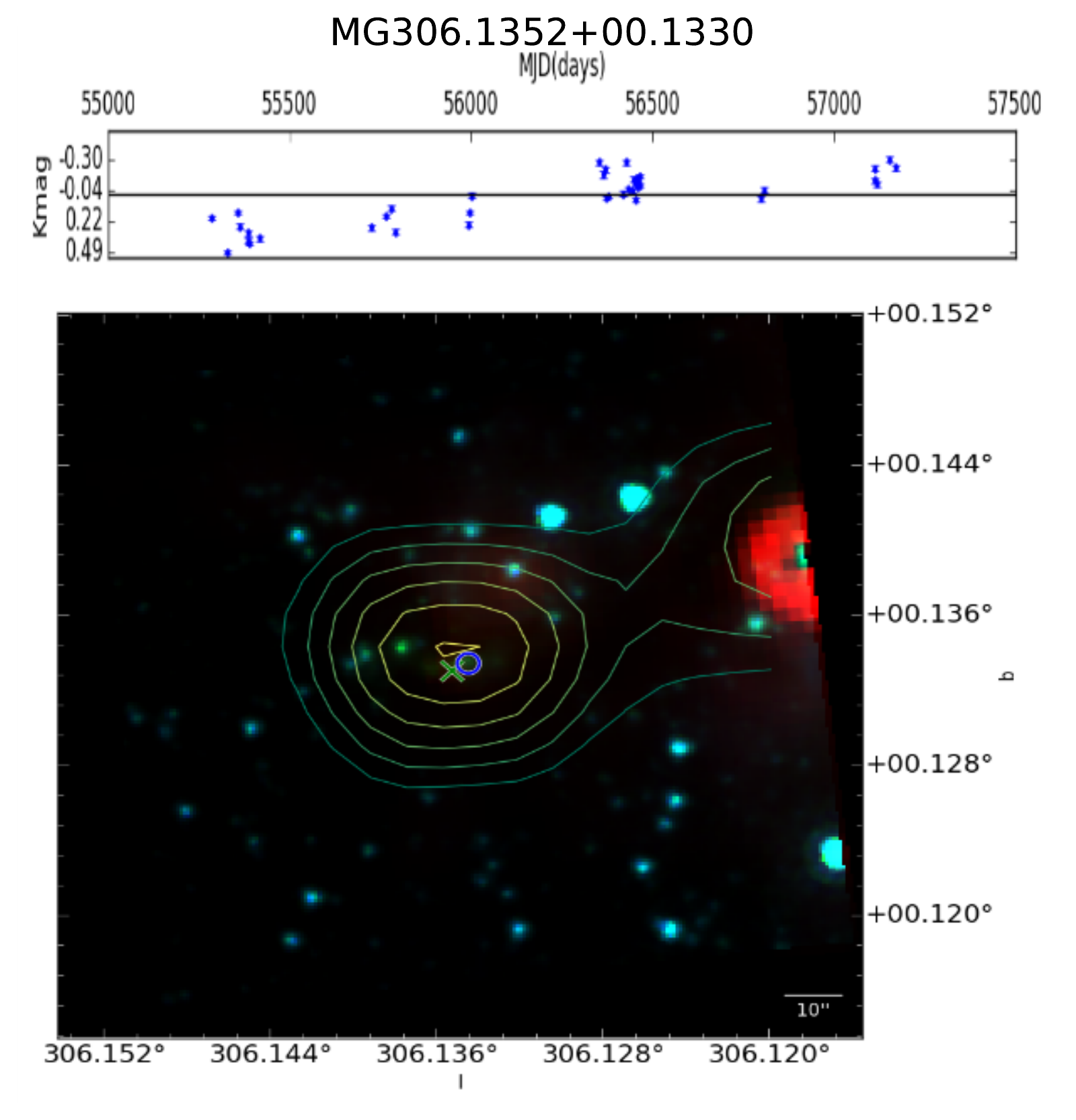}\\
        \end{tabular}
        \caption{Continuation of Fig. \ref{fig:testgrid}.}
\end{figure}

\begin{figure}
        \begin{tabular}{ccc}
                
                \includegraphics[width=0.31\textwidth]{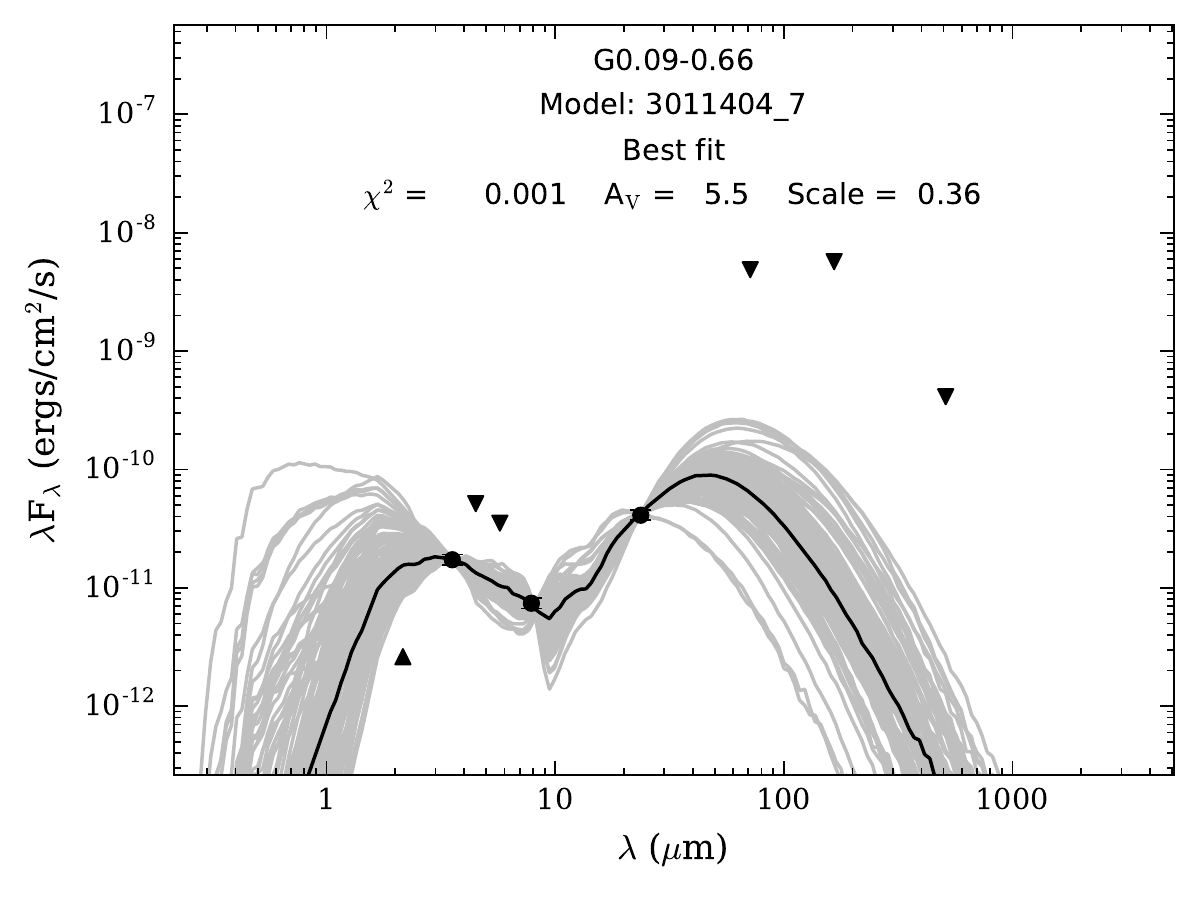}&
                \includegraphics[width=0.31\textwidth]{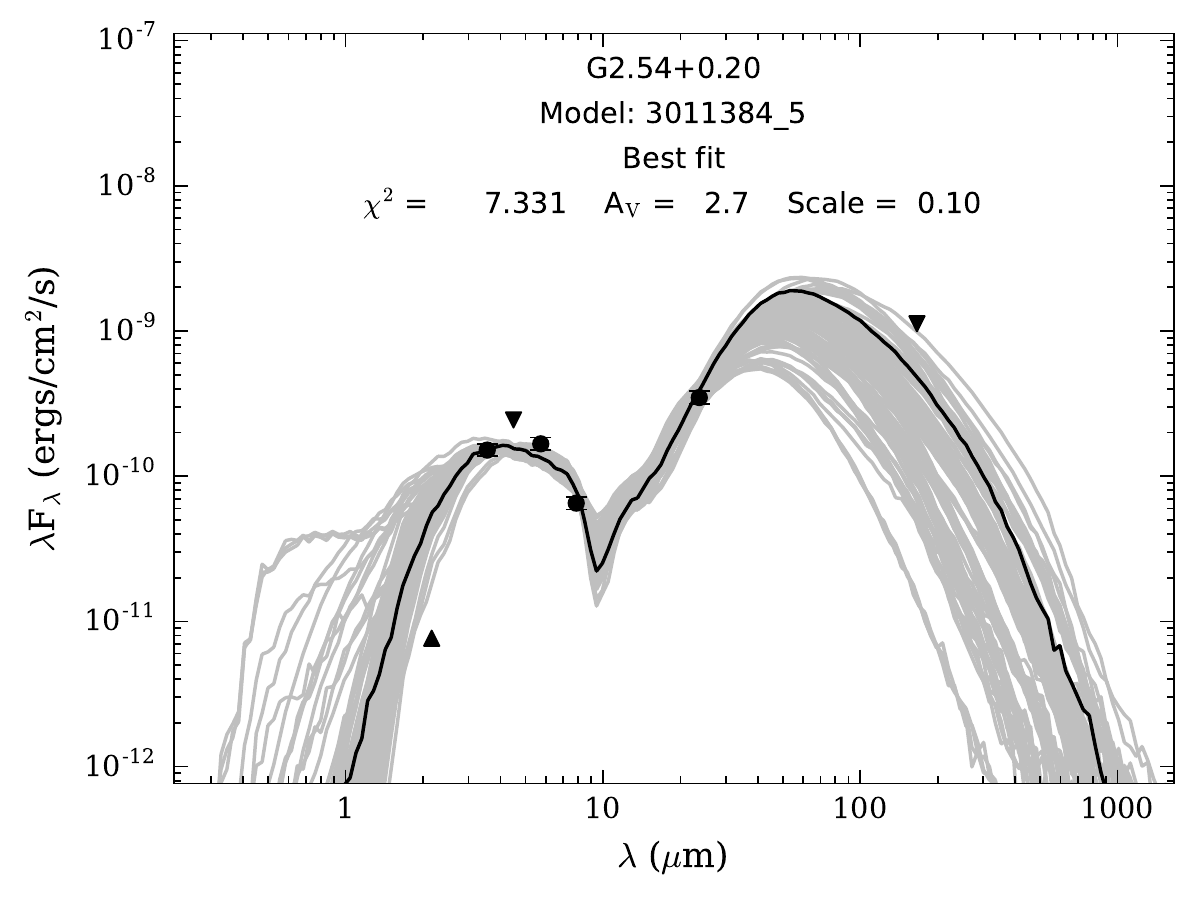}&
                \includegraphics[width=0.31\textwidth]{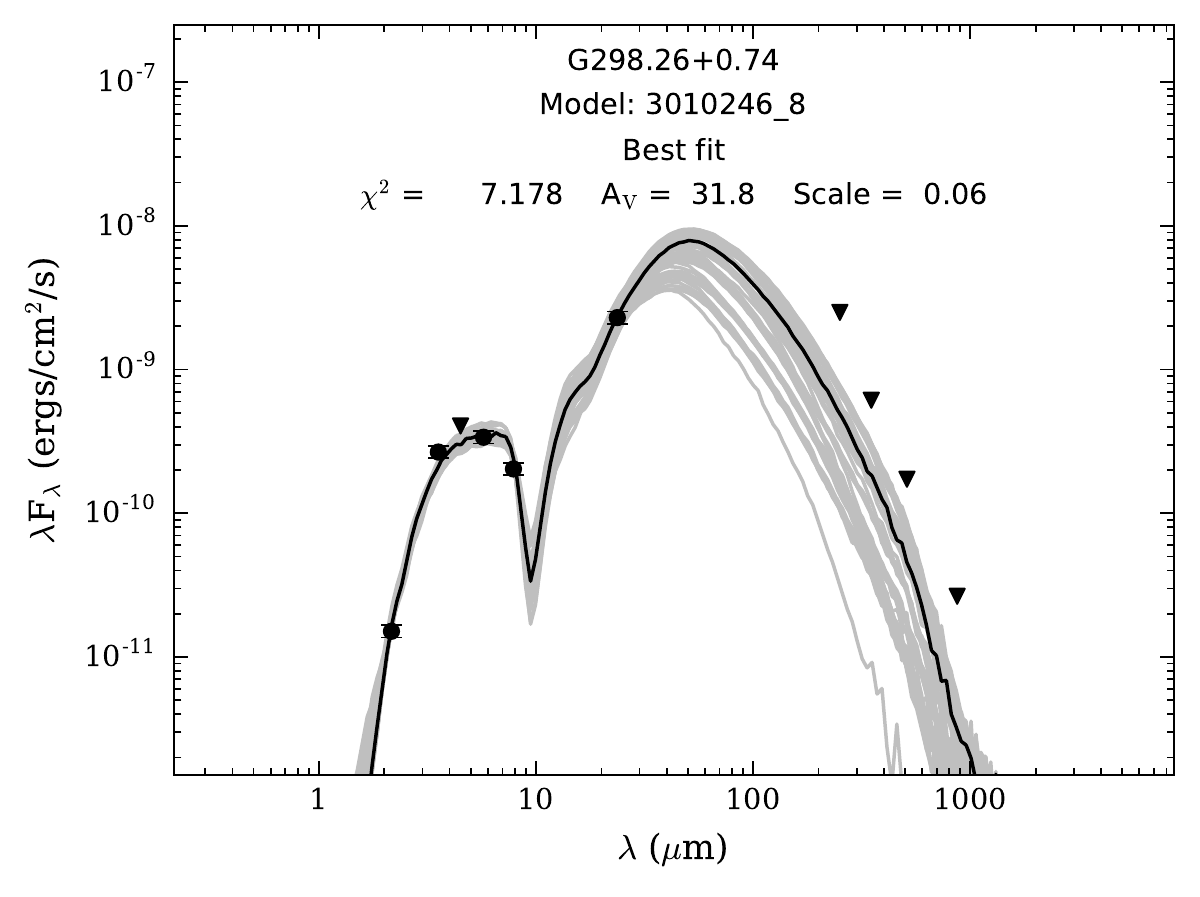}\\
                \includegraphics[width=0.31\textwidth]{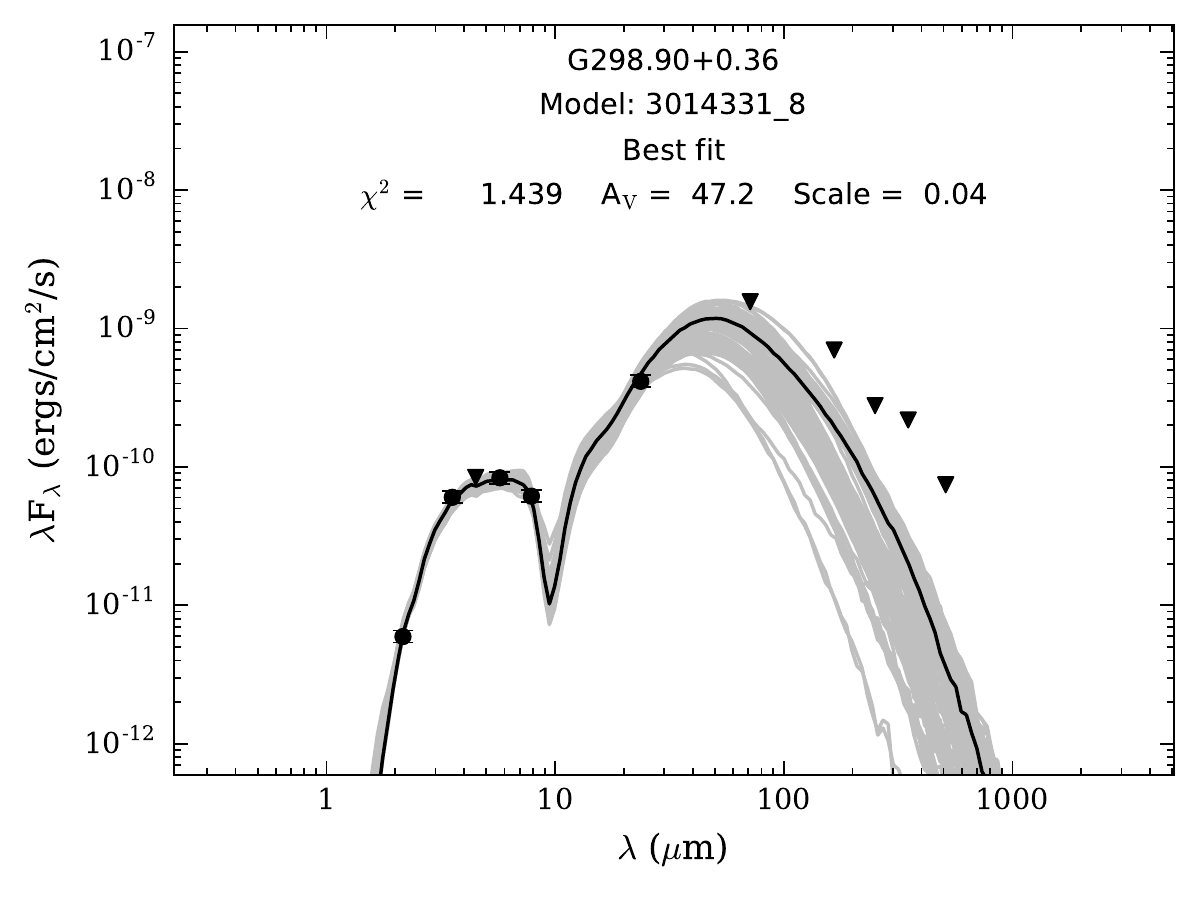}&
                \includegraphics[width=0.31\textwidth]{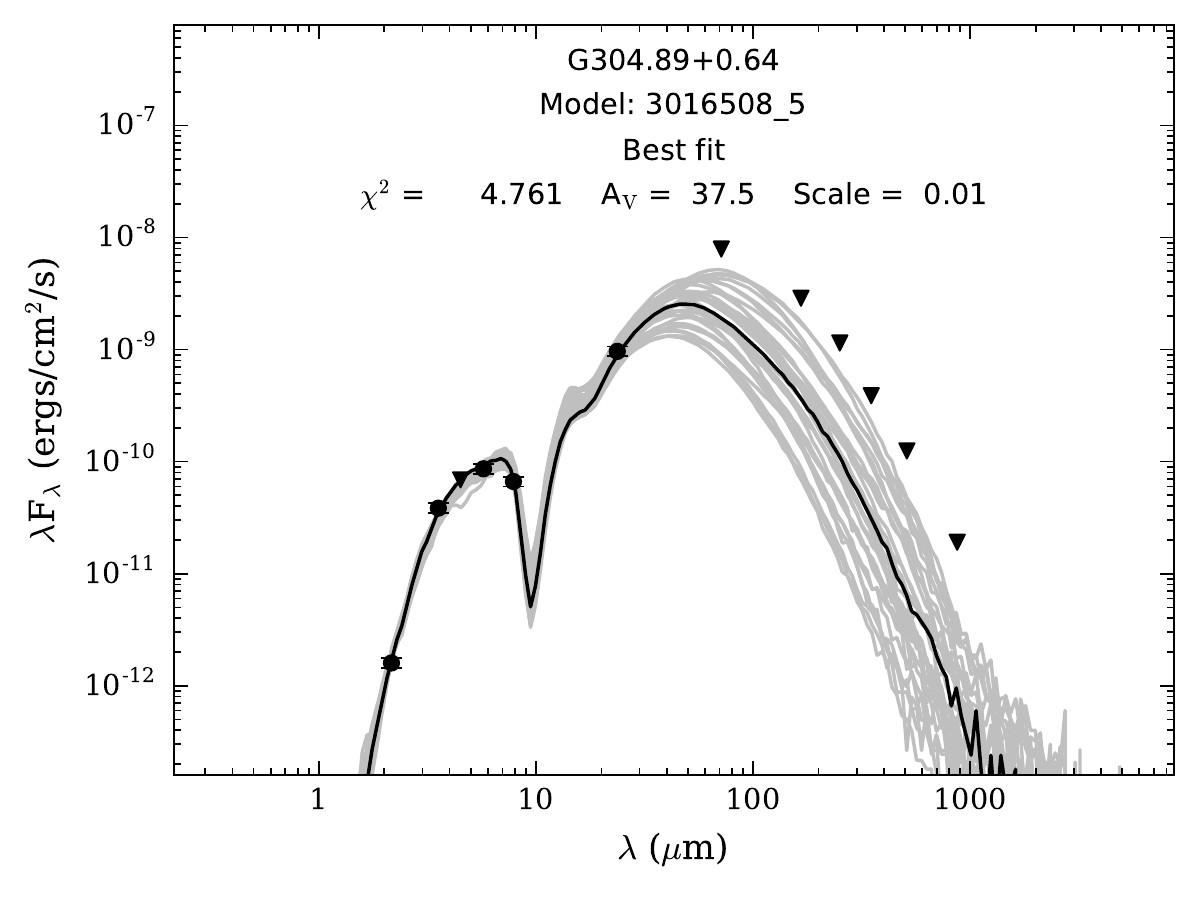}&
                \includegraphics[width=0.31\textwidth]{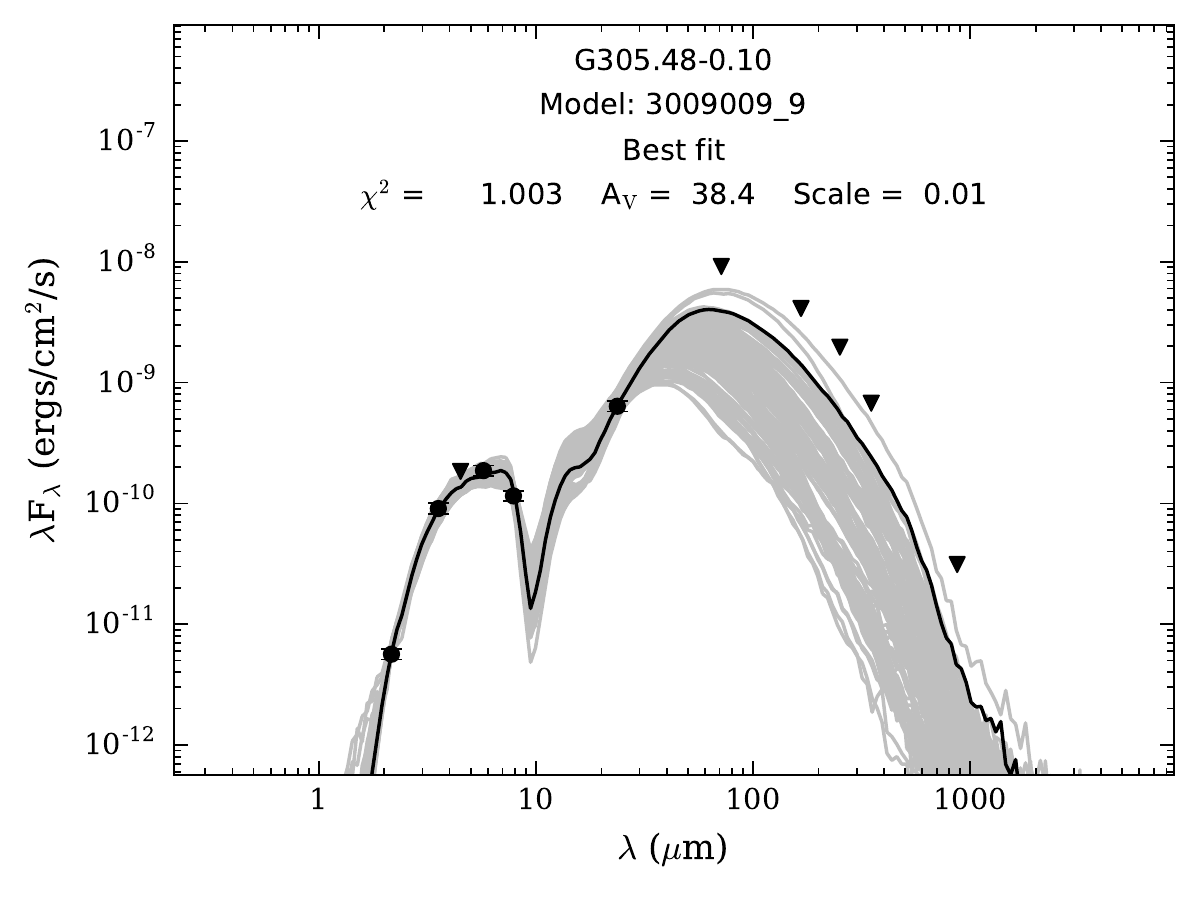}\\
                \includegraphics[width=0.31\textwidth]{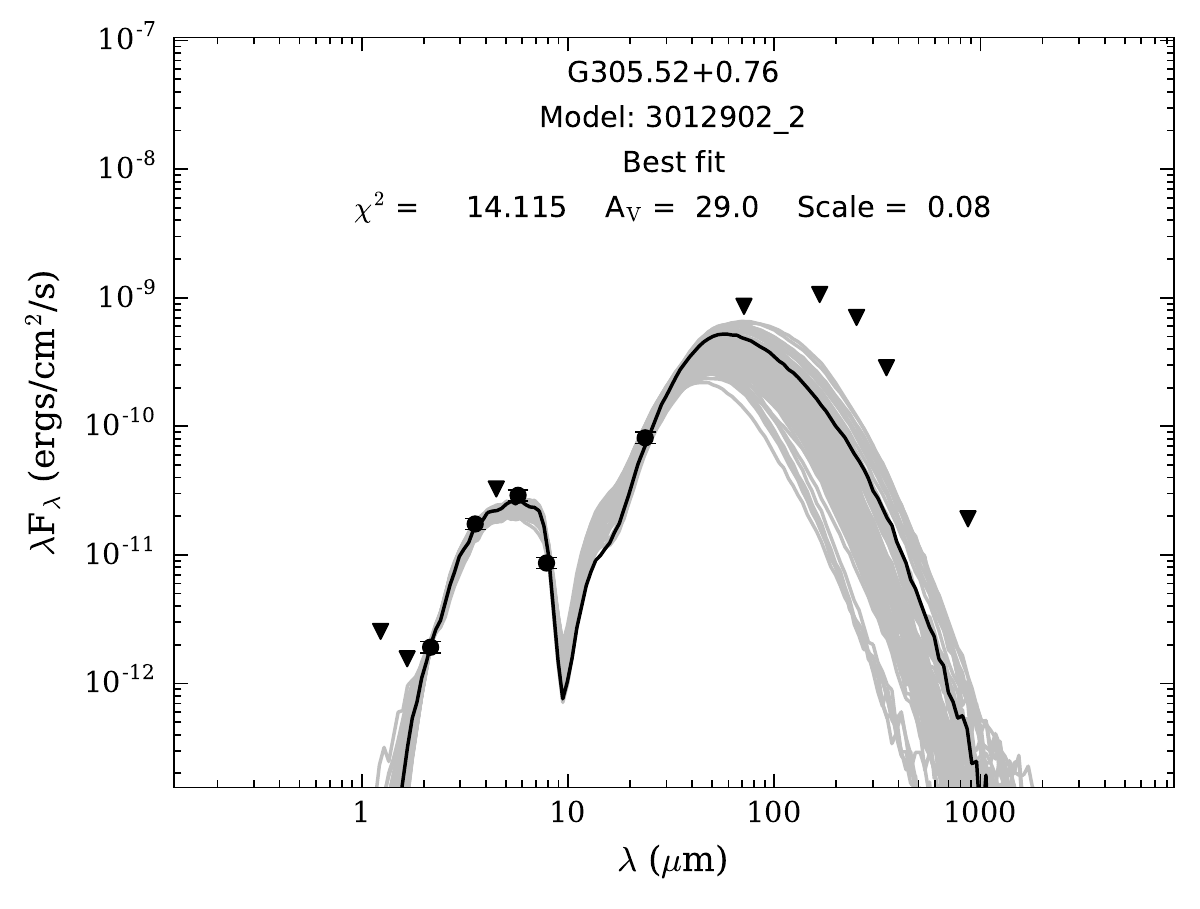}&
                \includegraphics[width=0.31\textwidth]{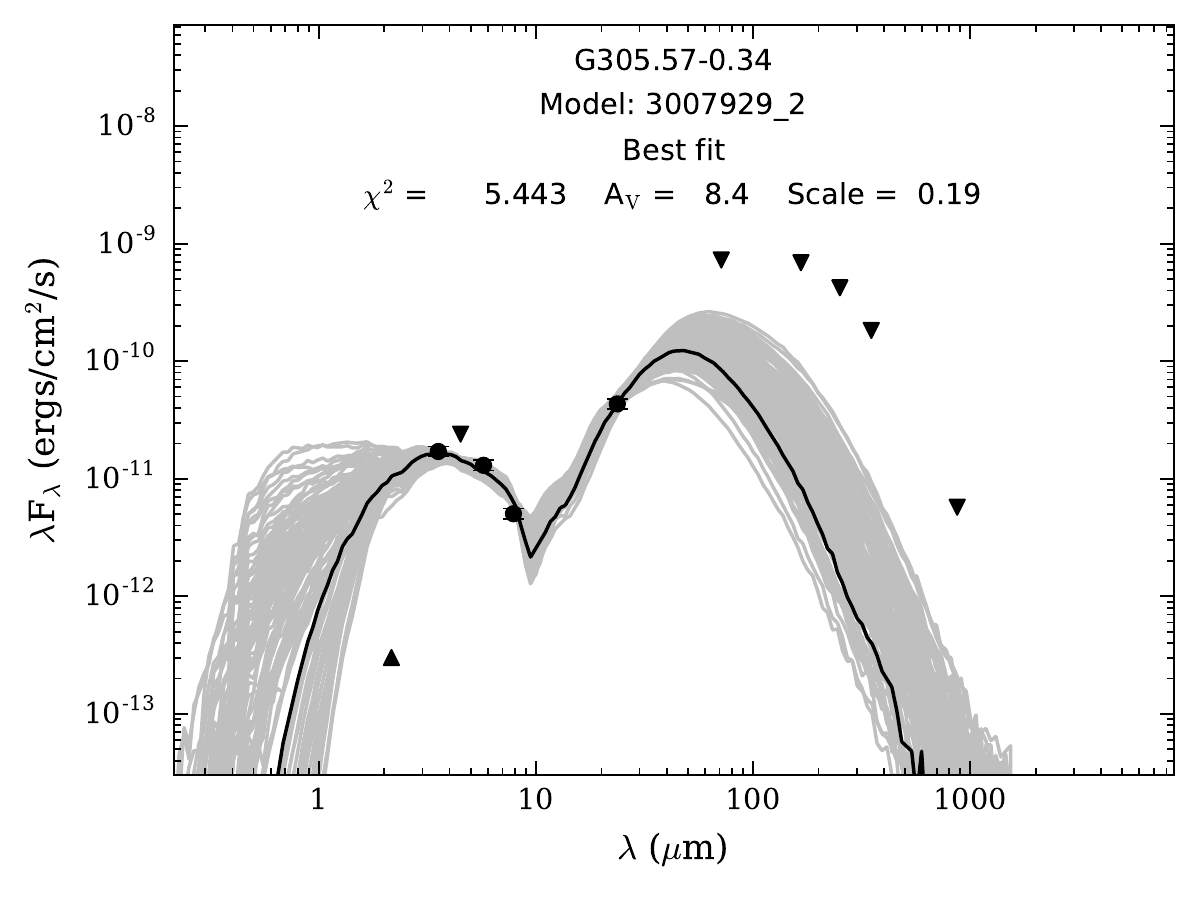}&
                \includegraphics[width=0.31\textwidth]{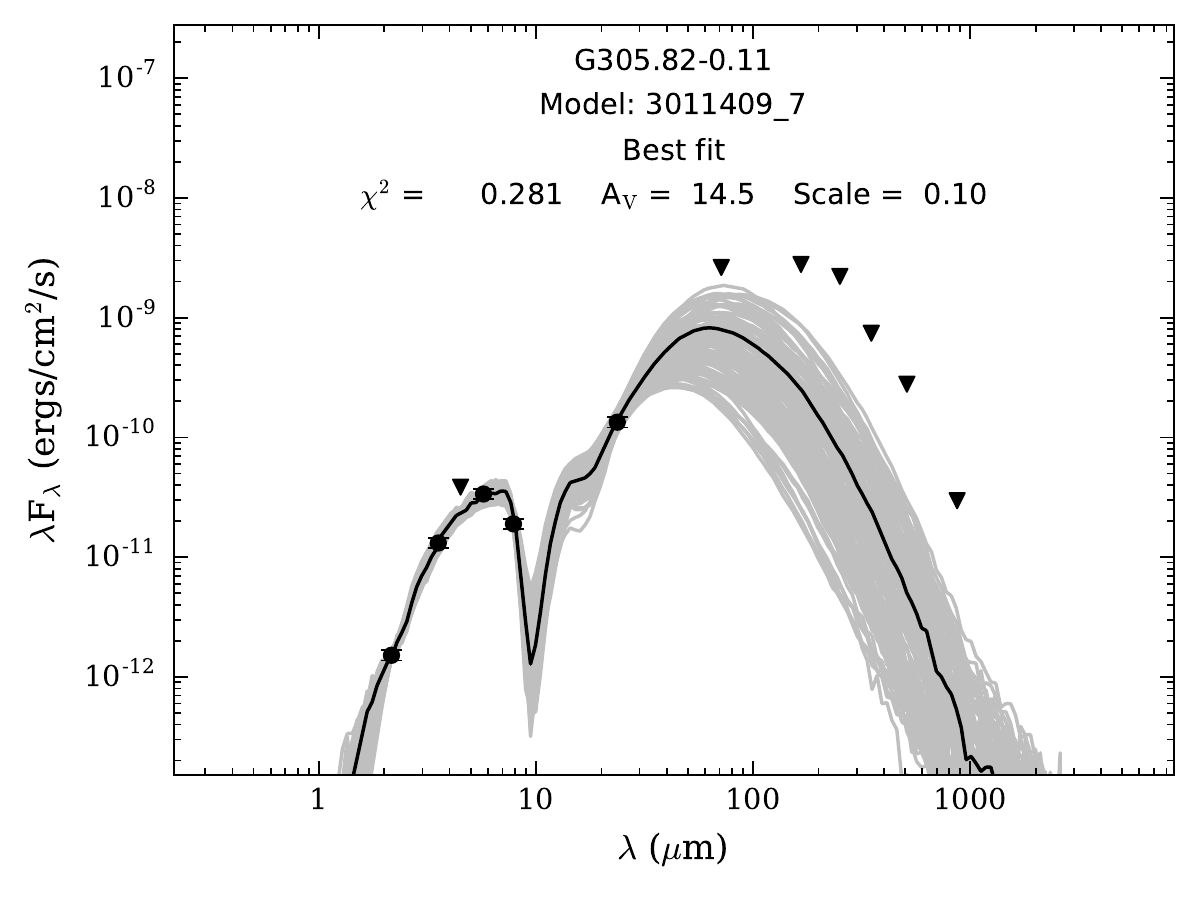}\\
                \includegraphics[width=0.31\textwidth]{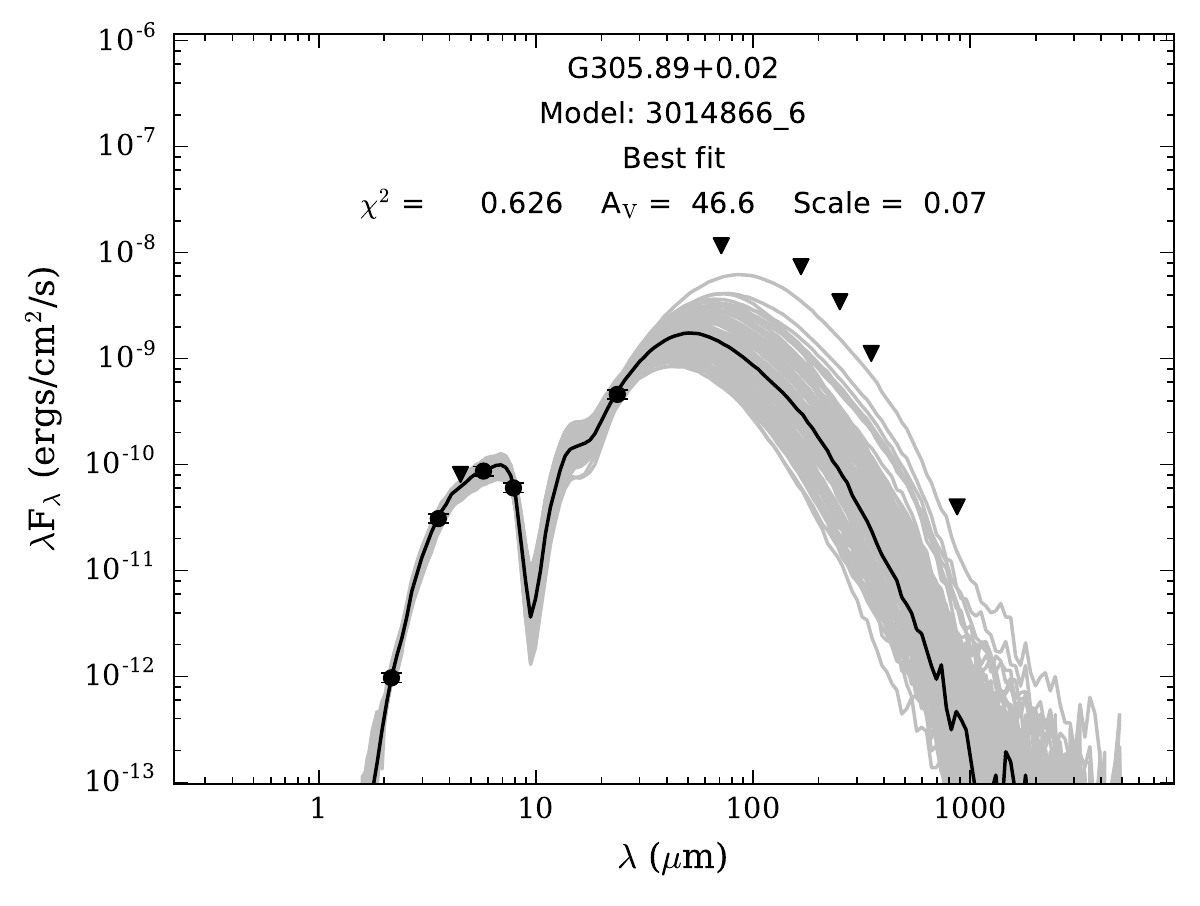}&
                \includegraphics[width=0.31\textwidth]{{./G309.91+0.32}.pdf}&
                \includegraphics[width=0.31\textwidth]{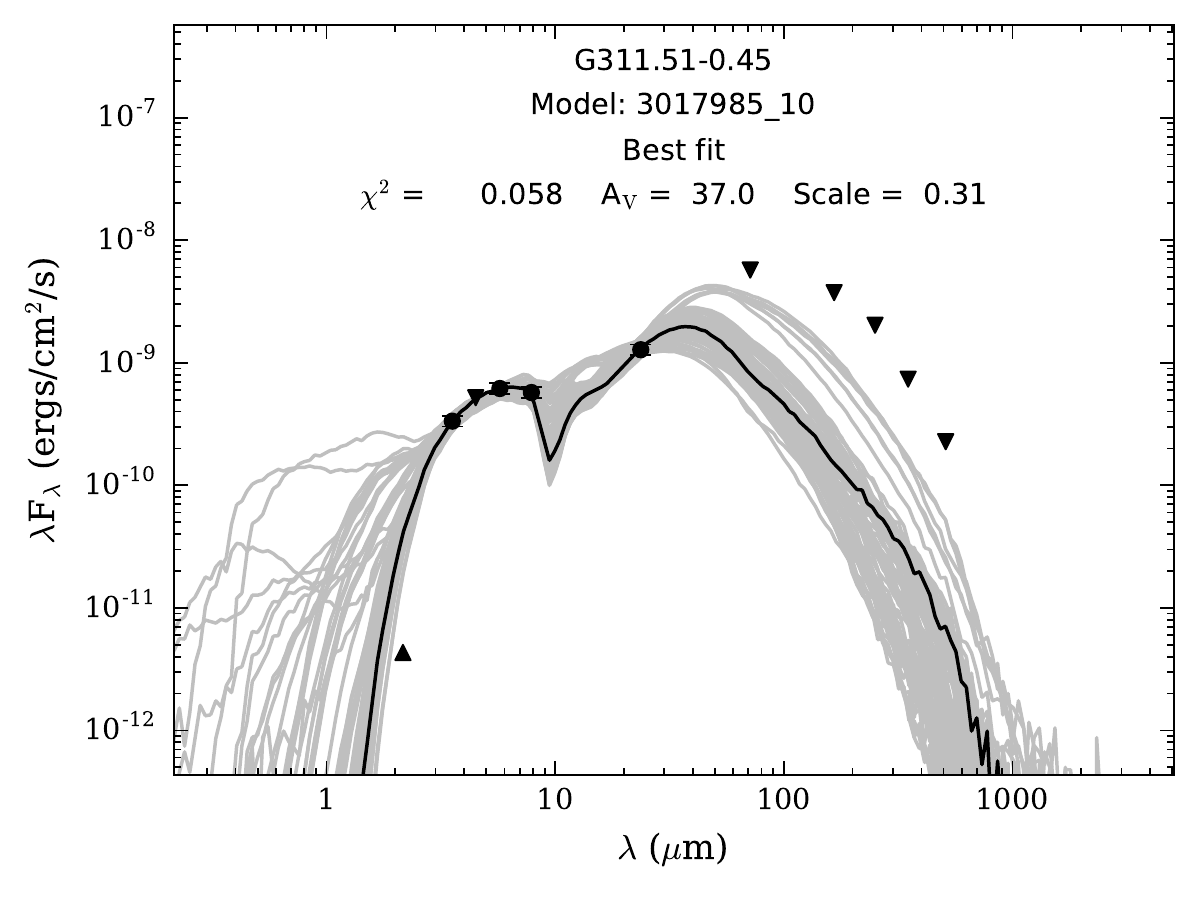}\\
                \includegraphics[width=0.31\textwidth]{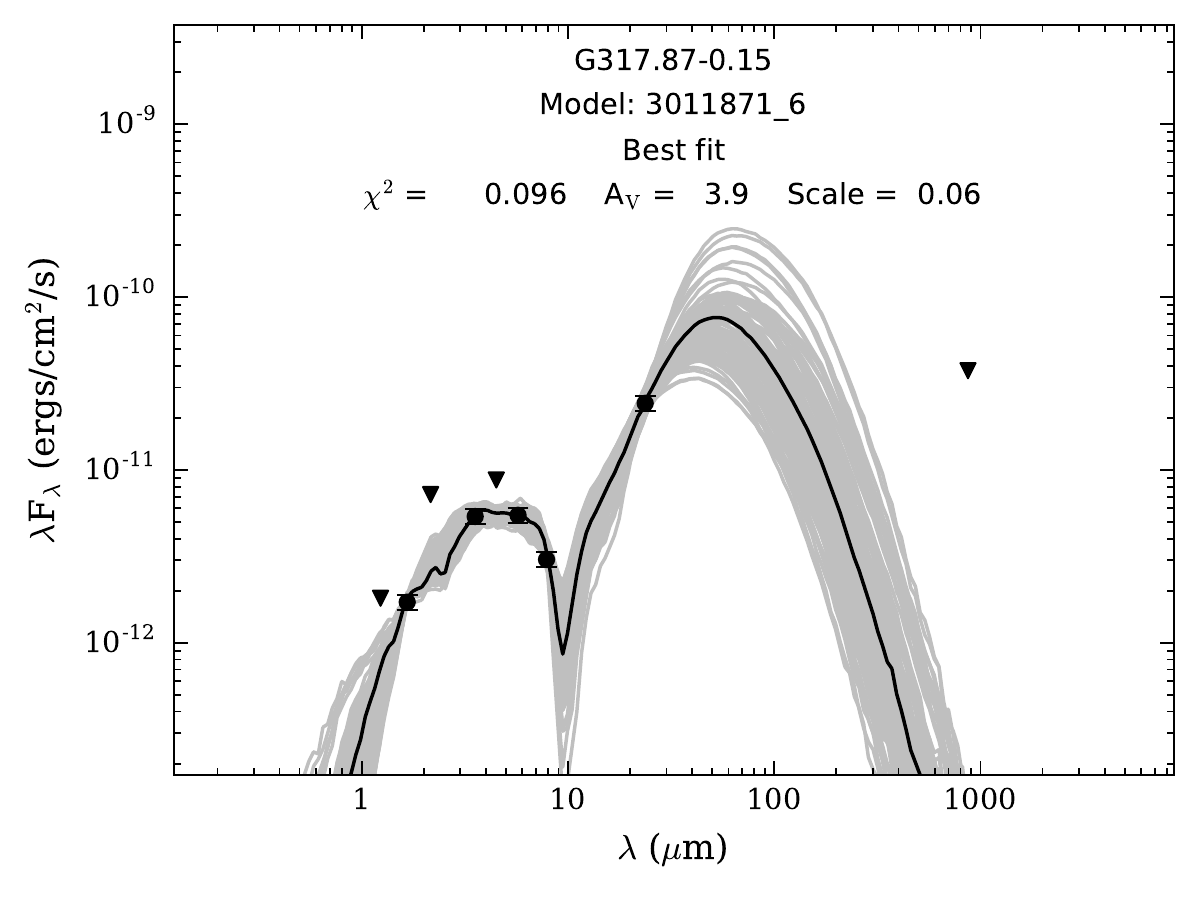}&
                \includegraphics[width=0.31\textwidth]{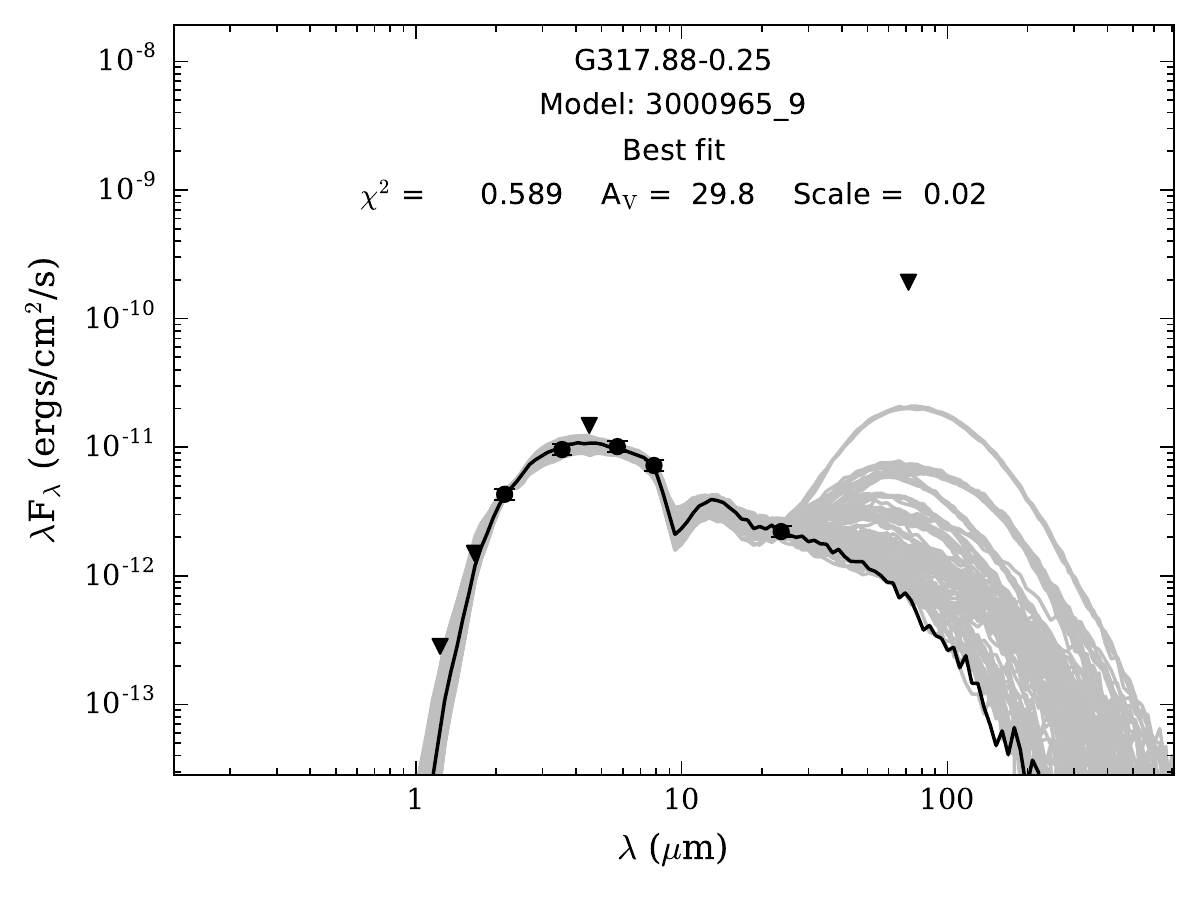}&
                \includegraphics[width=0.31\textwidth]{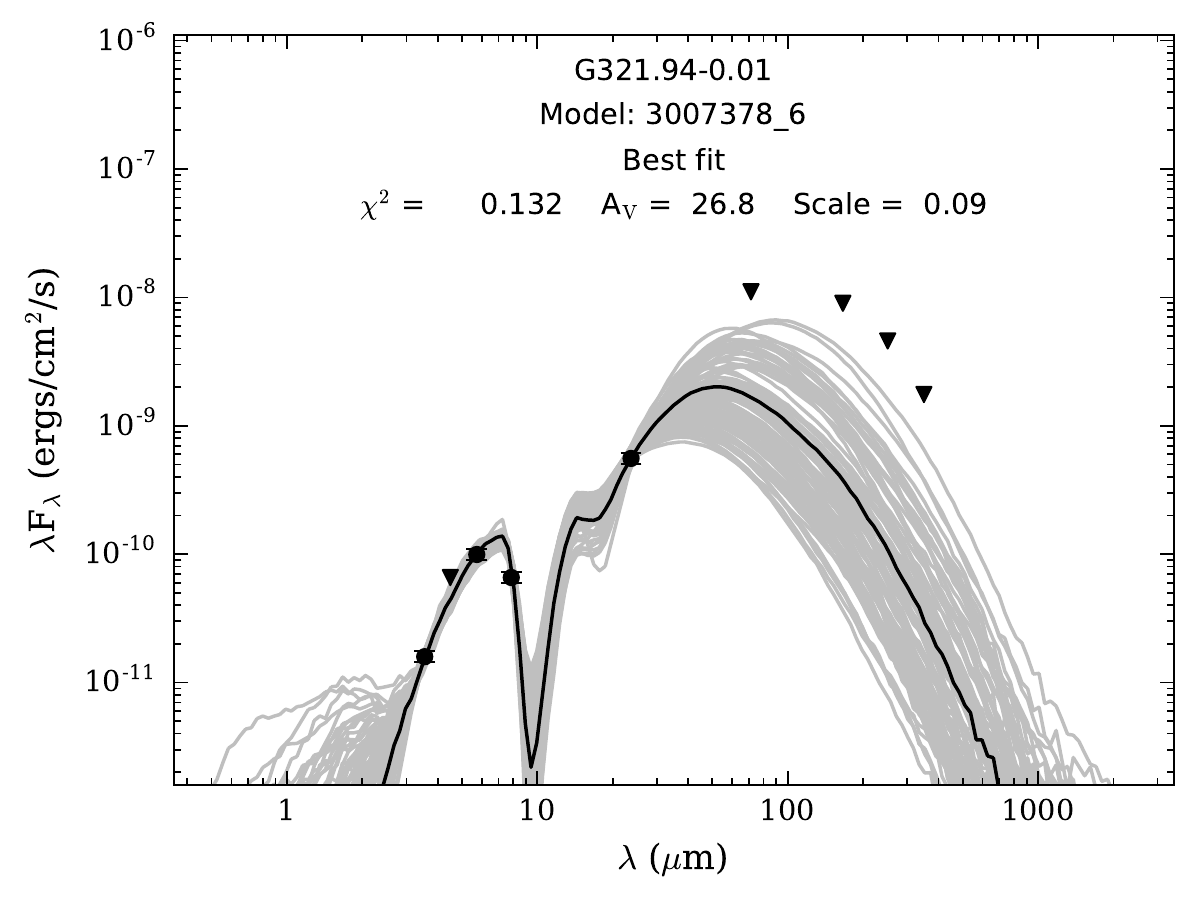}\\
        \end{tabular}
        \caption{The fitted SEDs with the best fit model in the black line, the grey lines are other $\chi^2 - \chi_{best}^2 <3$ models, data points, upper, and lower limits are, respectively, circles, inverted triangles, and triangles.}
        \label{fig:testgridsed}
\end{figure}

\begin{figure}
        \begin{tabular}{ccc}
                
                \includegraphics[width=0.31\textwidth]{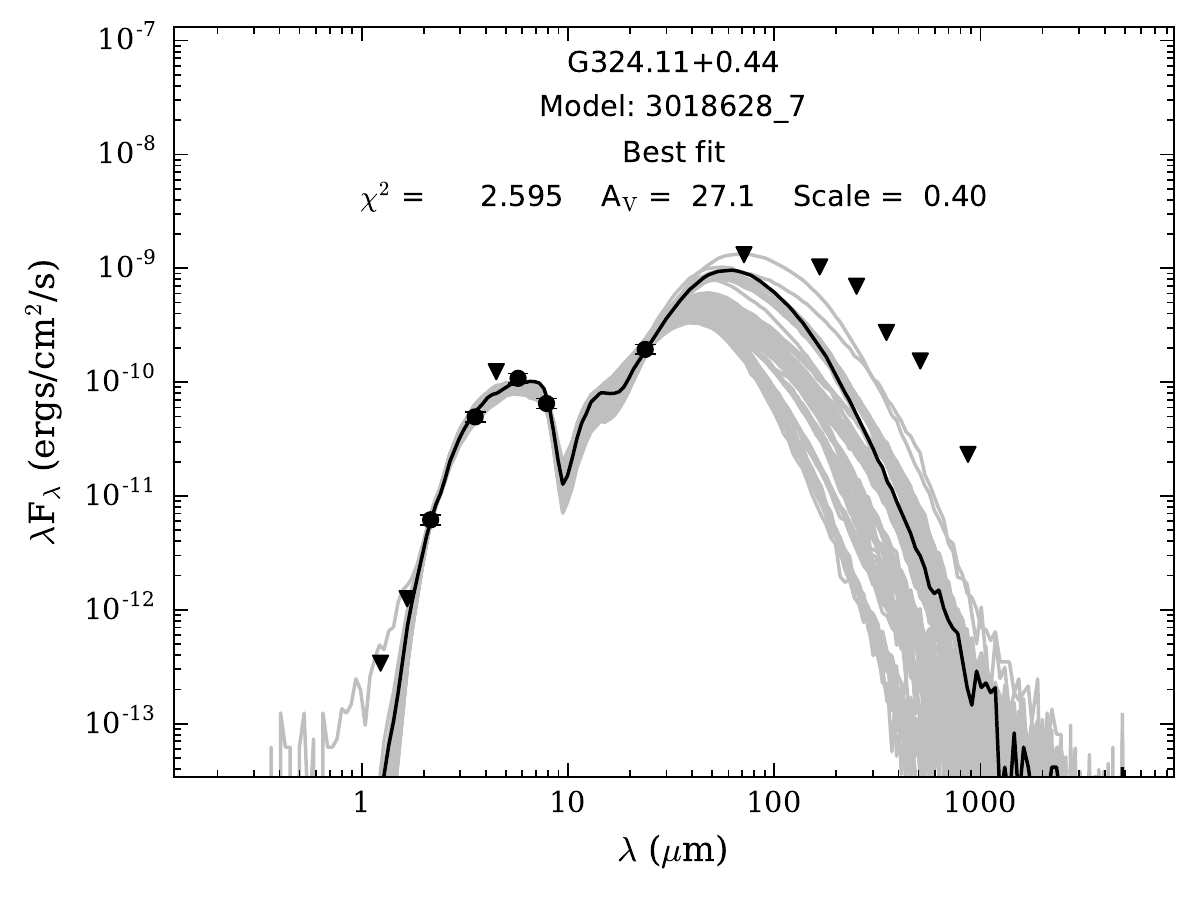}&
                \includegraphics[width=0.31\textwidth]{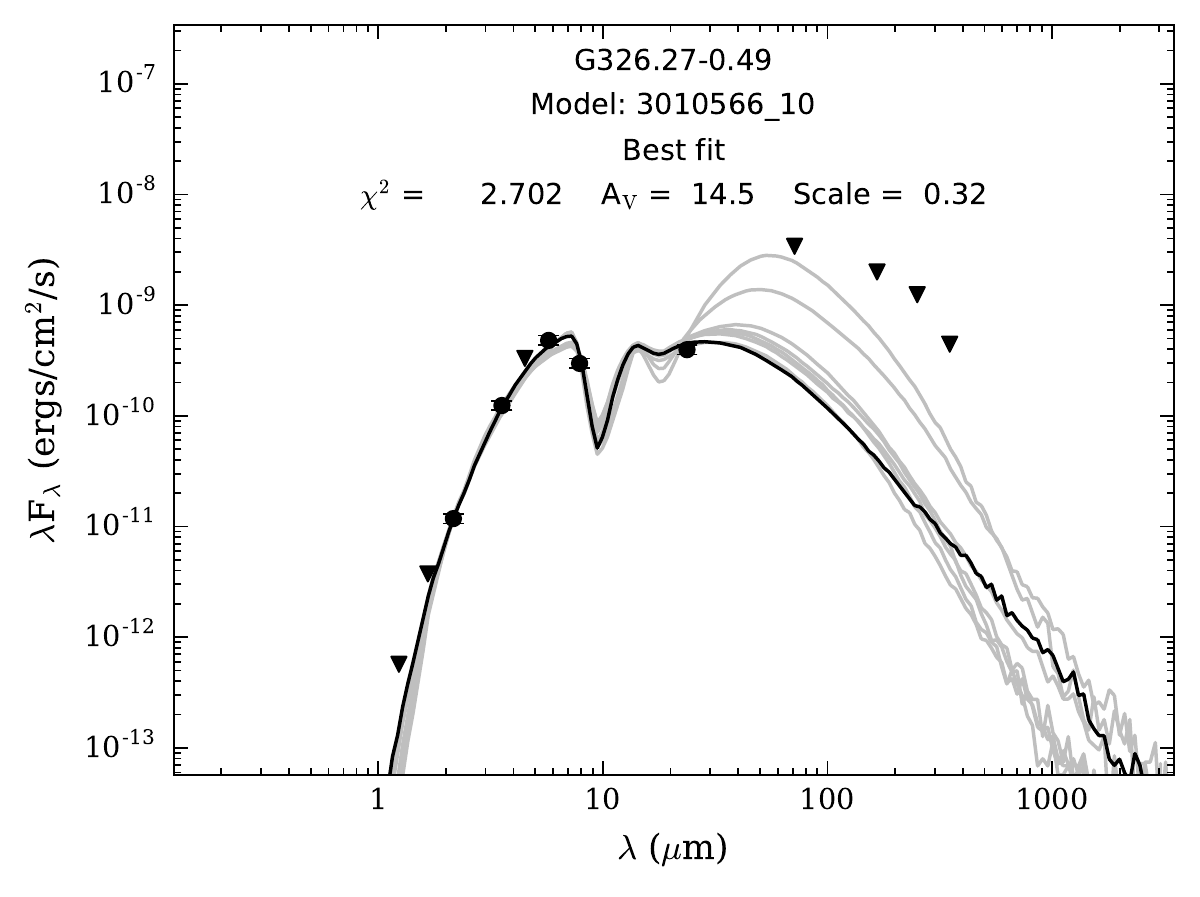}&
                \includegraphics[width=0.31\textwidth]{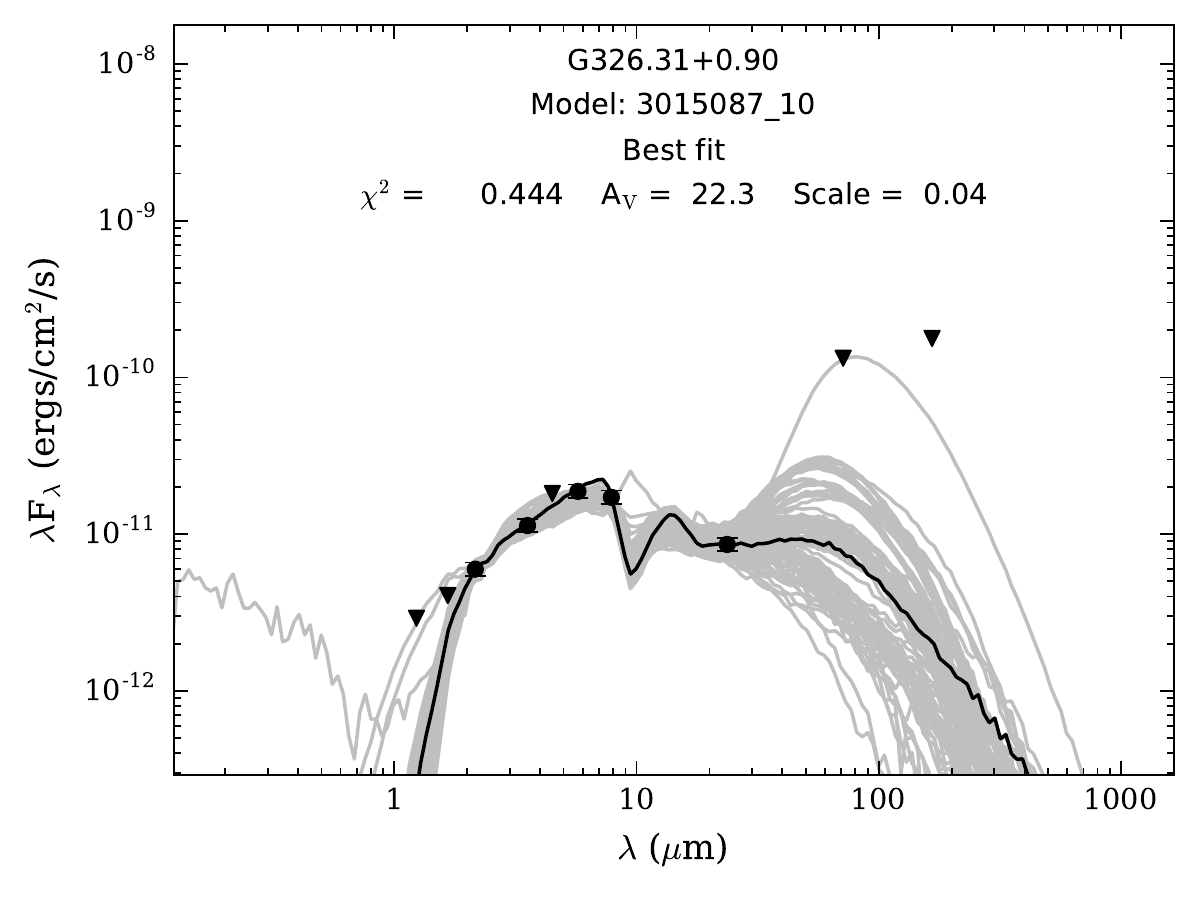}\\
                \includegraphics[width=0.31\textwidth]{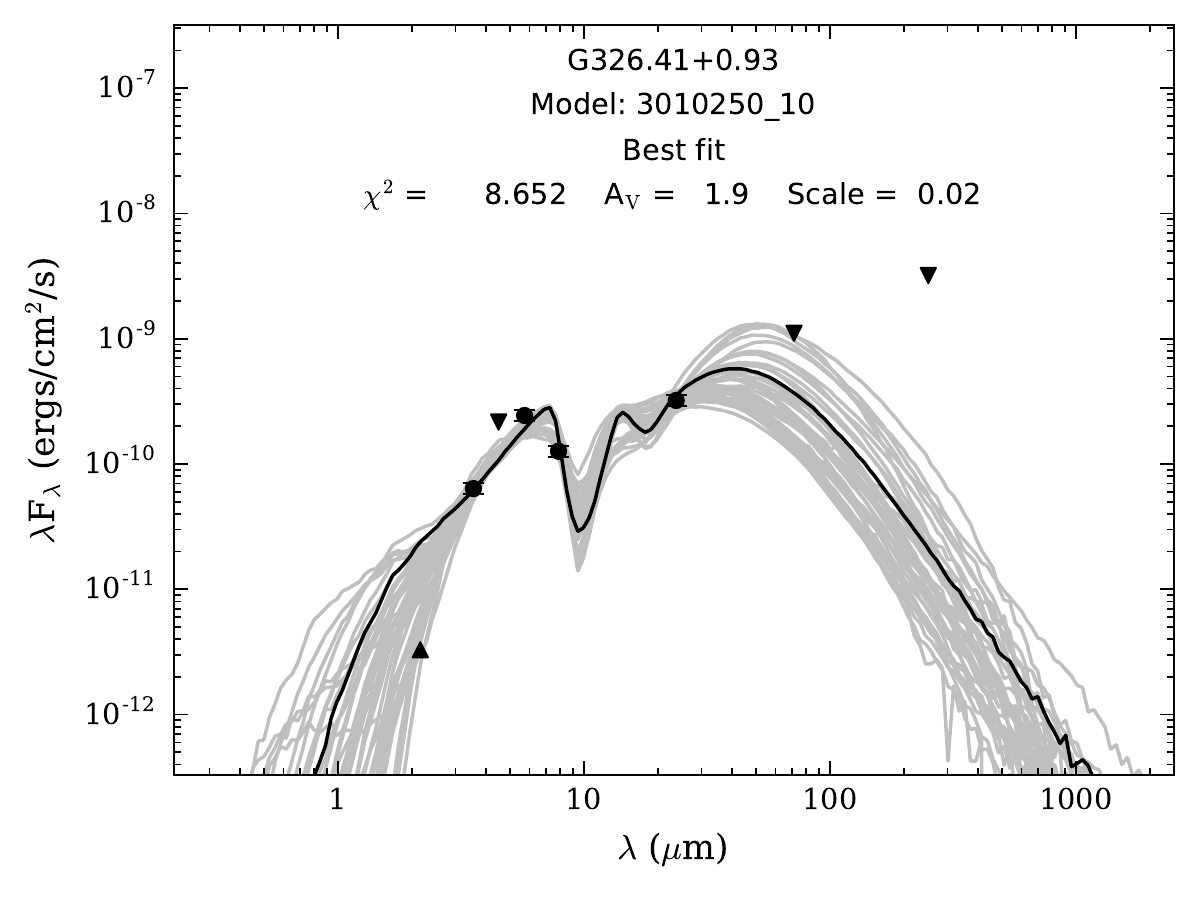}&
                \includegraphics[width=0.31\textwidth]{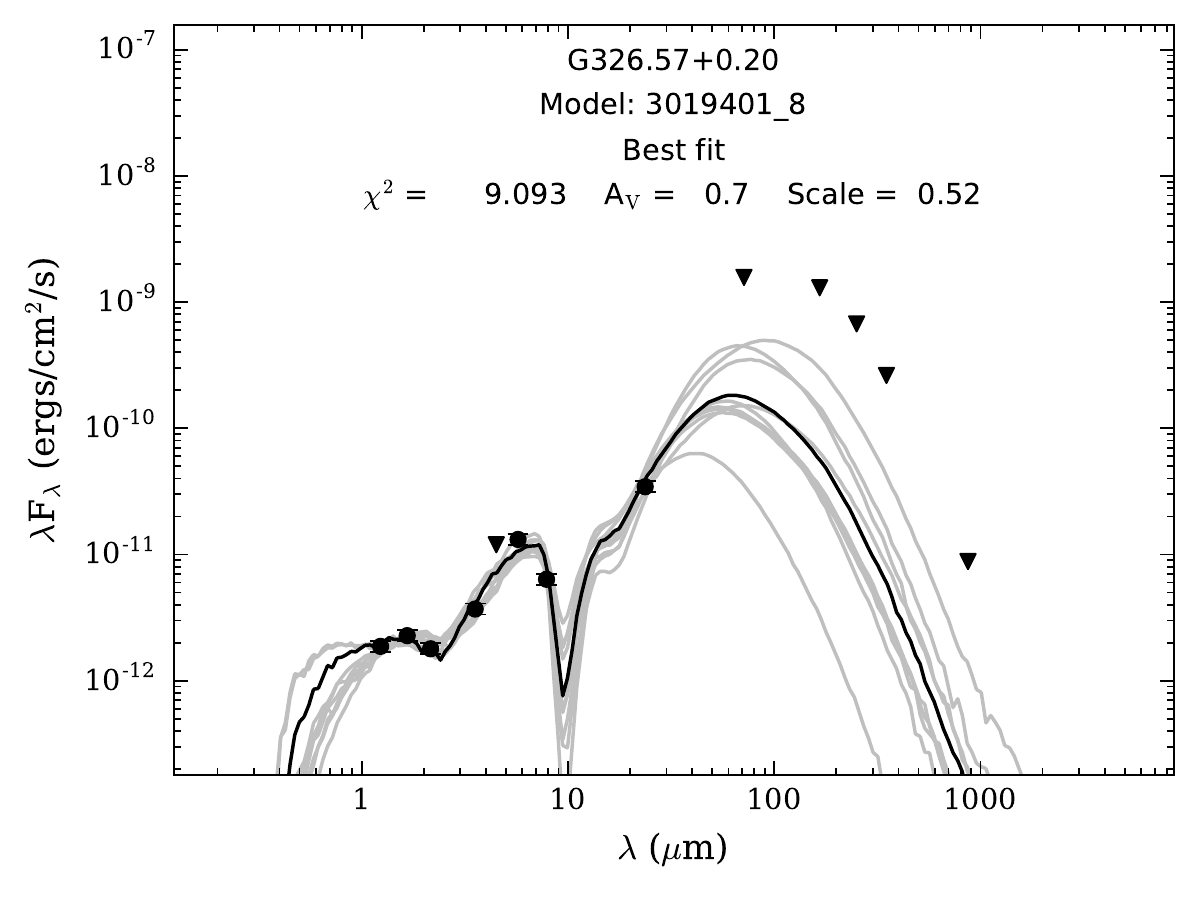}&
                \includegraphics[width=0.31\textwidth]{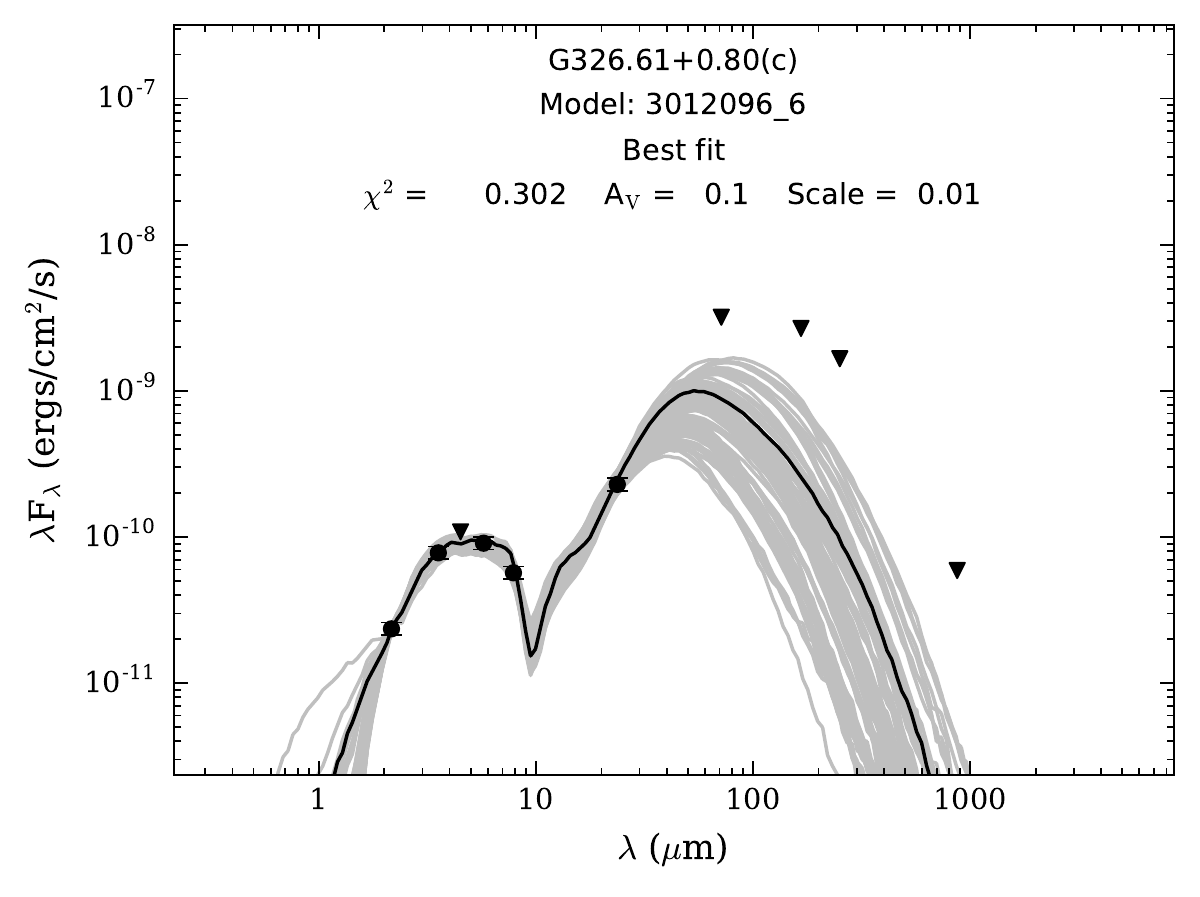}\\
                \includegraphics[width=0.31\textwidth]{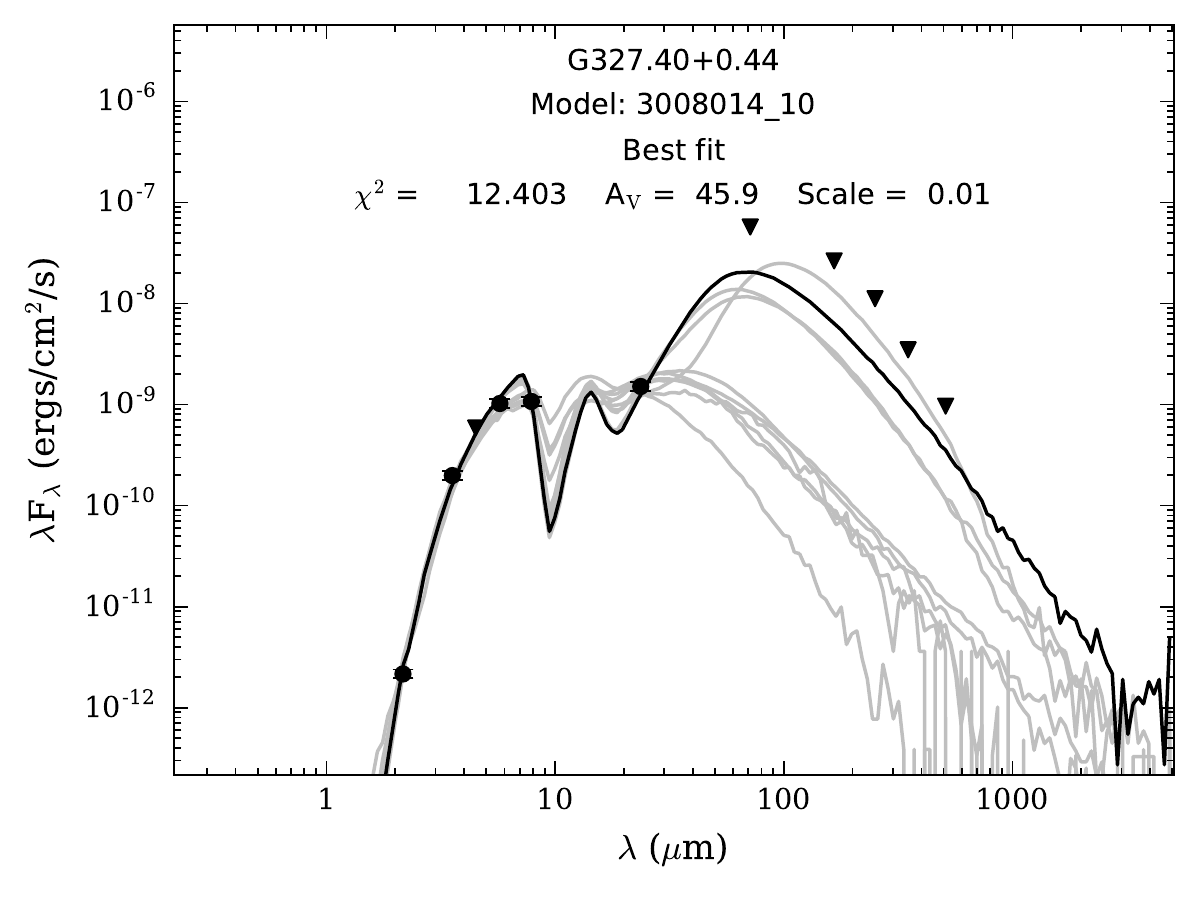}&
                \includegraphics[width=0.31\textwidth]{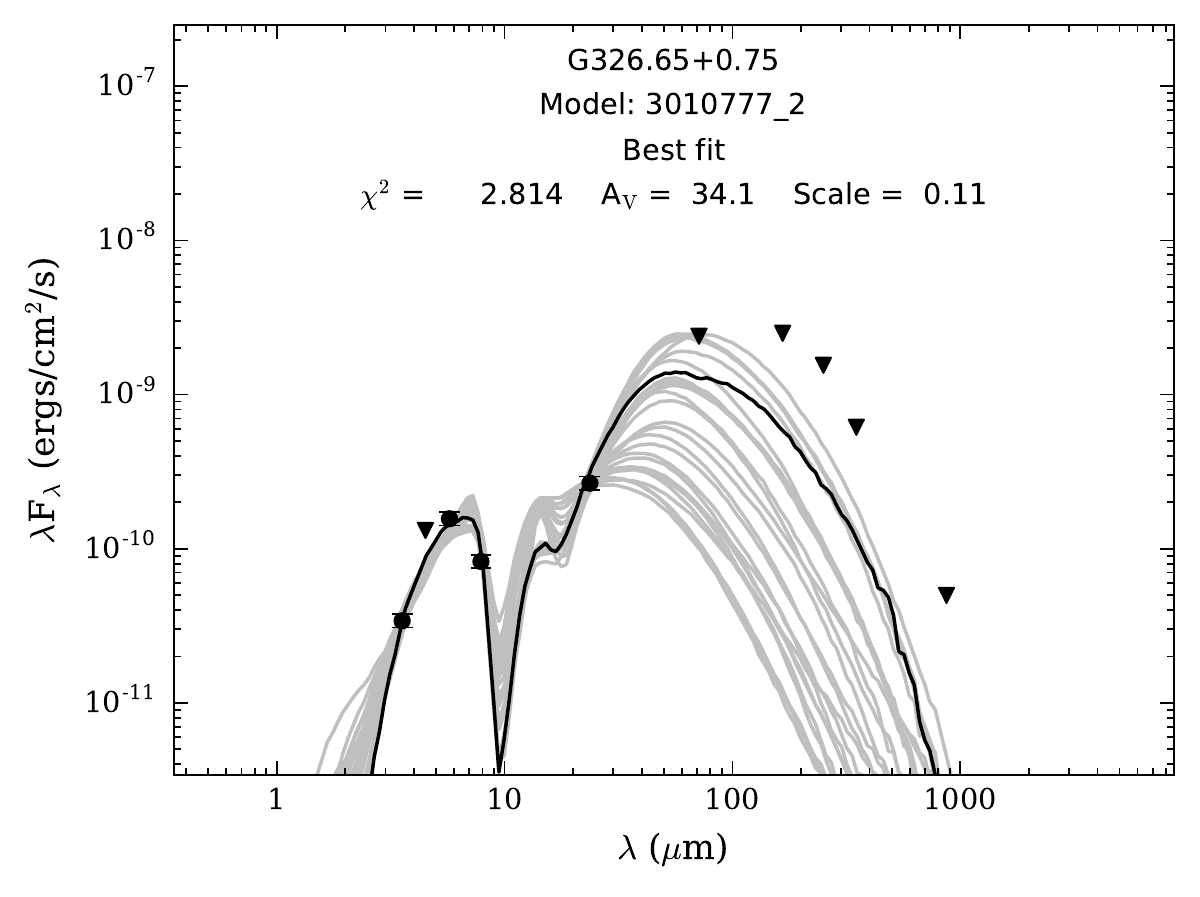}&
                \includegraphics[width=0.31\textwidth]{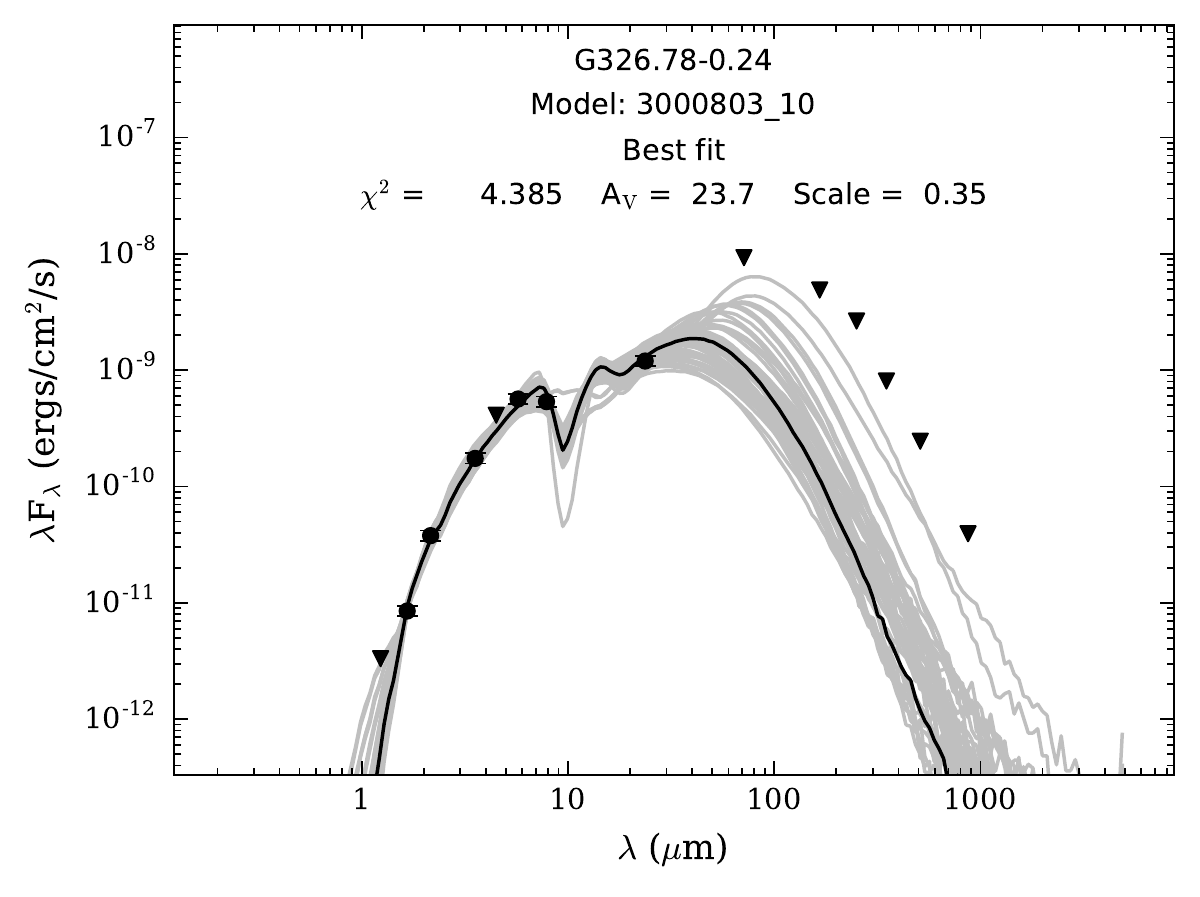}\\
                \includegraphics[width=0.31\textwidth]{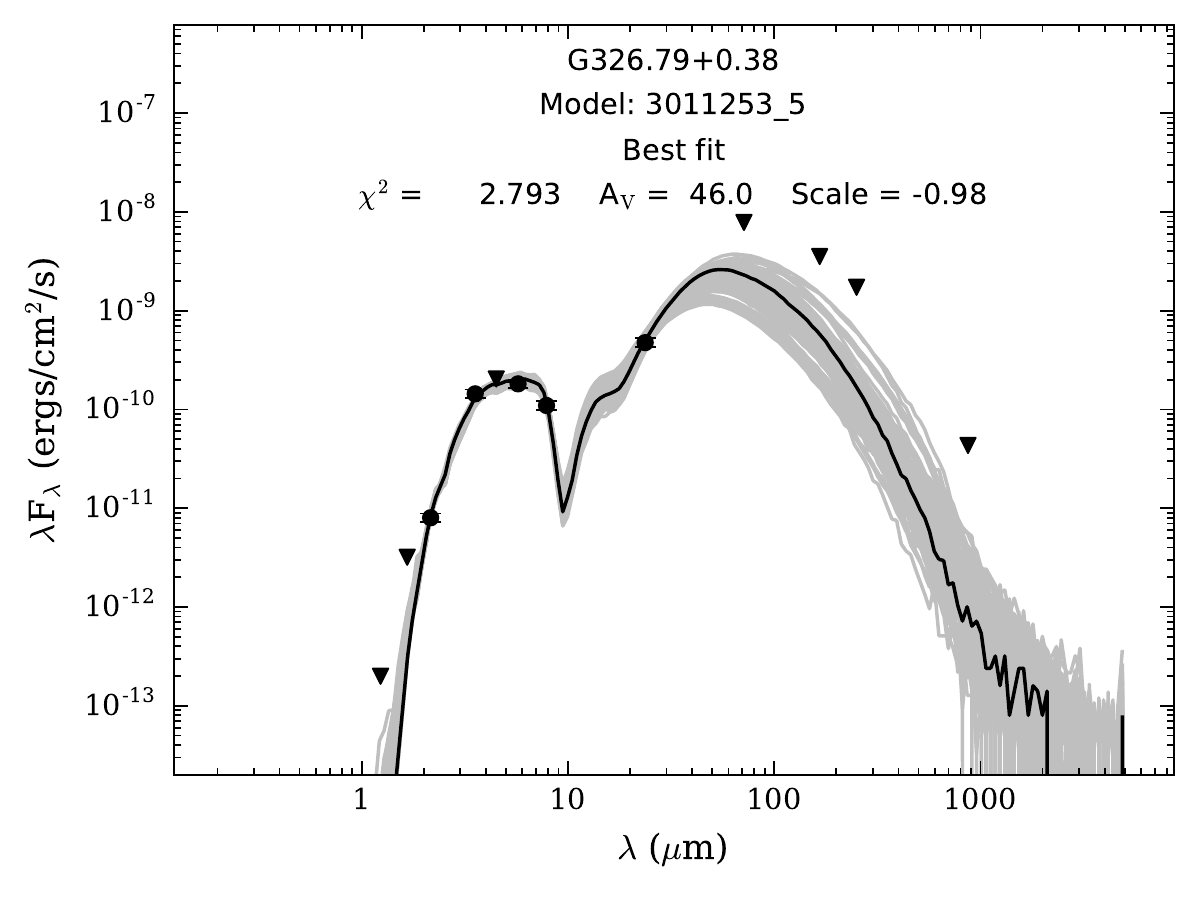}&
                \includegraphics[width=0.31\textwidth]{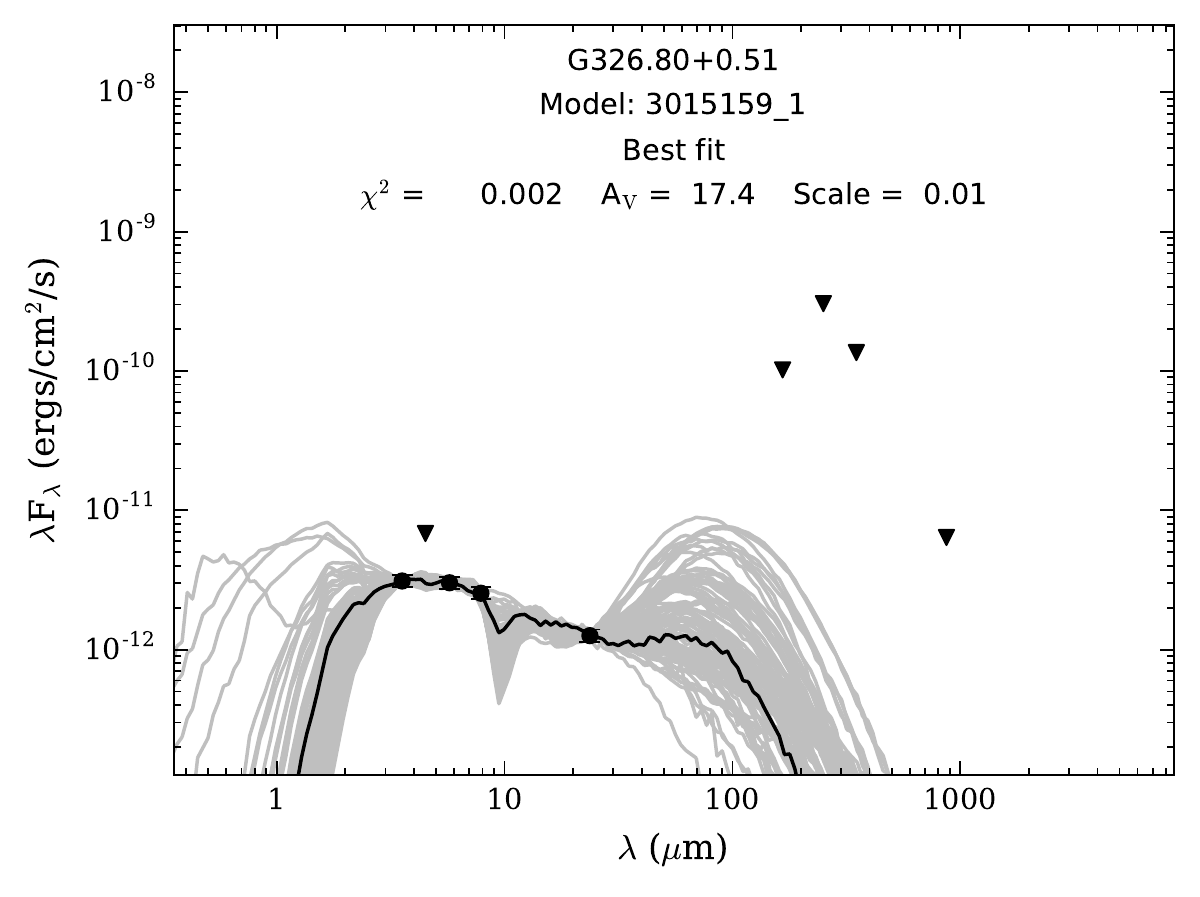}&
                \includegraphics[width=0.31\textwidth]{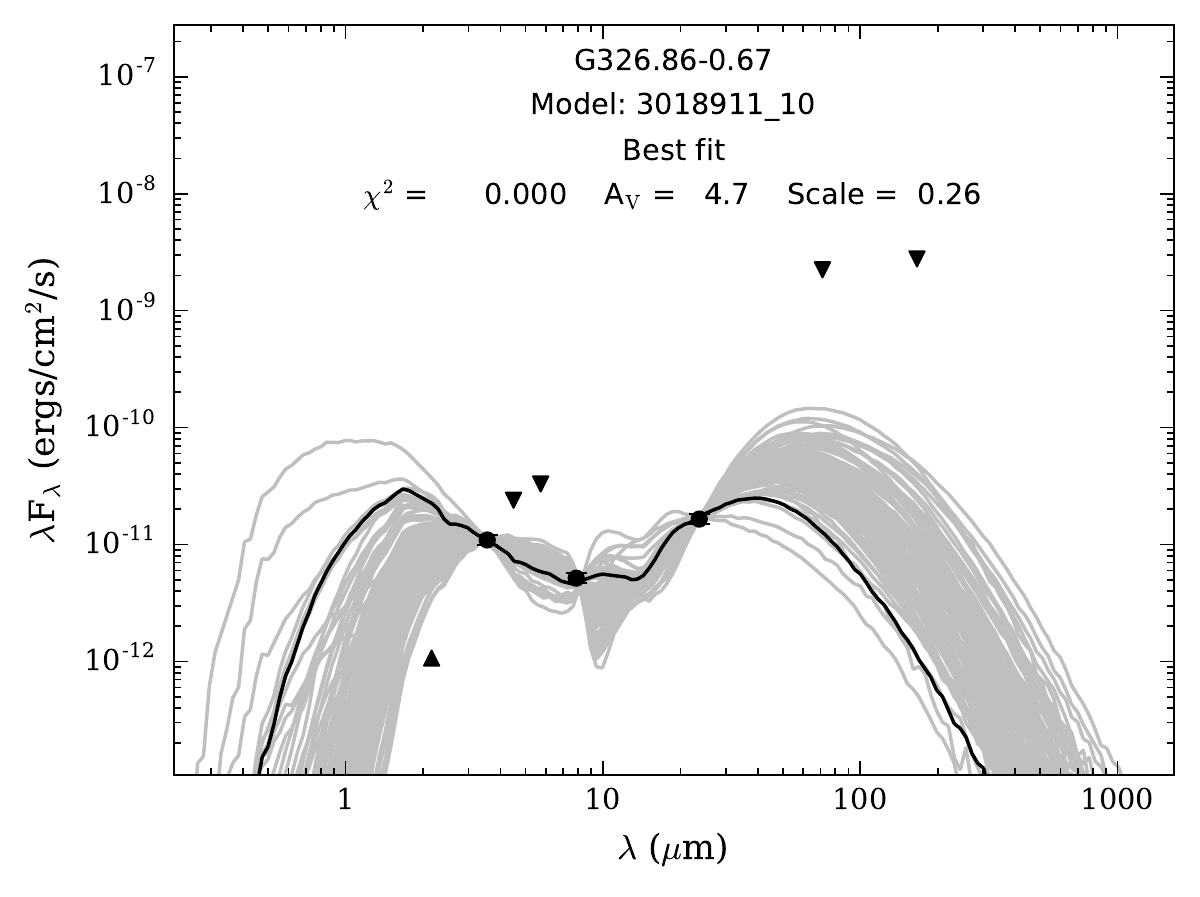}\\
                \includegraphics[width=0.31\textwidth]{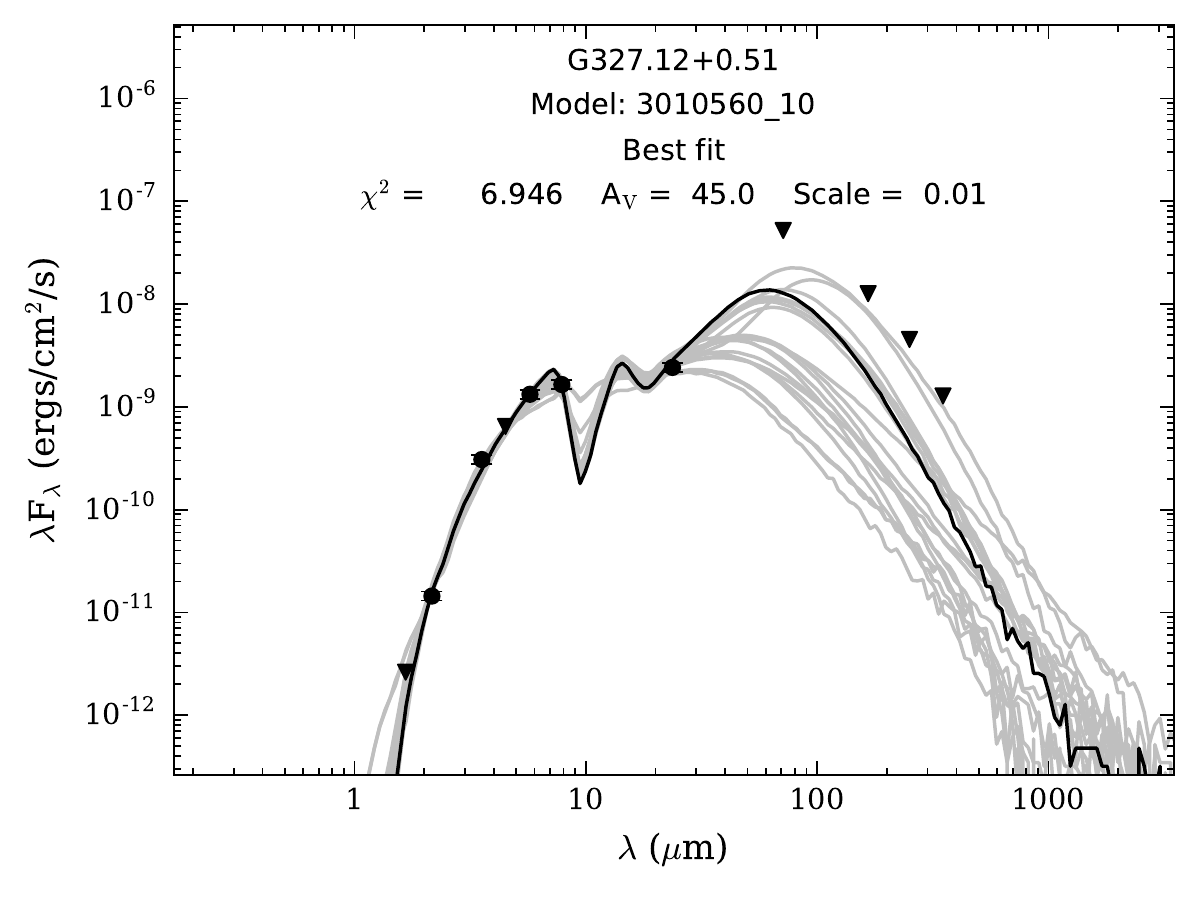}&
                \includegraphics[width=0.31\textwidth]{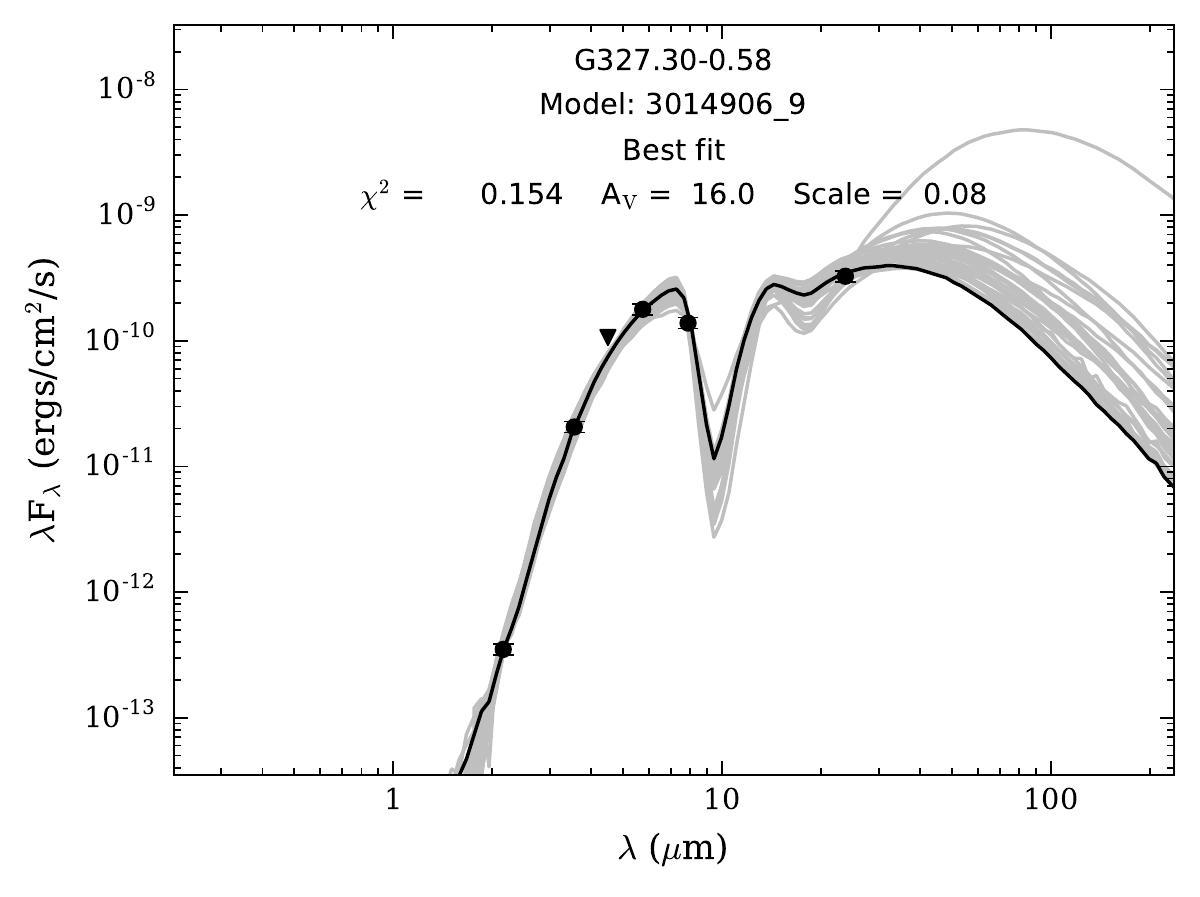}&
                \includegraphics[width=0.31\textwidth]{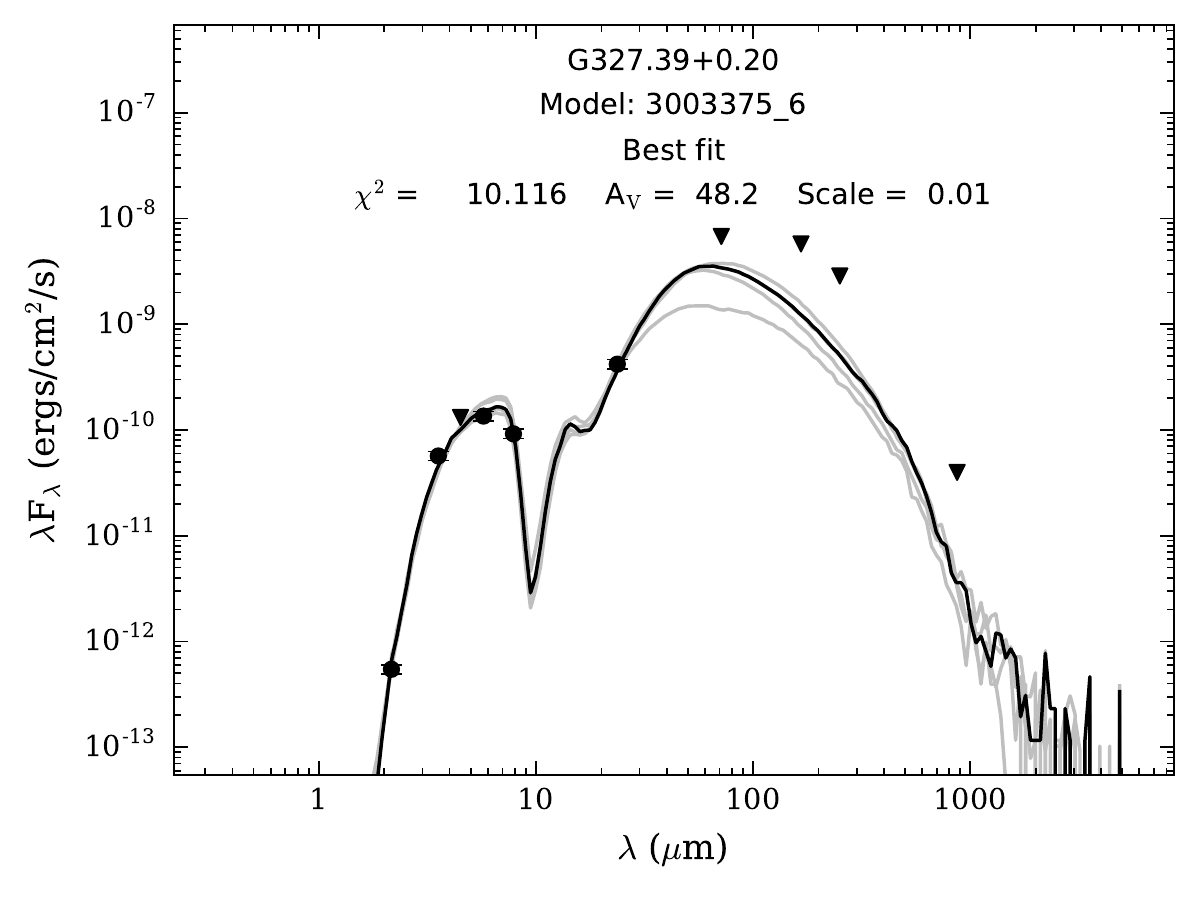}\\
        \end{tabular}
        \caption{Continuation of Fig. \ref{fig:testgridsed}.}
\end{figure}

\begin{figure}
        \begin{tabular}{ccc}
                
                \includegraphics[width=0.31\textwidth]{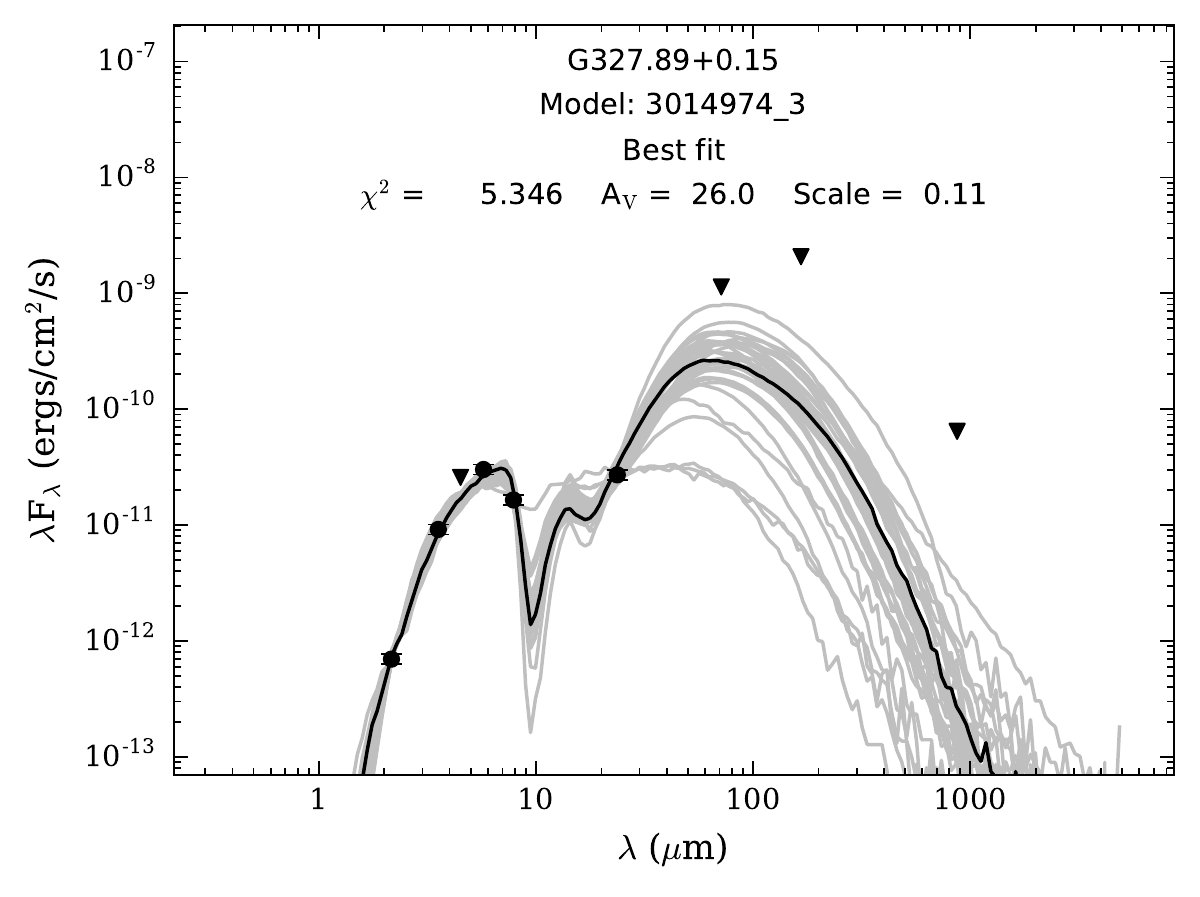}&
                \includegraphics[width=0.31\textwidth]{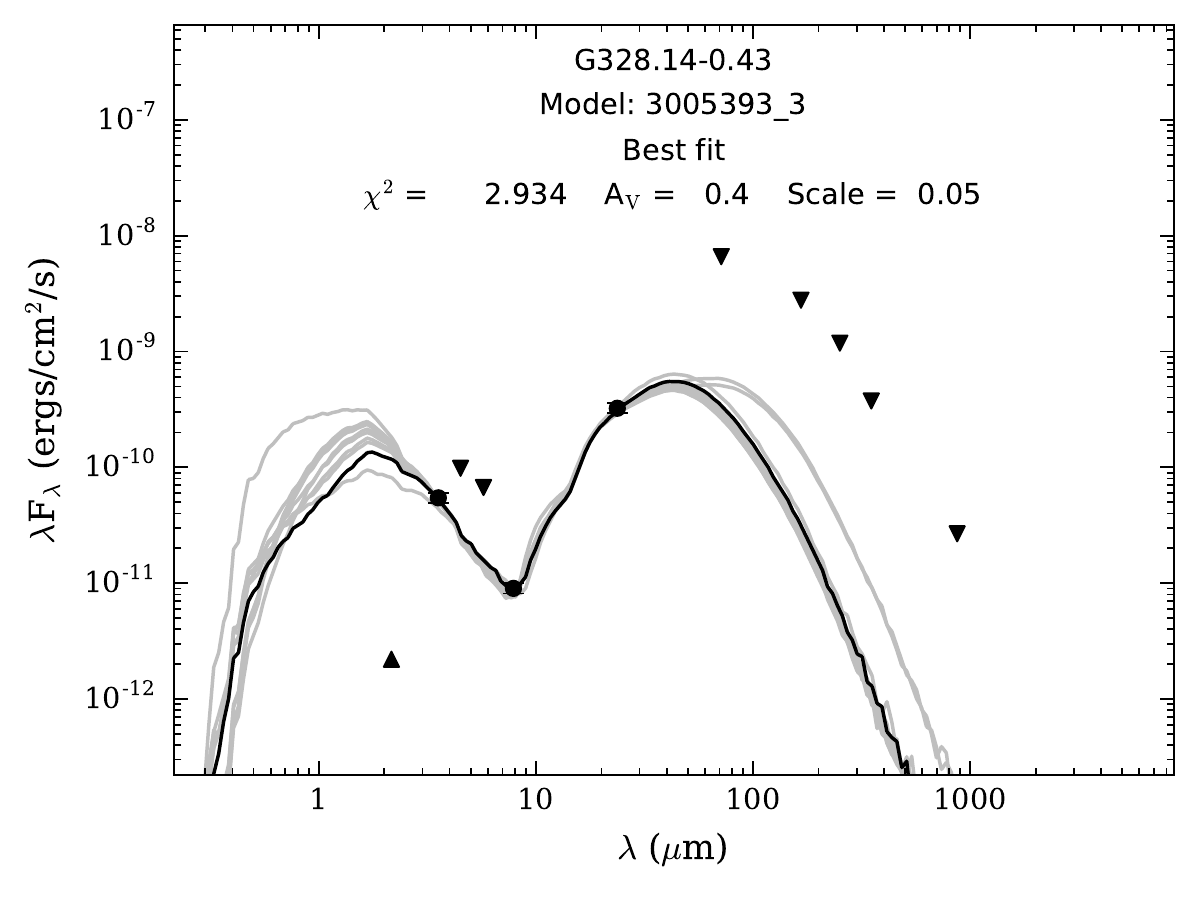}&
                \includegraphics[width=0.31\textwidth]{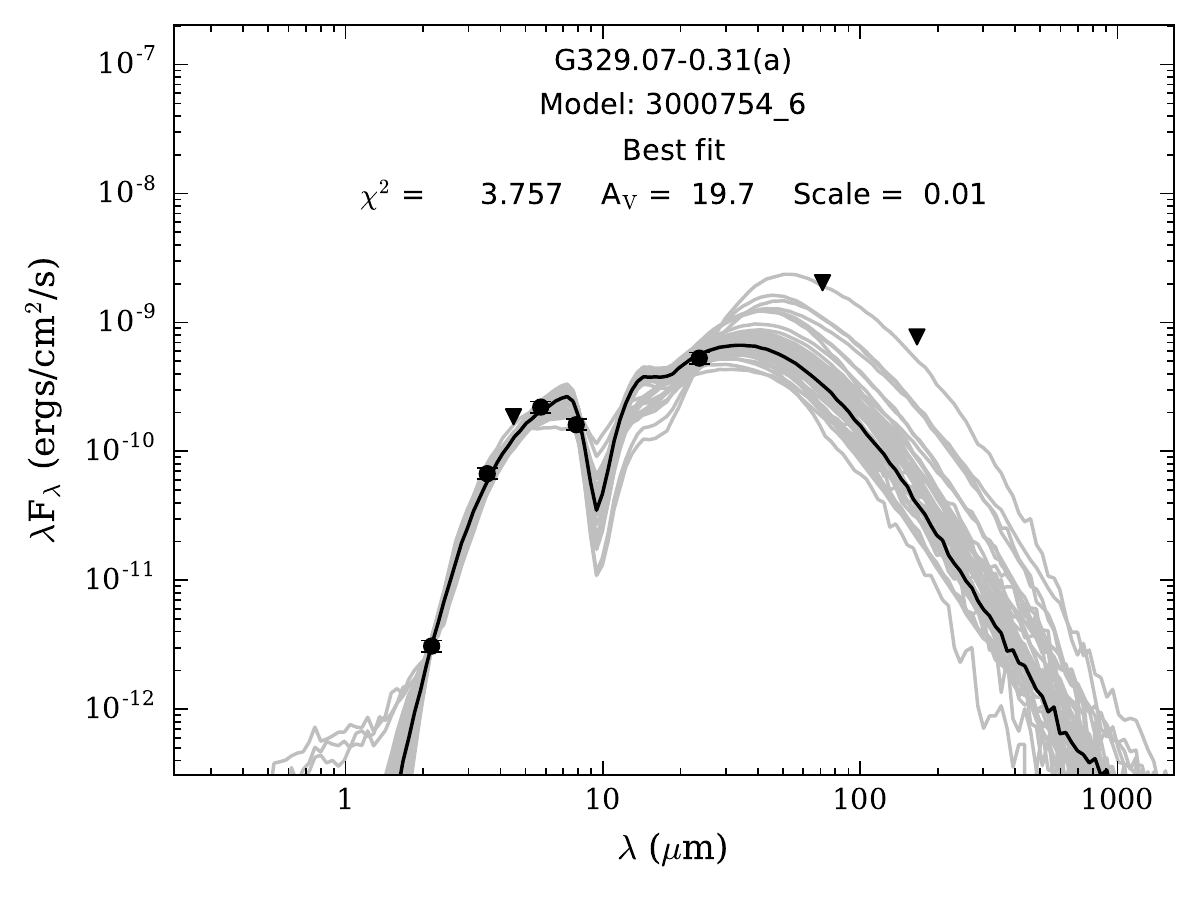}\\
                \includegraphics[width=0.31\textwidth]{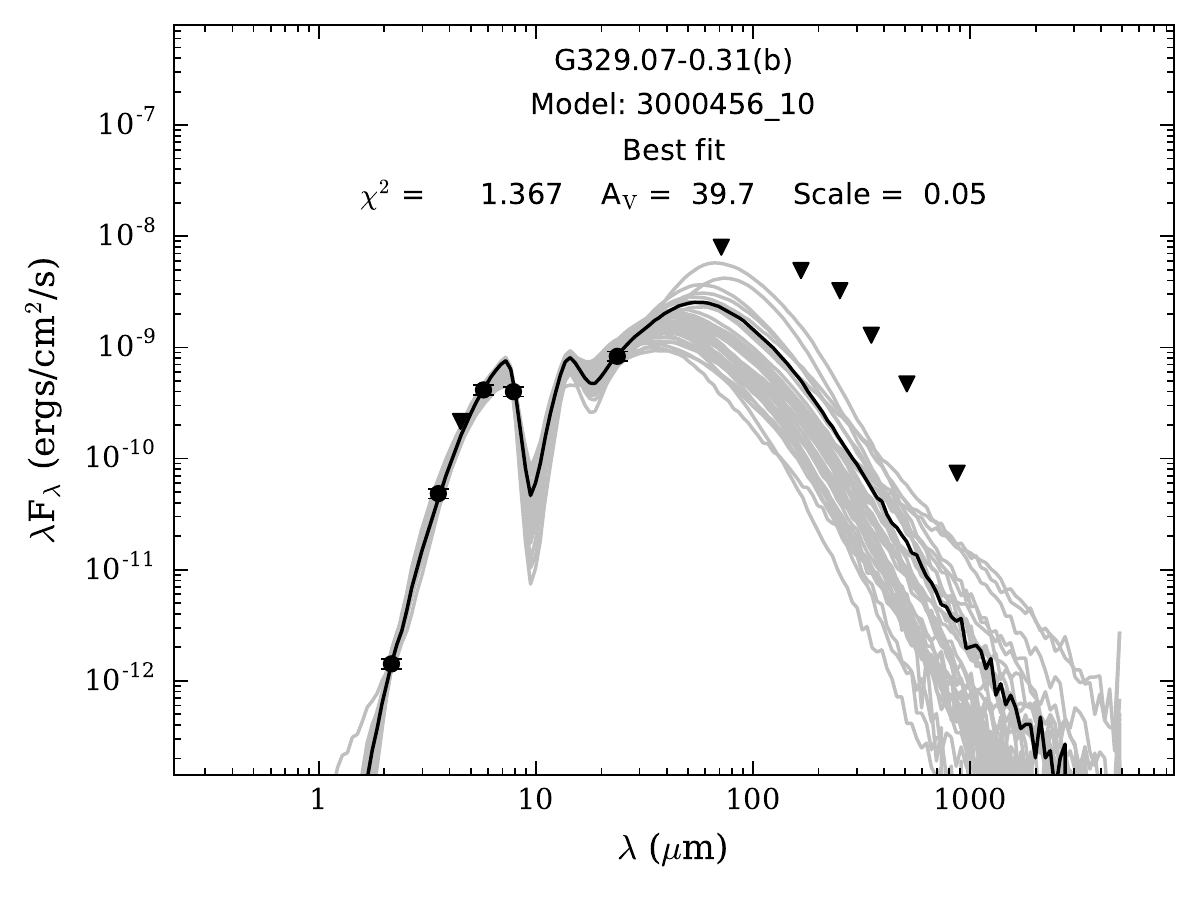}&
                \includegraphics[width=0.31\textwidth]{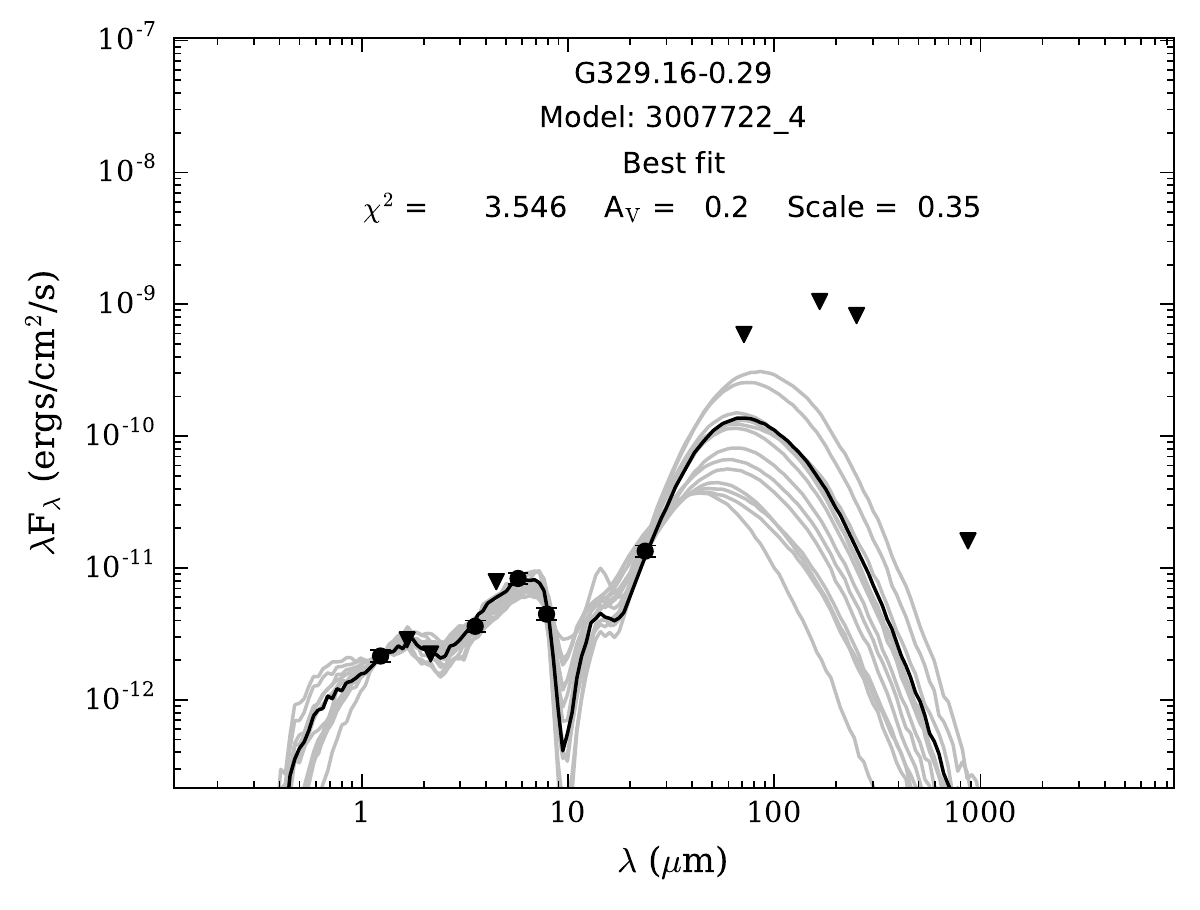}&
                \includegraphics[width=0.31\textwidth]{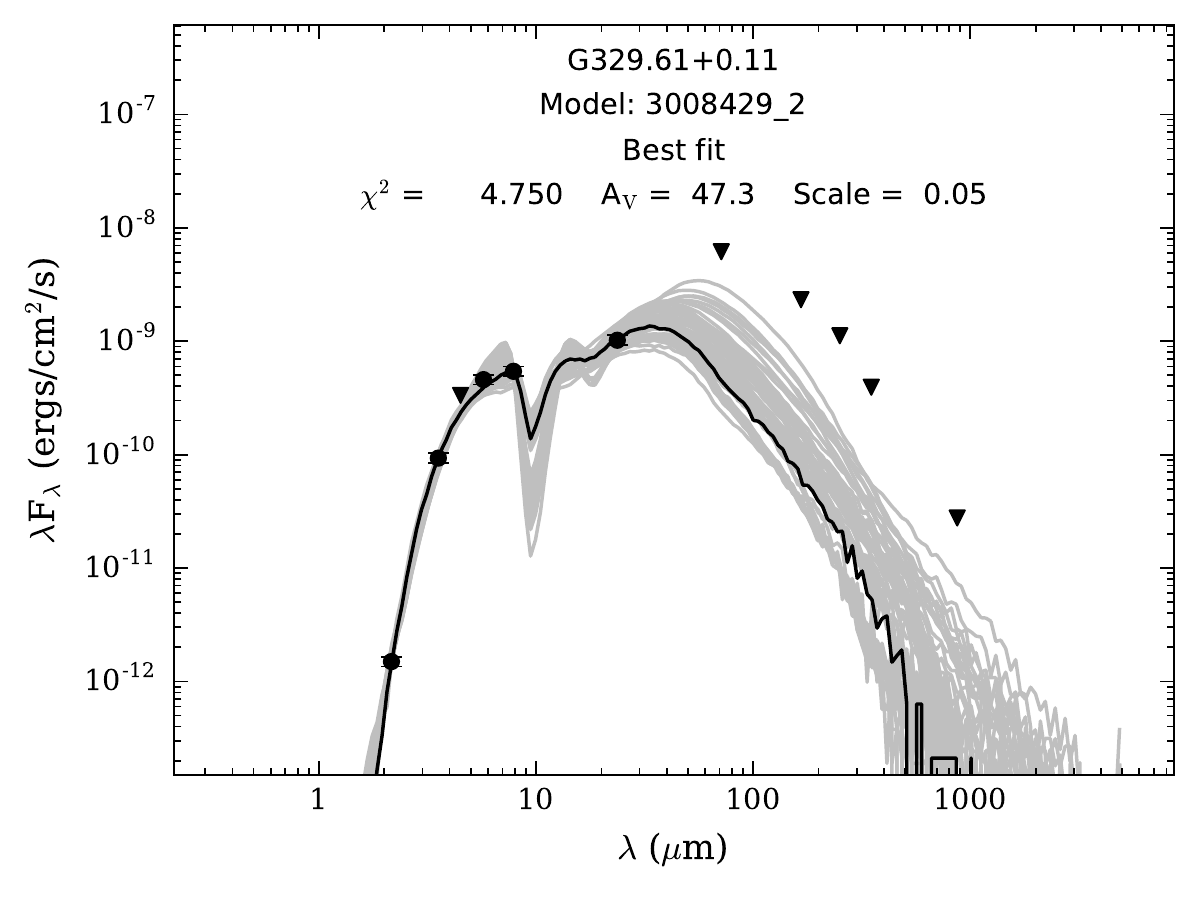}\\
                \includegraphics[width=0.31\textwidth]{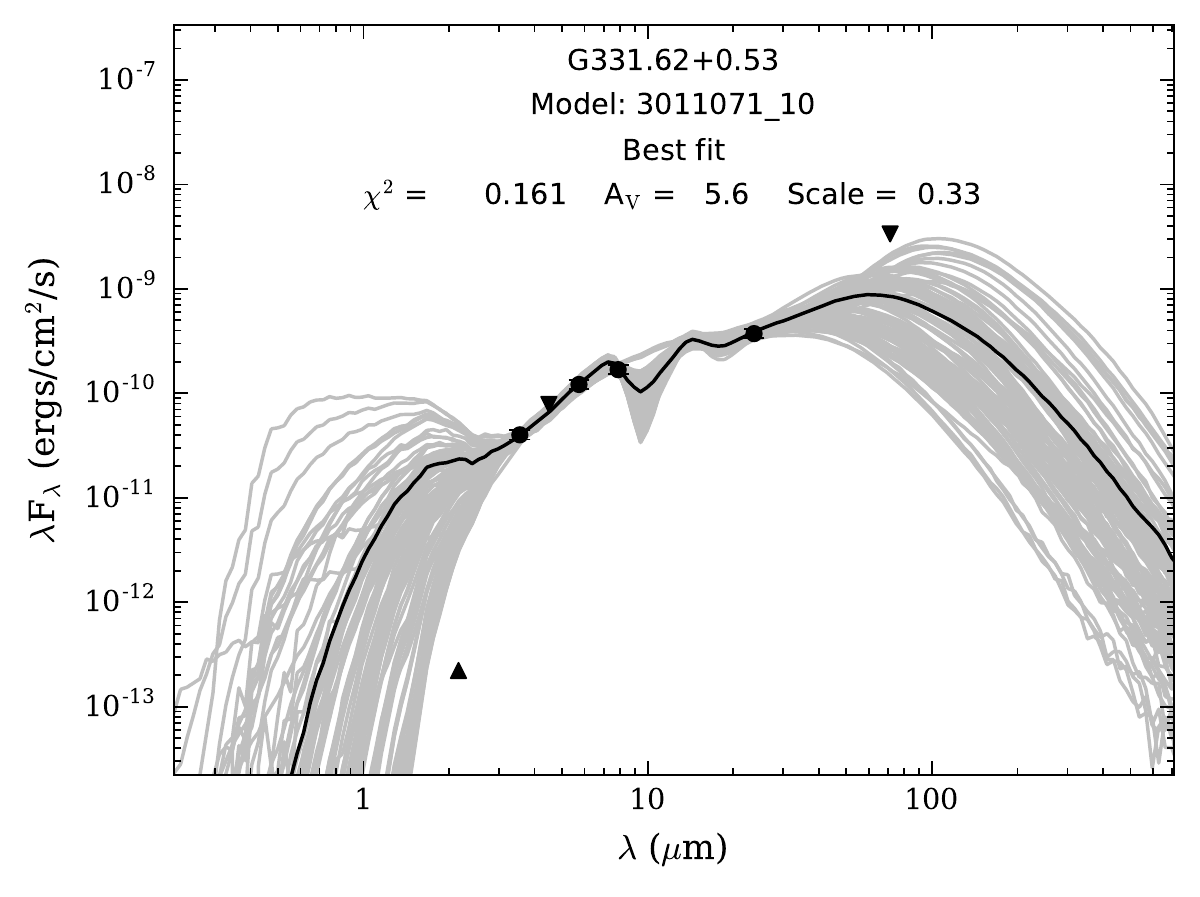}&
                \includegraphics[width=0.31\textwidth]{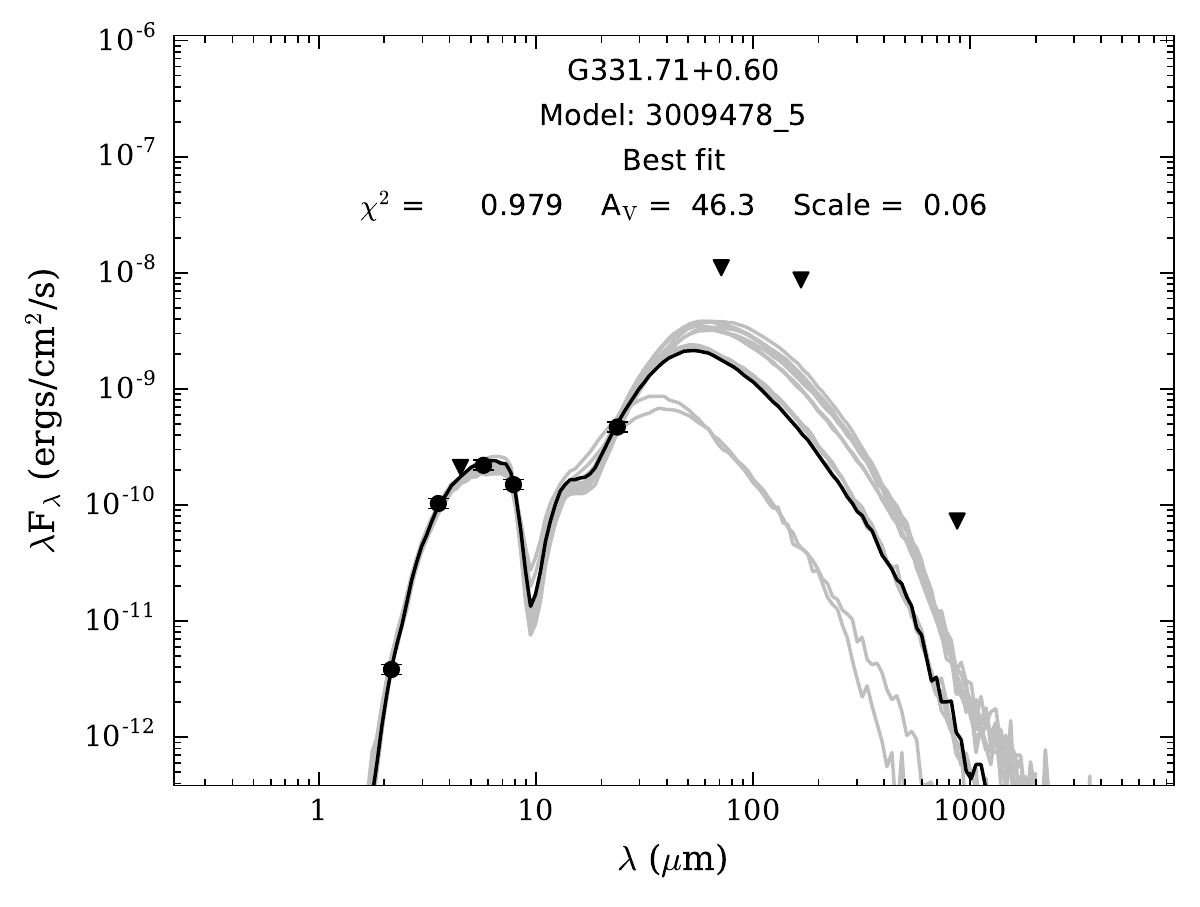}&
                \includegraphics[width=0.31\textwidth]{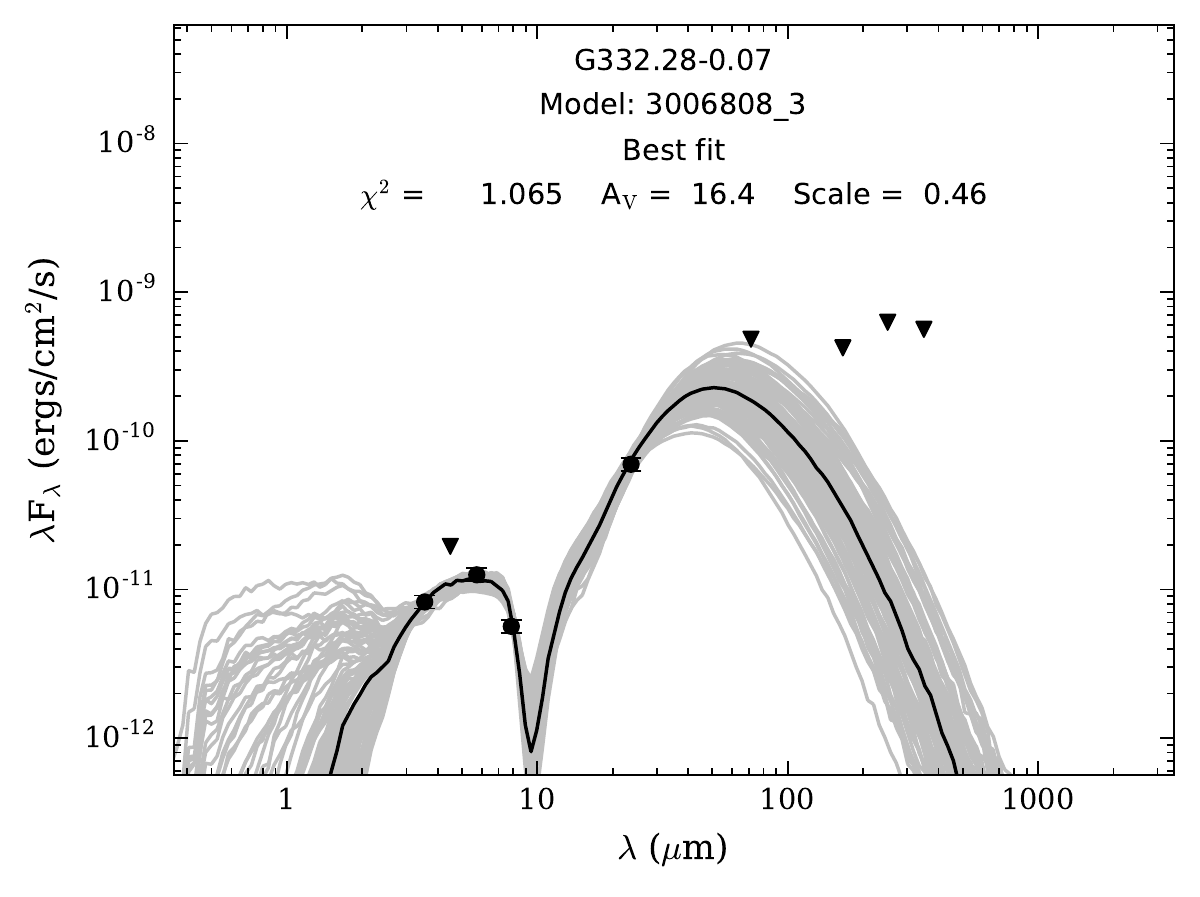}\\
                \includegraphics[width=0.31\textwidth]{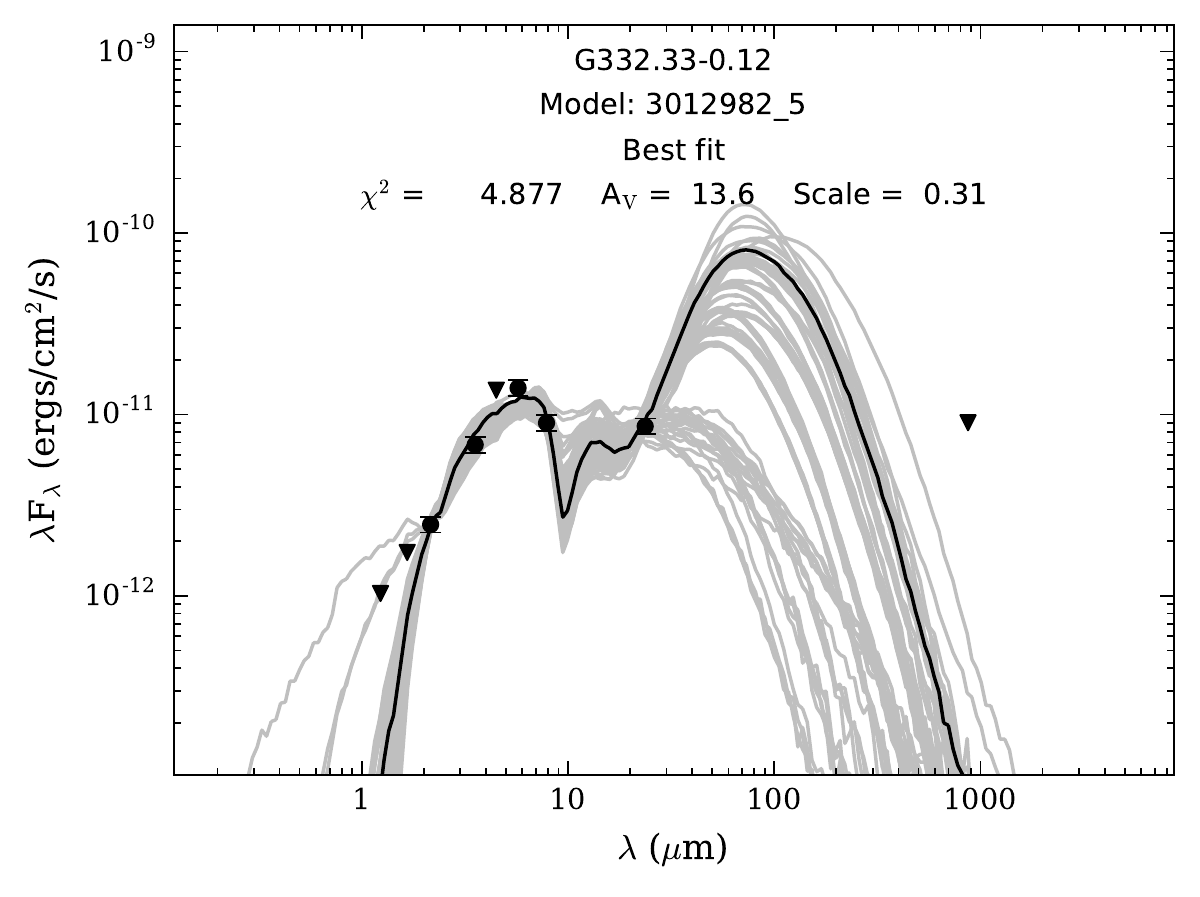}&
                \includegraphics[width=0.31\textwidth]{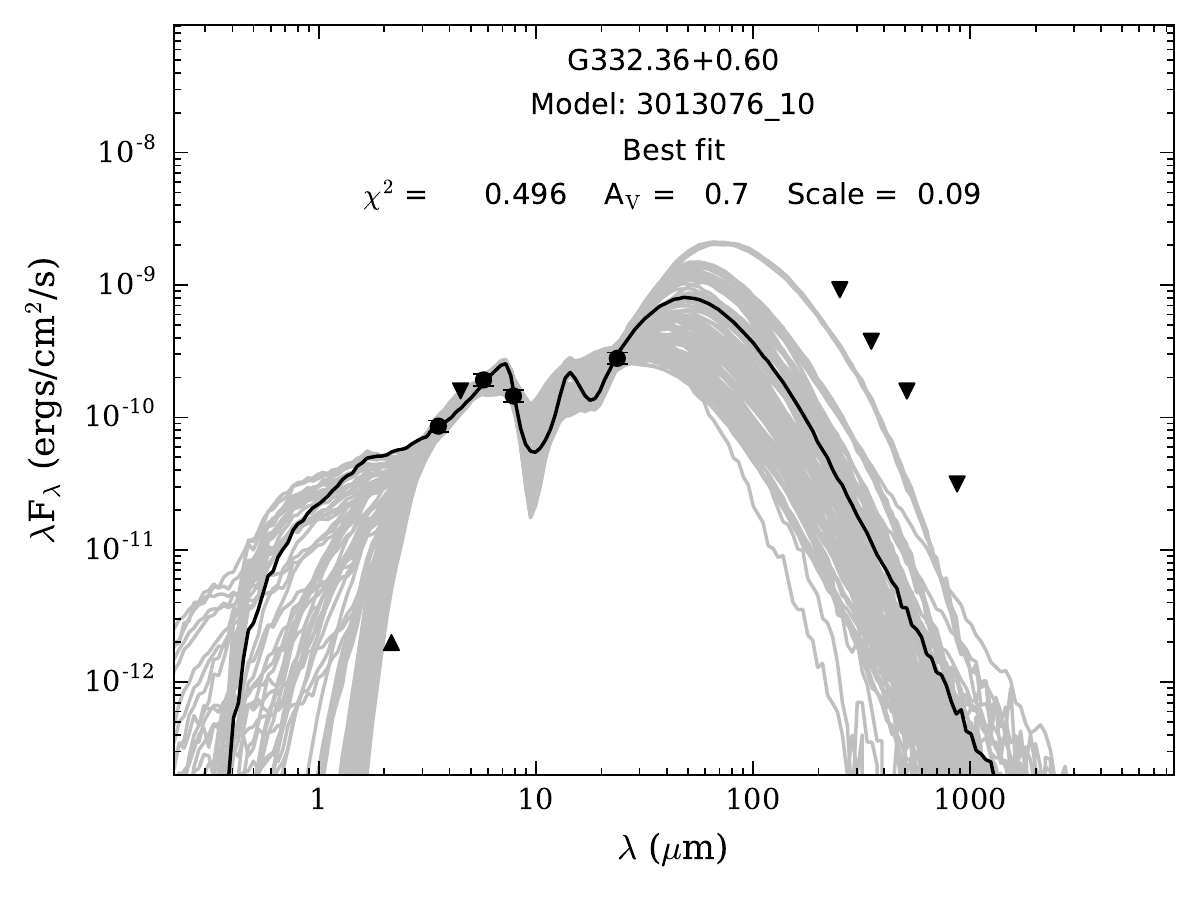}&
                \includegraphics[width=0.31\textwidth]{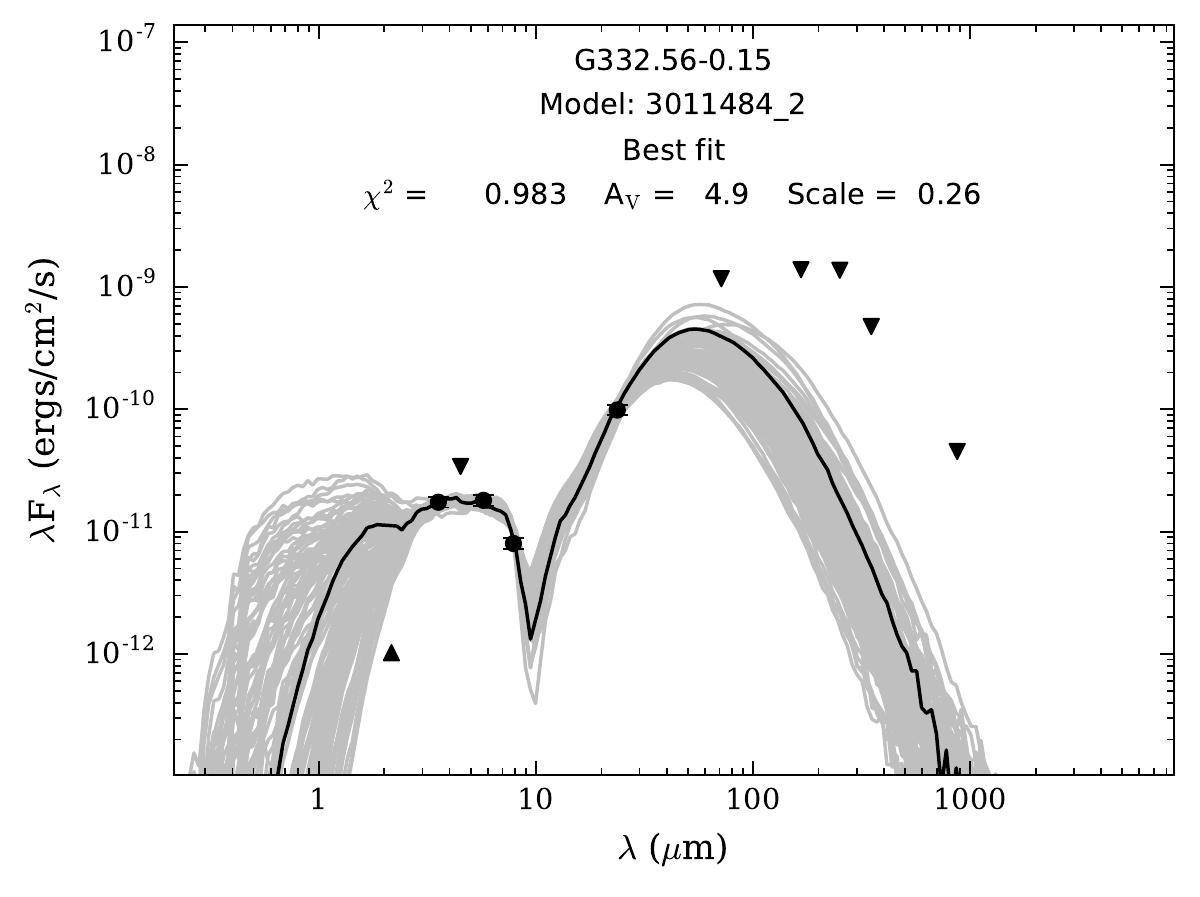}\\
                \includegraphics[width=0.31\textwidth]{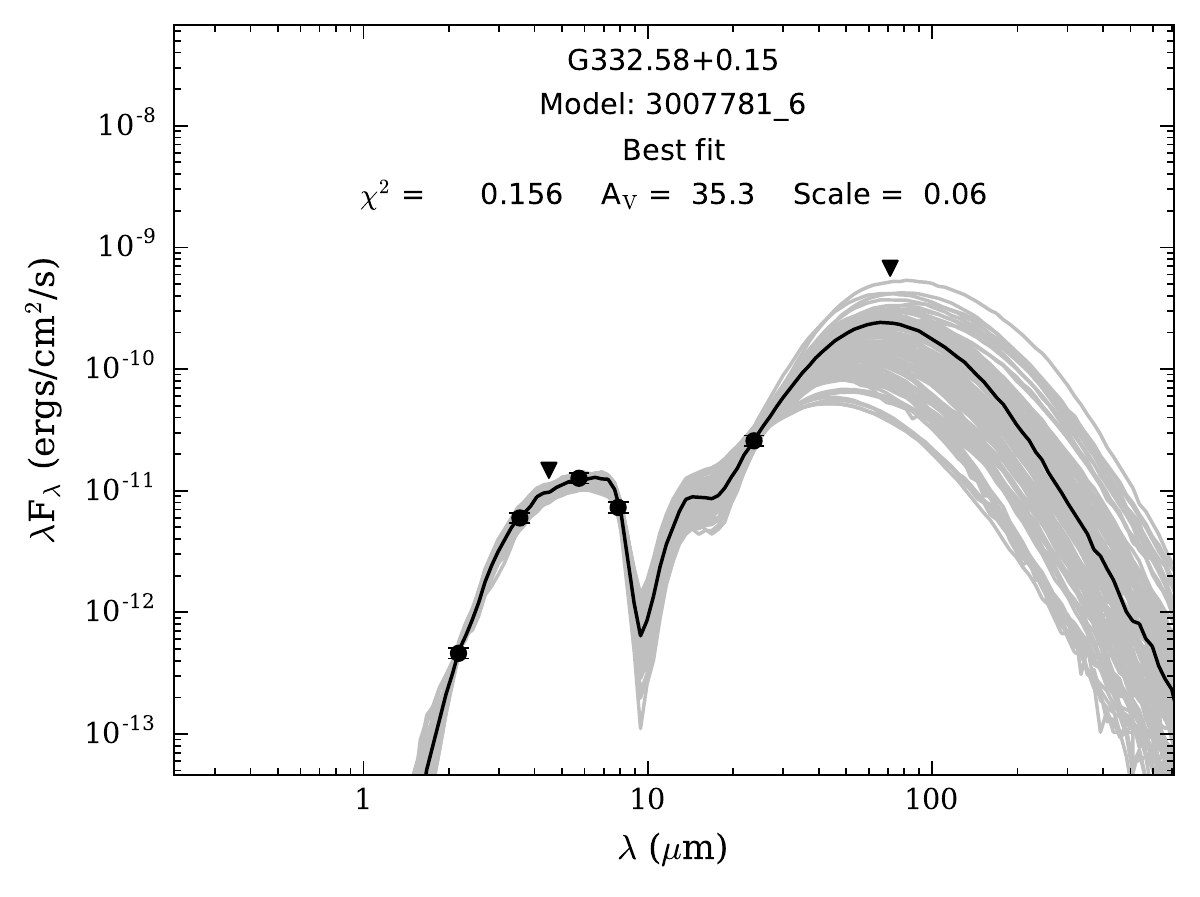}&
                \includegraphics[width=0.31\textwidth]{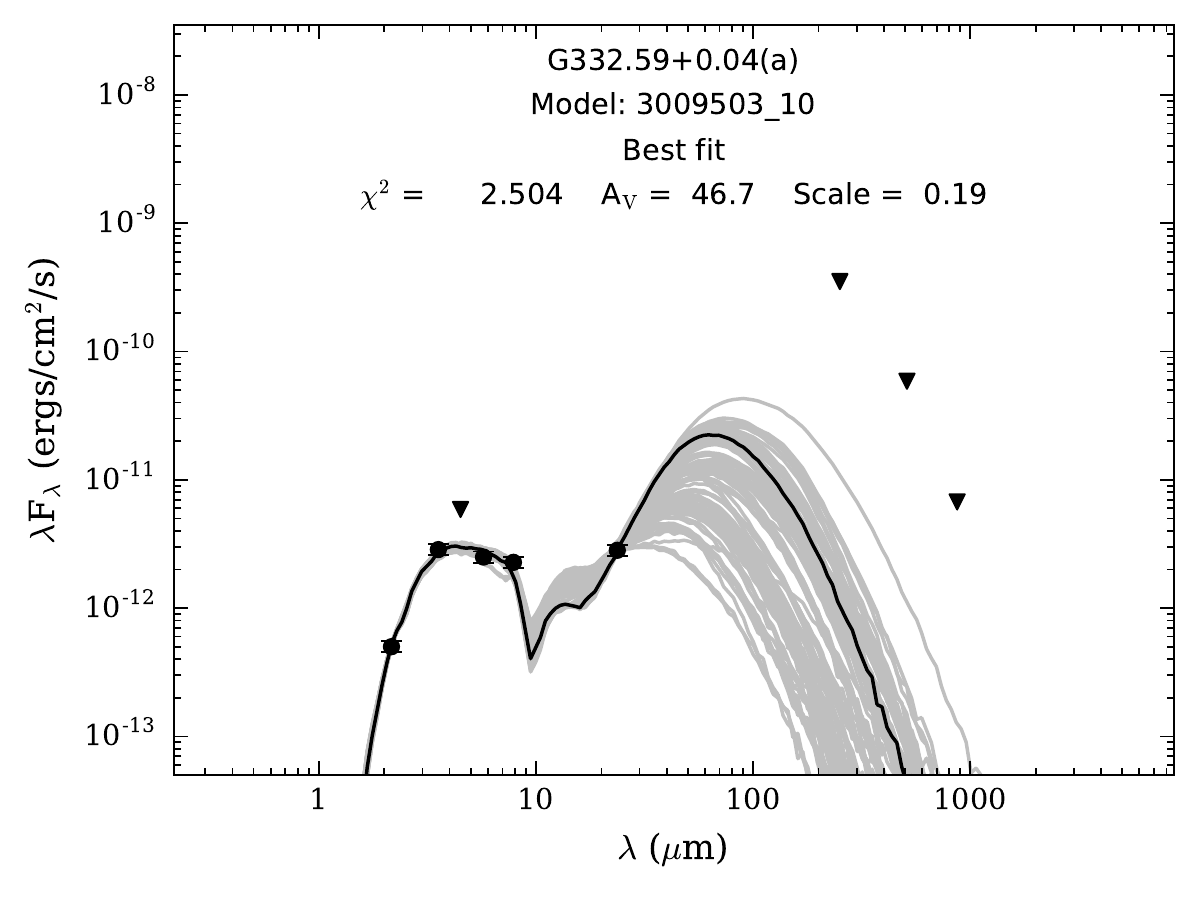}&
                \includegraphics[width=0.31\textwidth]{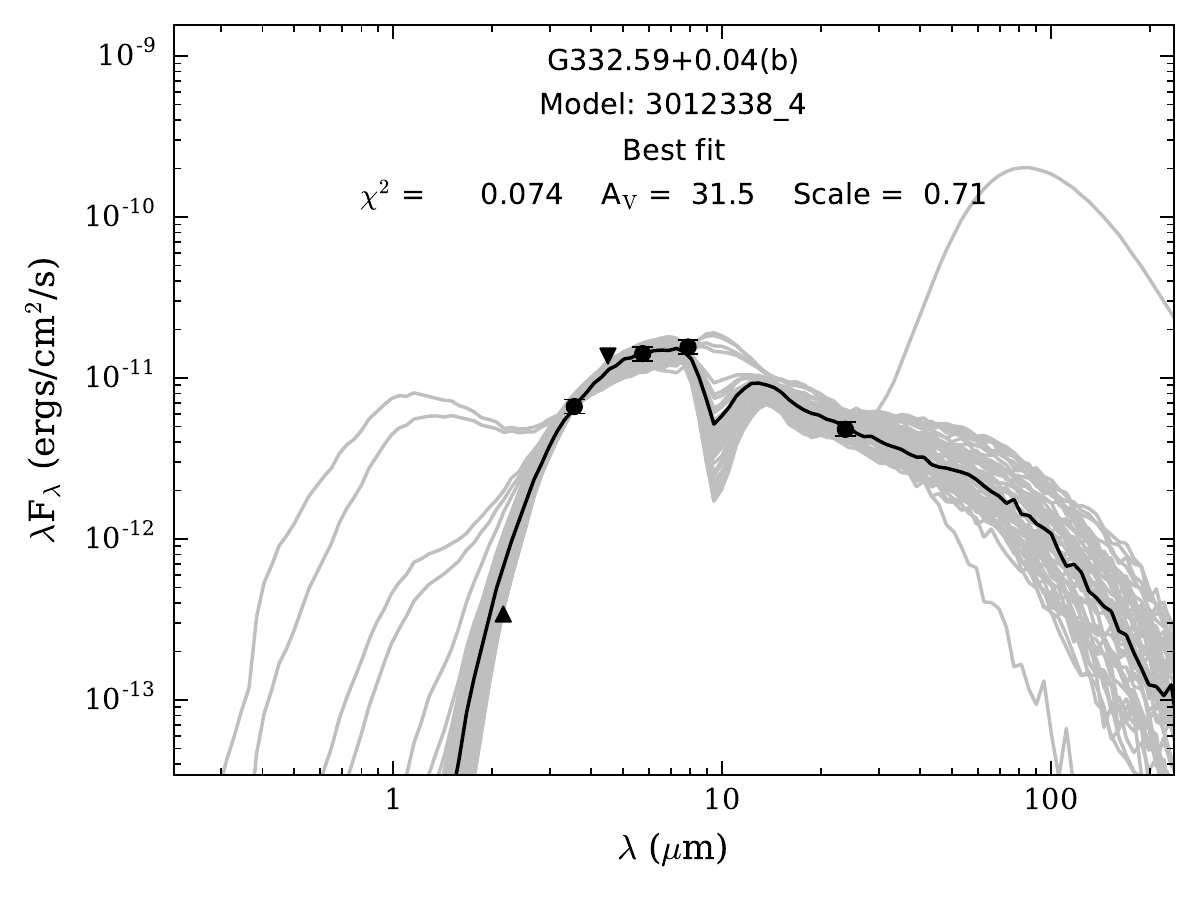}\\
        \end{tabular}
        \caption{Continuation of Fig. \ref{fig:testgridsed}.}
\end{figure}

\begin{figure}
        \begin{tabular}{ccc}
                \includegraphics[width=0.31\textwidth]{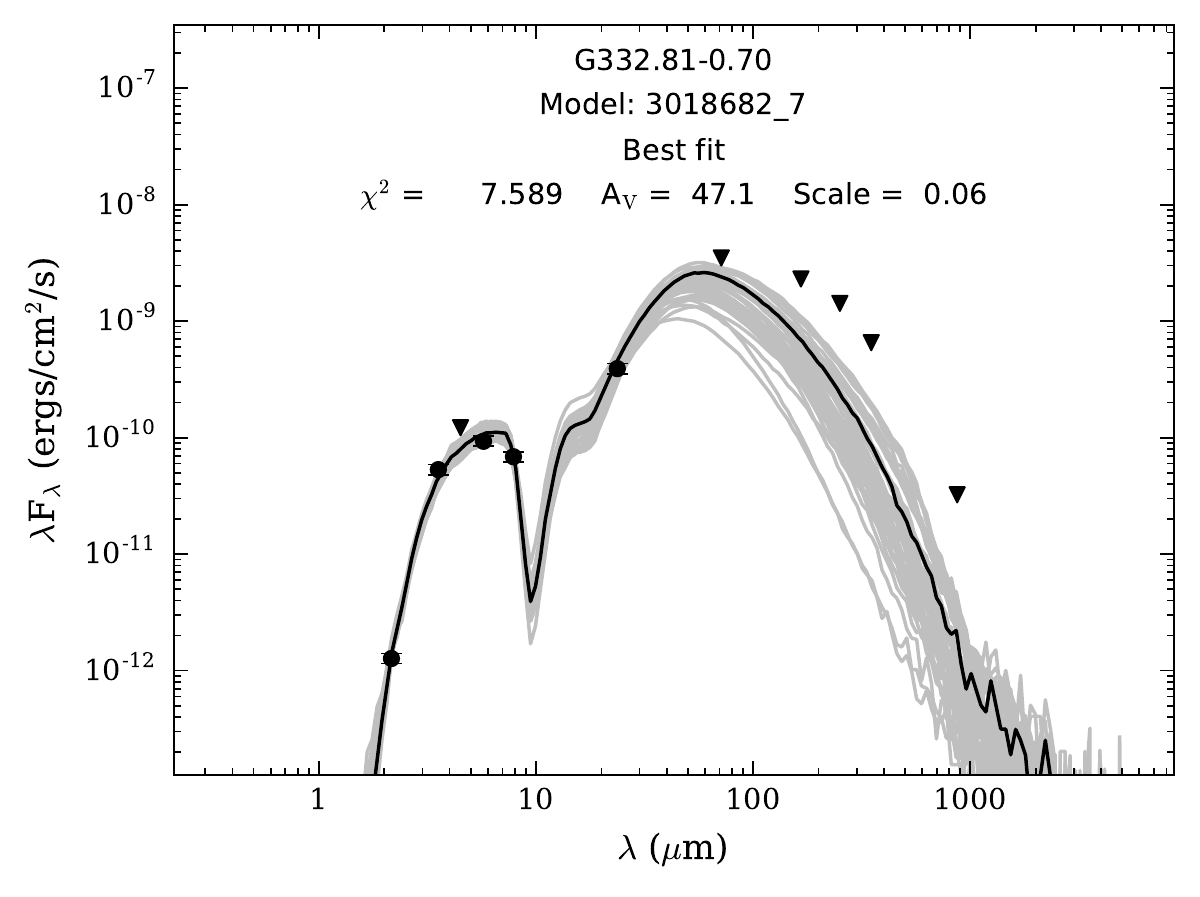}&
                \includegraphics[width=0.31\textwidth]{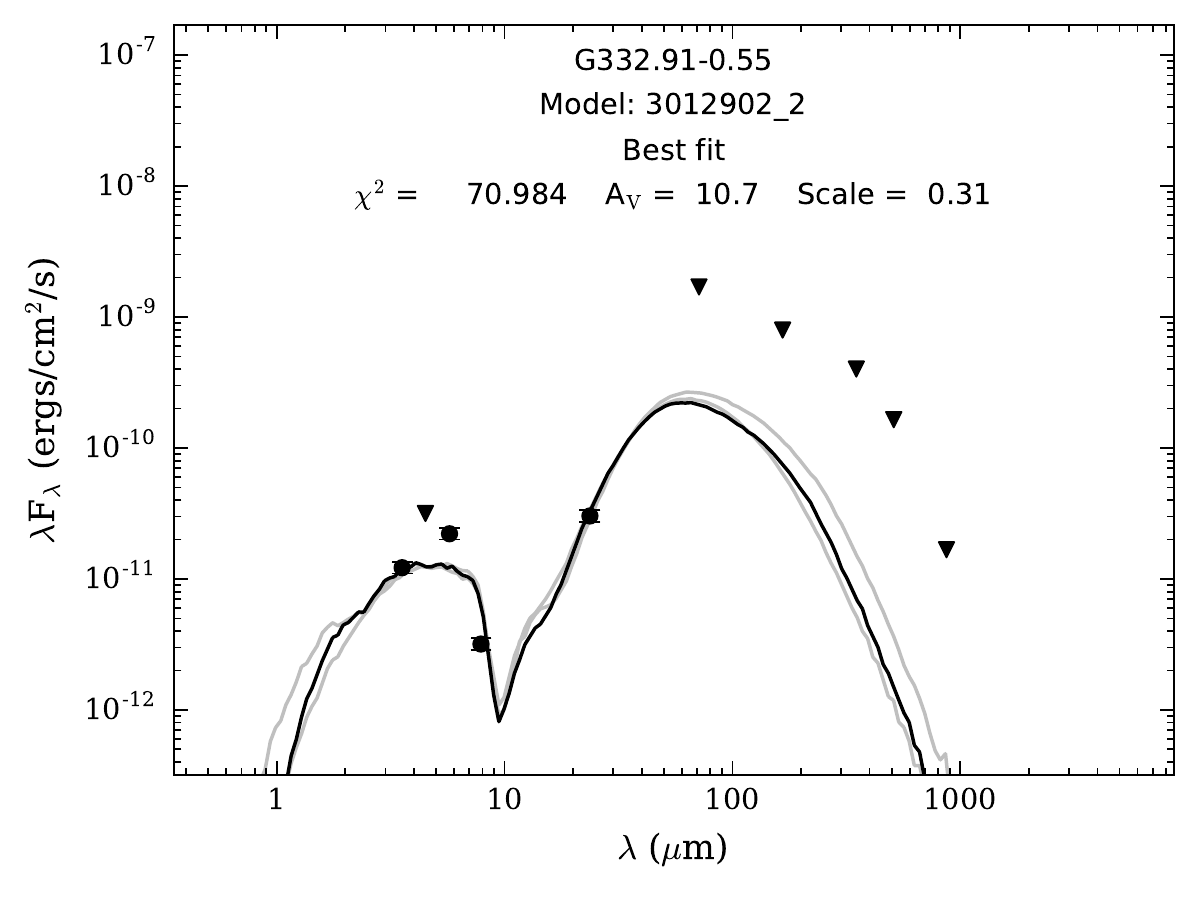}&
                \includegraphics[width=0.31\textwidth]{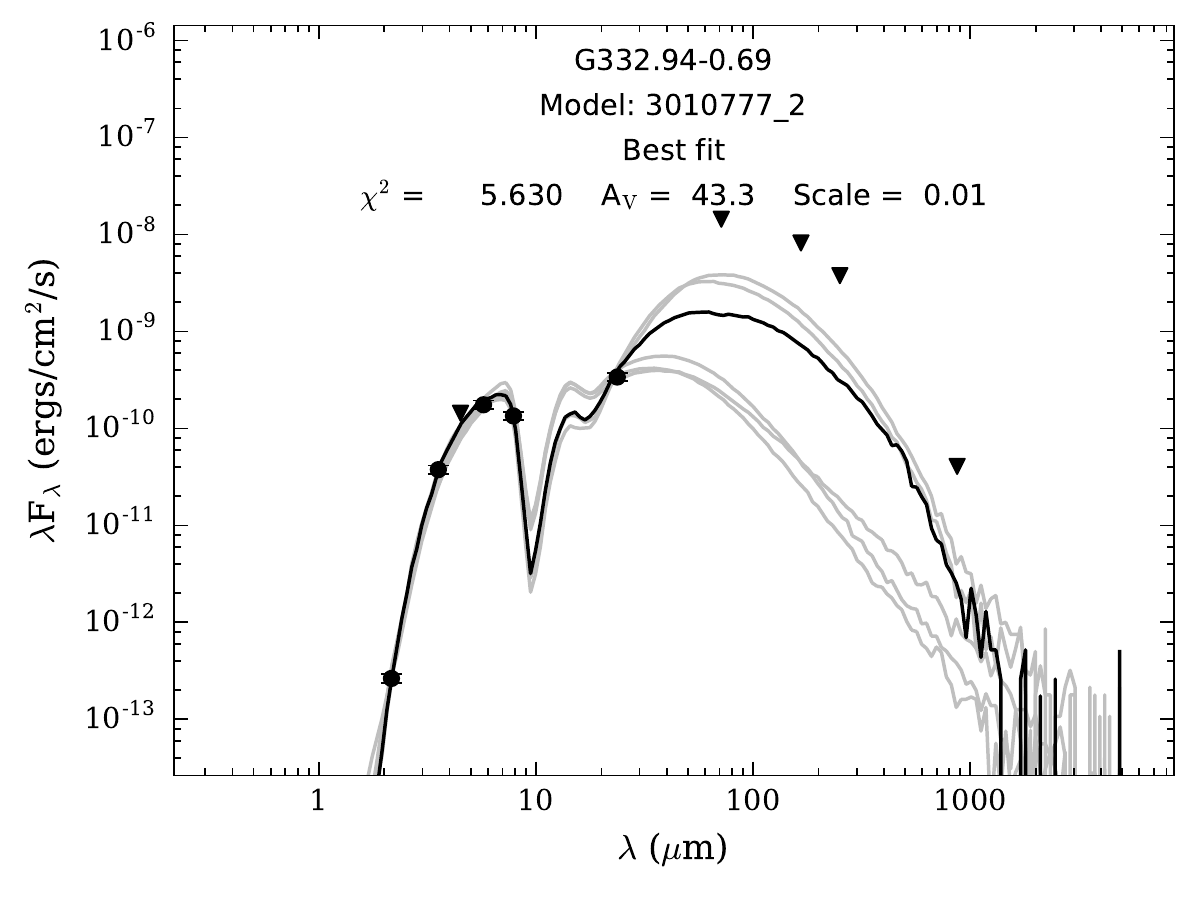}\\
                
                \includegraphics[width=0.31\textwidth]{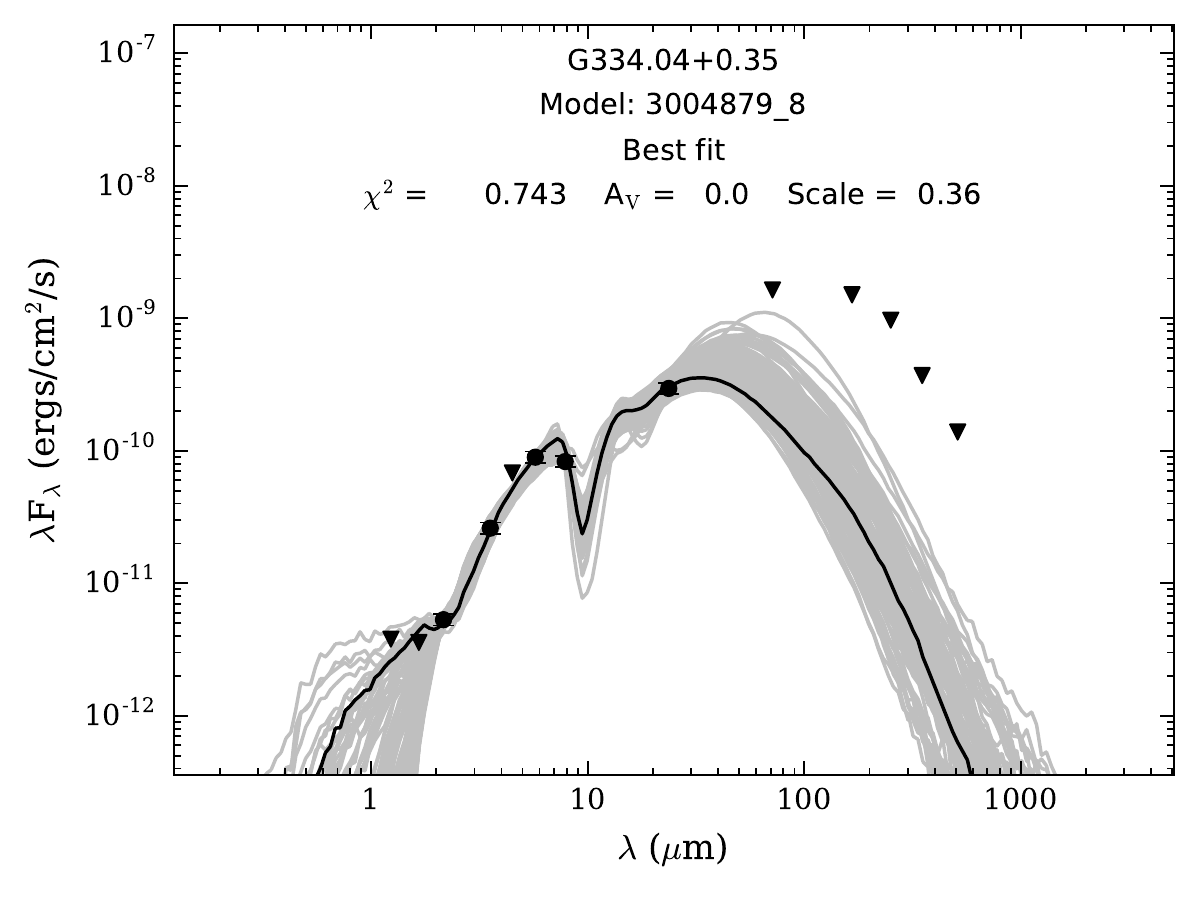}&
                \includegraphics[width=0.31\textwidth]{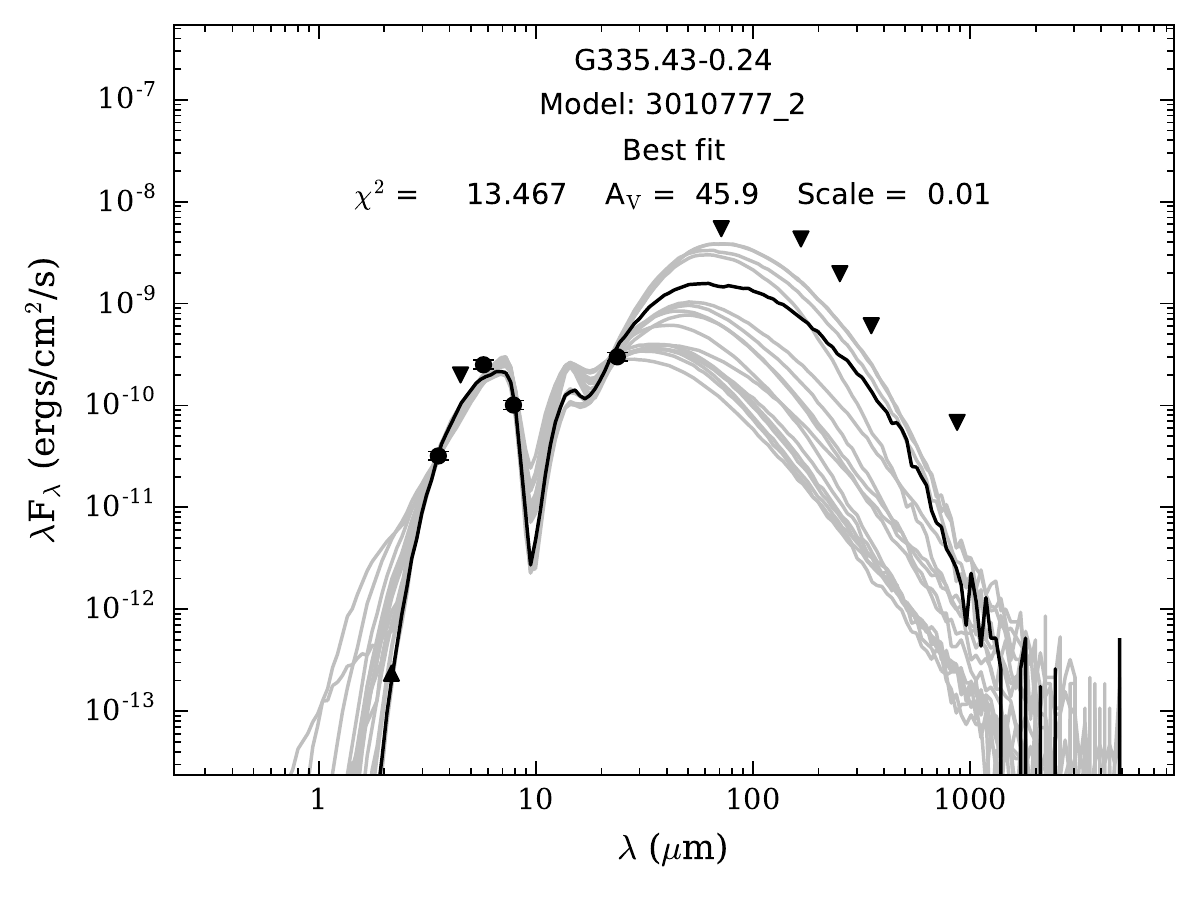}&
                \includegraphics[width=0.31\textwidth]{{./G335.59-0.29}.pdf}\\
                \includegraphics[width=0.31\textwidth]{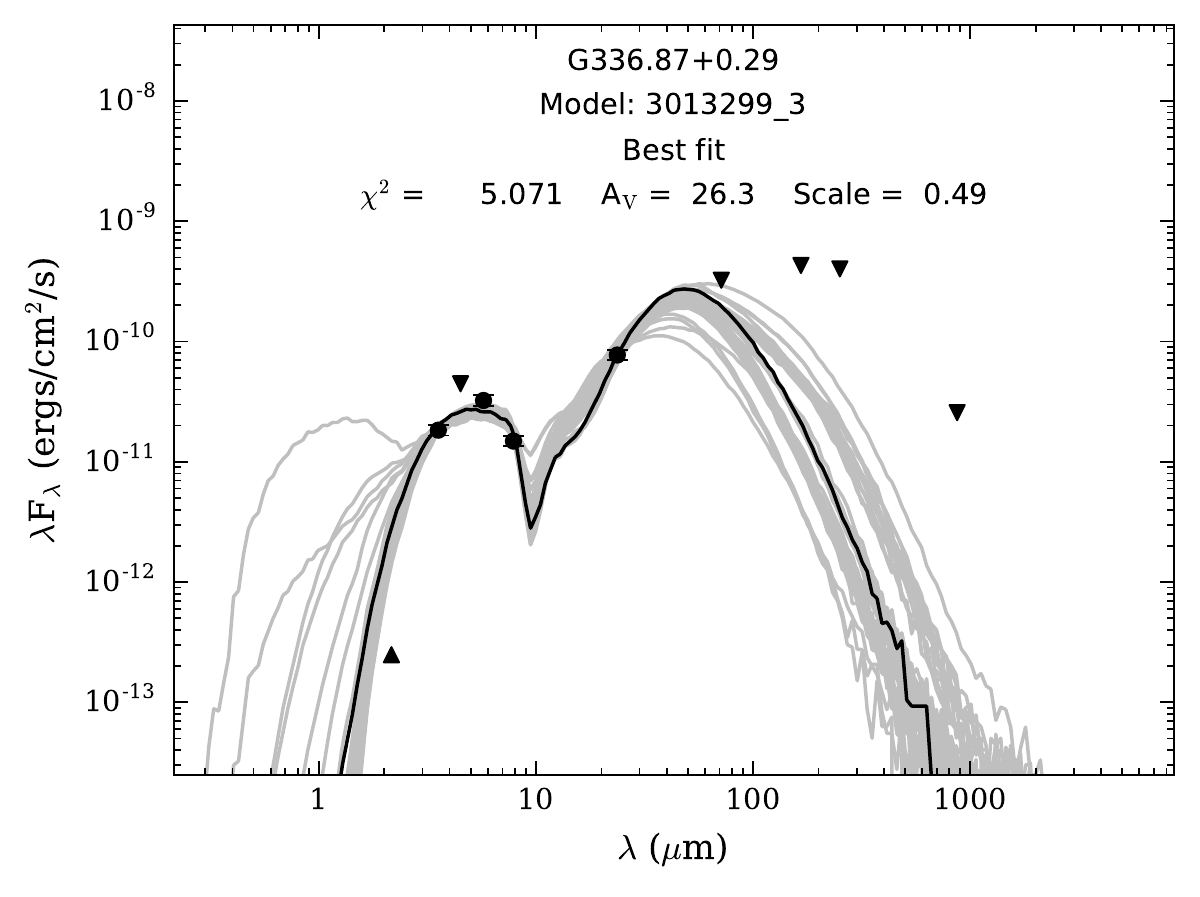}&
                \includegraphics[width=0.31\textwidth]{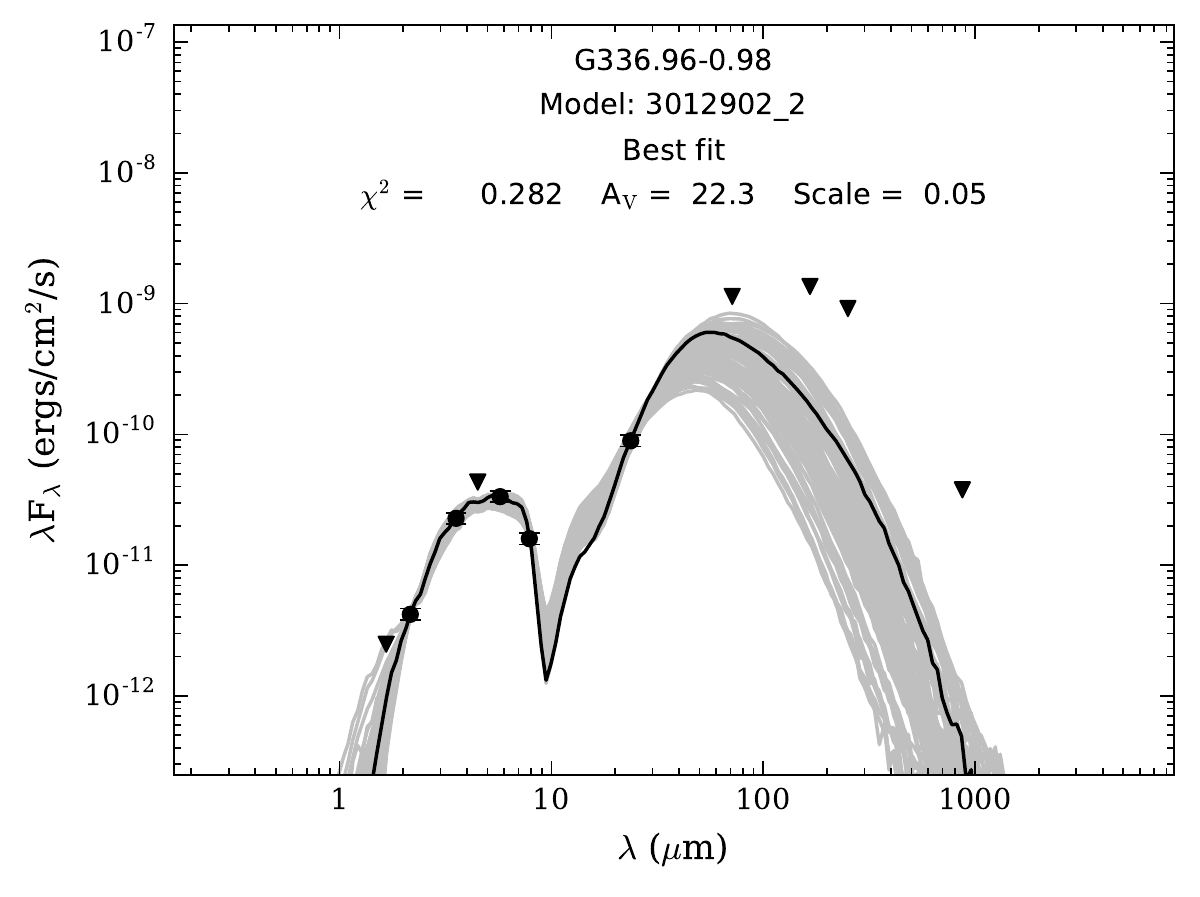}&
                \includegraphics[width=0.31\textwidth]{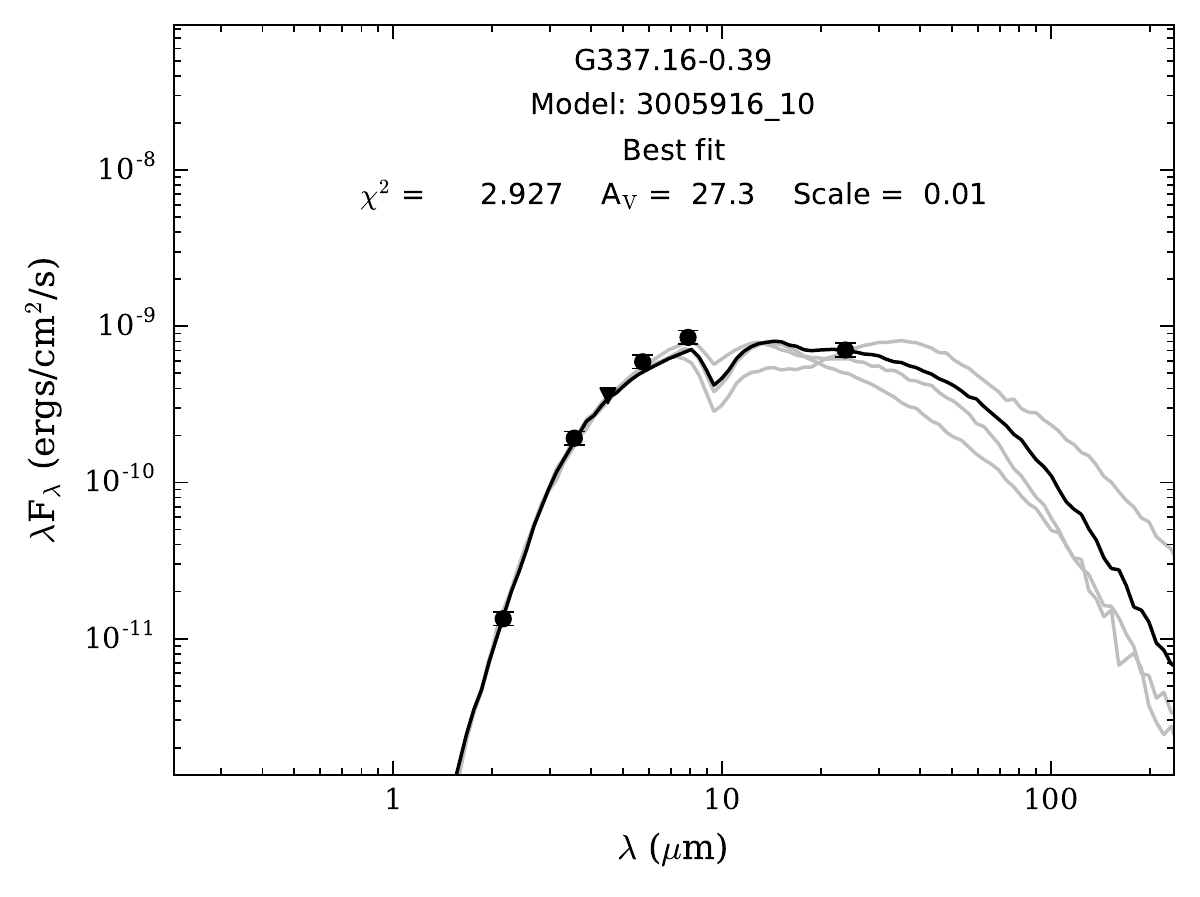}\\
                \includegraphics[width=0.31\textwidth]{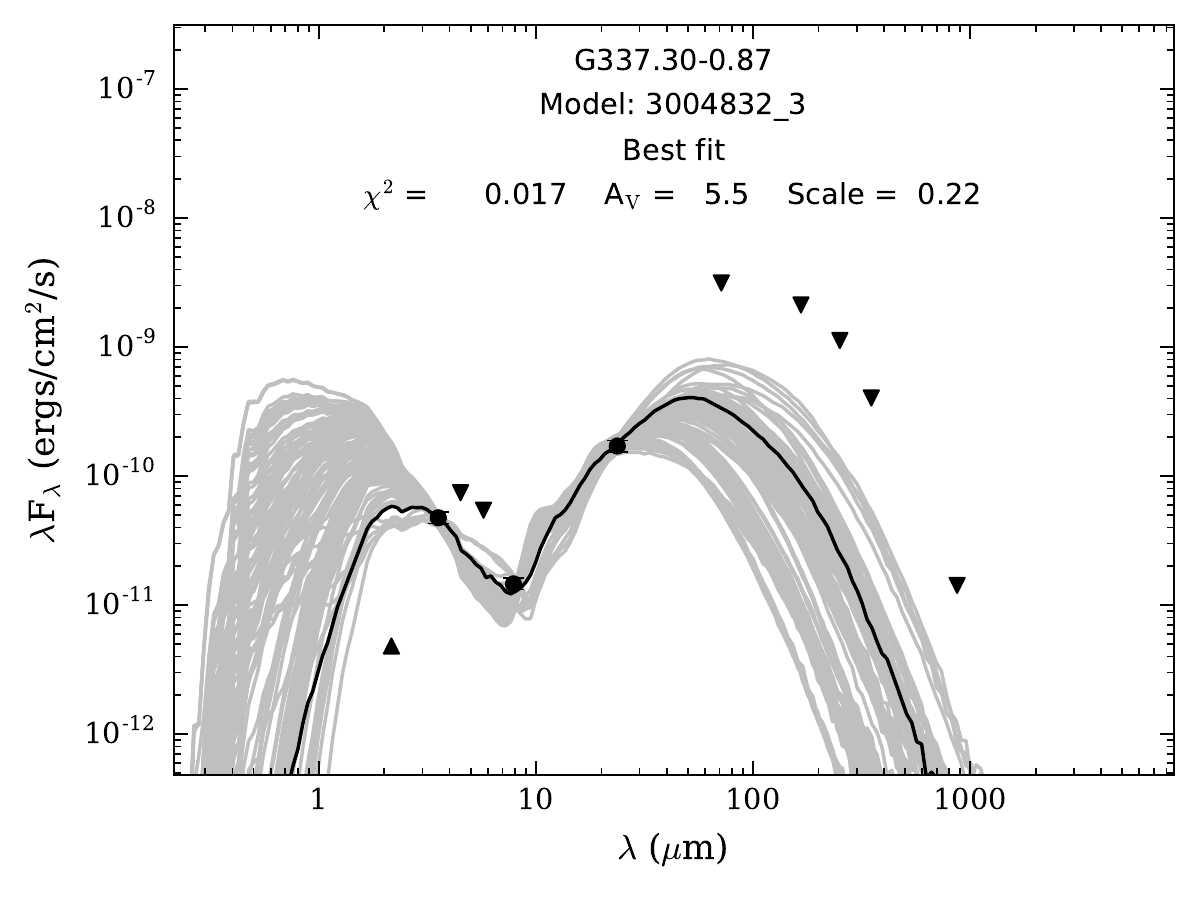}&
                \includegraphics[width=0.31\textwidth]{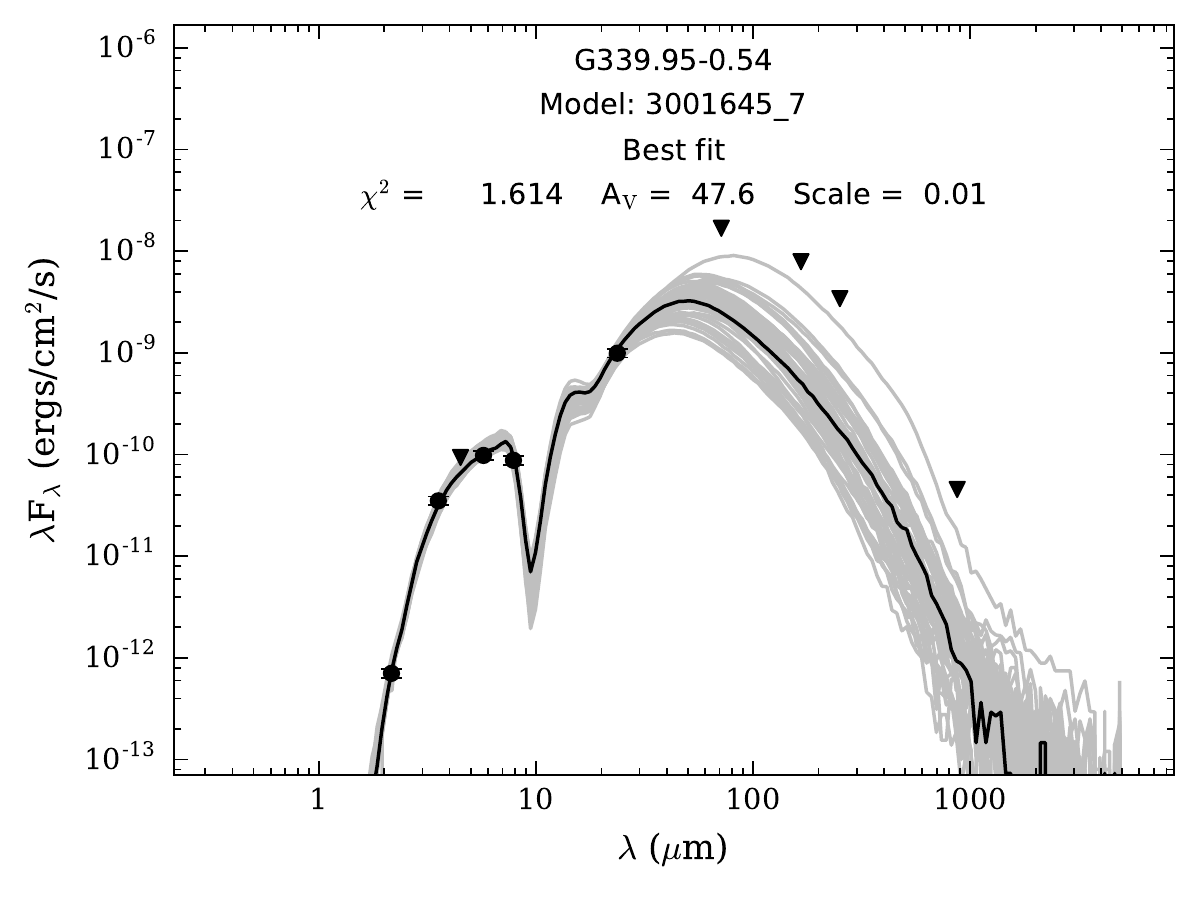}&
                \includegraphics[width=0.31\textwidth]{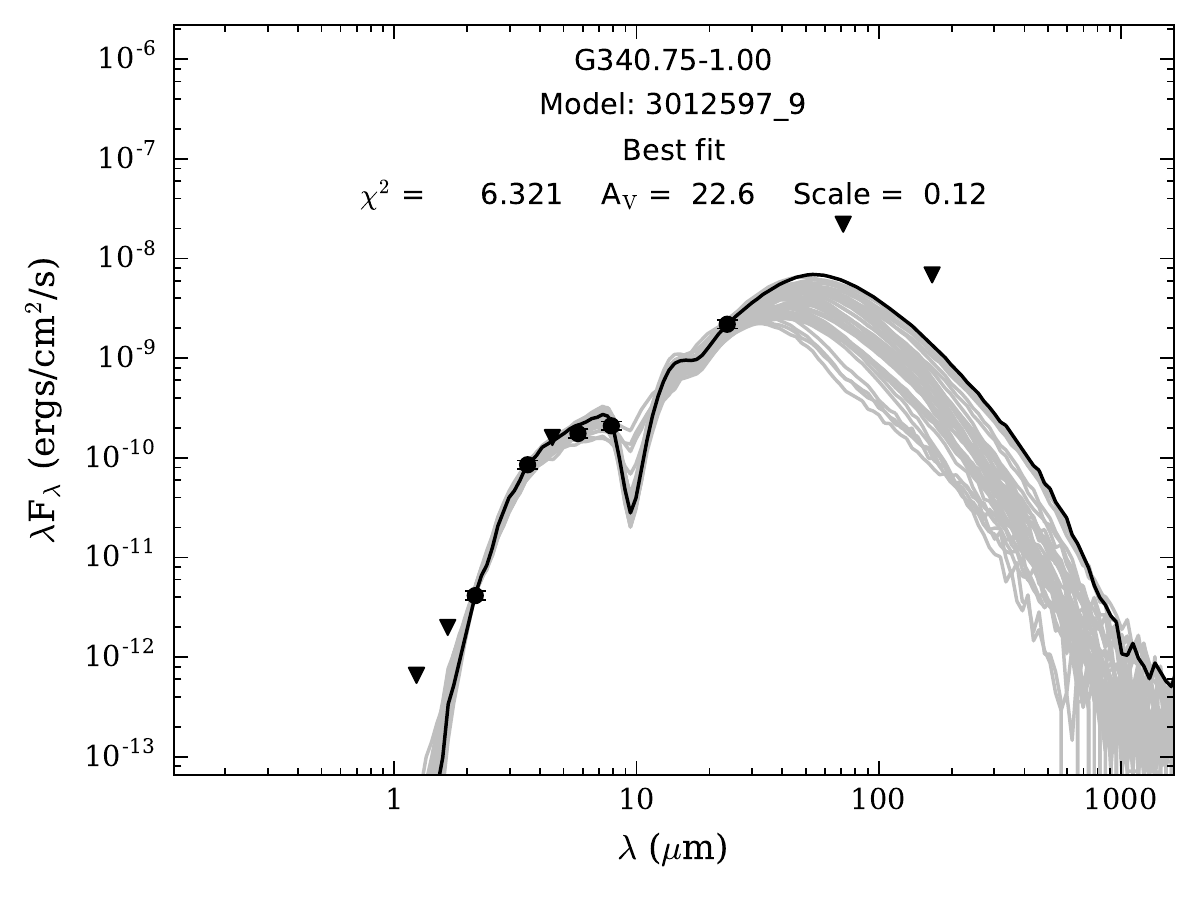}\\
                \includegraphics[width=0.31\textwidth]{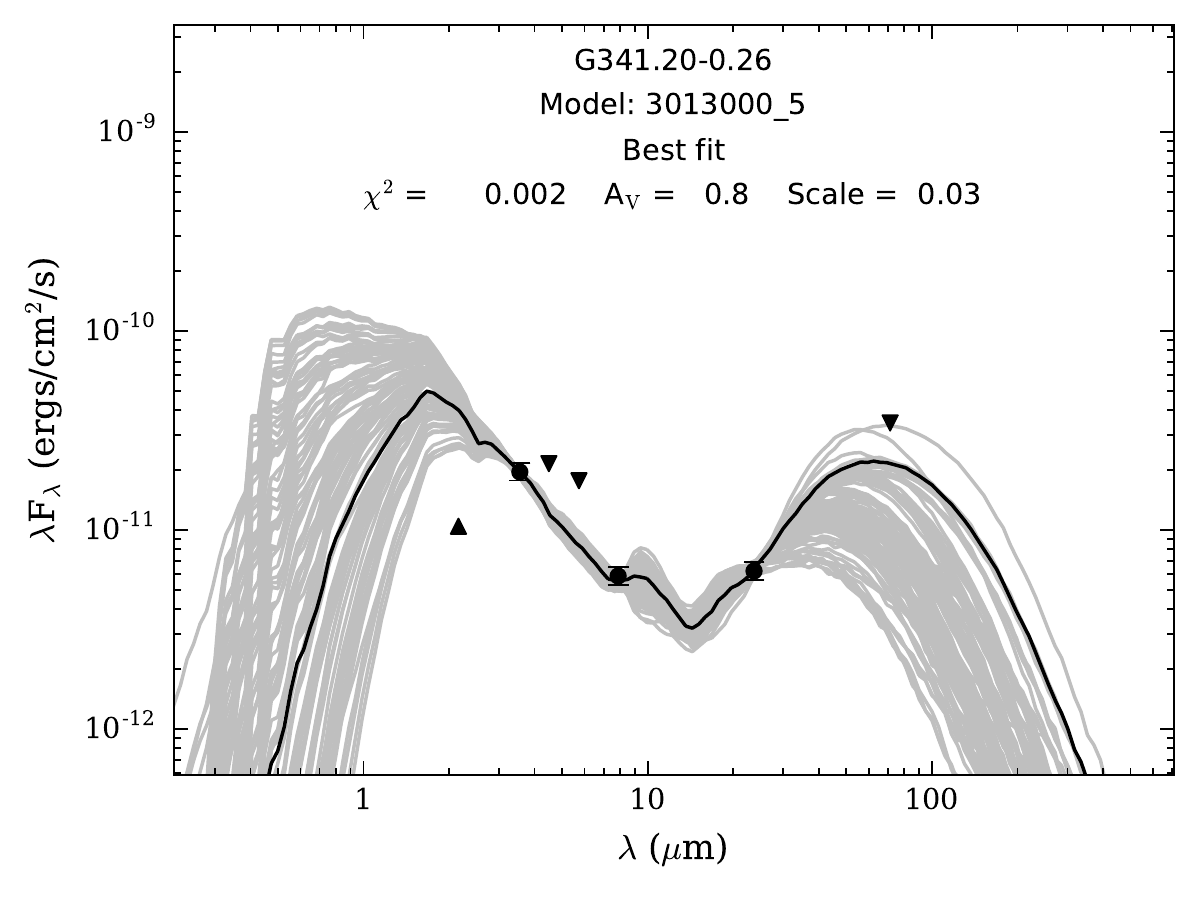}&
                \includegraphics[width=0.31\textwidth]{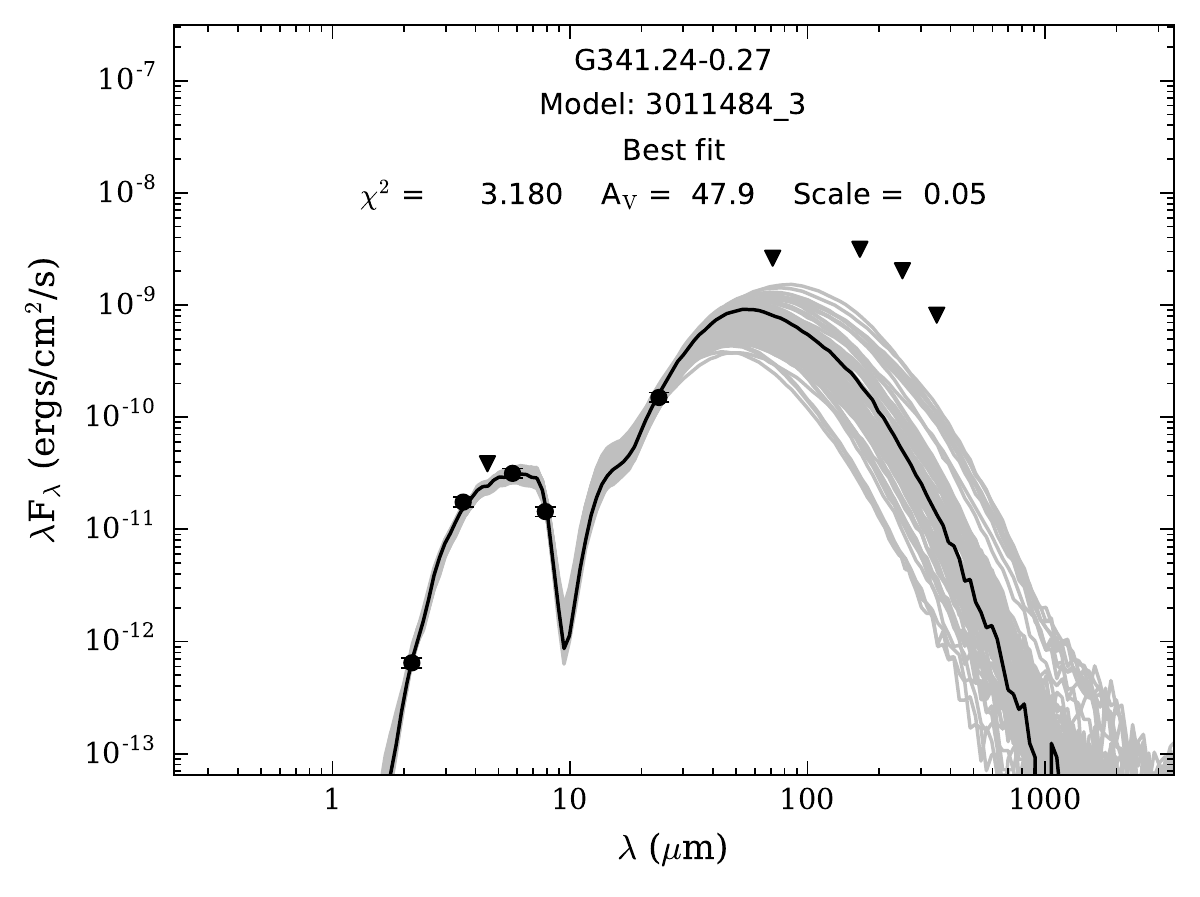}&
                \includegraphics[width=0.31\textwidth]{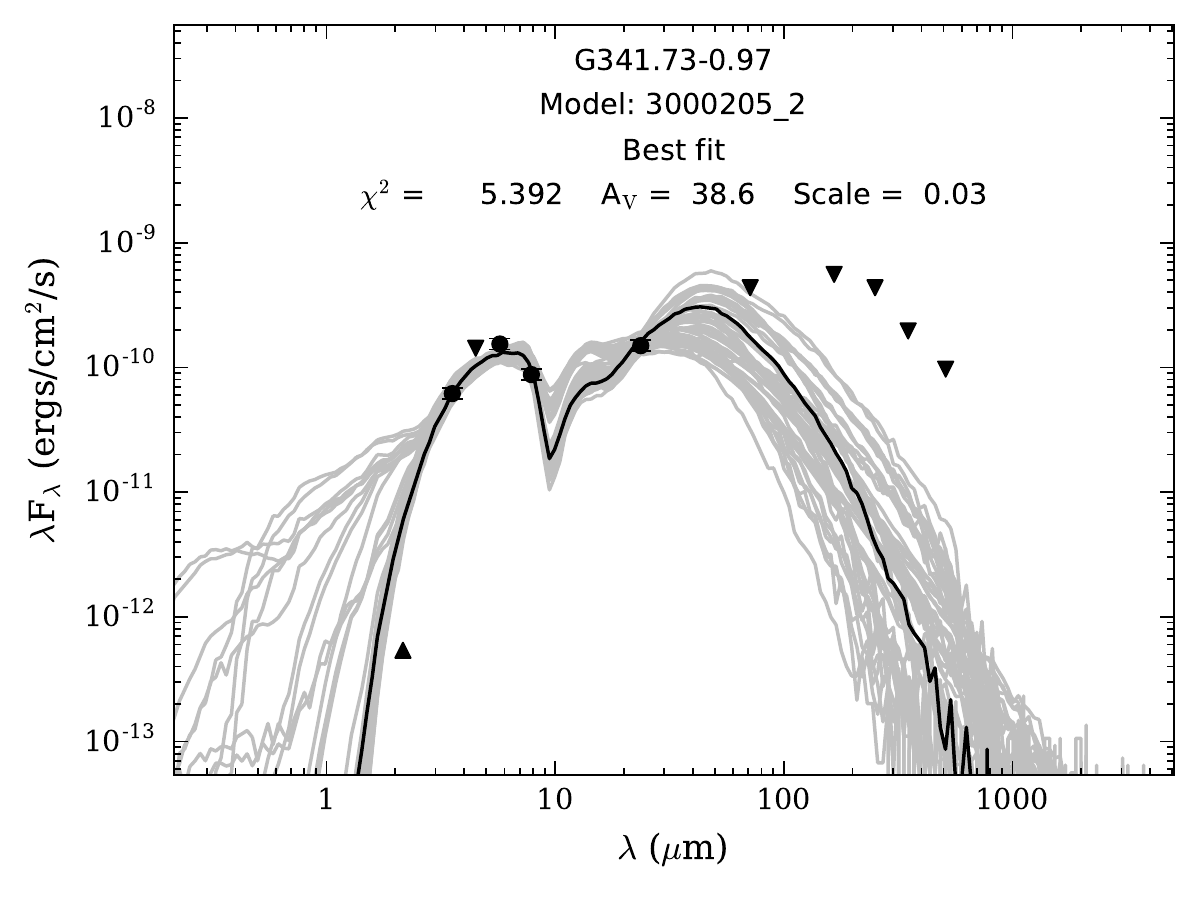}\\
        \end{tabular}
        \caption{Continuation of Fig. \ref{fig:testgridsed}.}
\end{figure}

\begin{figure}
        \begin{tabular}{ccc}
                
                \includegraphics[width=0.31\textwidth]{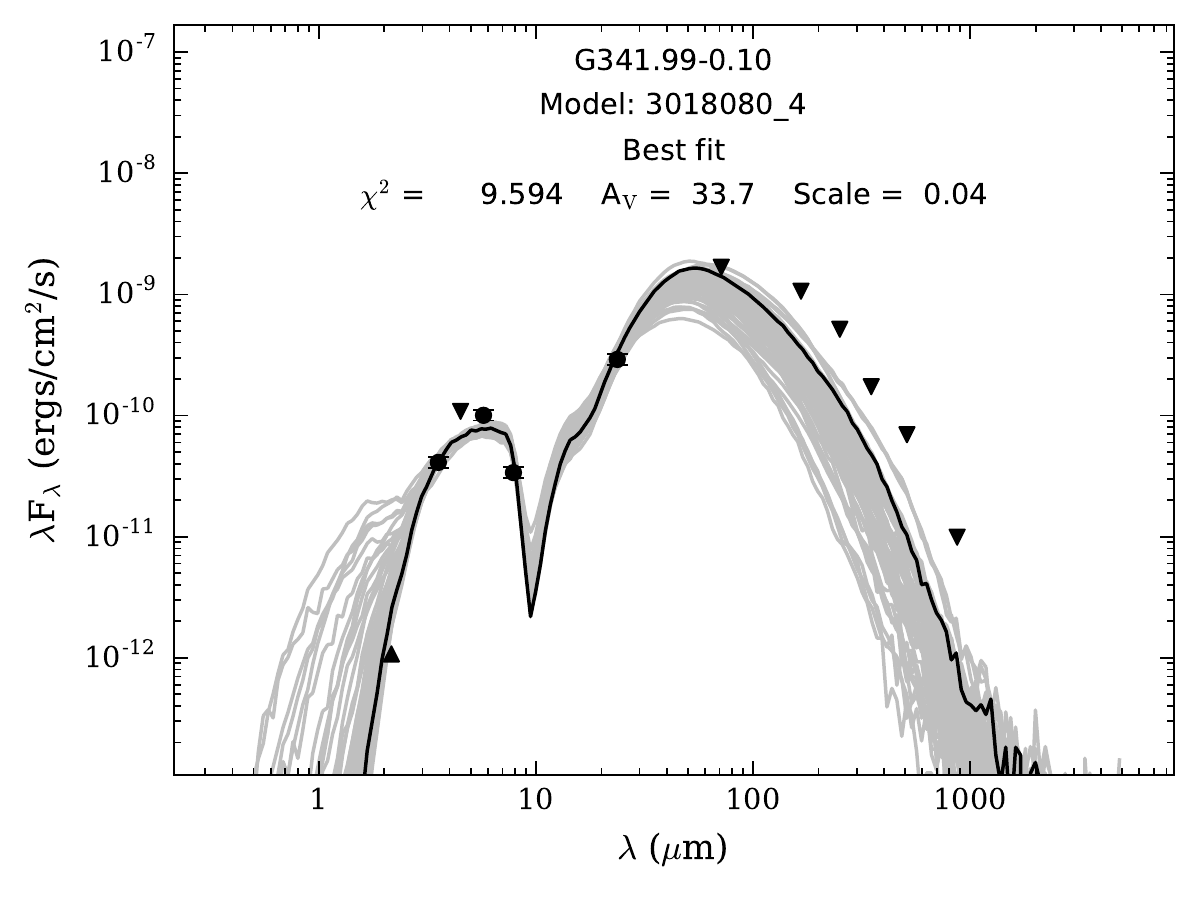}&
                \includegraphics[width=0.31\textwidth]{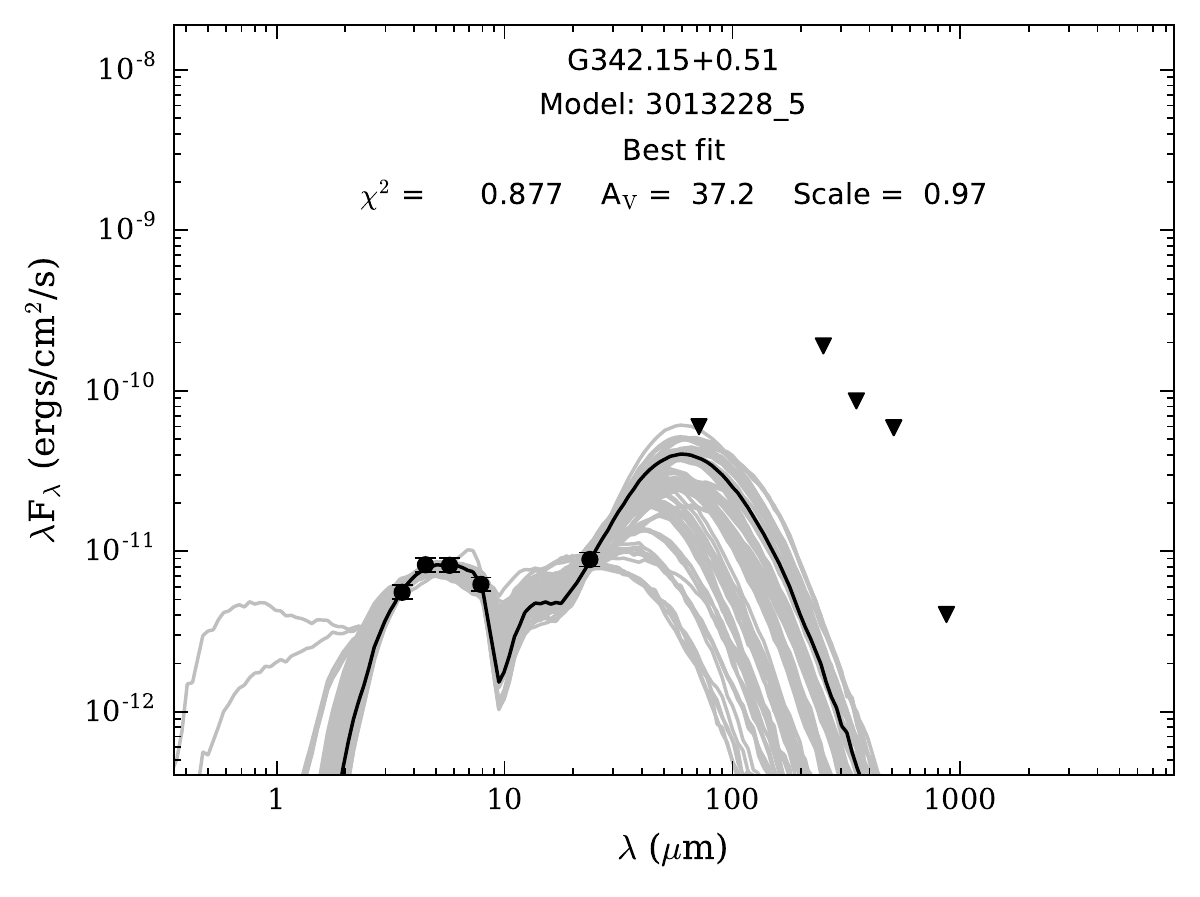}&
                \includegraphics[width=0.31\textwidth]{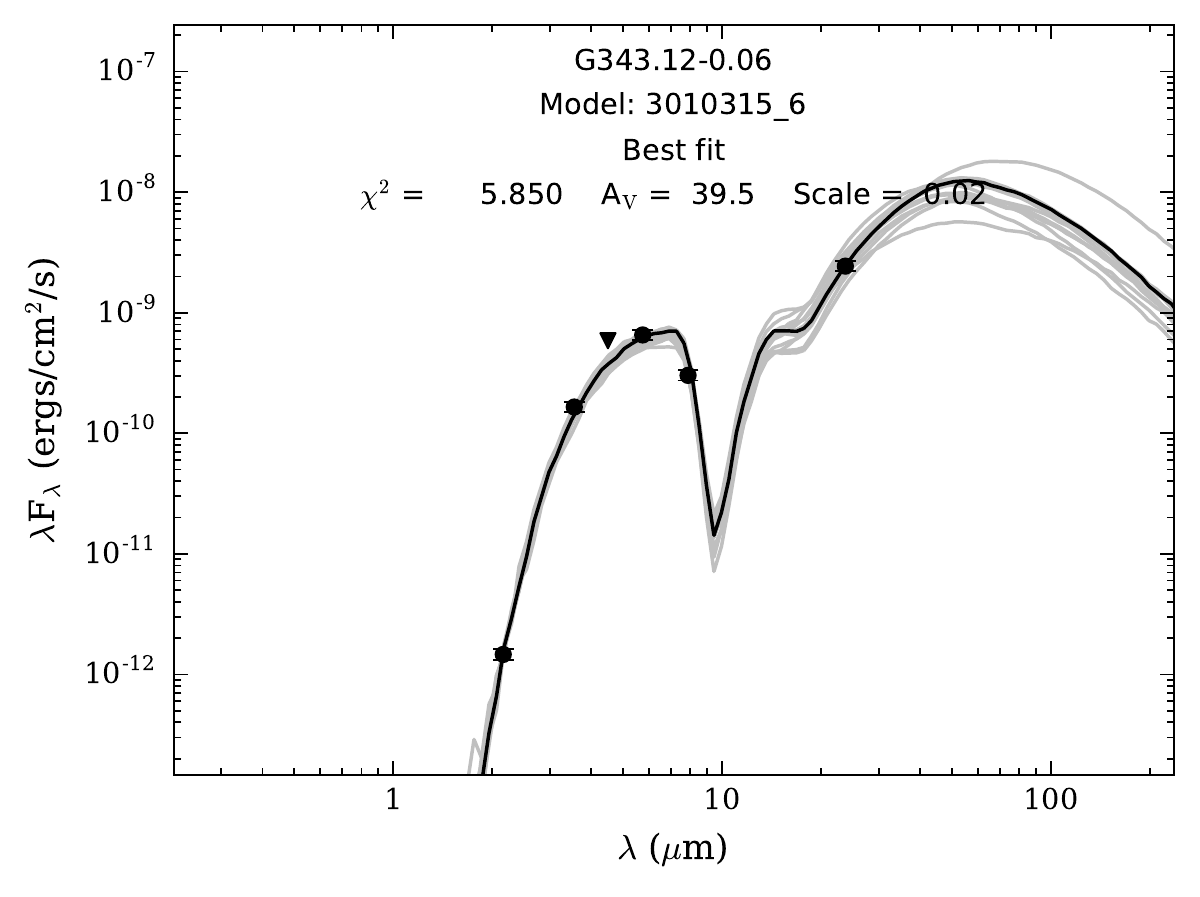}\\
                \includegraphics[width=0.31\textwidth]{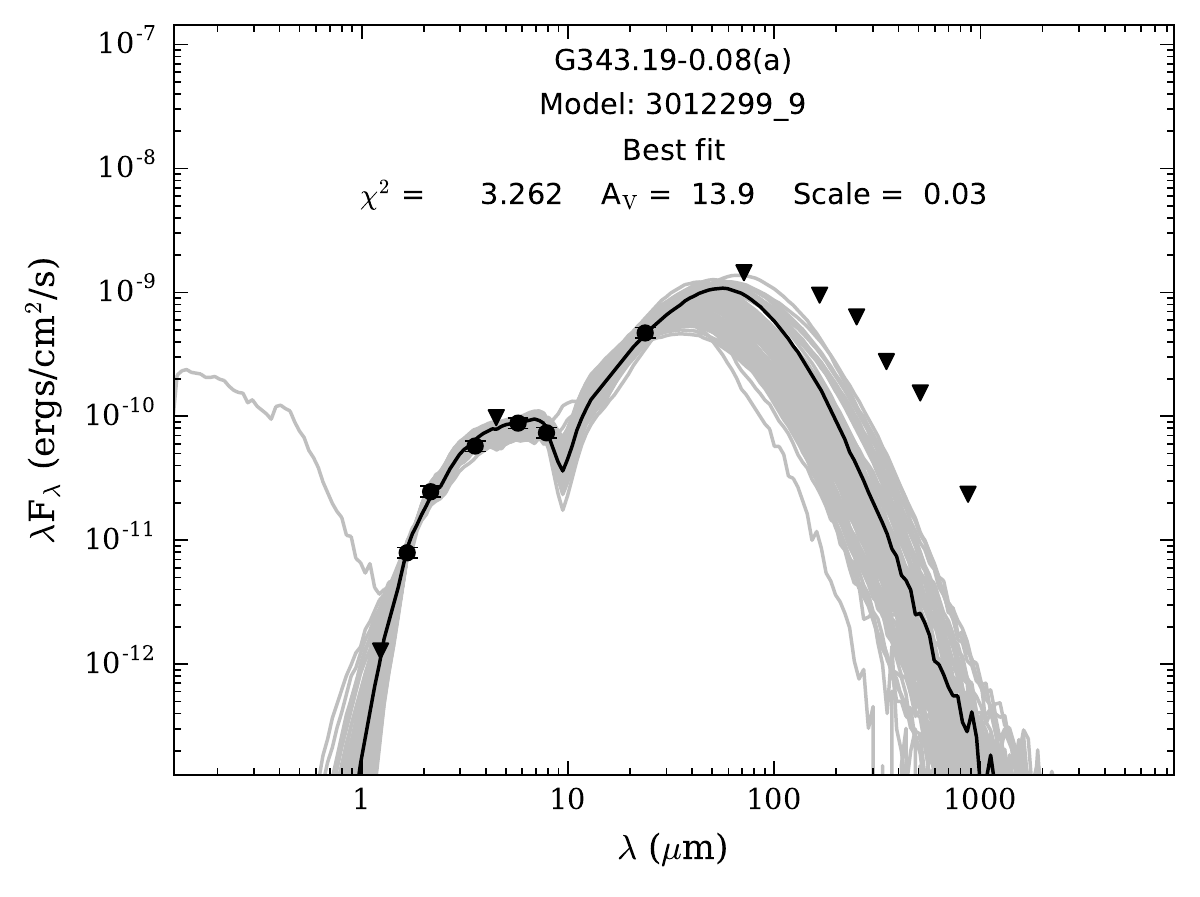}&
                \includegraphics[width=0.31\textwidth]{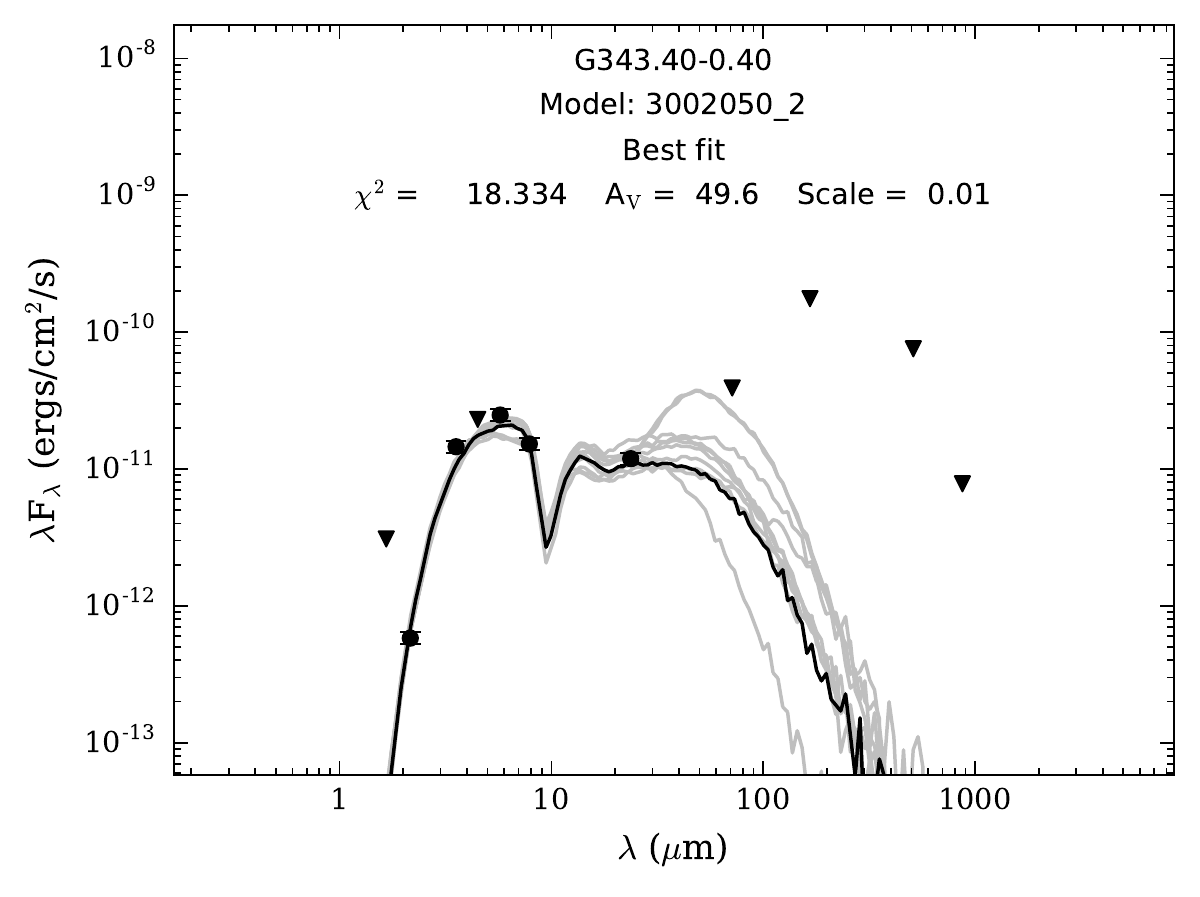}&
                \includegraphics[width=0.31\textwidth]{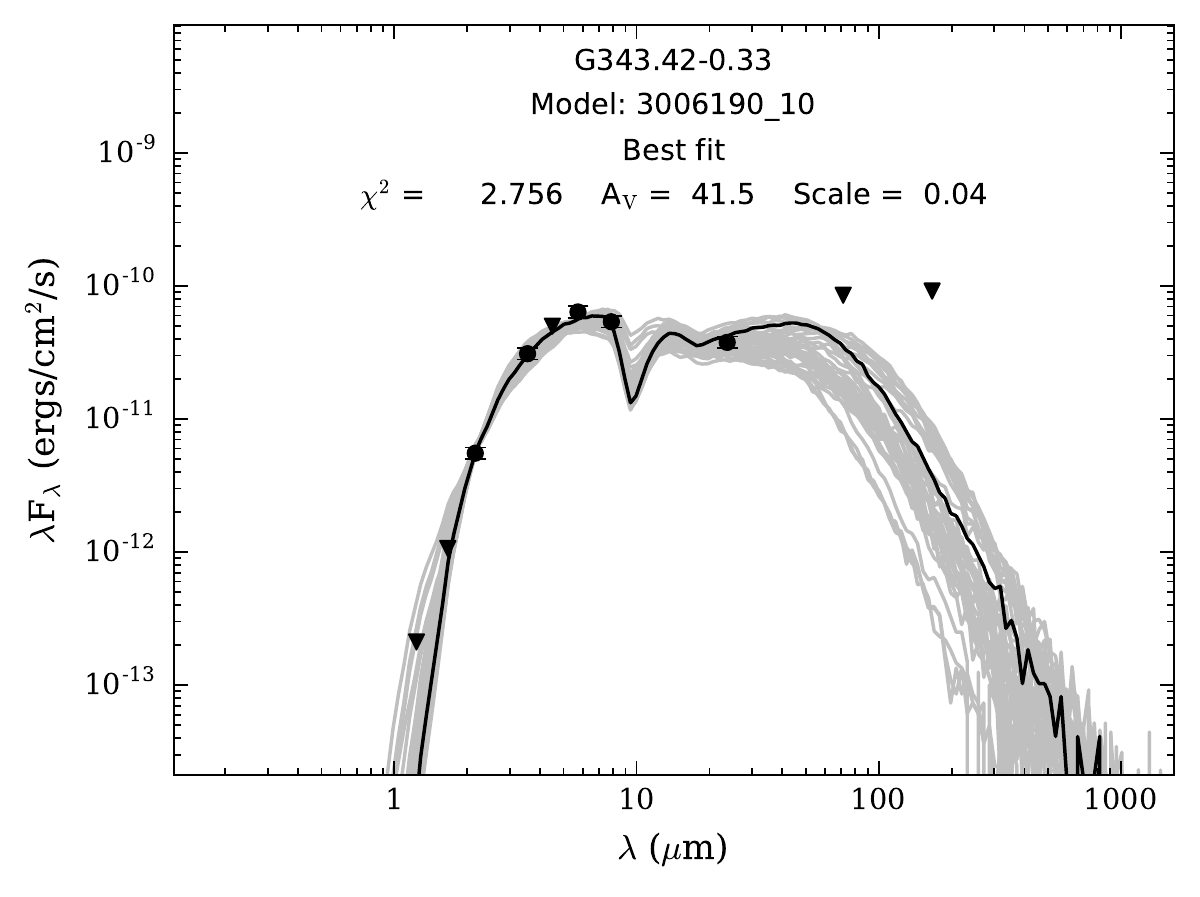}\\
                \includegraphics[width=0.31\textwidth]{{./G343.50-0.47}.pdf}&
                \includegraphics[width=0.31\textwidth]{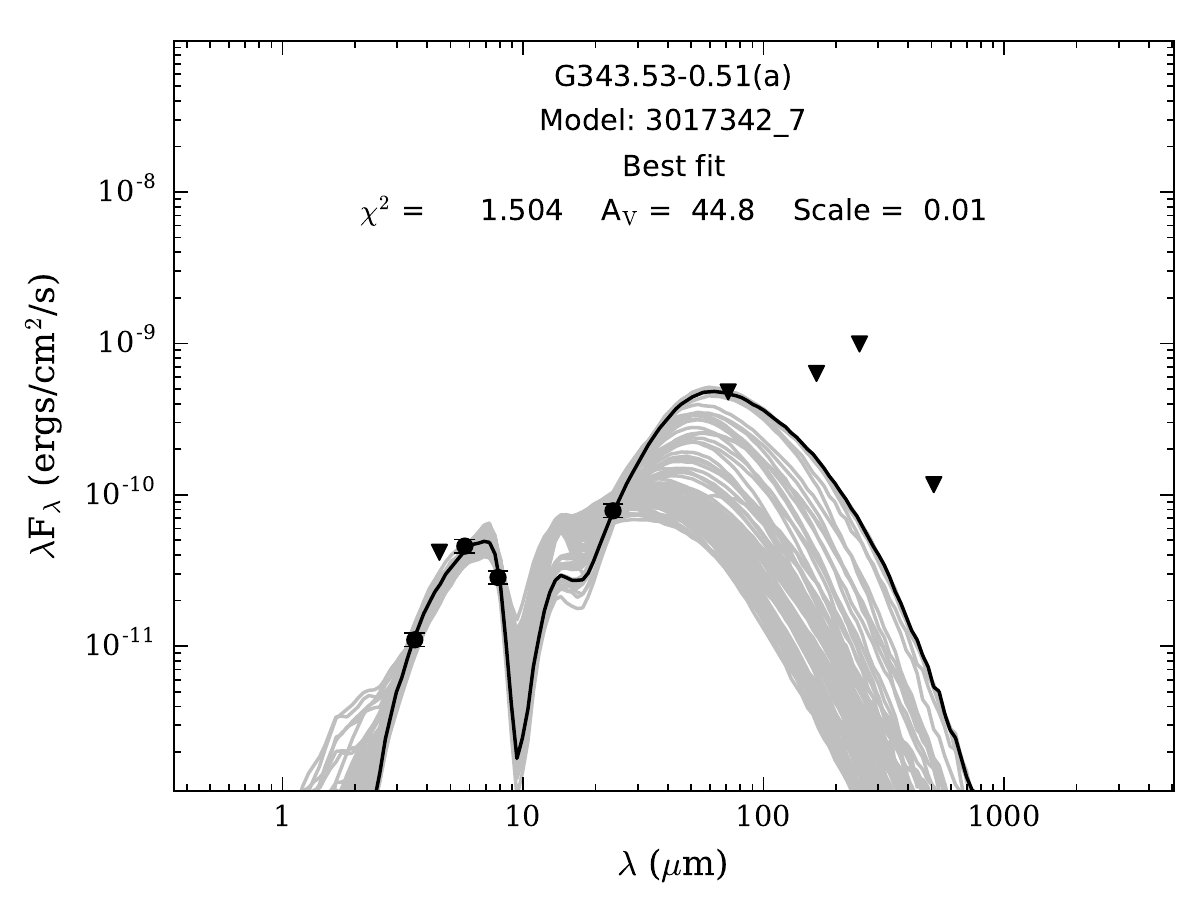}&
                \includegraphics[width=0.31\textwidth]{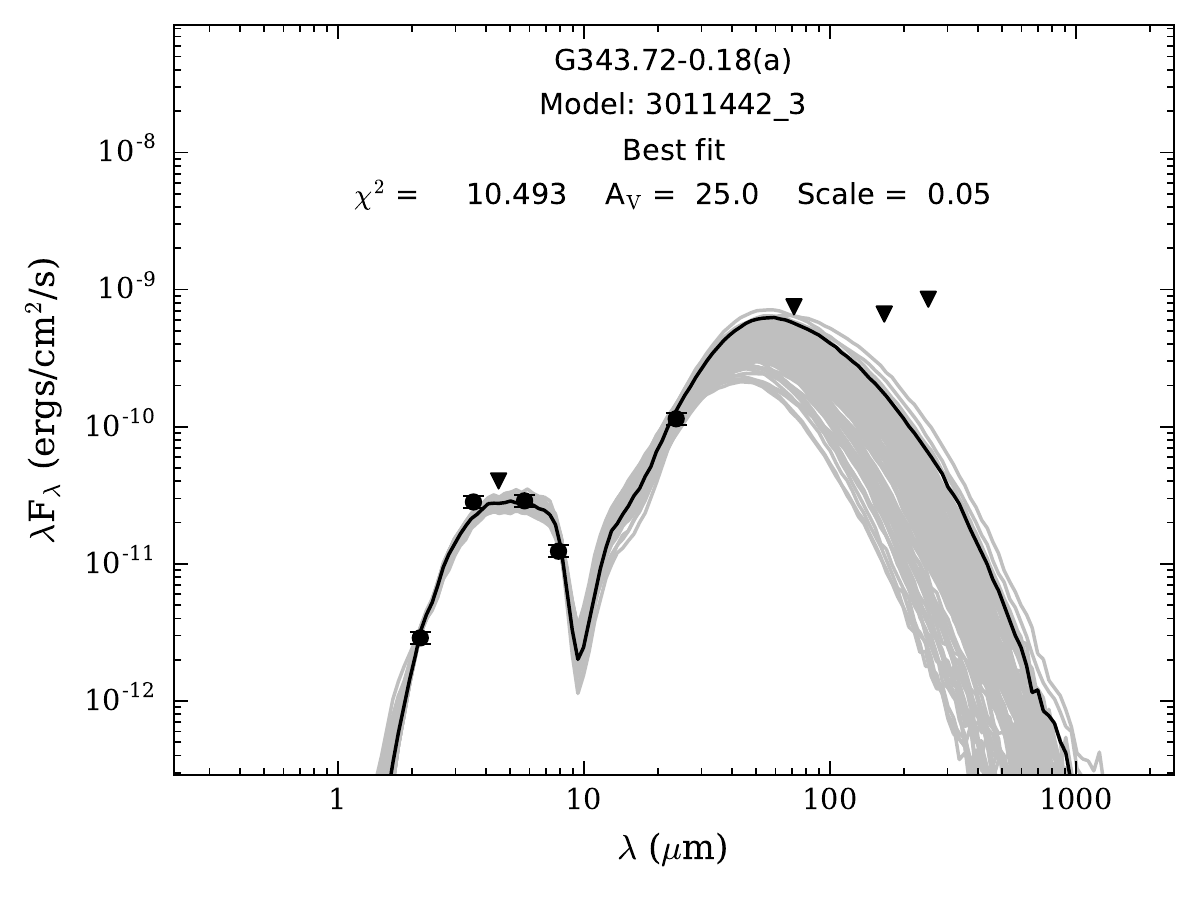}\\
                \includegraphics[width=0.31\textwidth]{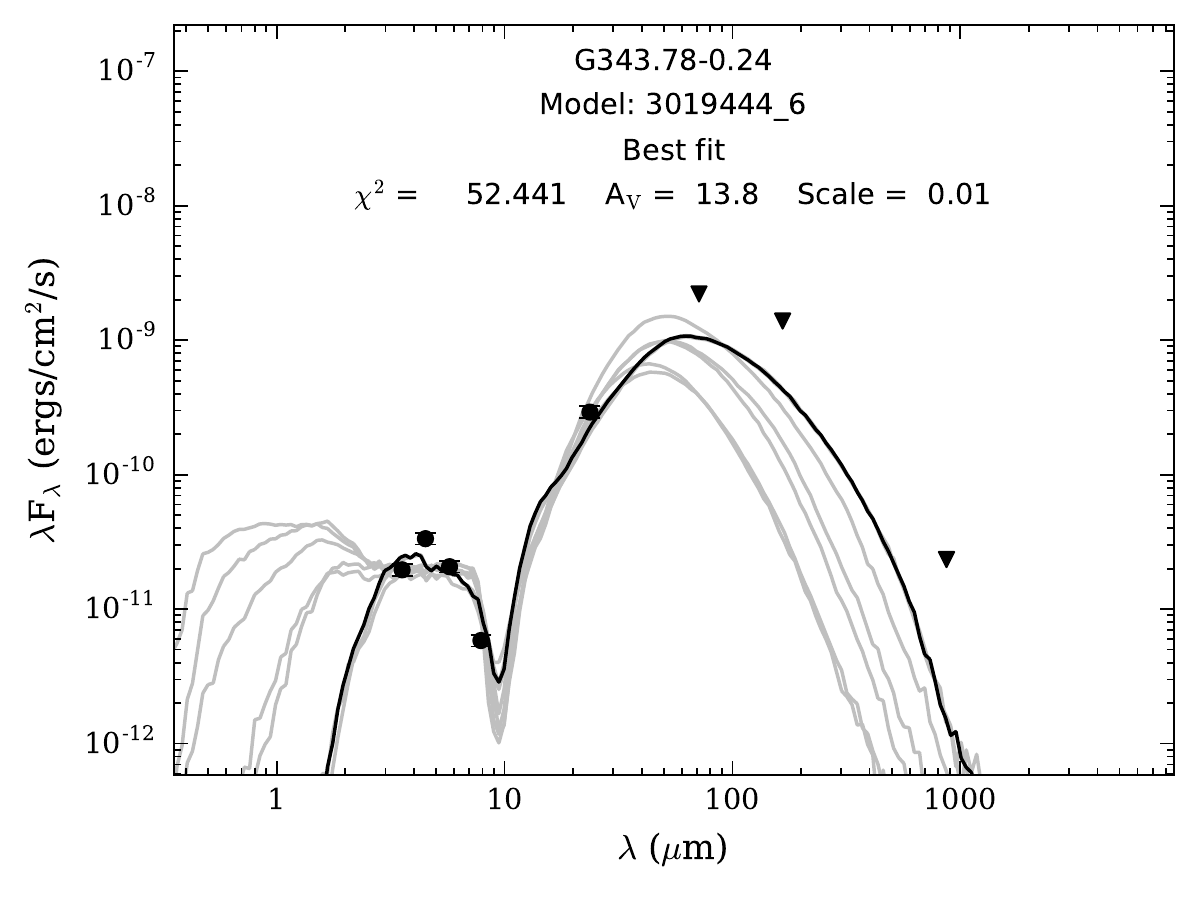}&
                \includegraphics[width=0.31\textwidth]{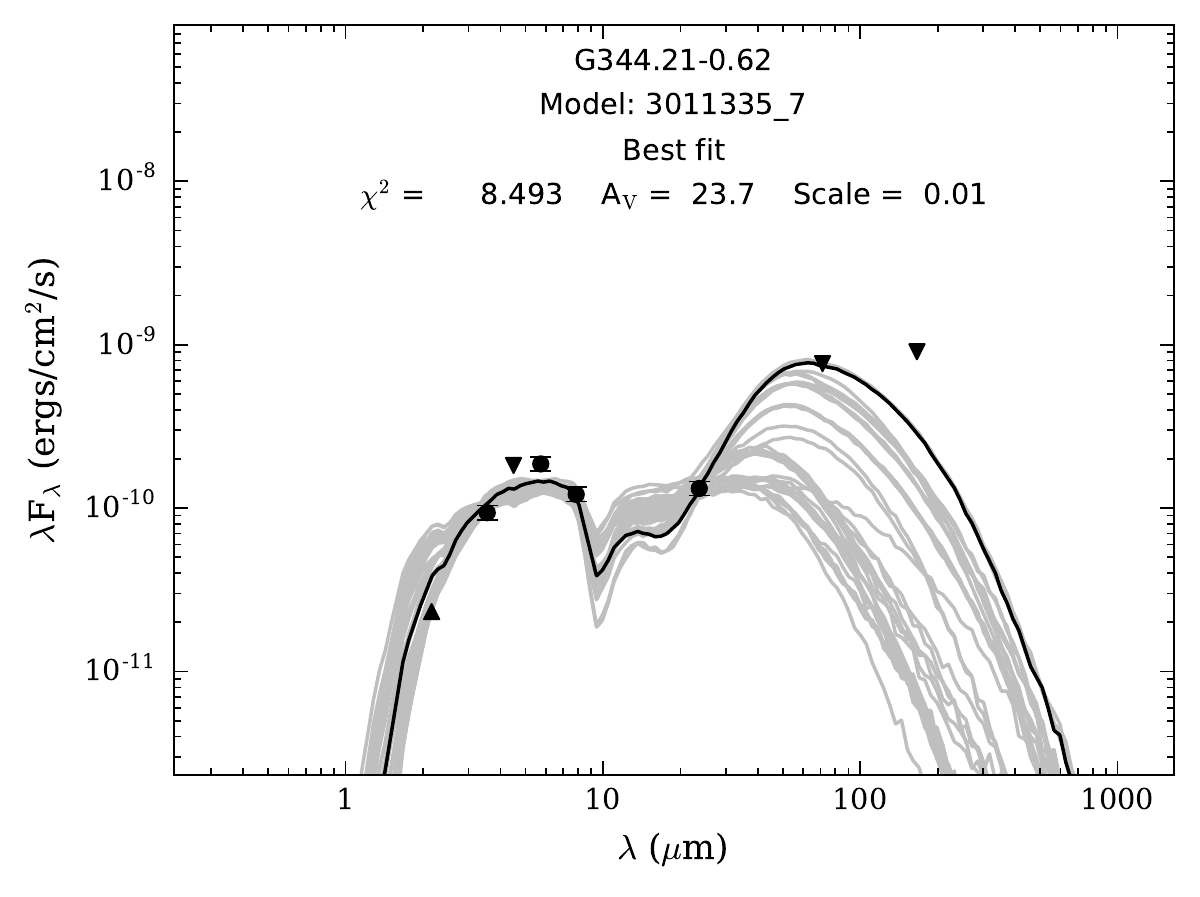}&
                \includegraphics[width=0.31\textwidth]{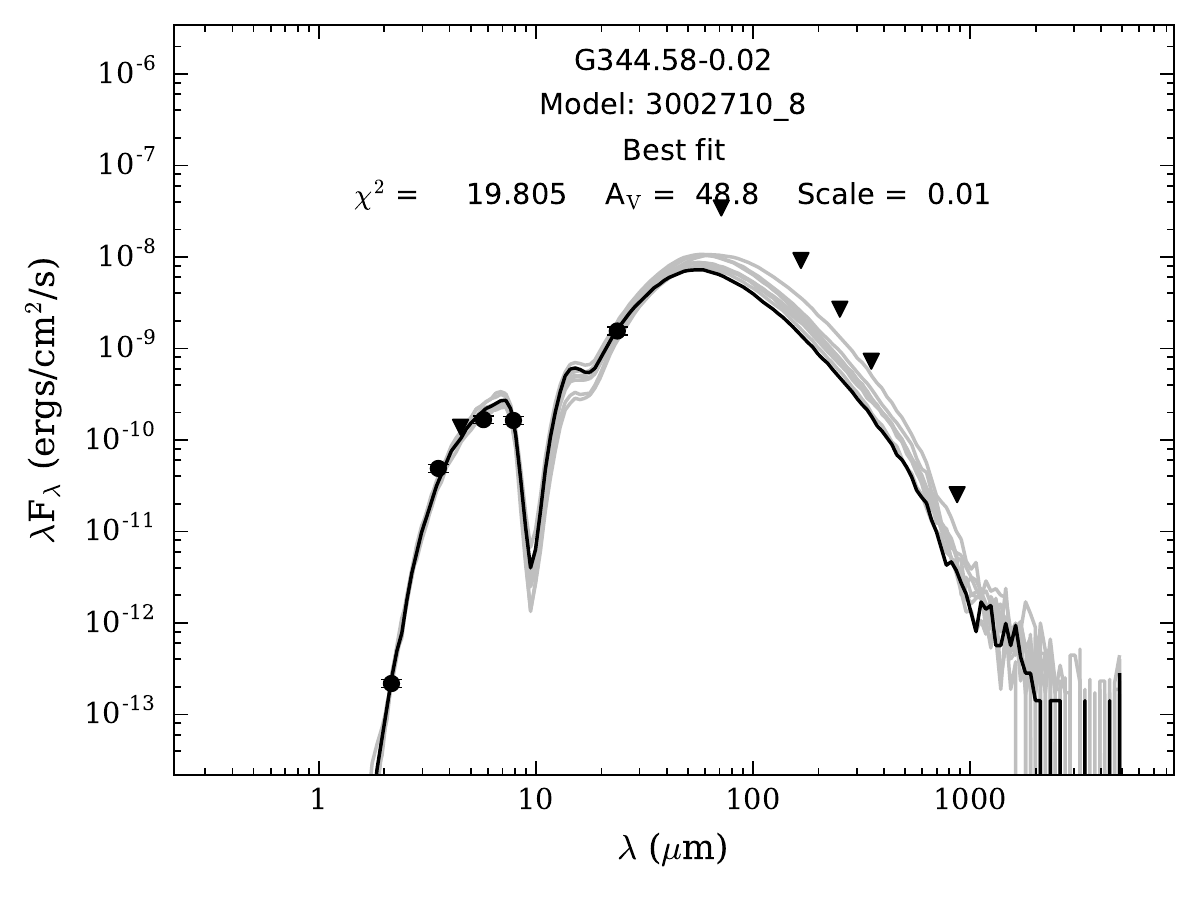}\\
                \includegraphics[width=0.31\textwidth]{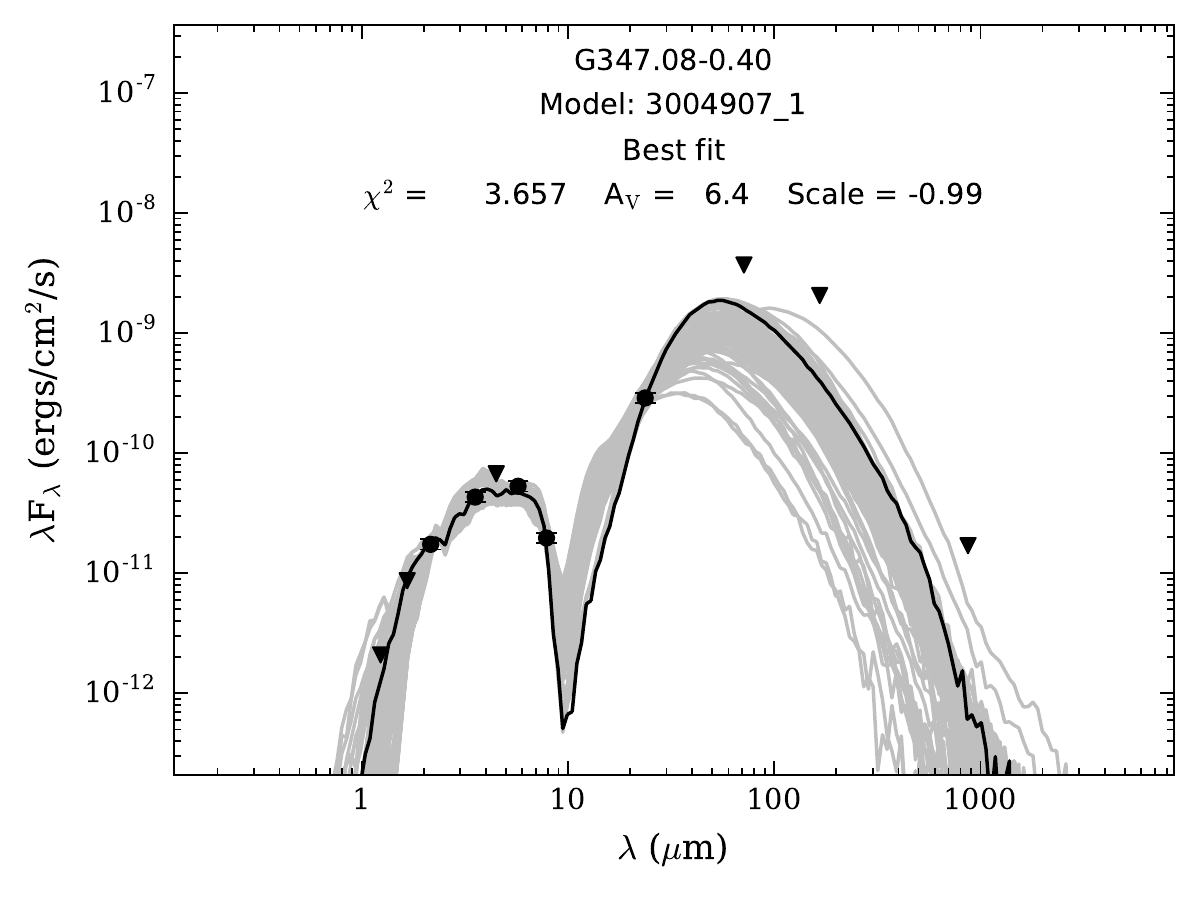}&
                \includegraphics[width=0.31\textwidth]{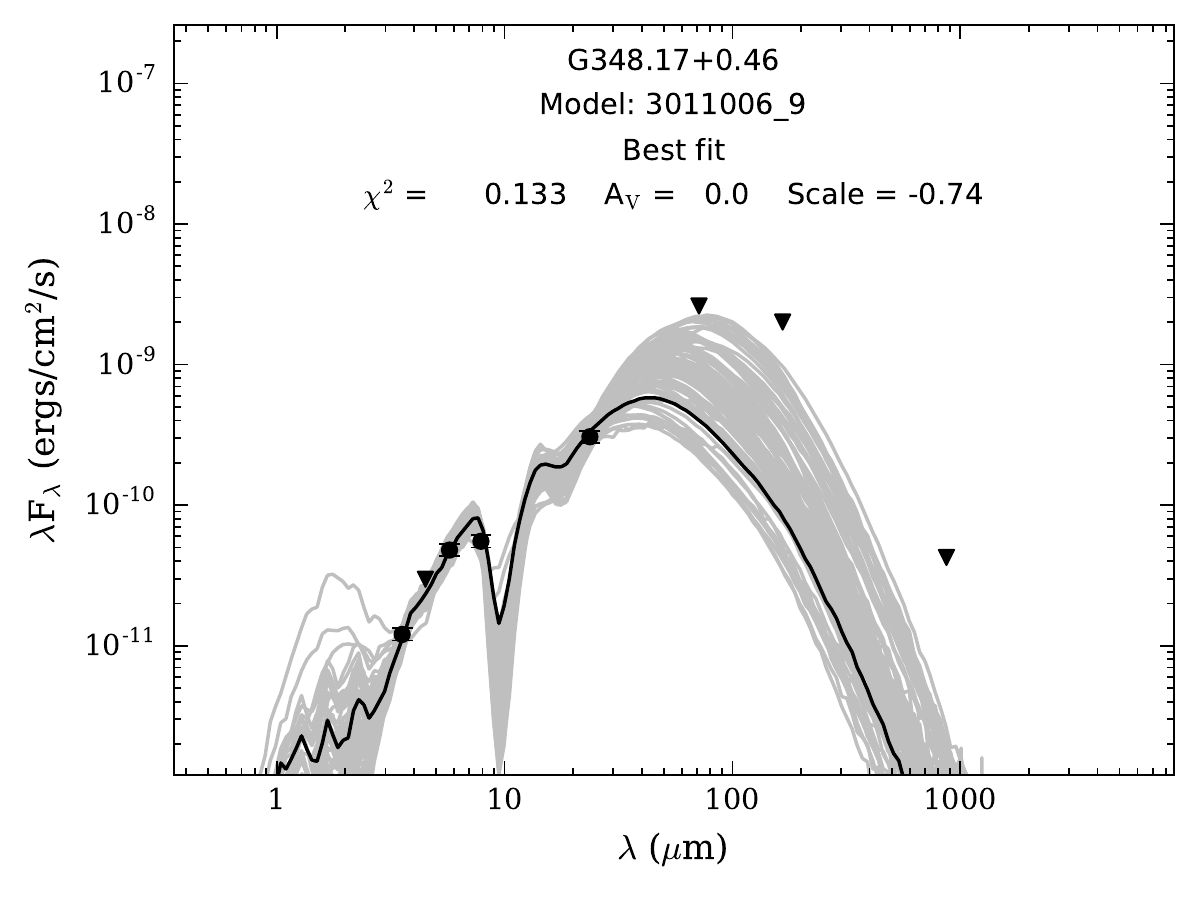}&
                \includegraphics[width=0.31\textwidth]{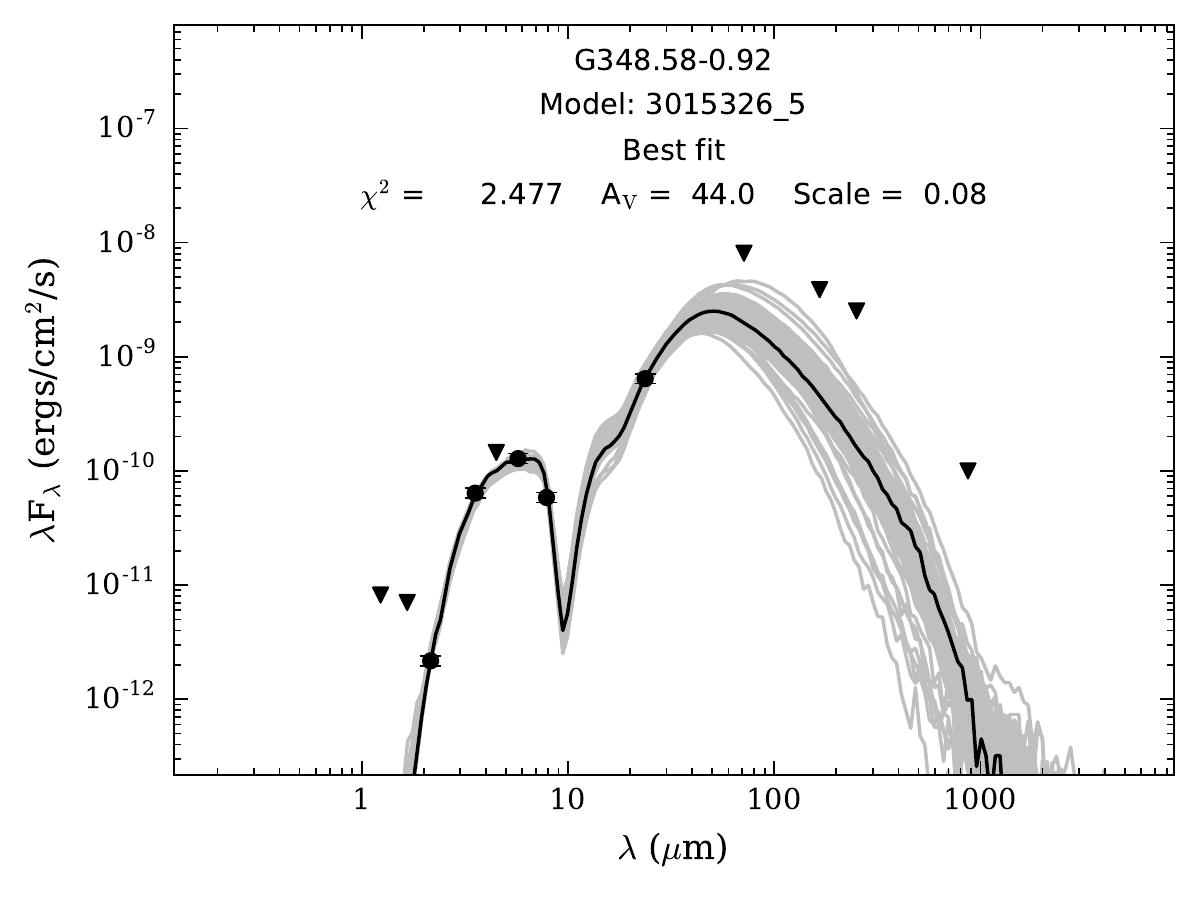}\\
        \end{tabular}
        \caption{Continuation of Fig. \ref{fig:testgridsed}.}
\end{figure}

\begin{figure}
        \begin{tabular}{ccc}
                
                \includegraphics[width=0.31\textwidth]{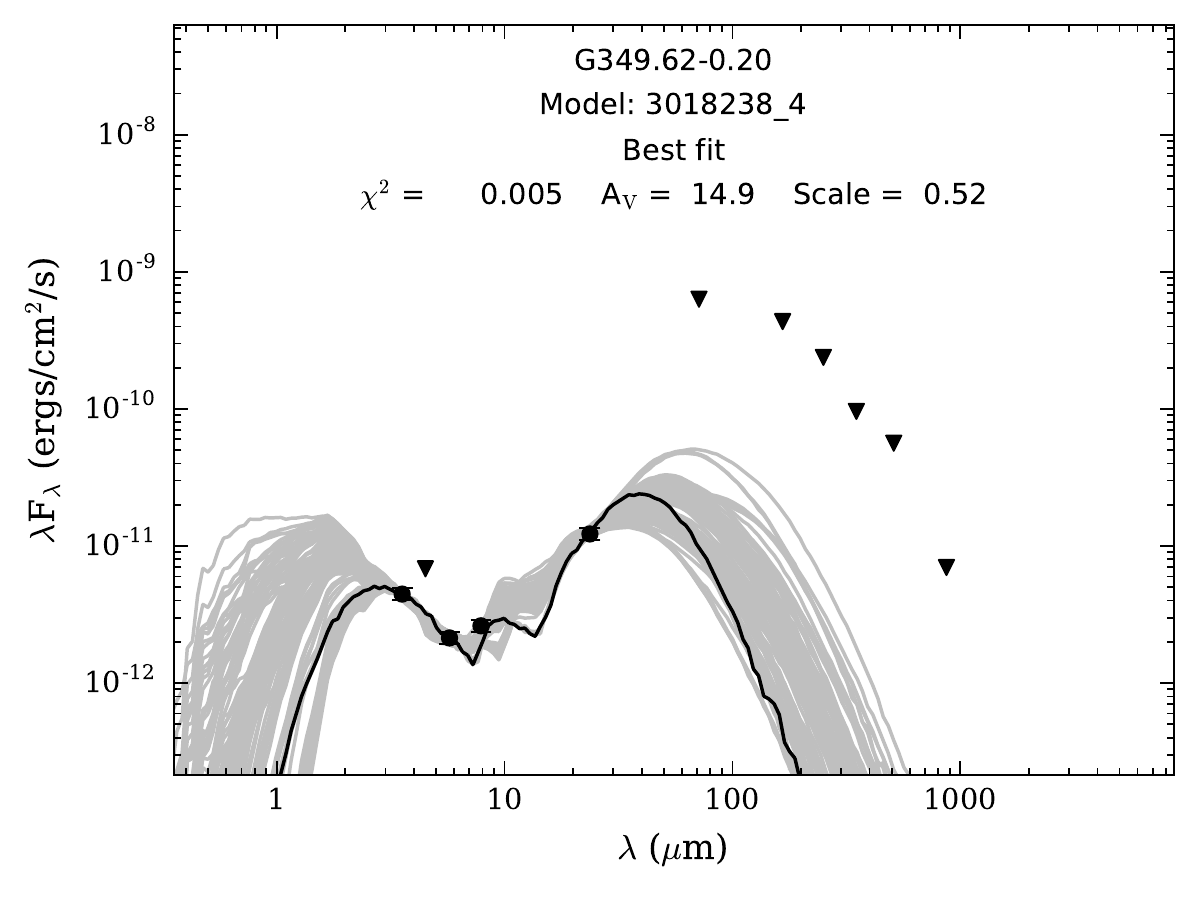}&
                \includegraphics[width=0.31\textwidth]{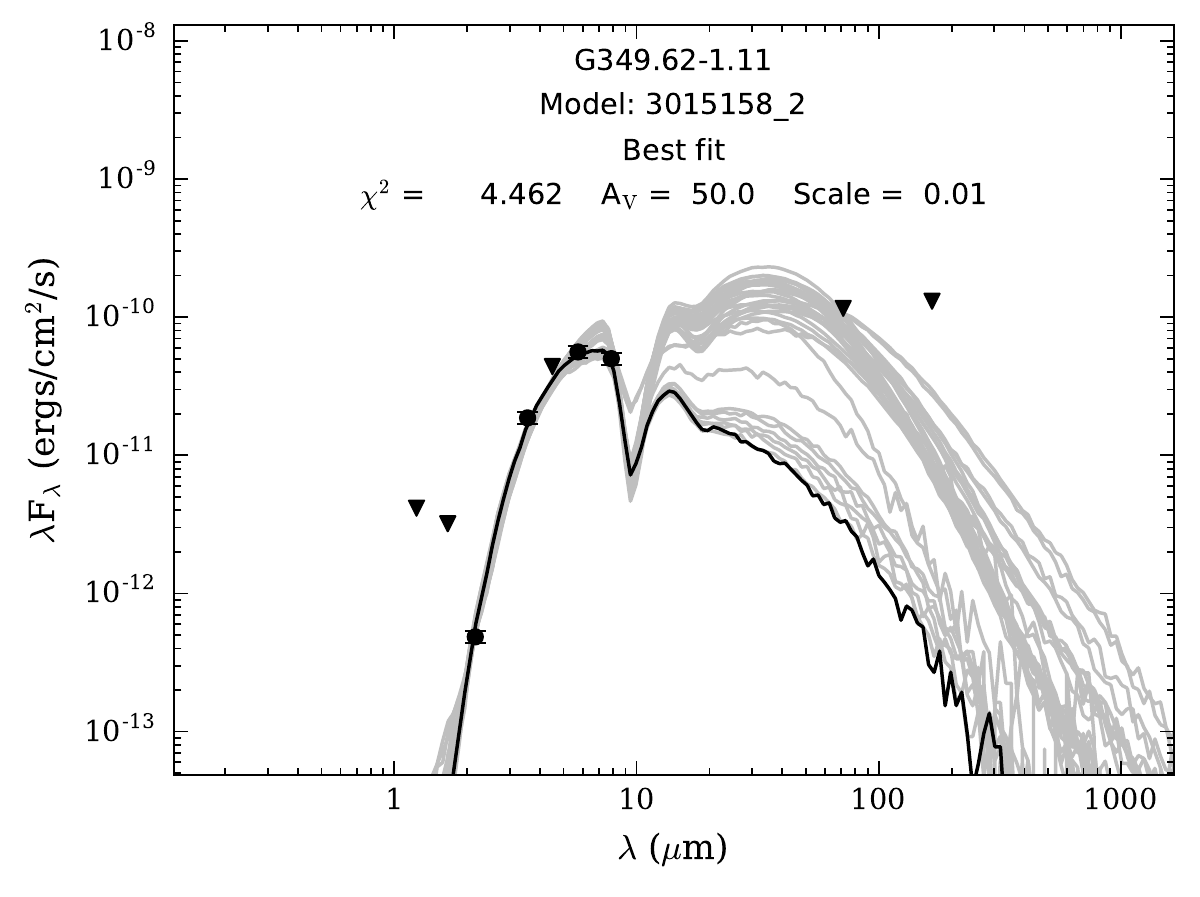}&
                \includegraphics[width=0.31\textwidth]{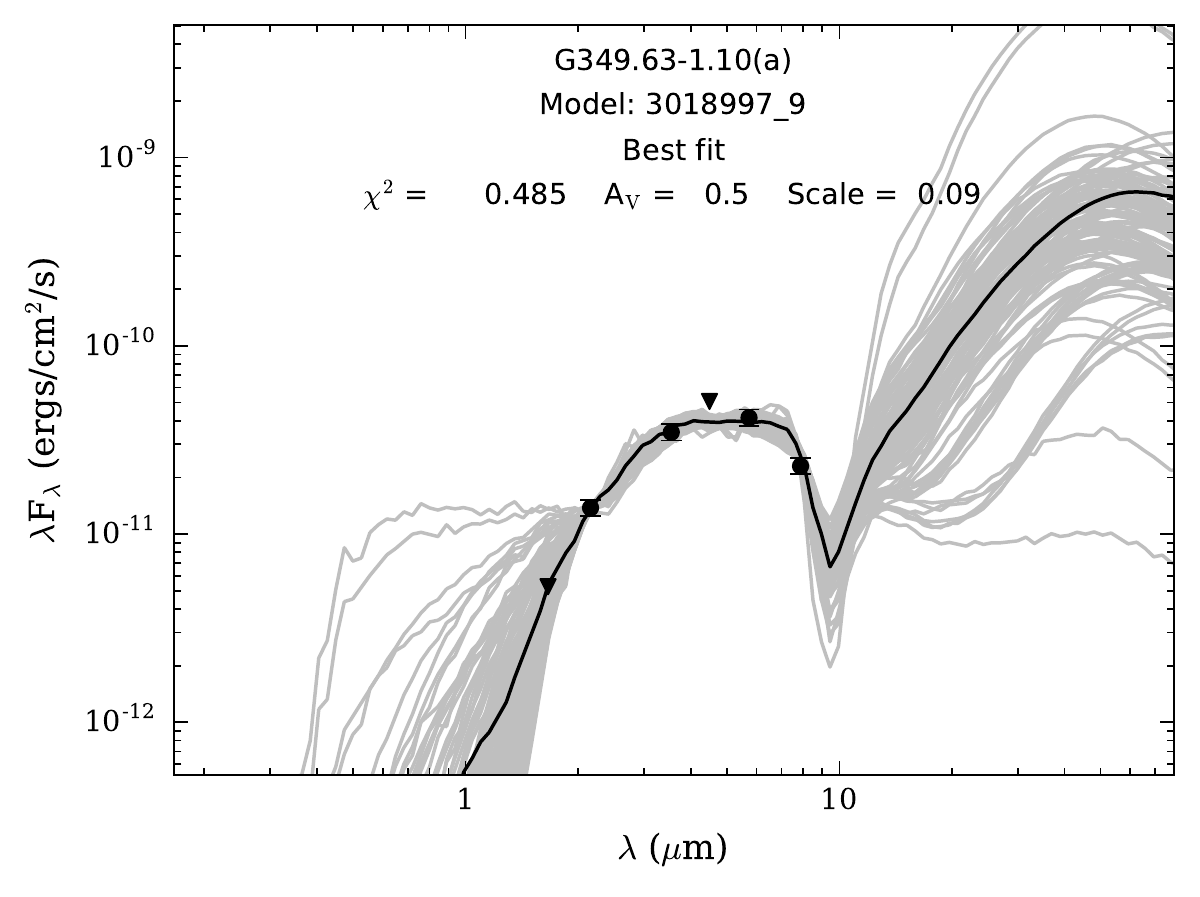}\\
                \includegraphics[width=0.31\textwidth]{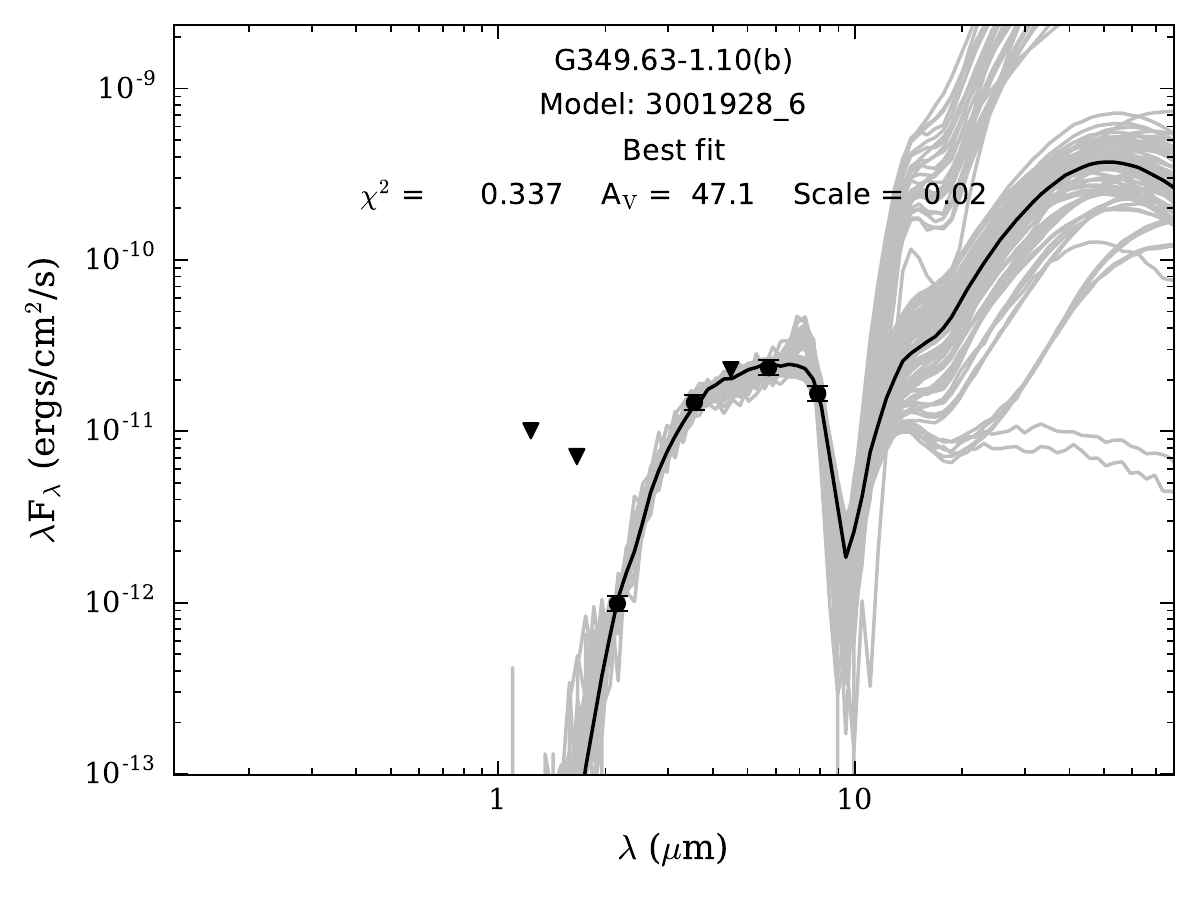}&
                \includegraphics[width=0.31\textwidth]{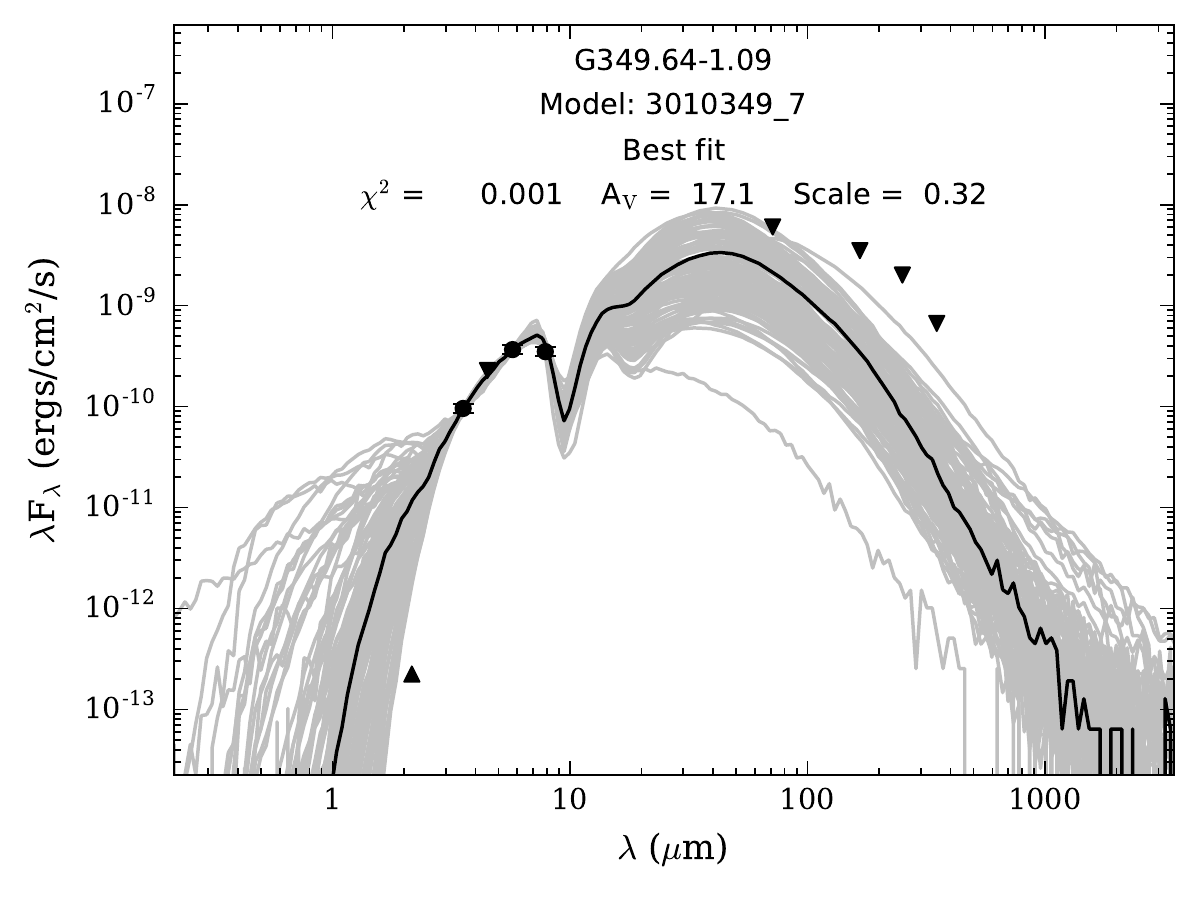}&
                \includegraphics[width=0.31\textwidth]{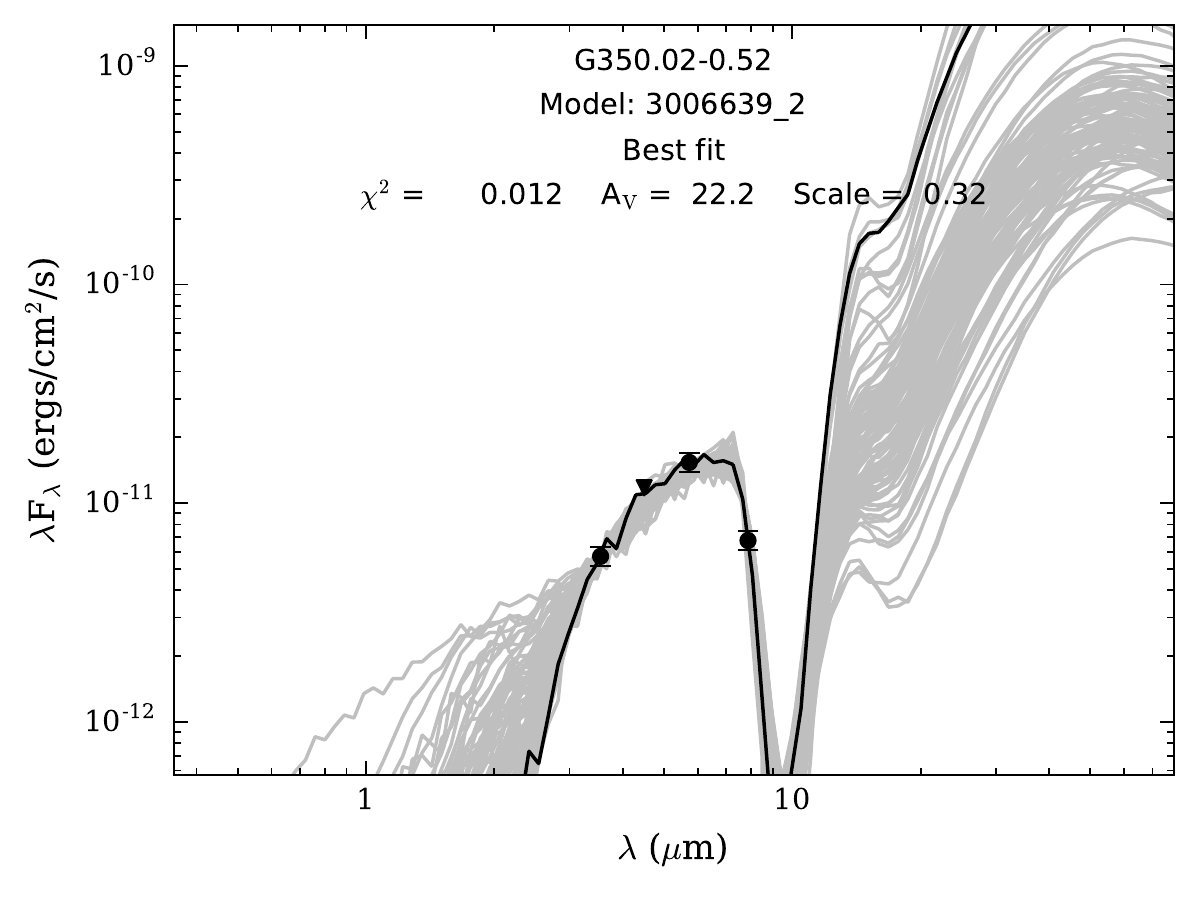}\\
                \includegraphics[width=0.31\textwidth]{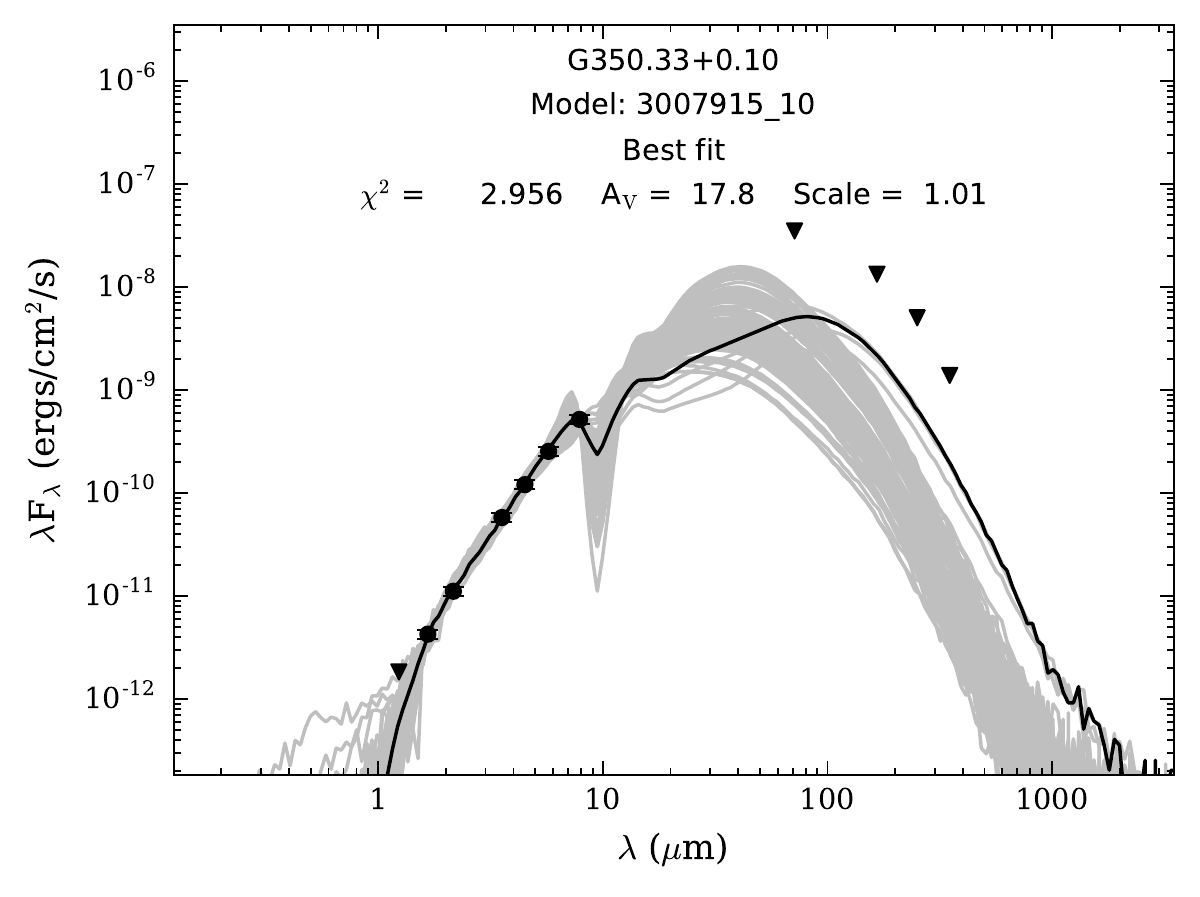}&
                \includegraphics[width=0.31\textwidth]{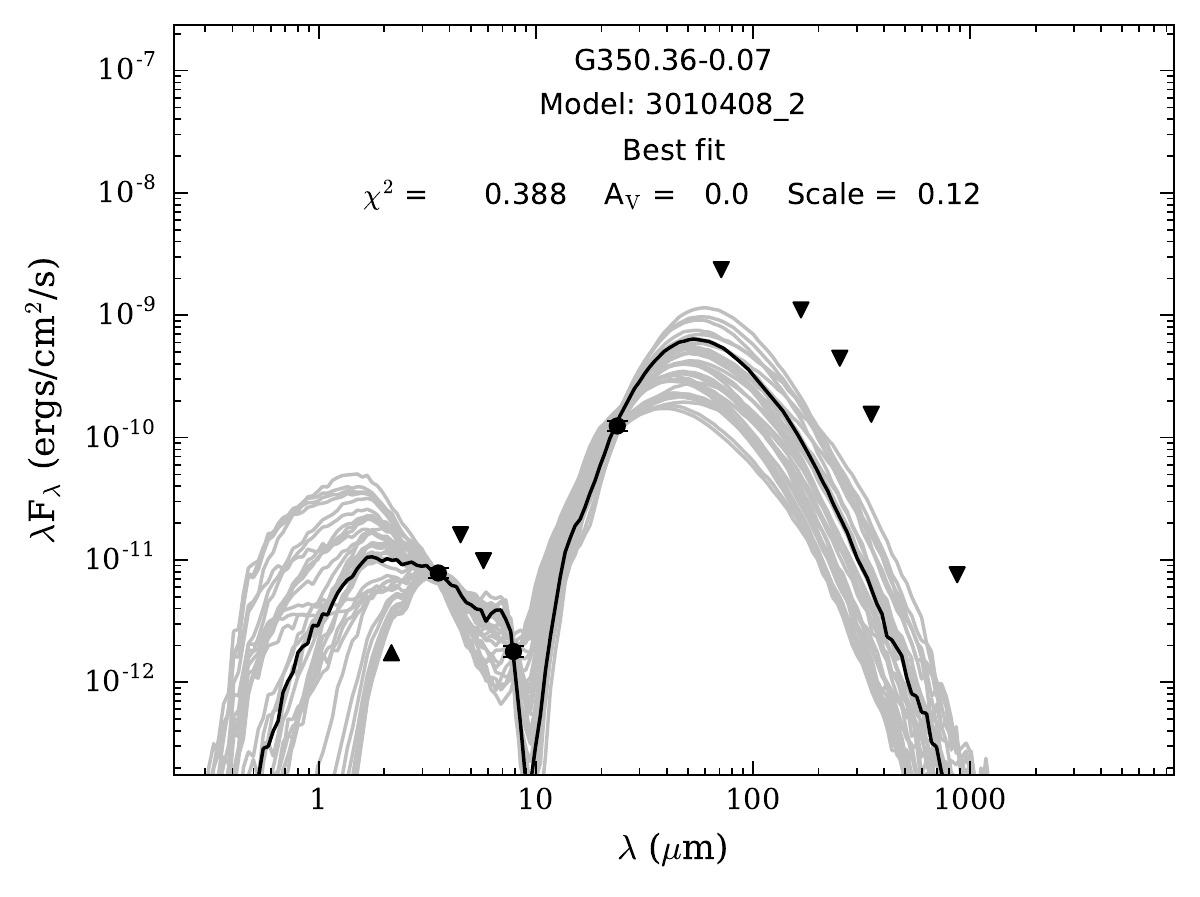}&
                \includegraphics[width=0.31\textwidth]{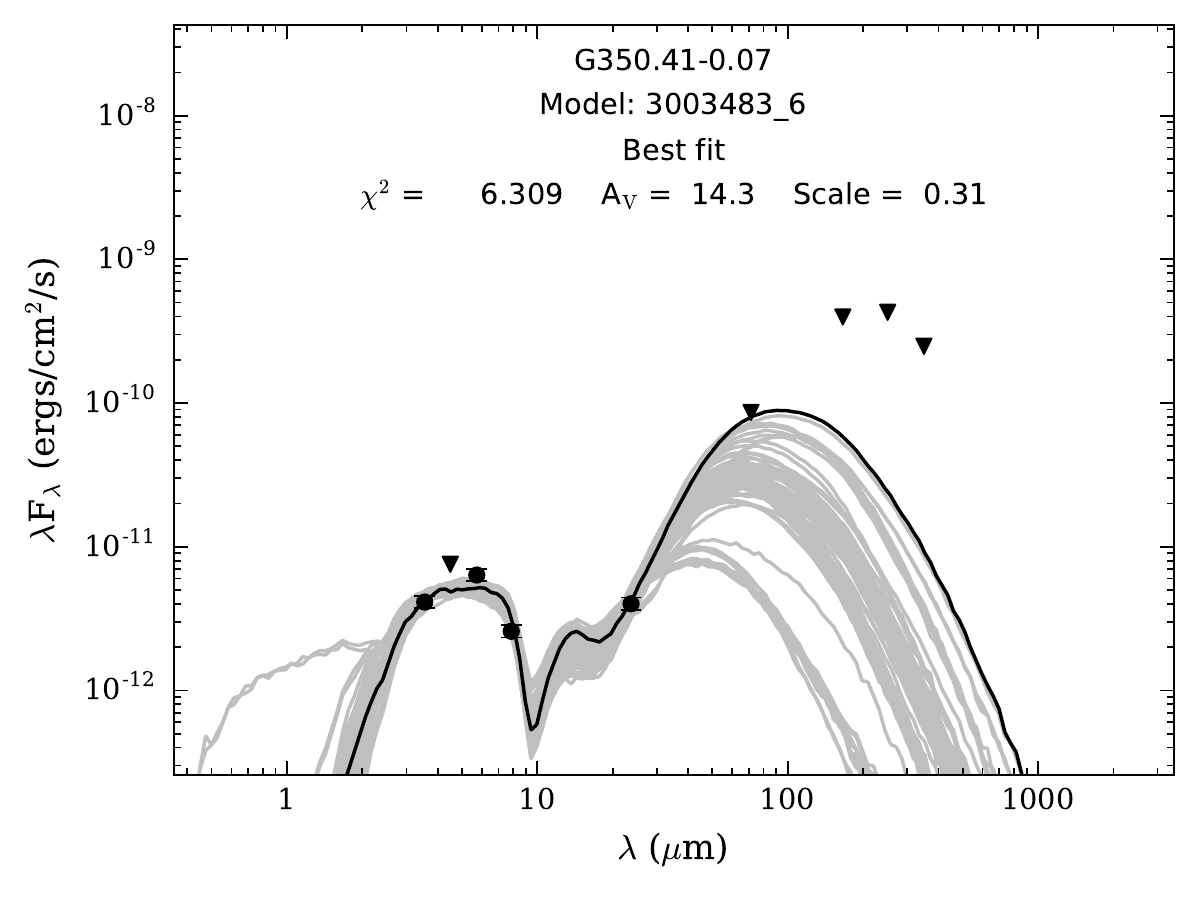}\\
                \includegraphics[width=0.31\textwidth]{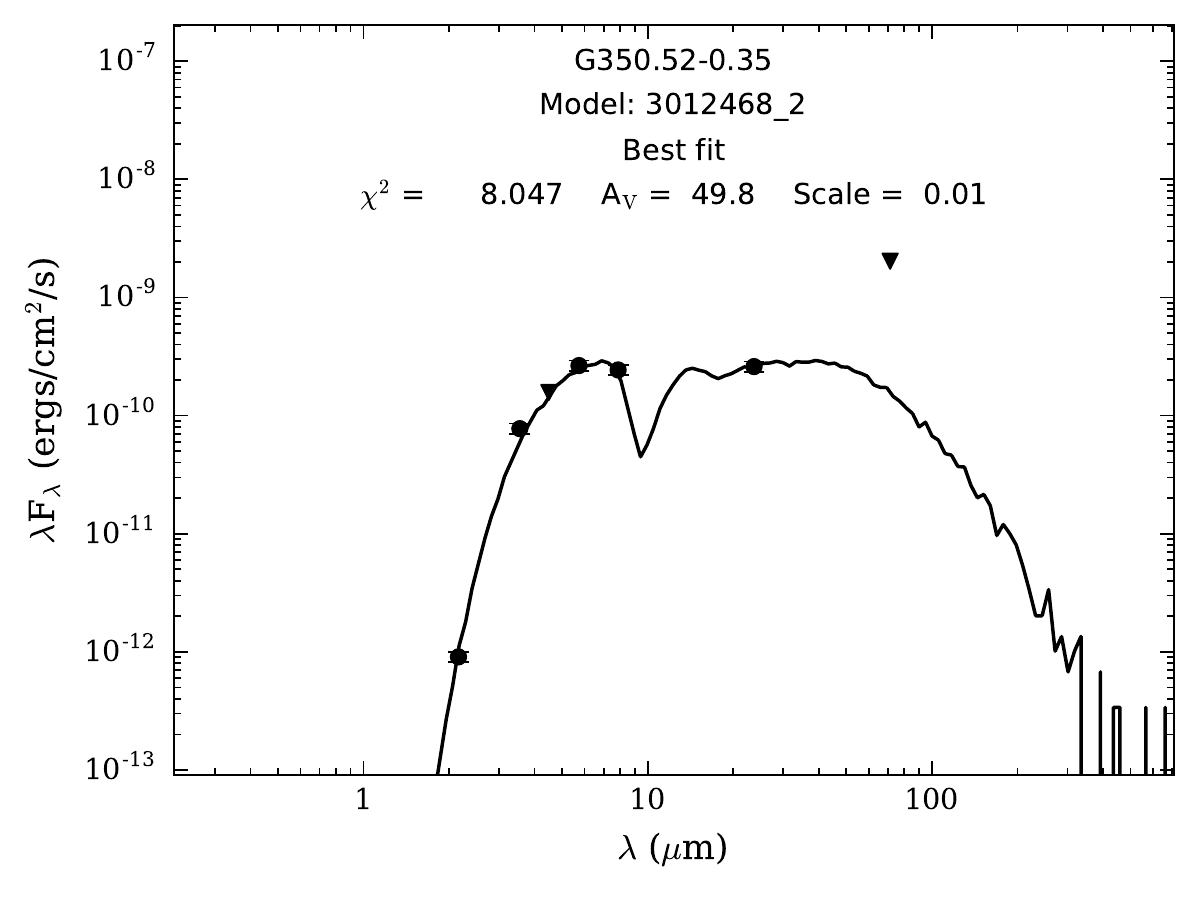}&
                \includegraphics[width=0.31\textwidth]{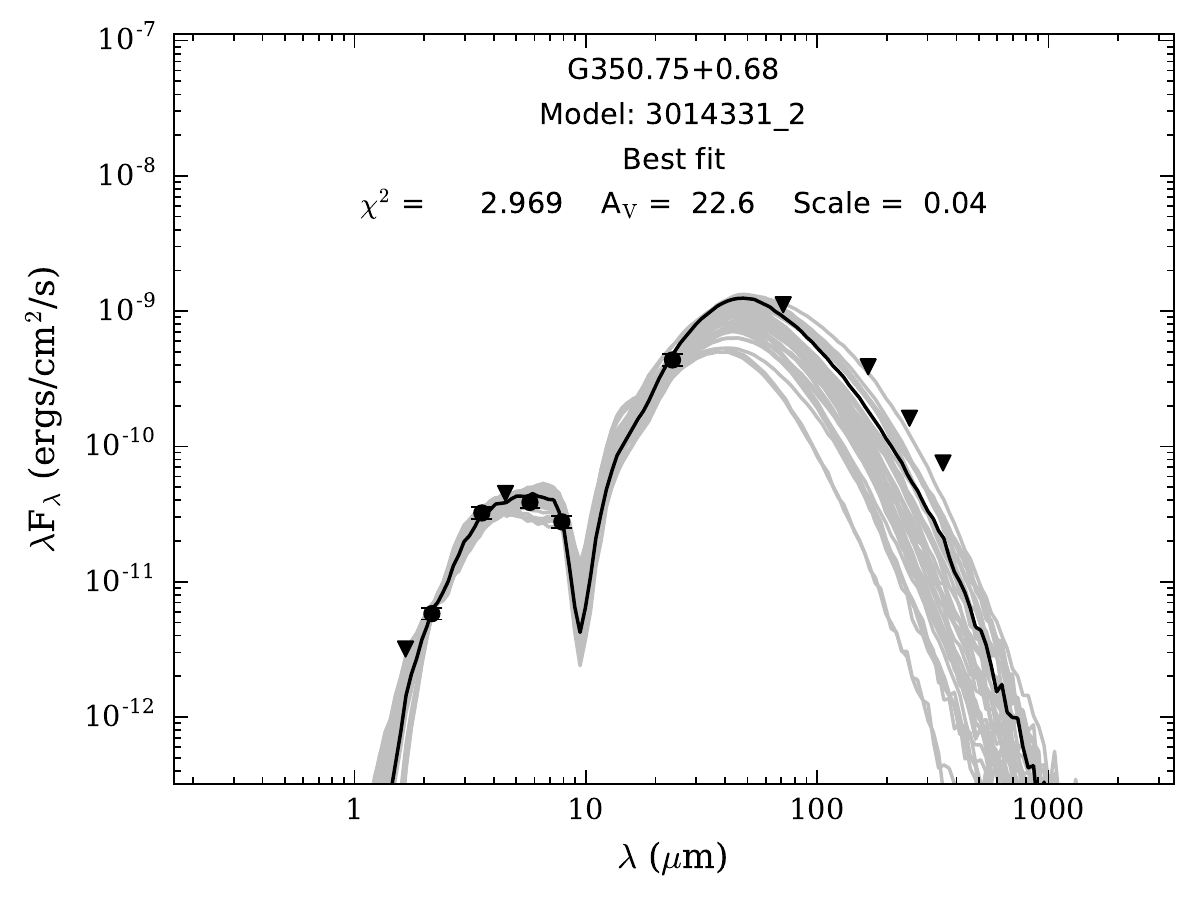}&
                \includegraphics[width=0.31\textwidth]{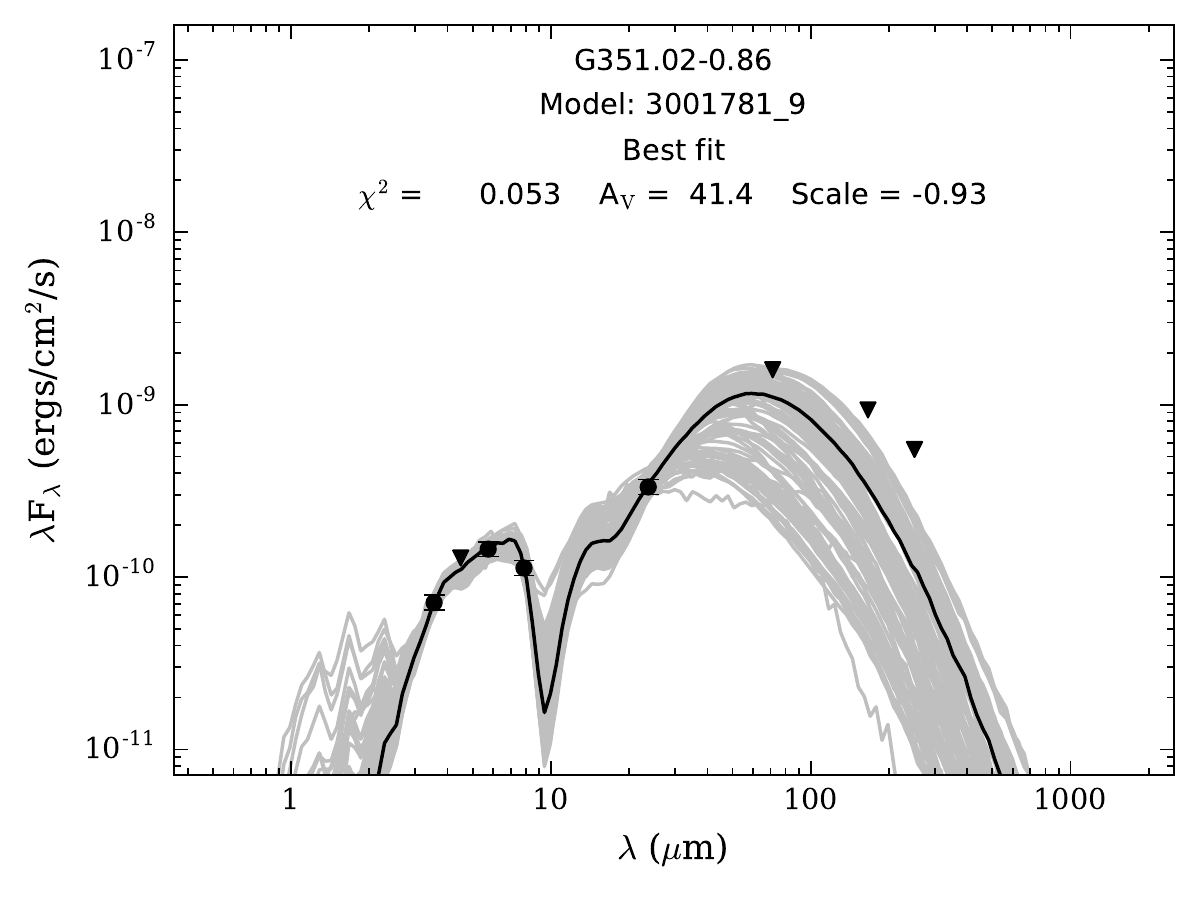}\\
                \includegraphics[width=0.31\textwidth]{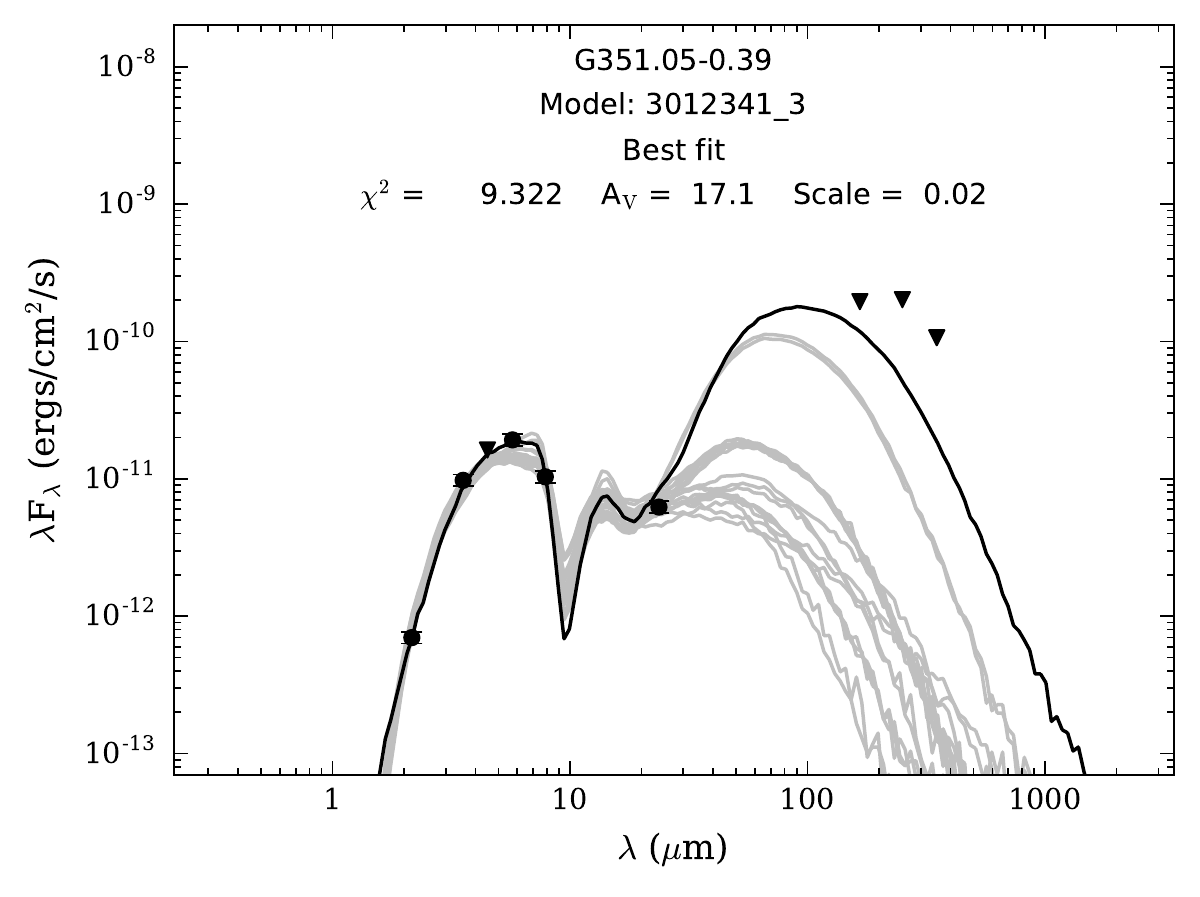}&
                \includegraphics[width=0.31\textwidth]{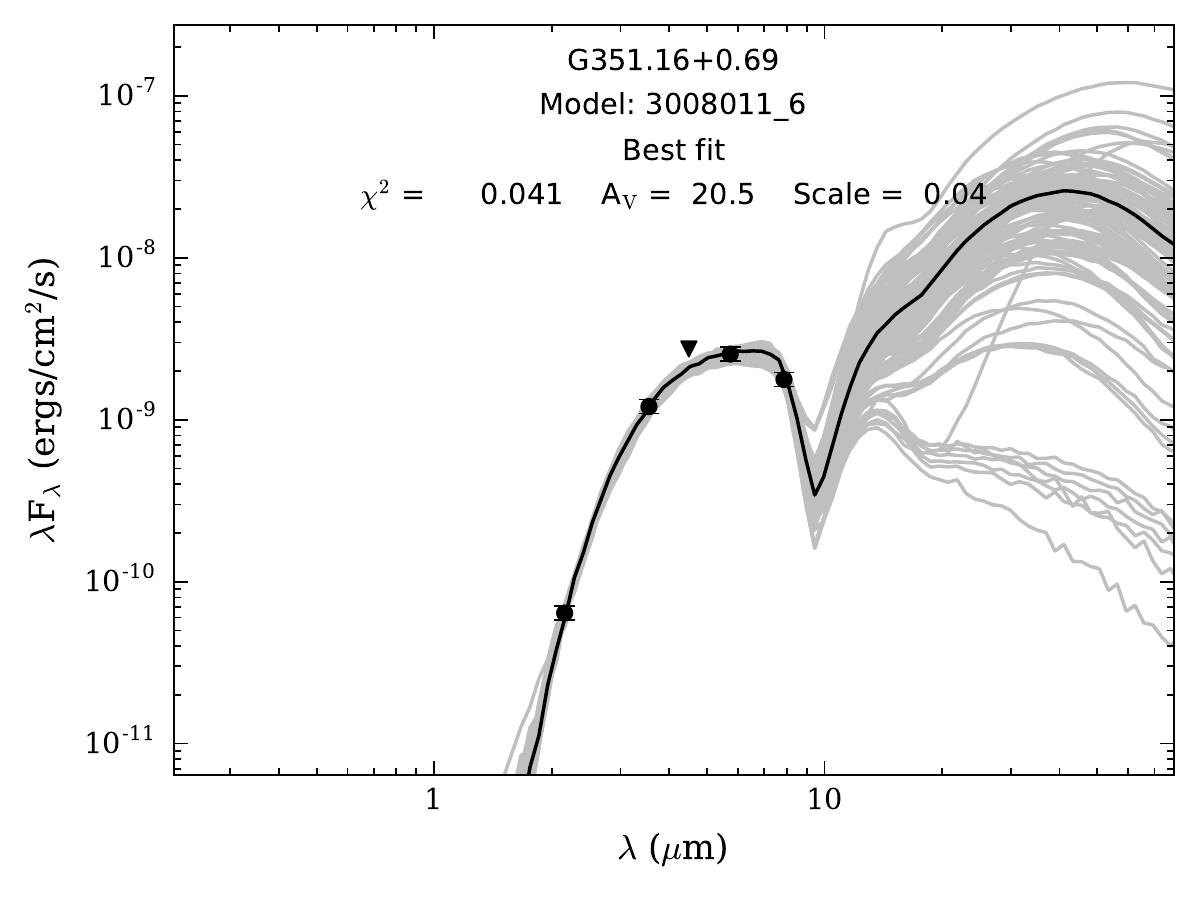}&
                \includegraphics[width=0.31\textwidth]{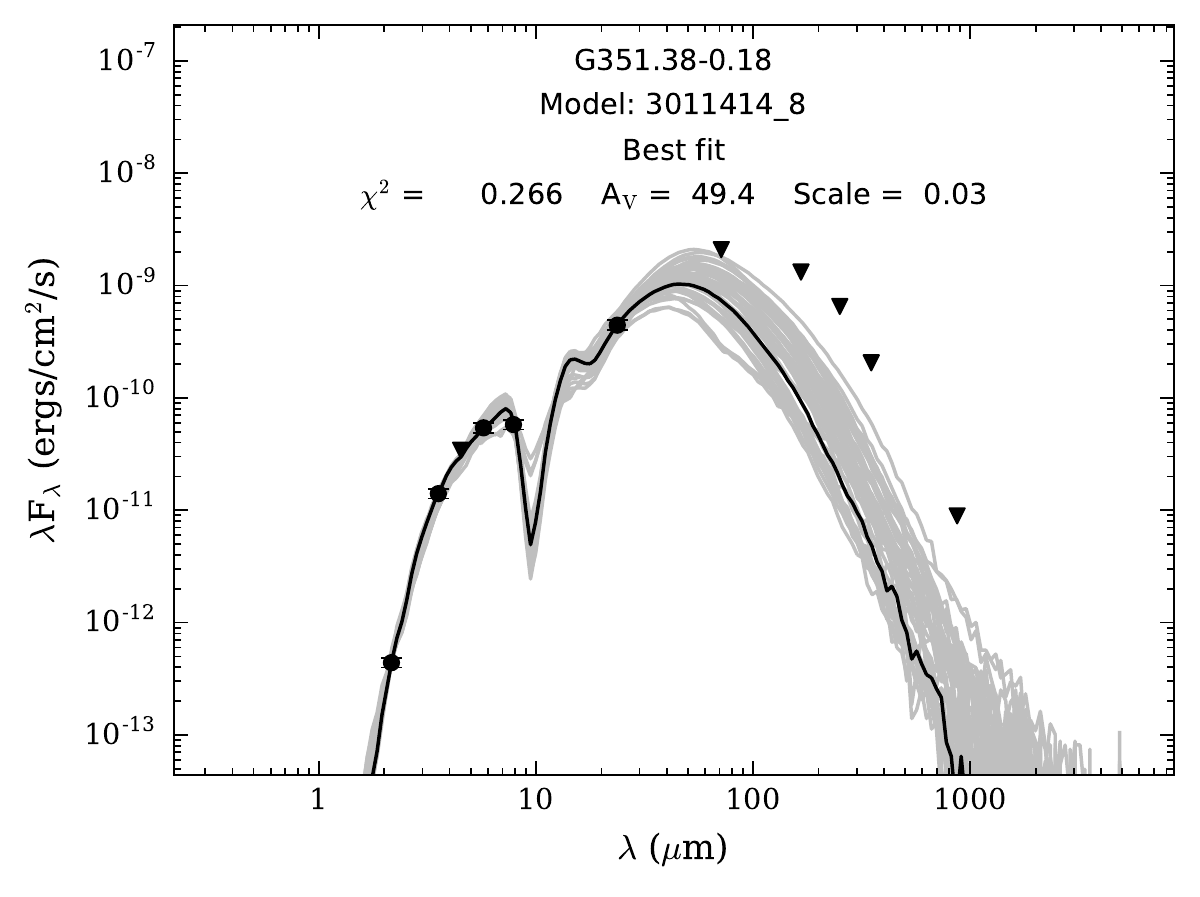}\\
        \end{tabular}
        \caption{Continuation of Fig. \ref{fig:testgridsed}.}
\end{figure}

\begin{figure}
        \begin{tabular}{ccc}
                
                \includegraphics[width=0.31\textwidth]{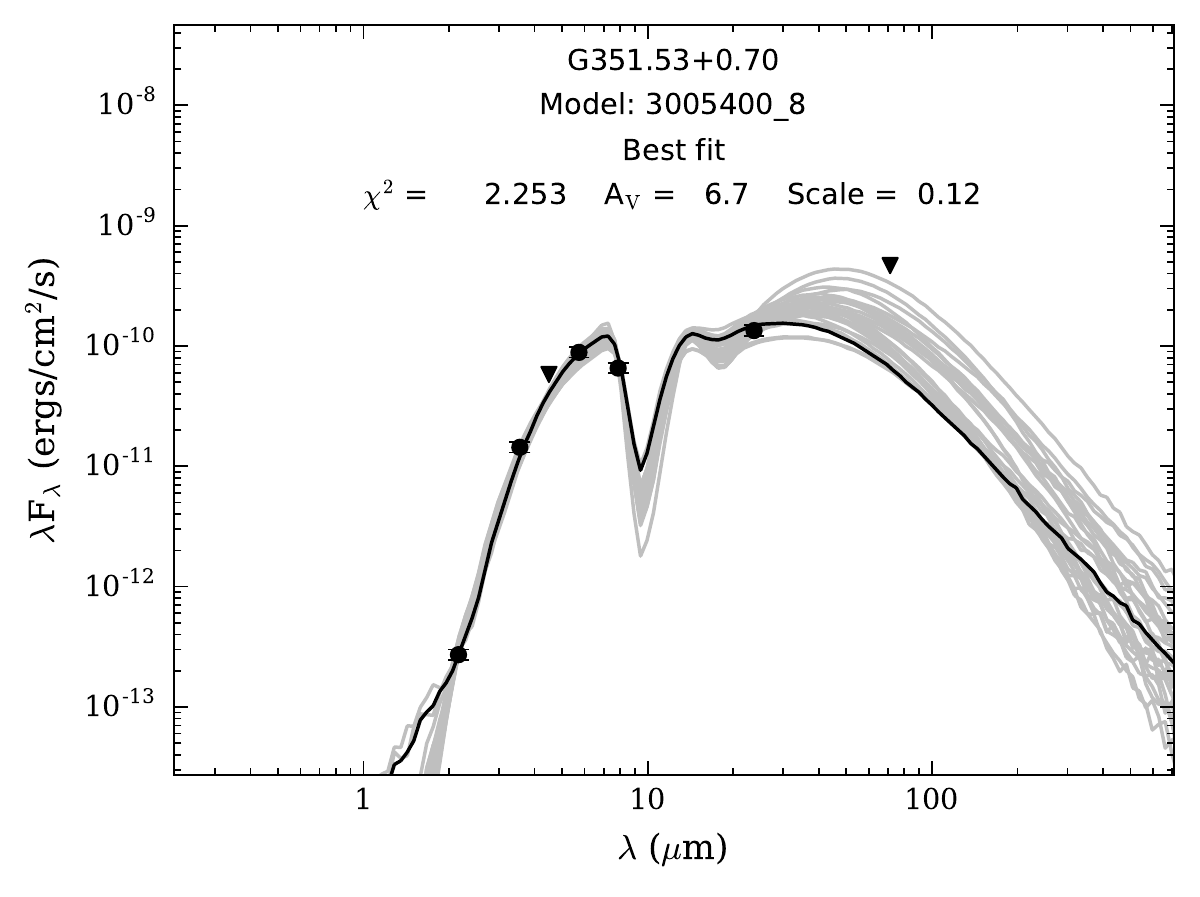}&
                \includegraphics[width=0.31\textwidth]{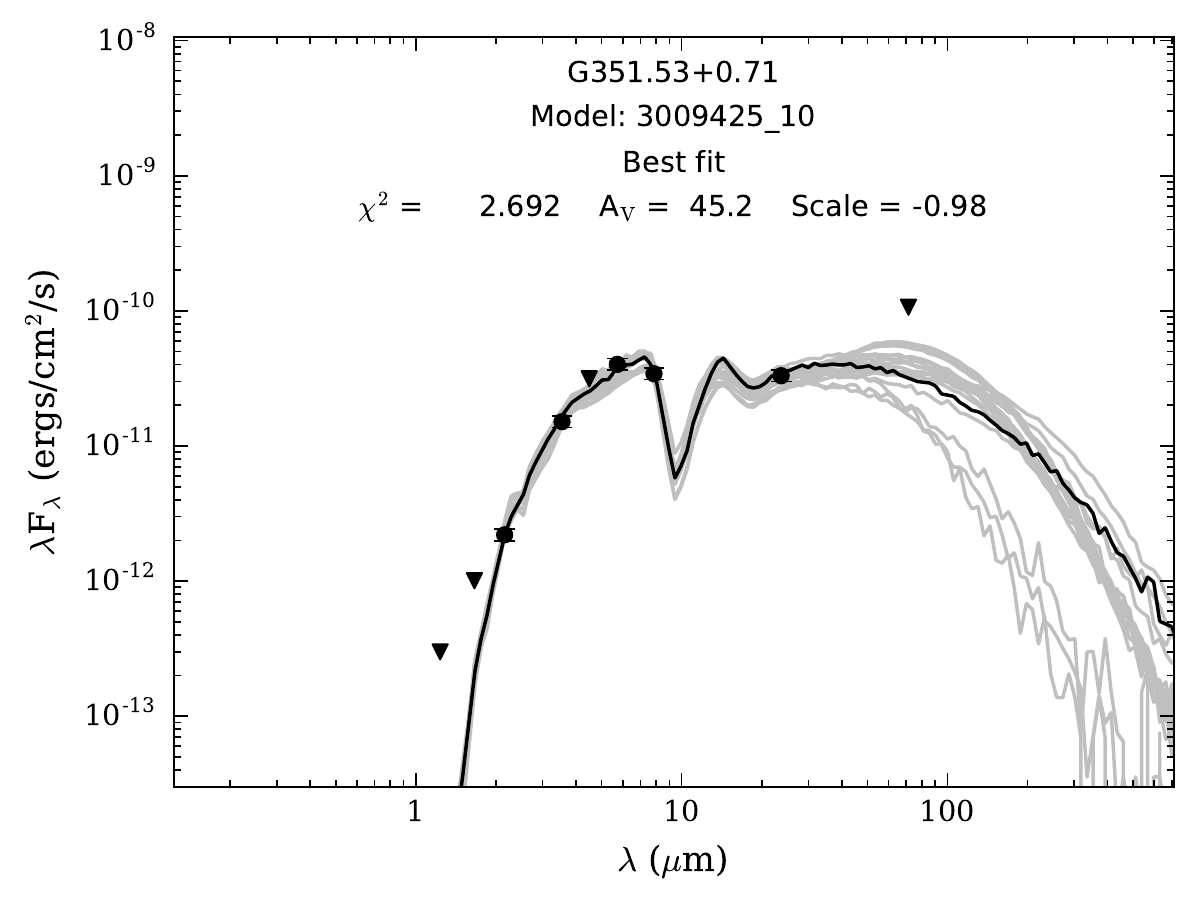}&
                \includegraphics[width=0.31\textwidth]{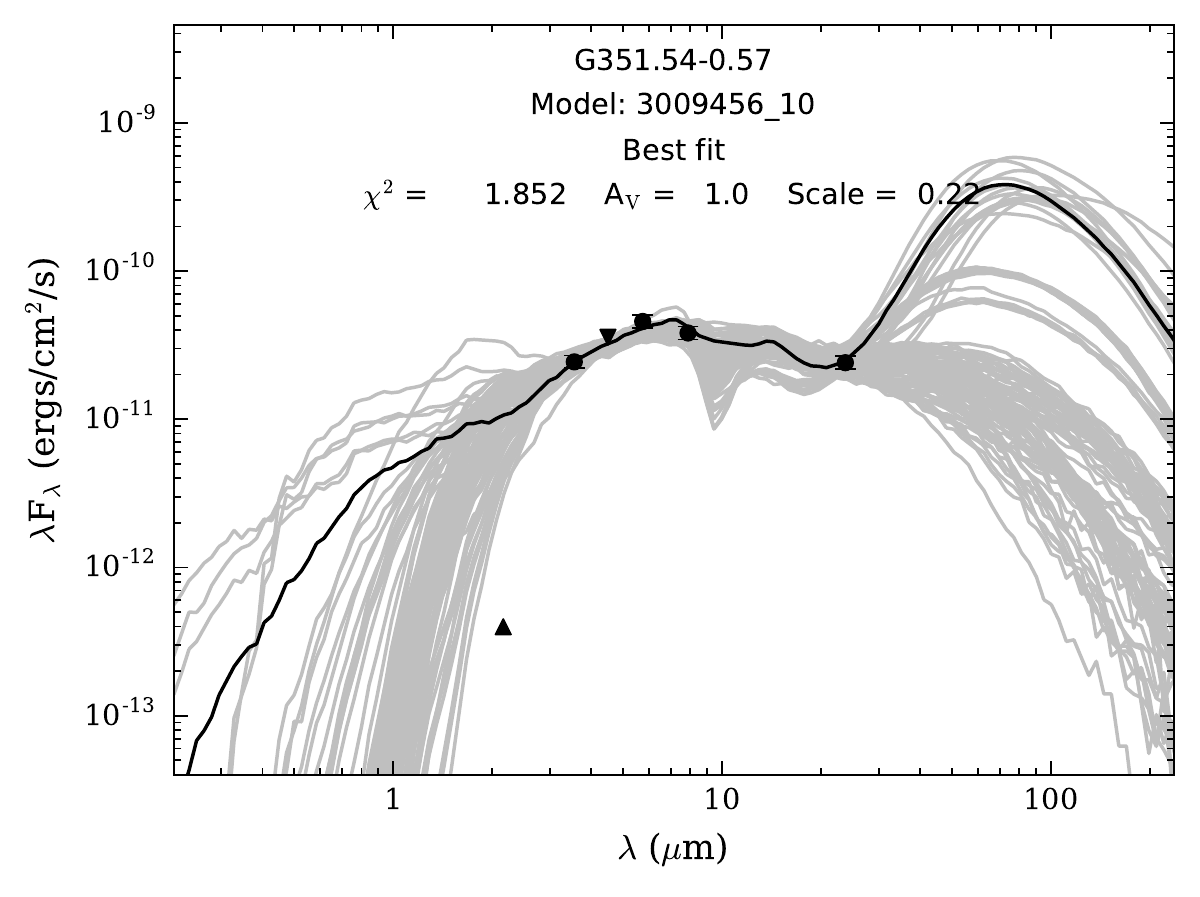}\\
                \includegraphics[width=0.31\textwidth]{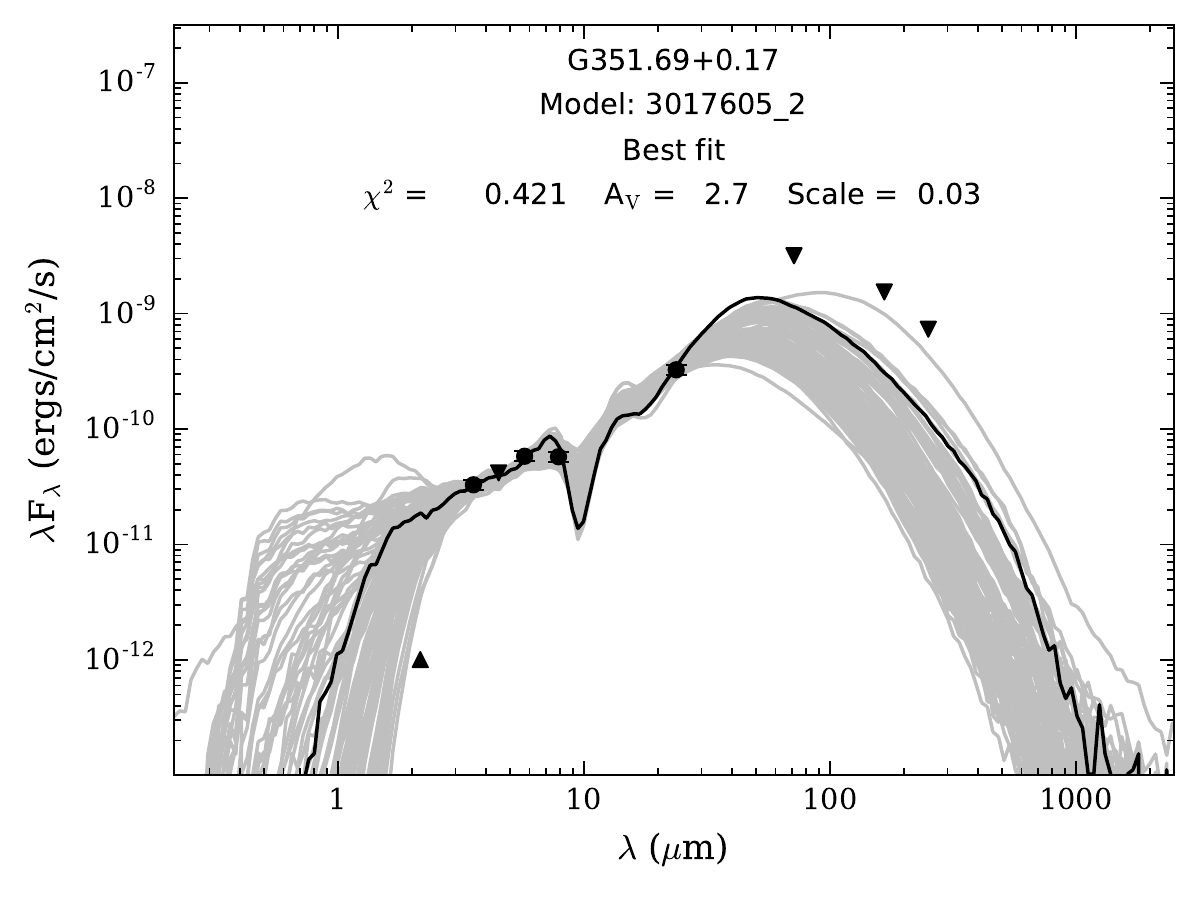}&
                \includegraphics[width=0.31\textwidth]{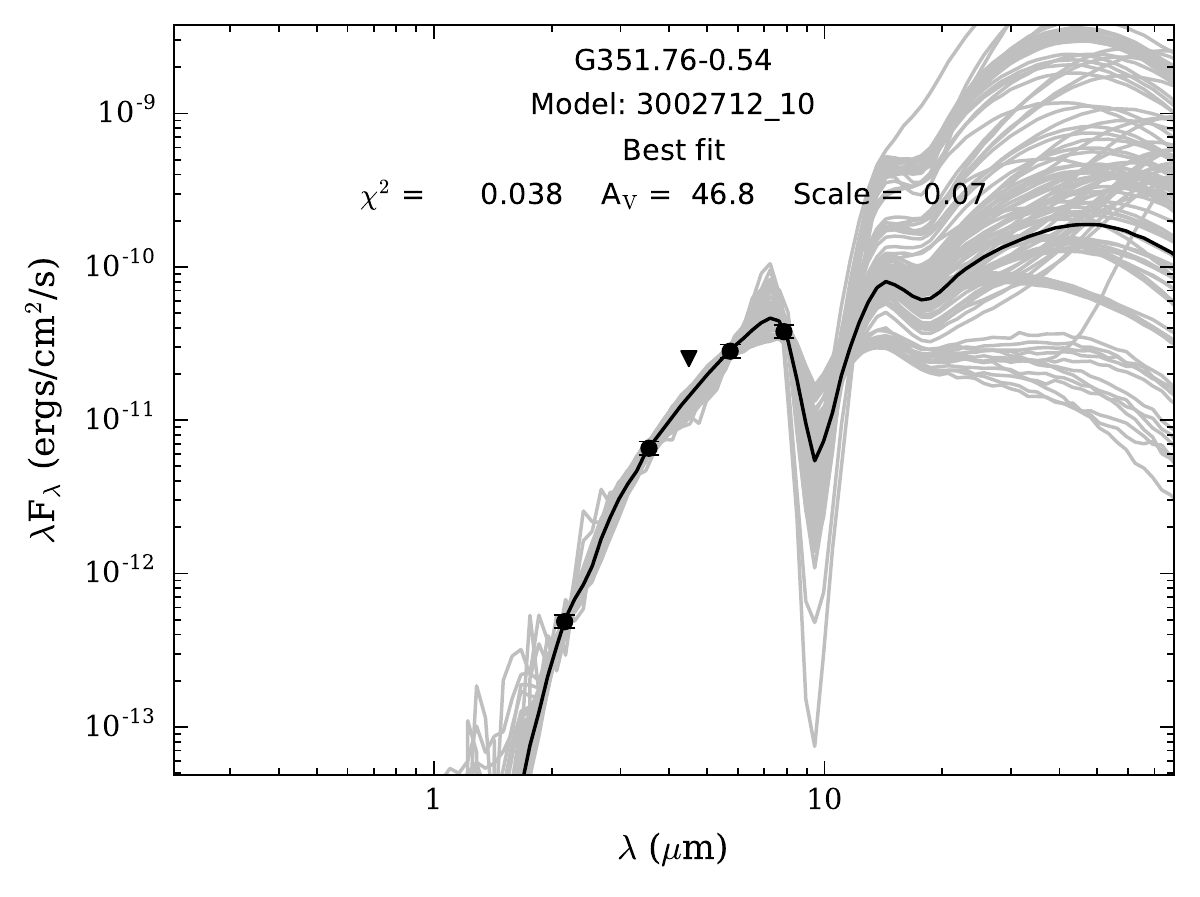}&
                \includegraphics[width=0.31\textwidth]{{./G351.78-0.54}.pdf}\\
                \includegraphics[width=0.31\textwidth]{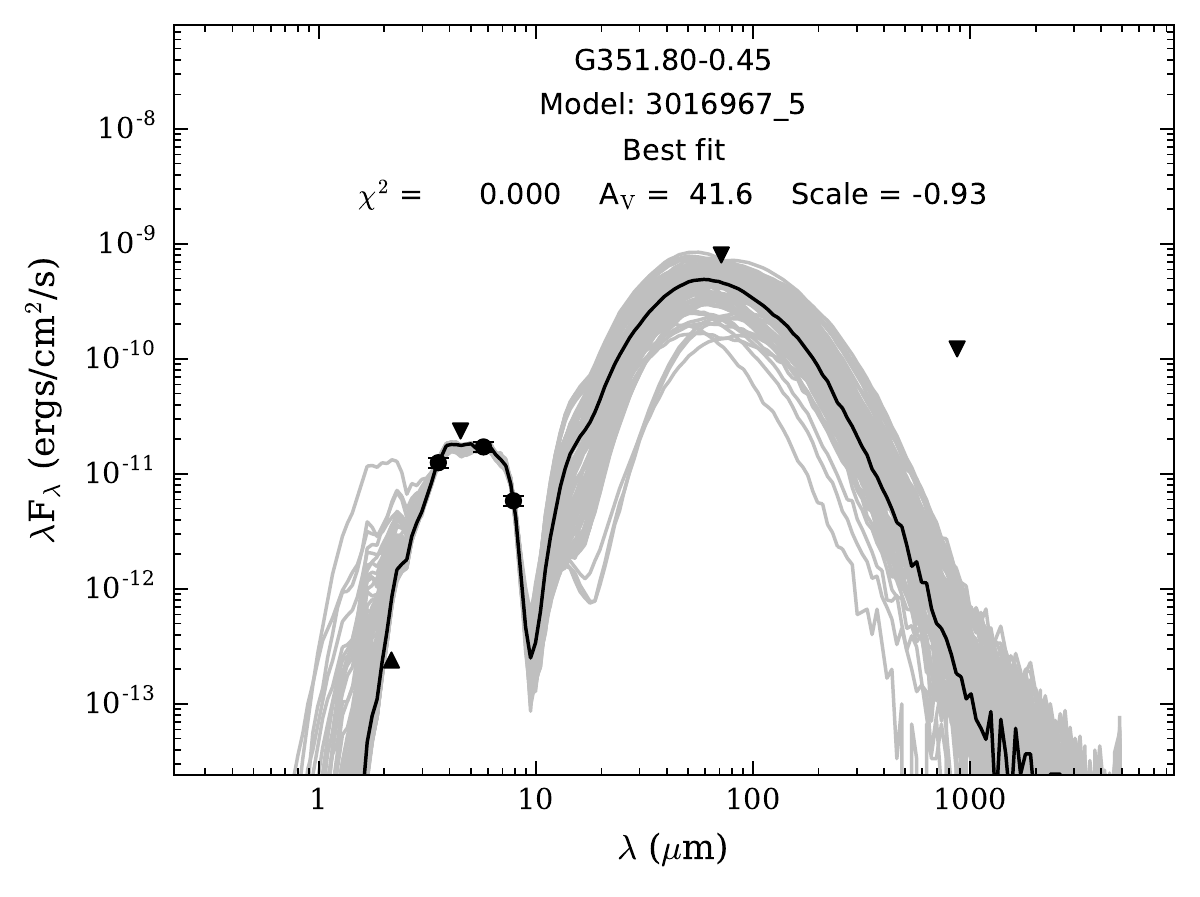}&
                \includegraphics[width=0.31\textwidth]{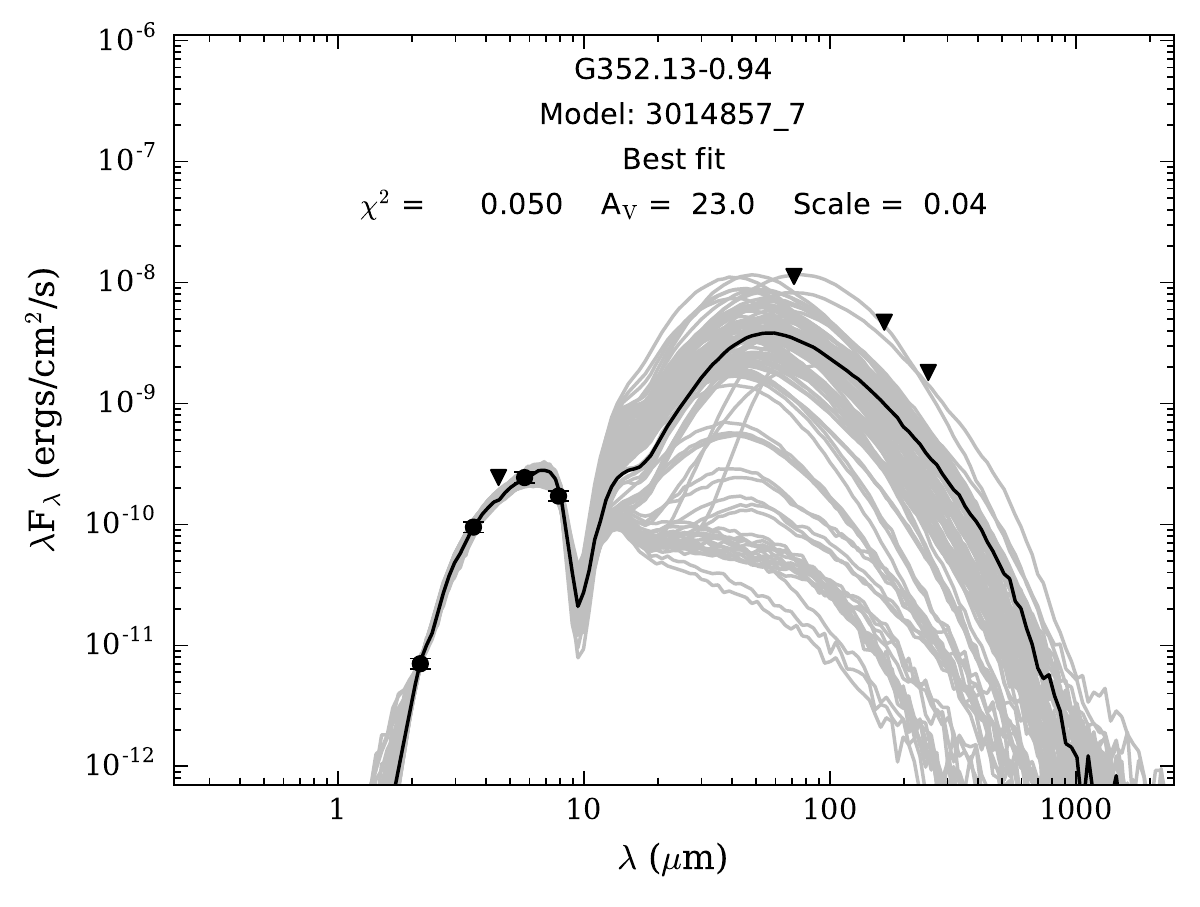}&
                \includegraphics[width=0.31\textwidth]{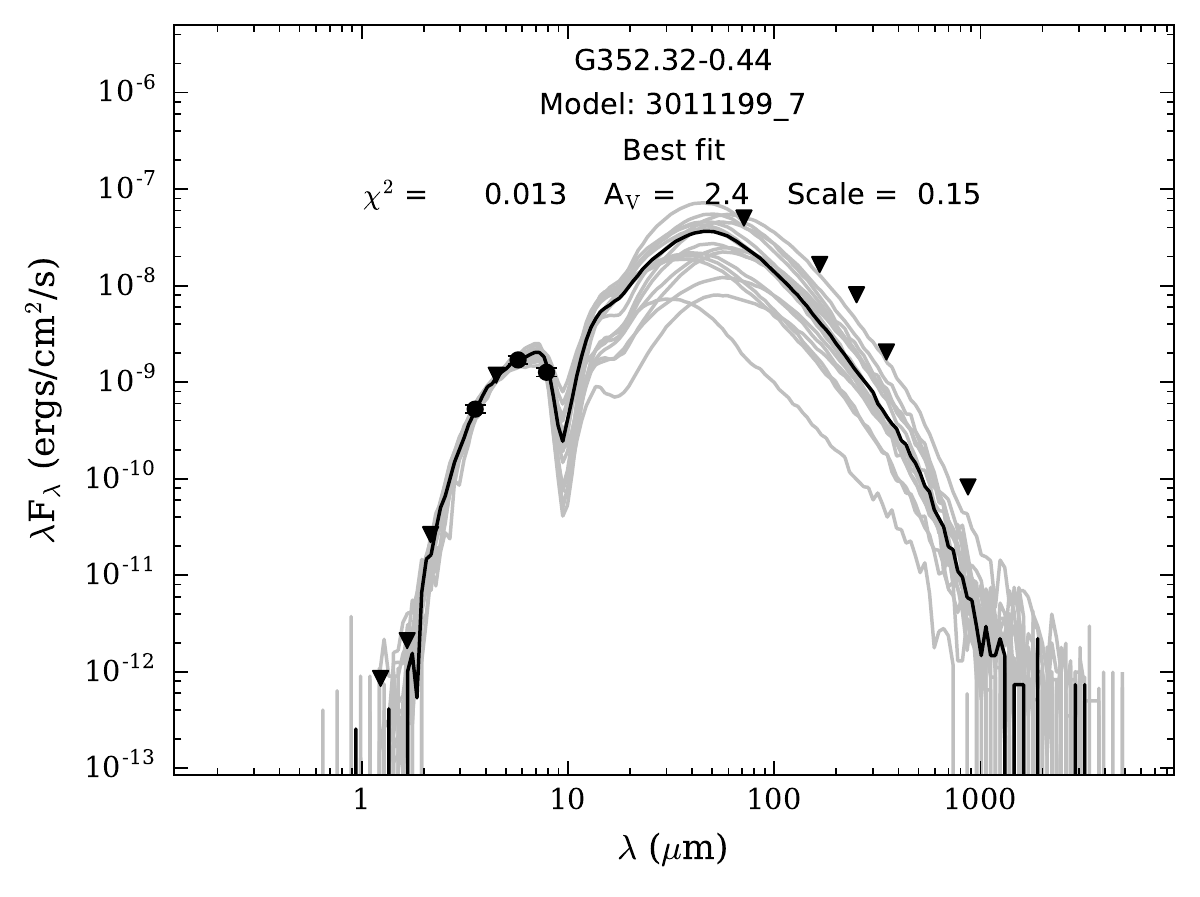}\\
                \includegraphics[width=0.31\textwidth]{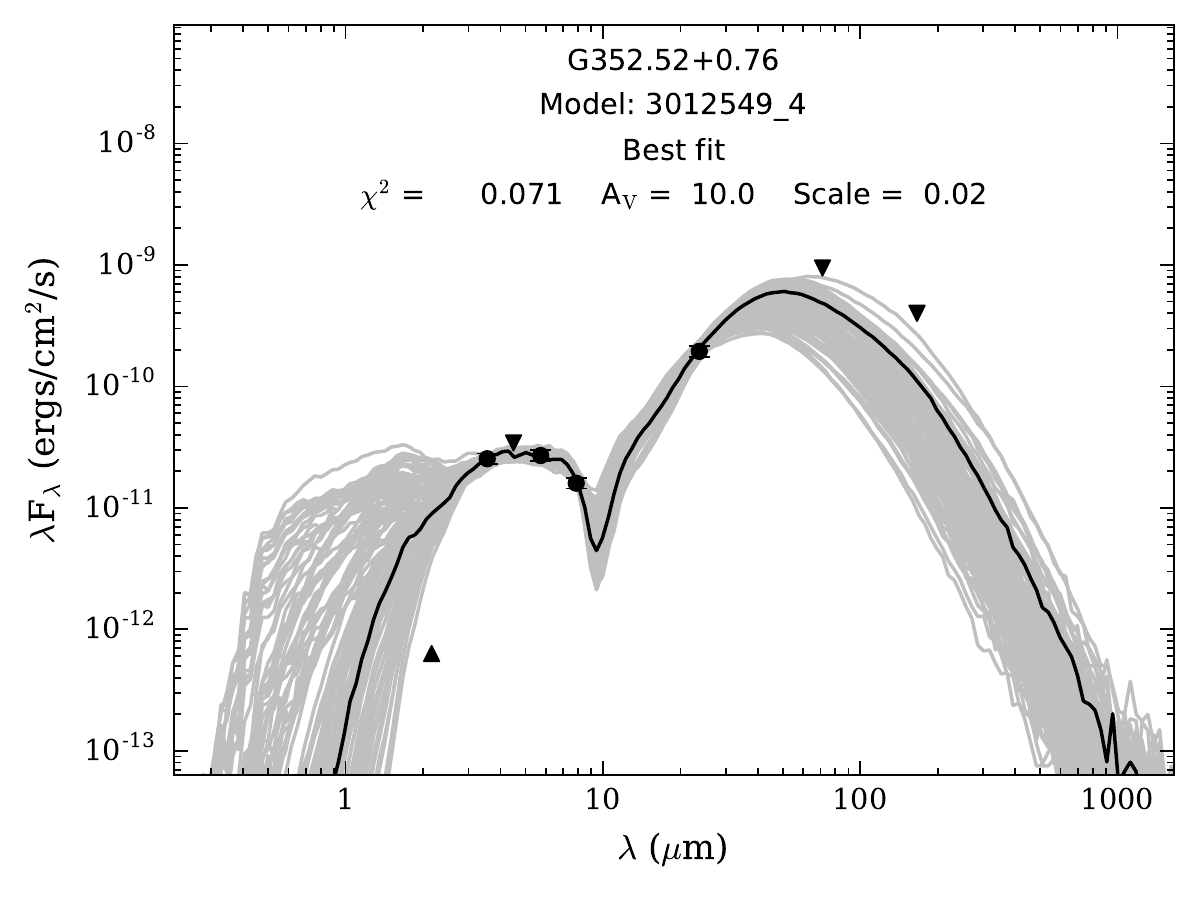}&
                \includegraphics[width=0.31\textwidth]{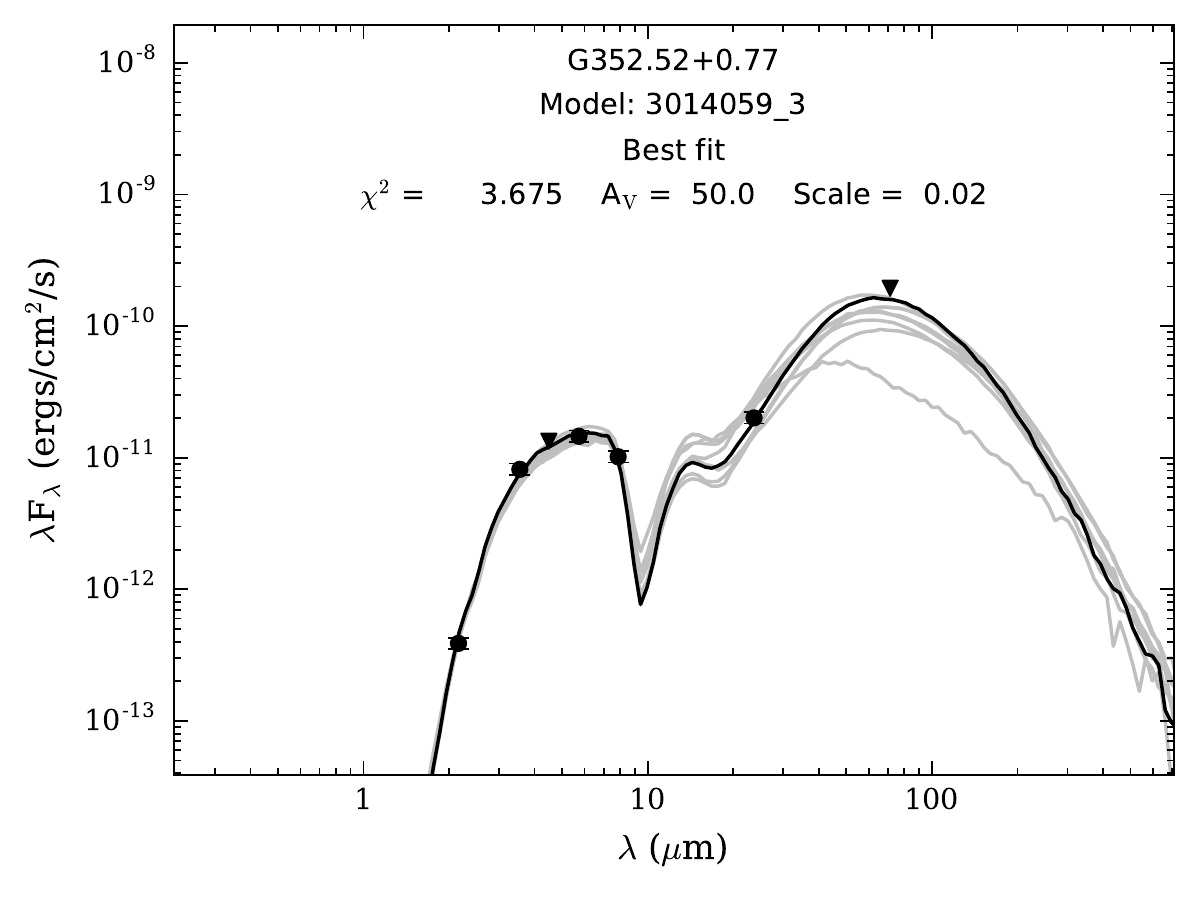}&
                \includegraphics[width=0.31\textwidth]{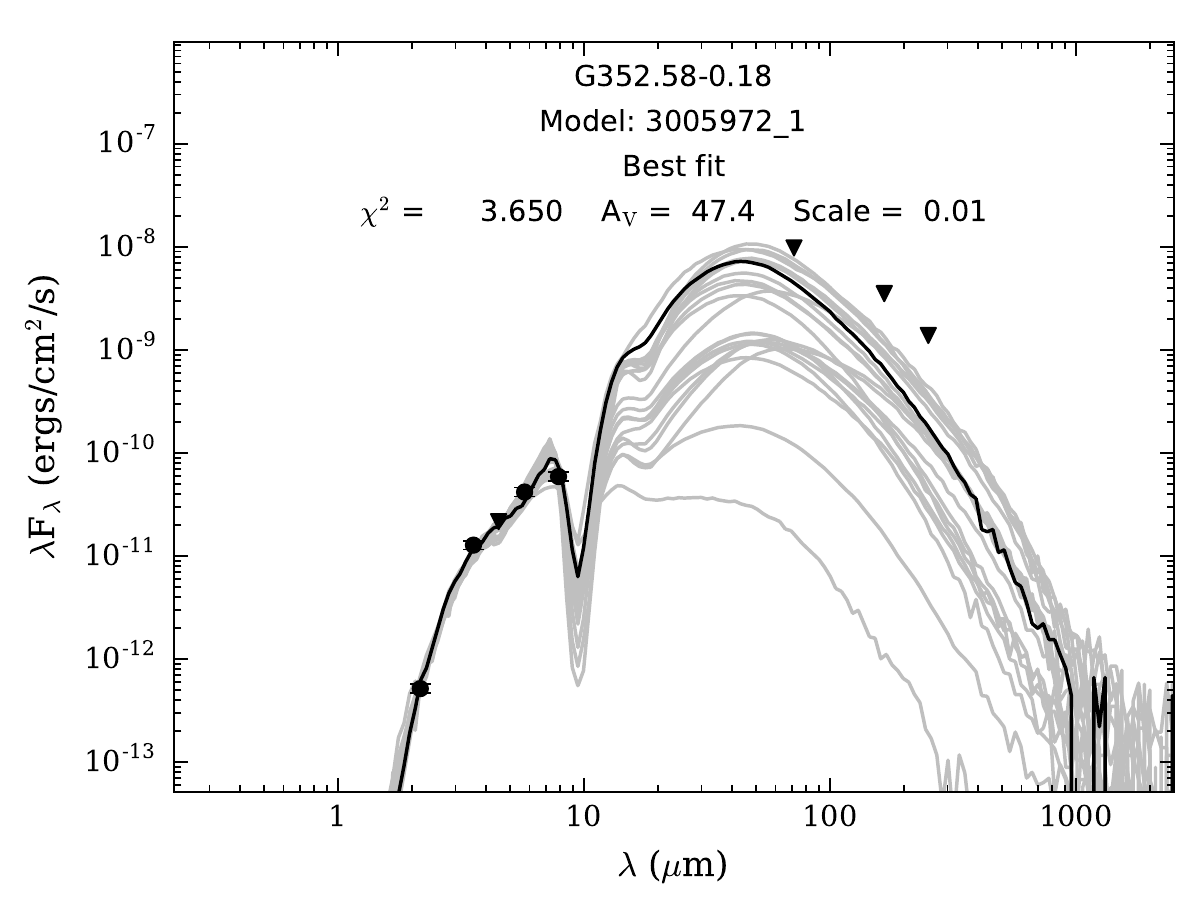}\\
                \includegraphics[width=0.31\textwidth]{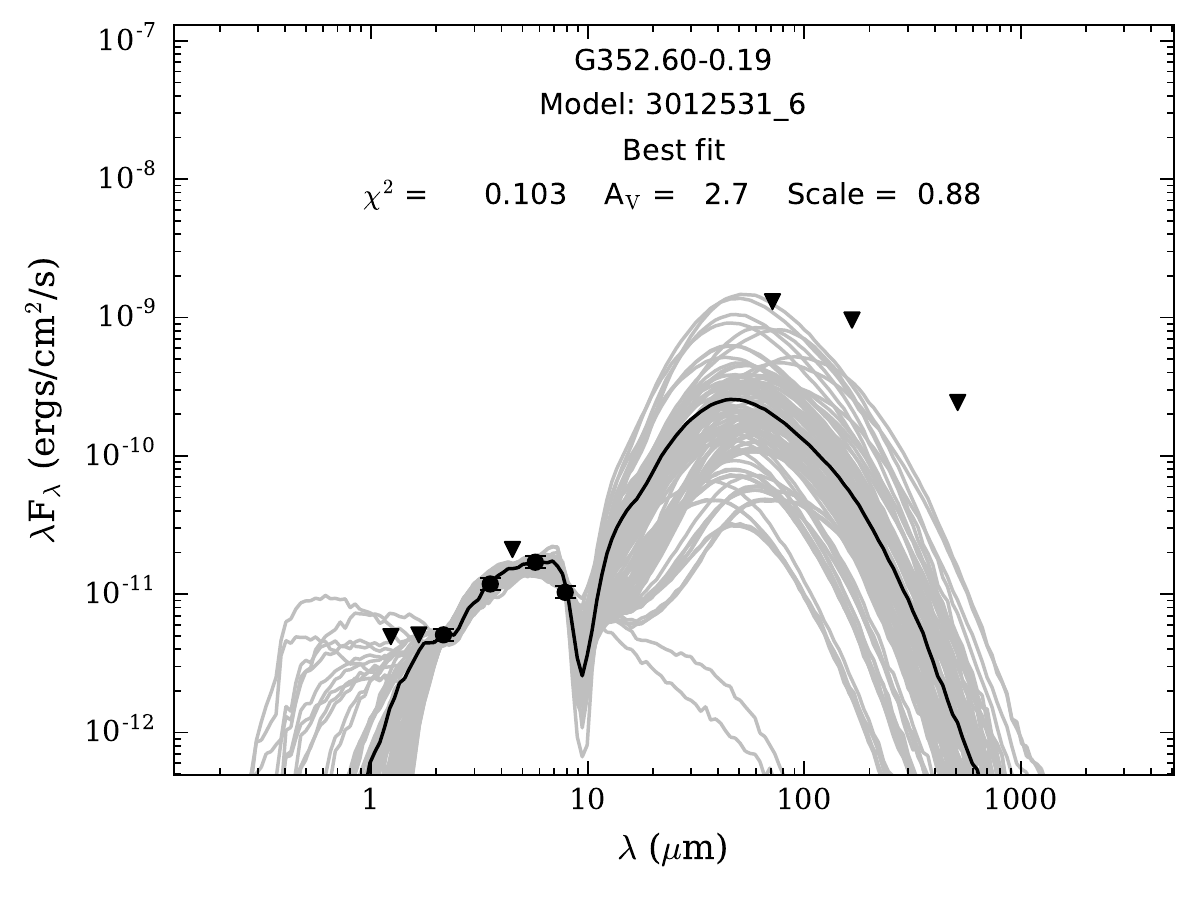}&
                \includegraphics[width=0.31\textwidth]{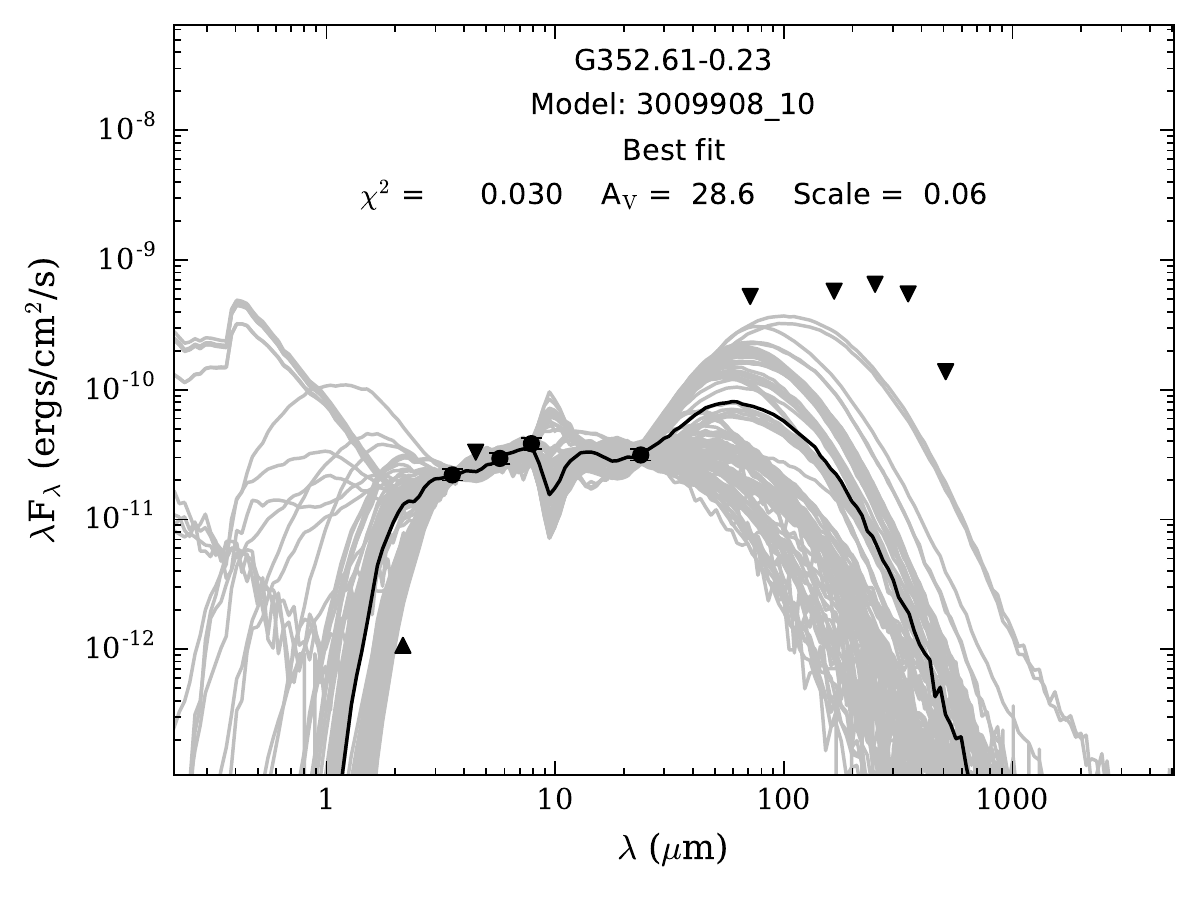}&
                \includegraphics[width=0.31\textwidth]{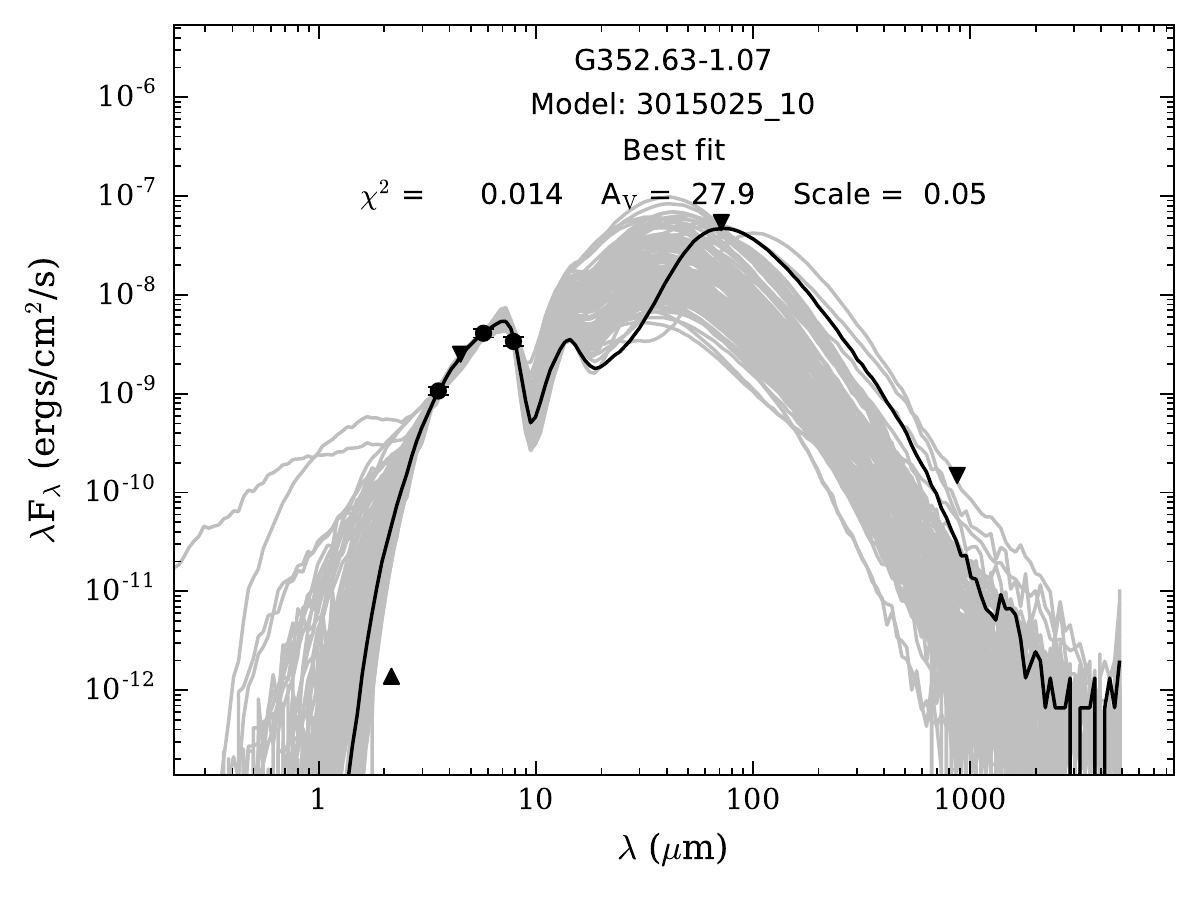}\\
        \end{tabular}
        \caption{Continuation of Fig. \ref{fig:testgridsed}.}
\end{figure}

\begin{figure}
        \begin{tabular}{ccc}
                \includegraphics[width=0.31\textwidth]{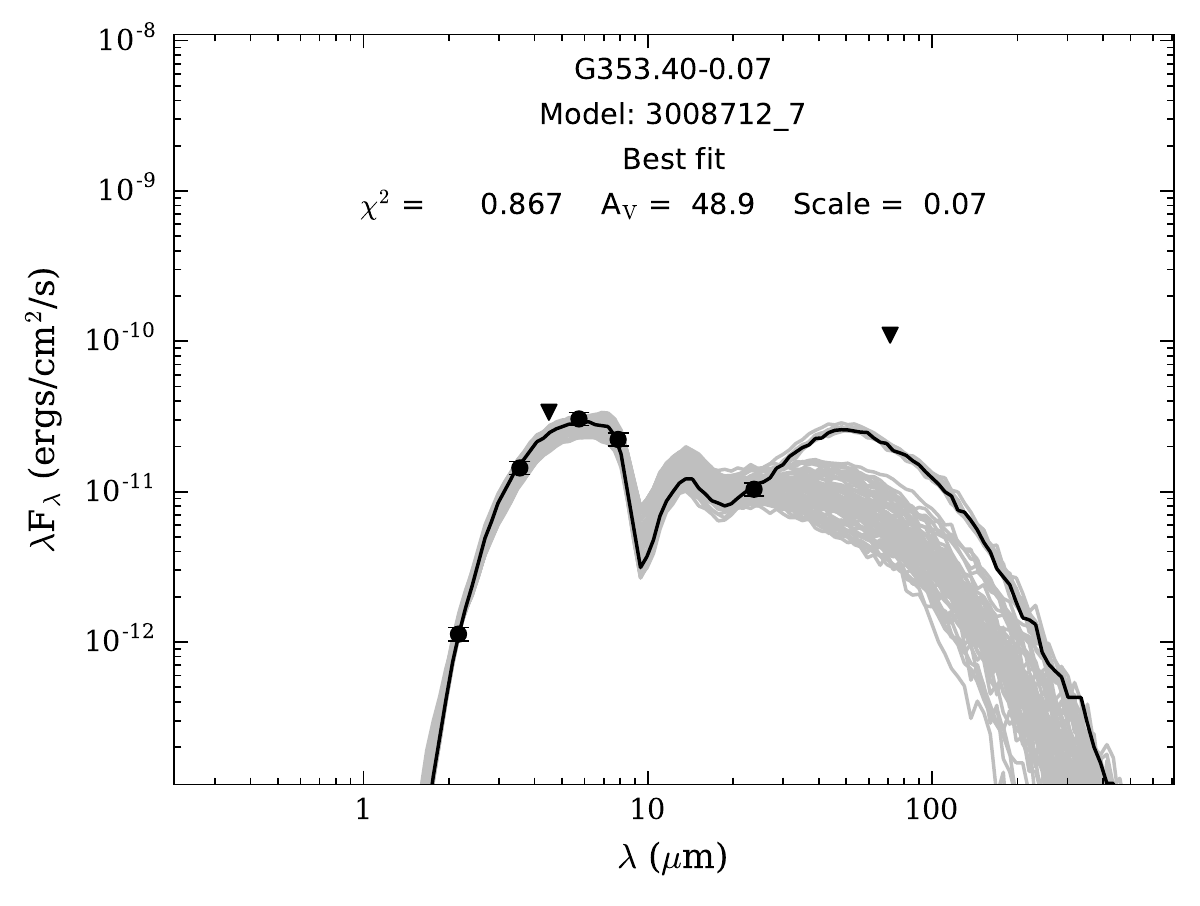}&
                \includegraphics[width=0.31\textwidth]{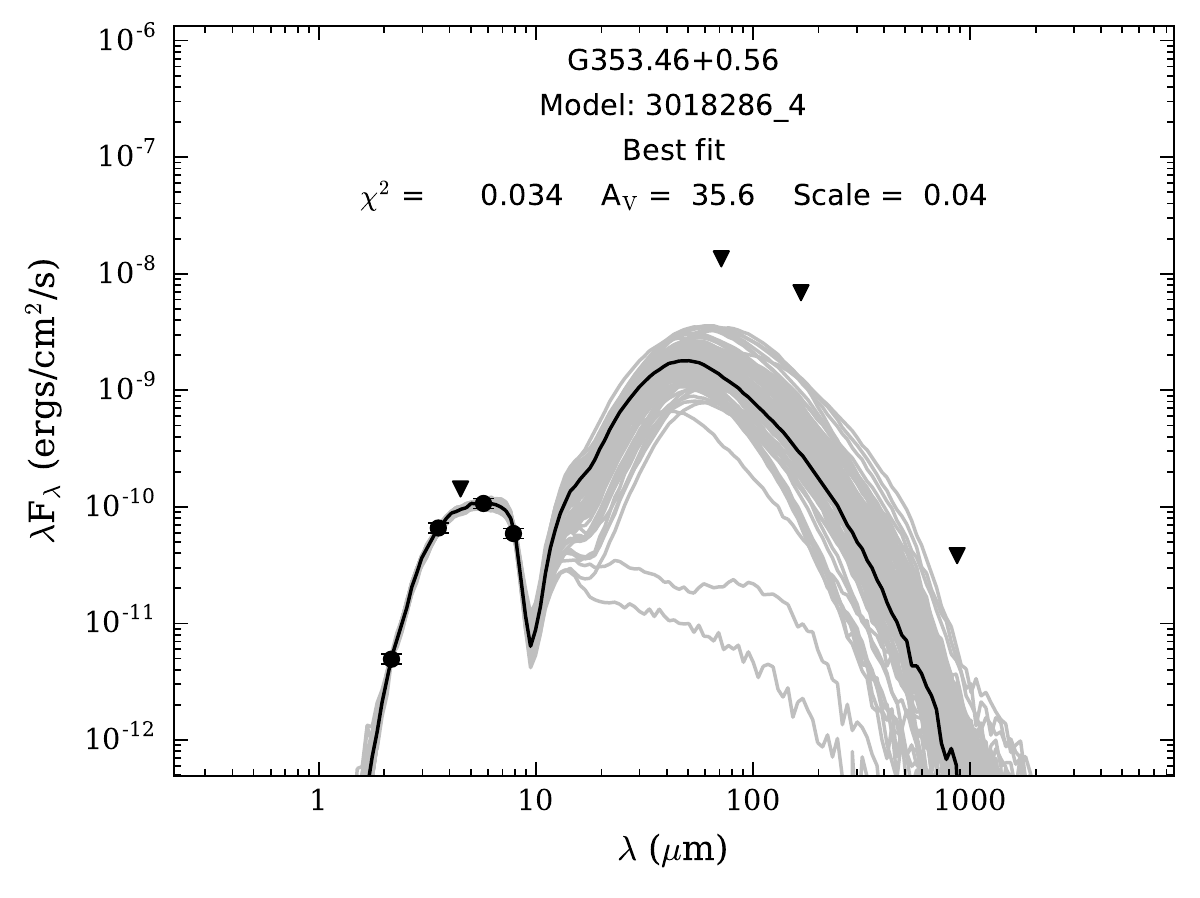}&
                \includegraphics[width=0.31\textwidth]{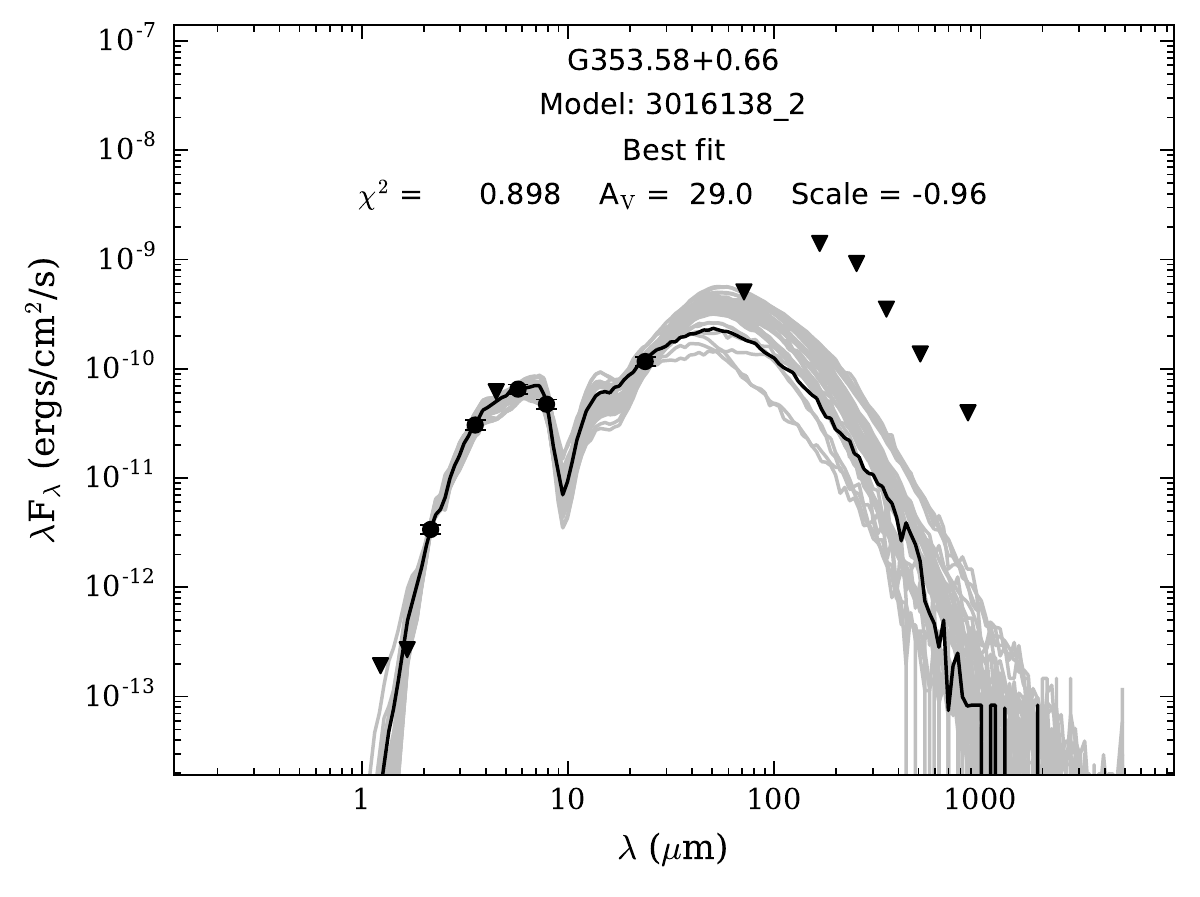}\\
                
                \includegraphics[width=0.31\textwidth]{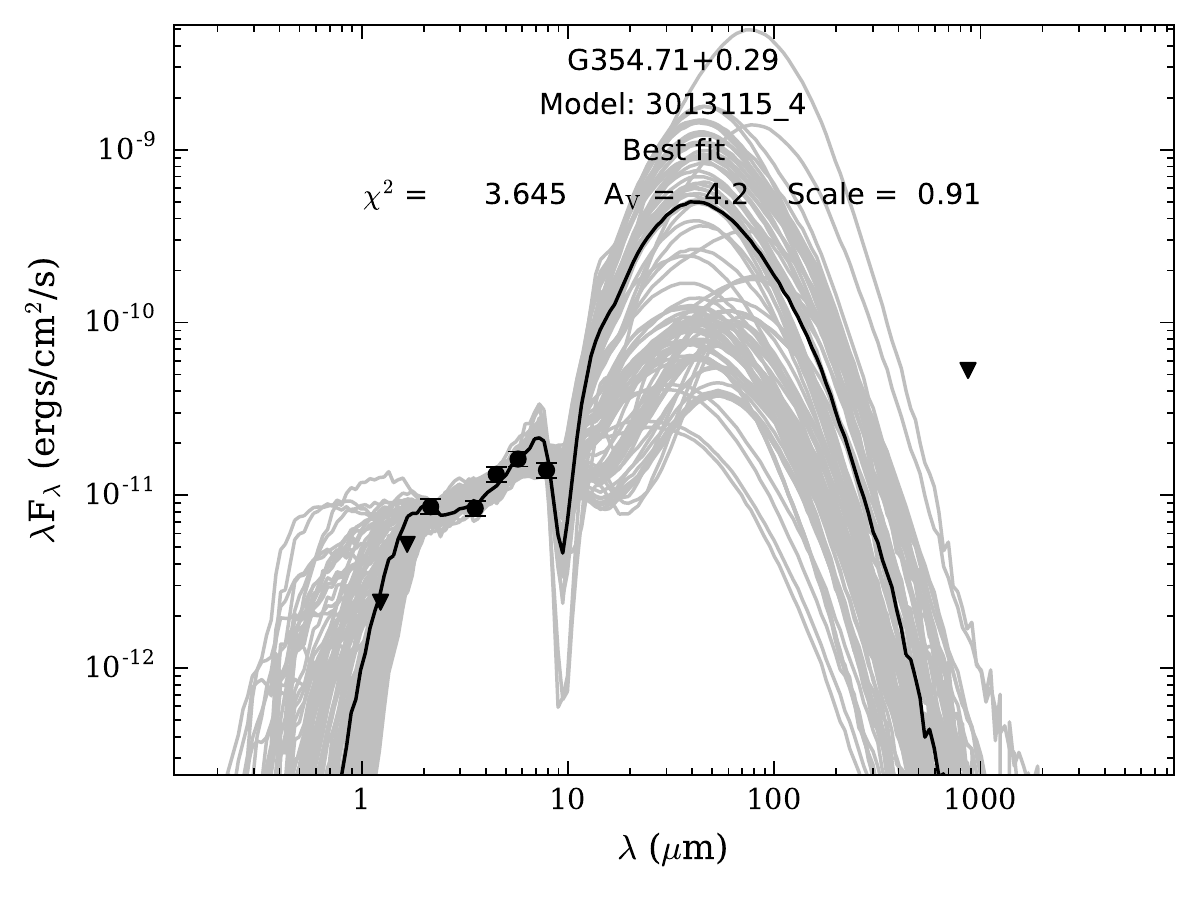}&
                \includegraphics[width=0.31\textwidth]{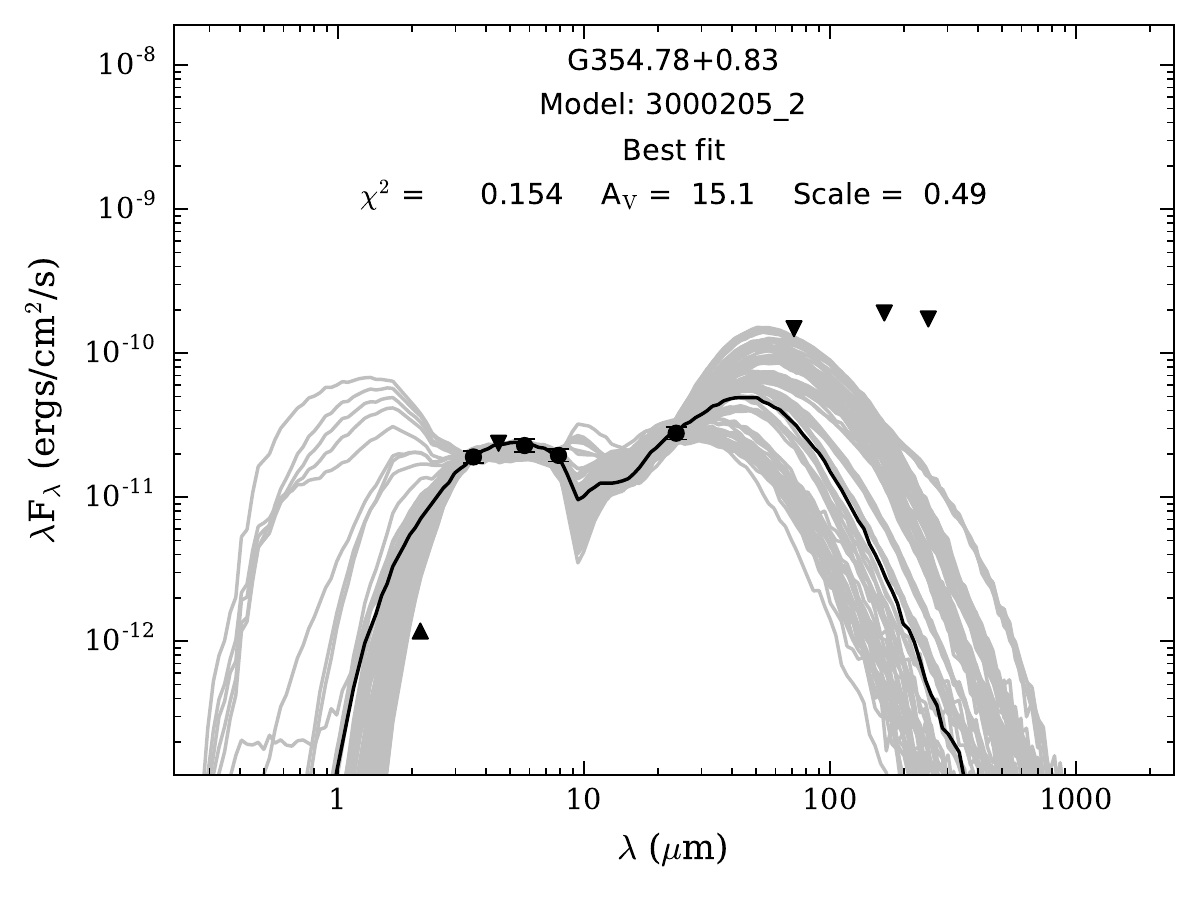}&
                \includegraphics[width=0.31\textwidth]{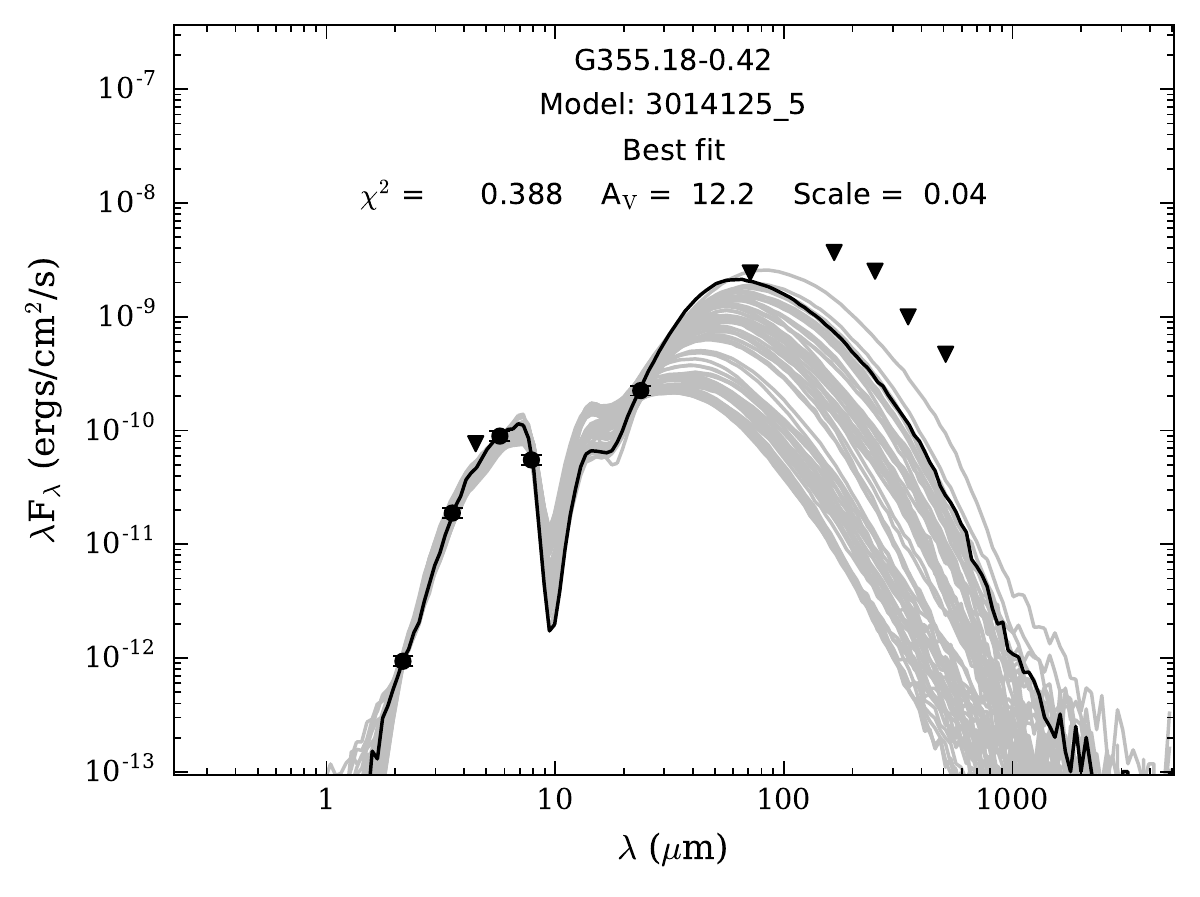}\\
                \includegraphics[width=0.31\textwidth]{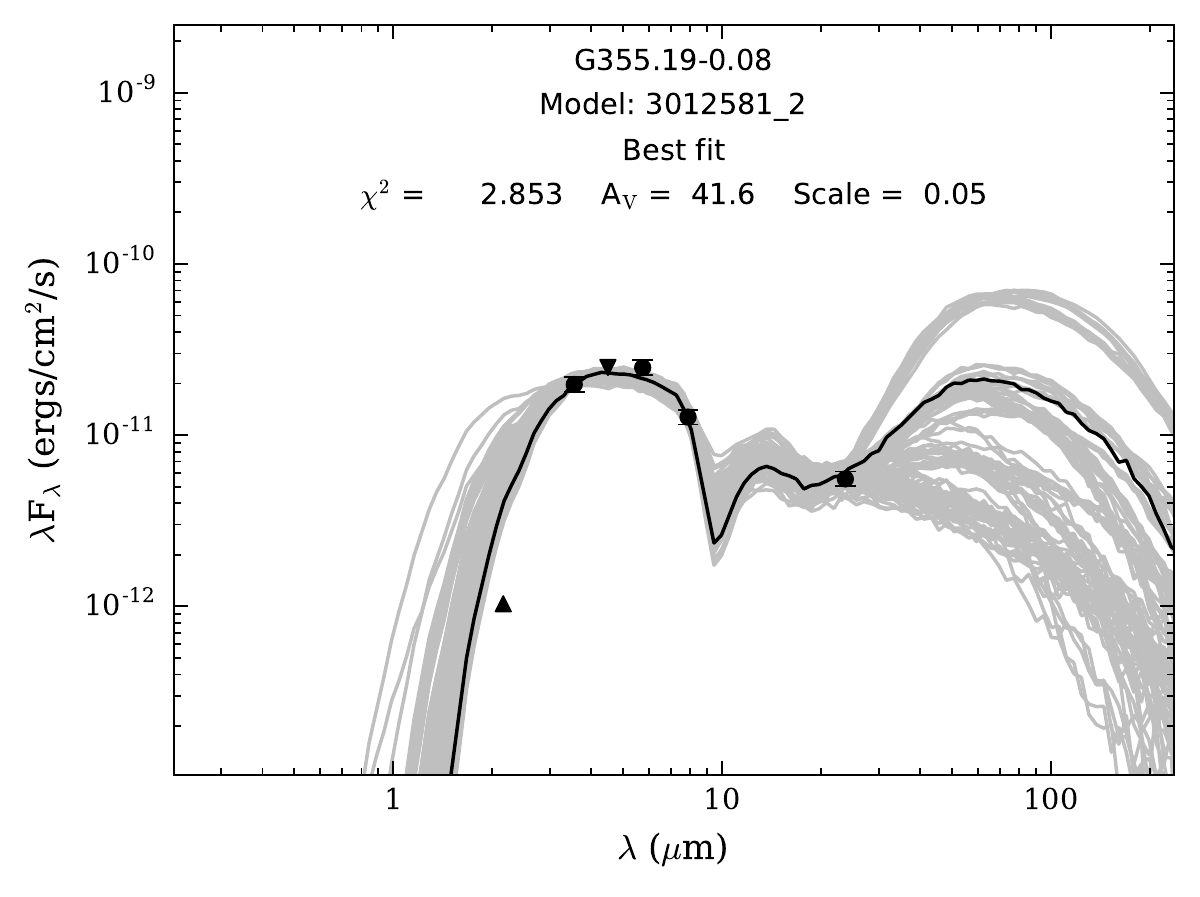}&
                \includegraphics[width=0.31\textwidth]{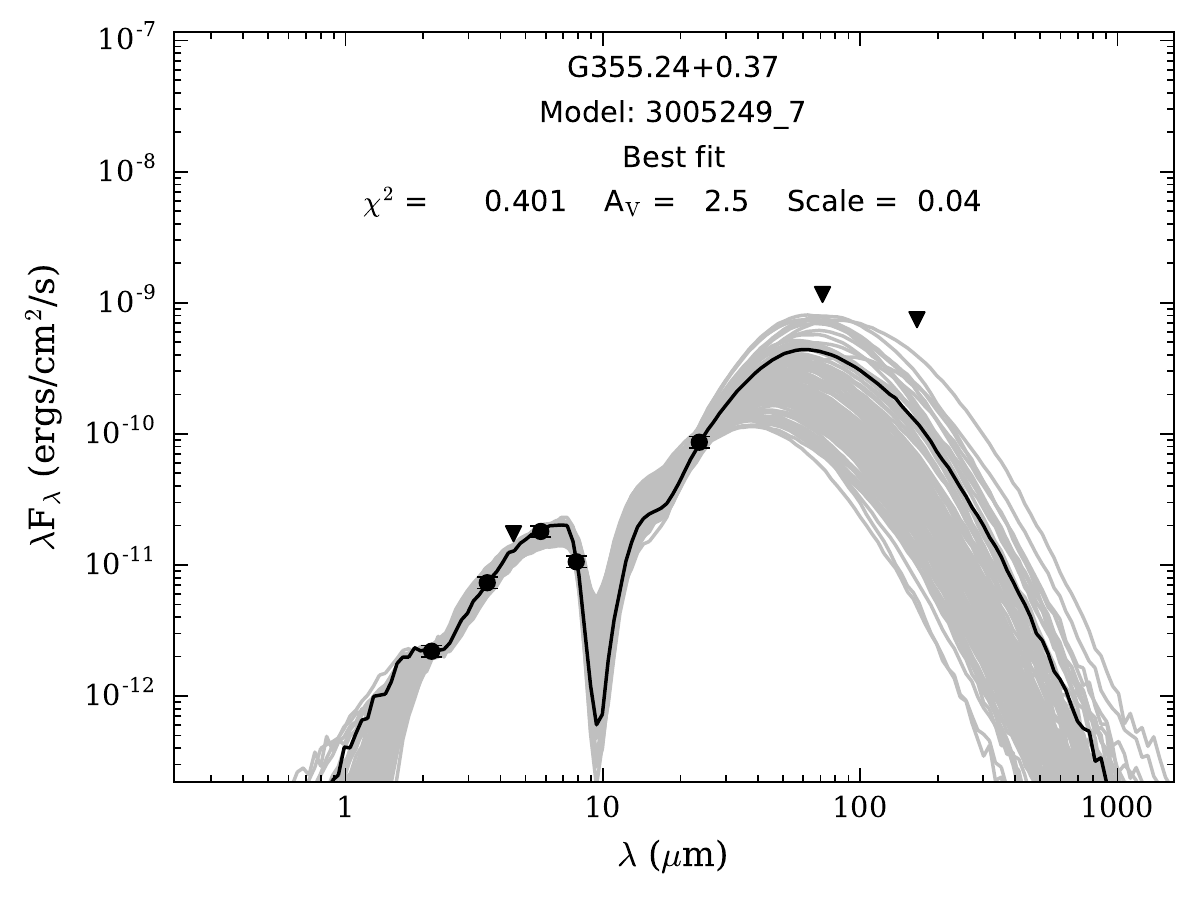}&
                \includegraphics[width=0.31\textwidth]{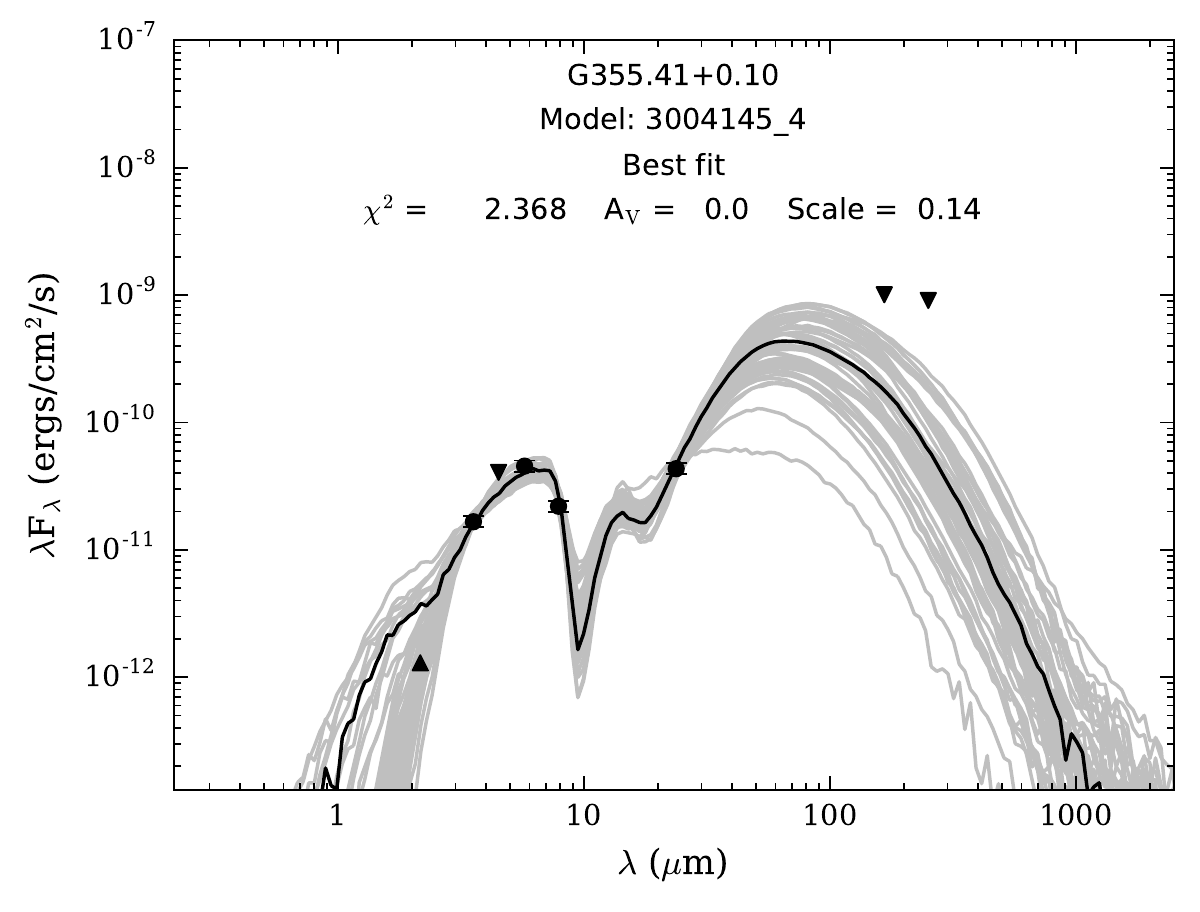}\\
                \includegraphics[width=0.31\textwidth]{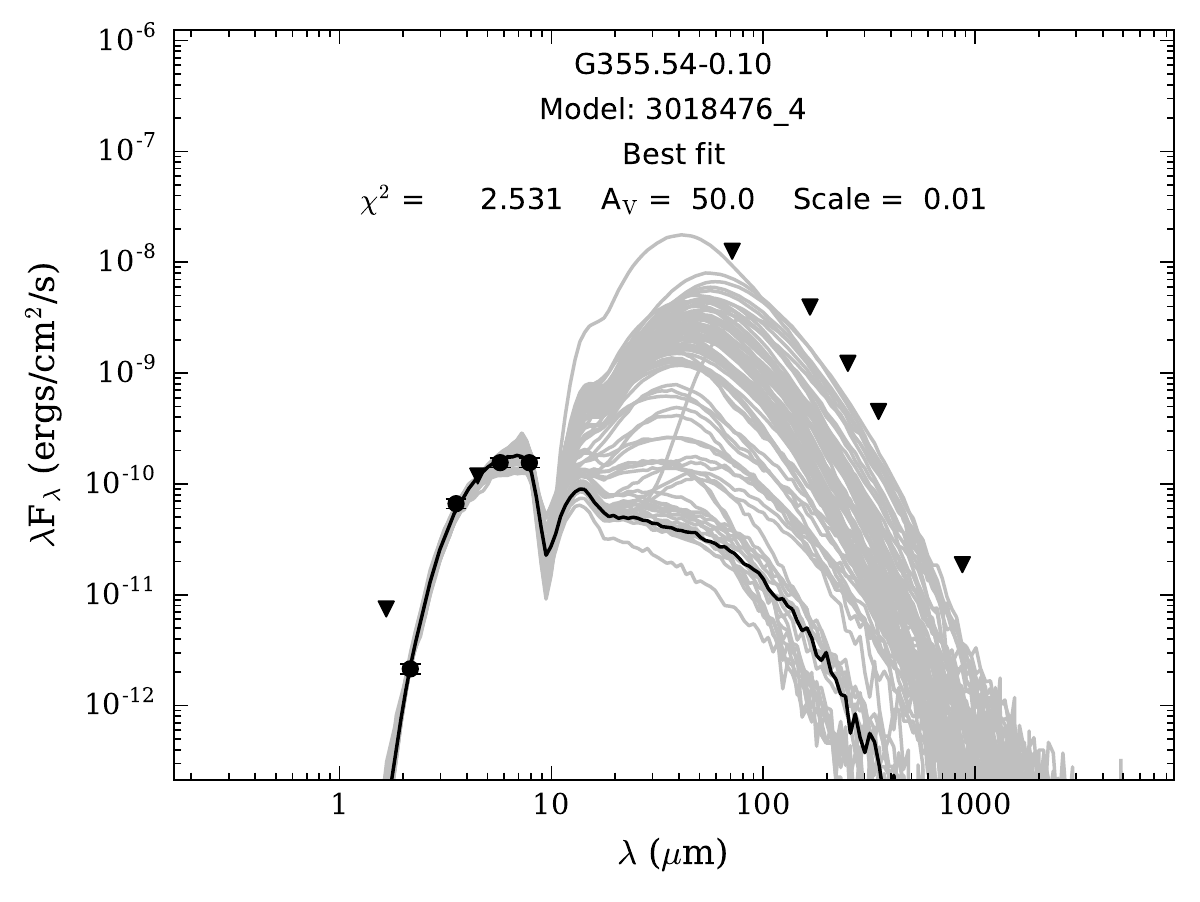}&
                \includegraphics[width=0.31\textwidth]{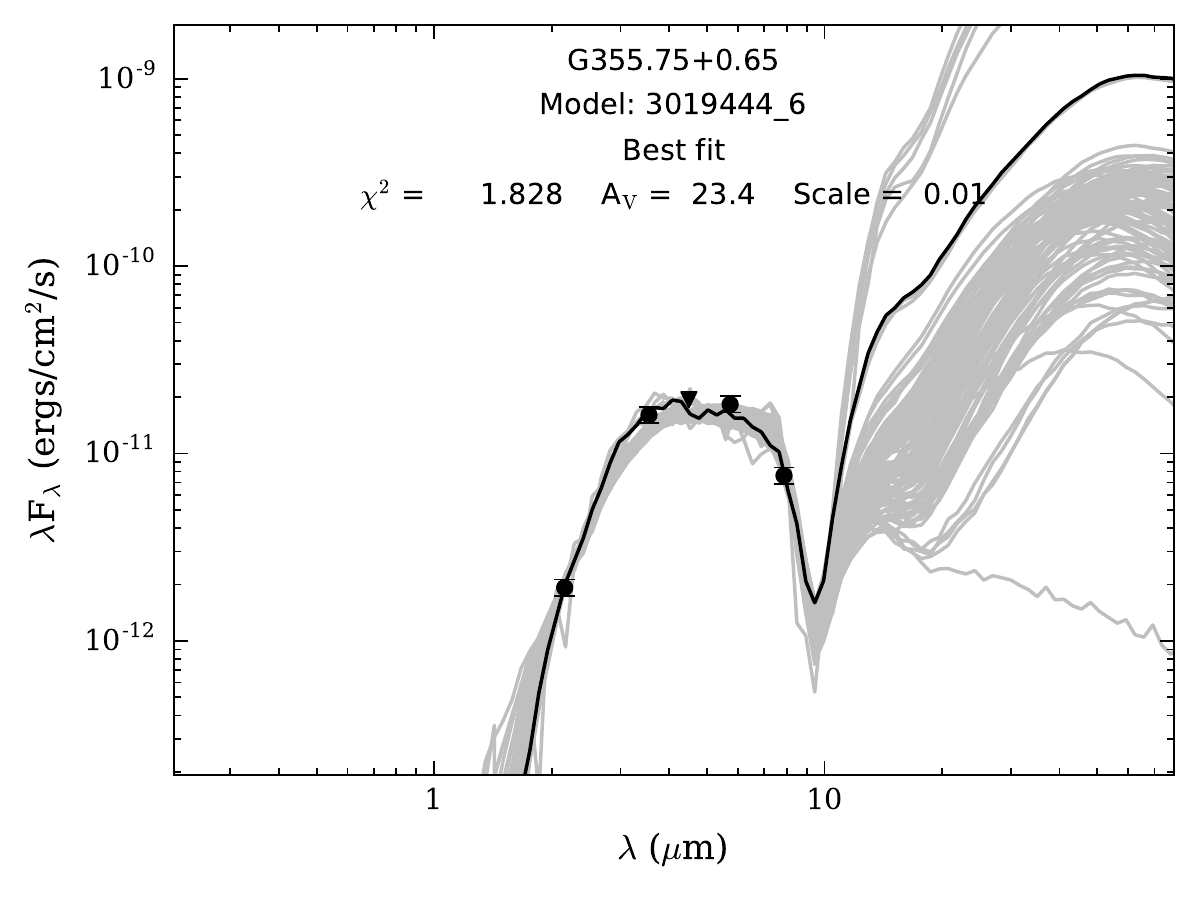}&
                \includegraphics[width=0.31\textwidth]{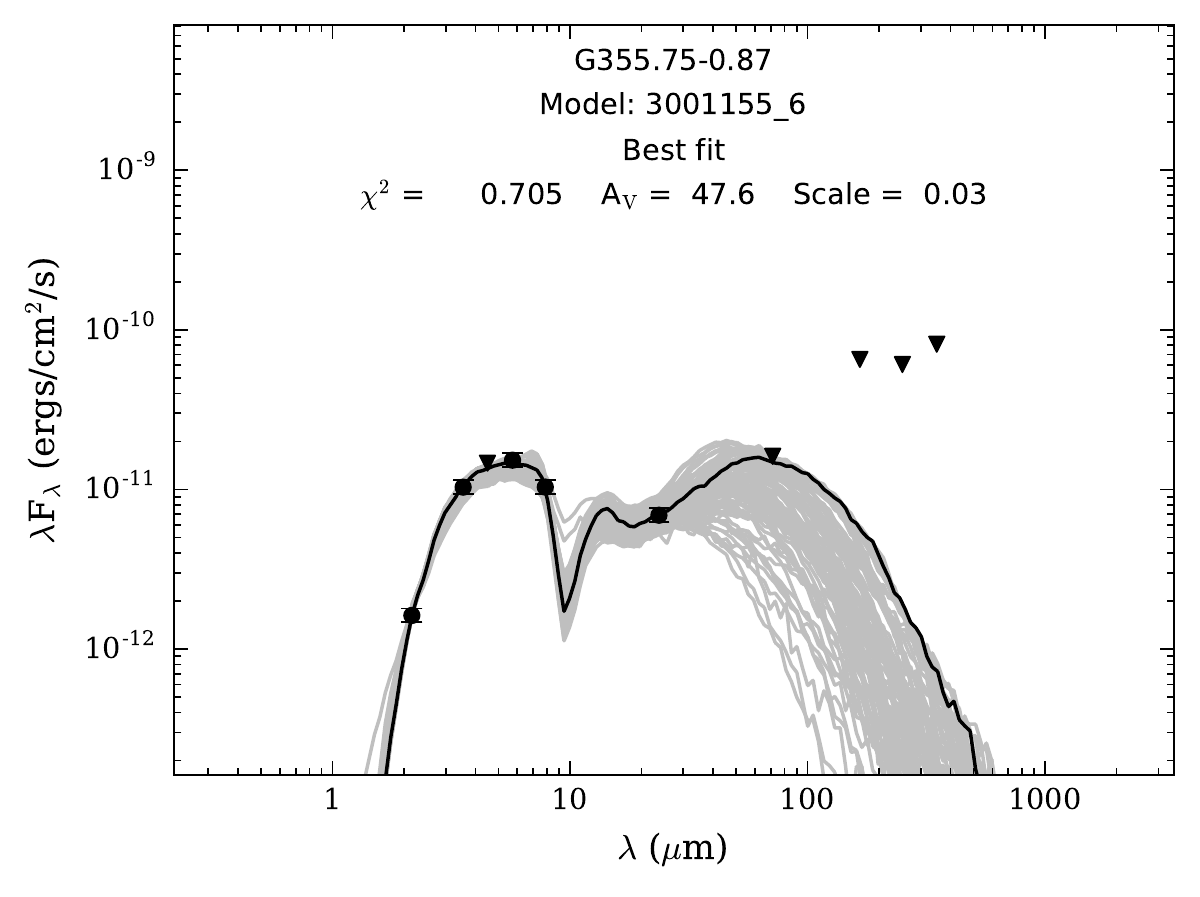}\\
                \includegraphics[width=0.31\textwidth]{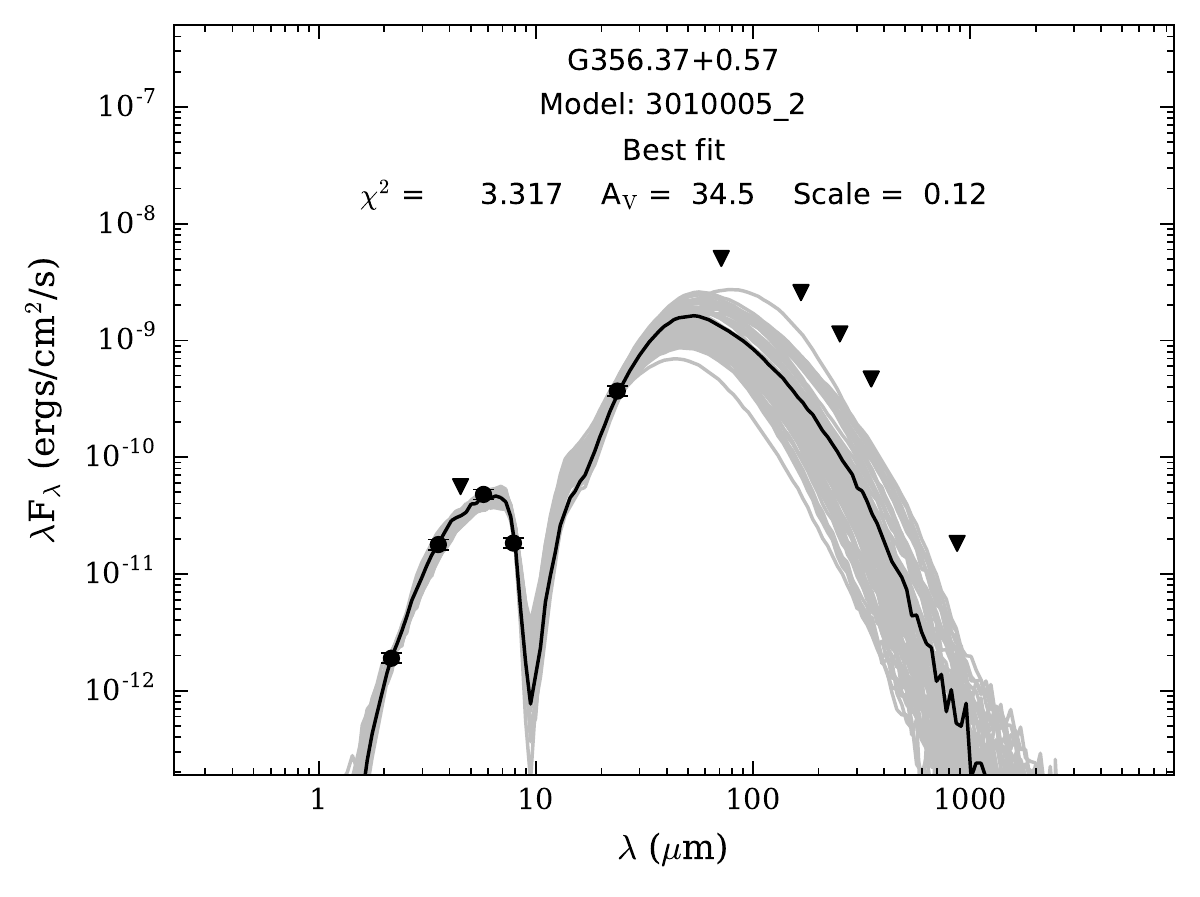}&
                \includegraphics[width=0.31\textwidth]{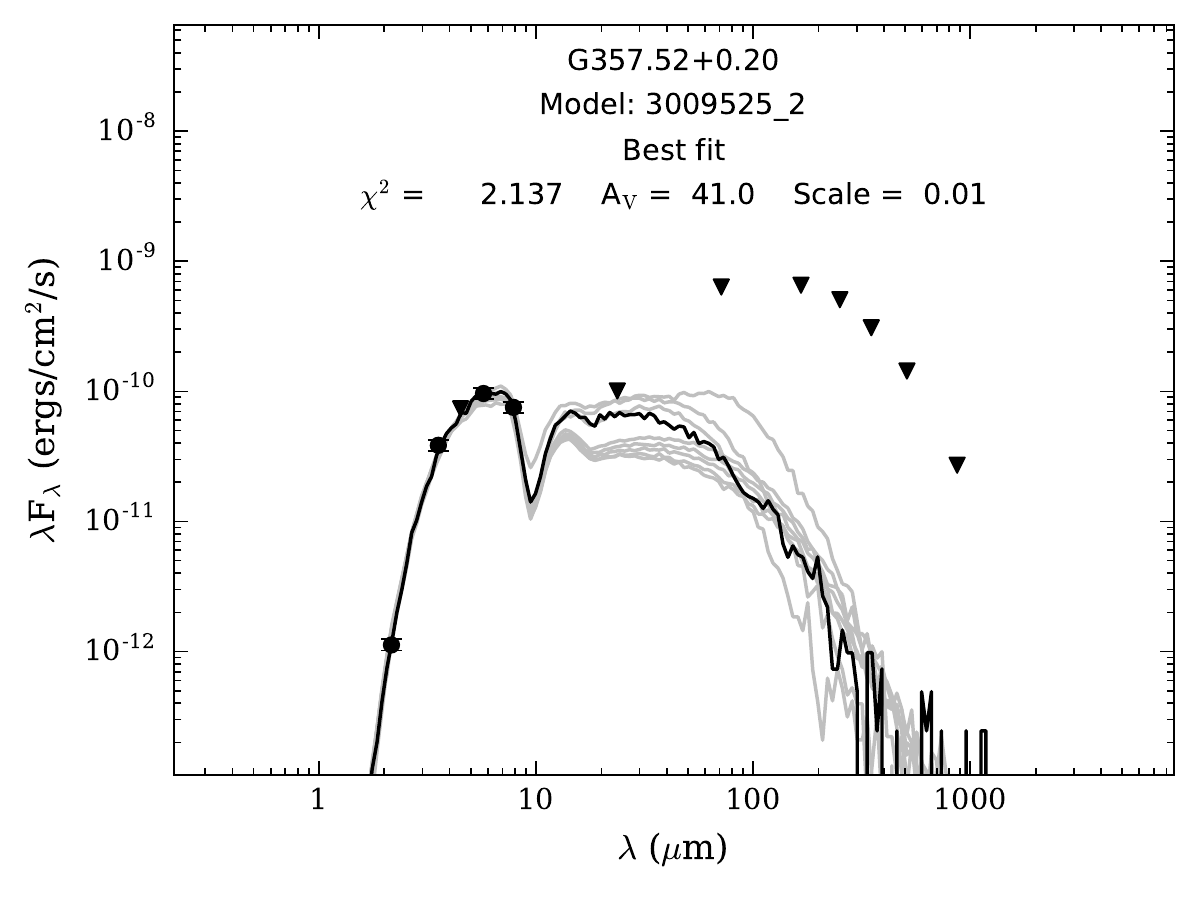}&
                \includegraphics[width=0.31\textwidth]{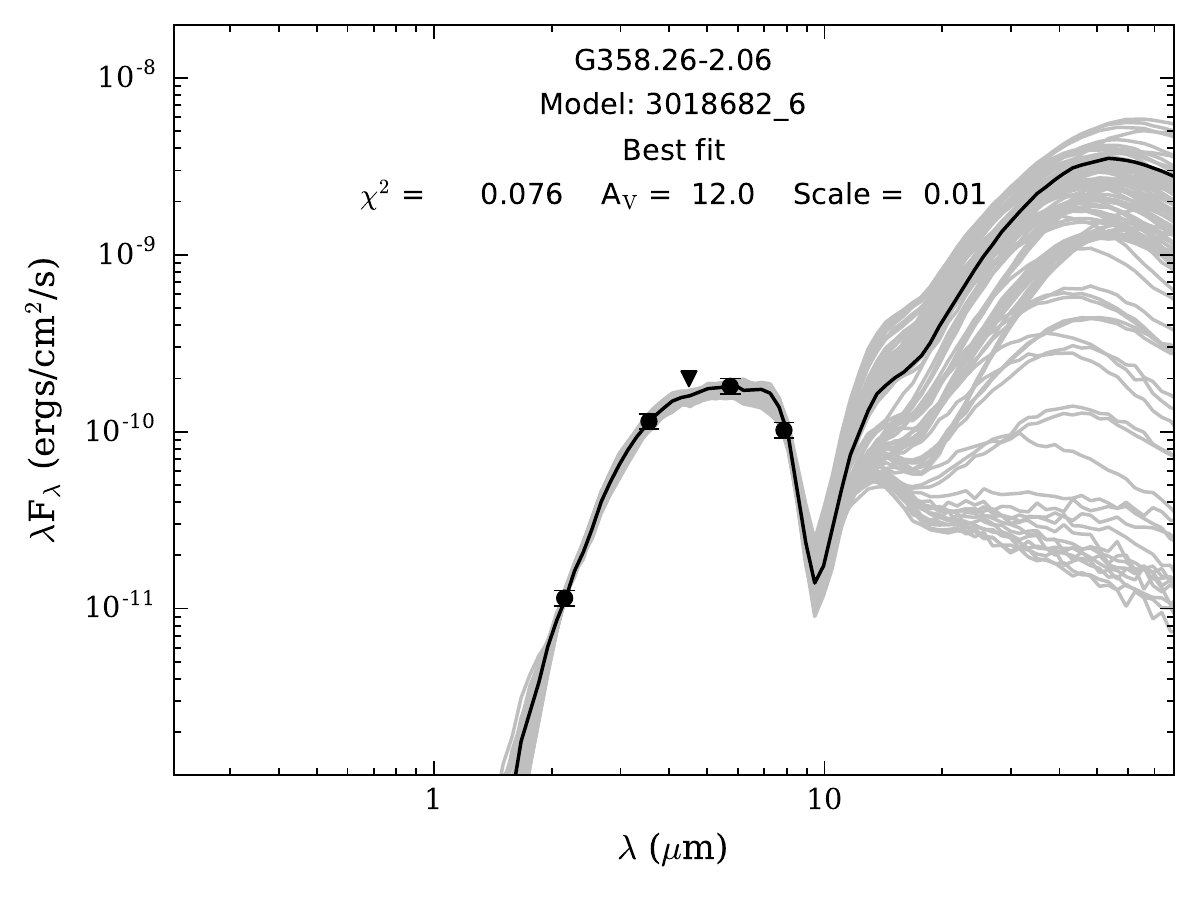}\\
        \end{tabular}
        \caption{Continuation of Fig. \ref{fig:testgridsed}.}
\end{figure}

\begin{figure}
        \begin{tabular}{ccc}
                
                \includegraphics[width=0.31\textwidth]{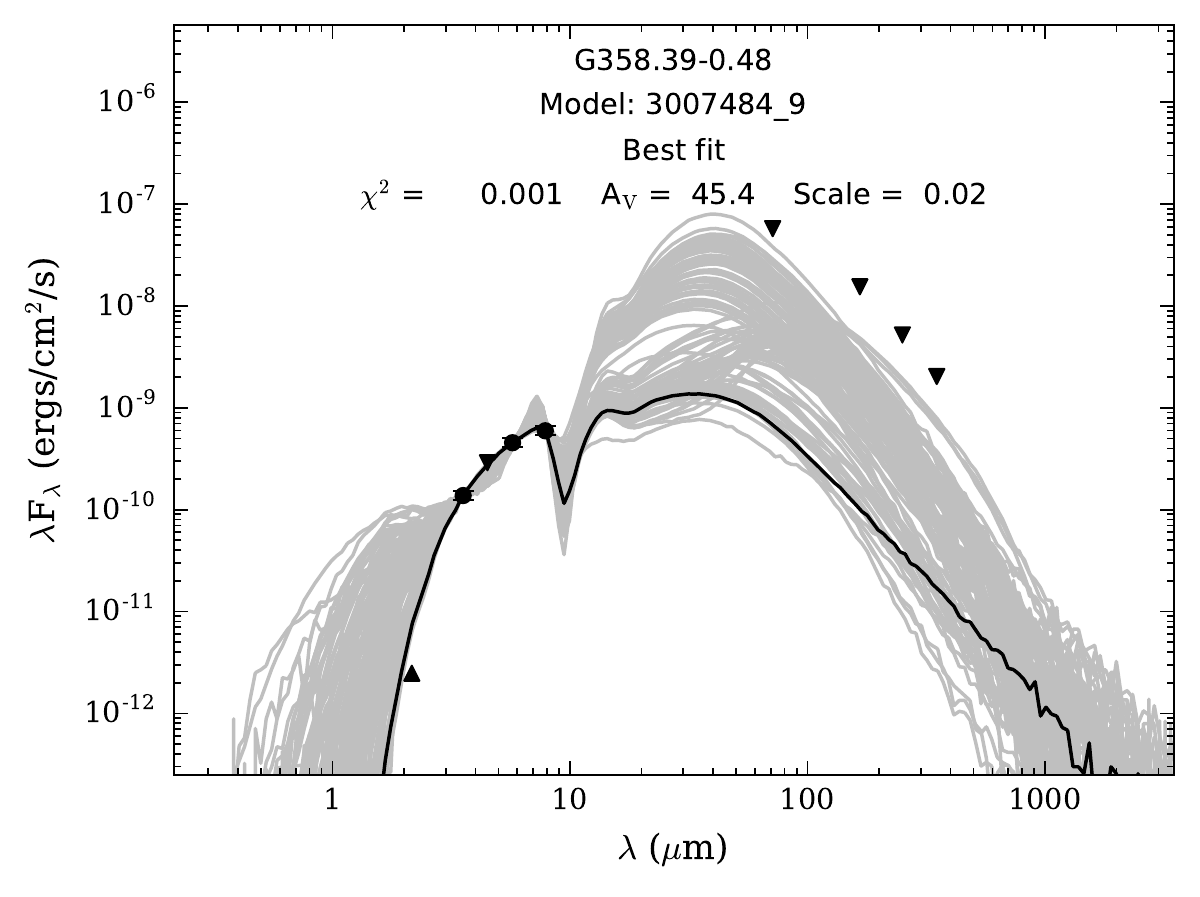}&
                \includegraphics[width=0.31\textwidth]{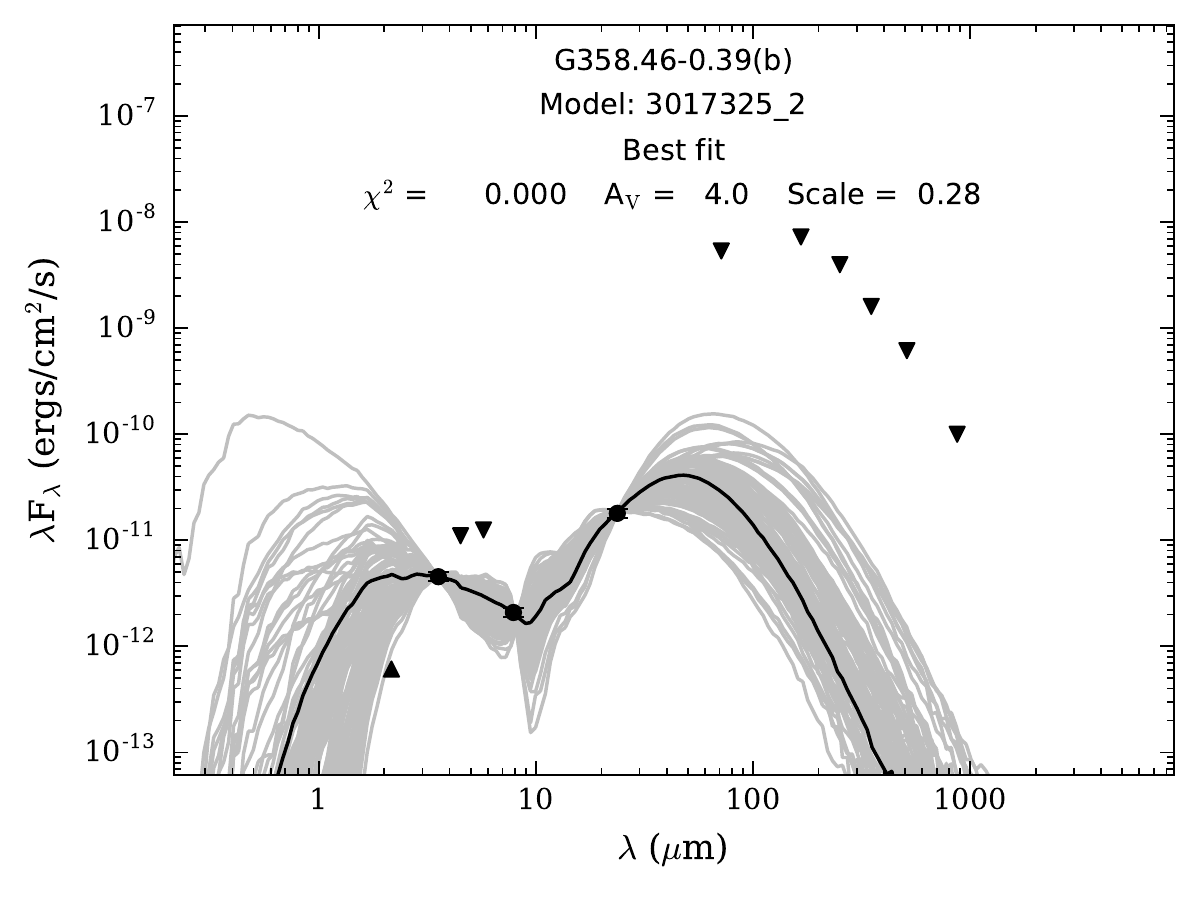}&
                \includegraphics[width=0.31\textwidth]{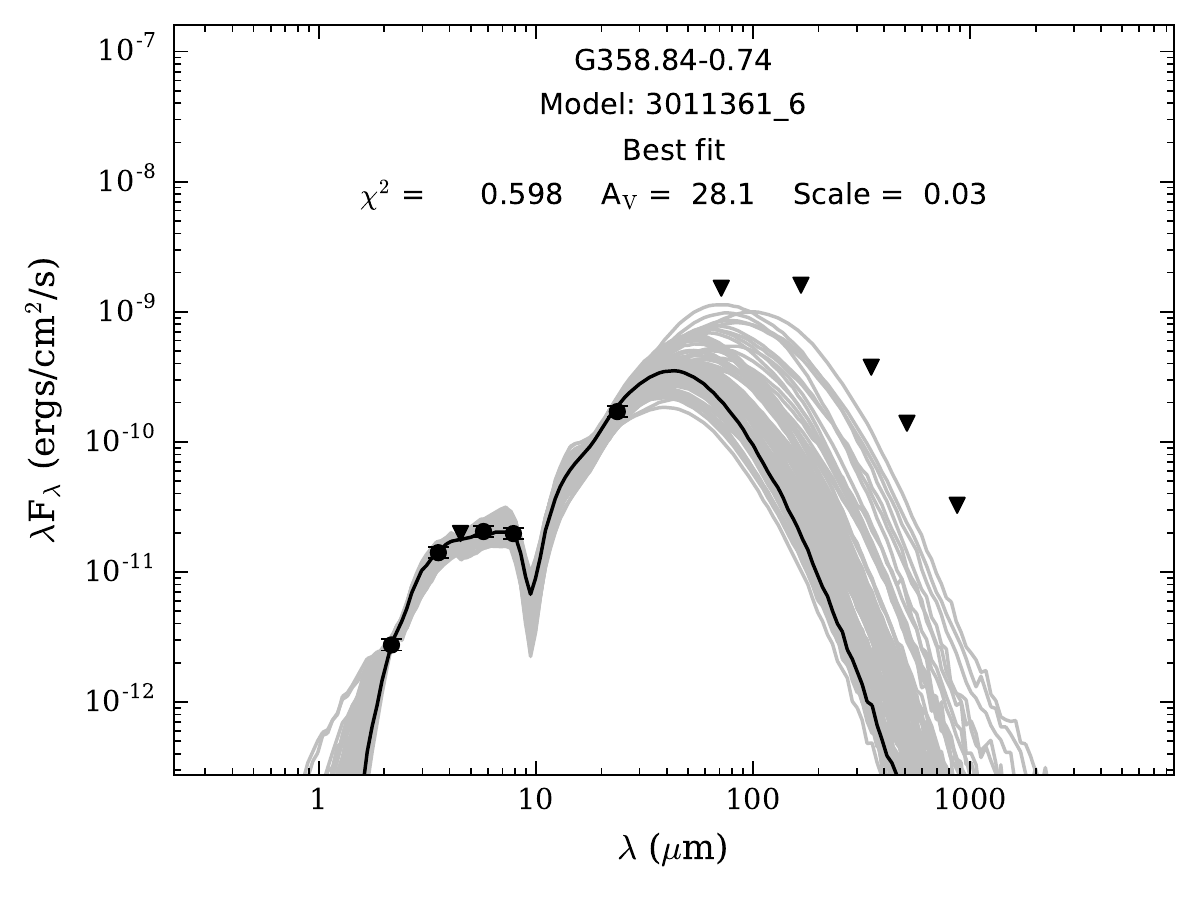}\\
                \includegraphics[width=0.31\textwidth]{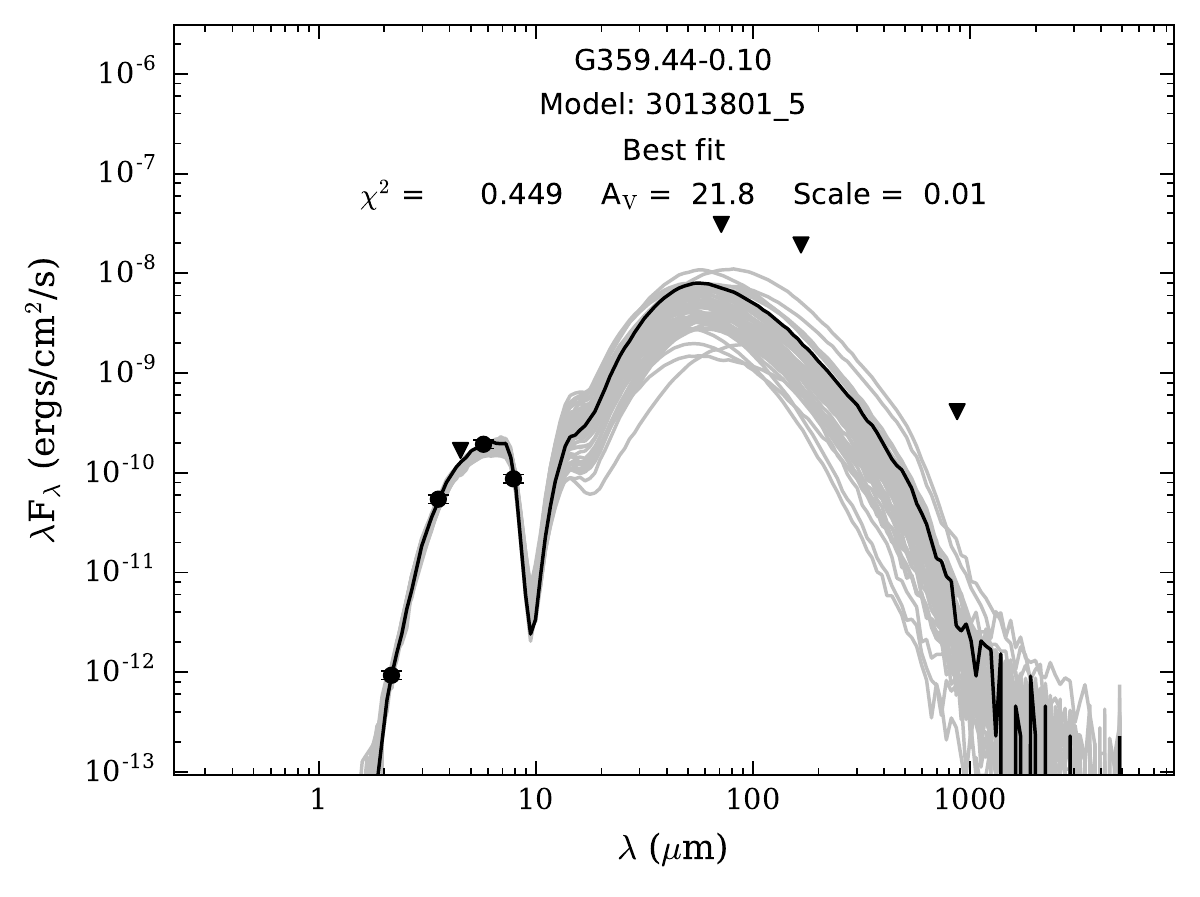}&
                \includegraphics[width=0.31\textwidth]{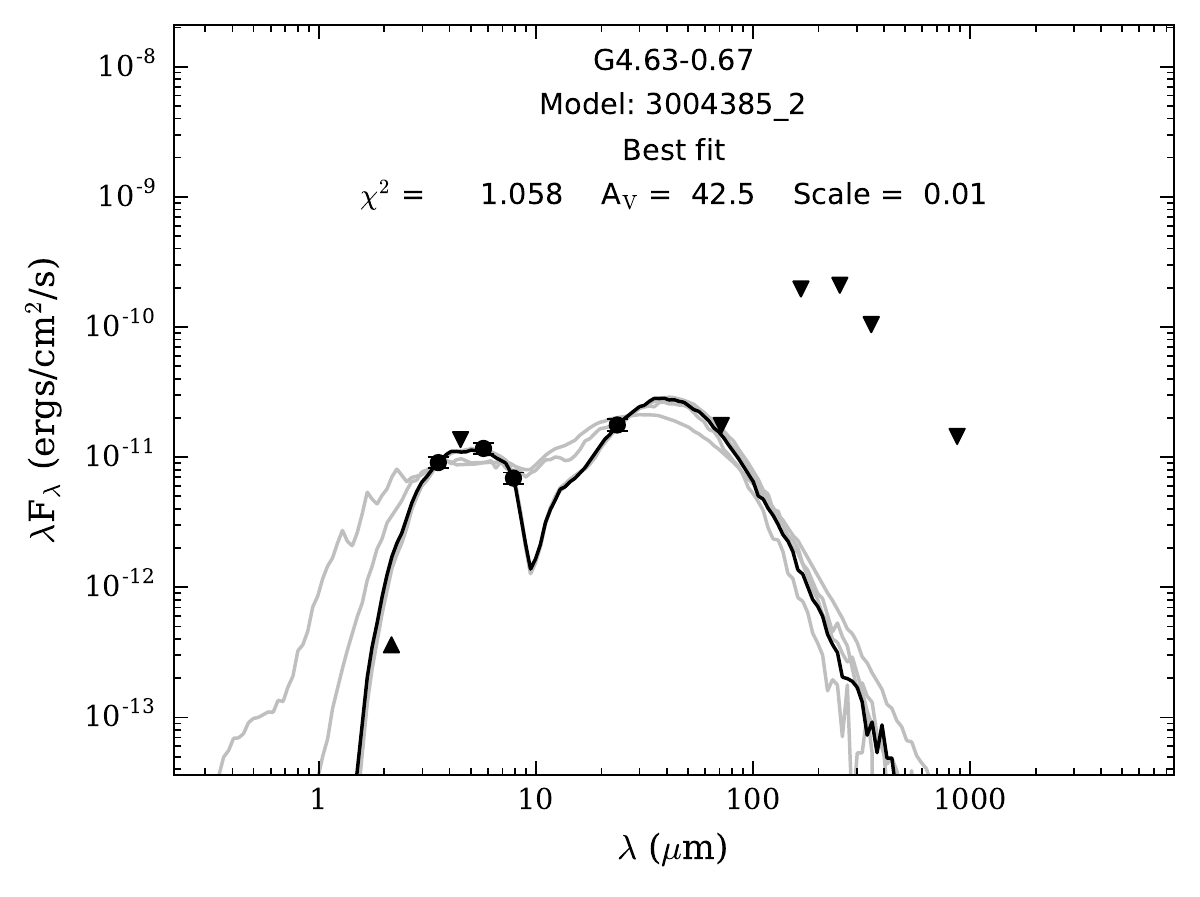}&
                \includegraphics[width=0.31\textwidth]{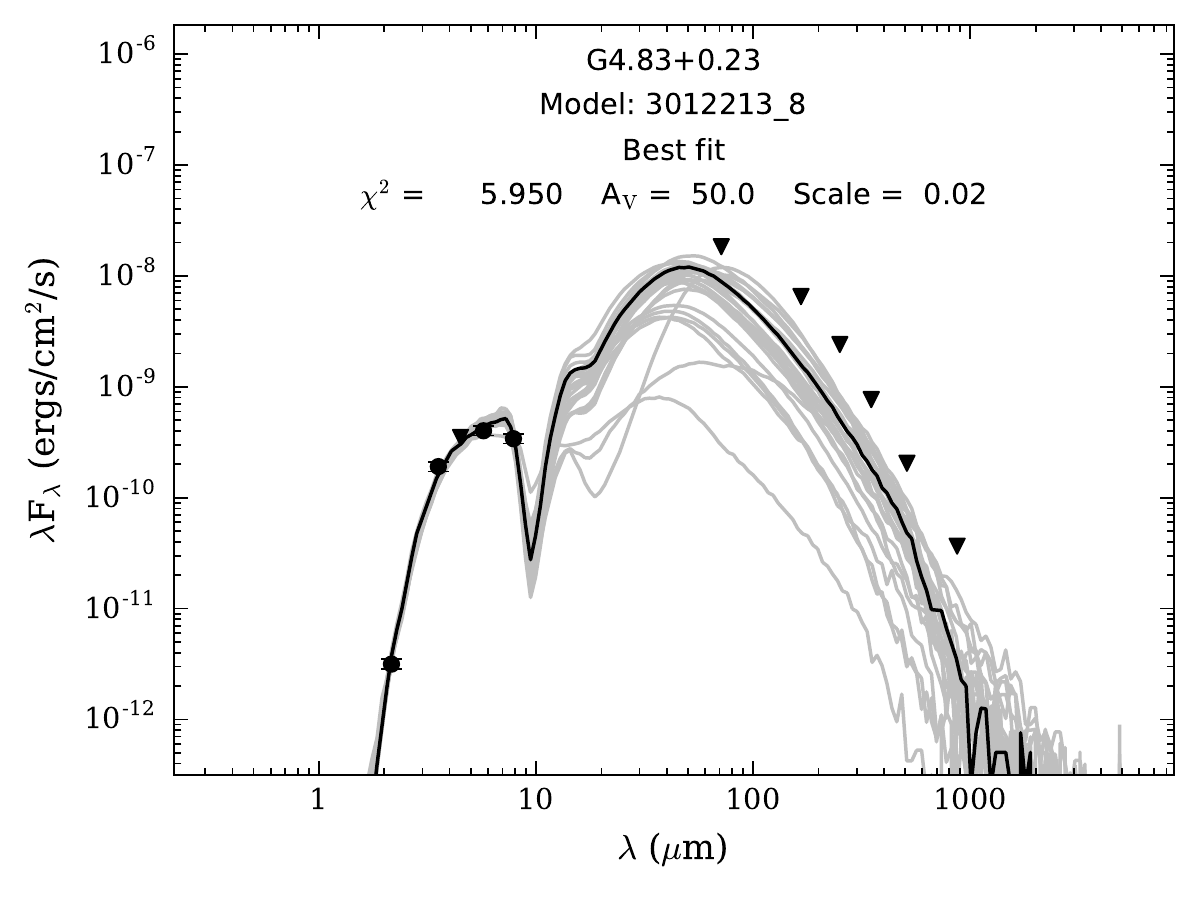}\\
                \includegraphics[width=0.31\textwidth]{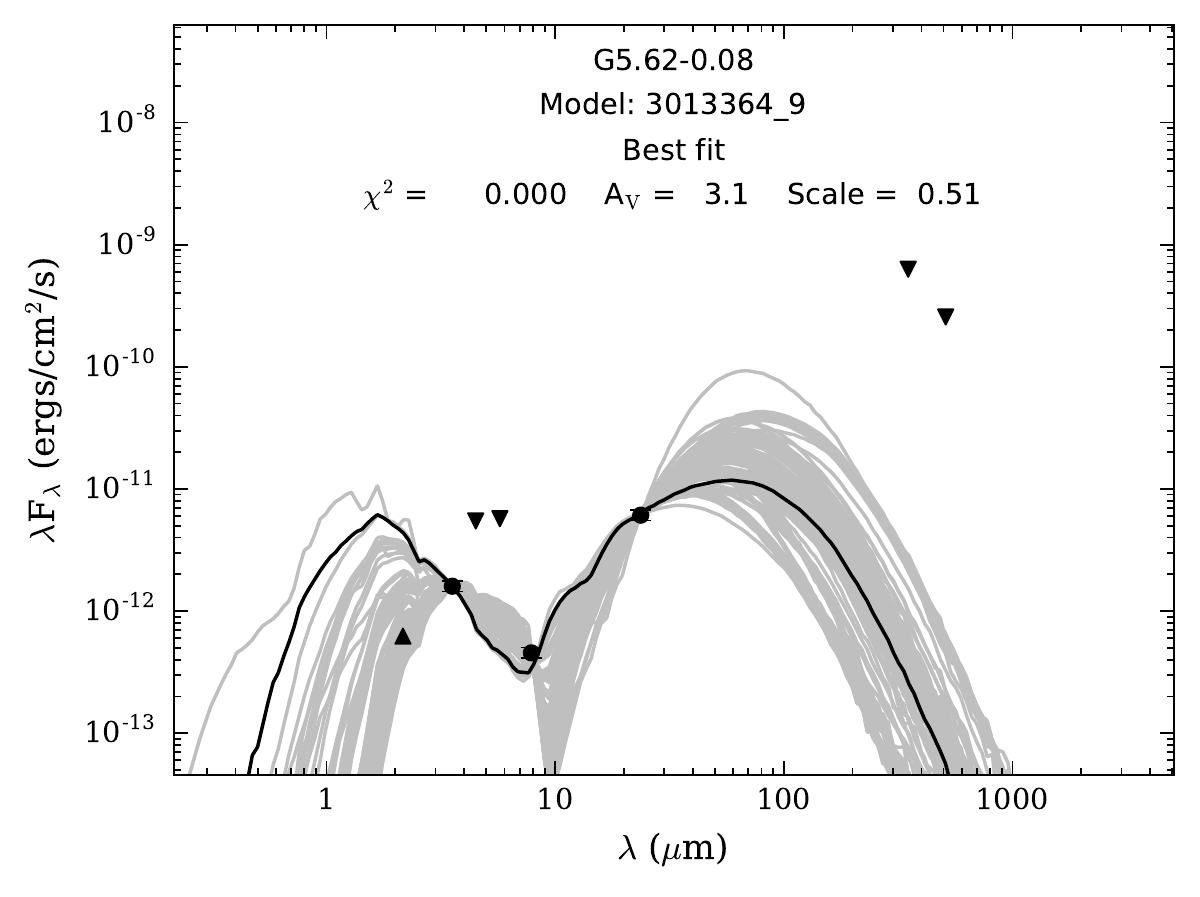}&
                \includegraphics[width=0.31\textwidth]{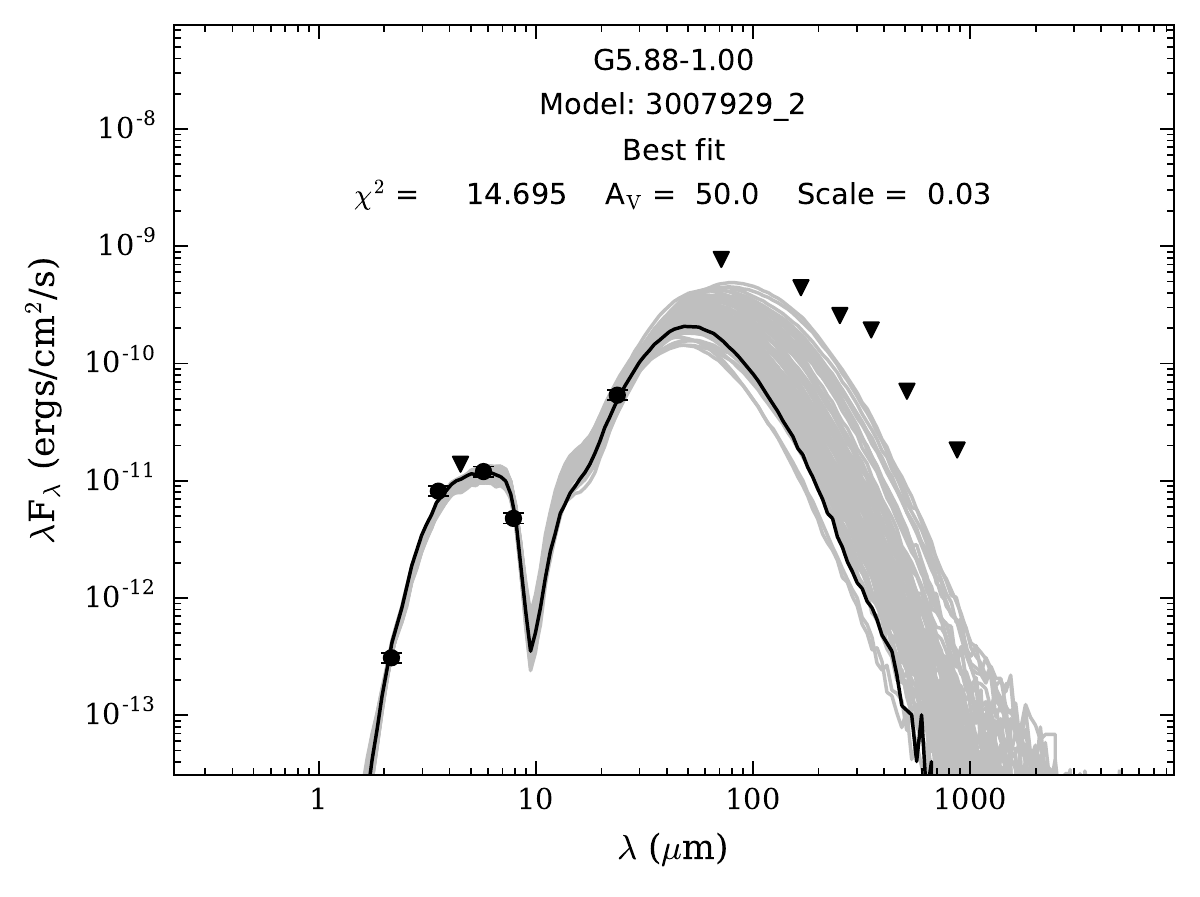}&
                \includegraphics[width=0.31\textwidth]{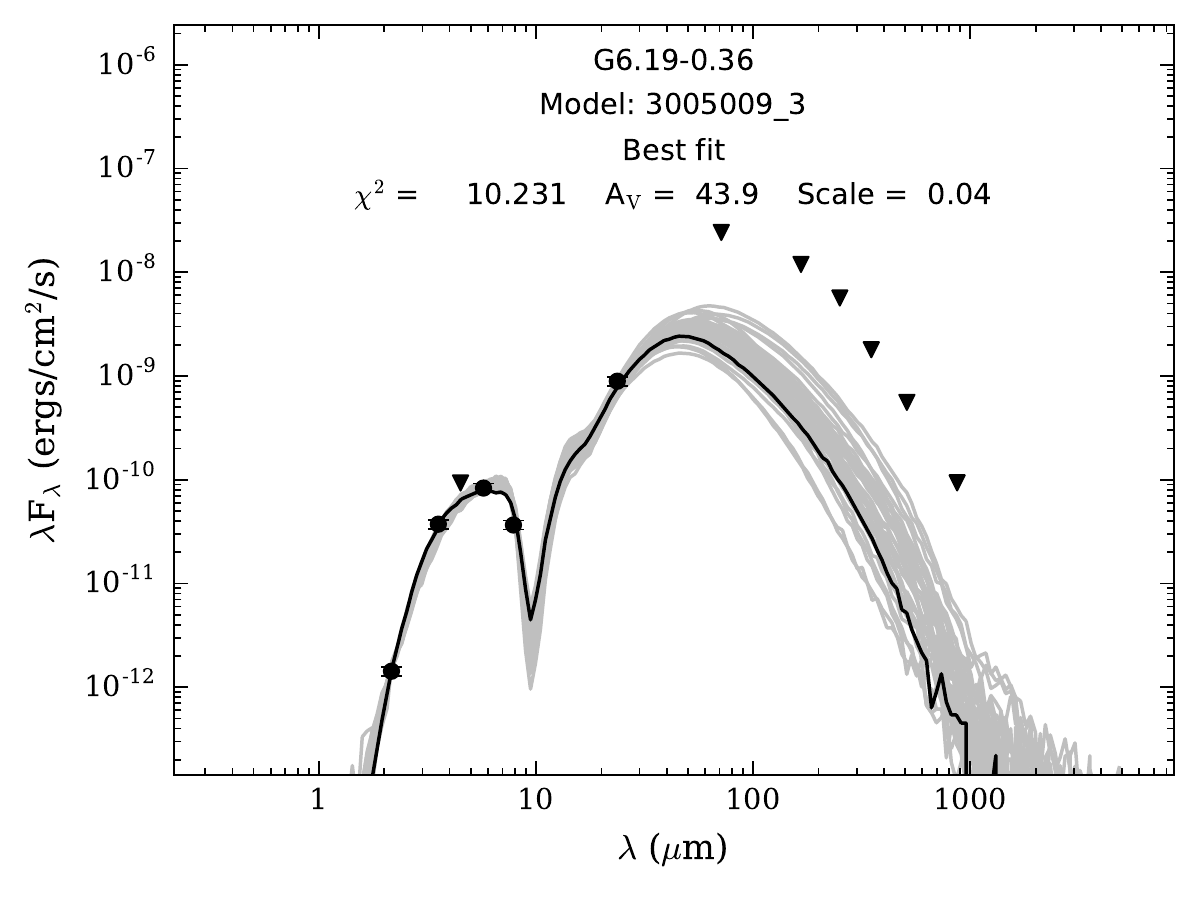}\\
                \includegraphics[width=0.31\textwidth]{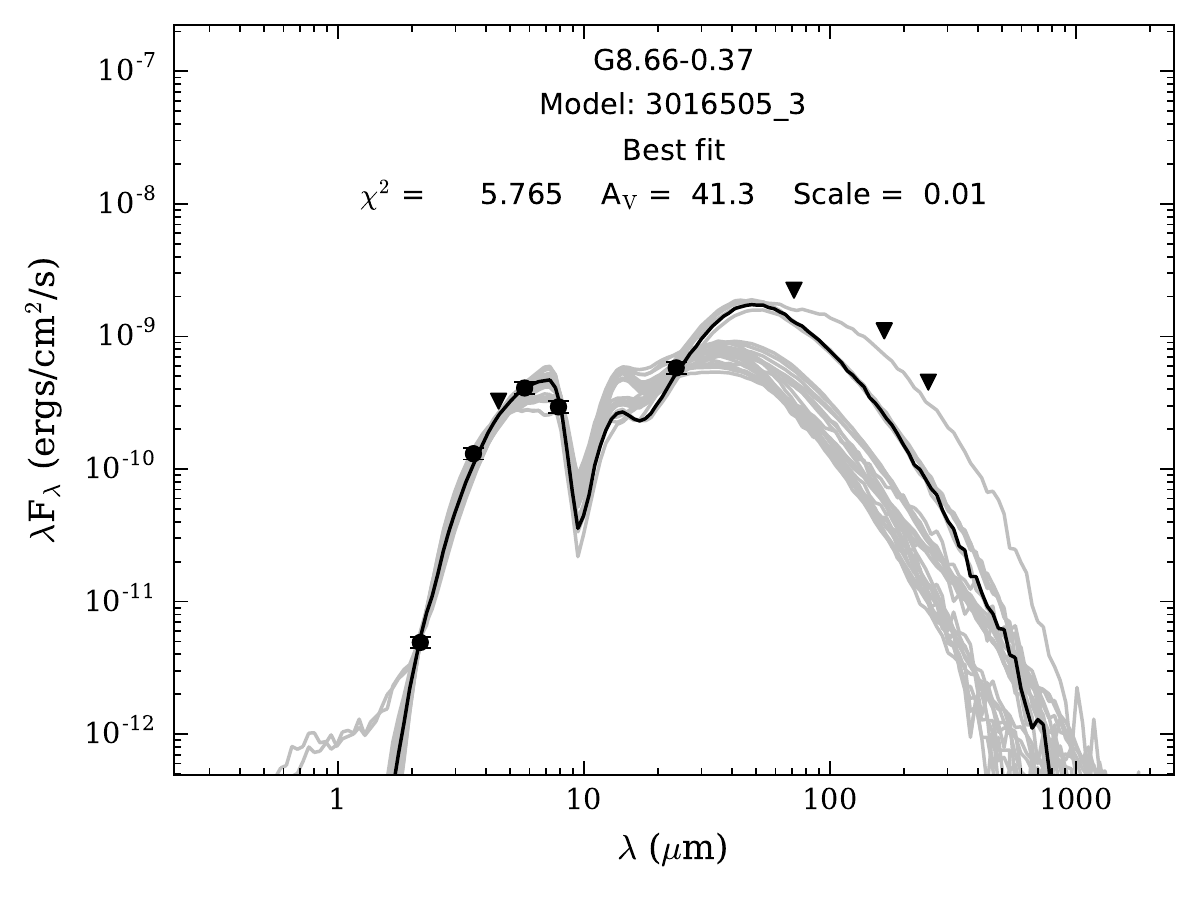}&
                \includegraphics[width=0.31\textwidth]{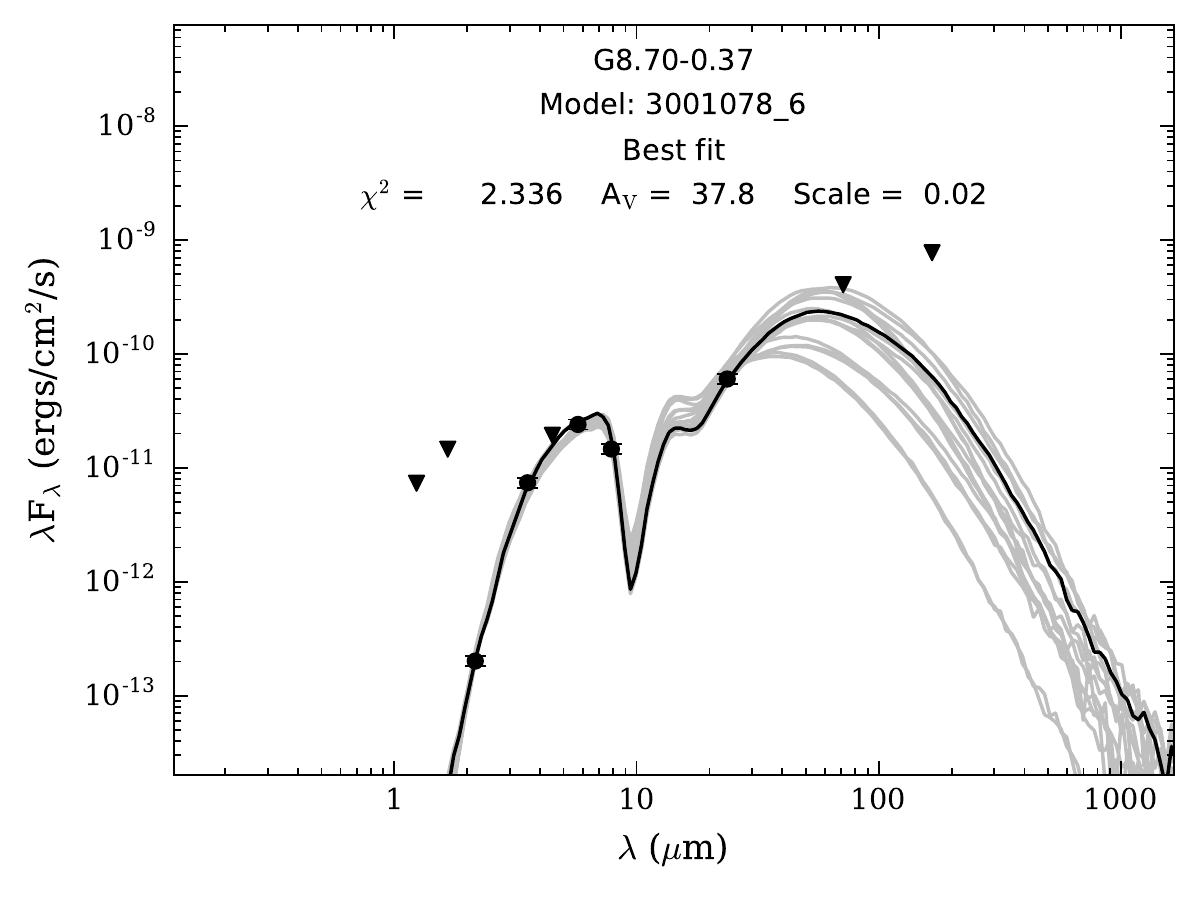}&
                \includegraphics[width=0.31\textwidth]{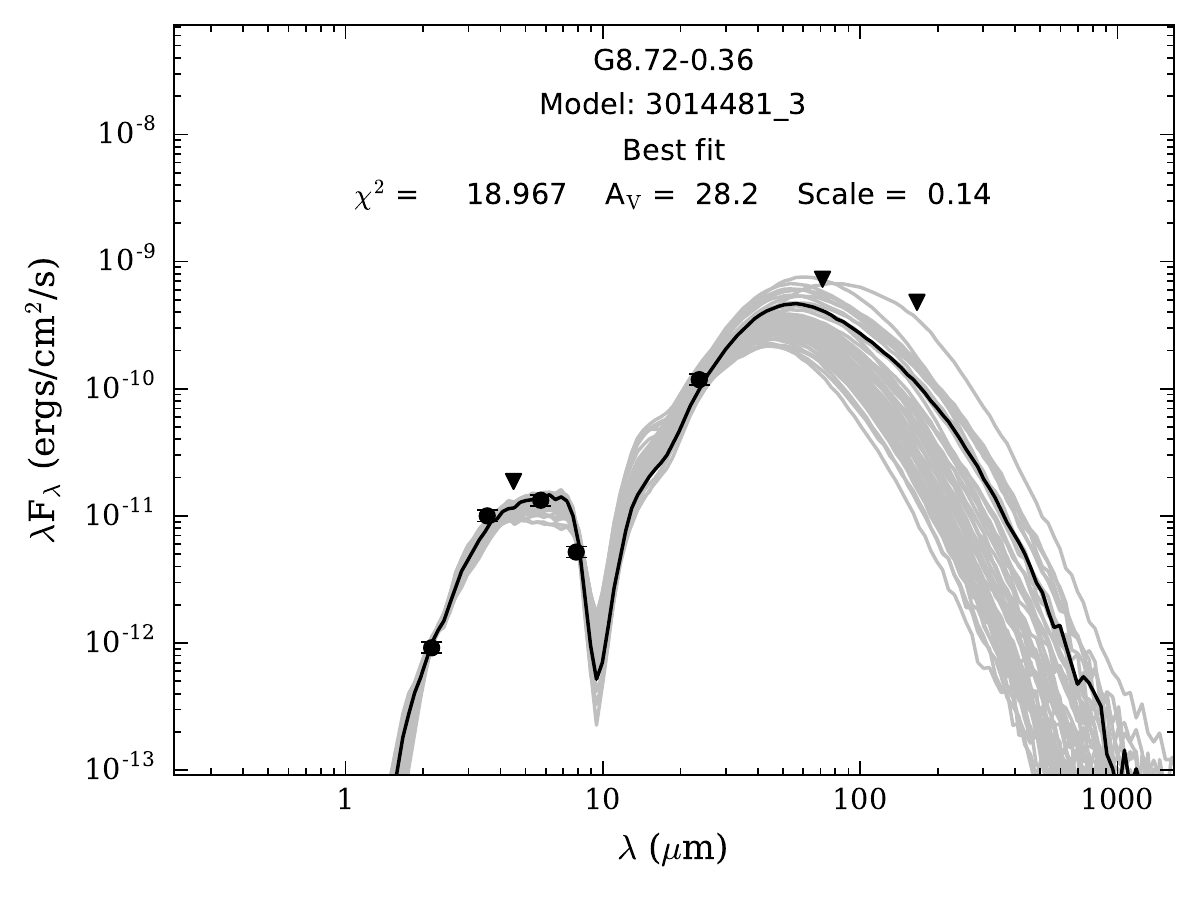}\\
                \includegraphics[width=0.31\textwidth]{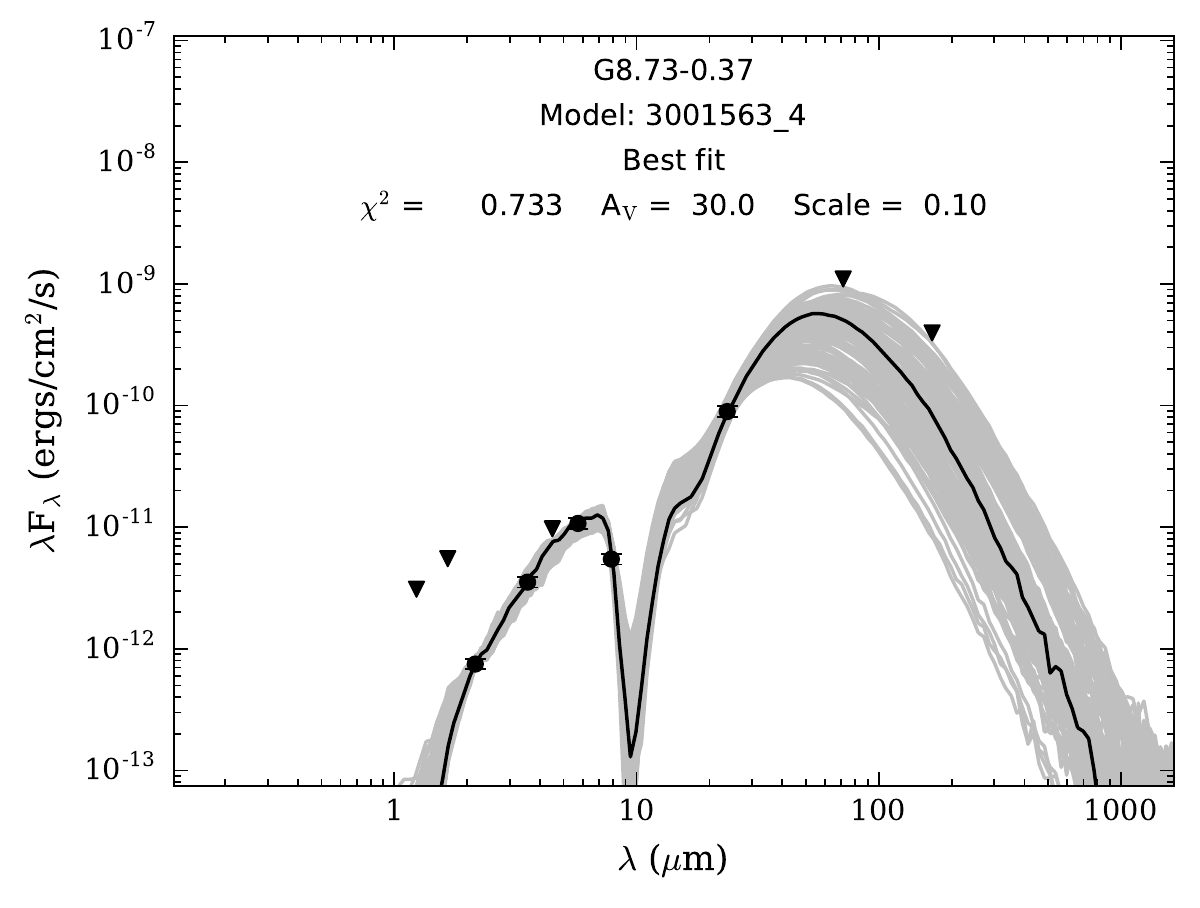}&
                \includegraphics[width=0.31\textwidth]{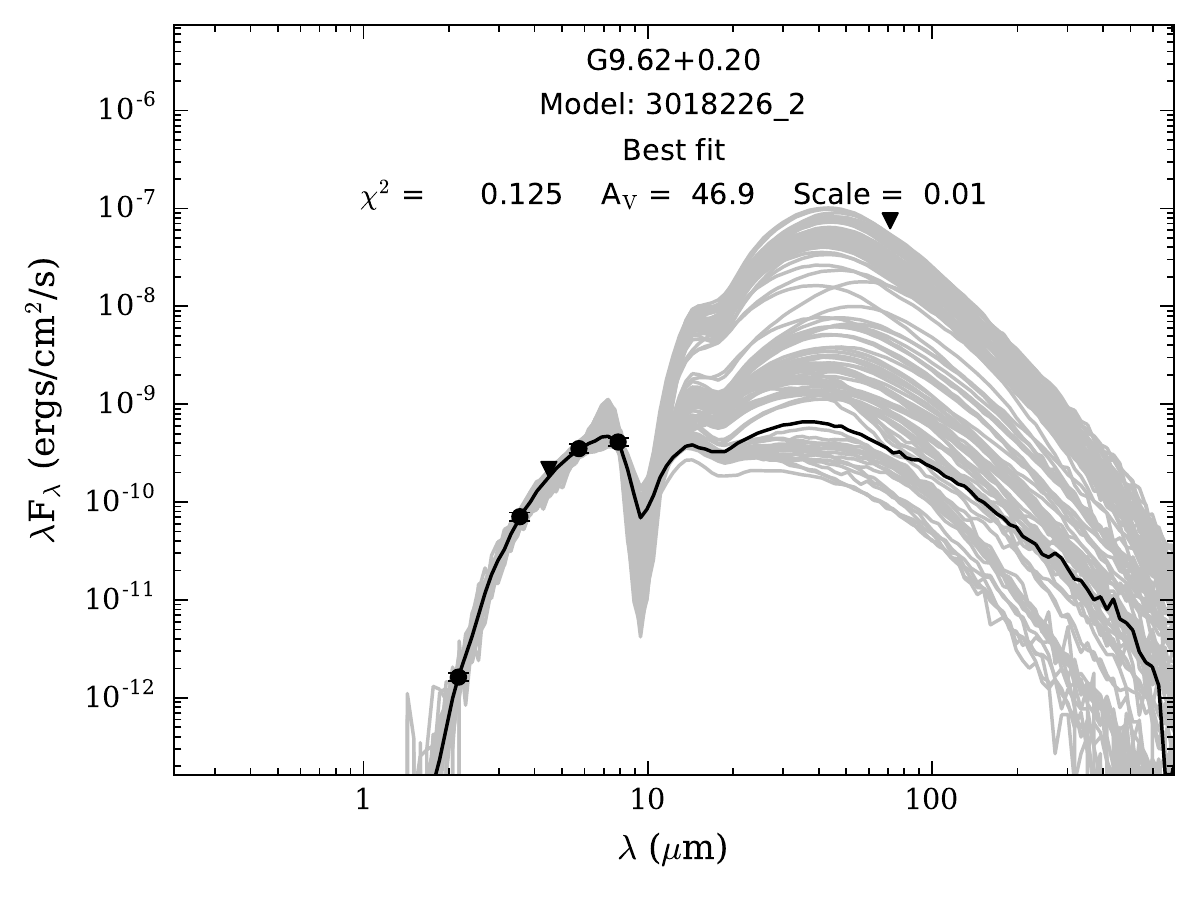}&
                \includegraphics[width=0.31\textwidth]{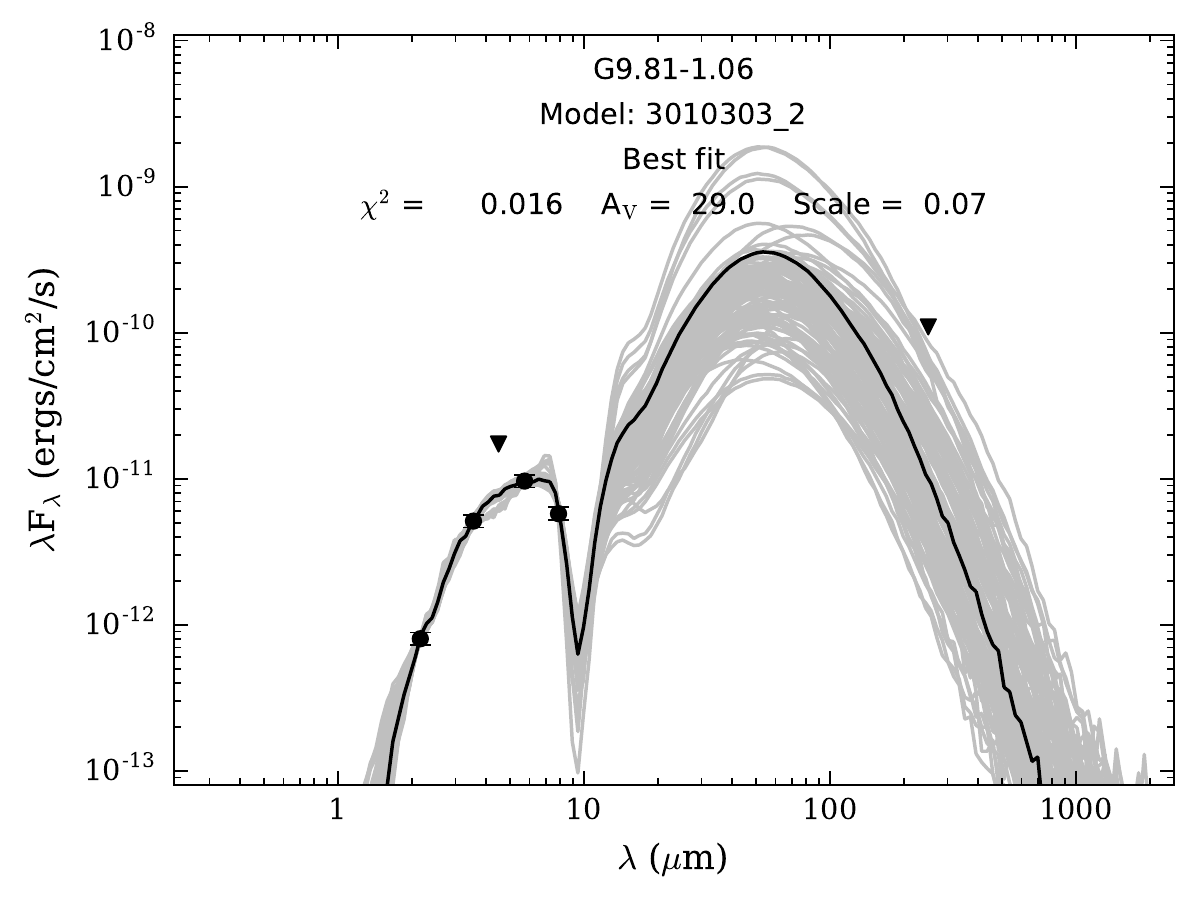}\\
        \end{tabular}
        \caption{Continuation of Fig. \ref{fig:testgridsed}.}
\end{figure}

\begin{figure}
        \begin{tabular}{ccc}
                
                \includegraphics[width=0.31\textwidth]{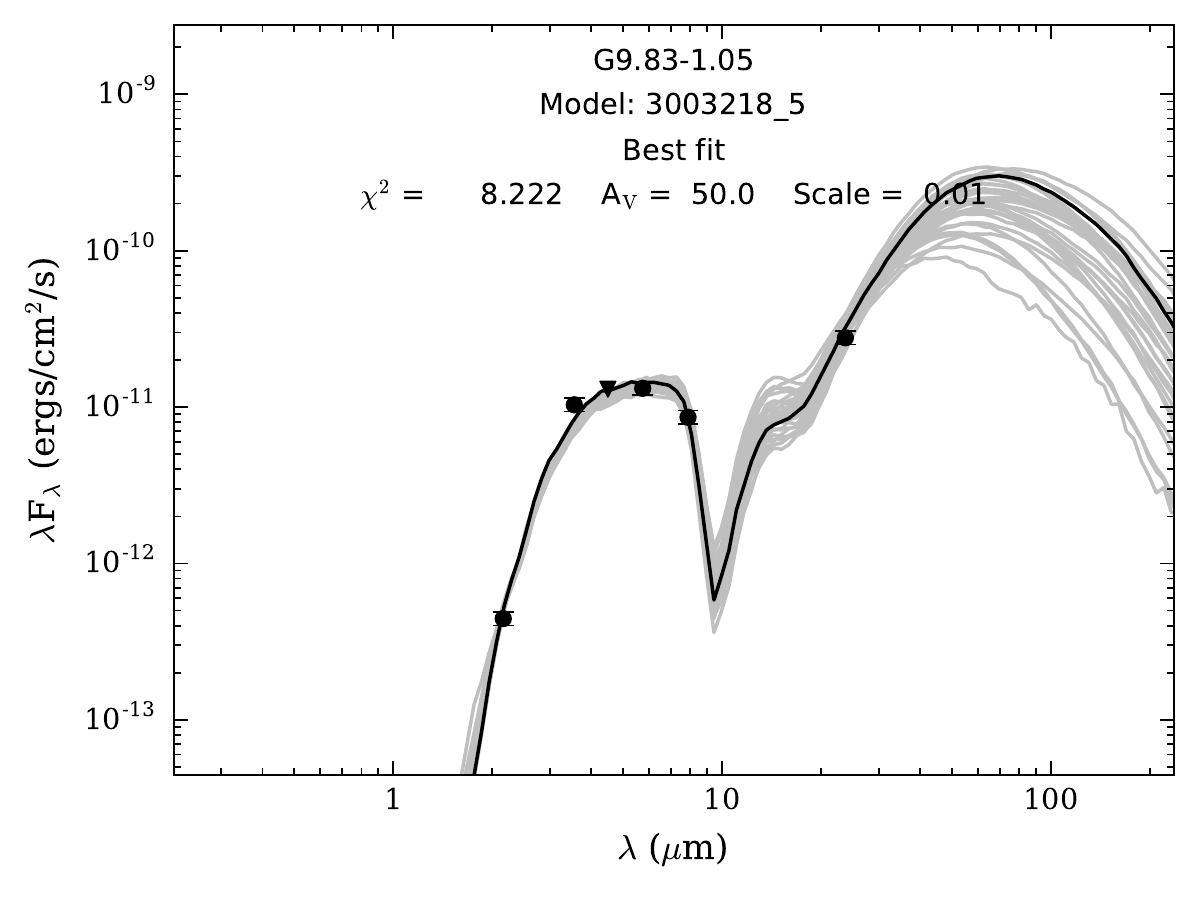}&
                \includegraphics[width=0.31\textwidth]{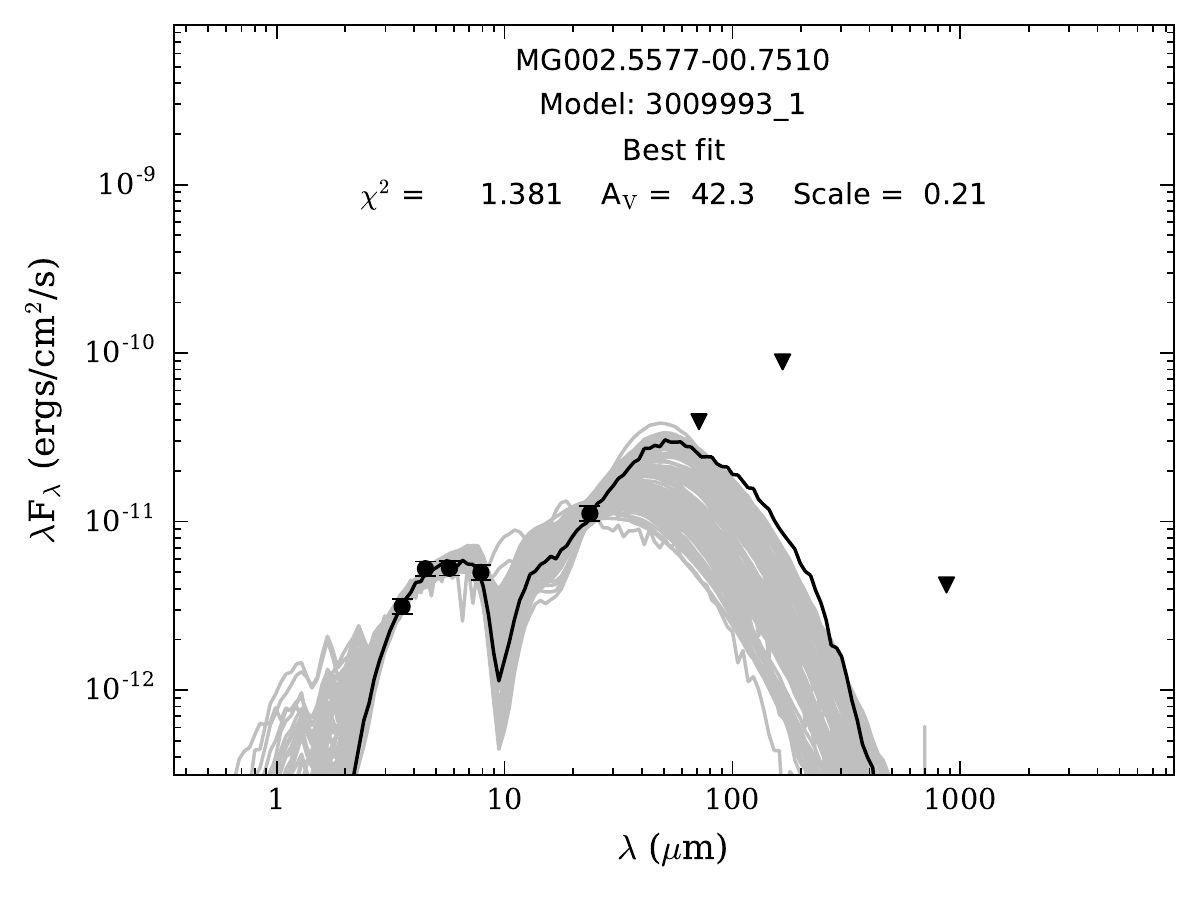}&
                \includegraphics[width=0.31\textwidth]{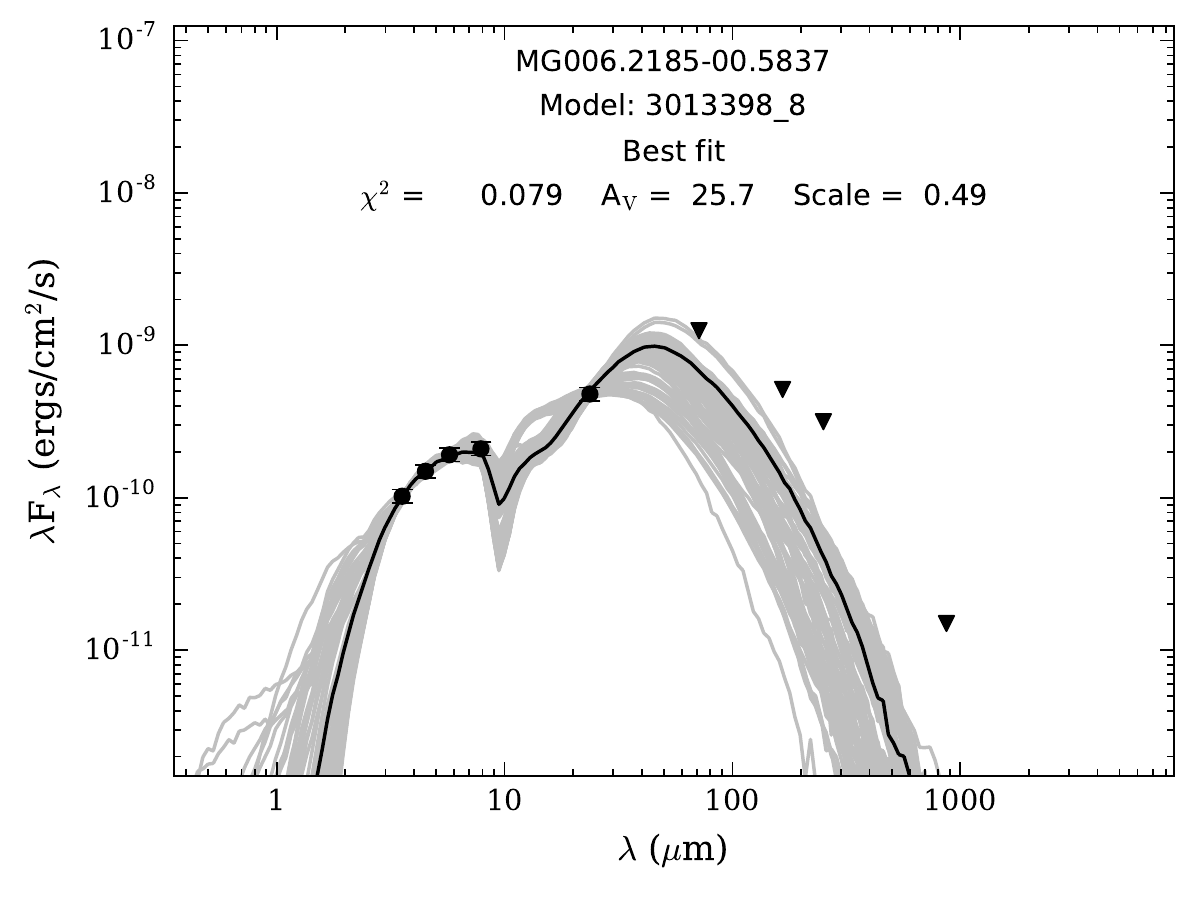}\\
                \includegraphics[width=0.31\textwidth]{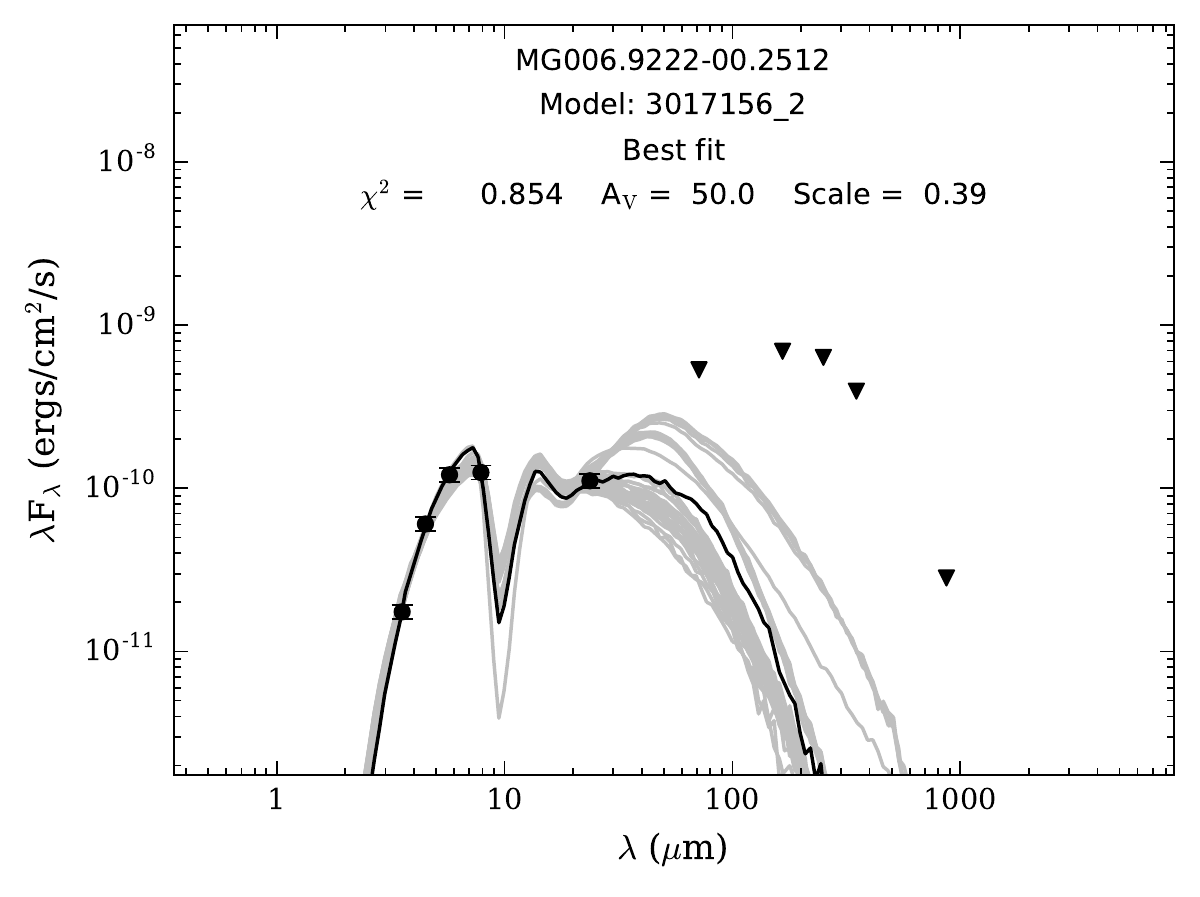}&
                \includegraphics[width=0.31\textwidth]{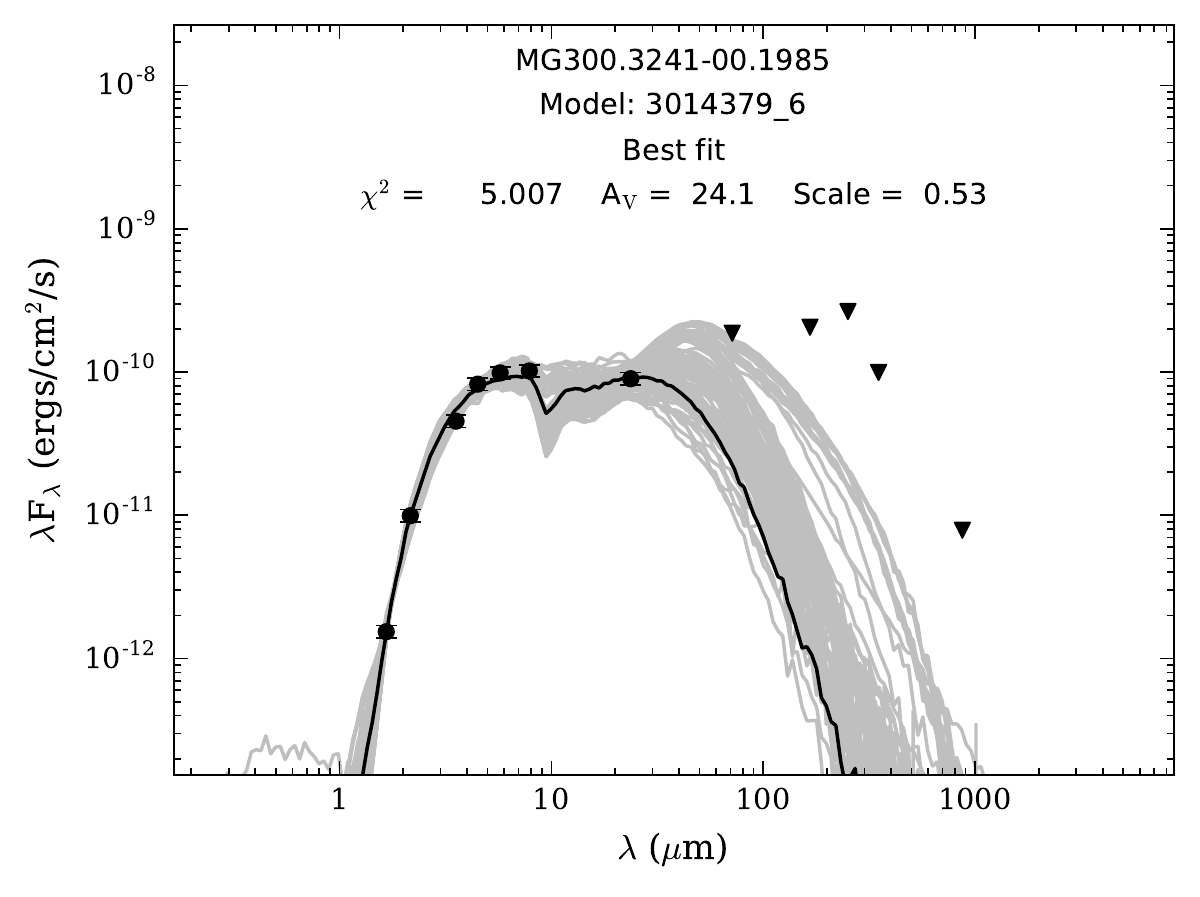}&
                \includegraphics[width=0.31\textwidth]{{./MG303.9304-00.6879}.pdf}\\
                \includegraphics[width=0.31\textwidth]{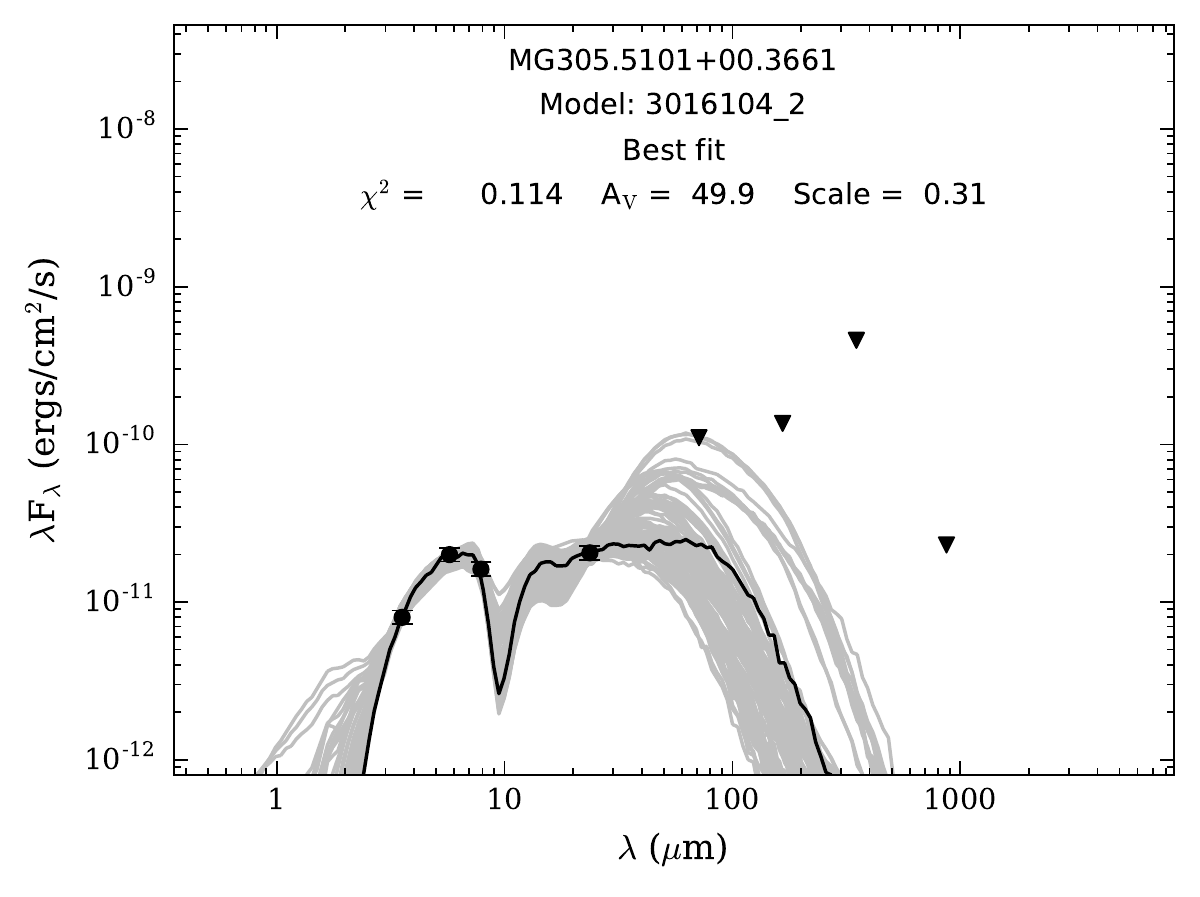}&
                \includegraphics[width=0.31\textwidth]{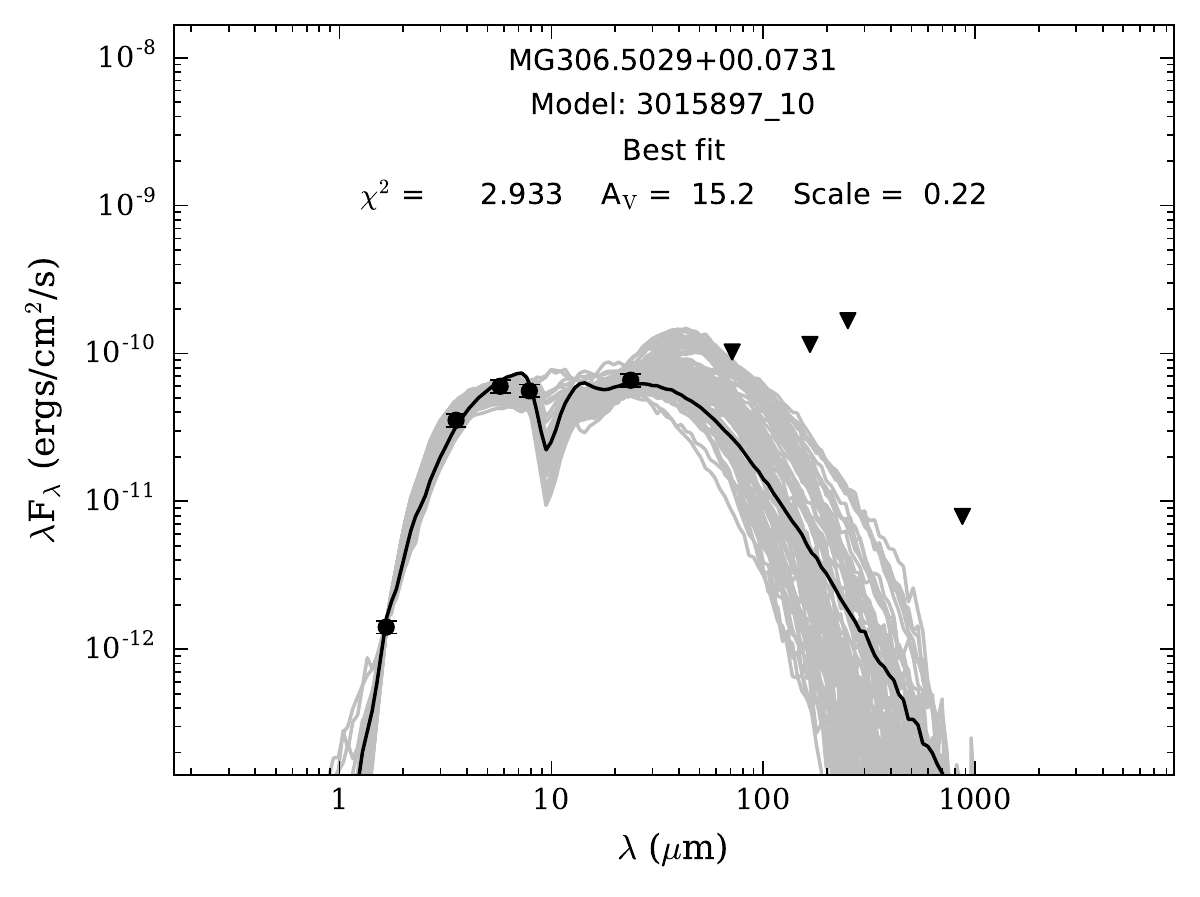}&
                \includegraphics[width=0.31\textwidth]{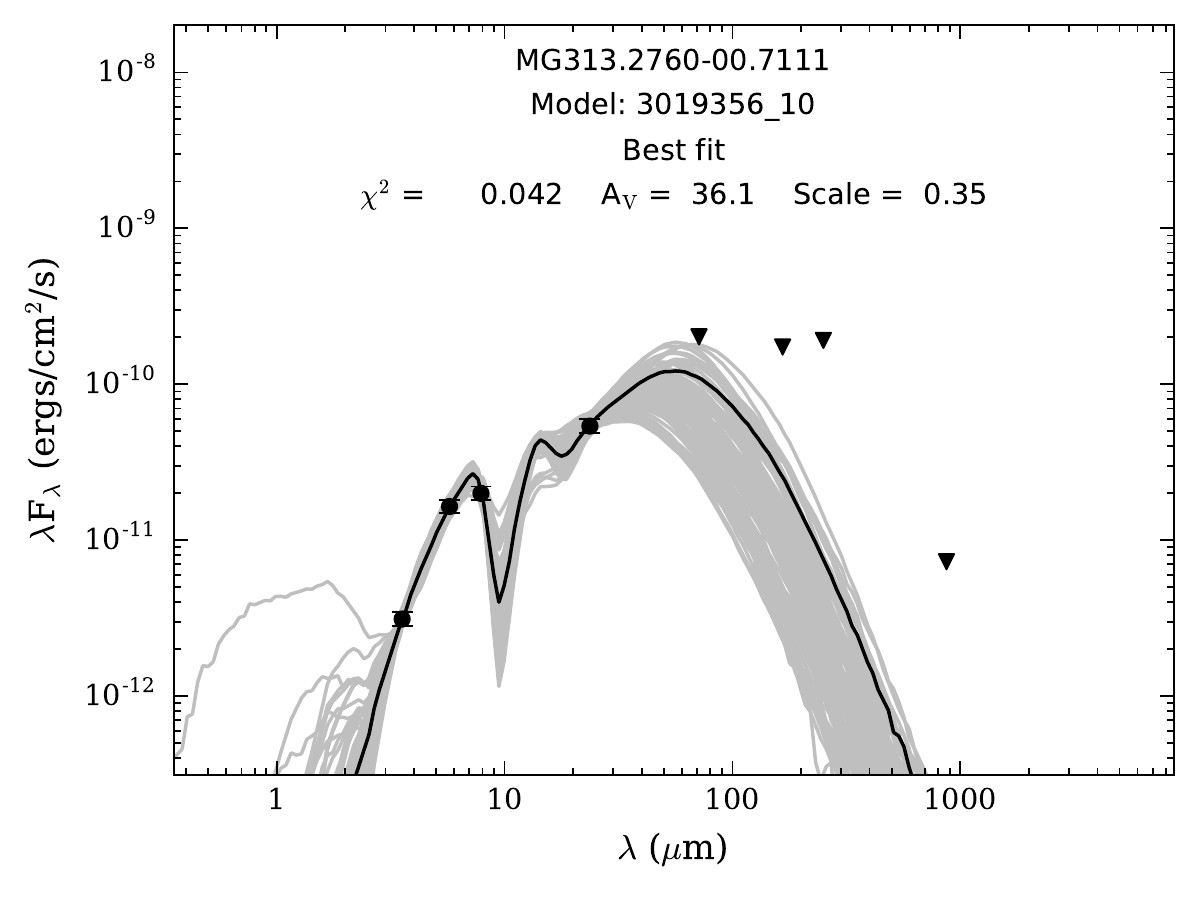}\\
                \includegraphics[width=0.31\textwidth]{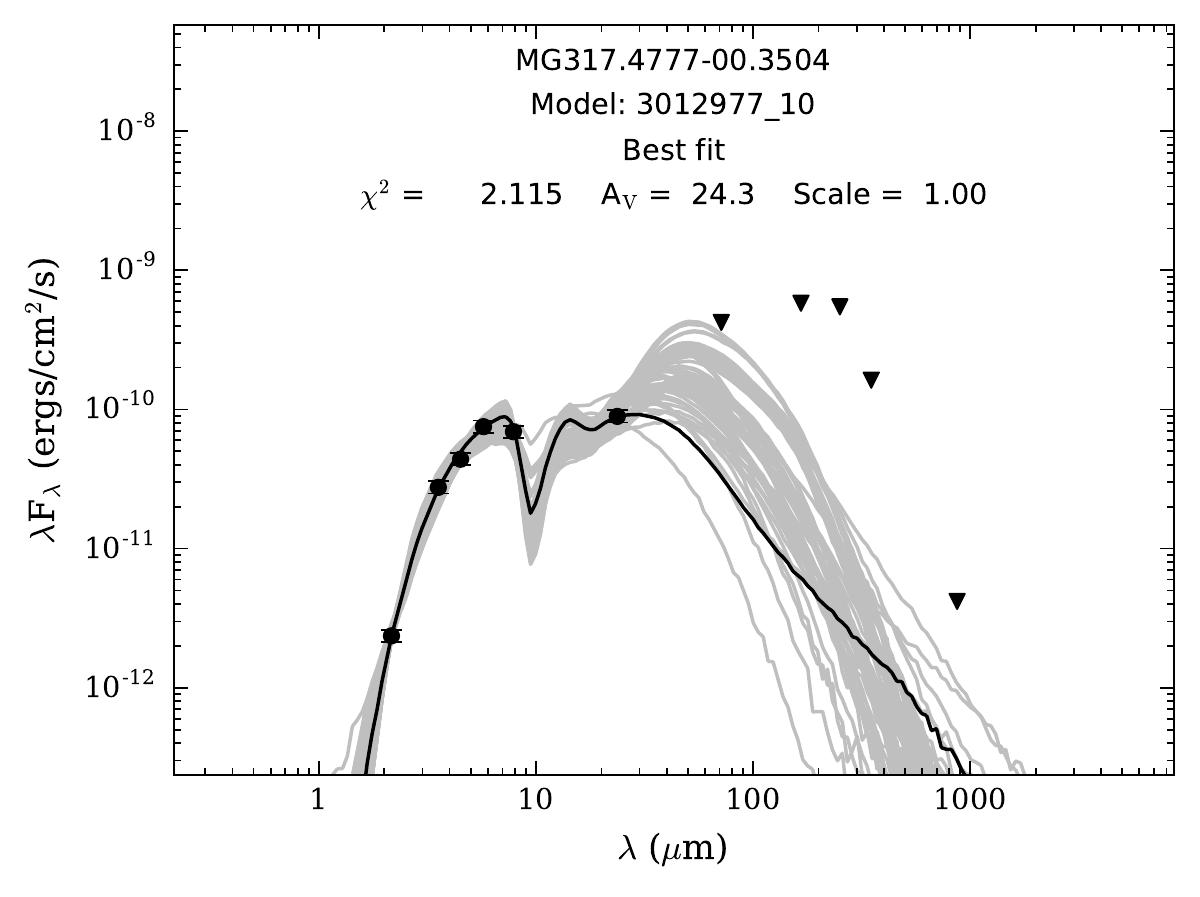}&
                \includegraphics[width=0.31\textwidth]{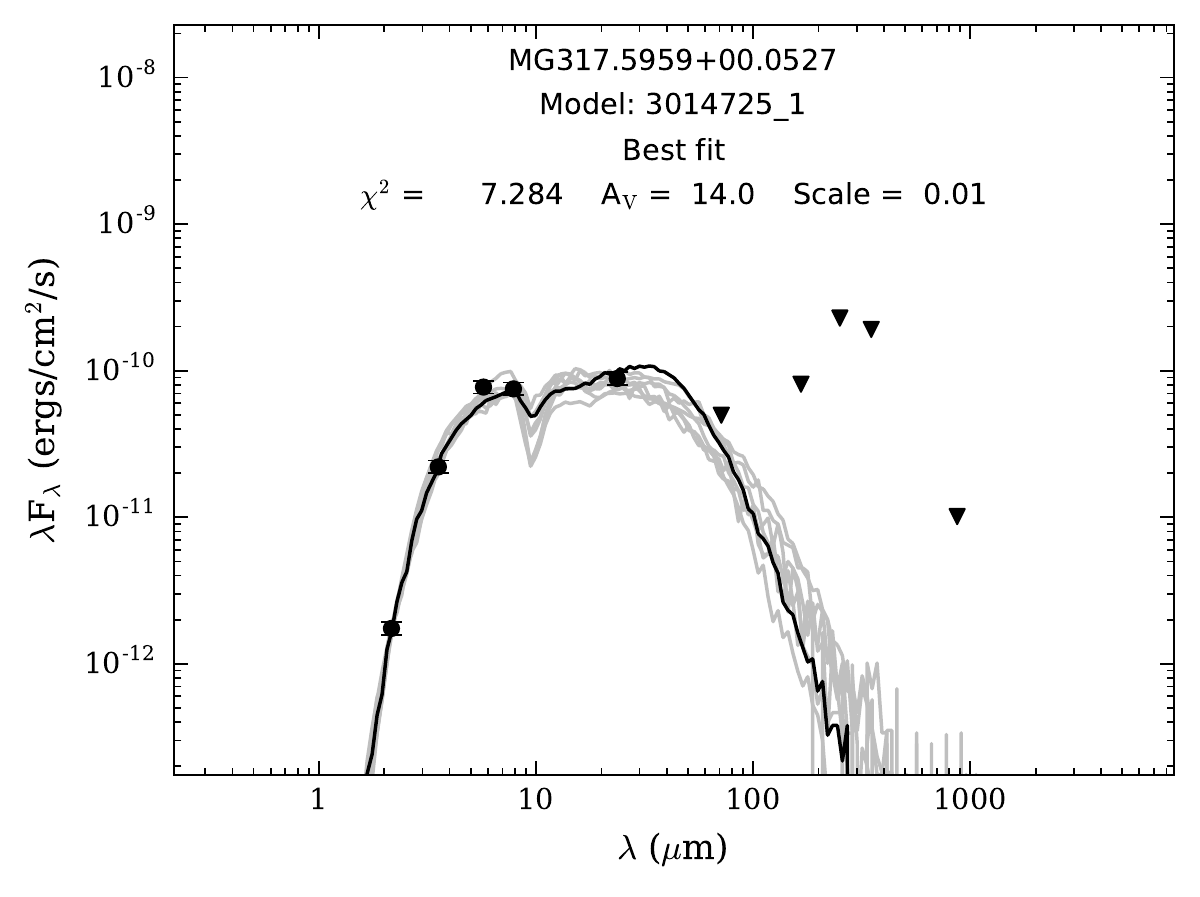}&
                \includegraphics[width=0.31\textwidth]{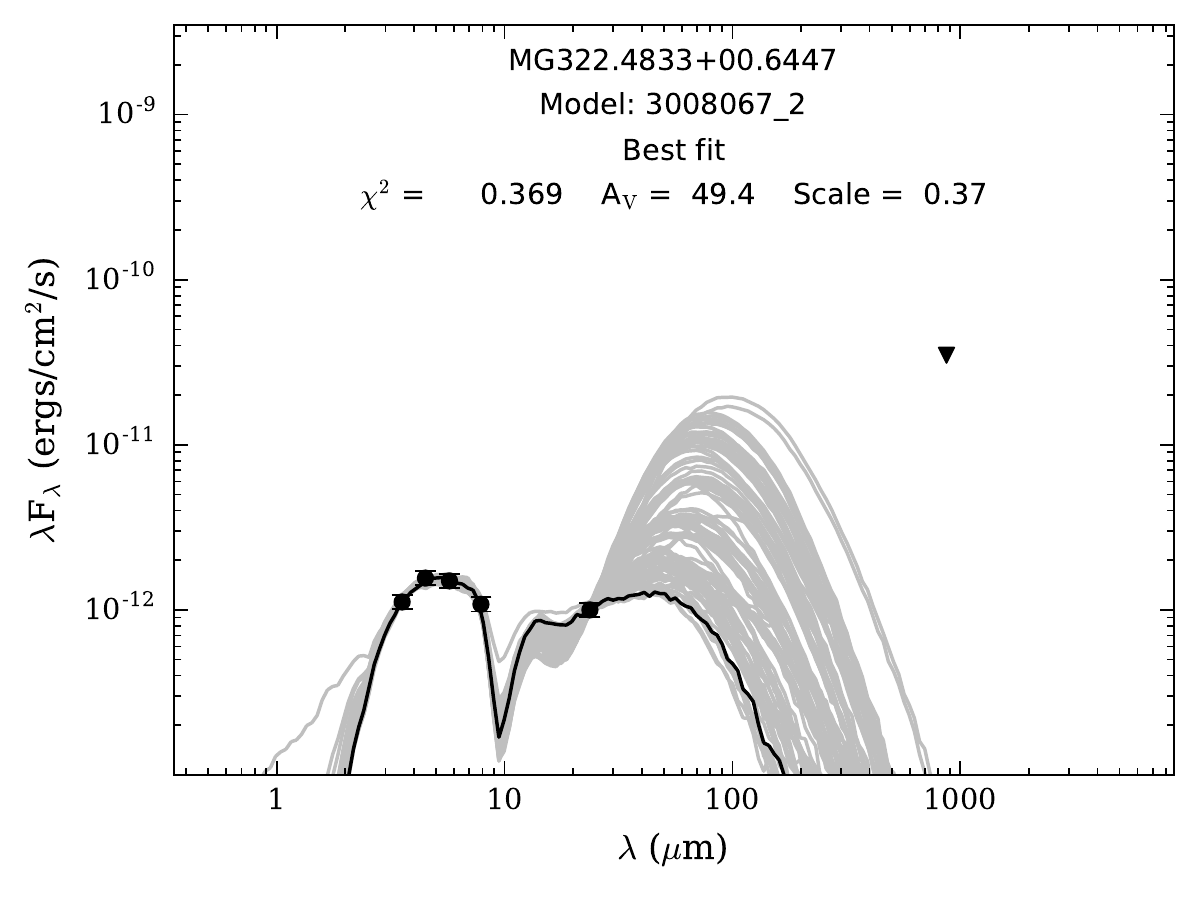}\\
                \includegraphics[width=0.31\textwidth]{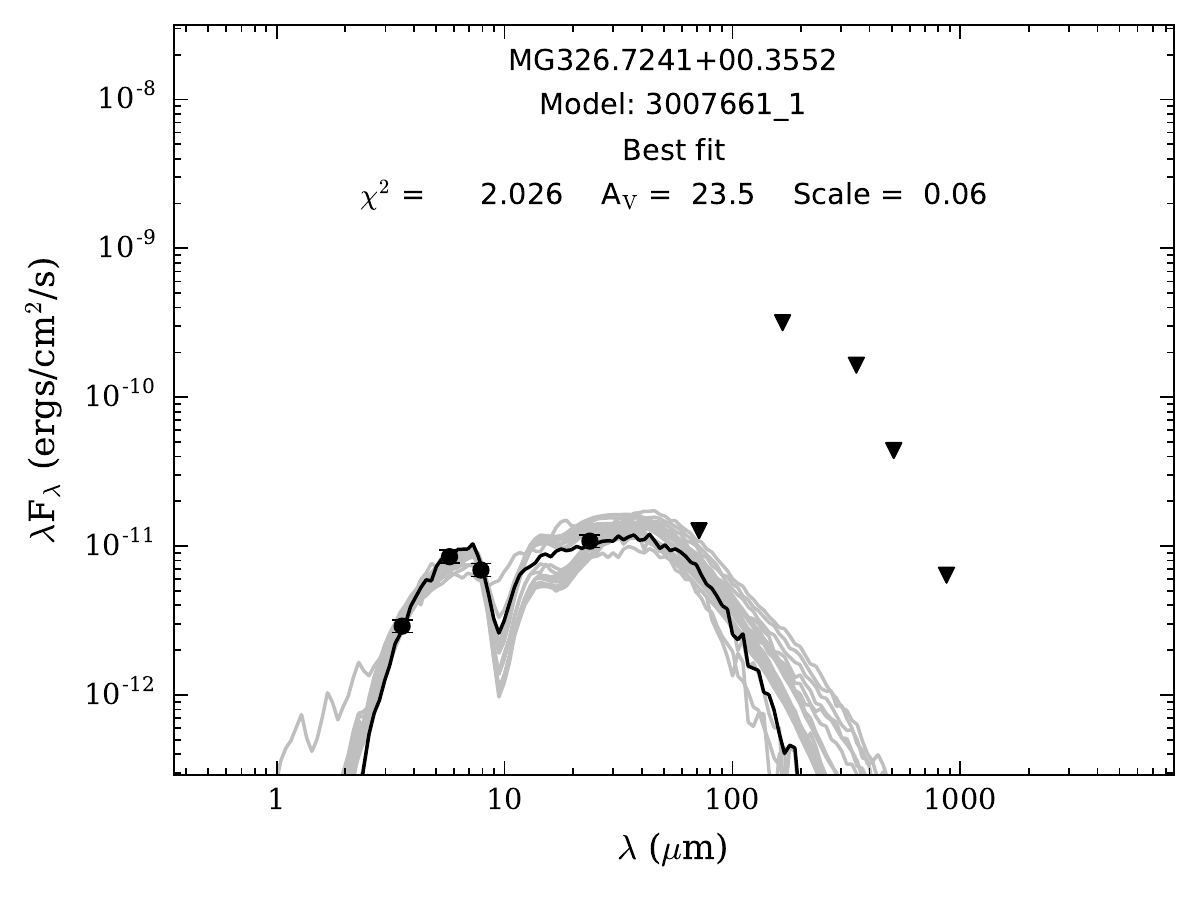}&
                \includegraphics[width=0.31\textwidth]{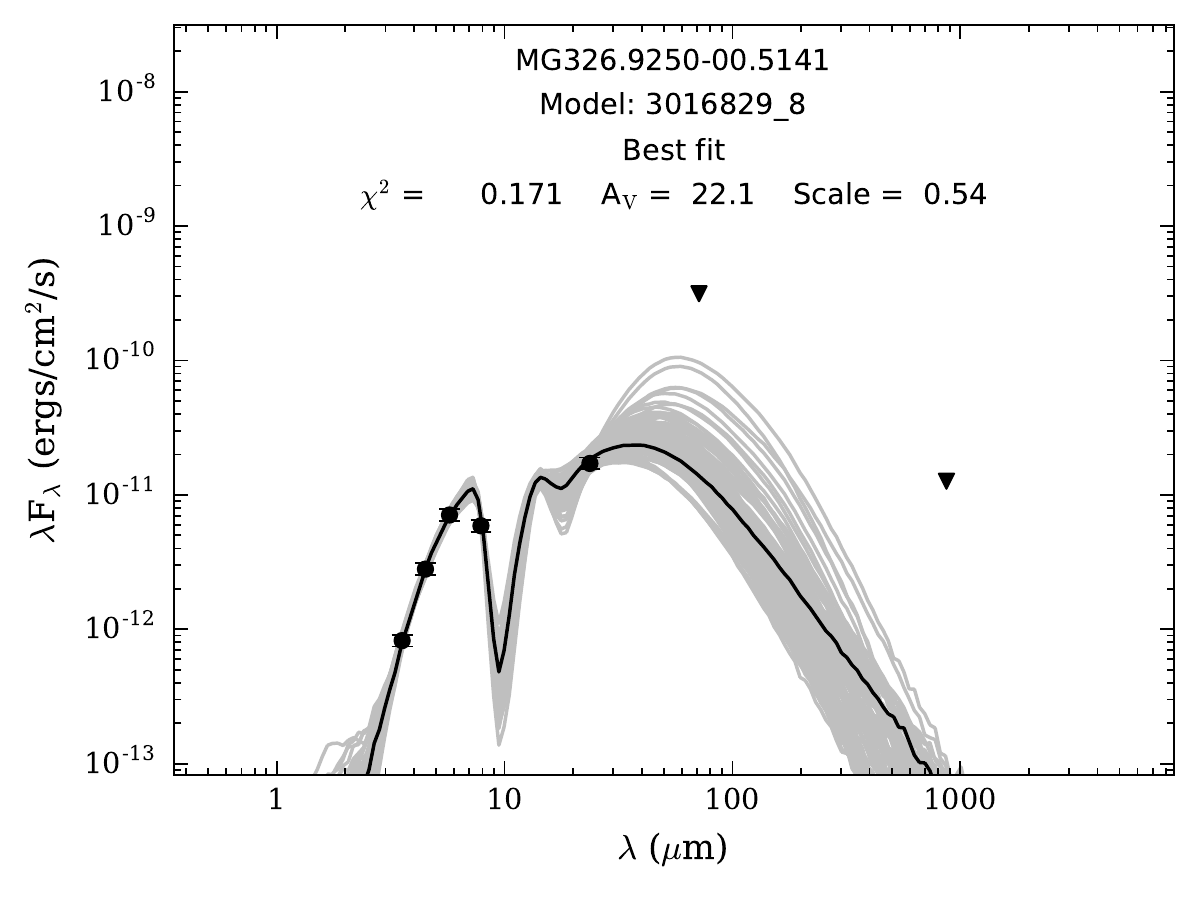}&
                \includegraphics[width=0.31\textwidth]{{./MG328.0494-00.0487}.pdf}\\
        \end{tabular}
        \caption{Continuation of Fig. \ref{fig:testgridsed}.}
\end{figure}

\begin{figure}
        \begin{tabular}{ccc}
                
                \includegraphics[width=0.31\textwidth]{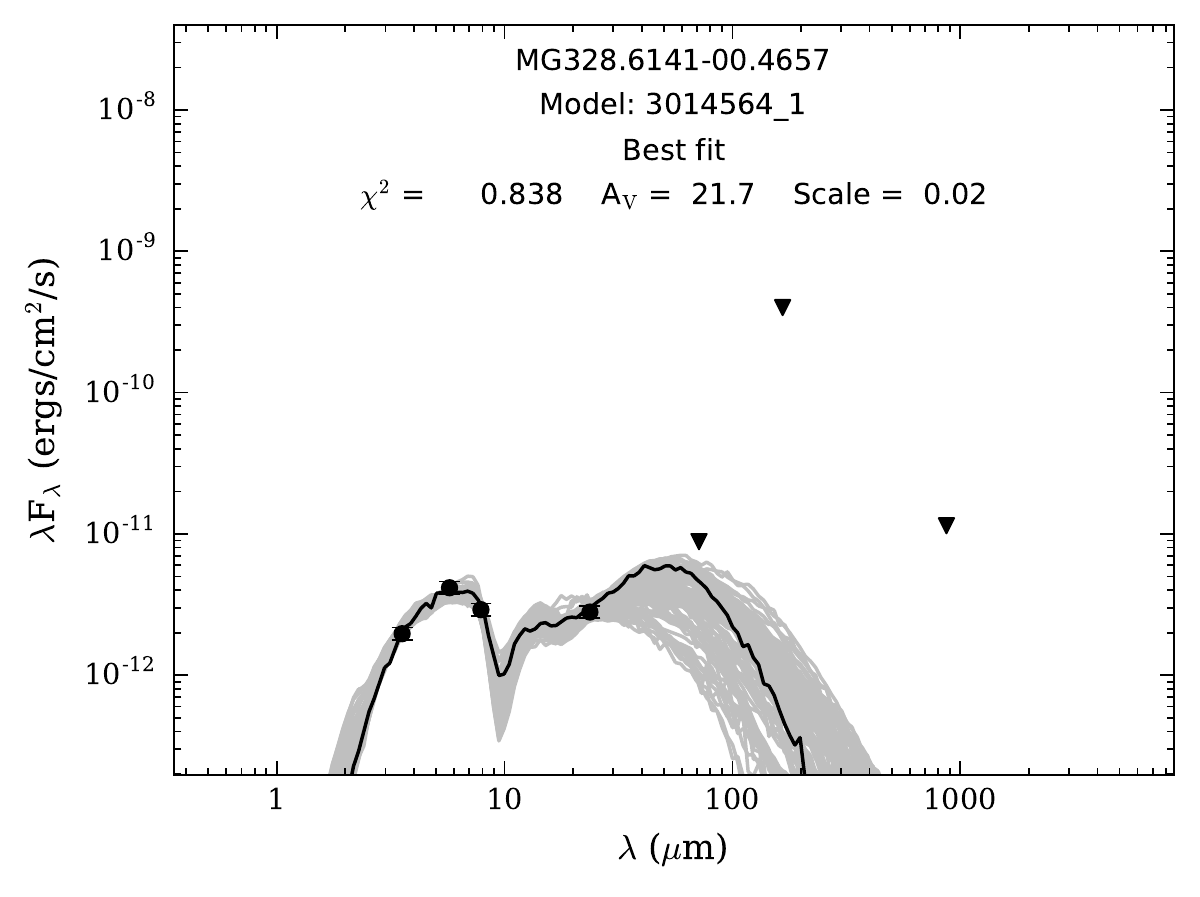}&
                \includegraphics[width=0.31\textwidth]{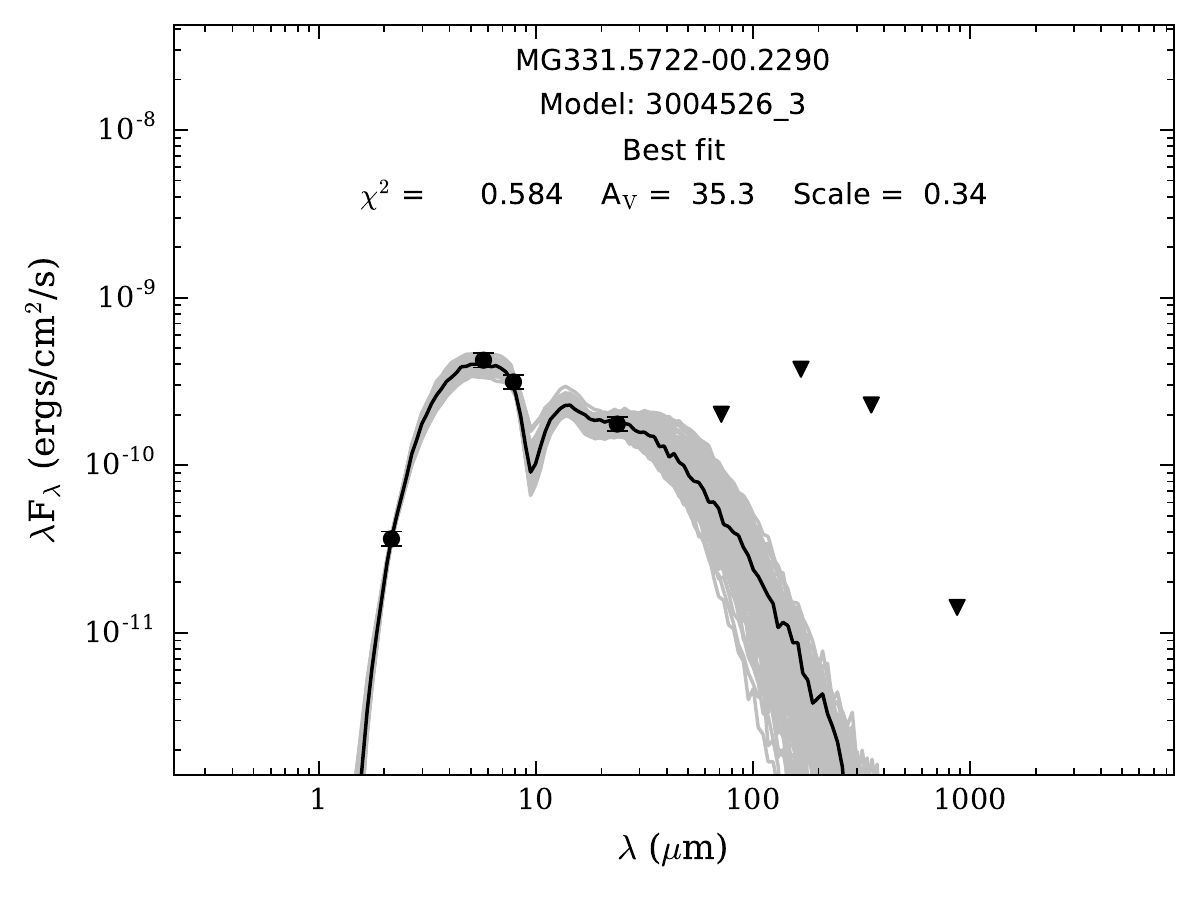}&
                \includegraphics[width=0.31\textwidth]{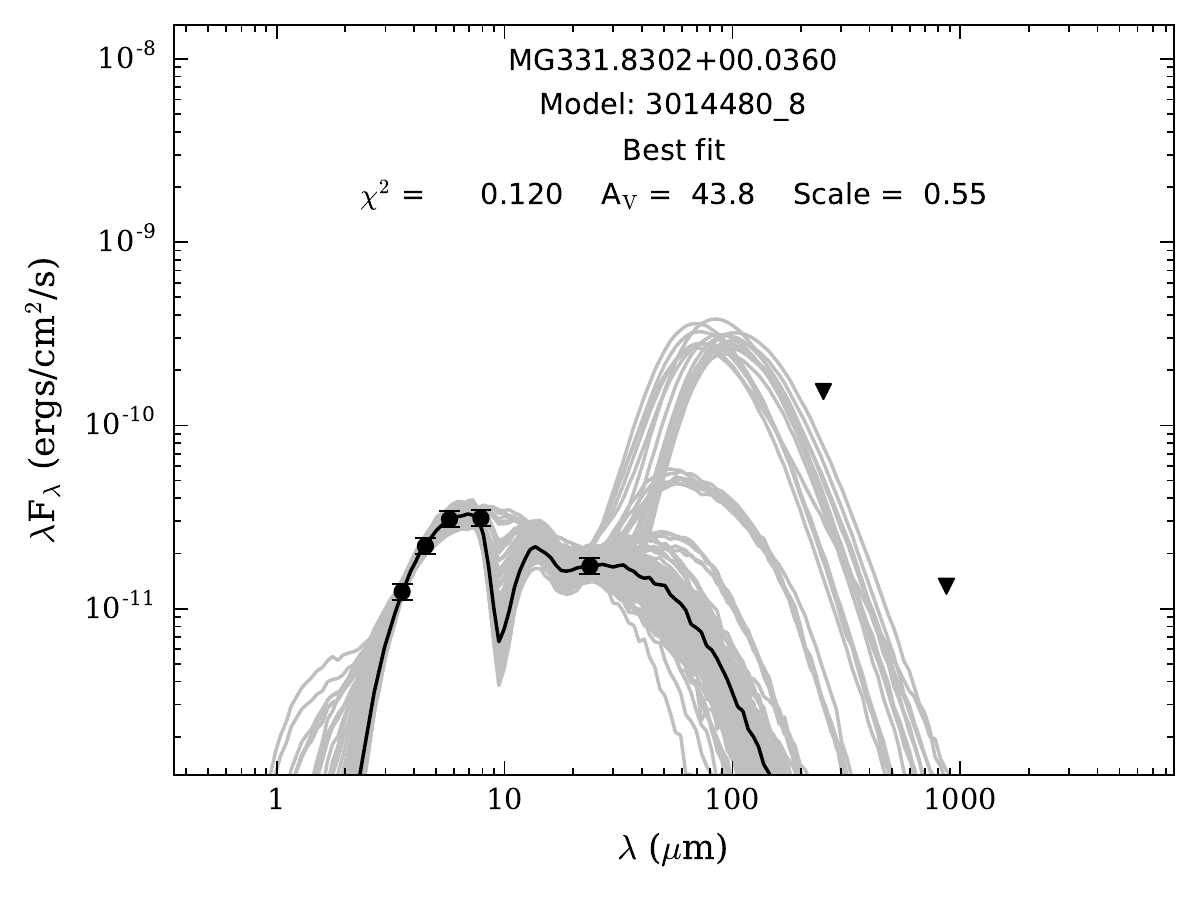}\\
                \includegraphics[width=0.31\textwidth]{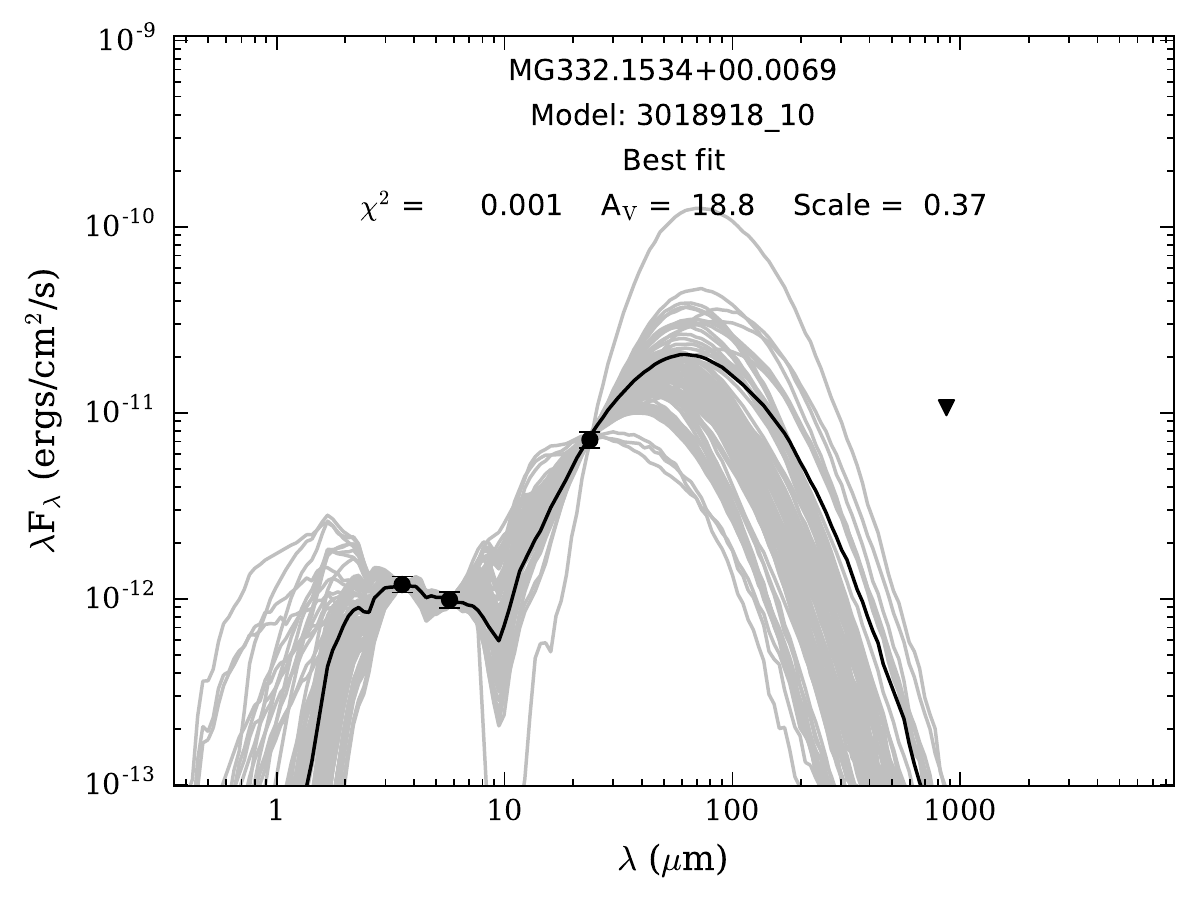}&
                \includegraphics[width=0.31\textwidth]{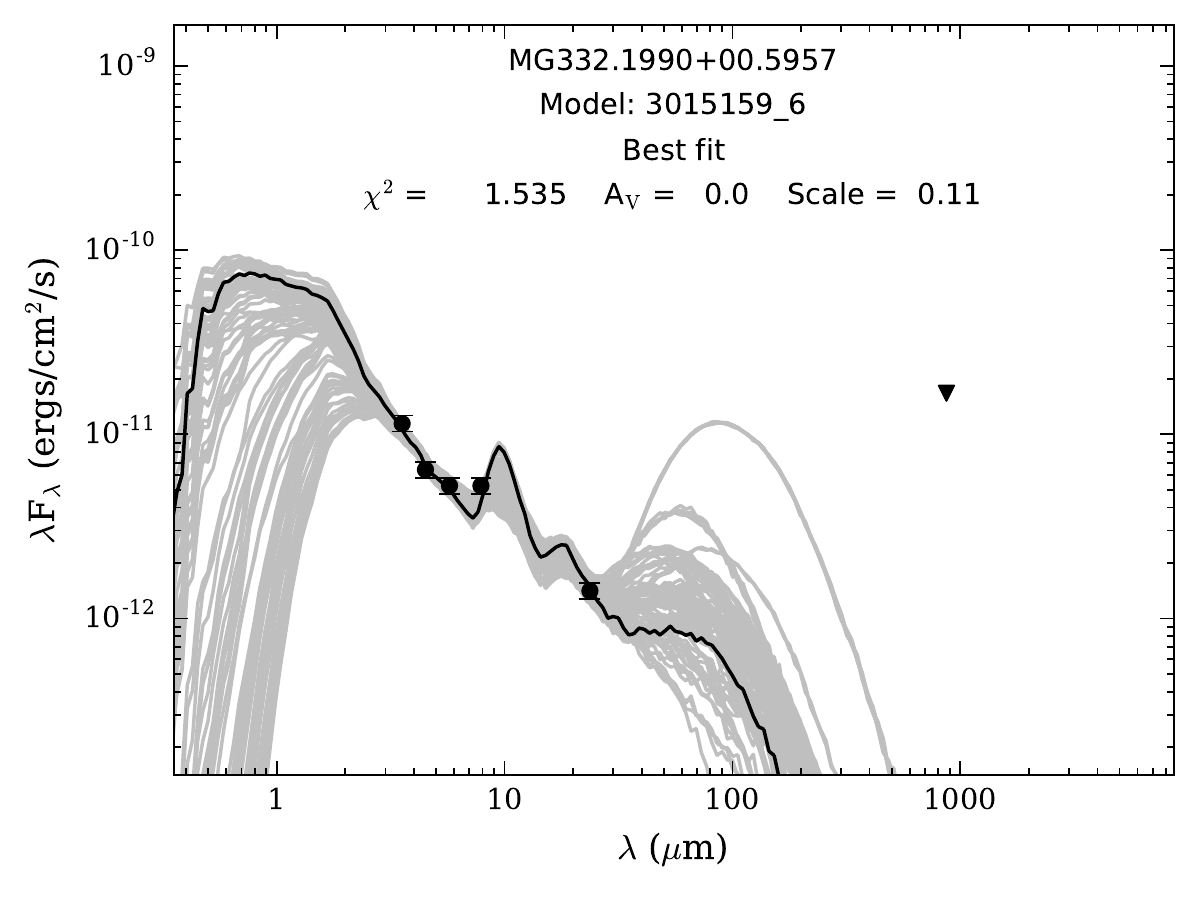}&
                \includegraphics[width=0.31\textwidth]{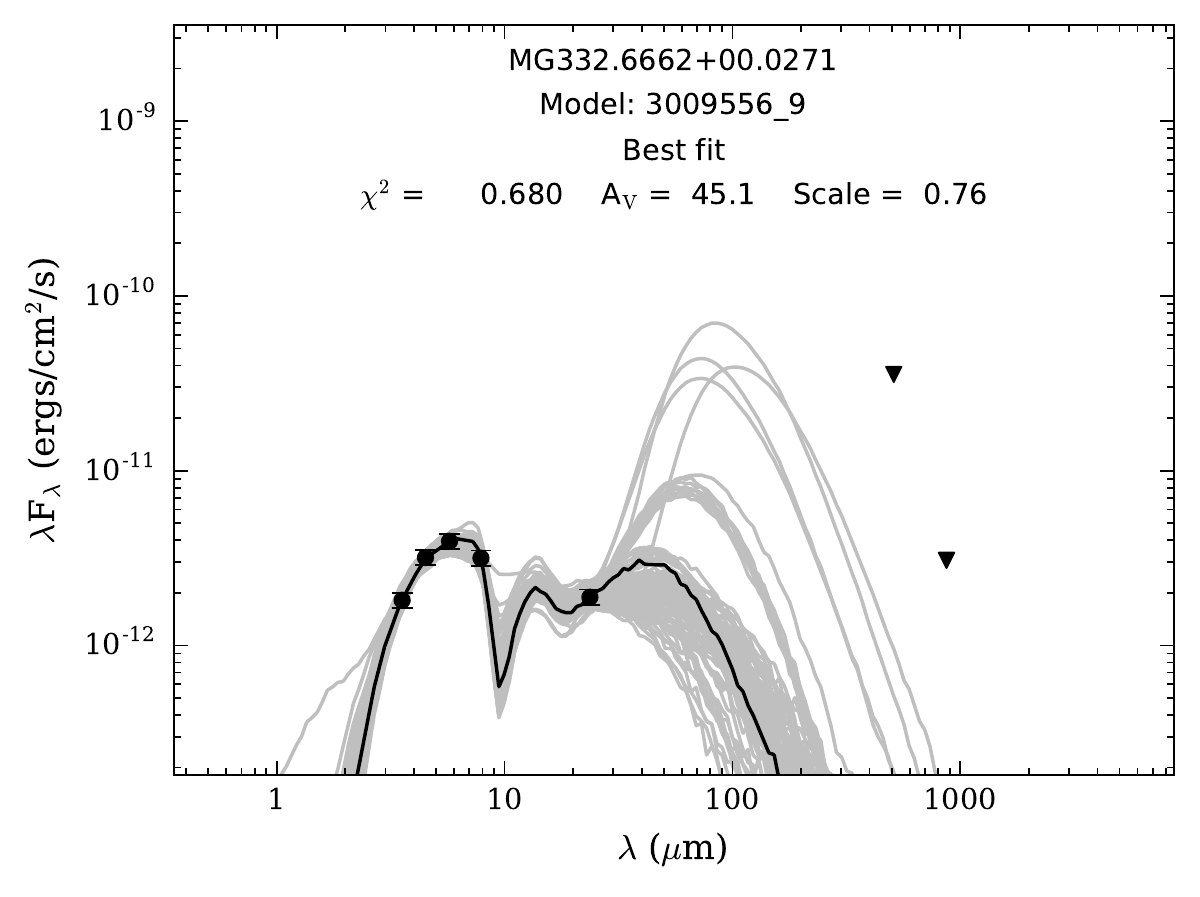}\\
                \includegraphics[width=0.31\textwidth]{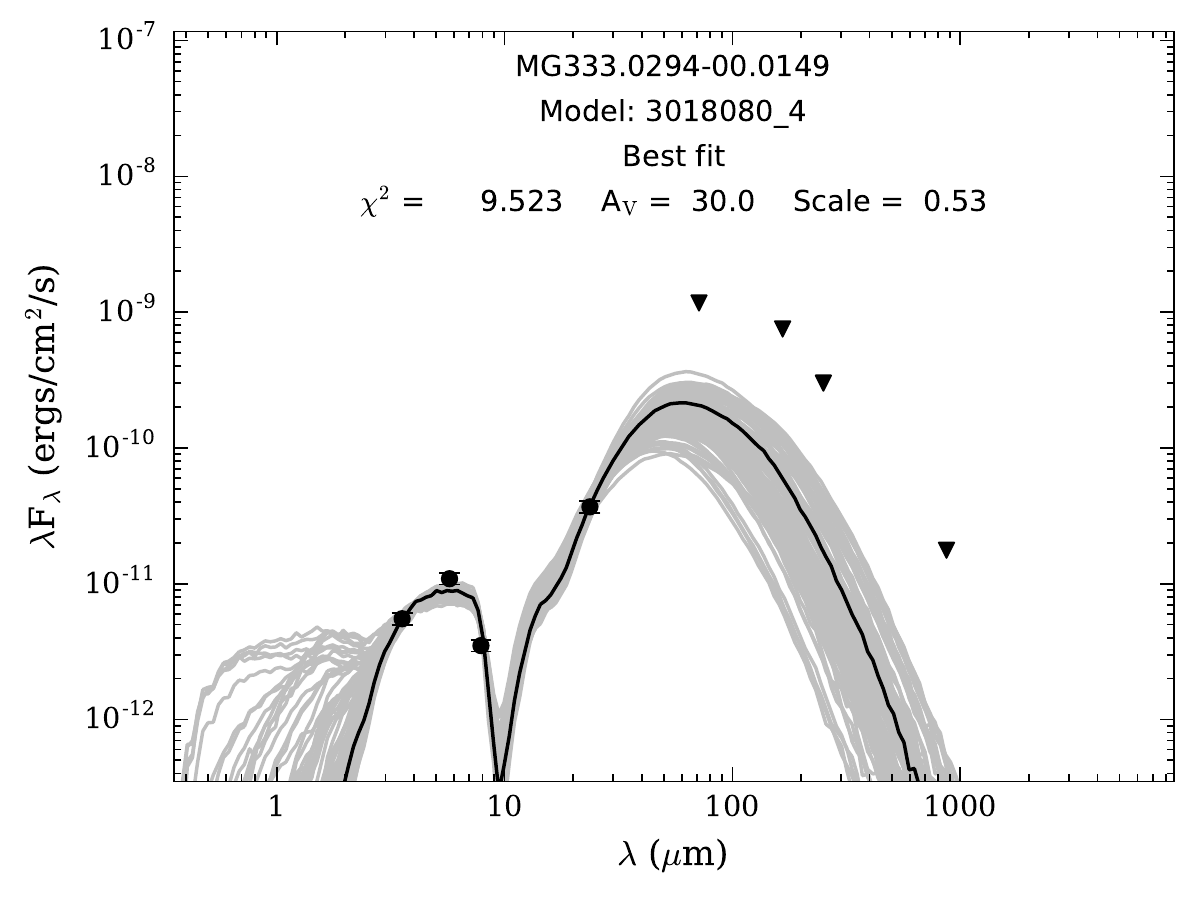}&
                \includegraphics[width=0.31\textwidth]{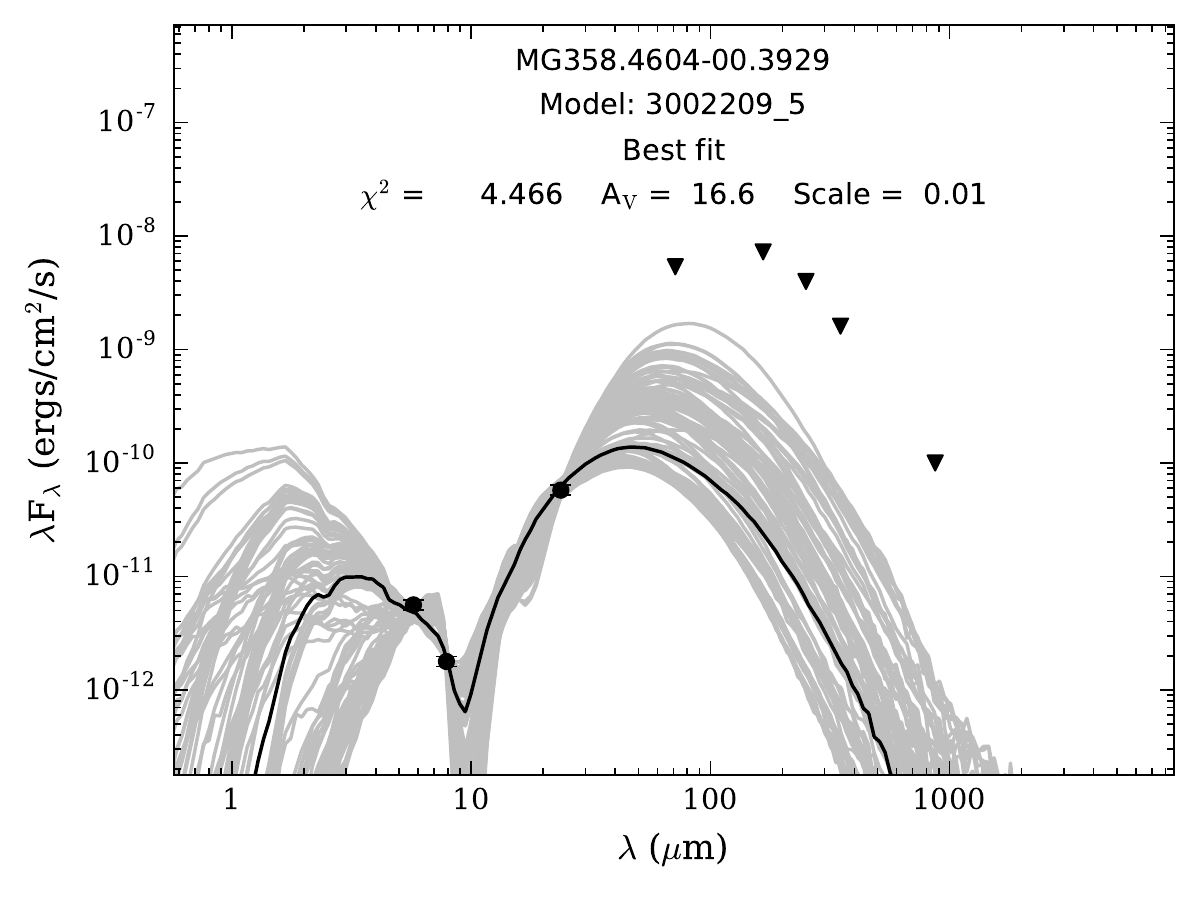}&
                \includegraphics[width=0.31\textwidth]{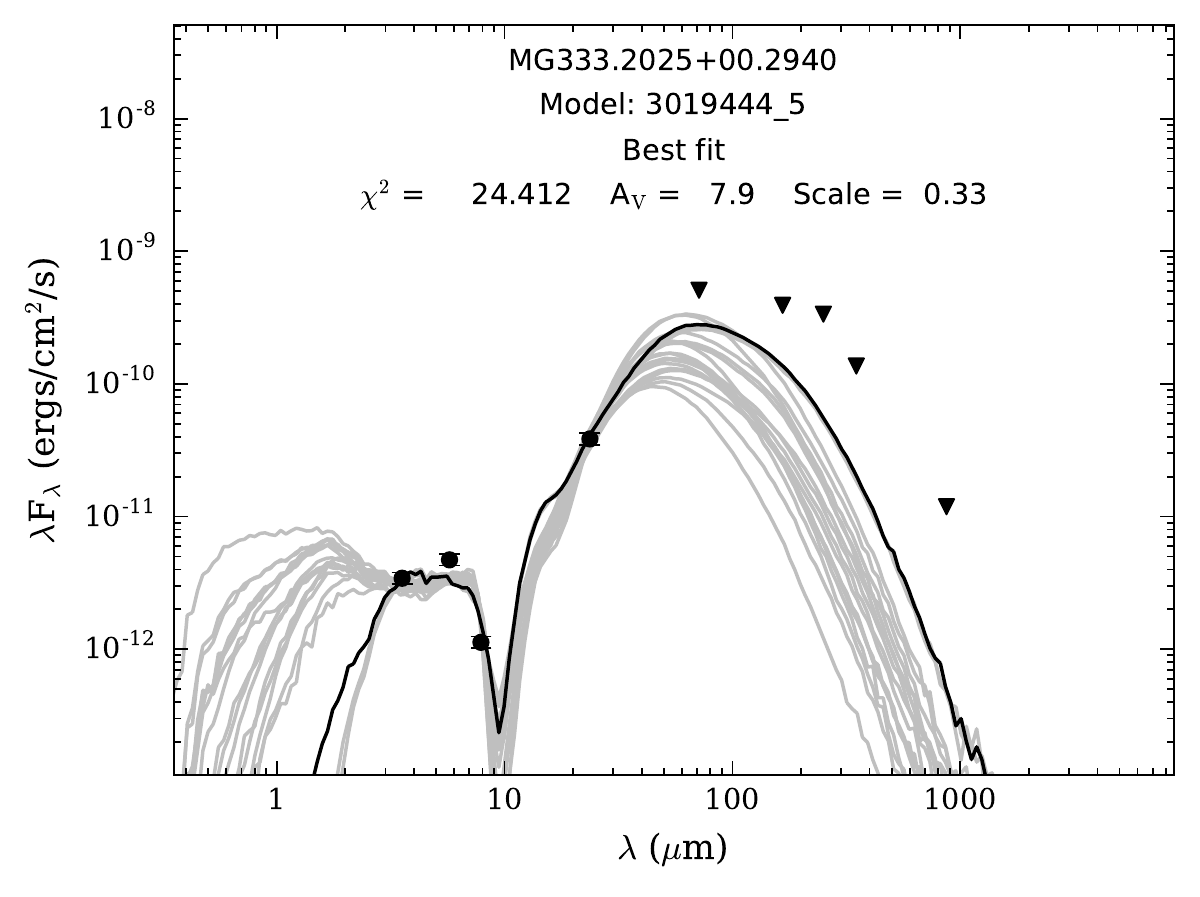}\\
                \includegraphics[width=0.31\textwidth]{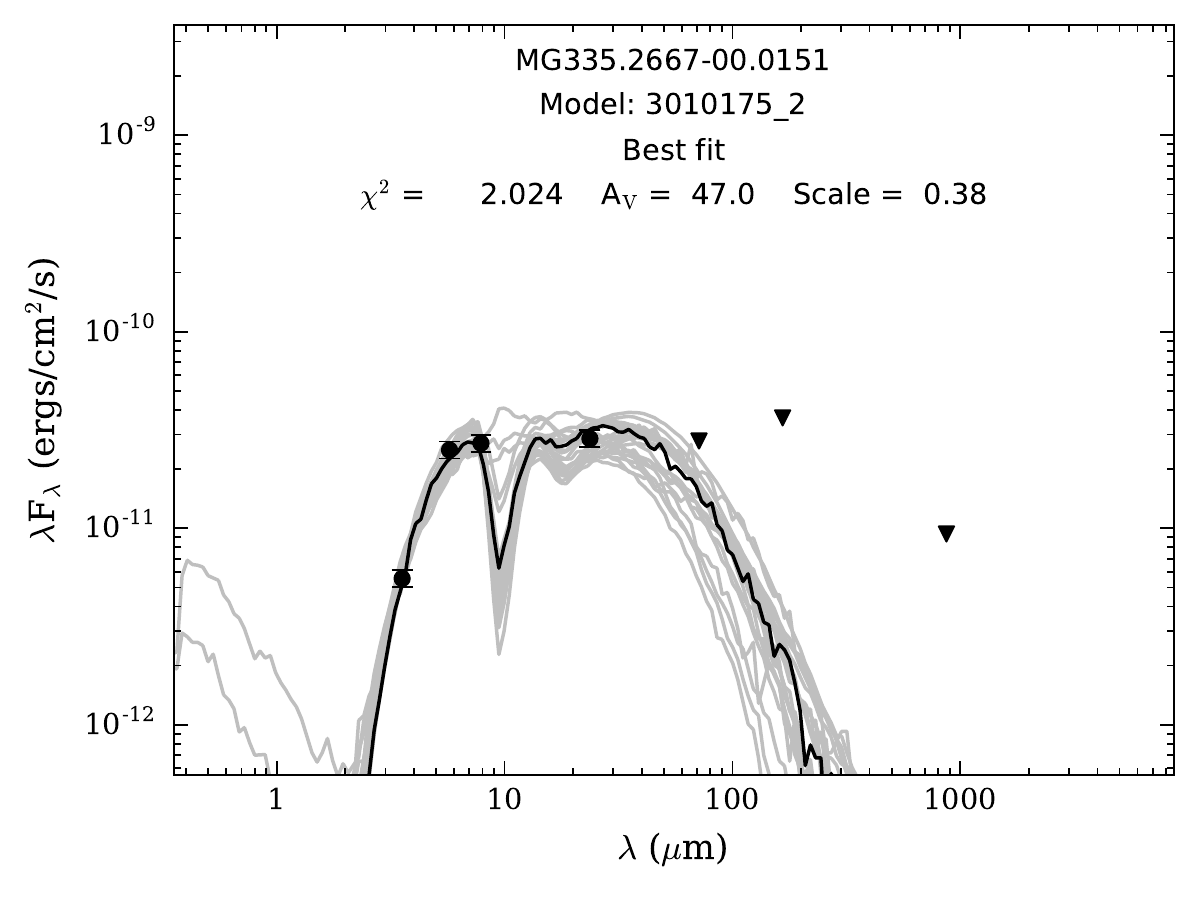}&
                \includegraphics[width=0.31\textwidth]{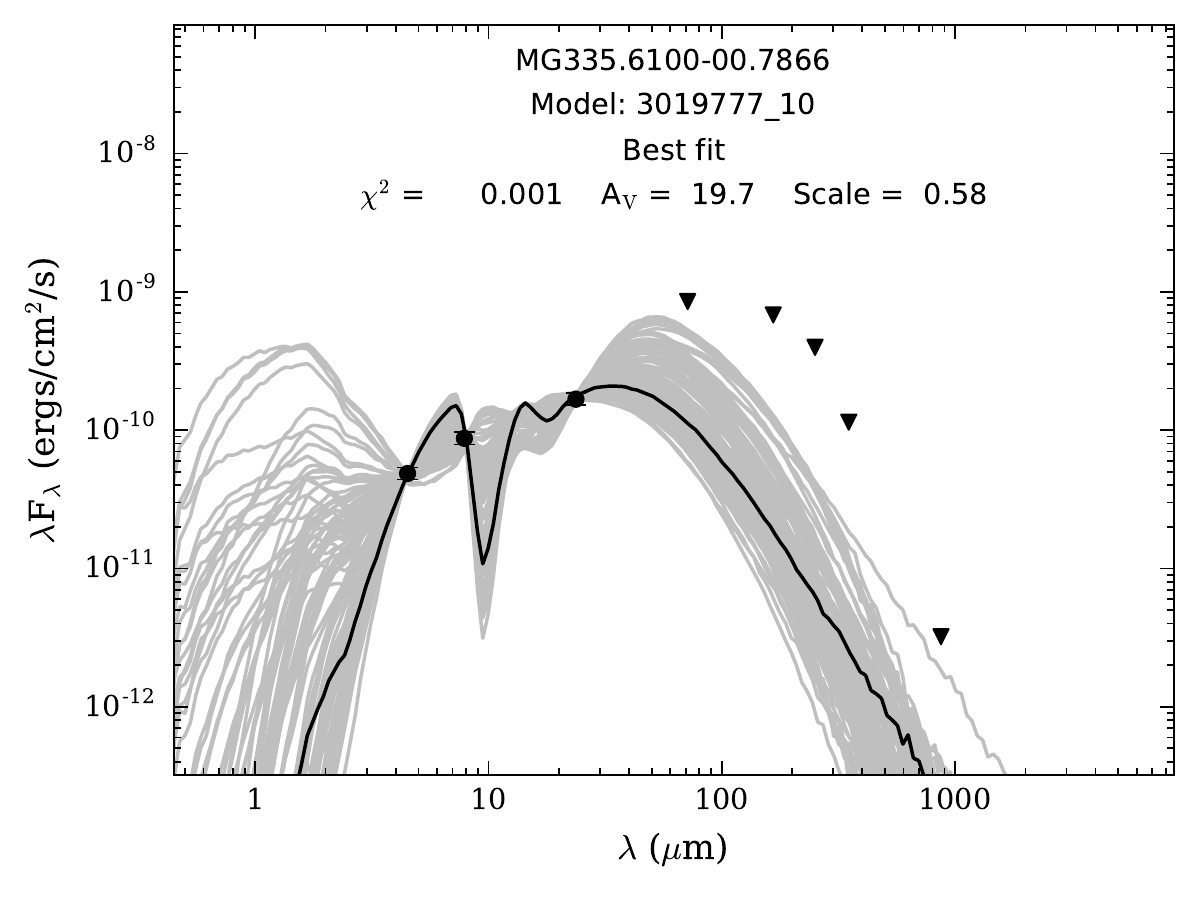}&
                \includegraphics[width=0.31\textwidth]{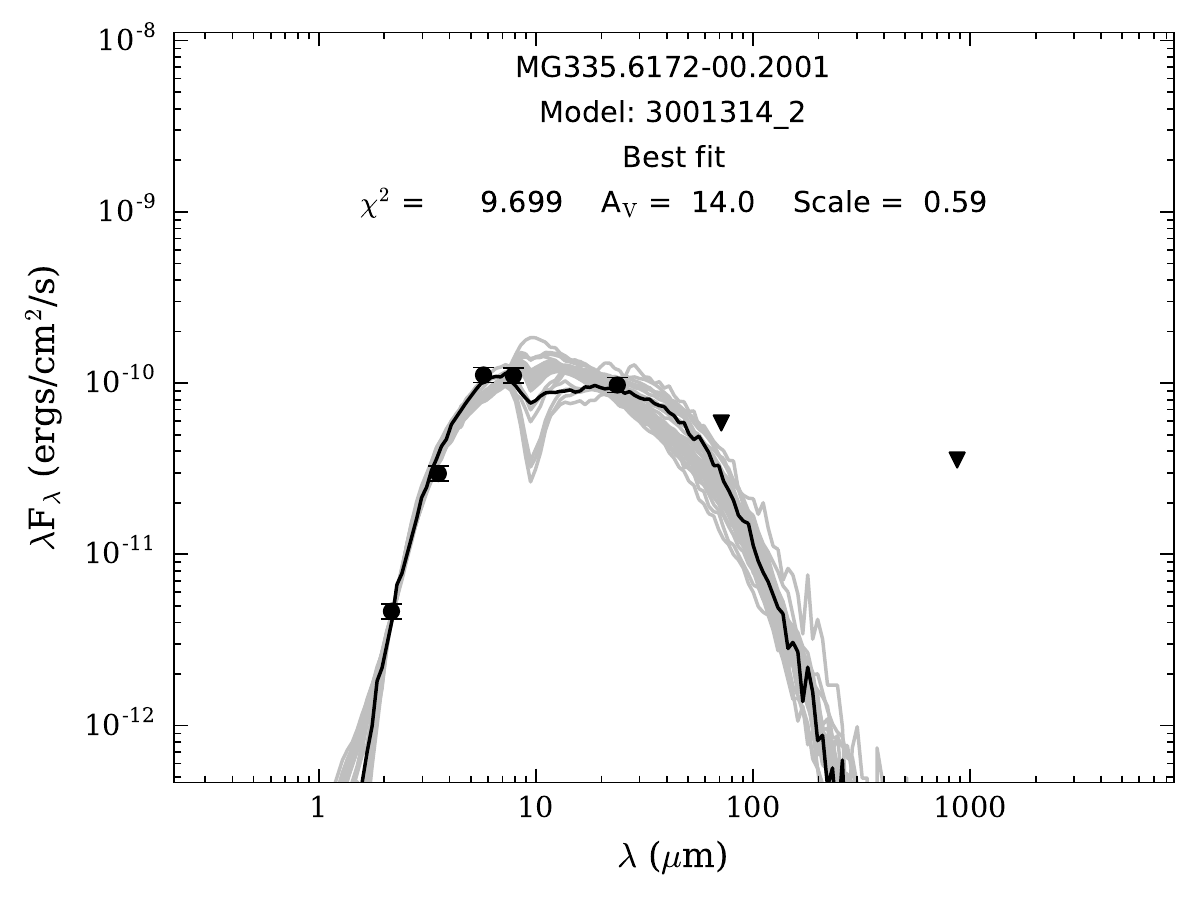}\\
                \includegraphics[width=0.31\textwidth]{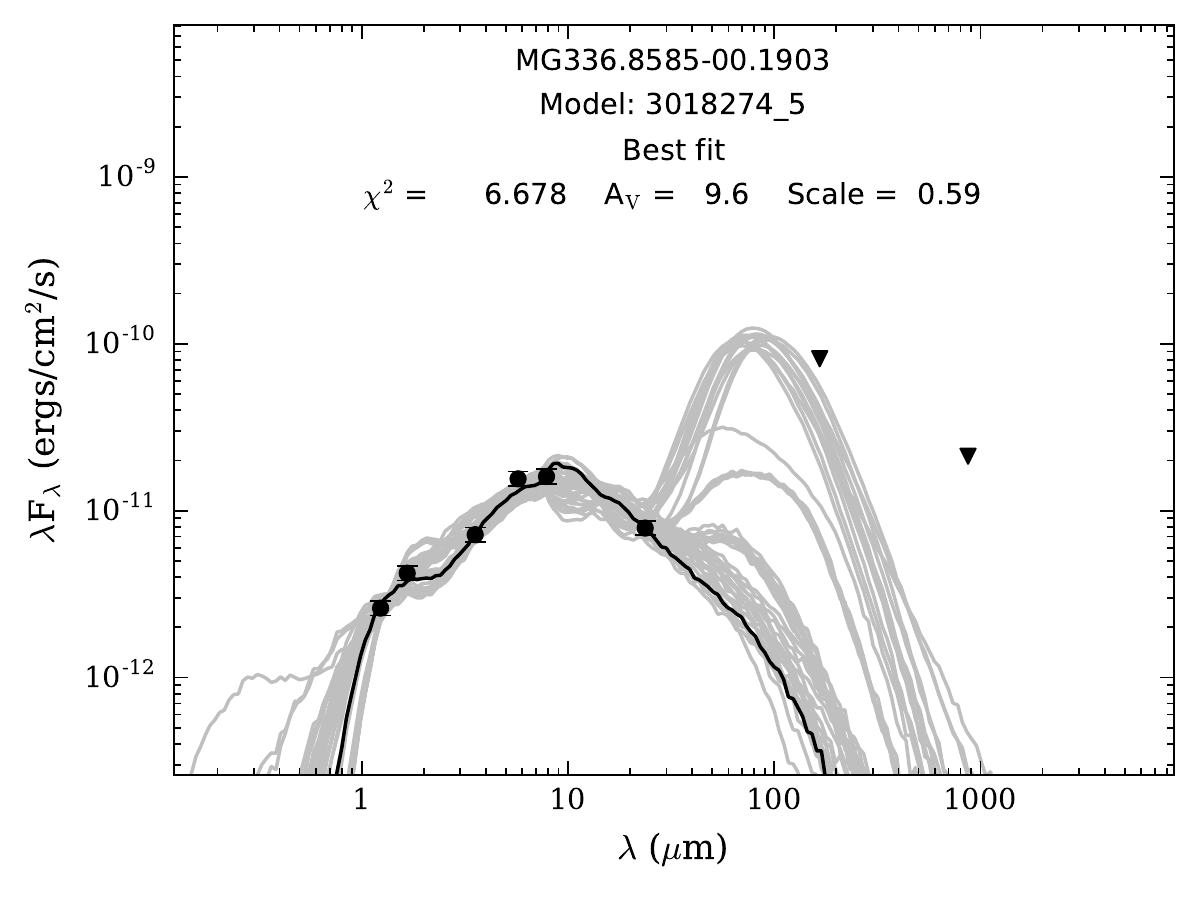}&
                \includegraphics[width=0.31\textwidth]{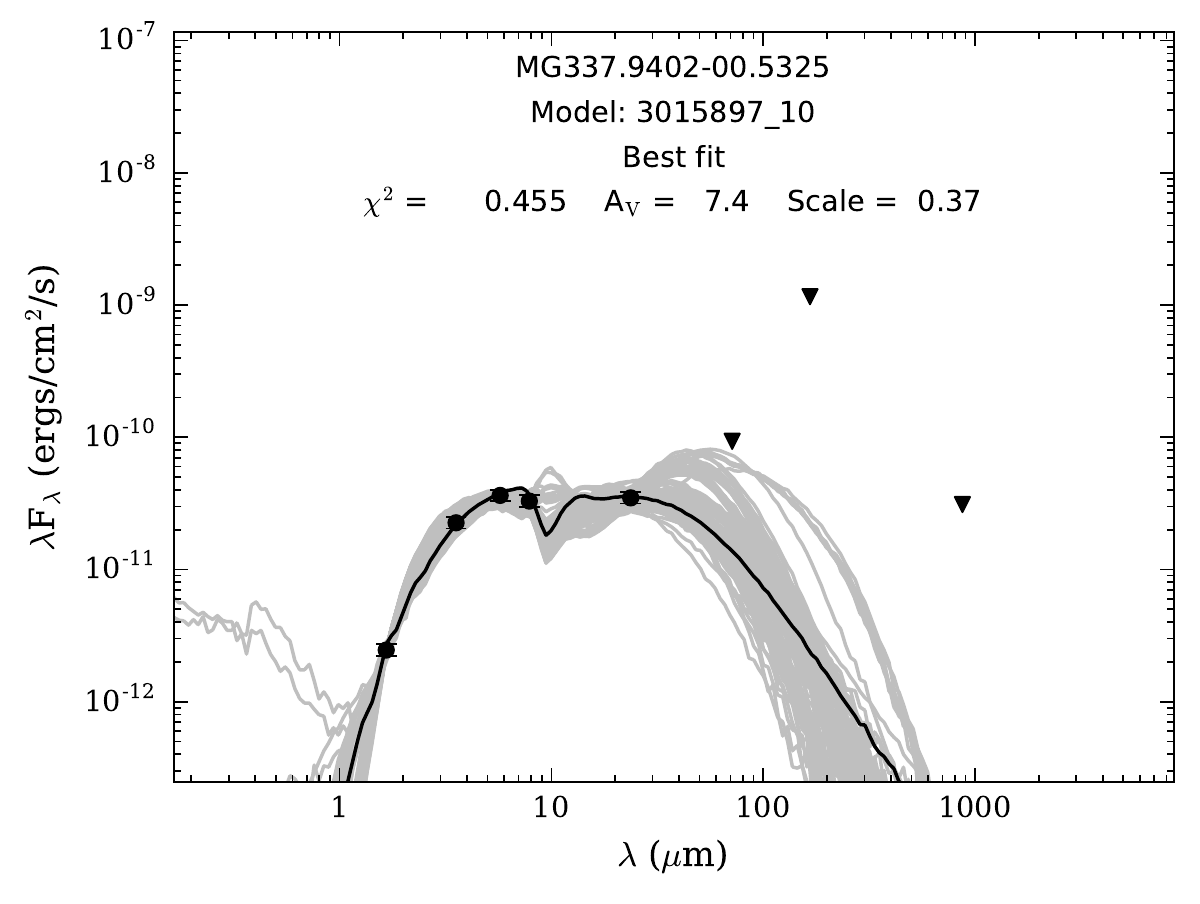}&
                \includegraphics[width=0.31\textwidth]{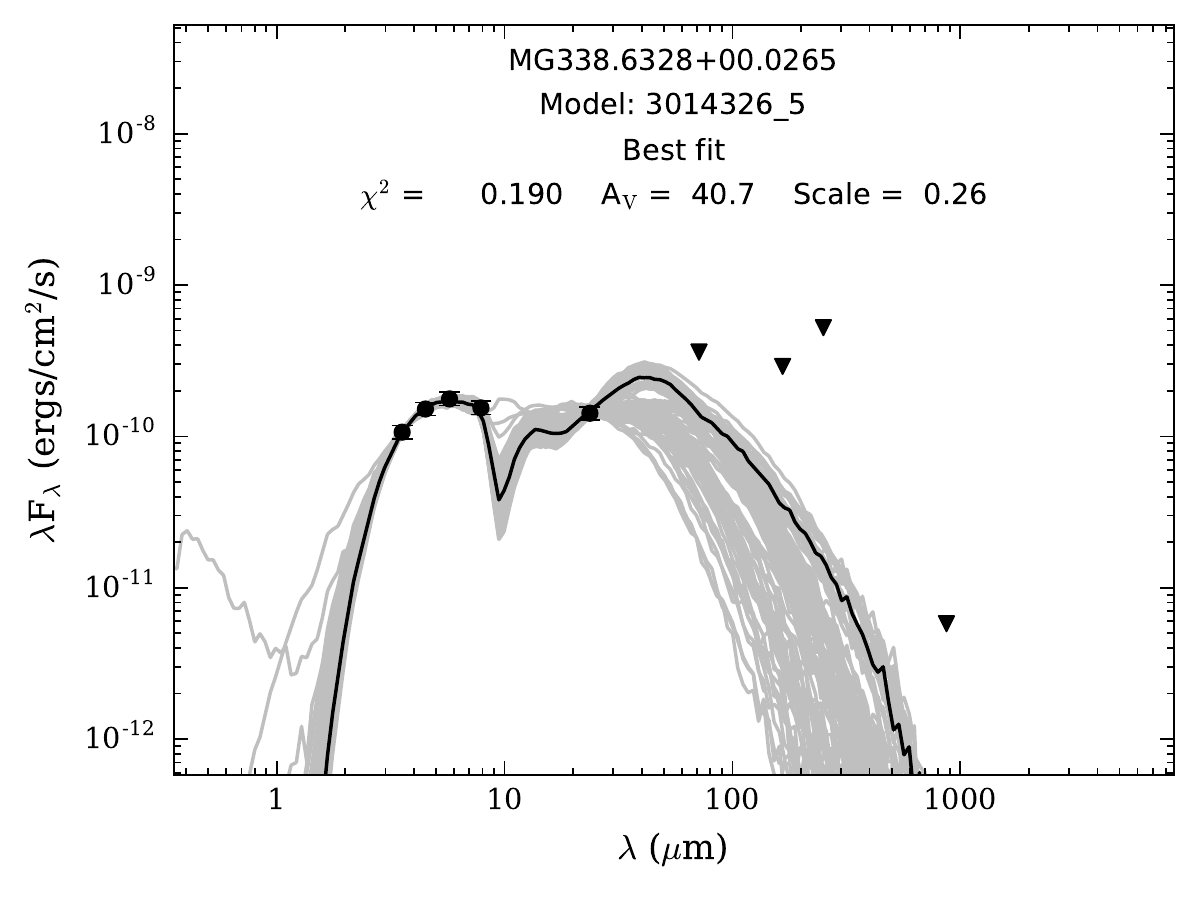}\\
        \end{tabular}
        \caption{Continuation of Fig. \ref{fig:testgridsed}.}
\end{figure}

\begin{figure}
        \begin{tabular}{ccc}
                \includegraphics[width=0.31\textwidth]{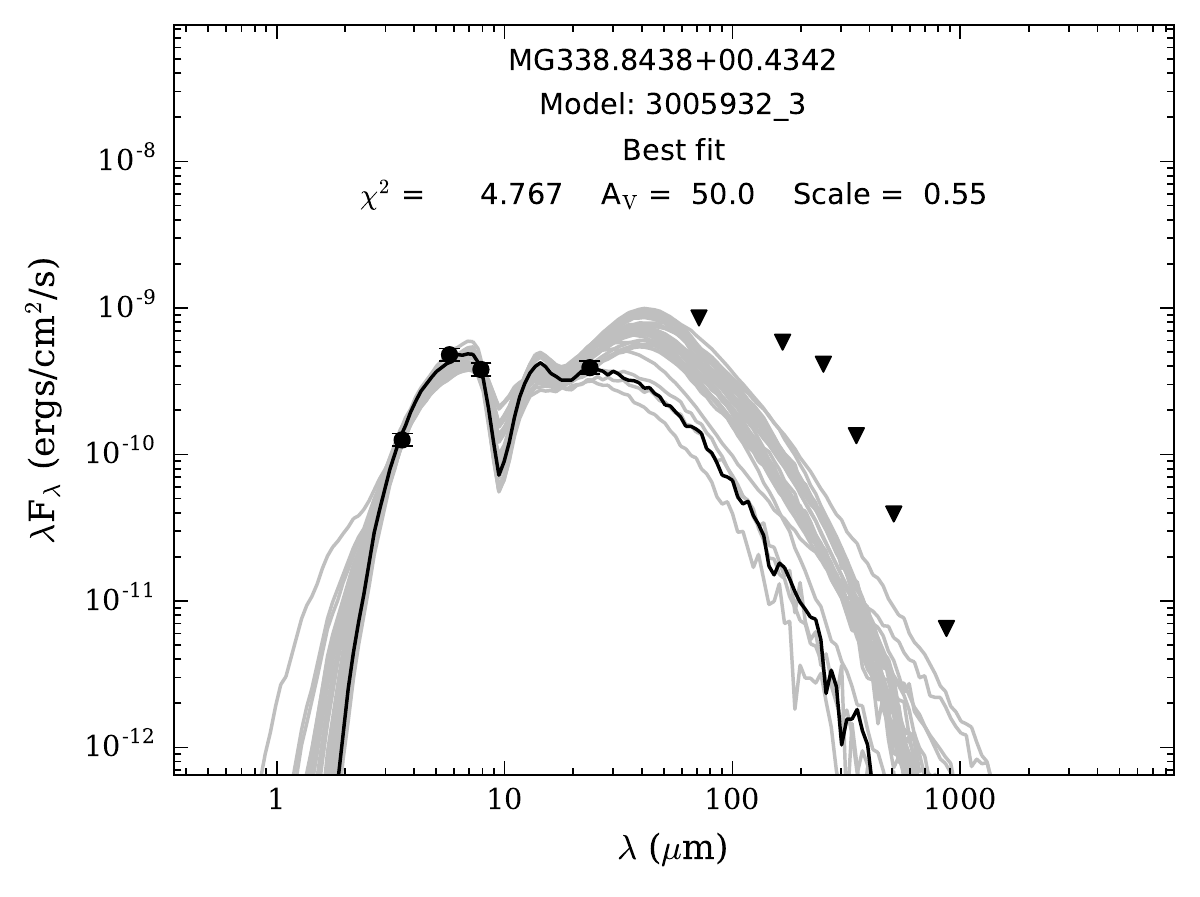}&
                \includegraphics[width=0.31\textwidth]{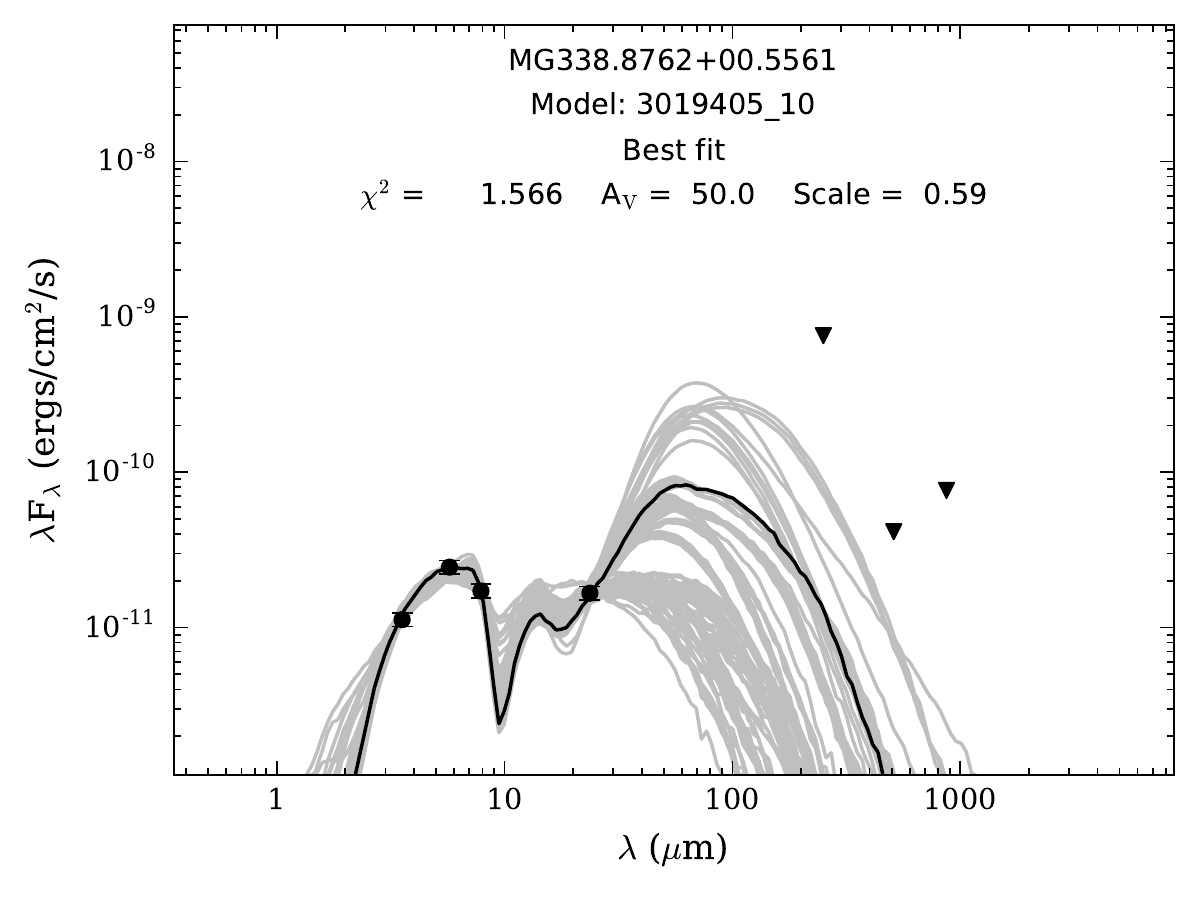}&
                \includegraphics[width=0.31\textwidth]{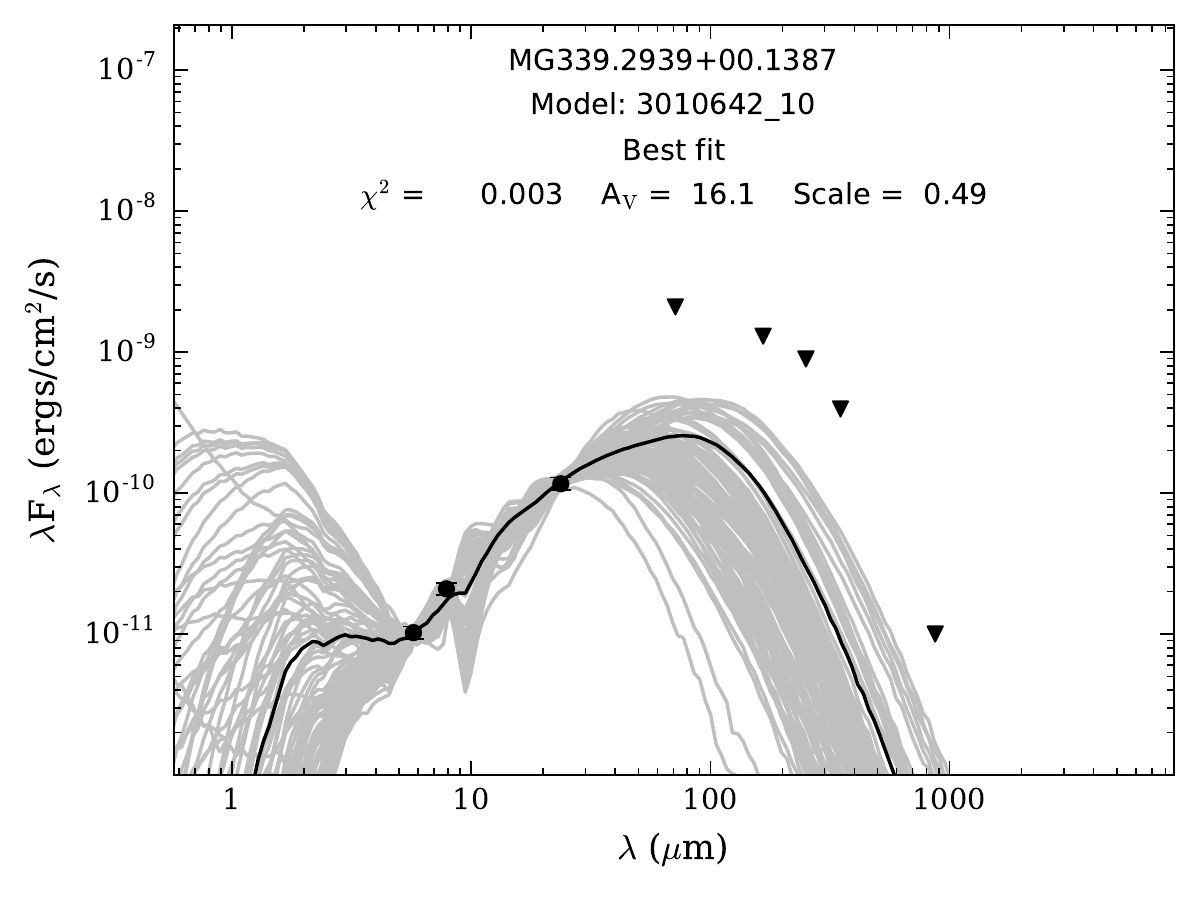}\\
                
                \includegraphics[width=0.31\textwidth]{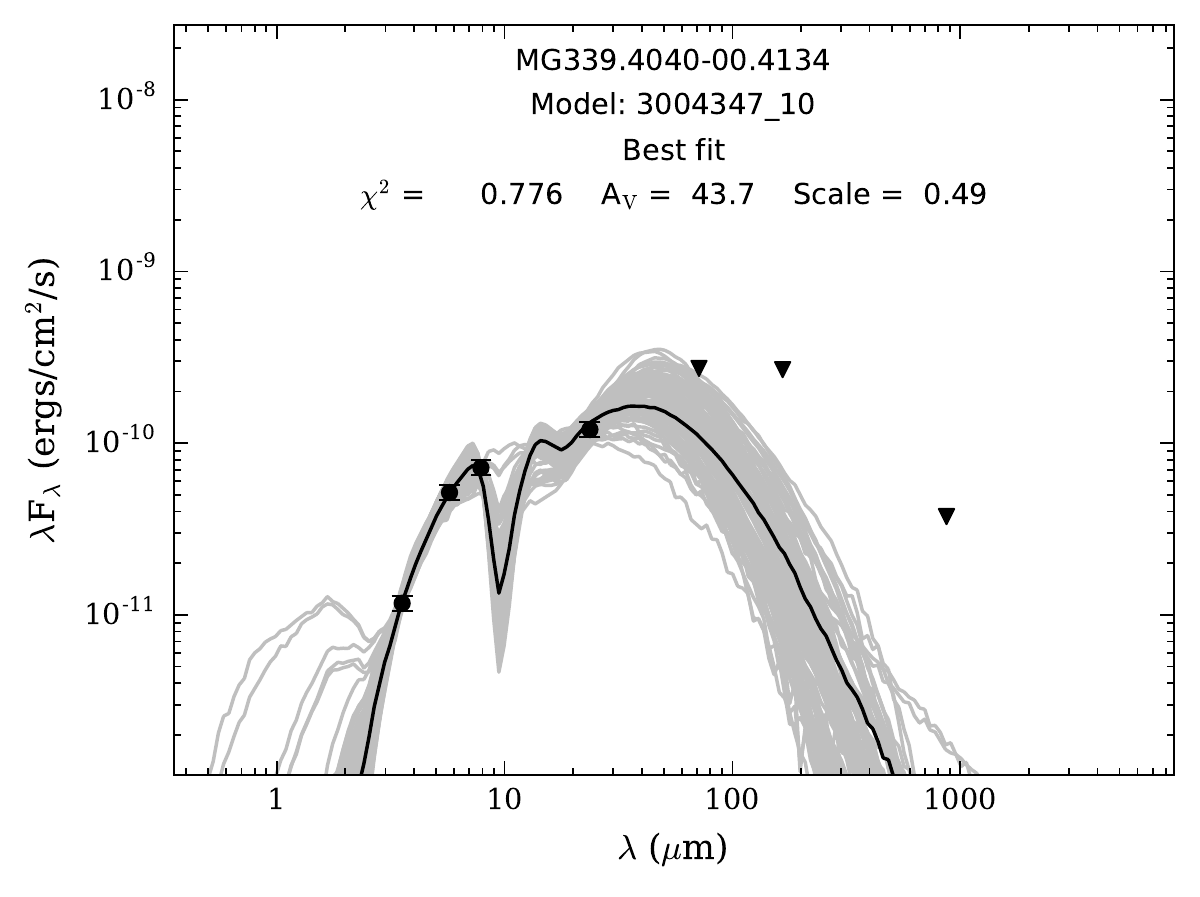}&
                \includegraphics[width=0.31\textwidth]{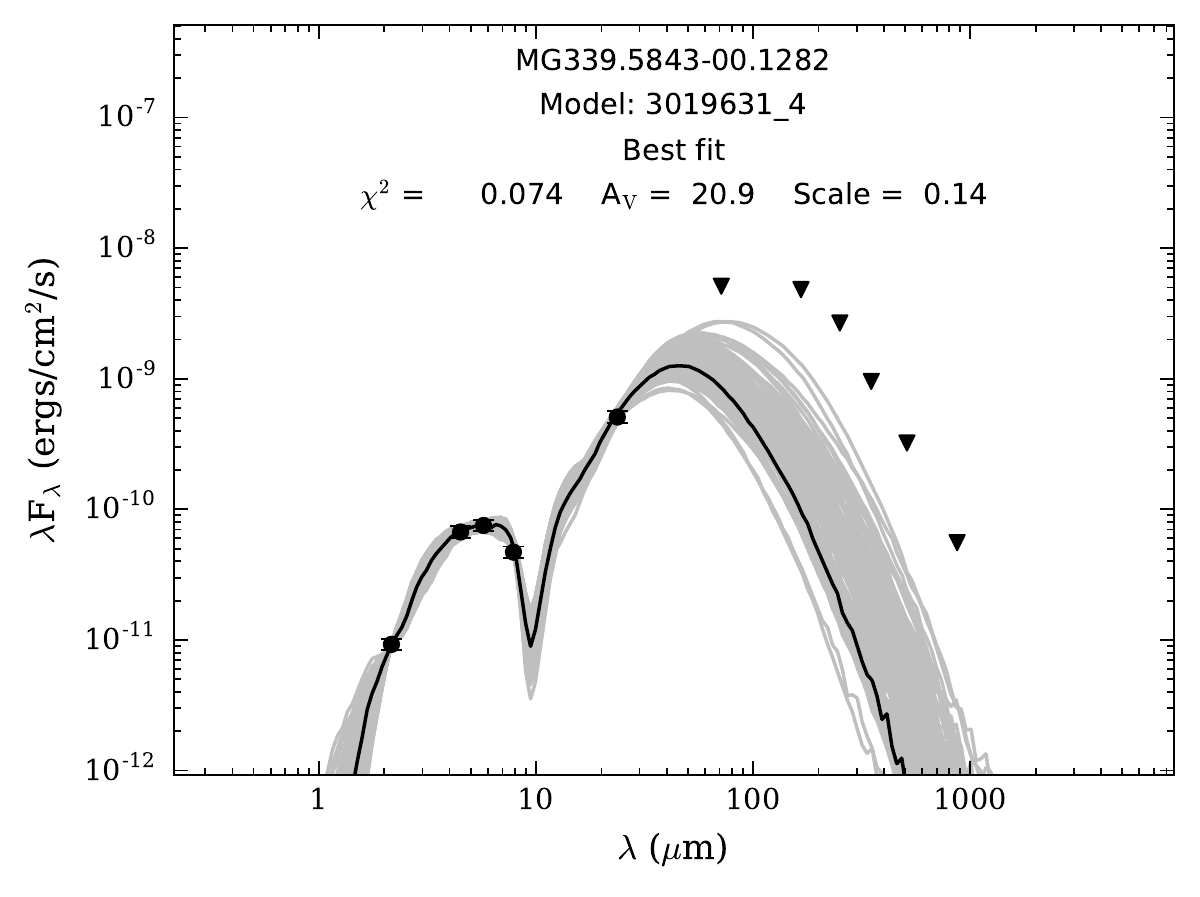}&
                \includegraphics[width=0.31\textwidth]{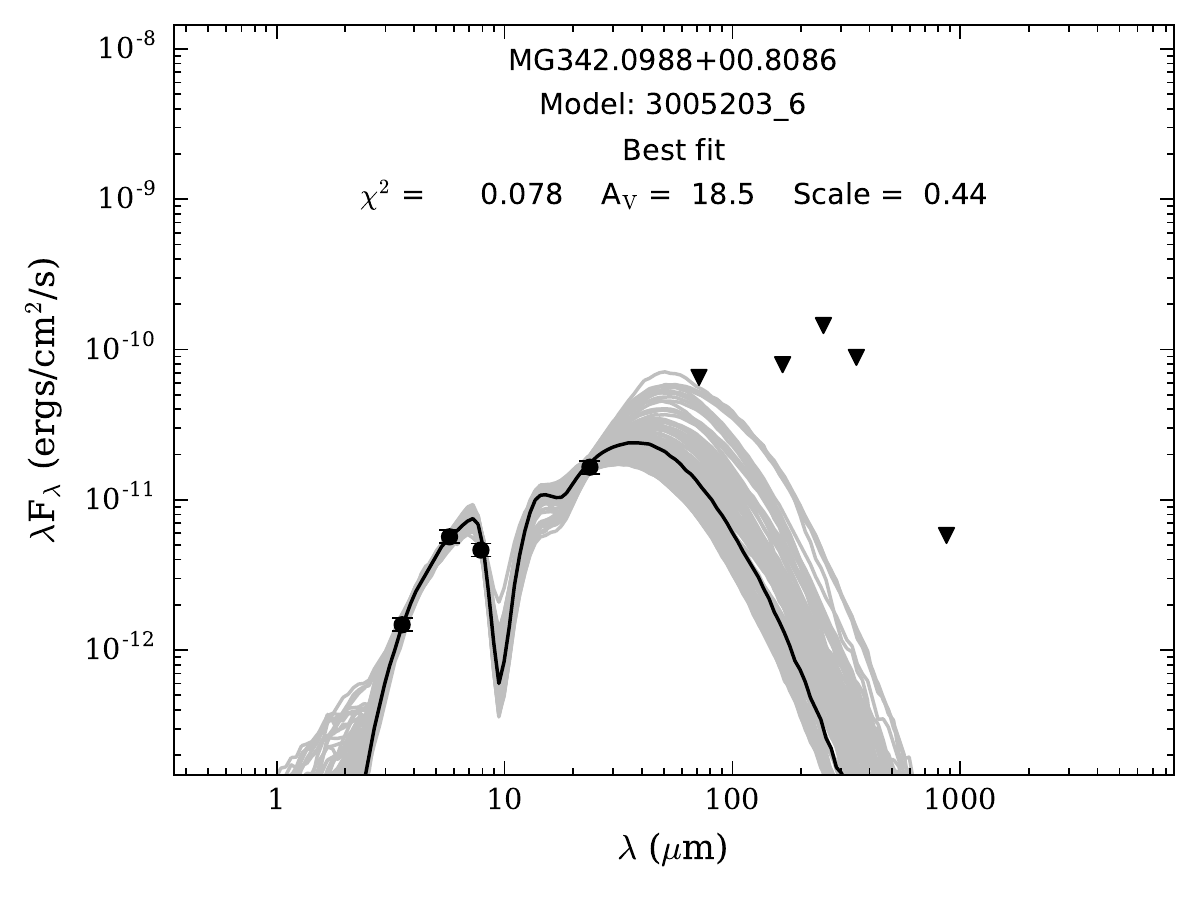}\\
                \includegraphics[width=0.31\textwidth]{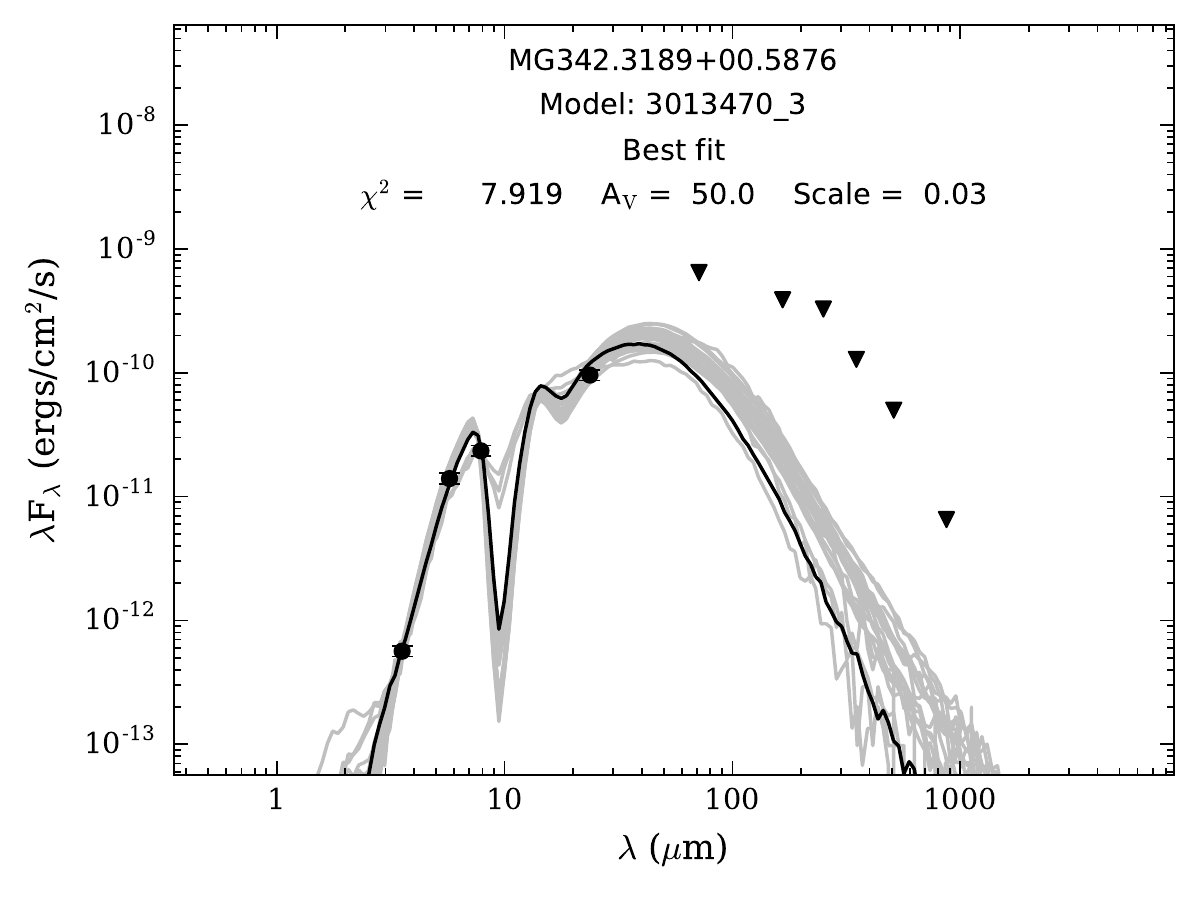}&
                \includegraphics[width=0.31\textwidth]{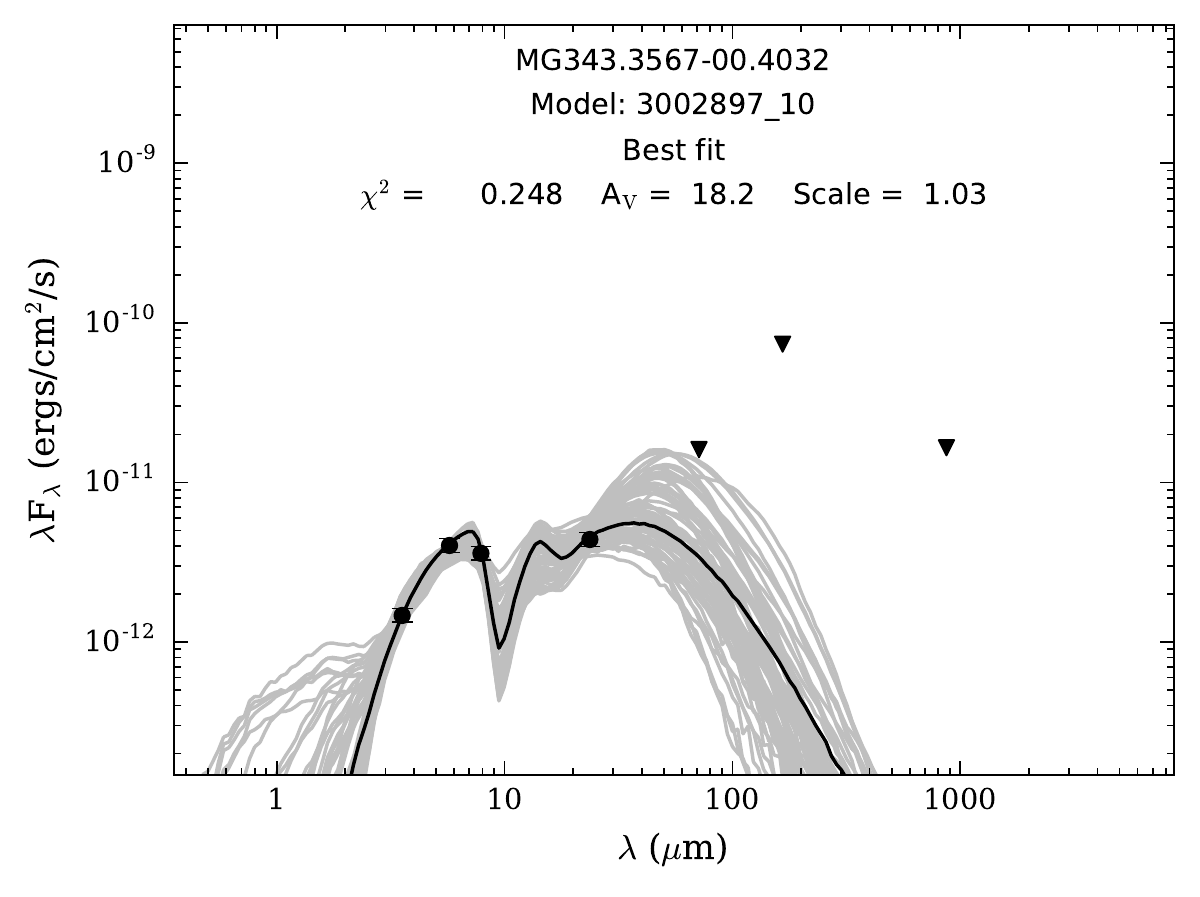}&
                \includegraphics[width=0.31\textwidth]{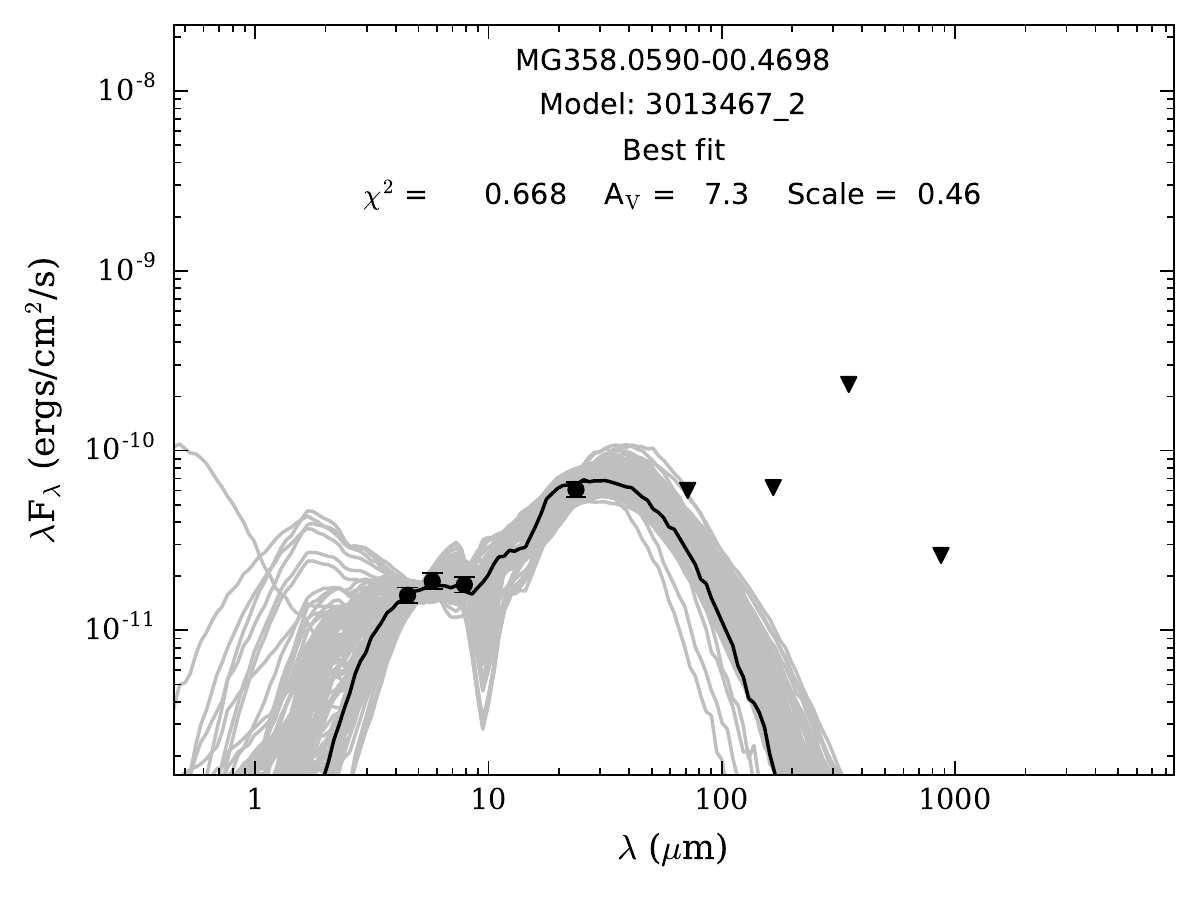}\\
                \includegraphics[width=0.31\textwidth]{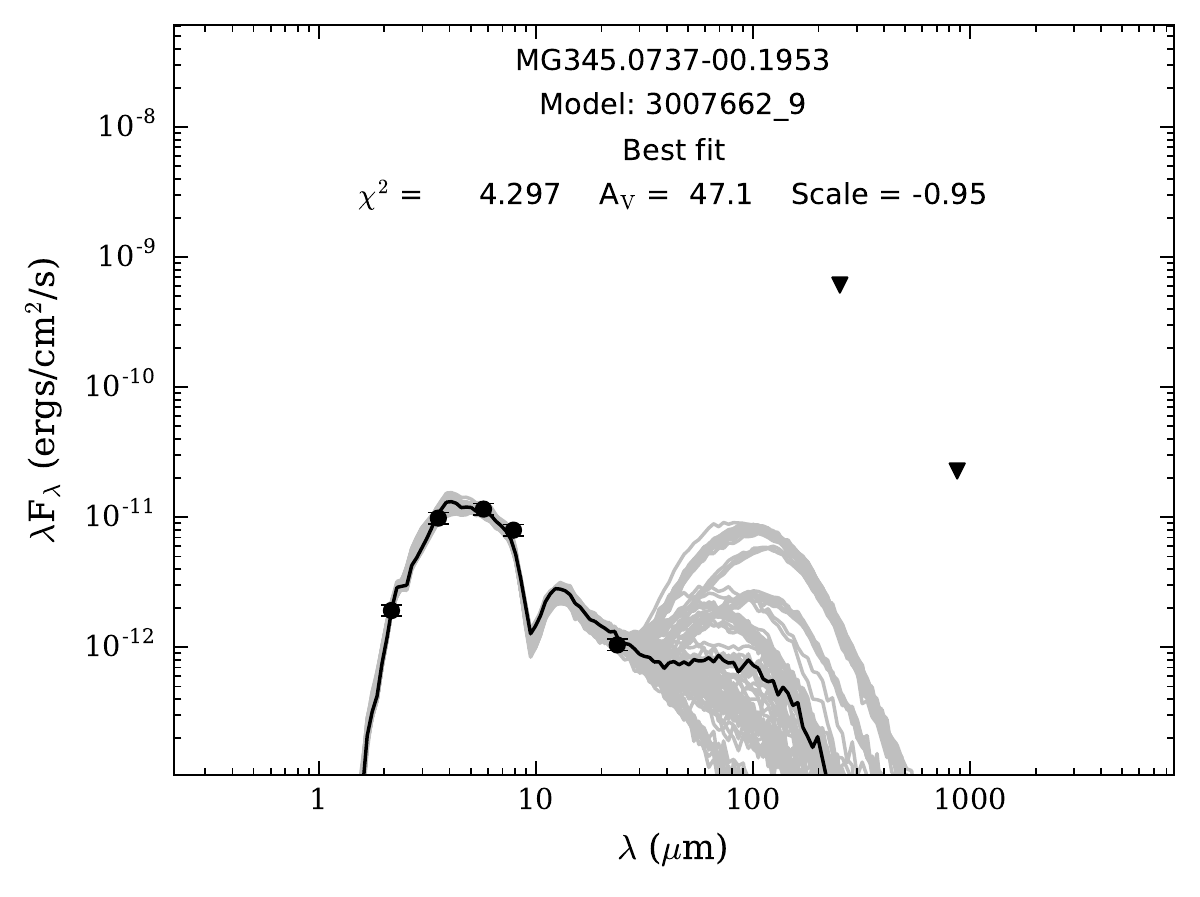}&
                \includegraphics[width=0.31\textwidth]{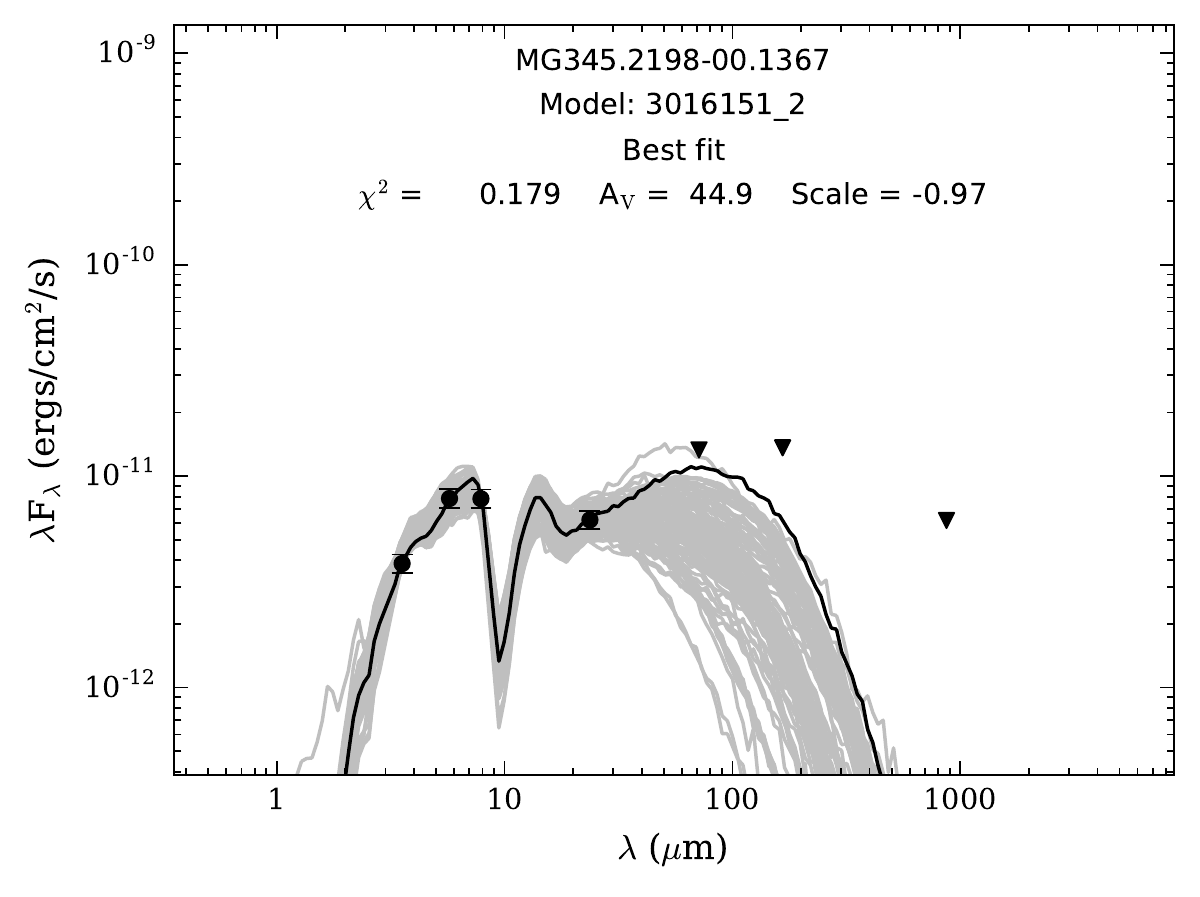}&
                \includegraphics[width=0.31\textwidth]{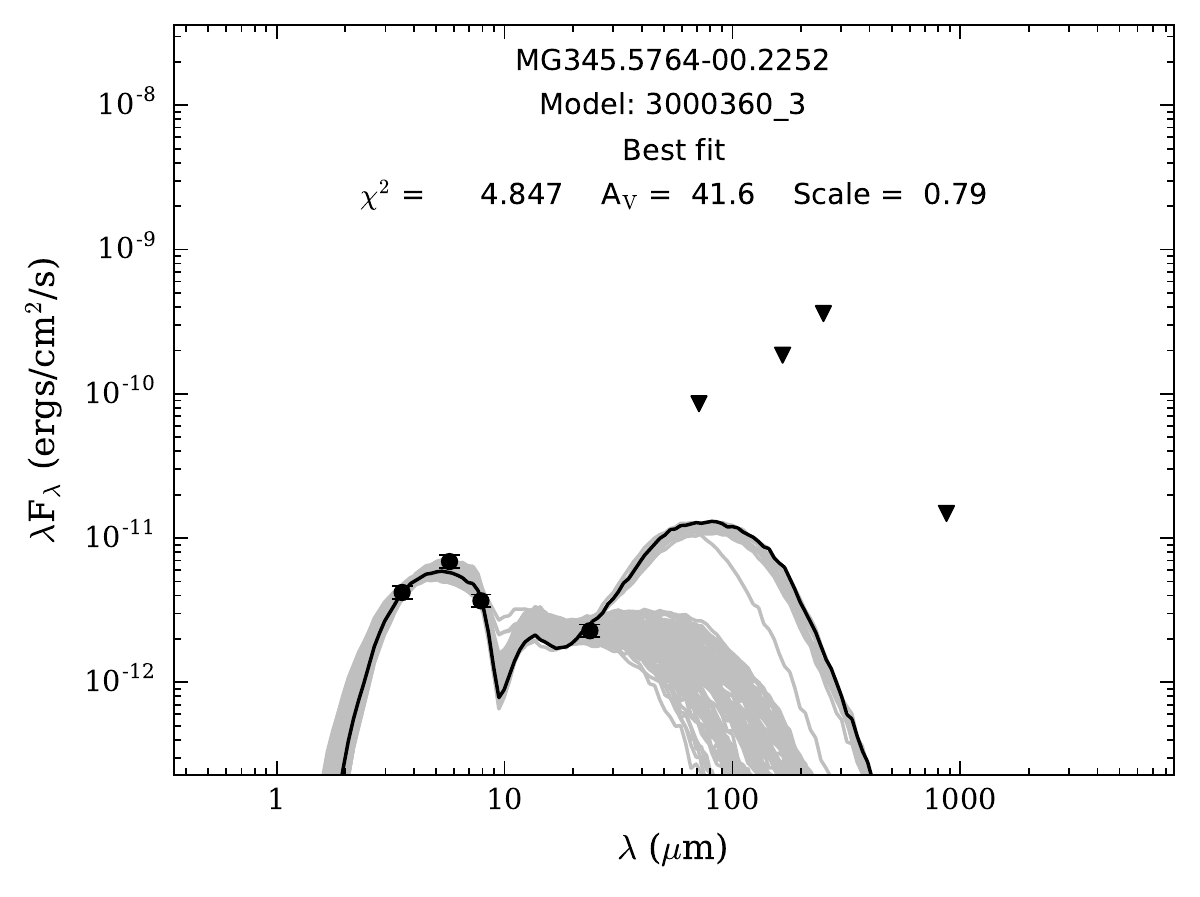}\\
                \includegraphics[width=0.31\textwidth]{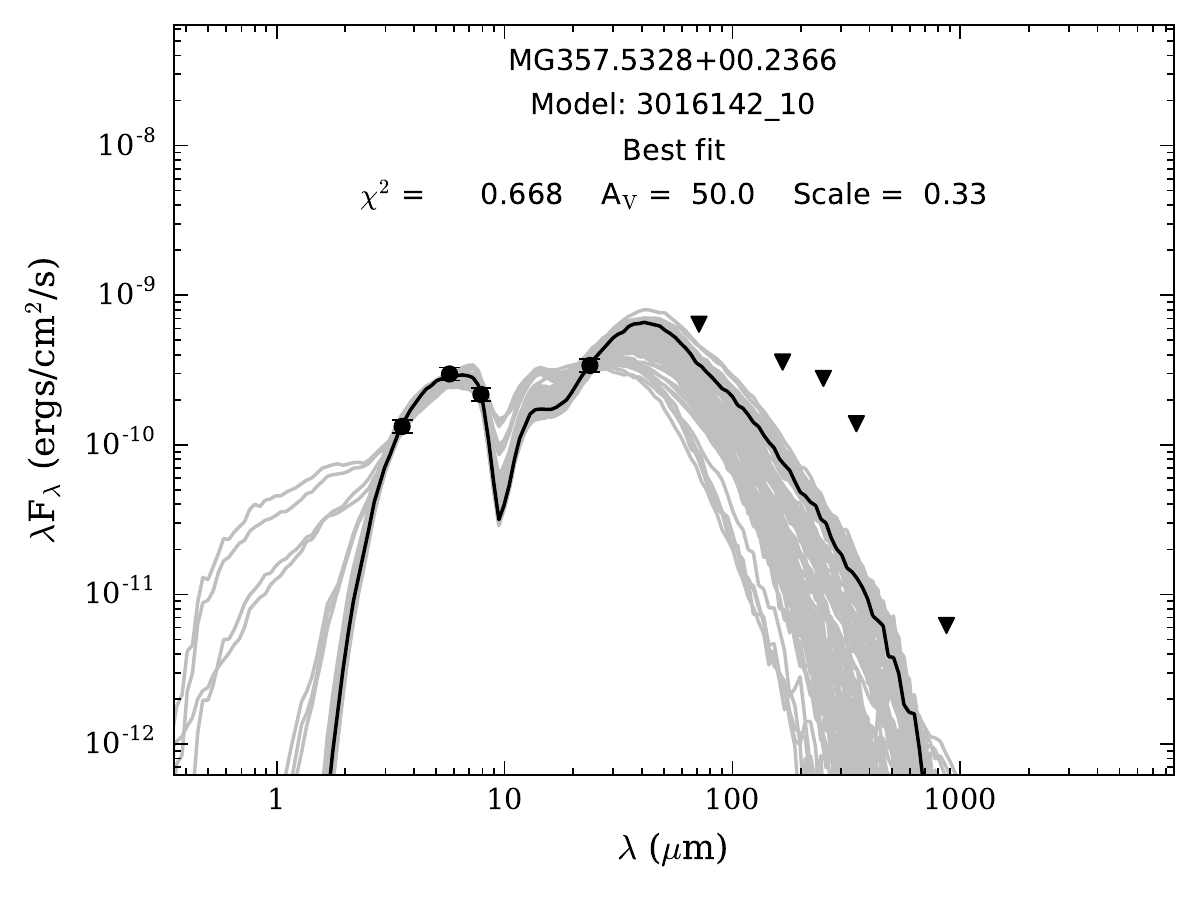}&
                \includegraphics[width=0.31\textwidth]{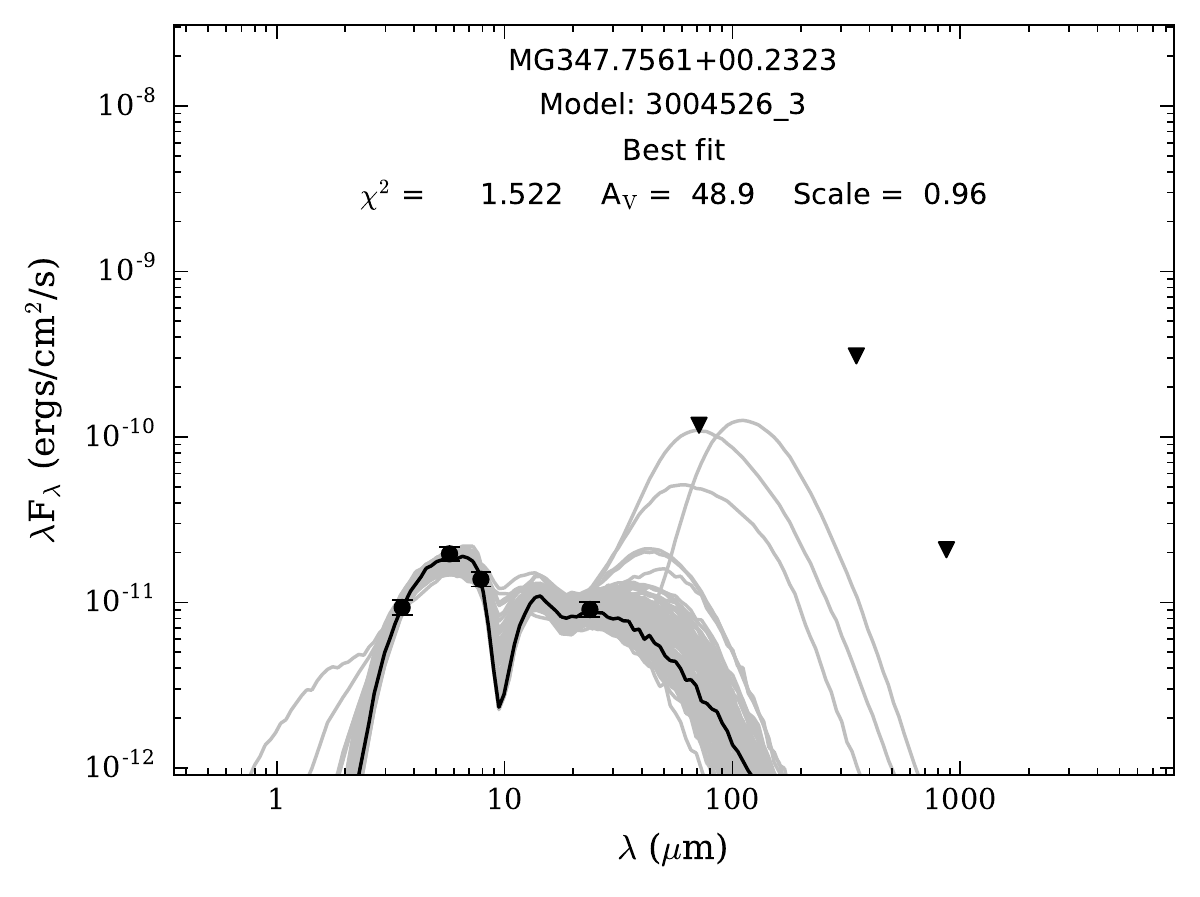}&
                \includegraphics[width=0.31\textwidth]{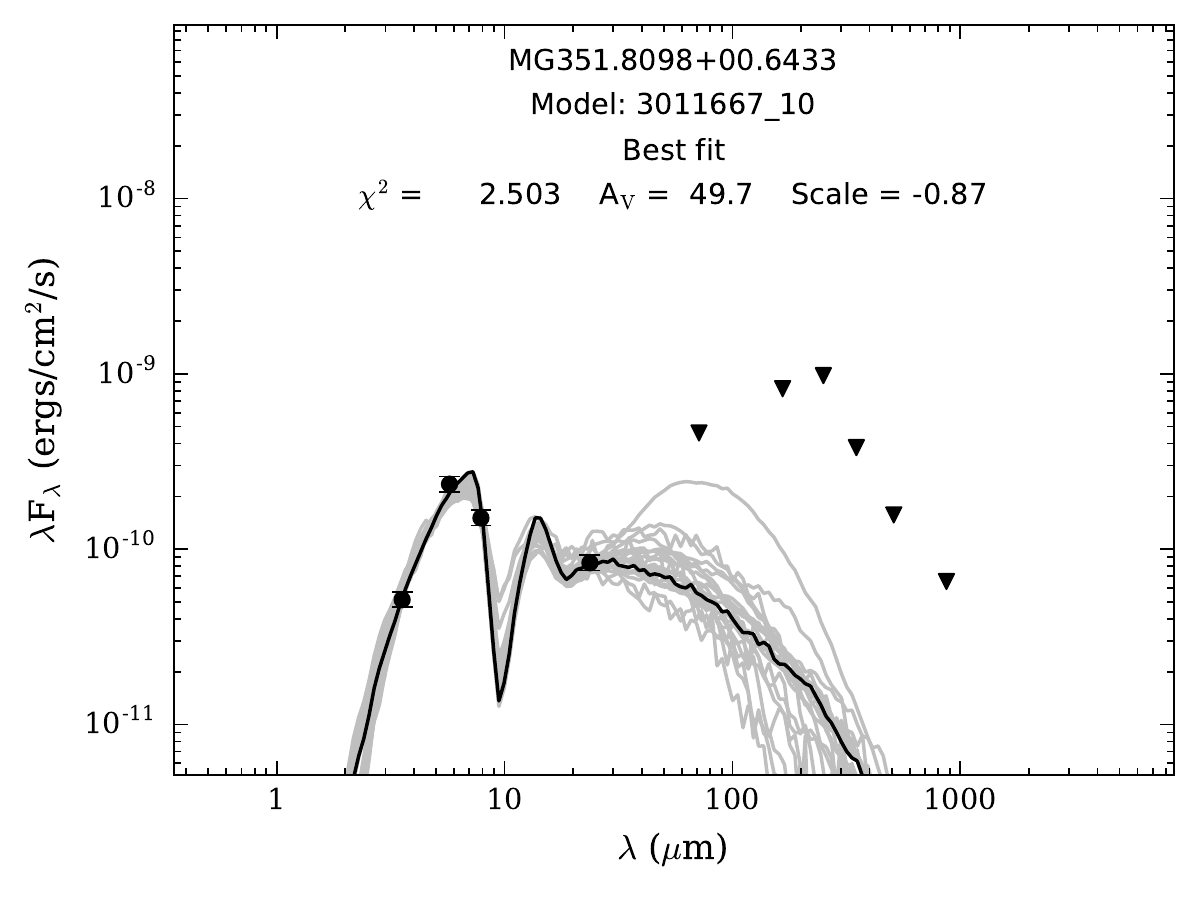}\\
        \end{tabular}
        \caption{Continuation of Fig. \ref{fig:testgridsed}.}
\end{figure}

\begin{figure}
        \begin{tabular}{ccc}
                
                \includegraphics[width=0.31\textwidth]{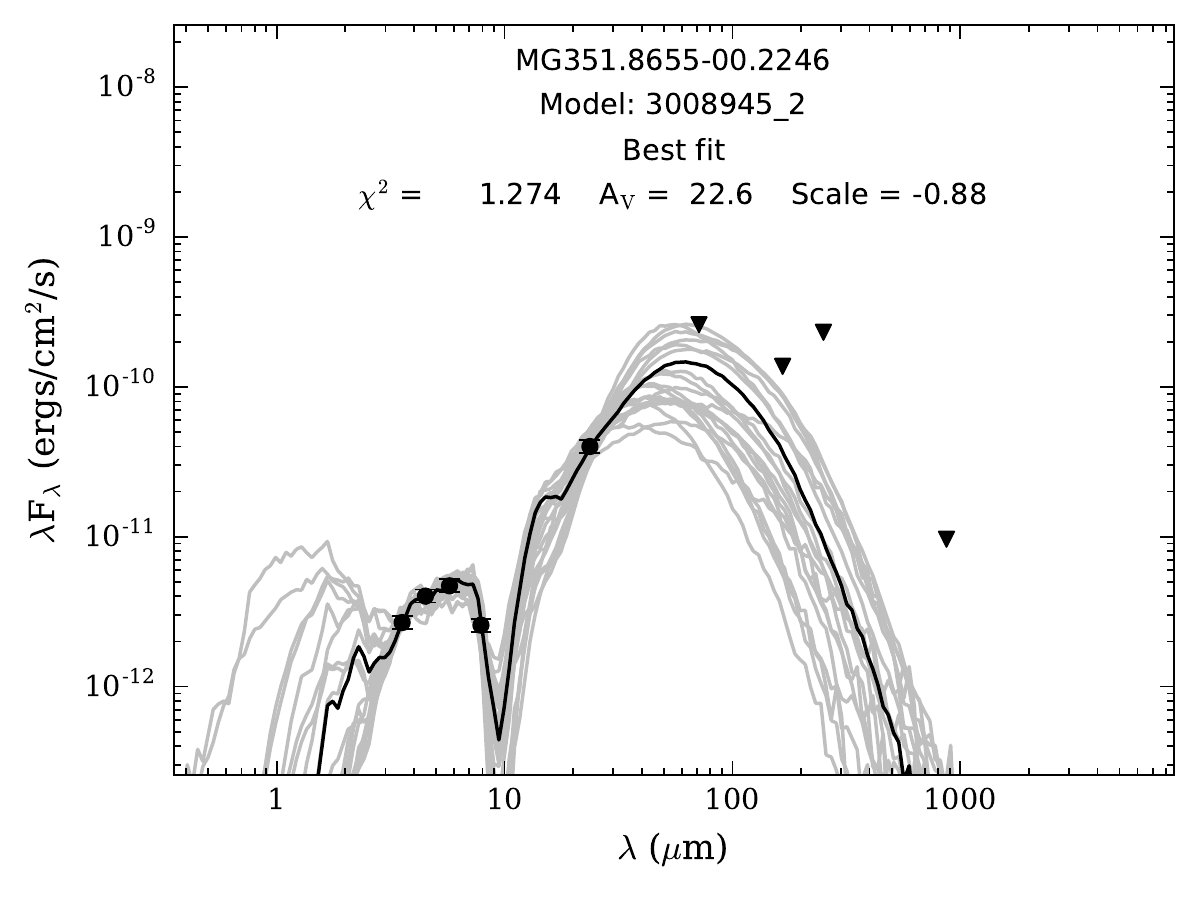}&
                \includegraphics[width=0.31\textwidth]{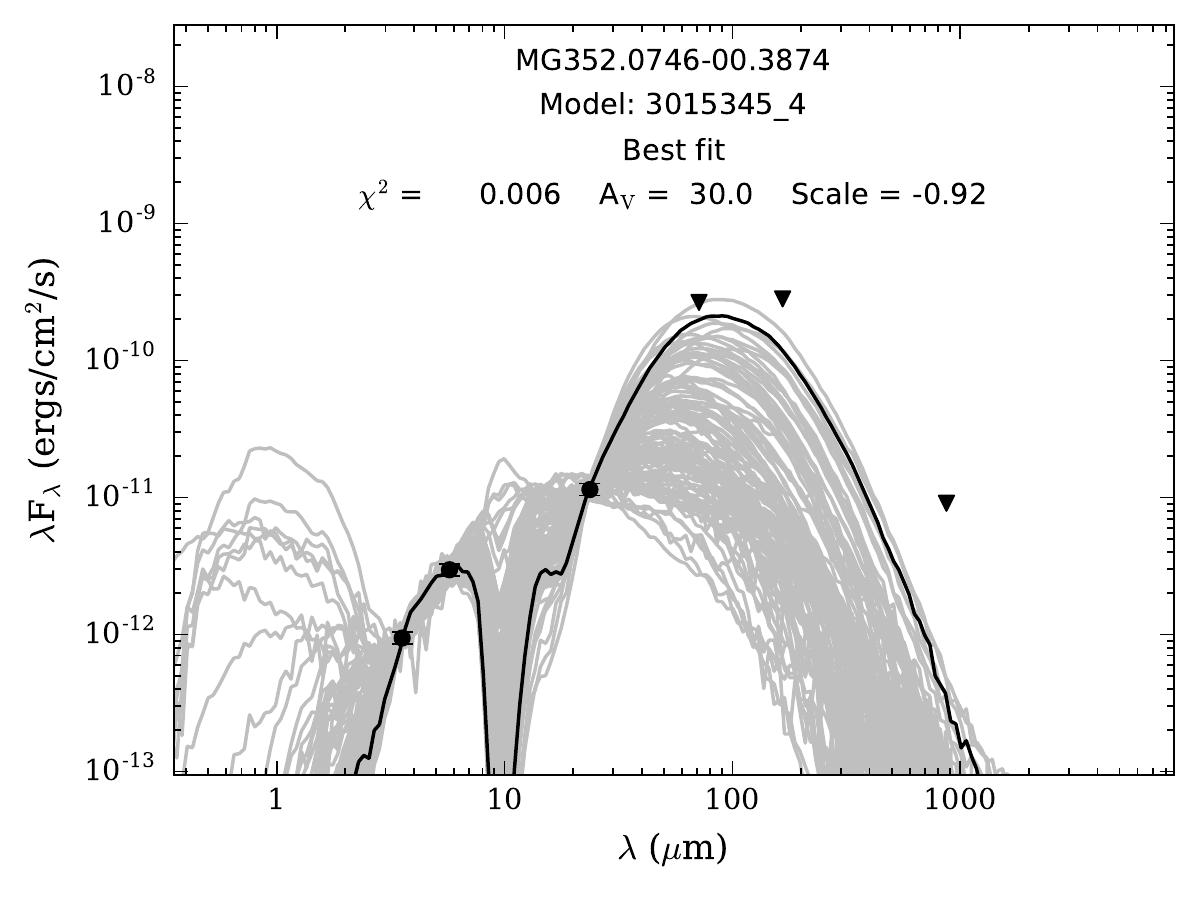}&
                \includegraphics[width=0.31\textwidth]{{./MG352.2452-00.0636}.pdf}\\
                \includegraphics[width=0.31\textwidth]{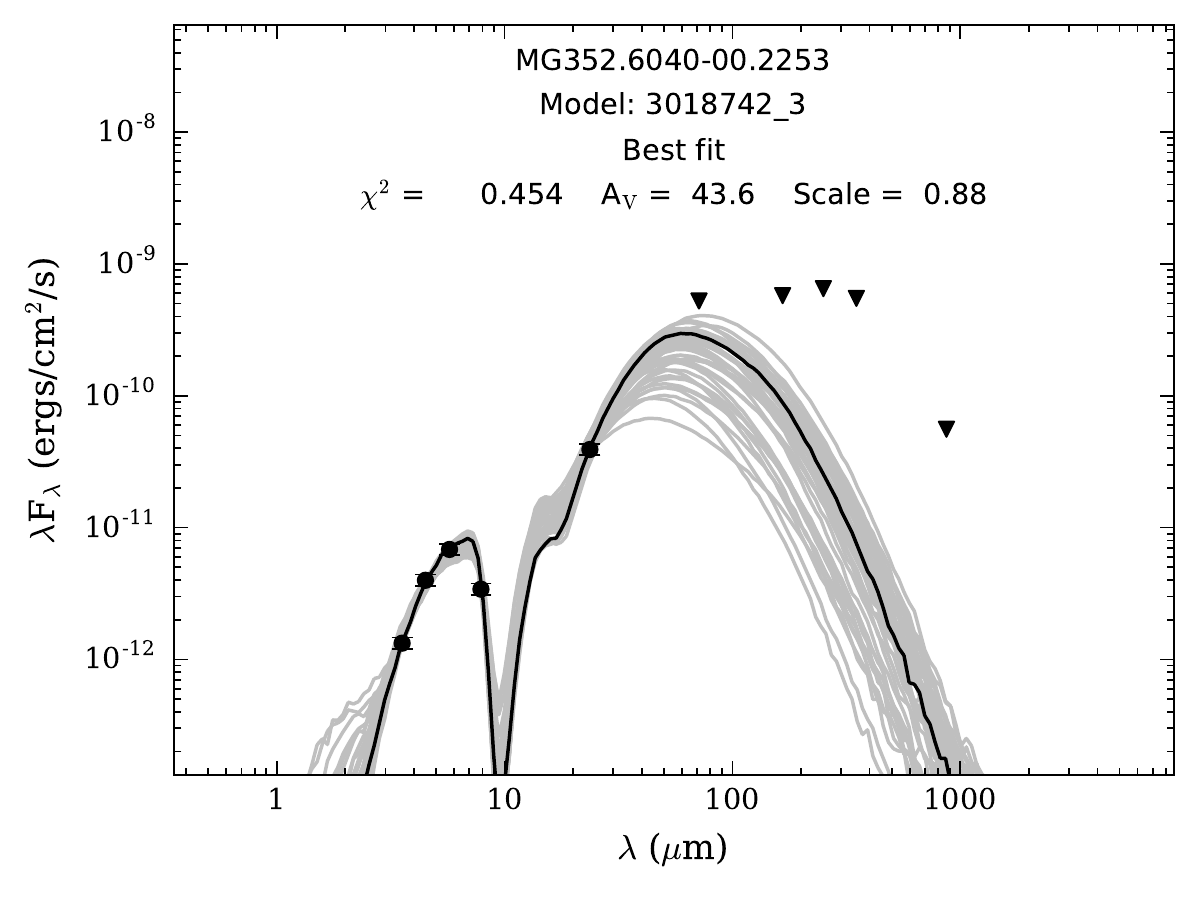}&
                \includegraphics[width=0.31\textwidth]{{./MG354.4384+00.4185}.pdf}&
                \includegraphics[width=0.31\textwidth]{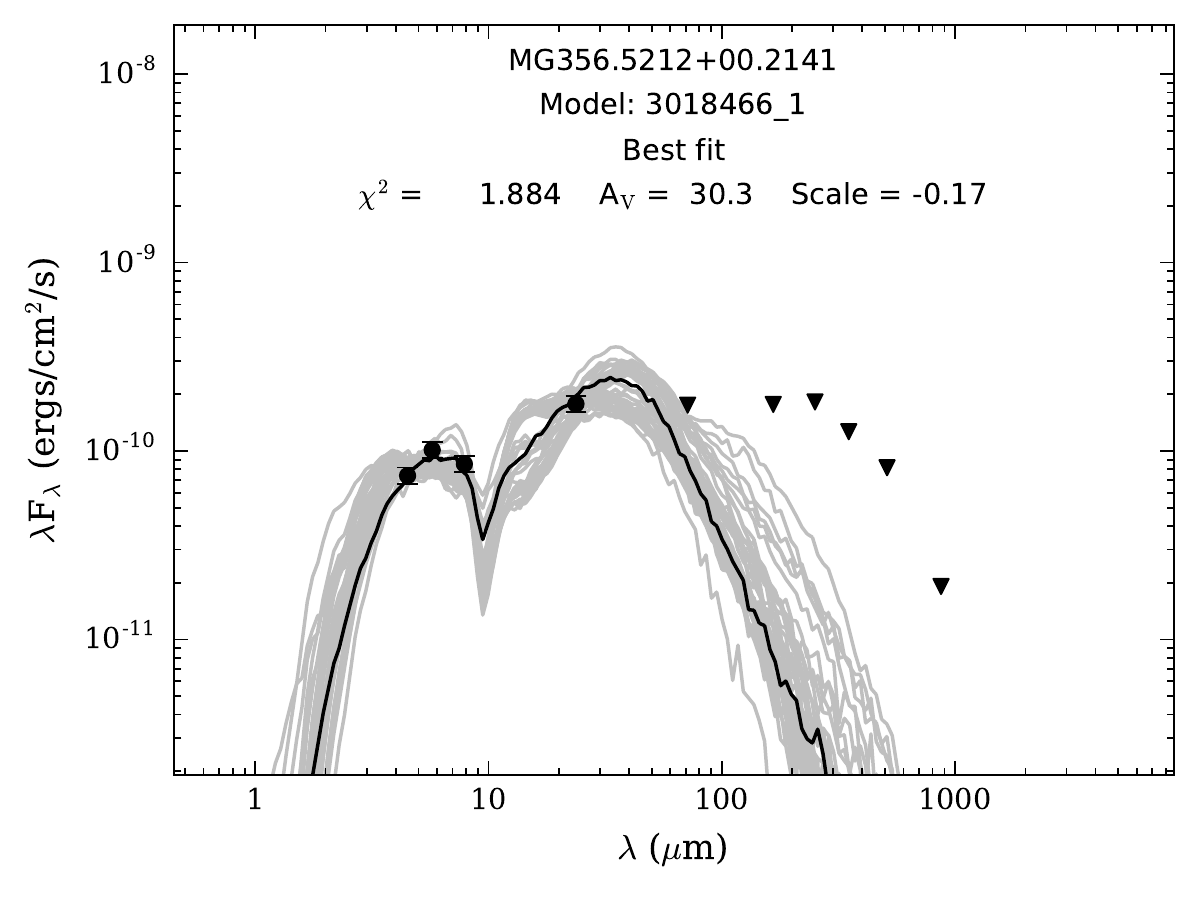}\\
        \end{tabular}
        \caption{Continuation of Fig. \ref{fig:testgridsed}.}
\end{figure}


\begin{thebibliography}{}
        \bibitem[Audard et al.(2014)]{audar14} Audard, M., \'Abrah\'am, P., Dunham, M.~M., et al.\ 2014, Protostars and Planets VI, 387
        
        \bibitem[Bally et al.(2010)]{bally10} Bally, J., Aguirre, J., Battersby, C., et al.\ 2010, \apj, 721, 137-163
        
        \bibitem[Bernasconi \& Maeder(1996)]{bernas96} Bernasconi, P.~A., \& Maeder, A.\ 1996, \aap, 307, 829
        
        \bibitem[Bouvier et al.(2003)]{bouv03} Bouvier, J., Grankin, K.~N., Alencar, S.~H.~P., et al.\ 2003, \aap, 409, 169
        
        \bibitem[Broos et al.(2013)]{mystix} Broos, P.~S., Getman, K.~V., Povich, M.~S., et al.\ 2013, \apjs, 209, 32
        
        \bibitem[Caratti o Garatti et al.(2015)]{garat15} Caratti o Garatti, A., Stecklum, B., Linz, H., Garcia Lopez, R., \& Sanna, A.\ 2015, \aap, 573, A82
        
        \bibitem[Caratti o Garatti et al.(2017)]{garat17} Caratti o Garatti, A., Stecklum, B., Garcia Lopez, R., et al.\ 2017, Nature Physics, 13, 276
        
        \bibitem[Carey et al.(2009)]{car09} Carey, S.~J., Noriega-Crespo, A., Mizuno, D.~R., et al.\ 2009, \pasp, 121, 76
        
        
        \bibitem[Caswell et al.(2010)]{cas10} Caswell, J.~L., Fuller, G.~A., Green, J.~A., et al.\ 2010, \mnras, 404, 1029
        
        \bibitem[Chambers et al.(2014)]{chambers2014} Chambers, E.~T., Yusef-Zadeh, F., \& Ott, J.\ 2014, \aap, 563, A68
        
        \bibitem[Chen et al.(2013a)]{chen2013} Chen, X., Gan, C.-G., Ellingsen, S.~P., et al.\ 2013, \apjs, 206, 9
        
        \bibitem[Chen et al.(2013b)]{chen2013pt2} Chen, X., Gan, C.-G., Ellingsen, S.~P., et al.\ 2013, \apjs, 206, 22
        
        \bibitem[Contreras Pe{\~n}a(2015)]{contreras2015PhDT} Contreras Pe{\~n}a, C.~E.\ 2015, Ph.D.~Thesis
        
        \bibitem[Contreras Pe{\~n}a et al.(2017)]{contreras17} Contreras Pe{\~n}a, C., Lucas, P.~W., Minniti, D., et al.\ 2017, \mnras, 465, 3011
        
        \bibitem[Contreras Pe{\~n}a et al.(2017)]{contrerasspec} Contreras Pe{\~n}a, C., Lucas, P.~W., Kurtev, R., et al.\ 2017, \mnras, 465, 3039 
        
        
        \bibitem[Contreras et al.(2013)]{contreras13} Contreras, Y., Schuller, F., Urquhart, J.~S., et al.\ 2013, \aap, 549, A45
        
        \bibitem[Csengeri et al.(2016)]{cseng16} Csengeri, T., Leurini, S., Wyrowski, F., et al.\ 2016, \aap, 586, A149
        
        \bibitem[Cyganowski et al.(2008)]{cyg08} Cyganowski, C.~J., Whitney, B.~A., Holden, E., et al.\ 2008, \aj, 136, 2391-2412
        
        \bibitem[de Villiers et al.(2015)]{devill15} de Villiers, H.~M., Chrysostomou, A., Thompson, M.~A., et al.\ 2015, \mnras, 449, 119 
        
        
        \bibitem[Dunham \& Vorobyov(2012)]{dun12} Dunham, M.~M., \& Vorobyov, E.~I.\ 2012, \apj, 747, 52 
        
        \bibitem[Dunham et al.(2014)]{dun14} Dunham, M.~M., Stutz, A.~M., Allen, L.~E., et al.\ 2014, Protostars and Planets VI, 195 
        
        \bibitem[Eiroa et al.(2002)]{eiro02} Eiroa, C., Oudmaijer, R.~D., Davies, J.~K., et al.\ 2002, \aap, 384, 1038
        
        \bibitem[Goedhart et al.(2014)]{goed2014} Goedhart, S., Maswanganye, J.~P., Gaylard, M.~J., \& van der Walt, D.~J.\ 2014, \mnras, 437, 1808 
        
        \bibitem[Grave \& Kumar(2009)]{grave09} Grave, J.~M.~C., \& Kumar, M.~S.~N.\ 2009, \aap, 498, 147
        
        \bibitem[Gutermuth \& Heyer(2015)]{gut15} Gutermuth, R.~A., \& Heyer, M.\ 2015, \aj, 149, 64
        
        \bibitem[Hampel (1974)]{hamp74} Hampel, F.~R.\ 1974 The influence curve and its role in robust estimation, Journal of the American Statistical Association, 69 (346) , pp. 383-393
        
        \bibitem[Hartmann et al.(1993)]{hart93} Hartmann, L., Kenyon, S.~J., \& Calvet, N.\ 1993, \apj, 407, 219
        
        \bibitem[Hayashi(1966)]{hayas66} Hayashi, C.\ 1966, \araa, 4, 171
        
        \bibitem[Herbst \& Shevchenko(1999)]{herbst99} Herbst, W., \& Shevchenko, V.~S.\ 1999, \aj, 118, 1043
        
        \bibitem[Heyer et al.(2016)]{hey16} Heyer, M., Gutermuth, R., Urquhart, J.~S., et al.\ 2016, \aap, 588, A29
        
        \bibitem[Hosokawa et al.(2010)]{hosok10} Hosokawa, T., Yorke, H.~W., \& Omukai, K.\ 2010, \apj, 721, 478 

\bibitem[Huchra et al.(2012)]{masssurv} Huchra, J.~P., Macri, L.~M., Masters, K.~L., et al.\ 2012, \apjs, 199, 26 
        
        \bibitem[Hunter et al.(2017)]{hunter17} Hunter, T.~R., Brogan, C.~L., MacLeod, G., et al.\ 2017, \apjl, 837, L29
        
        \bibitem[Inayoshi et al.(2013)]{ina13} Inayoshi, K., Hosokawa, T., \& Omukai, K.\ 2013, \mnras, 431, 3036 
        
        \bibitem[Kenyon et al.(1990)]{keny90} Kenyon, S.~J., Hartmann, L.~W., Strom, K.~M., \& Strom, S.~E.\ 1990, \aj, 99, 869 

    \bibitem[Kenyon \& Hartmann(1995)]{ken95} Kenyon, S.~J., \& Hartmann, L.\ 1995, \apjs, 101, 117 
        
        \bibitem[Kesseli et al.(2016)]{kesseli16} Kesseli, A.~Y., Petkova, M.~A., Wood, K., et al.\ 2016, \apj, 828, 42
        
        \bibitem[K{''o}nig et al.(2017)]{koen17} K{''o}nig, C., Urquhart, J.~S., Csengeri, T., et al.\ 2017, \aap, 599, A139
        
        \bibitem[Kumar \& Grave(2007)]{kumar07} Kumar, M.~S.~N., \& Grave, J.~M.~C.\ 2007, \aap, 472, 155
        
        \bibitem[Kumar et al.(2016)]{kumar2016} Kumar, M.~S.~N., Contreras Pe{\~n}a, C., Lucas, P.~W., \& Thompson, M.~A.\ 2016, \apj, 833, 24
        
        \bibitem[Larson(1969)]{larson69} Larson, R.~B.\ 1969, \mnras, 145, 271
        
        \bibitem[Lewis et al.(2010)]{lewis2010} Lewis, J.~R., Irwin, M., \& Bunclark, P.\ 2010, Astronomical Data Analysis Software and Systems XIX, 434, 91
        
        \bibitem[Li et al.(2016)]{li2016} Li, G.-X., Urquhart, J.~S., Leurini, S., et al.\ 2016, \aap, 591, A5
        
        \bibitem[Makin \& Froebrich(2018)]{makin18} Makin, S.~V., \& Froebrich, D.\ 2018, \apjs, 234, 8
        
        \bibitem[McKee \& Offner(2010)]{mcke10} McKee, C.~F., \& Offner, S.~S.~R.\ 2010, \apj, 716, 167 
        
        \bibitem[Medina et al.(2018)]{med2018} Medina, N., Borissova, J., Bayo, A., et al.\ 2018, arXiv:1806.04061 
        
        \bibitem[Meyer et al.(2017)]{meyerMNRAS2017} Meyer, D.~M.-A., Vorobyov, E.~I., Kuiper, R., \& Kley, W.\ 2017, \mnras, 464, L90
        
        \bibitem[Meyer et al.(2017)]{meyerproceeds2017} Meyer, D.~M.-A., Vorobyov, E.~I., Kuiper, R., \& Kley, W.\ 2017, arXiv:1710.02320
        
        \bibitem[Minniti et al.(2010)]{vvv2010} Minniti, D., Lucas, P.~W., Emerson, J.~P., et al.\ 2010, \na, 15, 433

\bibitem[Morales-Calder{\'o}n et al.(2011)]{moralesysovar} Morales-Calder{\'o}n, M., Stauffer, J.~R., Hillenbrand, L.~A., et al.\ 2011, \apj, 733, 50 
        
        \bibitem[Myers(2010)]{myer10} Myers, P.~C.\ 2010, \apj, 714, 1280 
        
        \bibitem[Ochsenbein et al.(2000)]{viziercit} Ochsenbein, F., Bauer, P., \& Marcout, J.\ 2000, \aaps, 143, 23

        \bibitem[Pilbratt et al.(2010)]{herschel10} Pilbratt, G.~L., Riedinger, J.~R., Passvogel, T., et al.\ 2010, \aap, 518, L1 
        
        \bibitem[Robitaille et al.(2006)]{robmodels} Robitaille, T.~P., Whitney, B.~A., Indebetouw, R., Wood, K., \& Denzmore, P.\ 2006, \apjs, 167, 256 
        
        \bibitem[Robitaille et al.(2007)]{rob10} Robitaille, T.~P., Whitney, B.~A., Indebetouw, R., \& Wood, K.\ 2007, \apjs, 169, 328
        
        \bibitem[Robitaille et al.(2008)]{robspitzer} Robitaille, T.~P., Meade, M.~R., Babler, B.~L., et al.\ 2008, \aj, 136, 2413 
        
        
        
        \bibitem[Scargle(1982)]{scargle82} Scargle, J.~D.\ 1982, \apj, 263, 835
        
        \bibitem[Siess et al.(2000)]{siess00} Siess, L., Dufour, E., \& Forestini, M.\ 2000, \aap, 358, 593 
        
        \bibitem[Schuller et al.(2009)]{schull09} Schuller, F., Menten, K.~M., Contreras, Y., et al.\ 2009, \aap, 504, 415
        
        \bibitem[Shu(1977)]{shu77} Shu, F.~H.\ 1977, \apj, 214, 488
        
        \bibitem[Smith et al.(2018)]{smith2017} Smith, L.~C., Lucas, P.~W., Kurtev, R., et al.\ 2018, \mnras, 474, 1826
        
        \bibitem[Sokolovsky et al.(2017)]{soko17} Sokolovsky, K.~V., Gavras, P., Karampelas, A., et al.\ 2017, \mnras, 464, 274
        
        \bibitem[Takami et al.(2012)]{taka12} Takami, M., Chen, H.-H., Karr, J.~L., et al.\ 2012, \apj, 748, 8 
        
        \bibitem[Upton \& Cook (1996)]{upton96} Upton, G.; Cook, I.\ 1996, Understanding Statistics, Oxford University Press. p. 55. ISBN 0-19-914391-9.
        
        \bibitem[Urquhart et al.(2013)]{urqu2016mas} Urquhart, J.~S., Moore, T.~J.~T., Schuller, F., et al.\ 2013, \mnras, 431, 1752
        
        \bibitem[Urquhart et al.(2014)]{urqu14} Urquhart, J.~S., Moore, T.~J.~T., Csengeri, T., et al.\ 2014, \mnras, 443, 1555
        
        \bibitem[Urquhart et al.(2014)]{urquhartcsc14} Urquhart, J.~S., Csengeri, T., Wyrowski, F., et al.\ 2014, \aap, 568, A41
        
        \bibitem[Urquhart et al.(2018)]{urqu17} Urquhart, J.~S., K{''o}nig, C., Giannetti, A., et al.\ 2018, \mnras, 473, 1059
        
        \bibitem[VanderPlas(2017)]{vanderplas17} VanderPlas, J.~T.\ 2017, arXiv:1703.09824
        
        \bibitem[Vorobyov \& Basu(2006)]{vorbas06} Vorobyov, E.~I., \& Basu, S.\ 2006, \apj, 650, 956
        
        
        \bibitem[Vorobyov \& Basu(2008)]{vorbas08} Vorobyov, E.~I., \& Basu, S.\ 2008, \apjl, 676, L139 
        
        \bibitem[Vorobyov \& Basu(2015)]{vorbas15} Vorobyov, E.~I., \& Basu, S.\ 2015, \apj, 805, 115
        
        \bibitem[Wenger et al.(2000)]{simbadtoolref} Wenger, M., Ochsenbein, F., Egret, D., et al.\ 2000, \aaps, 143, 9
        
        \bibitem[Wienen et al.(2015)]{wienen15} Wienen, M., Wyrowski, F., Menten, K.~M., et al.\ 2015, \aap, 579, A91
        
        \bibitem[Zapata et al.(2008)]{zapa2008} Zapata, L.~A., Leurini, S., Menten, K.~M., et al.\ 2008, \aj, 136, 1455 
        
        \bibitem[Zhu et al.(2009)]{zhu09} Zhu, Z., Hartmann, L., \& Gammie, C.\ 2009, \apj, 694, 1045
        
        
        
\end{thebibliography}
\end{document}